\documentclass[a4paper]{article}
\usepackage[french,english]{babel}
\usepackage[latin1]{inputenc}
\usepackage[T1]{fontenc}
\usepackage{amssymb}
\usepackage{enumitem}
\usepackage{footmisc}
\usepackage{graphics}
\usepackage{epsfig}
\usepackage[all,cmtip]{xy}

\let\ts\textstyle

\def\urltilde{\raise.3ex\hbox{\footnotesize$\!\sim$}}
\def\anneecours#1{\ifcase #1\or première\or deuxième\or troisi\`eme\fi}

\def\montitrecours{}
\def\monnom{{Gilles {\sc Leborgne}, www.isima.fr/leborgne\par\medskip}}
\def\montitre#1{
  \begin{center}
  \normalsize
  \LARGE\montitrecours\par\bigskip
  \normalsize\monnom
  \normalsize\today
  \end{center}
            }

\catcode`@=11
\newcommand{\boxed}[1]{\fbox{\m@th$\displaystyle#1$}}

\@addtoreset{equation}{section} 
\@addtoreset{figure}{section} 
\def\eqalign#1{\null\,\vcenter{\openup1\jot \m@th
   \ialign{\strut \hfil$\displaystyle{##}$ & $\displaystyle{{}##}$\hfil
      \crcr#1\crcr}}\,}
\def\eqalignrll#1{\null\,\vcenter{\openup1\jot \m@th
   \ialign{\strut \hfil$\displaystyle{##}$ & $\displaystyle{{}##}$\hfil
    & $\displaystyle{{}##}$\hfil
      \crcr#1\crcr}}\,}
\def\eqalignrllq#1{\null\,\vcenter{\openup1\jot \m@th
   \ialign{\strut \hfil$\displaystyle{##}$ & $\displaystyle{{}##}$\hfil
    & \quad$\displaystyle{{}##}$\hfil
      \crcr#1\crcr}}\,}

\def\binrel@#1{\begingroup
  \setboxz@h{\thinmuskip0mu
    \medmuskip\m@ne mu\thickmuskip\@ne mu
    \setbox\tw@\hbox{$#1\m@th$}\kern-\wd\tw@
    ${}#1{}\m@th$}%
  \edef\@tempa{\endgroup\let\noexpand\relax
    \ifdim\wdz@<\z@ \mathbin
    \else\ifdim\wdz@>\z@ \mathrel
    \else \relax\fi\fi}%
  \@tempa
}
\newcommand{\overset}[2]{\binrel@{#2}%
    \relax{\mathop{\kern\z@#2}\limits^{#1}}}
\def\overbigdot#1{\overset{\hbox{\tiny$\bullet$}}{#1}}
\def\overbbigdot#1{\overset{\hbox{\tiny$\bullet\bullet$}}{#1}}

\DeclareRobustCommand{\twodots}{\t@urdots}
\def\t@urdots{\mbox{\kern1\p@\vbox to 1ex{\hbox{.}\vss\vss\hbox{.}}}}
\DeclareRobustCommand{\treedots}{\th@urdots}
\def\th@urdots{\mbox{\kern1\p@\vbox to 1.3ex{\hbox{.}\vss \hbox{.}\vss\hbox{.}}}}

\def\runtitle{Objectivity in mechanics}
\def\ps@my{\let\@mkboth\markboth
      \def\@oddhead{\footnotesize\it\thepage\hfill\leftmark}
      \def\@evenhead{\footnotesize\hfill\it\rightmark\hfill}
      \def\@oddfoot{\rlap{\small\sl\runtitle}\hfil\thepage\hfil}
      \let\@evenfoot\@oddfoot
      \def\sectionmark##1{\markboth{\thesection. \ ##1}  % rappel du
                                   {\thesection. \ ##1}  % titre de la
                          \markright{\thesection. \ ##1}    % subsection
                         }
           }
\def\ps@myfin{\let\@mkboth\markboth
      \def\@oddhead{\footnotesize\it\thepage\hfill\rightmark}
      \def\@evenhead{\footnotesize\it\thepage\hfill\leftmark}
      \def\@oddfoot{\hfil\thepage\hfil} %\rlap{\small\sl\runtitle}\llap{\small\today}
      \let\@evenfoot\@oddfoot
      \let\@mkboth\markboth
      \def\subsectionmark##1{\markright{\thesubsection. \ ##1}}%
           }
\catcode`@=12

\def\textslbf#1{\textsl{\textbf{#1}}}

\def\vR{{\vec R}}
\def\PKtzt{{P\!\!\!K^\tz_t}}

\def\tPK{{\widetilde {P\!\!\!K}}}
\def\tSK{{\widetilde {S\!\!K}}}

\def\SKtzt{{S\!\!K^\tz_t}}

\def\piai{{\pi_{ai}}}
\def\piaj{{\pi_{aj}}}
\def\piak{{\pi_{ak}}}
\def\pibi{{\pi_{bi}}}
\def\pibj{{\pi_{bj}}}
\def\pibk{{\pi_{bk}}}

\def\tcalJ{{\tilde{{\cal J}}}}
\def\Mtu{\calM_{31}}

\def\Ftpt{{F^t_\pt}}%

\def\fEs{{f_{\calE*}}}
\def\pEs{{p_{\calE*}}}
\def\qEs{{q_{\calE*}}}

\def\pFb{{p_\calF{}^*}}

\def\fFb{{f_\calF{}^*}}

\def\thetatzs{{\theta_{\tz*}}}

\def\curveE{c_\calE}
\def\curveEp{c_\calE{}'}

\def\vag{\vec a_g}

\def\vbh{\vec b_h}
\def\vuE{\vec u_\calE}
\def\betaF{\beta_\calF}
\def\Phitmht{{\Phi^\tmh_t}}%
\def\Phittmh{{\Phi^t_\tmh}}%

\def\graph{\hbox{\rm graph}}

\def\Ltzt{{L^\tz_t}}

\def\tF{\tilde F}
\def\vbu{{\vb_1}}
\def\vbd{{\vb_2}}

\def\vVtzptz{{\vec V^\tz_\ptz}}
\def\vAtzptz{{\vec A^\tz_\ptz}}

\def\valphag{\vec a_g}

\def\equalref#1{\mathbin{\mathop{=}^{\eref{#1}}}}

\def\clanot{\hbox{clas{.}}}
\def\clanots{{\hbox{\footnotesize clas{.}}}}
\def\eqclas{\,\mathop{=}^{\clanots}}
\def\dualnot{\hbox{dual notation}}
\def\clasnot{\hbox{clas. not.}}
\def\dualnot{\hbox{dual not.}}

\def\duanot{\hbox{dual}}
\def\duanots{{\hbox{\footnotesize dual}}}
\def\eqdual{\mathop{=}^{\duanots}}
\def\underclas#1{\underbrace{#1}_\clanots}
\def\underdual#1{\underbrace{#1}_\duanots}

\def\mope{\mathop{=}}

\def\PK{{P\!\!\!K}}
\def\SK{{S\!\!K}}
\def\SKtzt{{S\!\!K^\tz_t}}

\def\vwtz{{\vw_\tz}}
\def\vwptz{{\vw_{\ptz}}}
\def\vwtzptz{{\vw_\tz(\ptz)}}

\def\curveFb{{c_\calF{}^*}}
\def\curveEs{{c_{\calE*}}}

\def\ain{{a^i_{new}}}

\def\aio{{a^i_{old}}}

\def\ajn{{a^j_{new}}}

\def\ajo{{a^j_{old}}}

\def\akn{{a^k_{new}}}

\def\ako{{a^k_{old}}}

\def\alphaE{{\alpha_\calE}}
\def\alphaEs{{\alpha_{\calE*}}}
\def\alphaF{{\alpha_{\!\calF}}}
\def\alphaFb{{\alpha_{\!\calF}}{}^*}

\def\bcdot{\raisebox{1pt}{\hbox{\tiny $\;\bullet\;$}}}
\def\bcdotG{\raisebox{1pt}{\hbox{\tiny $\;\bullet\;$}}_{\raisebox{-3pt}{\hbox{\scriptsize $\!\!\!G$}}}\,}
\def\bcdotg{\raisebox{1pt}{\hbox{\tiny $\;\bullet\;$}}_{\raisebox{-1pt}{\hbox{\scriptsize $\!\!\!g$}}}\,}
\def\bigC{\hbox{\large$\cal C$}}
\def\bin{{b^i_{new}}}
\def\bio{{b^i_{old}}}
\def\bjn{{b^j_{new}}}
\def\bjo{{b^j_{old}}}

\def\blo{{b^\ell_{old}}}

\def\bOmega{{\overline\Omega}}

\def\Btzt{{B^{t_0}_t}}
\def\calA{{\cal A}}

\def\calD{{\cal D}}
\def\calE{{\cal E}}
\def\calF{{\cal F}}
\def\calI{{\cal I}}

\def\calJ{{\cal J}}
\def\calL{{\cal L}}
\def\calM{{\cal M}}
\def\calO{{\cal O}}
\def\calP{{\cal P}}
\def\calQ{{\cal Q}}
\def\calR{{\cal R}}
\def\calRA{{\calR_{\!A}}}
\def\calRB{{\calR_{\!B}}}
\def\calT{{\cal T}}
\def\calU{{\cal U}}

\def\cf{cf{.}}
\def\Cf{Cf{.}}
\def\Cof{{\rm Cof}}

\def\CtzP{{C^{t_0}_P}}
\def\Ctzt{{C^{t_0}_t}}

\def\cwedge{{\ds\mathop{\wedge}^\circlearrowleft}}
\def\cwr{{\ds\mathop{\omega}^\circlearrowleft}}
\def\dd{(\cdot,\cdot)}
\def\demi{{1\over 2}}
\def\dvg{{\rm div}}
\def\eg{{e{.}g{.}}}
\def\Eg{{E{.}g{.}}}
\def\ein{{e^i_\new}}
\def\eio{{e^i_\old}}

\def\ejo{{e^j_{old}}}

\def\eqdef{:=}
\let\eqnamed\eqcalled
\def\eqnote{{\;\mathop{=}^{\hbox{\footnotesize noted}}\;}}
\def\Es{{E^*}}
\def\Ess{{E^{**}}}

\def\EtzP{{E^{t_0}_P}}
\def\Etzt{{E^{t_0}_{t}}}
\def\eul{{\calE\! ul}}
\def\fA{{f_A}}
\def\fAt{{f_{At}}}
\def\Fatzpatz{{F^\tz_{\pAtz}}}
\def\fB{{f_B}}
\def\fBt{{f_{Bt}}}

\def\fE{{f_\calE}}
\def\fEs{{f_{\calE*}}}

\def\fF{{f_\calF}}
\def\FiJ{{F^i_{\;J}}}
\def\foot{{\rm foot}}
\def\Fs{{F^*}}
\def\Fss{{F^{**}}}
\def\Ft{{F^t}}

\def\FTIj{(F^T)^I{}_{j}}
\def\Fttau{F^t_\tau}
\def\Fttph{{F^t_\tph}}
\def\Ftztzph{{F^\tz_\tzph}}
\def\Ftu{{F^t_u}}
\def\Ftz{{F^{t_0}}}
\def\FtzP{{F^{t_0}_P}}
\def\Ftzptz{{F^\tz_{p_{t_0}}}}
\def\Ftzt{{F^{t_0}_t}}

\def\gammaE{{c_\calE}}
\def\gammaF{{c_\calF}}
\def\Gammat{{\Gamma_t}}

\def\Htz{{H^{t_0}}}

\def\Htzt{{H^{t_0}_t}}
\def\ie{i{.}e{.}}
\def\Ie{I{.}e{.}}
\def\Im{{\rm Im}}
\def\Ker{{\rm Ker}}
\def\Lag{{{\cal L}ag}}
\def\Lagtz{{\Lag^\tz}}
\def\Lagtzt{{\Lag^\tz_t}}
\def\Lagtzptz{{\Lag^\tz_\ptz}}

\def\tLagtz{{\tilde\Lag^\tz}}

\def\Lij{{L^i{}_j}}
\def\Lji{{L^j{}_i}}
\def\Ldo{{L^2(\Omega)}}
\def\Lis{{L_{i\!s}}}
\def\metre{{\rm metre}}
\def\matrarrow{\mathop{\longrightarrow}}
\def\mrar{\mathop{\longrightarrow}}
\def\mn{\medskip\noindent}
\def\mn{\medskip\noindent}
\def\new{{new}}
\def\NN{\mathbb{N}}
\def\NNs{{\NN^*}}
\def\OA{{O_{\!A}}}

\def\OB{{O_{\!B}}}
\def\Obj{{\it O\!b\!j}}

\def\odd{\mathop{\;\raise-1.55pt\hbox{\large$0$}\mkern-9.5mu \twodots\;}}
\def\old{{old}}

\def\omegat{{\omega_t}}
\def\Omegat{{\Omega_t}}
\def\Omegatau{{\Omega_\tau}}
\def\Omegatz{{\Omega_{t_0}}}
\def\otd{\mathop{\;\raise-1.65pt\hbox{\Large$0$}\mkern-10.2mu \treedots}\;}
\def\paomegat{{\pa\omega_t}}

\def\pE{{p_{\!\calE}}}
\def\pF{{p_{\!\calF}}}
\def\Phit{{\Phi^t}}

\def\Phittau{{\Phi^t_\tau}}
\def\Phittaub{{\Phi^{t*}_\tau}}
\def\Phittph{{\Phi^t_\tph}}
\def\Phitz{{\Phi^{t_0}}}
\def\PhitzP{{\Phi^{t_0}_P}}
\def\Phitzptz{{\Phi^{t_0}_{p_{t_0}}}}
\def\Phitzt{{\Phi^{t_0}_t}}
\def\Phitztb{{\Phi^{\tz*}_t}}
\def\Phitzts{{\Phi^\tz_{t\,*}}}

\def\Phitztz{{\Phi_{t_0}^{t_0}}}
\def\Pij{{P^i{}_j}}

\def\piei{\pi_{ei}}
\def\piej{\pi_{ej}}

\def\Pobj{{P_{\!\Obj}}}

\def\Psit{{\Psi_t}}

\def\Psitz{{\Psi_\tz}}

\def\pt{{p_t}}
\def\ptau{{p_\tau}}
\def\ptmh{{p(\tmh)}}
\def\ptph{{p(t{+}h)}}
\def\ptu{{p_\tu}}
\def\ptz{{p_{t_0}}}
\def\ptzph{{p(\tz{+}h)}}
\def\pu{{p_u}}
\def\qand{\quad\hbox{and}\quad}

\def\qE{{q_{\!\calE}}}
\def\qF{{q_{\!\calF}}}
\def\qie{\quad\hbox{i.e.}\quad}
\def\qiff{\quad\hbox{iff}\quad}
\def\qinshort{\quad\hbox{in short}\quad}
\def\Qij{{Q^i{}_j}}

\def\qor{\quad\hbox{or}\quad}

\def\qso{\quad\hbox{so}\quad}
\def\qst{\quad\hbox{s{.}t{.}}\quad}
\def\qt{{q_t}}
\def\qtz{{q_{t_0}}}
\def\qtu{{q_{t_1}}}

\def\qti{{q_{t_i}}}
\def\qthen{\quad\hbox{then}\quad}
\def\qthus{\quad\hbox{thus}\quad}
\def\qwhile{\quad\hbox{while}\quad}
\def\qwith{\quad\hbox{with}\quad}
\def\qwhen{\quad\hbox{when}\quad}
\def\qwhere{\quad\hbox{where}\quad}
\def\qwritten{\quad\hbox{written}\quad}
\def\RR{\mathbb{R}}

\def\RRn{{\RR^n}}
\def\RRns{{\RRn^*}}
\def\RRnt{{\vec\RR^n_t}}
\def\RRnts{{\vec\RR^{n*}_t}}
\def\RRntzs{{\vec\RR^{n*}_\tz}}

\def\RRntz{{\vec\RR^n_\tz}}

\def\RRt{{\RR^3}}

\def\st{s{.}t{.}}
\def\Stzt{{S^{t_0}_{t}}}

\def\sumijkmn{{\sum_{i,j,k,m=1}^n}}
\def\sumijkn{{\sum_{i,j,k=1}^n}}

\def\sumijn{{\sum_{i,j=1}^n}}

\def\sumijm{{\sum_{i,j=1}^m}}
\def\sumiJn{{\sum_{i,J=1}^n}}

\def\sumikln{{\sum_{i,k,\ell=1}^n}}

\def\sumikn{{\sum_{i,k=1}^n}}
\def\sumim{{\sum_{i=1}^m}}
\def\sumin{{\sum_{i=1}^n}}
\def\sumIn{{\sum_{I=1}^n}}
\def\sumKn{{\sum_{K=1}^n}}
\def\sumjkn{{\sum_{j,k=1}^n}}

\def\sumjm{{\sum_{j=1}^m}}

\def\sumjn{{\sum_{j=1}^n}}
\def\sumJn{{\sum_{J=1}^n}}
\def\sumkm{{\sum_{k=1}^m}}
\def\sumkmn{{\sum_{k,m=1}^n}}
\def\sumkln{{\sum_{k,\ell=1}^n}}
\def\sumkn{{\sum_{k=1}^n}}
\def\sumln{{\sum_{\ell=1}^n}}
\def\sumlm{{\sum_{\ell=1}^m}}

\def\tdet{\mathop{\tilde\det}}
\def\Tduo{{T^2_1(\Omega)}}
\def\Tduo{{T^2_1(\Omega)}}

\def\tdvg{{\tilde\dvg}}
\def\Tdzo{{T^2_0(\Omega)}}
\def\Tdzo{{T^2_0(\Omega)}}
\def\Tdzu{{T^2_0(U)}}
\def\Tdzu{{T^2_0(U)}}
\def\Thetat{{\Theta_t}}
\def\Thetat{{\Theta_t}}

\def\Thetatz{{\Theta_\tz}}

\def\tJ{{\tilde J}}
\def\tL{{\tilde L}}
\def\tmh{{t{-}h}}
\def\tph{{t{+}h}}
\def\tPhi{{\widetilde\Phi}}
\def\tPhiPobj{{\widetilde\Phi_{\!\Pobj}}}

\def\tPsi{{\tilde\Psi}}
\def\tPsi{{\widetilde\Psi}}

\def\TptzOmegatz{T_{\!\ptz}\!(\Omegatz)}
\def\TptOmegat{T_{\!\pt}\!(\Omegat)}
\def\Tr{{\rm Tr}}

\def\Trsu{{T^r_s(U)}}
\def\Trsu{{T^r_s(U)}}
\def\ts{{\textstyle}}
\def\tu{{t_1}}
\def\Tudo{{T^1_2(\Omega)}}
\def\Tudo{{T^1_2(\Omega)}}
\def\Tudu{{T^1_2(U)}}
\def\Tuuo{{T^1_1(\Omega)}}
\def\Tuuot{{T^1_1(\Omegat)}}
\def\Tuuu{{T^1_1(U)}}
\def\Tuzo{{T^1_0(\Omega)}}

\def\Tuzu{{T^1_0(U)}}
\def\tvphi{{\widetilde\vphi}}
\def\tz{{t_0}}
\def\tzph{{\tz{+}h}}

\def\Tzdo{{T^0_2(\Omega)}}
\def\Tzdu{{T^0_2(U)}}

\def\Tzuo{{T^0_1(\Omega)}}
\def\Tzuu{{T^0_1(U)}}

\def\Tzzu{{T^0_0(U)}}
\def\UE{{{\cal U}_{\!\calE}}}
\def\UF{{{\cal U}_{\!\calF}}}
\def\uu#1{{\underline{\underline{#1}}}}
\def\uua{{\underline{\underline{a}}}}

\def\uuatz{{\underline{\underline{a}}^{t_0}}}

\def\uuatzt{{\underline{\underline{a}}^{t_0}_t}}
\def\uub{{\underline{\underline{b}}}}

\def\uubtz{{\underline{\underline{b}}^\tz}}
\def\uubtzt{{\underline{\underline{b}}^\tz_t}}
\def\uudelta{{\underline{\underline{\delta}}}}

\def\uueps{{\underline{\underline{\eps}}}}
\def\tuueps{{\tilde\uueps}}
\def\uuepstz{{{\underline{\underline{\eps}}^{t_0}}}}
\def\uuepstzt{{{\underline{\underline{\eps}}^{t_0}_t}}}
\def\uukappa{{\underline{\underline{\kappa}}}}
\def\uusigma{{\underline{\underline{\sigma}}}}
\def\uusigmatzt{{\uusigma^\tz_t}}

\def\uutau{{\underline{\underline{\tau}}}}
\def\va{{\vec a}}
\def\vA{{\vec A}}
\def\vain{{\va_{new,i}}}

\def\vaio{{\va_{old,i}}}

\def\vajo{{\va_{old,j}}}

\def\vako{{\va_{old,k}}}

\def\vajn{{\va_{new,j}}}

\def\vajtz{{\vec a_{j\tz}}}
\def\vakn{{\va_{new,k}}}

\def\vako{{\va_{old,k}}}

\def\van{{\va_{new}}}

\def\vao{{\va_{old}}}

\def\vAtz{{\vA^{t_0}}}

\def\vAtzt{{\vA^{t_0}_{t}}}
\def\vB{{\vec B}}
\def\vb{{\vec b}}
\def\vbin{{\vb_{new,i}}}
\def\vbio{{\vb_{old,i}}}
\def\vbjn{{\vb_{new,j}}}
\def\vbjo{{\vb_{old,j}}}
\def\vblo{{\vb_{old,\ell}}}

\def\vbko{{\vb_{old,k}}}

\def\vbn{{\vb_{new}}}
\def\vbo{{\vb_{old}}}
\def\vc{{\vec c}}
\def\vcalA{{\vec\calA}}
\def\vcalU{{\vec \calU}}
\def\vcalUtz{\vcalU^\tz}

\def\vcalUtzptz{{\vcalU^\tz_{\ptz}}}
\def\vcalUtzt{{\vcalU^\tz_t}}

\def\vcp{{\vec c\,'}}
\def\vd{{\vec d}}
\def\vDelta{{\vec\Delta}}

\def\vDeltatzt{{\vec\Delta^\tz_t}}

\def\vE{{\vec E}}
\def\Vect{{\rm Vect}}
\def\ve{{\vec e}}

\def\vein{{\ve_{new,i}}}
\def\veio{{\ve_{old,i}}}
\def\vejn{{\ve_{new,j}}}
\def\vejo{{\ve_{old,j}}}
\def\vekn{{\ve_{new,k}}}
\def\veko{{\ve_{old,k}}}
\def\vell{{\vec\ell}}

\def\ven{{\ve_{new}}}

\def\veo{{\ve_{old}}}

\def\vf{{\vec f}}

\def\vg{{\vec g}}

\def\vgamma{{\vec\gamma}}
\def\vGamma{{\vec\Gamma}}

\def\vgammaEp{{\vec{c_\calE}'}}
\def\vgammaEpp{{\vec{c_\calE}''}}
\def\vgammaFp{{\vec{c_\calF}'}}
\def\vgammaFpp{{\vec{c_\calF}''}}
\def\vgammap{{\vec\gamma\,'}}

\def\vGammatz{{\vec\Gamma^{\,t_0}}}
\def\vGammatzt{{\vec\Gamma_{t}^{\,t_0}}}
\def\vgrad{{\vec{\rm grad}}}
\def\vgradg{{\vec{\rm grad}_{\!g}}}

\def\vn{{\vec n}}
\def\vN{{\vec N}}

\def\vomega{{\vec\omega}}
\def\vp{{\vec p}}
\def\vphi{{\vec\phi}}
\def\vphitz{{\vphi^{\,t_0}}}

\def\vq{{\vec q}}
\def\vr{{\vec r}}
\def\vrot{{\vec{\rm rot}}}
\def\vrp{{\vr\,'}}
\def\vrpp{{\vr\,''}}
\def\vRRd{{\vec{\RR^2}}}
\def\vRRm{{\vec{\RR^m}}}
\def\vRRn{{\vec{\RR}^n}}

\def\vRRnt{{\vec{\RR^n_t}}}
\def\vRRntz{{\vec{\RR^n_\tz}}}
\def\vRRt{{\vec\RRt}}
\def\vS{{\vec S}}
\def\vu{{\vec u}}
\def\vU{{\vec U}}
\def\vuA{{\vu_A}}
\def\vuAt{{\vu_{At}}}

\def\vuB{{\vu_B}}
\def\vuBt{{\vu_{Bt}}}
\def\vuBts{{\vu_{Bt*}}}

\def\vv{{\vec v}}
\def\vV{{\vec V}}
\def\vvA{{\vv_{\!A}}}
\def\vvAt{{\vv_{\!At}}}

\def\vvB{{\vv_{\!B}}}
\def\vvBt{{\vv_{\!Bt}}}
\def\vvBts{{\vv_{Bt*}}}
\def\vgammaBts{{\vgamma_{Bt*}}}

\def\vvDt{{\vv_{\!Dt}}}

\def\vvTheta{{\vv_{\Theta}}}

\def\vvThetat{{\vv_{\Theta t}}}
\def\vVtz{{\vV^{t_0}}}

\def\vVtzt{{\vV^{t_0}_{t}}}
\def\vw{{\vec w}}
\def\vW{{\vec W}}
\def\vwA{{\vw_{\!A}}}
\def\vwAt{{\vw_{\!At}}}

\def\vwB{{\vw_B}}
\def\vwBt{{\vw_{Bt}}}
\def\vwBts{{\vw_{Bt*}}}
\def\vWd{{\vW_2}}
\def\vwd{{\vw_2}}
\def\vWdP{{\vW_{2P}}}

\def\vwE{{\vec w_{\calE}}}
\def\vwEs{{\vec w_{\calE*}}}
\def\vwF{{\vec w_{\calF}}}
\def\vwFb{{\vec w_{\calF}{}^*}}

\def\vWP{{\vW_P}}

\def\vWtz{\vec W^\tz}
\def\vwtzs{\vw_{\tz*}}
\def\vWu{{\vW_1}}
\def\vwu{{\vw_1}}

\def\vWuP{{\vW_{1P}}}

\def\vx{{\vec x}}
\def\vX{{\vec X}}
\def\vy{{\vec y}}

\def\vZ{{\vec Z}}
\def\vz{{\vec z}}

\let\ds\displaystyle
\let\eps\varepsilon
\let\la\langle
\let\lrar\longrightarrow
\let\ora\overrightarrow
\let\pa\partial
\let\phi\varphi
\let\ra\rangle
\let\rar\rightarrow
\let\tilde\widetilde
\let\ts\textstyle
\long\def\comment#1{}

\def\be{\begin{equation}}
\def\ee{\end{equation}}
\def\eref#1{(\ref{#1})}

\newtheorem{lemme}{Lemma}[section]  
\newtheorem{corollary}[lemme]{Corollary}
\newtheorem{definition}[lemme]{Definition}
\newtheorem{example}[lemme]{Example}
\newtheorem{exercise}[lemme]{Exercice}
\newtheorem{prop}[lemme]{Proposition}  
\newtheorem{remark}[lemme]{Remark}
\newtheorem{theorem}[lemme]{Theorem}
\newenvironment{reponse}{\par\medskip\noindent\small {\bf Answer}. }{}

   \def\boxending{\leavevmode\nolinebreak\hfill{$\tblackbox$}}

\def\debcor{\begin{corollary}\sl }
   \def\fincor{\end{corollary}}
\def\debdef{\begin{definition}\rm }
   \def\findef{\end{definition}}
\def\debexa{\begin{example}\rm}
   \def\finexa{\boxending\end{example}}
\def\debexe{\begin{exercise}\rm}
   \def\finexe{\boxending\end{exercise}}
\def\deblem{\begin{lemme}\sl }
   \def\finlem{\end{lemme}}
\def\debprop{\begin{prop}\sl }
   \def\finprop{\end{prop}}
\def\debrem{\begin{remark}\rm}
   \def\finrem{\boxending\end{remark}}
\def\debrep{\begin{reponse}\rm\small}
   \def\finrep{\end{reponse}}
\def\debthm{\begin{theorem}\sl }
   \def\finthm{\end{theorem}}

\def\debproof{\medskip\par\noindent{\bf Proof.\ }}
   \def\finproof{\boxending\par\medskip}
\let\debdem\debproof
   \let\findem\finproof
\def\tbox{\leavevmode\vrule height 3pt width 3pt depth 0pt\relax}
\def\tblackbox{\tbox\kern-.75pt\raise4pt\hbox{\tbox}\kern-.75pt\tbox}

\def\montitrecours{
\LARGE Objectivity in continuum mechanics, an introduction  %: Basic knowledge, objectivity and Lie derivatives.
\\ \bigskip
\large
Motions, Eulerian and Lagrangian variables and functions, deformation gradient,
\\
Lie derivatives,
velocity-addition formula, Coriolis.
}

\begin{document}
%%%%%%%%%%%%%%%%%%%%%%%%%%%%%%%%%%%%%%%%%%%%%%%%%%%%%%%%%%
\thispagestyle{empty}
\pagestyle{myfin}

\null

\bigskip
\bigskip
\bigskip

\montitre{3}

\bigskip
\bigskip
\bigskip
\bigskip

In classical mechanics, there are two objectivities: 1- The covariant objectivity concerns the universal laws of physics required to be observer independent (true in any reference frame);
This is a main topic in this manuscript. 
2- The isometric objectivity concerns the constitutive laws of materials once expressed in a reference frame.

Covariant objectivity in continuum mechanics follows Maxwell's requirements, \cf~\cite{maxwell} page~1:
``2. (...) The formula at which we arrive must be such that a person of any nation, by substituting for the different symbols the numerical value of the quantities as measured by his own national units, would arrive at a true result. (...) 10. (...) The introduction of coordinate axes into geometry by Des Cartes was one of the greatest steps in mathematical progress, for it reduced the methods of geometry to calculations performed on numerical quantities. The position of a point is made to depend on the length of three lines which are always drawn in determinate directions (...)
But for many purposes in physical reasoning, as distinguished from calculation, it is desirable to avoid explicitly introducing the Cartesian coordinates, and to fix the mind at once on a point of space instead of its three coordinates, and on the magnitude and direction of a force instead of its three components. This mode of contemplating geometrical and physical quantities is more primitive and more natural than the other,...''

And see the (short) historical note given in the introduction of Abraham and Marsden book ``Foundations of Mechanics''~\cite{abraham-marsden}, about qualitative versus quantitative theory:
``Mechanics begins with a long tradition of qualitative investigation
culminating with {\sc Kepler} and {\sc Galileo}. Following this is the period of
quantitative theory (1687-1889) characterized by concomitant developments
in mechanics, mathematics, and the philosophy of science that are epitomized
by the works of {\sc Newton}, {\sc Euler}, {\sc Lagrange}, {\sc Laplace}, {\sc Hamilton}, and {\sc Jacobi}.
(...) For celestial mechanics (...) resolution we owe to the
genius of {\sc Poincar\'e}, who resurrected the qualitative point of view (...) One advantage (...) %of this model 
is that by
suppressing unnecessary coordinates the full generality of the theory becomes
evident.''

After having defined motions, Eulerian and Lagrangian variables and functions,
we give the definition of the deformation gradient as a function.
% (and then its representation with bases as defined by mechanical engineers).
We then obtain a simple understanding of the Lie derivatives of vector fields which meet the needs of engineers.
Then we get the velocity addition formula and verify that the Lie derivatives are objective. 
Note that Cauchy would certainly have used the Lie derivatives if they had existed during his lifetime: To get a stress, Cauchy had to compare two vectors, whereas one vector is enough when using the derivatives of~Lie.

We systematically start with qualitive definitions (observer independent), before quantifying with bases and/or Euclidean dot products (observer dependent). 
A fairly long appendix tries to give in one manuscript the definitions, properties and interpretations, usually scattered across several books (and not always that easy to find).

\newpage
\tableofcontents

%\bigskip

\goodbreak

%%%%%%%%%%%%%%%%%%%%%%%%%%%%%%%%%%%%%%%%%%%%%%%%%%%%%%%%%%%%%%%%%%%%%%%%%%%%%%%%%%%
%%%%%%%%%%%%%%%%%%%%%%%%%%%%%%%%%%%%%%%%%%%%%%%%%%%%%%%%%%%%%%%%%%%%%%%%%%%%%%%%%%%

\newpage

A quantity $f$ being given then: $g$ defined by << $g$ equals $f$ >> is noted $g:=f$.

\part{Motions, Eulerian and Lagrangian descriptions, flows}

%%%%%%%%%%%%%%%%%%%%%%%%%%%%%%%%%%%%%%%%%%%%%%%%%%%%%%%%%%%%%%%%%%%%%%%%%%%%%%%%%%%
%%%%%%%%%%%%%%%%%%%%%%%%%%%%%%%%%%%%%%%%%%%%%%%%%%%%%%%%%%%%%%%%%%%%%%%%%%%%%%%%%%%

\section{Motions}

%To define Eulerian and Lagrangian functions, we first need to define a motion of an object.
The framework is classical mechanics, time being decoupled from space.
$\RRt$ is the classical geometric affine space (the space we live in), and
$(\vRRt,+,.) = \{\vec{pq}: p,q \in\RRt\} \eqnote \vRRt$ is the associated vector space of bipoint vectors equipped with its usual rules.
We also consider $\RR$ and~$\RR^2$ as subspaces of~$\RRt$, \ie\ we consider $\RRn$ and~$\vRRn$, $n=1,2,3$.

%%%%%%%%%%%%%%%%%%%%%%%%%%%%%%%%%%%%%%%%%%%%%%%%%%%%%%%%%%%%%%%%%%%%%%%%%%%%%%%%%%%

\subsection{Referential}
\label{secreferentiel}

%%%%%%%%%%%%%%%%%%%%%%%%%%%%%%%%%%%%%%%%%%%%%%%%%%%%%%%%%%%%%%%%%%%%%%%%%%%%%%%%%%%

\paragraph{Origin:}

An observer chooses an origin $\calO\in\RRn$; Thus a point $p\in\RRn$ can be located by the observer thanks to the bipoint vector $\ora{\calO p} = \vx \in \vRRn$; Hence $p = \calO + \vx$, and $\vx=\ora{\calO p}\eqnote p-\calO$.

Another observer chooses an origin $\tilde\calO\in\RRn$; Thus the point $p$ can also be located by this observer with the bipoint vector $\ora{\tilde\calO p} = \tilde\vx \in \vRRn$; So $p = \calO + \vx = \tilde\calO + \tilde\vx$, and $\tilde\vx = \ora{\calO\tilde\calO}+\vx$.% (Chasles' theorem).

%%%%%%%%%%%%%%%%%%%%%%%%%%%%%%%%%%%%%%%%%%%%%%%%%%%%%%%%%%%%%%%%%%%%%%%%%%%%%%%%%%%

\paragraph{Cartesian coordinate system:}

\def\RCart{\calR_{\!\rm Cart}}
\def\RCartb{\calR_{\!\rm Cart,b}}
\def\Rtime{\calR_{\!\rm time}}

A Cartesian coordinate system in the affine space~$\RRn$
is  a set $\RCart = (\calO,(\ve_i)_{i=1,...,n})$,
where $\calO$ is an origin and $(\ve_i):=(\ve_i)_{i=1,...,n}$ is a basis in~$\vRRn$ chosen by the observer.
Thus the location of a point $p\in\RRn$ can quantified by the observer 
$\exists \vx\in\vRRn$ \st
%$\exists x_1,...,x_n\in\RR$ \st
%(there is as many bipoint vectors as there are origins):
\be
\label{eqxi00}
p = \calO + \vx
\qwith \vx = \sumin x_i\ve_i
,  \qie [\ora{\calO p}]_{|\ve} = [\vx]_{|\ve} = \pmatrix{x_1\cr\vdots\cr x_n},
\ee
$[\vx]_{|\ve}=[\ora{\calO p}]_{|\ve}$ being the column matrix containing the components $x_i\in\RR$ of $\ora{\calO p} = \vx$ in the basis~$(\ve_i)$.
Another observer with his origin~$O_b$ and his Cartesian basis $(\vb_i)_{i=1,...,n}$ make the Cartesian coordinate system  $\RCartb = (\calO_b,(\vb_i)_{i=1,...,n})$, and gets for the same position~$p$ in~$\RRn$,
\be
p = \calO_b + \vy
\qwith \vy = \sumin \tilde y_i \tilde\vb_i
,  \qie [\ora{\calO_b p}]_{|\vb} = [\vy]_{|\vb} = \pmatrix{ y_1\cr\vdots\cr  y_n},
\ee
$[\vy]_{|\vb}=[\ora{\calO_b p}]_{|\vb}$ being the column matrix containing the components $ y_i\in\RR$ of $\ora{\calO_b p} = \vy$ in the basis~$(\vb_i)$.
And $\ora{\calO_b p} = \ora{\calO_b \calO} +\ora{\calO p}$, \ie\ $\vy = \ora{\tilde\calO \calO} +\vx$, gives the relation
between $\vx$ and~$\vy$ (drawing).

%%%%%%%%%%%%%%%%%%%%%%%%%%%%%%%%%%%%%%%%%%%%%%%%%%%%%%%%%%%%%%%%%%%%%%%%%%%%%%%%%%%

\paragraph{Chronology:}

A chronology (or temporal coordinate system) is a set $\Rtime=(t_0,(\Delta t))$  chosen by an observer,
where $t_0\in \RR$ is the time origin,
and $(\Delta t)$ is the time unit (a basis in~$\vec\RR$).

%%%%%%%%%%%%%%%%%%%%%%%%%%%%%%%%%%%%%%%%%%%%%%%%%%%%%%%%%%%%%%%%%%%%%%%%%%%%%%%%%%%

\paragraph{Referentiel:}

A referential $\calR$ is the set
\be
\calR = (\Rtime,\RCart)=\hbox{$(t_0,(\Delta t),\calO,(\ve_i)_{i=1,...,n})$ = (``chronologie'',``Cartesian coordinate system'')},
\ee
made of a chronology and a Cartesian coordinate system, chosen by an observer.
\medskip

In the following% (framework of classical mechanics)
, %when one deals mainly with spatial considerations, 
to simplify the writings, the same implicit chronology is used by all observers, and a referential $\calR = (\Rtime,\RCart)$ will simply be noted as the reference frame $\calR= (\calO,(\ve_i))$ (so $:=\RCart$).
% (the embedded Cartesian coordinate system).

\comment{
\debrem
%\label{rembpv}
The bipoint vectors $\vx{}'$ and $\vx$ are observer dependent: for one point~$p$ we have
$\vx{}' = \ora{\calO' p} = \ora{\calO' \calO} + \ora{\calO p}= \ora{\calO' \calO} + \vx$
(the bipoint vectors $\vx{}'$ and $\vx$ depend on the choice of the origin by an observer).
%A bi-point vector is not a vector: 
%So a bi-point vector depends on an observer (depends on an origin chosen by an observer).
\finrem
}

%%%%%%%%%%%%%%%%%%%%%%%%%%%%%%%%%%%%%%%%%%%%%%%%%%%%%%%%%%%%%%%%%%%%%%%%%%%%%%%%%%%

\subsection{Einstein's convention (duality notation)}
%\label{secEinteinc}
\label{secEC}

%We will also use Einstein's convention (duality notation, details see~\S~\ref{secEconv}):
Starting point: The classical notation $x_i$ for the components of a vector~$\vx$ relative to a basis, \cf~\eref{eqxi00}.
Then the duality notion is introduced: $x_i \eqnote x^i$ (enables to see the difference between a vector and
a function when using components).
So
\be
\label{eqxi}
\vx = \underbrace{\sumin x_i\ve_i}_{\hbox{\footnotesize classic not.}} = \underbrace{\sumin x^i\ve_i}_{\hbox{\footnotesize duality not.}}, \qand
[\vx]_{|\ve} \eqclas \pmatrix{x_1\cr\vdots\cr x_n} \eqdual \pmatrix{x^1\cr\vdots\cr x^n}.
\ee
The duality notation is part of the Einstein's convention; Moreover Einstein's convention uses the notation $\sumin x^i\ve_i \eqnote x^i\ve_i$, \ie\
the sum sign~$\sumin$ can be omitted when an index ($i$~here) is used twice, once up and once down, details at~\S~\ref{secEconv}. However this omission of the sum sign~$\sum$ will not be made in this manuscript (to avoid ambiguities): The \TeX-\LaTeX\ program makes it easy to print~$\sumin$.

\comment{
Einstein's convention aims to distinguish covariance (function) from contravariance (vector),
when coordinates relative to a basis are used:
%This aims to distinguish a (linear) function from a vector.
\Eg, if you see $x^i$ then you deal with a vector $\vx\in\vRRn$ (contravariant), and if you see $\ell_i$ then you deal with a (linear) function $\ell\in\RRns$ (covariant).
}

\debexa
\label{exa11}
The height of a child is represented on a wall by a vertical bipoint vector $\vx$ starting from the ground up to a pencil line.
%The vector $\vx$ is objective: It is the same for any observer.
Question: What is the size of the child ?

Answer: It depends... on the observer (quantitative value = subjective result). %\Eg, all observers choose the same origin = ``the ground''.
% with the Einstein convention and observers who choose the same origin = ``the floor'':
\Eg, an English observer chooses a vertical basis vector $\va_1$ which length is one English foot~(ft).
So he writes $\vx = x_1\va_1$,
and for him the size of the child (size of~$\vx$) is $x_1$ in~\foot. \Eg\ $x_1 = 4$ means the child is 4 ft tall.
A French observer chooses a vertical basis vector $\vb_1$ which length is one metre~(m).
So he writes $\vx = y_1\vb_1$,
and for him the size of the child (size of~$\vx$) is  $y^1$~\metre.
\Eg, if $x_1 = 4$ then $y_1 \simeq 1.22$, since 1~ft $:=$ 0.3048~$m$:
The child is both $4$ and $1.22$ tall... in foot or metre. 
This quantification is written $\vx= 4$~ft $=1.22$~m, where ft means~$\va_1$ and m means~$\vb_1$ here.
NB: The qualitative vector $\vx$ is the same vector for all observers, not the quantitative values 4 or 1.22 (depends on a choice of a unit of measurement).

With duality notation: 
$\vx = x^1\va_1 = y^1\vb_1$, % with $y^1=0.3048\,x^1$, 
so if $x^1 = 4$ then $y^1 \simeq 1.22$.
\finexa

\comment{
\debrem
See~\S~\ref{secRfc}.
\finrem
}

This manuscript insists on covariant objectivity; Thus an English engineer (and his foot) and a French engineer (and his metre) will be able to work together ... and be able to avoid crashes like that of the Mars Climate Orbiter probe, see remark~\ref{remMCOC}. And they will be able to use the results of Galileo, Descartes,  Newton, Euler... who used their own unit of length, and knew nothing about the metre %(which replaced the Toise) 
defined in 1793 %(one ten-millionth of the distance from the North Pole to the Equator) 
and adopted in 1799 in France (after 6 years of measurements), and considered by the scientific community at the end of the ninetieth century... and couldn't explicitly use the ``Euclidean dot products'' either (which seems to have been defined mathematically by Grassmann around 1844). %, who named it the linear product). % (the metre did not exist at the time).
%(To compare with isometric objectivity: same metric for everyone.)

%\medskip See the detailed appendix, \S~\ref{secann1}, \ref{secrefE} and \ref{secRRT},  to possibly remove the doubts.

%%%%%%%%%%%%%%%%%%%%%%%%%%%%%%%%%%%%%%%%%%%%%%%%%%%%%%%%%%%%%%%%%%%%%%%%%%%%%%%%%%%

\subsection{Motion of an object}

Let $\Obj$ be a ``real object'', or ``material object'', made of particles (\eg, the Moon: Exists independently of an observer). Let $t_1,t_2\in\RR$, $t_1<t_2$.

\debdef
\label{deftPhi}
The motion of~$\Obj$ in~$\RRn$ is the map
\be
\label{eqdeftPhi0}
\tPhi : 
\left\{\eqalign{
[t_1,t_2] \times \Obj &\rar \RRn \cr
(t,\underbrace{
\Pobj)
}_{\hbox{\footnotesize particle}}
 & \rar 
\underbrace{
p
= \tPhi(t,\Pobj) %= \calO + \sumin x^i\ve_i = \calO + \vx
}_{\hbox{\footnotesize its position at $t$ in the Universe}}
%}_{\makebox[1cm]{\hbox{\scriptsize its position at $t$}}} 
%= \hbox{position of $\Pobj$ at~$t$ in~$\RRn$}.
.
}\right.
\ee
And $t$ is the time variable, $p$ is the space variable, and $(t,p)\in\RR\times \RRn$ is the time-space variable.
And $\tPhi$~is supposed to be $C^2$ in time.
\findef

With an origin $\calO$ (observer dependent),
the motion can be described with the bi-point vector
\be
\label{eqdeftvphi}
%\tvphi : 
%\left\{\eqalign{
%[t_1,t_2] \times \Obj &\rar \vRRn \cr
%(t,\Pobj) & \rar 
\vx %= \tvphi(t,\Pobj)  
= \ora{\calO \tPhi(t,\Pobj)} = \ora{\calO p} \eqnote \tvphi(t,\Pobj).
%}\right.
\ee
But then, two observers with different origins $\calO$ and~$\calO_b$ have different description of the motion. Therefore, in the following we won't use $\tvphi$.
%Moreover, in a non-planar surface considered on its own (a manifold), a bi-point vector is meaningless (it is not tangent to the surface: it goes ``through'').
Then (quantification) with a Cartesian basis $(\ve_i)$ to make a referential~$\calR$, we get~\eref{eqxi00}.

\comment{
\debrem
%\label{remdiffsObj}
The motion of an object~$\Obj$ (\eg\ a planet) has been described before the definition of groups, rings, vector spaces, algebra (19th century) (Copernicus 1473-1543, Descartes 1596-1650).
\finrem
}

%%%%%%%%%%%%%%%%%%%%%%%%%%%%%%%%%%%%%%%%%%%%%%%%%%%%%%%%%%%%%%%%%%%%%%%%%%%%%%%%%%%

\subsection{Virtual and real motion}

\debdef
\label{defmouvP}
A virtual (or possible) motion of $\Obj$ is a function $\tPhi$ ``regular enough for the calculations to be meaningful''.
%: In the following, the parametric trajectories $\tPhiPobj$ are at least $C^2$ for velocities and accelerations to exist.
Among all the virtual motions, the observed motion is called the real motion.
\findef

%%%%%%%%%%%%%%%%%%%%%%%%%%%%%%%%%%%%%%%%%%%%%%%%%%%%%%%%%%%%%%%%%%%%%%%%%%%%%%%%%%%

\subsection{Hypotheses (Newton and Einstein)} % and spatial (Eulerian) variables} %, , Eulerian and Lagrangian variables
\label{sechypNE}

{\bf Hypotheses of Newtonian mechanics (Galileo relativity) and general relativity (Einstein)}:

1- You can describe a phenomenon only at the actual time~$t$ and from the location~$p$ you are at
(you have \textbf{no} gift of ubiquity in time or space);
%And $t$ and~$p$ willconstitute the Eulerian variables;

2- You don't know the future; 

3- You can use your memory, so use some past time~$\tz$ and some past position~$\ptz$; 
%And $\tz$ and~$\ptz$ constitute the Lagrangian variables;

4- You can use someone else memory (results of measurements) if you can communicate objectively. %, to use a past observation.

%%%%%%%%%%%%%%%%%%%%%%%%%%%%%%%%%%%%%%%%%%%%%%%%%%%%%%%%%%%%%%%%%%%%%%%%%%%%%%%%%%%

\subsection{Configurations} % and spatial (Eulerian) variables}

%Let $\tPhi$ be a motion, \cf~\eref{eqdeftPhi0}.
Fix $t\in[t_1,t_2]$, and define
$
%\label{eqconfig0}
\tPhi_t : 
\left\{\eqalign{
\Obj &\rar \RRn \cr
\Pobj & \mapsto p %= \pt(\Pobj) 
= \tPhi_t(\Pobj) := \tPhi(t,\Pobj). \cr
}\right.
$

\debdef
\label{defconfig0}
The ``configuration at~$t$'' of~$\Obj$ is the range (or image) of~$\tPhi_t$, \ie\ is the subset of~$\RRn$ (affine space) defined by% is the range (= image) of~$\tPhi_t$, that is, it is
\be
\label{eqconfig}
\Omegat := \{p\in\RRn : \exists \Pobj\in\Obj \hbox{ \st\ } p = \tPhi_t(\Pobj)\}
\eqnote \tPhi_t(\Obj) \eqnote \Im(\tPhi_t) .
\ee

If $t$ is the actual time then $\Omegat$ is the actual (or current or Eulerian) configuration.

If $t_0$ is a time in the past then $\Omegatz$ is the past (or initial or Lagrangian) configuration.
\comment{
And $p = \tPhi_t(\Pobj)\in\Omegat$ is the spatial variable (at~$t$), or Eulerian variable, relative to~$\Pobj$ at~$t$.
%If an origin~$\calO$ has been chosen by an observer, then $\ora{\calO \pt}$ is the spatial location of~$\Pobj$ in~$\vRRn$ at~$t$ relative to $\calO$.

And if a Cartesian referential $\calR=(\calO,(\ve_i))$ has been chosen, then $\ora{\calO \pt}=\sumin x_i\ve_i$ is called a vectorial spatial variable, or vectorial Eulerian variable, relative to~$\Pobj$ at~$t$ and relative to the referential~$\calR$.
}
\findef

\noindent
{\bf Hypothesis:} At any time $t$, $\Omegat$ is supposed to be a ``smooth domain'' in~$\RRn$, and the map $\tPhi_t$ is assumed to be one-to-one (= injective): $\Obj$ does not crash onto itself.
%, \ie, the closure of a ``regular open set'' in~$\RRt$, or of a 2-D differentiable surface in~$\RRt$, or of a 1-D differentiable curve in~$\RRt$ (continuum mechanics).

%%%%%%%%%%%%%%%%%%%%%%%%%%%%%%%%%%%%%%%%%%%%%%%%%%%%%%%%%%%%%%%%%%%%%%%%%%%%%%%%%%%

\subsection{Definition of the Eulerian and Lagrangian variables}

%\debdef\ 

$\bullet$ If $t$ is the actual time, then $\pt = \tPhi_t(\Pobj)\in\Omegat$ is called the Eulerian variable
relative to $\Pobj$ and~$t$.

\mn
$\bullet$ If $\tz$ is a time in the past, then $\ptz = \tPhi_\tz(\Pobj)\in\Omegatz$ is called the Lagrangian variable relative to $\Pobj$ and~$\tz$.
(A Lagrangian variable is a ``past Eulerian variable''). (Two observers with two different origin of time $\tz$ and~$\tz'$ get two different Lagrangian variable while they have the same Eulerian variable.)
%\findef

%So you can consider a Lagrangian variable if it first was a Eulerian variable considered by the observer who made the measurement at~$\tz$. In other words, a Lagrangian variable is a past Eulerian variable.

%%%%%%%%%%%%%%%%%%%%%%%%%%%%%%%%%%%%%%%%%%%%%%%%%%%%%%%%%%%%%%%%%%%%%%%%%%%%%%%%%%%

\subsection{Trajectories}

Let $\tPhi$ be a motion of~$\Obj$, \cf~\eref{eqdeftPhi0},
and $\Pobj\in\Obj$ (a particle in $\Obj$ = \eg\ the Moon).

\debdef
\label{deftPhiPobj}
The (parametric) trajectory of~$\Pobj$  %between $t_1$ and $t_2$ 
is the function
\be
\label{eqtPhiPobj}
\tPhiPobj : 
\left\{\eqalign{
[t_1,t_2] &\rar \RRn, \cr
t & \mapsto p(t) = \tPhiPobj(t) \eqdef \tPhi(t,\Pobj) %\eqnote p_\Pobj(t) 
\quad (\hbox{position of $\Pobj$ at $t$ in the Universe}). \cr
}\right.
\ee
Its geometric trajectory is the range (image) of $\tPhiPobj$, \ie\
\be
\hbox{geometric trajectory of $\Pobj$} 
:= \{q\in\RRn : \exists t\in[t_1,t_2] \hbox{ \st\ } q=\tPhiPobj(t)\}
= \Im(\tPhiPobj) = \tPhiPobj([t_1,t_2]).
\ee
\findef

%%%%%%%%%%%%%%%%%%%%%%%%%%%%%%%%%%%%%%%%%%%%%%%%%%%%%%%%%%%%%%%%%%%%%%%%%%%%%%%%%%%

\subsection{Pointed vector, tangent space, fiber, vector field, bundle}
\label{secRRnt}

\def\tvw{{\tilde\vw}}

(See \eg\ Abraham--Marsden~\cite{abraham-marsden}.)
To deal with surfaces $S$ in~$\RRt$, \eg\ with $S =$ a sphere (and more generally with manifolds in~$\RRn$),
a vector cannot simply be a ``bi-point vector connecting two points of~$S$'' (would get ``through the surface''). A vector
is defined to be tangent to~$S$: %, to point toward a direction tangent to~$S$. 
%Thus, a vector will be defined to be a ``tangent vector'':
 Consider a ``regular'' curve $c:s\in ]-\eps,\eps[ \rar c(s)\in S$ where $S$ is a surface in an affine space, and the vector tangent to~$S$ at $c(0)$ is $\vw(c(0))=\lim_{h\rar0}{c(h) - c(0)\over h}$ (it is defined with a parametrization of~$c$ in a general manifold); Considering all the possible curves, we get ``all possible vectors on~$S$''.
%In~$\RRn$, let $m\in[1,n]_\NN$ and let $S$ be a ``regular $m$-surface'' (a $m$-differentiable manifold in~$\RRn$).That is, in~$\RRn=\RRt$, a $3$-surface $S$ is an open set $\Omega$ in~$\RRt$, a $2$-surface $S$ is a ``usual surface'', a $1$-surface $S$ is a ``usual curve''.

\medskip
\noindent
{\bf Notation:}
%The tangent space at~$p$ relative to~$S$ is
\be
\label{eqTpS}
T_p S:
= \{\hbox{tangent vectors $\vw_p$ at $S$ at~$p$}\}
=\hbox{The tangent space at~$p \in S$} .
\ee
%In particular if $\tPhi:(t,\Pobj)\rar \RRn$ is a motion and $S=\Omegat = \tPhi_t(\Obj)$ is an open set in~$\RRn$, then $\ds T_p \Omegat \eqnote \RRnt$.

\Eg, if $S$ is a sphere in~$\RR^3$ and $p\in S$, then $T_pS$ is its usual tangent plane at~$p$ at~$S$.

\Eg, particular case: If $S=\Omega$ is an open set in~$\RRn$, then $T_p S = T_p\Omega=\vRRn$ is independent of~$p$.

\debdef
\be
\label{eqfiber}
\hbox{The fiber at~$p$}
:=\{p\}\times T_p S
=\{\underbrace{(p,\vw_p)}_{\hbox{\footnotesize pointed vector}}\in \{p\} \times T_p S\},
\ee
\ie, the fiber at~$p$ is the set of ``pointed vectors at~$p$'', 
a pointed vector being the couple $(p,\vw_p)$ made of the ``base point'' $p$ and the vector $\vw_p$ defined at~$p$.

Drawing: A vector in~$\vRRn$ can be drawn anywhere in~$\RRn$; While a ``pointed vectors at~$p$'' has to be drawn at the point~$p$ in~$\RRn$. %(And a regular set of pointed vectors will define a ``vector field''.)

If the context is clear, a pointed vector is simply noted $\tvw(p) \eqnote \vw(p)$ (lighten the writing).
\findef

Particular case: If $S=\Omega$ is an open set in~$\RRn$, then the fiber at~$p$ is $T_p\Omega=\{p\}\times \vRRn$.

\debdef
\be
\label{eqbundle}
\hbox{The tangent bundle } TS
:= \bigcup_{p\in S} (\{p\}\times T_p S),
\ee
that is, is the union of the fibers.
\findef

\debdef
A vector field $\tvw$ in~$S$ is a~$C^\infty$ function (or at least $C^2$ in the following)
\be
\tvw: 
\left\{\eqalign{
S & \rar TS \cr
p  & \rar \tvw(p)=(p,\vw(p)). \cr
}\right.
\ee
If the context is clear, a vector field is simply noted $\tvw \eqnote \vw$ (lighten the writing).
\findef

%Thus the image (the range) of a vector field~$\tvw$ is included in the tangent bundle: $\Im(\tvw)\subset TS$.

%%%%%%%%%%%%%%%%%%%%%%%%%%%%%%%%%%%%%%%%%%%%%%%%%%%%%%%%%%%%%%%%%%%%%%%%%%%%%%%%%%%
%%%%%%%%%%%%%%%%%%%%%%%%%%%%%%%%%%%%%%%%%%%%%%%%%%%%%%%%%%%%%%%%%%%%%%%%%%%%%%%%%%%

\section{Eulerian description (spatial description at actual time $t$)}
\label{seceul}

%%%%%%%%%%%%%%%%%%%%%%%%%%%%%%%%%%%%%%%%%%%%%%%%%%%%%%%%%%%%%%%%%%%%%%%%%%%%%%%%%%%

\subsection{The set of configurations}

%The geometric space is $\RRn=\RRt$.
Let $\tPhi$ be a motion of~$\Obj$, \cf~\eref{eqdeftPhi0},
and $\Omega_t = \tPhi_t(\Obj) \subset\RR^n$ be the configuration at~$t$, \cf~\eref{eqconfig}.
The set of configurations is the subset $\bigC \subset \RR\times \RRn$ (the ``time-space'') defined by
\be
\label{eqbigU}
\eqalign{
\bigC
:= &\bigcup_{t\in[t_1,t_2]}(\{t\}\times\Omega_t)  \quad(=\hbox{set in which you find particles in ``time-space''}) \cr
% \RR\times \RRn
= &\{(t,p)\in\RR\times \RRn : \exists (t,\Pobj)\in[t_1,t_2] \times\Obj,\; p=\tPhi(t,\Pobj)\},
}
\ee

\medskip
Question: Why don't we simply use $\bigcup_{t\in[t_1,t_2]} \Omegat$ instead of $\bigC = \bigcup_{t\in[t_1,t_2]}(\{t\}\times\Omega_t)$?

Answer: $\bigC$ gives the film of the life of~$\Obj$ = the succession of the photos~$\Omegat$ taken at each~$t$;
And $\Omegat$ is obtained from~$\bigC$ thanks to the pause feature at~$t$.
Whereas $\bigcup_{t\in[t_1,t_2]} \Omegat \subset \RRn$ is the superposition of all the photos on the image $\bigcup_{t\in[t_1,t_2]}\Omegat$... %: The film is superimposed on one photo... 
and we don't distinguish the past from the present.

%%%%%%%%%%%%%%%%%%%%%%%%%%%%%%%%%%%%%%%%%%%%%%%%%%%%%%%%%%%%%%%%%%%%%%%%%%%%%%%%%%%

\subsection{Eulerian variables and functions}

\debdef
In short:
A Eulerian function relative to~$\Obj$ is a function, with $m\in\NNs$,
\be
\label{eqdeffspa20}
\eul : 
\left\{\eqalign{
\bigC & \rar \vRRm \hbox{ (or more generally a suitable set of tensors)}\cr
(t,p) & \rar \eul(t,p) %\quad(=\hbox{value of $\eul$ at $(t,p)$})
, %\qst \forall t,\; \eul_t(p) \in \calL^r_s(\RRn).
}\right.
\ee
the spatial variable $p$ being the Eulerian variable.

In details: A function $\eul$ being given as in~\eref{eqdeffspa20},
the associated Eulerian function $\widehat\eul$ is the function %defined by
\be
\label{eqdeffspa2}
\widehat\eul : 
\left\{\eqalign{
\bigC & \rar \bigC \times \vRRm \hbox{ (or $\bigC \times$ some suitable set of tensors)} \cr
(t,p) & \rar \widehat\eul(t,p)=((t,p),\eul(t,p)) =(\hbox{time-space position\ ,\ value}),
}\right.
\ee
and is called ``a field of functions''. So $\widehat\eul(t,p)$ is the ``pointed $\eul(t,p)$'' at $(t,p)$ (in time-space). 

So, the range $\Im(\widehat\eul) = \widehat\eul(\bigC)$ of an Eulerian function~$\widehat\eul$ is the graph of~$\eul$.
(Recall: The graph of a function $f:x\in A \rar f(x)\in B$ is the subset $\{(x,f(x))\in A\times B\} \subset A\times B$: gives the ``drawing of~$f$'').

If there is no ambiguity, $\widehat\eul \eqnote \eul$ for short.
\findef

At~$t$, the Eulerian vector field at~$t$ is
$
%\label{eqdeffspa3}
\widehat\eul_t :
\left\{\eqalign{
\Omegat & \rar  \Omegat \times \vRRn \cr
p & \rar \widehat\eul_t(p) \eqdef (p,\eul_t(p)) =(\hbox{position\ ,\ value}). 
}\right.
$

\debexa
\label{exaeul1}
$\eul(t,p)=\theta(t,p) \in \RR=$  temperature
of the particle $\Pobj$ which is at~$t$ at~$p=\tPhi(t,\Pobj)$;
\finexa

\debexa
\label{exaeul12}
$\eul(t,p)=\vu(t,p)\in \vRRn=$ force applied on the particle $\Pobj$
which is at~$t$ at~$p$.
\finexa

\debexa
$\eul(t,p)=d\vu(t,p)\in \calL(\vRRn:\vRRn)=$ the differential at~$t$ at~$p$ of a Eulerian function~$\vu$.
\finexa

%\medskip
{\bf Question:} Why introduce $\widehat\eul$? Isn't $\eul$ sufficient?

{\bf Answer:} 
%With $p_+=(t,p) \in \RR\times \RRn$, a value $\eul(p_+)=\eul(p_+)\in\RRn$ is drawn on ``the vertical axis'' (the image set), when 
The ``pointed value'' $\widehat\eul(t,p) = ((t,p),\eul((t,p)))$ is drawn on the graph of~$\eul$.

%\Eg, the temperature $\theta(t,p) \in \RR$ is a value on the ``$y$-axis'', while the ``pointed value'' $\widehat\theta(t,p)=((t,p),\theta(t,p))$ is drawn at $(t,p)$ on the graph of~$\theta$ in $(\RR\times \RRn)\times \RR$. %: The drawing gives the temperature field (usually drawn in colors).

\Eg, at $t$ at~$p$ the velocity vector $\vv(t,p)\in \vRRt$ can be drawn anywhere, while the ``pointed vector'' $\widehat\vv(t,p)=((t,p);\vv(t,p))$ is $\vv(t,p)$ drawn at~$t$ at~$p$ (and $\widehat\vv$ is called the velocity field).

Moreover~\eref{eqdeffspa2} emphasizes the difference between a Eulerian vector field and a Lagrangian vector function, see~\eref{eqdefLag}.

\debrem
\Eg, the initial framework of Cauchy for his description of forces is Eulerian: The Cauchy stress vector $\vec t=\uusigma.\vn$ is considered at the actual time~$t$ at a point $p\in\Omegat$.
(It is not Lagrangian.)
\finrem

%%%%%%%%%%%%%%%%%%%%%%%%%%%%%%%%%%%%%%%%%%%%%%%%%%%%%%%%%%%%%%%%%%%%%%%%%%%%%%%%%%%

\subsection{Eulerian velocity (spatial velocity) and speed}

\debdef
In short:
Consider a particle $\Pobj$ and its (regular) trajectory $\tPhiPobj: t\rar p(t) = \tPhiPobj(t)$, \cf~\eref{eqtPhiPobj}.
Its Eulerian velocity at~$t$ at $p(t) = \tPhiPobj(t)$ is
%is the vectorial valued map defined on $\bigC= \bigcup_{t\in[t_1,t_2]}(\{t\}\times\Omega_t)$ by
\be
\label{eqdefve}
\vv(t,p(t)) := \tPhiPobj{}'(t) \eqnote {\pa\tPhi \over \pa t}(t,\Pobj), \qwhen p(t) = \tPhiPobj(t),
\ee
\ie, $\vv(t,p(t))$ is the tangent vector at~$t$ at $p(t) = \tPhiPobj(t)$ to the trajectory $\tPhiPobj$.
This defines the vector field (in short)
$\vv:
\left\{\eqalign{
\bigC & \rar \vRRn \cr
(t,\pt) & \rar\vv(t,\pt)
}\right\}
$.

In details: \cf~\eref{eqdeffspa2}, the Eulerian velocity is 
the function
$ \widehat\vv:
\left\{\eqalign{
\bigC & \rar \bigC \times \vRRm \cr
(t,p) & \rar \widehat\vv(t,p)=((t,p),\vv(t,p))
}\right\}
$
(pointed vector) where $\vv(t,p)$ is given by~\eref{eqdefve}.
\findef

\debrem
${d\tPhiPobj \over dt}(t)  = \vv(t,\tPhiPobj(t))$, with $p(t)=\tPhiPobj(t)$, is often written
\be
\label{eqed00}
{dp \over dt}(t) = \vv(t,p(t)), \qor {d\vx \over dt}(t) = \vv(t,\vx(t)), \qor {d\vx \over dt} = \vv(t,\vx),
\ee
the two last notations when an origin $O$ is chosen and $\vx(t) = \ora{\calO p(t)}$. %, \cf~\eref{eqdeftvphi}.
Such an equation is the prototype of an ODE (ordinary differential equation)
solved with the Cauchy--Lipschitz theorem, see~\S~\ref{secflot}. % and remark~\ref{remhypNE}.
(A~Lagrangian velocity does not produce an ODE, see~\eref{eqdefVl00}.)
\finrem

\debdef
If an observer chooses a Euclidean dot product~$\dd_g$ (\eg\ foot or metre built), the associated norm being~$||.||_g$,
then the length $||\vv(t,p)||_g$ is the speed (or scalar velocity) of~$\Pobj$ 
(\eg\ in ft/s or in m/s).
And the context must remove the ambiguities: the ``velocity'' is
either the vector velocity $\vv(t,p)=\tPhiPobj{}'(t)$ %(depends on the time unit),
or the speed (the scalar velocity) $||\vv(t,p)||_g$.
\findef

\debexe
\def\vxp{\vx\,'}
\def\vxpp{\vx\,''}
%Let $\vr(t)=\tPhi(t,\Pobj)=$  trajectory of a~$\Pobj$ (so $\vrp(t) = \vv(t,\vr(t))=$ Eulerian velocity),
Euclidean dot product~$\dd_g$,
$\vx(t)= \ora{\calO p(t)}$,
$\vec T(t)= {\vxp(t) \over ||\vxp(t)||_g}$,
and $f(t) = ||\vxp(t)||_g$ (speed).
Prove : ${ d f \over dt}(t) = (\vxpp(t),\vec T(t))_g \eqnote \vxpp(t) \bcdot \vec T(t)$ (= tangential acceleration).

\debrep 2-D and Euclidean basis:
$\vx(t)=\pmatrix{x(t) \cr y(t)}$ gives
$f(t)=(x'(t)^2 + y'(t)^2)^\demi$, thus
$f'(t)= {x'(t)x''(t) + y'(t)y''(t) \over f(t)}
={\vrp(t) \bcdot \vrpp(t) \over ||\vrp(t)||}$.
Idem in $n$-D.
\finrep
\finexe

%%%%%%%%%%%%%%%%%%%%%%%%%%%%%%%%%%%%%%%%%%%%%%%%%%%%%%%%%%%%%%%%%%%%%%%%%%%%%%%%%%%

\subsection{Spatial derivative of the Eulerian velocity}

$t\in [t_1,t_2]$ is fixed, $\eul$ is a given Eulerian function, and 
$\eul_t:
\left\{\eqalign{
\Omegat &\rar\vRRm \cr
p &\rar \eul_t(p) :=\eul(t,p) \cr
}\right\}$ is~$C^1$.

%%%%%%%%%%%%%%%%%%%%%%%%%%%%%%%%%%%%%%%%%%%%%%%%%%%%%%%%%%%%%%%%%%%%%%%%%%%%%%%%%%%

\subsubsection{Definition}

Recall: If $\Omega$ is an open set in~$\RRn$ and if $f:\Omega \rar\RR$ is differentiable at $p$, then its differential at $p$ is the linear form $df(p)\in\calL(\vRRn;\RR)$ (linear map with real values) defined by, for all $\vu\in\vRRn$ (vector at~$p$),
\be
\label{eqconv0}
df(p).\vu = \lim_{h\rar0} {f(p{+}h\vu) - f(p) \over h} .
\ee
This expression is the same for all observers (English, French...: There is no inner dot product here).

\debdef
The space derivative of $\eul$ at~$(t,p)$ is the differential $d\eul_t$ at~$p$, \ie,
%$d\eul(t,p) := d\eul_t(p)$; %\cf~\eref{eqdeffspa3},
for all $t\in [t_1,t_2]$, all $p\in\Omegat$ and all $\vw_p \in \RRnt$ (vector at $p$),
\be
\label{eqspd}
 (d\eul_t(p).\vw_p =) \quad
\boxed{d\eul(t, p).\vw_p
= \lim_{h\rar0} {\eul(t,p{+}h \vw_p) - \eul(t,p) \over h} } \eqnote {\pa\eul\over \pa p}(t,p).\vw_p.
\ee
%Interpretation: 
In $\Omegat$ (the photo at~$t$), $d\eul(t, p).\vw_p$ gives the rate of variations of~$\eul_t$ at~$p$ in the direction~$\vw_p$.
\findef

\Eg, at~$t$, the space derivative $d\vv$ of the Eulerian velocity field is defined by
\be
\label{eqspdv}
d\vv(t, p).\vw_p = \lim_{h\rar0} {\vv(t,p{+}h \vw_p) - \vv(t,p) \over h} \quad (=d\vv_t(p).\vw_p).
%\quad ( = \lim_{h\rar0} {\vv_t(p+h \vw) - \vv_t(p) \over h}).
\ee

\debrem
In differential geometry, \eref{eqconv0} is also written $\vu(f)(p) = {d\over dh}f(p{+}h\vu)_{|h=0}$; Don't use this notation if you are not at ease with differential geometry (where a vector is defined to be a derivation, so $\vu[f]$ is the derivation of~$f$ by~$\vu$).
\finrem

%%%%%%%%%%%%%%%%%%%%%%%%%%%%%%%%%%%%%%%%%%%%%%%%%%%%%%%%%%%%%%%%%%%%%%%%%%%%%%%%%%%

\subsubsection{The convective derivative $d\eul.\vv$}

\debdef
If $\vv$ is the Eulerian velocity field, % and $f=\eul=$ a Eulerian function, 
then $d\eul.\vv$ is called the convective derivative of~$\eul$.
\findef

\comment{
\noindent
{\bf Quantification:} With a Cartesian basis $(\ve_i)$ in~$\vRRn$, \eref{eqconv01} gives
\be
d\eul.\vv = (\vv.\vgrad)\eul
= \sumin  v_i {\pa \eul \over \pa x_i}= \hbox{the convective derivative in a basis}.
\ee
(With duality notations, $d\eul.\vv = (\vv.\vgrad)\eul
= \sumin  v^i {\pa \eul \over \pa x^i}$.)
}

%%%%%%%%%%%%%%%%%%%%%%%%%%%%%%%%%%%%%%%%%%%%%%%%%%%%%%%%%%%%%%%%%%%%%%%%%%%%%%%%%%%

\subsubsection{Quantification in a basis: $df.\vu$ is written $(\vu.\vgrad)f$}

{\bf Quantification:}
Let $f:p\in \RRn \rar f(p) \in \RR$ be~$C^1$.
Let $(\ve_i)$ be a basis in~$\vRRn$.
Let (usual definition)
\be
\label{eqconv02}
{\pa f \over \pa x_i}(p) := df(p).\ve_i \qand 
[df(p)]_{|\ve} = \pmatrix{{\pa f \over \pa x_1}(p)& ... & {\pa f \over \pa x_n}(p)} \;\hbox{ (line matrix)}.
\ee
(Recall: The matrix which represents a linear form is a line matrix.) And $[df(p)]_{|\ve}$ is the Jacobian matrix of $f$ at~$p$ relative to~$(\ve_i)$.
\comment{
, and, with $(dx_i)$ the (covariant) dual basis of~$(\ve_i)$,
\be
df(p)=\sumin {\pa f \over \pa x_i}(p)\,dx_i
\ee
(unavoidable in thermodynamics).
}
So, with $\vu=\sumin u_i\ve_i$ a vector at~$p$, and with the usual matrix multiplication rule, we have
\be
\label{eqconv01}
df(p).\vu =[df(p)]_{|\ve}.[\vu]_{|\ve}
= \sumin {\pa f \over \pa x_i}(p) u_i
= \sumin  u_i {\pa f \over \pa x_i}(p)
\eqnote (\vu.\vgrad)_{|e} f(p),
\ee
where $(\vu.\vgrad)_{|e} : C^1(\Omega;\RR) \rar C^0(\Omega;\RR)$ is the differential operator defined relative to a basis~$(\ve_i)$ by
\be
\label{eqconv03}
(\vu.\vgrad)_{|e} (f)=\sumin  u_i {\pa f \over \pa x_i} . %\quad (=df.\vu).
\ee
If the basis $(\ve_i)$ is unambiguously imposed, then $(\vu.\vgrad)_{|e} \eqnote \vu.\vgrad$

For vector valued functions $\vf:\Omega\rar\vRRm$, 
the above steps apply to the components of $\vf$ in a basis $(\vb_i)$ in~$\vRRm$: If $\vf = \sumim f_i\vb_i$,
\ie\ $\vf(p) = \sumim f_i(p)\vb_i$,
then 
\be
\label{eqconv04}
(\vu.\vgrad)_{|e} (\vf)
=\sumim (df_i.\vu) \vb_i = \sumim ((\vu.\vgrad)_{|e} f_i)\vb_i
=\sumim\sumjn (u_j.{\pa f_i \over \pa x_j}) \vb_i. %\quad ( = d\vf.\vu),
\ee
%\ie\ $[(\vu.\vgrad)_{|e} (\vf)]_{\vb} = [df]_{|\ve,\vb}.[\vu]_{|\ve} = [df.\vu]_{|\vb}$ where $[d\vf]_{|\ve,\vb} = [{\pa f_i \over \pa x_j}]_{i=1,...,m \atop j=1,...,n}$ is the Jacobian matrix of~$\vf$.
%And with tensorial notations and the contraction rule, $d\vf \eqnote \sumim \vb_i \otimes df_i$ (the calculation of $df.\vu$ has to be done as $(\sumim \vb_i \otimes df_i).\vu = \sumim \vb_i(df_i.\vu)$).

\comment{
With duality notations ${\pa f \over \pa x_i} \eqnote {\pa f \over \pa x^i}$,
and $df(p)=\sumin {\pa f \over \pa x^i}(p)\,dx^i$,
thus $df.\vu = \sumin  u^i {\pa f \over \pa x^i}$,
and $\vu.\vgrad = \sumin  u^i {\pa \over \pa x^i}$.
And $d\vf.\vu=\sumim\sumjn (u^j.{\pa f^i \over \pa x^j}) \vb_i$ and
$[d\vf]_{|\ve,\vb} = [{\pa f^i \over \pa x^j}]$. % and $d\vf \eqnote \sumim \vb_i \otimes df^i$.
}

\comment{
\medskip
\noindent
{\bf Application:}
Consider a motion $\tPhiPobj$ of a particle $\Pobj\in\Obj$, \cf~\eref{eqtPhiPobj}, let $t\in\RR$, let $p=\tPhiPobj(t)$,
let $\vv(t,p)=\tPhiPobj{}'(t)$ (the Eulerian velocity at~$t$ at~$p$).
And consider a differentiable Eulerian function $\eul$, \cf~\eref{eqdeffspa20},
and  let $\eul(t,p) \eqnote \eul_t(p)$.
Then, with $f=\eul_t$ and $\vu=\vv(t,p)$ in~\eref{eqconv0} we define:
}

%%%%%%%%%%%%%%%%%%%%%%%%%%%%%%%%%%%%%%%%%%%%%%%%%%%%%%%%%%%%%%%%%%%%%%%%%%%%%%%%%%%

\subsubsection{Representation relative to a Euclidean dot product: $\vgrad f$}

An observer chooses a distance unit (foot, metre...) and uses the associated Euclidean dot product~$\dd_g$.
%(the following results depend on~$\dd_g$, \ie\ on the observer).

Let $\Omega$ be an open set in~$\RRn$, $f \in C^1(\Omega;\RR)$ (scalar valued function), and $p\in\Omega$.
Then the $\dd_g$-Riesz representation vector of the differential form $df(p)$ is called the gradient of~$f$ at~$p$ relative to~$\dd_g$, and named $\vgrad_g f(p)\in\vRRn$: It is defined by %, \cf~\eref{thmRiesz},
\be
\label{eqdfgf0}
\forall \vu\in\vRRn, \quad (\vgrad_g f(p),  \vu)_g = df(p).\vu, \qwritten \vgrad f \bcdot \vu = df.\vu,
\ee
the last notation iff a Euclidean dot product $\dd_g$ is imposed to all observer (quite subjective: foot, metre ?).
%(Also see~\eref{eqdifff2}.)
%($\vgrad_g f$ depends on~$\dd_g$, cf~\eref{eqsecemp2}, whereas $df$ does not.)

(The first order Taylor expansion $f(p{+}h\vu) = f(p) + h\, df(p).\vu + o(h)$ can therefore, after a choice of an Euclidean dot product, be written $f(p{+}h\vu) = f(p) + h\, \vgrad_g f(p) \bcdotg  \vu + o(h)$.)

\medskip
\noindent
{\bf Quantification:} Let $(\ve_i)$ be a Cartesian basis in~$\RRn$. 
Then~\eref{eqdfgf0} gives
$[df].[\vu] = [\vgrad f]^T. [g].[\vu],$ for all $\vu\in\RRnt$
(more precisely $[df]_{|\ve}.[\vu]_{|\ve} = [\vgrad_g f]_{|\ve}^T. [g]_{|\ve}.[\vu]_{|\ve}$),
thus (since $[g]_{|\ve}$ is symmetric)
\be
\label{eqdfgf01}
[\vgrad f] = [g].[df]^T \quad\hbox{(column matrix)}.
\ee
\Ie, if $\vgrad f=\sumin a_i\ve_i$ then $a_i=\sumjn g_{ij} {\pa f \over \pa x_j}$ for all~$i$.
In particular, if $(\ve_i)$ is a $\dd_g$-orthonormal basis then $[\vgrad f] =[df]^T$.

With duality notations, $\vgrad f=\sumin a^i\ve_i$ and \eref{eqdfgf01} gives $a^i=\sumjn g_{ij} {\pa f \over \pa x^j}$: The Einstein convention is \textslbf{not} satisfied (the index~$j$ is twice bottom), which is expected since the definition of $\vgrad_g f$ depends on a subjective choice (unit of length). In comparison, $df = \sumin {\pa f \over \pa x^i}dx^i$ satisfies the Einstein convention (a differential is objective).

\medskip
\noindent
{\bf Mind the notations:} The gradient $\vgrad_g f\eqnote \vgrad f$ depends on~$\dd_g$, \cf~\eref{eqdfgf0}-\eref{eqdfgf01}, while
$(\vu.\vgrad) f$ does not (only depends on a basis), \cf~\eref{eqconv03} (historical notations...).
%However in thermodynamics, \eg\ for the internal energy~$U$ where no inner dot product makes sense, you don't use the gradient notation $(\vu.\vgrad) f$: you use $df.\vu$, \eg\  $dU.\vX = {\pa U \over \pa T}(\vX)\,dT + {\pa U \over \pa T}(\vX)\,dP$ where $\vX=(T,P)$.)

%%%%%%%%%%%%%%%%%%%%%%%%%%%%%%%%%%%%%%%%%%%%%%%%%%%%%%%%%%%%%%%%%%%%%%%%%%%%%%%%%%%

\subsubsection{Vector valued functions}

For vector valued functions $\vf:\Omega\rar\vRRm$, the above steps apply to the components $f_i$ of~$\vf$
relative to a basis $(\vb_i)$ in~$\vRRm$... But, depending on the book you read:

%11- Non ambiguous: For the differential $d\vf$ there is just one ``Jacobian matrix'' relative to a given basis, \cf~\eref{eqconv04},

1- Ambiguous: $d\vf$, the differential of~$\vf$, is unfortunately also sometimes called the ``gradient matrix'' (although no Euclidean dot product is required).

2- Ambiguous: It could mean the differential...
or the Jacobian matrix... or its transposed... because an orthonormal basis relative to an imposed Euclidean dot product is chosen (which one?) and then $[\vgrad f_i] = [df_i]^T$... And calculations confuses $[.]$ and~$[.]^T$...
% And the use of some Euclidean dot product (which one?) may be required... or not... 

3- Non ambiguous: In the objective framework of this manuscript, we will use the differential $d\vf$ (objective) to begin with; And only after an explicit choice of bases~$(\ve_i)$ for quantitative purposes, the Jacobian matrix, which is 
$[df]_{|\ve}$, will be used.

\comment{
In particular, if $(\ve_i)$ is a $\dd_g$-orthonormal basis, then $\vgrad_g w^k = \sumin {\pa w^k\over\pa x^i}\ve_i$, and for all $\vu=\sumin u^i\ve_i\in\RRnt$, we get
\be
\label{eqdfgf2}
\eqalign{
d\vw.\vu 
= & \sumijn {\pa w^i \over \pa x^j} u^j\,\ve_i
%= \sumin  (\sumjn  u^j{\pa \over \pa x^j})\,w^i\,\ve_i
= (\sumjn  u^j{\pa \over \pa x^j})(\sumin  \,w^i\,\ve_i)
 \eqnote  (\vu.\vgrad) \vw.
}
\ee
}

\debexe
A Euclidean framework being chosen, prove:
$
(\vv.\vgrad)\vv = \demi \vgrad (||\vv||^2) + \vrot\vv\wedge\vv.
$

\debrep
Euclidean basis $(\vE_i)$,
Euclidean dot product $\dd_g\eqnote \dd$, associated norm $||.||_g\eqnote ||.||$.
Thus $\vv=\sumin v_i\vE_i$ gives $\ds||\vv||^2=\sum_i v_i^2$, thus
$\ds{\pa||\vv||^2\over\pa x_k}=\sum_i 2v_i{\pa v_i\over\pa x_k}$, for any $k=1,2,3$.
And, the first component of~$\vrot\vv$ is $\ds(\vrot\vv)_1={\pa v_3\over\pa x_2}-{\pa v_2\over\pa x_3}$, idem for $(\vrot\vv)_2$ and $(\vrot\vv)_3$
(circular permutation). Thus (first component)
$\ds (\vrot\vv\wedge\vv)_1
=({\pa v_1\over\pa x_3}-{\pa v_3\over\pa x_1})v_3
-({\pa v_2\over\pa x_1}-{\pa v_1\over\pa x_2})v_2$, idem for $(\vrot\vv\wedge\vv)_2$
and $(\vrot\vv\wedge\vv)_2$.
Thus $(\demi \vgrad (||\vv||^2) + \vrot\vv\wedge\vv)_1
=v_1{\pa v_1\over\pa x_1} + v_2{\pa v_2\over\pa x_1} + v_3{\pa v_3\over\pa x_1}
+ {\pa v_1\over\pa x_3}v_3 - {\pa v_3\over\pa x_1}v_3
-{\pa v_2\over\pa x_1}v_2 + {\pa v_1\over\pa x_2}v_2
=v_1{\pa v_1\over\pa x_1}+v_2{\pa v_1\over\pa x_2}+v_3{\pa v_1\over\pa x_3}
=(\vv.\vgrad)v_1$. Idem for the other components.
\finrep
\finexe

%%%%%%%%%%%%%%%%%%%%%%%%%%%%%%%%%%%%%%%%%%%%%%%%%%%%%%%%%%%%%%%%%%%%%%%%%%%%%%%%%%%

\subsection{Streamline (current line)}

%%%%%%%%%%%%%%%%%%%%%%%%%%%%%%%%%%%%%%%%%%%%%%%%%%%%%%%%%%%%%%%%%%%%%%%%%%%%%%%%%%%

%\subsubsection{Lignes de courant}

Fix $t\in \RR$, and consider the photo $\Omega_t = \tPhi_t(\Obj)$. 
Let $\pt\in\Omegat$, $\eps>0$, and consider the spatial curve in~$\Omegat$ at~$\pt$ defined by:
\be
\label{eqcpt}
c_\pt : 
\left\{\eqalign{
]-\eps,\eps[ & \rar \Omega_t \cr
s & \rar q = c_\pt(s) \cr
}\right\} \qst %q(0) = 
c_\pt(0)=\pt.
\ee
So $s$ is a curvilinear spatial coordinate (dimension of a length), and the graph of $c_\pt$ is drawn in the photo $\Omegat$ at~$t$.
%and $\eps$ is small enough for $\Im(c_\pt)$ to be in~$\Omegat$.

\debdef
$\vv: (t,p)\rar \vv(t,p)$ being the Eulerian velocity field of~$\Obj$,
a streamline through a point $\pt\in\Omegat$ is a  (parametric) spatial curve $c_\pt$ solution of the differential equation
\be
\label{eqlc1}
{dc_\pt\over ds}(s)=\vv_t(c_\pt(s)) \qwith c_\pt(0)=\pt.
\ee
%\ie\ ${dc_\pt\over ds}(s)=\vv(t,c_\pt(s))$ with $c_\pt(0)=\pt$.
And $\Im(c_\pt)$ is the geometric associated streamline ($ \subset\Omegat$).
\findef

NB: \eref{eqlc1} cannot be confused with~\eref{eqed00}:
In~\eref{eqed00} the variable is the time variable~$t$,
while in~\eref{eqlc1} the variable is the space variable~$s$. % (its solution is not related to a trajectory in general).

\mn
{\bf Usual notation:}
If an origin $\calO$ is chosen at~$t$ by an observer and $\vx(s) := \ora{\calO c_\pt(s)} $ %= \ora{\calO q(s)}$
, then~\eref{eqlc1} is written
\be
\label{eqlc1b}
%{dq \over ds}(s)=\vv_t(q(s))\qwith q(0)=\pt, \qor 
{d\vx \over ds}(s)=\vv_t(\vx(s)) \qwith\vx(0)=\ora{O\pt}.
\ee
Moreover, with a Cartesian basis $(\ve_i))$ chosen at~$t$ by the observer, with
$\vx(s) = \sumin x_i(s)\ve_i$ we get
${d\vx \over ds}(s) = \sumin {dx_i \over ds}(s)\ve_i$, and \eref{eqlc1b} reads as the differential system of $n$ equations in~$\vRRn$
\be
\label{eqlignec1}
\forall i=1,...,n,\quad
{dx_i\over ds}(s) = v_i(t,x_1(s),...,x_n(s)) \qwith x_i(0)=(\ora{O\pt})_i
\ee
(the $n$ functions $x_i : s \rar x_i(s)$ are the unknown).
Also written
\be
\label{eqlignec2}
{dx_1\over v_1}=...={dx_n\over v_n}=ds,
\ee
which means: It is the differential system~\eref{eqlignec1} of $n$ equations and $n$ unknowns which must be solved.
% (at each~$t$), that is, ${dx_i\over ds}(s)=v_i(x_1(s),...,x_n(s))$ with $x_i(0)=\ora{O\pt}_i$, where $c_\pt(s) = O + \sumin x_i(s)\ve_i$ and $\vv = \sumin v_i\ve_i$.

(With duality notations, 
${dx^i \over ds}(s) = v^i(t,x^1(s),...,x^n(s))$ and $ x^i(0)=(\ora{O\pt})^i$ for all~$i$.)

%%%%%%%%%%%%%%%%%%%%%%%%%%%%%%%%%%%%%%%%%%%%%%%%%%%%%%%%%%%%%%%%%%%%%%%%%%%%%%%%%%%

\subsection{Material time derivative (dériv\'ees particulaires)}

%%%%%%%%%%%%%%%%%%%%%%%%%%%%%%%%%%%%%%%%%%%%%%%%%%%%%%%%%%%%%%%%%%%%%%%%%%%%%%%%%%%

\subsubsection{Usual definition}

Goal: To compute the variations of a Eulerian function~$\eul$ along the trajectory $\tPhiPobj$ of a particle~$\Pobj$
(\eg\ the temperature of a particle along its trajectory).
%(\eg\ $\eul(t,p(t)=$ the temperature $\theta(t,p(t))$ of~$\Pobj$). 
So consider the function $g_\Pobj$ giving the values of~$\eul$ relative to a $\Pobj$ along its trajectory:
%Consider one particle $\Pobj\in\Obj$ and consider a regular Eulerian map~$\eul$, \cf~\eref{eqdeffspa2 0}.And let
\be
\label{eqdefdfdt00}
g_\Pobj(t) :=  \eul(t,p(t)) \qwhen p(t) := \tPhiPobj(t). %\eul(t,\tPhiPobj(t)) \eqnote
\ee

\debdef
The Material time derivative of $\eul$ at $(t,p(t))$ is $g_\Pobj{}'(t)\eqnote {D\eul \over Dt}(t,p(t))$.
\findef

So:
\be
\label{eqdmat}
{D\eul \over Dt}(t,p(t)) \eqdef  g_\Pobj{}'(t) \quad (= \lim_{h\rar 0} {\eul(t{+}h,p(t{+}h)) - \eul(t,p(t)) \over h}).
\ee

Since $g_\Pobj(t) := \eul(t,\tPhiPobj(t))$ we get
$
g_\Pobj{}'(t)
=  {\pa \eul\over\pa t}(t,\tPhiPobj(t)) + d \eul(t,\tPhiPobj(t)).\tPhiPobj'(t)
%\eqnote {D\eul \over Dt}(t,\tPhiPobj(t))
$, thus, having $\tPhiPobj'(t) = \vv(t,p(t))$ (Eulerian velocity),
${D\eul \over Dt}(t,p(t)) = {\pa \eul\over\pa t}(t,p(t)) + d \eul(t,p(t)).\vv(t,p(t))$:
%in $\bigC= \bigcup_{t\in [t_1,t_2]}(\{t\}\times \Omegat)$,
\be
\label{eqdefdfdt2}
\boxed{ {D\eul \over Dt} \eqdef {\pa \eul\over\pa t} + d \eul.\vv }.
\ee

\debexe
Prove, if $\eul$ is $C^2$:
\be
\label{eqdfddtd}
\eqalign{
{D^2 \eul\over D t^2}
= & {\pa^2 \eul \over \pa t^2} + 2d{\pa \eul \over \pa t}.\vv + d\eul.{\pa \vv \over \pa t}
+ d^2\eul(\vv,\vv) + d\eul.d\vv.\vv\cr
%= & {\pa^2 \eul\over \pa t^2} + d{\pa \eul\over \pa t}.\vv + d^2\eul(\vv,\vv) + d\eul.{D\vv \over Dt}. \cr
}
\ee

\debrep
$$
\eqalign{
{D^2 \eul \over D t^2}
= {D  {D \eul \over D t}\over D t}
= g_\Pobj''(t)
= & {\pa ({\pa \eul\over \pa t} + d\eul.\vv) \over \pa t} + d( {\pa \eul\over \pa t} + d\eul.\vv).\vv \cr
= & {\pa^2 \eul \over \pa t^2} + {\pa (d\eul) \over \pa t}.\vv + d\eul.{\pa \vv \over \pa t}
+  d{\pa \eul \over \pa t}.\vv + d^2\eul(\vv,\vv) + d\eul.d\vv.\vv, \cr
}
$$
and $\eul$~$C^2$ gives ${\pa \over \pa t}\circ d = d\circ{\pa \over \pa t}$ (Schwarz theorem),
hence~\eref{eqdfddtd}. %(Or use~\eref{eqdfddtd2} and~\eref{eqdfddtd0}.)
\finrep
\finexe

\debexe
Prove, if $\eul$ is $C^2$, for any vector field $\vw$,
\be
\label{eqdfddtd0}
{D (d\eul.\vw) \over Dt}
= d{\pa \eul \over \pa t}.\vw + d\eul.{\pa \vw \over \pa t} + d^2\eul(\vv,\vw) + d\eul.d\vw.\vv.
\ee

\debrep
$\ds {D (d\eul.\vw) \over D t}
= {\pa (d\eul.\vw) \over \pa t} + d(d\eul.\vw).\vv
= {\pa d\eul \over \pa t}.\vw + d\eul.{\pa \vw \over \pa t} + (d(d\eul).\vv).\vw + d\eul.d\vw.\vv
$. And the Schwarz theorem gives ${\pa (d \eul) \over \pa t} = d ({\pa \eul\over \pa t})$ since $\eul\in C^2$.
Hence~\eref{eqdfddtd0}.
\finrep
\finexe

%%%%%%%%%%%%%%%%%%%%%%%%%%%%%%%%%%%%%%%%%%%%%%%%%%%%%%%%%%%%%%%%%%%%%%%%%%%%%%%%%%%

\subsubsection{Remark: About notations}

\leavevmode

$\bullet$ The notation ${d\over dt}$ (lowercase letters) concerns a function of one variable,
\eg\ ${dg_\Pobj\over dt}(t):=g_\Pobj{}'(t) := \lim_{h\rar 0} {g_\Pobj(t{+}h)) - g_\Pobj(t) \over h}$;

$\bullet$ The notation ${\pa \over \pa t}$ %(curved~$d$) 
concerns a function with more than one variable,
\eg\ ${\pa \eul \over \pa t}(t,p)= \lim_{h\rar 0} {\eul(t{+}h,p) - \eul(t,p) \over h}$;

$\bullet$ The notation ${D\over Dt}$ (capital letters) concerns a Eulerian function %(function of more than one variable) 
differentiated along a motion, \cf~\eref{eqdmat}.
%that is ${D\eul \over Dt}(t,p(t))  := \lim_{h\rar 0} {\eul(t{+}h,p(t{+}h)) - \eul(t,p(t)) \over h}$ when $p(\tau)=\tPhiPobj(\tau)$.

$\bullet$ Other notations, often practical but might be ambiguous if composed functions are considered:
\be
\label{eqdefdfdt1}
{d\eul(t,p(t)) \over dt} := {D\eul \over Dt}(t,p(t))
, \qand
{d\eul(t,p(t)) \over dt}_{|t=\tz} := {D\eul \over Dt}(\tz,p(\tz)) .
\ee
%(This improper notation is problematic if composite functions are considered.)

\comment{
%%%%%%%%%%%%%%%%%%%%%%%%%%%%%%%%%%%%%%%%%%%%%%%%%%%%%%%%%%%%%%%%%%%%%%%%%%%%%%%%%%%

\subsubsection{Quantification}

Let $(\ve_1,...,\ve_n)$ be a Cartesian basis and $(dx_i)$ its dual basis.
Then
\be
\label{eqdefdfdt12}
D\eul = {\pa \eul \over \pa t}dt + d\eul = {\pa \eul \over \pa t}dt + \sumin {\pa \eul \over \pa x_i}\,dx_i.
\ee
Unmissable in thermodynamics.
With duality notations: $D\eul = {\pa \eul \over \pa t}dt + \sumin {\pa \eul \over \pa x^i}\,dx^i$.
}

%%%%%%%%%%%%%%%%%%%%%%%%%%%%%%%%%%%%%%%%%%%%%%%%%%%%%%%%%%%%%%%%%%%%%%%%%%%%%%%%%%%

\subsubsection{Definition bis:  Time-space definition}
\label{secDET}

Consider a $C^1$ time-space function $f:(t,p)\in \RR\times \RRn \rar f(t,p)$ where $t=$ time and $p=$ space.

\debdef
The differential of $f: (t,p) \in \RR^{n+1} \rar f(t,p)$ considered as a function on the Cartesian (time$\times$space) product $\RR\times \RRn$ is called the ``total differential'', or ``total derivative'', and is written $Df$
(here time and space are of a different nature).
\findef

So if $p_+=(t,p)\in \RR\times \RRn$ and $\vw_+=(w_0,\vw)\in\vec\RR\times \vRRn$ (time$\times$space) then,
by definition of a differential,
\be
Df(p_+).\vw_+
:= \lim_{h\rar0}{f (p_++h\vw_+) - f(p_+) \over h},
\ee
\ie
\be
\label{eqDxi00}
Df(t,p).(w_0,\vw)
:= \lim_{h\rar0}{f (t{+}hw_0,p{+}h\vw) - f(t,p) \over h}.
\ee

Thus
\be
\label{eqDxi01}
Df(t,p) = {\pa f \over \pa t}(t,p)\, dt + df(t,p). % \quad \hbox{(called the total differential)}.
\ee

\medskip
\noindent
{\bf Along a trajectory $\tPhiPobj: t \rar p(t)=\tPhiPobj(t)$ with $f=\eul$ a Eulerian function:}
Consider the time-space trajectory %(the graph of~$\tPhi$)
\be
\label{eqDxi1}
\tilde\Psi_\Pobj :
\left\{\eqalign{
[t_1,t_2] & \rar \RR\times \RRn \cr
t & \rar \tilde\Psi_\Pobj(t) \eqdef (t,\tPhiPobj(t)) \quad ( = (t,p(t))).
}\right.
\ee
(So $\Im(\tilde\Psi_\Pobj) = $ graph$(\tPhiPobj)$.)
The tangent vector to this curve at~$t$ is %We get, with $\tPhiPobj{}'(t)=\vv(t,p(t))$ the Eulerian velocity,
\be
\tilde\Psi_\Pobj{}'(t)
= (1,\tPhiPobj{}'(t))
= (1,\vv(t,p(t)) \in \vec\RR\times \vRRn
\ee
where $\vv(t,p(t))$ the Eulerian velocity at $p_+ = (t,p(t))$.
And \eref{eqdefdfdt00} reads
\be
g_\Pobj(t) = (\eul \circ \tilde\Psi_\Pobj)(t)=\eul(\tPsi_\Pobj(t)),
\ee
thus
\be
g_\Pobj'(t)=D\eul(\tPsi(t)).\tilde\Psi_\Pobj{}'(t).
\ee
And we recover~\eref{eqdefdfdt2}:
$
g_\Pobj'(t) %=D\eul(\tPsi(t)).\tilde\Psi_\Pobj{}'(t)
\mope^{\eref{eqDxi01}} {\pa \eul\over\pa t}(t,p(t)).1 + d \eul(t,p(t)).\vv(t,p(t))
\eqnote {D\eul \over Dt}(t,p(t)):
$
The material time derivative is the ``total derivative'' $D\eul$
along the time-space trajectory~$\tilde\Psi_\Pobj$.

\comment{
\medskip
\noindent
{\bf Quantification:} If $\vec1$ is a (time) basis in~$\vec\RR$ with dual basis $dt$, and $(\ve_1,...,\ve_n)$ is a Cartesian basis in~$\vRRn$ with dual basis $(dx_1,...,dx_n)$, then we recover~\eref{eqdefdfdt12}.
}

\comment{
%%%%%%%%%%%%%%%%%%%%%%%%%%%%%%%%%%%%%%%%%%%%%%%%%%%%%%%%%%%%%%%%%%%%%%%%%%%%%%%%%%%

\subsubsection{Material time derivative and primitive}

\eref{eqdefdfdt0} and
$\int_{\Im(g_\Pobj)} dg_\Pobj = \int_{t_1}^{t_2} g_\Pobj{}'(t)\,dt = g_\Pobj({t_2}) - g_\Pobj({t_1})$
%$g_\Pobj{}'(t) = {D\eul \over Dt}(t,p(t))$,
give
%(le travail d'une forme différentielle exacte le long d'une courbe joignant deux points ne dépend pas du chemin suivi reliant ces deux points),
\be
\int_{t=t_1}^{t_2} {D\eul\over D t}(t,p(t)) \,dt
= \eul(t_2,p(t_2)) - \eul(t_1,p(t_1)) = \int_{(t_1,p(t_1))}^{(t_2,p(t_2))} D\eul.
\ee
}

%%%%%%%%%%%%%%%%%%%%%%%%%%%%%%%%%%%%%%%%%%%%%%%%%%%%%%%%%%%%%%%%%%%%%%%%%%%%%%%%%%%

\subsubsection{The material time derivative is a derivation}

\debprop
All the functions are Eulerian and supposed $C^1$.

$\bullet$ Linearity:
\be
\label{eqdpaw20}
{D(\eul_1+ \lambda \eul_2) \over Dt} = {D\eul_1 \over Dt} +\lambda {D\eul_2 \over Dt}.
\ee

$\bullet$ Product rules:
If $\eul_1,\eul_2$ are scalar valued functions then
\be
\label{eqdpaw21}
{D(\eul_1\eul_2) \over Dt} = {D\eul_1 \over Dt}\,\eul_2 + \eul_1\,{D\eul_2 \over Dt}.
\ee
And if $\vw$ is a vector field and $T$ a compatible tensor (so $T.\vw$ is meaningful) then
\be
\label{eqdpaw2}
{D(T.\vw) \over Dt} = {DT \over Dt}.\vw + T.{D\vw \over Dt}.
\ee
\finprop

\debdem
Let $i=1,2$, and $g_i$ defined by $g_i(t) := \eul_i(t,p(t))$ where $p(t) = \tPhiPobj(t)$.

$\bullet$ $(g_1 + \lambda g_2)' = g_1' + \lambda g_2'$ gives~\eref{eqdpaw20}.

$\bullet$ On the one hand ${D(T.\vw) \over Dt}
= {\pa (T.\vw) \over \pa t} + d(T.\vw).\vv
= {\pa T \over \pa t}.\vw + T. {\pa \vw \over \pa t} + (dT.\vv).\vw + T.(d\vw.\vv)
$,
and on the other hand
${DT \over Dt}.\vw + T.{D\vw \over Dt}
= ({\pa T\over \pa t} + dT.\vv).\vw + T.({\pa \vw\over \pa t} + d\vw.\vv)$. Thus~\eref{eqdpaw21}-\eref{eqdpaw2}.
\findem

%%%%%%%%%%%%%%%%%%%%%%%%%%%%%%%%%%%%%%%%%%%%%%%%%%%%%%%%%%%%%%%%%%%%%%%%%%%%%%%%%%%

\subsubsection{Commutativity issue}

The Schwarz theorem that, when $\eul$ is~$C^2$, the derivatives ${\pa\eul \over \pa t}$ and $d\eul$ commute.
% since they concern the independent variables $t$ and~$p$:
%${\pa\eul \over \pa t}(t,p) = \lim_{h\rar0} {\eul(t{+}h,p) - \eul(t,p) \over h}$ (here $p$ is fixed), and $d\eul(t,p).\vu = \lim_{h\rar0} {\eul(t,p{+}h\vu) - \eul(t,p) \over h}$ (here $t$ is fixed). 
But

\debprop
The material time derivative ${D\over Dt}$ does not commute with the temporal derivation ${\pa \over \pa t}$
or with the spatial derivation $d$: We have
${\pa ({D\eul \over Dt}) \over \pa t} \ne {D ({\pa \eul\over \pa t}) \over Dt}$
and $d ({D\eul \over Dt}) \ne {D (d\eul) \over Dt}$ in general (because the variables $t$ and $p$ are not independent along a trajectory). In facts:
\be
\label{eqdfddtd2}
\left.\eqalign{
{\pa ({D\eul \over Dt}) \over \pa t}
= & {D ({\pa \eul\over \pa t}) \over Dt} + d\eul.{\pa \vv\over \pa t} \cr
= & {\pa^2 \eul\over \pa t^2} +  d{\pa \eul\over \pa t}.\vv  + d\eul.{\pa \vv\over \pa t}
}\right\},
\qand
\left\{\eqalign{
d ({D\eul \over Dt})
= & {D (d\eul) \over Dt}  + d\eul.d\vv \cr
= & {\pa (d\eul)\over \pa t} +  d^2\eul.\vv  + d\eul.d\vv
}\right.
% = {\pa (d\eul) \over \pa t} + d^2\eul.\vv + d\eul.d\vv. \cr
\ee
\finprop

\debdem
$\ds
{\pa {D \eul \over D t} \over \pa t}
= {\pa ({\pa \eul \over \pa t} + d\eul.\vv) \over \pa t}
= {\pa ({\pa \eul \over \pa t}) \over \pa t} + {\pa  d\eul \over \pa t}.\vv + d\eul.{\pa  \vv \over \pa t}
\mathop{=}^{\rm Schwarz} {\pa ({\pa \eul \over \pa t})\over \pa t} +  d{\pa \eul\over \pa t}.\vv + d\eul.{\pa \vv\over \pa t}
$, thus~\eref{eqdfddtd2}$_1$.

$\ds
d {D \eul \over D t}
= d({\pa \eul \over \pa t} + d\eul.\vv)
%= d({\pa \eul \over \pa t}) + d(d\eul.\vv)
= {\pa (d\eul) \over \pa t} + d(d\eul).\vv + d\eul.d\vv
$,
thus~\eref{eqdfddtd2}$_2$.

So $\ds {\pa ({D\eul \over Dt}) \over \pa t} \ne {D ({\pa \eul\over \pa t}) \over Dt}$
and $\ds d ({D\eul \over Dt}) \ne {D (d\eul) \over Dt}$
in general.
\findem

\debexe
Prove~\eref{eqdfddtd2} with components.

\debrep
$(\ve_i)$ is a Cartesian basis.
${\pa {D \eul \over D t} \over \pa t}
= {\pa ({\pa \eul \over \pa t} + \sum_i {\pa \eul \over \pa x^i}.v^i ) \over \pa t}
= {\pa^2 \eul \over \pa t^2} + \sum_i {\pa^2 \eul \over \pa t \pa x^i}.v^i
 + \sum_i {\pa \eul \over \pa x^i}.{\pa v^i \over \pa t}
= {\pa^2 \eul \over \pa t^2} + \sum_i {\pa^2 \eul \over \pa t \pa x^i}.v^i
 + d\eul.{\pa \vv \over \pa t}
$.
And
${D ({\pa \eul\over \pa t}) \over Dt}
= {\pa^2 \eul \over \pa t^2}+ \sum_i{\pa {\pa \eul\over \pa t} \over \pa x^i}.v^i
= {\pa^2 \eul \over \pa t^2}+ \sum_i{\pa^2 \eul\over \pa t \pa x^i}.v^i
$.

And $d({D\eul \over Dt}).\vw
= \sum_j {\pa {D\eul \over Dt} \over \pa x^j} w^j
= \sum_j {\pa ({\pa\eul \over \pa t}+ \sum_i {\pa\eul \over \pa x^i} v^i)  \over \pa x^j} w^j
= \sum_j {\pa^2\eul \over \pa t \pa x^j} w^j + \sum_{ij} {\pa^2\eul \over \pa x^i \pa x^j} v^i w^j
+ \sum_{ij} {\pa\eul \over \pa x^i} {\pa v^i \over \pa x^j} w^j
= \sum_j {\pa^2\eul \over \pa t \pa x^j} w^j +d^2\eul(\vv,\vw) + d\eul.d\vv.\vw
$.
And ${D (d\eul) \over Dt}.\vw
= ({\pa (d\eul) \over \pa t} +d(d\eul).\vv).\vw
= {\pa (d\eul) \over \pa t}.\vw + d^2\eul(\vv,\vw)
= \sum_i {\pa^2 \eul \over \pa x^i \pa t}\,w^i  + d^2\eul(\vv,\vw)
$.
Thus $d ({D\eul \over Dt}).\vw
= {D (d\eul) \over Dt}.\vw  + d\eul.d\vv.\vw$ for all~$\vw$.
\finrep
\finexe

%%%%%%%%%%%%%%%%%%%%%%%%%%%%%%%%%%%%%%%%%%%%%%%%%%%%%%%%%%%%%%%%%%%%%%%%%%%%%%%%%%%

\subsection{Eulerian acceleration}

\debdef
In short: If $\tPhiPobj$ is $C^2$, then the Eulerian acceleration of the particle $\Pobj$ which is at~$t$ at $\pt = \tPhi(t,\Pobj)$ is
\be
\label{eqvaeul0}
\vgamma(t,\pt) \eqdef \tPhiPobj{}''(t) \eqnote {\pa^2 \tPhi\over\pa t^2}(t,\Pobj).
\ee
In details: as in~\eref{eqdeffspa2}, the Eulerian acceleration (vector) field $\widehat\vgamma$ is defined with~\eref{eqvaeul0} by
\be
\label{eqvaeul00}
\widehat\vgamma(t,\pt) = ((t,\pt),\vgamma(t,\pt)) \in \bigC\times\RRnt \quad\hbox{(pointed vector)}.
\ee
\findef

\debprop
%Let $\vv$ be the Eulerian velocity, that is, $\vv(t,p(t)) = \tPhiPobj{}'(t)$ when $p(t)=\tPhi(t,\Pobj)$, \cf~\eref{eqdefve}. Then $\vgamma(t,p) = {D\vv \over Dt}(t,p) = {\pa \vv\over \pa t}(t,p) + d\vv(t,p).\vv(t,p)$ for all $(t,p)\in\bigC$, \ie, in~$\bigC$,
\be
\label{eqvaeul}
\boxed{\vgamma = {D\vv \over Dt} = {\pa \vv\over \pa t} + d\vv.\vv}.
\ee
%NB: $\vgamma$ is a non linear function of~$\vv$. 
And if $\vv$ is~$C^2$ then
\be
\label{eqdfddtd2v}
\eqalign{
%& {\pa \vgamma \over \pa t} = {D ({\pa \vv\over \pa t}) \over D t} + d\vv.{\pa \vv \over \pa t}, \cr
& d \vgamma
= {\pa (d\vv) \over \pa t} + d^2\vv.\vv + d\vv.d\vv
=  {D (d\vv) \over Dt}  + d\vv.d\vv.
}
\ee
\finprop

\debdem
With $g(t)=\vv(t,p(t))=\tPhiPobj{}'(t)$ and~\eref{eqdefdfdt2} we get
$\vgamma(t,p(t)) = g'(t) = {D\vv\over Dt}(t,p(t))$.
And $\vv$ being~$C^2$, the Schwarz theorem gives $d{\pa \vv \over \pa t} = {\pa (d\vv) \over \pa t}$.
\findem

\comment{
\debexe
Prove:
\be
\label{eqdfddtd0}
{D (d\eul.\vw) \over Dt}
= d{\pa \eul \over \pa t}.\vw + d\eul.{\pa \vw \over \pa t} + d^2\eul(\vv,\vw) + d\eul.d\vw.\vv.
\ee
And with $g_\Pobj{}''(t) = (g_\Pobj{}')'(t)
= {D({D \eul\over D t})\over D t}(t,p(t)) \eqnote {D^2 \eul\over D t^2}(t,p(t))$, prove:
\be
\label{eqdfddtd}
\eqalign{
{D^2 \eul\over D t^2}
= & {\pa^2 \eul \over \pa t^2} + 2d{\pa \eul \over \pa t}.\vv + d\eul.{\pa \vv \over \pa t}
+ d^2\eul(\vv,\vv) + d\eul.d\vv.\vv\cr
= & {\pa^2 \eul\over \pa t^2} + d{\pa \eul\over \pa t}.\vv + d^2\eul(\vv,\vv) + d\eul.{D\vv \over Dt}. \cr
}
\ee

\debrep
$\ds {D (d\eul.\vw) \over D t}
= {\pa (d\eul.\vw) \over \pa t} + d(d\eul.\vw).\vv
= {\pa d\eul \over \pa t}.\vw + d\eul.{\pa \vw \over \pa t} + (d(d\eul).\vv).\vw + d\eul.d\vw.\vv
$. And the Schwarz theorem gives ${\pa (d \eul) \over \pa t} = d ({\pa \eul\over \pa t})$ since $\eul\in C^2$.
Hence~\eref{eqdfddtd0}.
Then,
$$
\eqalign{
{D^2 \eul \over D t^2}
= {D  {D \eul \over D t}\over D t}
= & {\pa ({\pa \eul\over \pa t} + d\eul.\vv) \over \pa t} + d( {\pa \eul\over \pa t} + d\eul.\vv).\vv \cr
= & {\pa^2 \eul \over \pa t^2} + d{\pa \eul \over \pa t}.\vv + d\eul.{\pa \vv \over \pa t}
+  d{\pa \eul \over \pa t}.\vv + d^2\eul(\vv,\vv) + d\eul.d\vv.\vv, \cr
}
$$
hence~\eref{eqdfddtd}. (Or use~\eref{eqdfddtd2} and~\eref{eqdfddtd0}.)
\finrep
\finexe
}

\debdef
If an observer chooses a Euclidean dot product $\dd_g$ (based on a foot, a metre...), the associated norm being~$||.||_g$,
then the length $||\vgamma(t,\pt)||_g$ is the (scalar) acceleration of~$\Pobj$.
\findef

\comment{
Quantification: If a basis $(\ve_i)$ is chosen, then \eref{eqconv03} gives
\be
\vgamma = {\pa \vv\over \pa t} + (\vv.\vgrad).\vv \quad (={\pa \vv\over \pa t}+d\vv.\vv).
\ee
}

%%%%%%%%%%%%%%%%%%%%%%%%%%%%%%%%%%%%%%%%%%%%%%%%%%%%%%%%%%%%%%%%%%%%%%%%%%%%%%%%%%%%

\subsection{Time Taylor expansion of~$\tPhi$}

Let $\Pobj\in\Obj$ and $t\in]t_1,t_2[$. Suppose $\tPhiPobj \in C^2(]t_1,t_2[;\RRn)$. %Let $p(\tau)=\tPhiPobj(\tau)$.
Its second-order (time) Taylor expansion of~$\tPhiPobj$ is, in the vicinity of a $t\in ]t_1,t_2[$,
\be
\label{eqdl20}
\eqalign{
\tPhiPobj(\tau)
= & \tPhiPobj(t) + (\tau{-}t) \tPhiPobj'(t) + {(\tau{-}t)^2 \over 2} \tPhiPobj''(t) + o((\tau{-}t)^2),\cr
%= & \tPhiPobj(t) + (\tau{-}t) \vv(t,\pt) + {(\tau{-}t)^2 \over 2} \vgamma(t,\pt) + o((\tau{-}t)^2), \cr
}
\ee
\ie
\be
\label{eqdl20E}
p(\tau) = p(t) + (\tau{-}t) \vv(t,p(t)) + {(\tau{-}t)^2 \over 2} \vgamma(t,p(t)) + o((\tau{-}t)^2).
\ee
%(NB: $\vgamma$ is non linear in~$\vv$, \cf~\eref{eqvaeul}, thus~\eref{eqdl20E} is non linear in~$\vv$.)

%%%%%%%%%%%%%%%%%%%%%%%%%%%%%%%%%%%%%%%%%%%%%%%%%%%%%%%%%%%%%%%%%%%%%%%%%%%%%%%%%%%
%%%%%%%%%%%%%%%%%%%%%%%%%%%%%%%%%%%%%%%%%%%%%%%%%%%%%%%%%%%%%%%%%%%%%%%%%%%%%%%%%%%

\section{Motion from an initial configuration: Lagrangian description}
\label{secmoaic}

Instead of working on~$\Obj$, an observer may prefer to work with an initial configuration $\Omegatz=\tPhi(\tz,\Obj)$ of~$\Obj$ (essential for elasticity): This is the ``Lagrangian approach''. This Lagrangian approach is not objective: Two observers may choose two different initial (times and) configurations.

% And an observer $B$, with his unit of measurement, \eg~metre, may use the results of~$A$ who made the measurements at~$\tz$ and whose unit of measurement was \eg\ the~foot, .
% And another observer will have another initial configuration $\Omega_{\tz'}=\tPhi(\tz',\Obj)$. 
%It is then clear that the results will then be ``very observer dependent'':.

%And at~$t$ the observers have the same (qualitative) result (quantification depends on their unit of measurement).

%%%%%%%%%%%%%%%%%%%%%%%%%%%%%%%%%%%%%%%%%%%%%%%%%%%%%%%%%%%%%%%%%%%%%%%%%%%%%%%%%%%

\subsection{Initial configuration and Lagrangian ``motion''}

%%%%%%%%%%%%%%%%%%%%%%%%%%%%%%%%%%%%%%%%%%%%%%%%%%%%%%%%%%%%%%%%%%%%%%%%%%%%%%%%%%%

\subsubsection{Definition}

%Soit $\tz$ fixé (par un observateur), $\tz\in]t_1,t_2[$.
$\Obj$ is a material object, $\tPhi:[t_1,t_2[\times \Obj \rar\RRn$ is its motion, %\cf~\eref{eqdeftPhi0}.
$\Omegatau = \tPhi_\tau(\Obj)$ is its configuration at~$\tau$,
$\tz\in]t_1,t_2[$ is an ``initial time'', and $\Omegatz$ is the initial configuration for the observer who chose~$\tz$.

\debdef
The motion of~$\Obj$ relative to the initial configuration $\Omegatz = \tPhi(\tz,\Obj)$ is the function
%est la fonction $\Phitz : [t_1,t_2]\times\Omegatz$ décrivant les trajectoires des particules~$\Pobj$ à partir de~$\Omegatz$ :
\be
\label{eqdefPhi}
\Phitz :
\left\{\eqalign{
[t_1,t_2]\times\Omegatz& \rar \RRn \cr
(t, \ptz) & \mapsto \pt=\Phitz(t,\ptz) := \tPhi(t,\Pobj) \qwhen \ptz=\tPhi(\tz,\Pobj).
}\right.
\ee
So, $\pt=\Phitz(t,\ptz):=\tPhi(t,\Pobj)$ is the position at~$t$ of the particle $\Pobj$
which was at $\ptz$ at~$\tz$. In~particular $\ptz = \Phitz(\tz,\ptz):=\tPhi(\tz,\Pobj)$.
%(its initial position relative to the observer who chose~$\tz$),
\findef

Marsden and Hughes notations: Once an initial time~$\tz$ has been chosen by an observer,
then $\Phitz\eqnote\Phi$,
then $\ptz \eqnote P$ (capital letter for positions at~$\tz$)
and $\pt \eqnote p$ (lowercase letter for positions at~$t$), so
\be
\label{eqdefP}
p = \Phi(t,P) \; \in\Omegat.
\ee
(When objectivity is under concern, we need to switch back to the notations $\Phitz$, $\ptz$ and~$\pt$.)

\medskip
NB:
$\bullet$ Talking about the motion of a position~$\ptz$ is absurd: A position in~$\RRn$ does not move. Thus $\Phitz$ has no existence without the definition, at first, 
of the motion~$\tPhi$ of particles. % and of the definition of $\Omegatz:=\tPhi(\tz,\Obj)$.

$\bullet$ The domain of definition of $\Phitz$ depends on $\tz$ through $\Omegatz$:
The superscript ${}^\tz$ recalls it.
And a late observer with initial time $\tz'>\tz$ defines $\Phi^{\tz'}$
which  domain of definition is $[t_1,t_2]\times\Omega_{\tz'}$;
And $\Phi^{\tz'} \ne \Phitz$ in general because $\Omega_{\tz'} \ne \Omegatz$ in general.
%As far as objectivity is concerned, the $\tz$ in~$\Phitz$ must not be forgotten in the writing of~$\Phitz$.

$\bullet$ The following notation is also used:
\be
\Phitz(t,\ptz) = \Phi(t;\tz,\ptz).
\ee
(The couple $(\tz,\ptz)$ is ``the initial condition'', or $\tz$ and~$\ptz$ are the initial conditions, see the~\S\ on flows).

$\bullet$ If a origin $\calO\in\RRn$ is chosen by the observer, we may also use, with~\eref{eqdeftvphi},
\be
\label{eqdefPhiv}
\vx_\tz = \ora{\calO\ptz} = \vphitz(\tz,\vx_\tz) = \vX = \ora{\calO P} \qand
\vx_t = \ora{\calO\pt} = \vphitz(t,\vx_\tz) = \vx = \ora{\calO p}.
\ee

%%%%%%%%%%%%%%%%%%%%%%%%%%%%%%%%%%%%%%%%%%%%%%%%%%%%%%%%%%%%%%%%%%%%%%%%%%%%%%%%%%%

\subsubsection{Diffeomorphism between configurations}

%Let $\tz,t \in[t_1,t_2]$, $\Omegatz = \tPhi(\tz,\Obj)$ and $\Omegat = \tPhi(t,\Obj)$ (initial and actual configurations).
With~\eref{eqdefPhi}, define
\be
\label{eqPhit}
\Phitzt :
\left\{\eqalign{
\Omegatz & \rar \Omegat \cr
\ptz & \rar \pt = \Phitzt(\ptz) := \Phitz(t,\ptz)  .
%\ptz = \tPhi(\tz,\Pobj)) & \rar \pt = \Phitzt(\ptz) := \tPhi(t,\Pobj),\qie \Phitzt(\ptz) := \Phitz(t,\ptz)  .
}\right.
\ee

\mn
{\bf Hypothesis:} For all $\tz,t\in]t_1,t_2[$,  the map $\Phitzt : \Omegatz \rar \Omegat$ is a $C^k$ diffeomorphism
(a $C^k$ invertible function whose inverse is~$C^k$), where $k\in\NNs$ depends on the required regularity.

\medskip
%Then \eref{eqdefPhi} reads $\Phitz(t,\tPhi(\tz,\Pobj)) = \tPhi(t,\Pobj)$,
Thus~\eref{eqPhit} gives $\tPhi_t(\Pobj) = \Phitzt(\tPhi_\tz(\Pobj))$, true for all $\Pobj\in\Obj$,
thus $\Phitzt\circ\tPhi_\tz = \tPhi_t$, 
\ie
\be
\label{eqPhitb}
\boxed{\Phitzt := \tPhi_t\circ(\tPhi_{t_0})^{-1}}.
\ee
Thus, $\Phitztz  = I$ and $\Phi^t_\tz \circ \Phitzt
= (\tPhi_t\circ(\tPhi_{t_0})^{-1})\circ (\tPhi_{t_0}\circ(\tPhi_t)^{-1}) = I$ give
\be
\label{eqPhit2i}
\Phi^t_\tz = (\Phitzt)^{-1}.
\ee
%(NB: An Eulerian function does not need any initial time~$\tz$.)

%%%%%%%%%%%%%%%%%%%%%%%%%%%%%%%%%%%%%%%%%%%%%%%%%%%%%%%%%%%%%%%%%%%%%%%%%%%%%%%%%%%

\subsubsection{Trajectories}

Let $(\tz,\ptz)\in [t_1,t_2]\times \Omegatz$ (initial conditions) and with~\eref{eqdefPhi} define
\be
\label{eqPhiP}
\Phitzptz :
\left\{\eqalign{
[t_1,t_2] & \rar\RRn \cr
t & \mapsto  p(t) = \Phitzptz(t) := \tPhiPobj(t) = \Phitz(t,\ptz) \qwhen \ptz=\tPhiPobj(t_0).
}\right.
\ee

\debdef
$\Phitzptz$ is called the (parametric) ``trajectory of~$\ptz$'', which means: $\Phitzptz$ is the trajectory of the particle $\Pobj$ that is located at $\ptz=\tPhi(t,\Pobj)$ at~$\tz$. And the geometric ``trajectory of~$\ptz$'' is
\be
\Im(\Phitzptz)=\Phitzptz([t_1,t_2])
= \bigcup_{t\in[t_1,t_2]}\{\Phitzptz(t)\} \quad (=\Im(\tPhiPobj)).
\ee
%(Recall: A position~$\ptz$ does not move... It is $\Pobj$ that moves.)
\findef

NB: The terminology ``trajectory of~$\ptz$'' is awkward, since a position $\ptz$ does not move: It is indeed the trajectory $\tPhiPobj$ of a particle $\Pobj$ which is at $\ptz$ at~$\tz$ that must be understood.

%%%%%%%%%%%%%%%%%%%%%%%%%%%%%%%%%%%%%%%%%%%%%%%%%%%%%%%%%%%%%%%%%%%%%%%%%%%%%%%%%%%

\subsubsection{Streaklines (lignes d'émission)}

Take a film between $\tz$ and $T$ (start and end).

\debdef
Let $Q$ be a fixed point in~$\RRn$ (you see the point $Q$ on each photo that make up the film).
The streakline through~$Q$ is the set
\be
\eqalign{
E_{\tz,T}(Q)
= & \{ p \in \Omega : \exists \tau\in [\tz,T] : p = \Phi^\tau_T(Q) = (\Phi^T_\tau)^{-1}(Q)\} \cr
= & \{ p \in \Omega : \exists u\in [0,T{-}\tz] : p = \Phi^{T-u}_T(Q) = (\Phi^T_{T-u})^{-1}(Q)\} \cr
}
\ee
= the set at~$T$ of the positions (a line in~$\RRn$) of all the particles which were at $Q$ at a $\tau \in [\tz,T]$.
\findef

\debexa
Smoke comes out of a chimney.
Fix a camera nearby, choose a point $Q$ at the top of the chimney where the particles are colored, and make a film.
At~$T$ stop filming.
Then (at time~$T$) superimpose the photos in the film: The colored curve we see is the streakline.
\finexa

In other words
$
=  \bigcup_{\tau\in[\tz,T]} \{\Phi_Q^{\tau}(T)\}
=  \bigcup_{u\in[0,T{-}\tz]} \{\Phi_Q^{T{-}u}(T)\}.
$
%(It can be an empty set: depends on~$Q$.)

%%%%%%%%%%%%%%%%%%%%%%%%%%%%%%%%%%%%%%%%%%%%%%%%%%%%%%%%%%%%%%%%%%%%%%%%%%%%%%%%%%%
%%%%%%%%%%%%%%%%%%%%%%%%%%%%%%%%%%%%%%%%%%%%%%%%%%%%%%%%%%%%%%%%%%%%%%%%%%%%%%%%%%%

%\section{Lagrangian description}
%\label{seclag}

%%%%%%%%%%%%%%%%%%%%%%%%%%%%%%%%%%%%%%%%%%%%%%%%%%%%%%%%%%%%%%%%%%%%%%%%%%%%%%%%%%%

\subsection{Lagrangian variables and functions}
\label{seclag}

%%%%%%%%%%%%%%%%%%%%%%%%%%%%%%%%%%%%%%%%%%%%%%%%%%%%%%%%%%%%%%%%%%%%%%%%%%%%%%%%%%%

\subsubsection{Definition}

Consider a motion $\tPhi$, \cf~\eref{eqdeftPhi0}.
An observer chose (subjective) a $\tz \in[t_1,t_2]$ (``in the past''); So $\Omegatz = \tPhi(\tz,\Obj)$ is his initial configuration.
%and consider $\Phitz:\Omegatz \rar \RRn$ the associated Lagrangian motion, \cf~\eref{eqdefPhi}.
Let $m\in\NNs$.

\debdef
\label{defLag}
In short: A Lagrangian function, relative to~$\Obj$, $\tPhi$ and~$\tz$, is a function
\be
\label{eqdefLag0}
\Lagtz : 
\left\{\eqalign{
[t_1,t_2] \times \Omegatz & \rar \vRRm \  \hbox{ (or, more generally, some adequat set)} \cr
(t,\ptz) & \rar \Lagtz(t,\ptz),
}\right.
\ee
and $\ptz$ is called the Lagrangian variable relative to the (subjective) choice~$\tz$. 

(To compare with~\eref{eqdeffspa20}: A Eulerian function does not depend on~any~$\tz$.)
\findef

\debexa
Scalar values:
$\Lagtz(t,\ptz) = \Theta^\tz(t,\ptz) =$ temperature at~$t$ at $\pt = \Phitzt(\ptz) = \tPhi(t,\Pobj)$
of the particle $\Pobj$ that was at $\ptz$ at~$\tz$.
(So, continuing example~\ref{exaeul1}, $\Theta^\tz(t,\ptz) = \theta(t,\pt)$.)
\finexa

\debexa
Vectorial values:
$\Lagtz(t,\ptz) = \vU^\tz(t,\ptz)=$ force at~$t$ at $\pt = \Phitzt(\ptz) =  \tPhi(t,\Pobj)$
acting on the particle $\Pobj$ that was at $\ptz$ at~$\tz$.
(So, continuing example~\ref{exaeul12}, $\vU^\tz(t,\ptz) = \vu(t,\pt)$.)
\finexa

If $t$ is fixed or if $\ptz\in\Omegatz$ is fixed, then we define
\be
\Lagtzt :
\left\{\eqalign{
\Omegatz & \rar  \vRRm \hbox{ (or, more generally, some adequat set)}\quad \cr
\ptz & \rar \Lagtzt(\ptz) := \Lagtz(t,\ptz),
}\right.
\ee
\be
\Lagtzptz : 
\left\{\eqalign{
[t_1,t_2]  & \rar \vRRm \hbox{ (or, more generally, some adequat set)} \cr
t & \rar \Lagtzptz(t) := \Lagtz(t,\ptz).
}\right.
\ee

\debrem
The position $\ptz$ is also sometimes called a ``material point'', which is counter intuitive: 
$\Pobj$~(objective) is the material point, and $\ptz$ is just its spatial position at~$\tz$ (subjective); And a Eulerian variable $\pt$ is not called a ``material point'' at~$t$...
%Recall: The material particle is~$\Pobj$.

By the way, the variable~$\pt$ is also called the ``updated Lagrangian variable''...
%: Gives $\Lag^t(\tau,\pt)$ a value at $\tau$ at $\ptau{=}\Phittau(\pt))$).
\finrem

%%%%%%%%%%%%%%%%%%%%%%%%%%%%%%%%%%%%%%%%%%%%%%%%%%%%%%%%%%%%%%%%%%%%%%%%%%%%%%%%%%%

\subsubsection{A Lagrangian function is a two point tensor}
\label{secbrtpt}

%\label{remvVnoncv0}

\debdef
\label{defvVnoncv0}
In details: %With $\bigC=\bigcup_{t\in[t_1,t_2]}(\{t\} \times \Omegat)$, and 
$\Lagtz$ being defined in~\eref{eqdefLag0},
a Lagrangian function is a function
\be
\label{eqdefLag}
\tLagtz :
\left\{\eqalign{
[t_1,t_2] \times \Omegatz & \rar \bigC \times \vRRm \cr
(t,\ptz) & \rar \tLagtz(t,\ptz)  = ((t,\pt),\Lagtz(t,\ptz)) \qwhen \pt = \Phitzt(\ptz).
}\right.
\ee
\Ie\ $\tLagtz(t,\ptz)  = ((t, \Phitzt(\ptz)),\Lagtz(t,\ptz))$. (And $\vRRm$ can be replaced by some set.)
\findef

\debdef
(Marsden and Hughes~\cite{marsden-hughes}.) A Lagrangian function is a ``two point vector field'' (or more generally a ``two point tensor'') in reference to the points $\ptz\in\Omegatz$ (departure set) and $\pt\in\Omegat$ (arrival set) where the value $\Lagtz(t,\ptz)$ is considered.
\findef

\noindent
{\bf Interpretation:} \eref{eqdefLag} tells that $\Lagtz(t,\ptz)$ is \textsl{\textbf{not}} represented at~$(t,\ptz)$, but at $(t,\pt)$: That is, having
\be
\graph(\Lagtz) = \{((t,\ptz), \Lagtz(t,\ptz))
\qand
\Im(\tLagtz)=\{((t,\pt),\Lagtz(t,\ptz))\}
,
\ee
we have
\be
\label{eqLntg}
\Im(\tLagtz) \ne \graph(\Lagtz) :
\ee
So a Lagrangian function does \textsl{\textbf{not}} define a tensor in the usual sense. To compare with the Eulerian function $\eul$ which defines a tensor (in particular $\Im(\widehat\eul) = \graph(\eul)$), \cf~\eref{eqdeffspa2}. 
%This is why Marsden and Hughes called it a ``two point vector field'' (or more generally a ``two point tensor'').

%%%%%%%%%%%%%%%%%%%%%%%%%%%%%%%%%%%%%%%%%%%%%%%%%%%%%%%%%%%%%%%%%%%%%%%%%%%%%%%%%%%

\subsection{Lagrangian function associated with a Eulerian function}

%%%%%%%%%%%%%%%%%%%%%%%%%%%%%%%%%%%%%%%%%%%%%%%%%%%%%%%%%%%%%%%%%%%%%%%%%%%%%%%%%%%

\subsubsection{Definition}

Let $\tPhi$ be a motion, \cf~\eref{eqdeftPhi0}.
Let $\eul$ be a Eulerian function, \cf~\eref{eqdeffspa2}. Let $\tz\in[t_1,t_2]$. 

\debdef
The Lagrangian function $\Lagtz$ associated with the Eulerian function~$\eul$ is defined by,
for all $(t,\Pobj)\in[t_1,t_2] \times \Obj$,
\be
\Lagtz(t,\tPhi(\tz,\Pobj)) := \eul(t,\tPhi(t,\Pobj)),
\ee
\ie, for all $(t,\ptz)\in[t_1,t_2]\times \Omegatz$,
\be
\label{eqrelel}
\Lagtz(t,\ptz) := \eul(t,\pt), \qwhen \pt=\tPhi(t,\Pobj) = \Phitzt(\pt)
\ee
\ie,
$\Lagtz(t,\ptz) \eqdef \eul(t,\pt)$ when $\ptz = (\Phitzt)^{-1}(\pt)$ for all $(t,\pt)\in\bigC$.
In other words:
\be
\label{eqrelel2}
\boxed{\Lagtzt := \eul_t\circ \Phitzt}.
\ee
\findef

%(Recall: a Eulerian function $\eul$ does not depend on any initial time~$\tz$.)

%%%%%%%%%%%%%%%%%%%%%%%%%%%%%%%%%%%%%%%%%%%%%%%%%%%%%%%%%%%%%%%%%%%%%%%%%%%%%%%%%%%

\subsubsection{Remarks}

$\bullet$
If you have a Lagrangian function, then you can associate the function
\be
\label{eqrelel4}
\eul_t^\tz:=\Lagtzt \circ (\Phitzt)^{-1} % = \Lag^{\tz'}_t \circ (\Phi^{\tz'}_t)^{-1}.
\ee
which thus a priori depends on~$\tz$.
But, a Eulerian function is independent of any initial time~$\tz$.

%However, $\eul_t^\tu = \Lag^\tu_t \circ (\Phi^\tu_t)^{-1}$

\medskip
\noindent
$\bullet$ For one measurement, there is only one Eulerian function~$\eul$,
%\cf~remark~\ref{remhypNE},
while there are as many associated Lagrangian function~$\Lagtz$ as they are~$\tz$ (as many as observers):
The Lagrangian function $\Lag^{\tz'}$ of a late observer who chooses $\tz' > \tz$
is different from $\Lagtz$ since the domains of definition $\Omegatz$ and~$\Omega_{\tz'}$
are different (in general).
%However, relative to one Eulerian function~$\eul$, the functions $\Lagtz$ and $\Lag^{\tz'}$ define the same result at~$t$ at $\pt$: $\Lagtz(t,\ptz) =\eul(t,\pt)=\Lag^{\tz'}(t,\ptz')$.

%\medskip With Marsden and Hughes notations, $\tz$ being fixed once for all, let $\ptz \eqnote P$.

%%%%%%%%%%%%%%%%%%%%%%%%%%%%%%%%%%%%%%%%%%%%%%%%%%%%%%%%%%%%%%%%%%%%%%%%%%%%%%%%%%%

\subsection{Lagrangian velocity}

%%%%%%%%%%%%%%%%%%%%%%%%%%%%%%%%%%%%%%%%%%%%%%%%%%%%%%%%%%%%%%%%%%%%%%%%%%%%%%%%%%%

\subsubsection{Definition}

\debdef
In short: The Lagrangian velocity at~$t$ at $\pt = \tPhi(t,\Pobj)$ of the particle $\Pobj$ is
the function
\be
\label{eqdefVl00}
\vVtz :
\left\{\eqalign{
\RR \times \Omegatz & \rar \vRRn \cr
(t,\ptz) & \rar \vVtz(t,\ptz) := \tPhiPobj{}'(t) \qwhen \ptz = \tPhi(\tz,\Pobj).
}\right.
\ee
In details: With~\eref{eqdefVl00}, the Lagrangian velocity is the two point vector field given by
\be
\label{eqdefV99}
\widehat\vVtz(t,\ptz) :
\left\{\eqalign{
\RR \times \Omegatz & \rar \bigC\times \vRRn \cr
(t,\ptz) & \rar \widehat\vVtz(t,\ptz) := ((t,\pt), \vVtz(t,\ptz)) %= ((t,\pt), \tPhiPobj{}'(t))
, \qwhen \pt = \Phitz(t,\ptz). %=\tPhi(t,\Pobj).
}\right.
\ee
\findef

Thus $\vVtz(t,\ptz) = \tPhiPobj{}'(t) = \vv(t,\pt)$ is the velocity at~$t$ at $\pt=\tPhi(t,\Pobj)$ of the particle~$\Pobj$
which was at $\ptz=\tPhi(\ptz,\Pobj)$ at~$\tz$; And $\vVtz(t,\ptz)$ is \textsl{\textbf{not}} tangent to $\graph(\vVtz)$, \cf~\eref{eqLntg}: It is tangent to $\graph(\vv)$ at~$(t,\pt)$. %: $\vVtz$ is a two-point tensor, \cf\ definition~\ref{defvVnoncv0}.

If $t$ is fixed, or if $\ptz\in\Omegatz$ is fixed, then we define
\be
\label{eqvVvv0}
\vVtzt(\ptz) := \vVtz(t,\ptz), \qor
\vV^\tz_\ptz(t) := \vVtz(t,\ptz).
\ee
{\bf Remark:} A usual definition is given without explicit reference to a particle; It is, instead of~\eref{eqdefVl00},
\be
\label{eqdefV98}
\vVtz(t,\ptz) :=  {\pa \Phitz\over\pa t}(t,\ptz), \quad \forall (t,\ptz) \in \RR \times \Omegatz.
\ee

%%%%%%%%%%%%%%%%%%%%%%%%%%%%%%%%%%%%%%%%%%%%%%%%%%%%%%%%%%%%%%%%%%%%%%%%%%%%%%%%%%%

\subsubsection{Lagrangian velocity versus Eulerian velocity}

\eref{eqdefVl00} and~\eref{eqdefve} give (alternative definition), with $\ptau= \tPhi(\tau,\Pobj)$,
\be
\label{eqremvVnoncv}
\vVtz(t,\ptz) = \vv(t,\pt) \quad (= {\pa \Phitz\over\pa t}(t,\ptz)= \tPhiPobj{}'(t) = \hbox{velocity at $t$ at $\pt$ of $\Pobj$}).
\ee
%(Recall: $\vv$ doesn't depend on any~$\tz$.)
In other words, %\cf~\eref{eqrelel2},
\be
\label{eqremvVnoncv2}
\boxed{\vVtzt  = \vv_t \circ \Phitzt}.
\ee

%%%%%%%%%%%%%%%%%%%%%%%%%%%%%%%%%%%%%%%%%%%%%%%%%%%%%%%%%%%%%%%%%%%%%%%%%%%%%%%%%%%

\subsubsection{Relation between differentials}

For $C^2$ motions \eref{eqremvVnoncv2} gives
\be
\label{eqvVvv}
%( {\pa d\Phitz\over\pa t}(t,\ptz)=)\quad 
d\vVtzt(\ptz)=d\vv_t(\pt).d\Phitzt(\ptz)\qwhen \pt = \Phitzt(\ptz).
\ee
\Ie, with
\be
\label{eqdg}
\Ftzt = d\Phitzt \eqnote \hbox{the deformation gradient relative to~$\tz$ and~$t$} ,
\ee
%we have
\be
\label{eqvVvv2}
\boxed{d\vVtzt(\ptz)=d\vv_t(\pt).\Ftzt(\ptz)} \qwhen \pt = \Phitzt(\ptz).
\ee
Abusively written (dangerous notation: At what points, relative to what times?)
\be
d\vV = d\vv.F.
\ee

%%%%%%%%%%%%%%%%%%%%%%%%%%%%%%%%%%%%%%%%%%%%%%%%%%%%%%%%%%%%%%%%%%%%%%%%%%%%%%%%%%%

\subsubsection{Computation of $d\vv$ called $L=\protect\overbigdot F.F^{-1}$ wih Lagrangian variables}
%\subsubsection{Computation of $\ds d\vv \eqnote L=\mathop{F}^{\bullet}.F^{-1}$ from Lagrangian variables}

The Lagrangian approach can be introduced before the Eulerian approach: $\vVtz$ being given, define
\be
\label{eqLtzt0}
%\vv^\tz(t,\pt) = \vv^\tz_t(\pt) := \vVtzt(\ptz) = \vVtz(t,\ptz) ,
\vv^\tz(t,\pt)  :=  \vVtz(t,\ptz) ,
  \qwhen  \pt=\Phitzt(\ptz),
%, \qie \Ltzt := \vVtzt \circ \Phitzt^{-1}.
\ee
\cf~\eref{eqrelel4}. (\Ie\ $\vv^\tz(t,\pt):=\vVtz(t,\Phitzt^{-1}(\pt))$).
%NB: With this definition $\vv$ does depend a priori on~$\tz$ since $\vVtz$ does.
So $\vv^\tz(t,\Phitzt(\ptz)) = {\pa\Phitz \over\pa t}(t,\ptz)$, thus
%, for $C^2$ motions and with $d\Phitz \eqnote \Ftz$ and $\pt=\Phitzt(\ptz)$,
\be
\label{eqLtzt}
d\vv^\tz(t,\pt).d\Phitz(t,\ptz)
= d({\pa \Phitz \over\pa t})(t,\ptz)
= {\pa (d\Phitz) \over\pa t}(t,\ptz)
= {\pa \Ftz \over\pa t}(t,\ptz),
%\eqnote \overbigdot\Ftzptz(t) .
\ee
when $\Phitz$  is $C^2$ and $\Ftz:=d\Phitz$. Thus
\be
\label{eqvVvv2b}
%d\vv^\tz(t,\pt) = \overbigdot\Ftzptz(t).\Ftzt(\ptz)^{-1}.
d\vv^\tz(t,\pt) = {\pa \Ftz \over\pa t}(t,\ptz).\Ftz(t,\ptz)^{-1},
\quad\hbox{written in short}\quad
d\vv = \overbigdot F.F^{-1} \quad(\hbox{points? times?}).
\ee
%(at what points, what times?).
And $d\vv^\tz_t$ can be written $L^\tz_t$ in classical mechanics books, so you can find
\be
\label{eqdefL}
\Ltzt(\pt):= \overbigdot\Ftzptz(t).\Ftzt(\ptz)^{-1},
\quad\hbox{written in short}\quad L=\overbigdot F.F^{-1} 
\quad(\hbox{at what points, what times?}).
\ee
Here it is not obvious that $\Ltzt(\pt)$ does \textsl{\textbf{not}} depend on~$\tz$, which is indeed the case, \cf~\eref{eqvVvv2}:
\be
\label{eqdefL2}
\Ltzt(\pt) = d\vv_t(\pt).
\ee
Reminder: if possible, use Eulerian quantities as long as possible%
\footnote{
To get Eulerian results from Lagrangian computations
can make the understanding of a Lie derivative quite difficult: To introduce the ``so-called'' Lie derivatives in classical mechanics you can find the following steps:
1- At~$t$ consider the Cauchy stress vector $\vec t$ (Eulerian),
2- then with a unit normal vector~$\vn$, define the associated Cauchy stress tensor $\uusigma$
(satisfying $\vec t = \uusigma.\vn$),
3- then use the virtual power and the change of variables in integrals to be back into~$\Omegatz$ to be able to work with Lagrangian variables,
4- then introduce the first Piola--Kirchhoff (two point) tensor~$\PK$,
5- then introduce the second Piola--Kirchhoff tensor~$\SK$ (endomorphism in~$\Omegatz$),
6- then differentiate~$\SK$ in~$\Omegatz$ (in the Lagrangian variables although the initials variables are the Eulerian variables in~$\Omegat$),
7- then back in~$\Omegat$ to get back to Eulerian functions (change of variables in integrals),
8- %together with some confusion between covariance and contravariance, 
then you get some Jaumann or Truesdell or other so called Lie derivatives type terms% (cf~\S~\ref{secWPi})
, the appropriate choice among all these derivatives being quite obscure because the covariant objectivity has been forgotten en route... % (and does not seem to give that satisfying results)...
%which moreover satisfies nobody...%
While, with simple Eulerian considerations, it requires a few lines
to understand the (real) Lie derivative (Eulerian concept) and its simplicity, see~\S~\ref{secdL}, and deduce second order covariant objective results.
%(The Lie derivative can be directly applied to Eulerian Cauchy stress type vectors, without reference to the machinery ``the Cauchy stress tensor'' used to represent the Cauchy vector thanks to a unit normal vector.)
\label{footrem}
}.

\comment{
Let $\tPhi$ be a motion, \cf~\eref{eqdeftPhi0}.
%, and $\vv$ be its Eulerian velocity field, \cf~\eref{eqdefve}.
Then, with the irrepressible urge to use an initial time~$\tz$ (observer dependent),
consider the initial configuration $\Omegatz = \tPhi(\tz,\Obj)$ and $\Phitz$, \cf~\eref{eqdefPhi}.
Then define the Lagrangian velocity $\vVtz(t,\ptz) = {\pa \Phitz \over \pa t}(t,\ptz)$, 
and then associate the Eulerian velocity
\be
\label{eqdefLdvv}
\vv^\tz(t,\pt) := \vVtz(t,\ptz) \qwhen \pt = \Phitz(t,\ptz).
\ee
(With this point of view, the Eulerian velocity depends a priori on~$\tz$.) Thus
$\vv^\tz_t(\Phitzt(\ptz)) := \vVtzt(\ptz)$ gives
\be
\label{eqdefLdv0}
d\vv^\tz_t(\pt).d\Phitzt(\ptz) = d\vVtz(t,\ptz) \qwhen \pt = \Phitz(t,\ptz).
\ee
Thus, with $d\vVtz(t,\ptz) = d{\pa \Phitz \over \pa t}(t,\ptz)
={\pa d\Phitz \over \pa t}(t,\ptz)
={\pa \Ftz \over \pa t}(t,\ptz)
={d\Ftzptz \over d t}(t)
\eqnote  \overbigdot{\Ftzptz}(t)
$, we obtain
\be
\label{eqdefLdv}
d\vv^\tz_t(\pt) = \overbigdot{\Ftzptz}(t).\Ftzt(\ptz)^{-1}  \qwhen \pt = \Phitz(t,\ptz). %\quad \hbox{written in short}\quad L=\overbigdot{F}.F,
\ee
This gives the differential $d\vv^\tz$ of the Eulerian velocity as a function of $\tz,t,\pt$.
Then define
\be
\label{eqdefL}
L^\tz(t,\pt) := \overbigdot{\Ftzptz}(t).\Ftzt(\ptz)^{-1}, 
\quad\hbox{to get}\quad L^\tz(t,\pt) = d\vv^\tz(t,\pt)  \qwhen \pt = \Phitz(t,\ptz).
\ee
}

%%%%%%%%%%%%%%%%%%%%%%%%%%%%%%%%%%%%%%%%%%%%%%%%%%%%%%%%%%%%%%%%%%%%%%%%%%%%%%%%%%%

\subsection{Lagrangian acceleration}

Let $\Pobj\in\Obj$, $\tz,t\in\RR$, $\ptz =\tPhiPobj(\tz)$ and $\pt=\tPhiPobj(t)$ (positions of~$\Pobj$ at~$\tz$ and~$t$).

\debdef
In short, the Lagrangian acceleration at $t$ at~$\pt$ of the particle $\Pobj$ is
\be
\label{eqvaci20b}
\vGammatz(t,\ptz) %:= {\pa^2 \tPhi\over\pa t^2}(t,\Pobj) 
:= \tPhiPobj{}''(t)  \qwhen \ptz = \tPhiPobj(\tz).
\ee
In other words
\be
\label{eqremvGnoncv}
%\tPhiPobj{}''(t) = 
\vGammatz(t,\ptz)  := \vgamma(t,\pt)   \qwhen \pt = \Phitz(t,\ptz),
\ee
where $\vgamma(t,\pt)=\tPhiPobj{}''(t)$ is the Eulerian acceleration at $t$ at $\pt=\tPhi(t,\Pobj)$, \cf~\eref{eqvaeul0}.

In details,  the Lagrangian acceleration is the ``two point vector field'' defined on $\RR\times \Omegatz$ by
\be
\label{eqvaci20}
\widetilde\vGammatz(t,\ptz) = ((t,\pt),\tPhiPobj{}''(t)),
\qwhen \pt = \Phitz(t,\ptz).
\ee
(To compare with~\eref{eqvaeul00}.) In particular $\vGammatz(t,\ptz)$ is not drawn on the graph of~$\vGammatz$ at $(t,\ptz)$,
but on the graph of~$\vgamma$ at $(t,\pt)$.
\findef
% (which is a usual vector field: $\vgamma$ doesn't depend on any~$\tz$, and isn't a ``two point vector field''.) 

If $t$ is fixed, or if $\ptz\in\Omegatz$ is fixed, then define
\be
\vGammatzt(\ptz) := \vGammatz(t,\ptz), \qand
\vGamma_\ptz^\tz(t) := \vGammatz(t,\ptz).
\ee
Thus
\be
\label{eqvAva}
\vGammatzt = \vgamma_t \circ \Phitzt, \qand
d\vGammatz_t(\ptz)=d\vgamma_t(\pt).\Ftzt(\ptz),
\ee
when $\pt = \Phitzt(\ptz)$ and $\Ftzt := d\Phitzt$ (the deformation gradient).

Risky notation: $d\vGamma = d\vgamma.F$ (points? times?).

%%%%%%%%%%%%%%%%%%%%%%%%%%%%%%%%%%%%%%%%%%%%%%%%%%%%%%%%%%%%%%%%%%%%%%%%%%%%%%%%%%%%

\subsection{Time Taylor expansion of~$\Phitz$}

Let $\ptz\in\Omegatz$. Then, at second order,
\be
\label{eqttexp1}
\eqalign{
\Phitzptz(\tau)
= & \Phitzptz(t) + (\tau{-}t) \Phitzptz'(t) + {(\tau{-}t)^2 \over 2} \Phitzptz''(t) + o((\tau{-}t)^2), \cr
}
\ee
that is, with $p(\tau)=\tPhiPobj(\tau)=\Phi^\tz_\tau(\ptz)$,

\be
\label{eqttexp2}
\eqalign{
p(\tau)
= & p(t) + (\tau{-}t) \vVtz(t,\ptz) + {(\tau{-}t)^2 \over 2} \vGammatz(t,\ptz) + o((\tau{-}t)^2). \cr
%(= & p(t) + (\tau{-}t) \vv(t,\pt) + {(\tau{-}t)^2 \over 2} \vgamma(t,\pt) + o((\tau{-}t)^2)). \cr
}
\ee
NB: There are \textsl{\textbf{three}} times involved: $\tz$ (observer dependent), $t$ and~$\tau$ (for the Taylor expansion).
To compare with~\eref{eqdl20}-\eref{eqdl20E}:
$p(\tau) = p(t) + (\tau{-}t) \vv(t,p(t)) + {(\tau{-}t)^2 \over 2} \vgamma(t,p(t)) + o((\tau{-}t)^2)$, independent of~$\tz$.

%%%%%%%%%%%%%%%%%%%%%%%%%%%%%%%%%%%%%%%%%%%%%%%%%%%%%%%%%%%%%%%%%%%%%%%%%%%%%%%%%%%

\subsection{A vector field that let itself be deformed by a motion}

Consider a $C^0$ Eulerian vector field $\vw : 
\left\{\eqalign{
\bigC &\rar \vRRn \cr
(t,\pt) &\rar \vw(t,\pt) \cr
}\right\}$.
Let $\tz\in[t_1,t_2[$ and let
$\vw_\tz : 
\left\{\eqalign{
\Omegatz &\rar \vRRn \cr
\ptz &\rar \vw_\tz(\ptz):=\vw(\tz,\ptz) \cr
}\right\}$
(vector field in~$\Omegatz$).
Then define the (virtual) vector field
\be
\label{eqdwdt05}
\vw_{\tz*} : 
\left\{\eqalign{
\bigC &\rar \vRRn \cr
(t,\pt) &\rar \vw_{\tz*}(t,\pt) := d\Phitz(t,\ptz).\vw_\tz(\ptz), \qwhen p(t) = \Phitz(t,\ptz). \cr
}\right.
\ee
(The push-forward = result of the deformation of $\vw_\tz$ by the motion, see figure~\ref{figpf}.)
%This vector field, considered as a flow, is associated with trajectories (= functions $\Psi$ solutions of ${d\Psi \over dt}(t)=\vw_{\tz*}(t,\Psi(t))$).

\debprop
\label{peqdwdt04}
For $C^2$ motions, we have (time variation rate along a virtual trajectory)
\be
\label{eqdwdt04}
{D\vw_{\tz*} \over Dt} = d\vv.\vw_{\tz*},
%\qie {D \vw_{\tz*} \over d t} - d\vv.\vw_{\tz*} = \vec0,
\ee
\ie\ $\calL_\vv \vw_{\tz*}=\vec0$,
where $\calL_\vv \vu := {D\vu \over Dt} - d\vv.\vu$ ($= {\pa \vu \over \pa t} + d\vu.\vv - d\vv.\vu$) is the Lie derivative of a (unsteady) vector field $\vu:\bigC \rar \vRRn$ along~$\vv$.

{\bf Interpretation:} We will see that
$\calL_\vv \vw(\tz,\ptz) = \lim_{t\rar\tz}{ \vw(t,p(t)) - \vw_{\tz*}(t,p(t)) \over h}$ measures the ``resistance of~$\vw$ to a motion'', see~\S~\ref{secdlresist};
Thus the result $\calL_\vv \vw_{\tz*}(\tz,\ptz)=\vec0$ is ``obvious'' ($=\lim_{t\rar\tz}{ \vw_{\tz*}(t,p(t)) - \vw_{\tz*}(t,p(t)) \over h}$):
If $\vw=\vwtzs$ then the vector (``force'') field $\vw$ does not oppose any resistance to the flow.
\finprop

\debdem
$\ptz$ being fixed, with $d\Phitz(t,\ptz) \eqnote F(t)$ we have
$\vw_{\tz*}(t,p(t)) \mope^{\eref{eqdwdt05}}F(t).\vw_\tz(\ptz)$, thus
$
{D\vw_{\tz*} \over Dt}(t,p(t))
= F'(t).\vw_\tz(\ptz)
= F'(t).F(t)^{-1}.\vw_{\tz*}(t,p(t))
\mathop{=}^{\eref{eqvVvv2b}} d\vv(t,p(t)).\vw_{\tz*}(t,p(t))
$, 
\ie~\eref{eqdwdt04}.
\findem

%%%%%%%%%%%%%%%%%%%%%%%%%%%%%%%%%%%%%%%%%%%%%%%%%%%%%%%%%%%%%%%%%%%%%%%%%%%%%%%%%%%
%%%%%%%%%%%%%%%%%%%%%%%%%%%%%%%%%%%%%%%%%%%%%%%%%%%%%%%%%%%%%%%%%%%%%%%%%%%%%%%%%%%

\section{Deformation gradient $F:=d\Phi$}
\label{secremF}

Consider a motion 
$\tPhi : 
\left\{\eqalign{
\RR\times \Obj & \rar\RRn \cr
(t,\Pobj) & \rar \pt=\tPhi(t,\Pobj)
}\right\}$, %\cf~\eref{eqdeftPhi0},
$\Omegat:=\tPhi(t,\Obj)$ the configuration of~$\Obj$ at any~$t$,
fix $\tz,t$ in~$\RR$, and let
$
\Phitzt:
\left\{\eqalign{
\Omegatz &\rar \Omegat \cr
\ptz=\tPhi(\tz,\Pobj) &\rar \pt=\Phitzt(\ptz):=\tPhi(t,\Pobj) \cr
}\right\}$,
%(the associated motion between $\tz$ and~$t$), \cf~\eref{eqPhit},
supposed to be a~$C^1$ diffeomorphism. 
Notations for calculations (quantification), to comply with practices:

1- Classical (unambiguous) notations as in Arnold, Germain: \Eg, $(\va_i)$ and $(\vb_i)$ are bases resp. in~$\RRntz$ and~$\RRnt$,
$\vw_\tz(\ptz)=\sum_i w_{\tz,i}(\ptz)\va_i\in\RRntz$, $\vw_{t,i}(\pt)=\sum_i w_{t,i}(\pt)\vb_i\in\RRnt$; And 
% $\vw_\tz(\ptz)=\sumin w_{\tz,i}(\ptz)\va_i\in\RRntz$, $\vw_{t,i}(\pt)=\sumin w_{t,i}(\pt)\vb_i\in\RRnt$; And 

2- Marsden--Hughes duality notations: Capital letters at~$\tz$, lower case letters at~$t$, duality notation, \eg\ $(\vE_I)$ and~$(\ve_i)$ are bases resp. in~$\RRntz$ and~$\RRnt$,
$\vW(P)=\sum_I W^I(P)\vE_I\in\RRntz$, $\vw(p)=\sum_i w^i(p)\ve_i\in\RRnt$.
%$\vW(P)=\sumIn W^I(P)\vE_I\in\RRntz$, $\vw(p)=\sumin w^i(p)\ve_i\in\RRnt$.

%%%%%%%%%%%%%%%%%%%%%%%%%%%%%%%%%%%%%%%%%%%%%%%%%%%%%%%%%%%%%%%%%%%%%%%%%%%%%%%%%%%

\subsection{Definitions}
\label{secremF0}

%%%%%%%%%%%%%%%%%%%%%%%%%%%%%%%%%%%%%%%%%%%%%%%%%%%%%%%%%%%%%%%%%%%%%%%%%%%%%%%%%%%

\subsubsection{Definition of the deformation gradient $F$}

\begin{figure}[!h]
\qquad\includegraphics[width=0.9\textwidth]{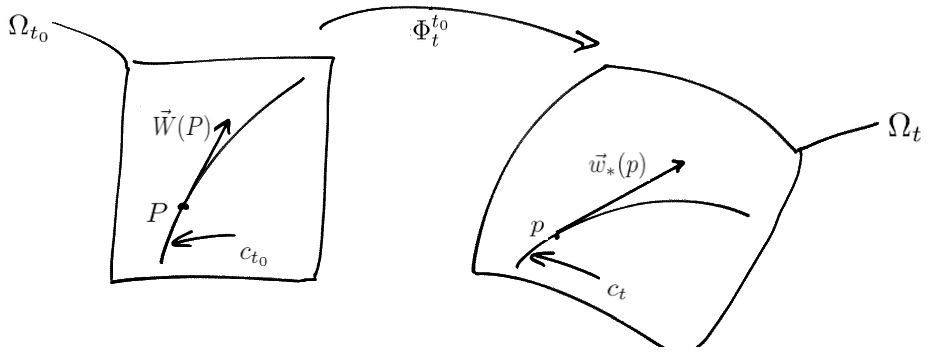}
\vspace{-10pt}
\caption{$\vw$ is a Eulerian vector field. At~$\tz$ define vector field $\vwtz$ in~$\Omegatz$ by $\vwtz(\ptz):=\vw(\tz,\ptz)$.
The (spatial) curve $c_\tz:s\rar \ptz=c_\tz(s)$ in~$\Omegatz$ is an integral curve of~$\vwtz$, \ie\ satisfies $c_\tz{}'(s) = \vwtz(c_\tz(s))$. It is transformed by $\Phitzt$ into the (spatial) curve $c_t = \Phitzt\circ c_\tz :
s\rar \pt=c_t(s) {=} \Phitzt(c_\tz(s))$ in~$\Omegat$;
Hence $c_t{}'(s) = d\Phitzt(\ptz).c_\tz{}'(s) = d\Phitzt(\ptz).\vwtz(\ptz) \eqnote \vwtzs(t,\pt)$ is the tangent vector at~$c_t$ at~$\pt$ (the push-forward of~$\vwtz$ by~$\Phitzt$). And $\vw(t,p(t))$ (actual value) is also drawn.
}\label{figpf}
\end{figure}

\debdef
The differential $d\Phitzt \eqnote \Ftzt : 
\left\{\eqalign{
\Omegatz &\rar \calL(\RRntz;\RRnt) \cr
\ptz & \rar \Ftzt(\ptz):=d\Phitzt(\ptz)
}\right\}$
is called ``the covariant deformation gradient between $\tz$ and~$t$'', or simply ``the deformation gradient''.
And ``the covariant deformation gradient at~$\ptz$ between $\tz$ and~$t$'', or in short ``the deformation gradient at~$\ptz$'' is the linear map $\Ftzt(\ptz) \in\calL(\RRntz;\RRnt)$, so defined by, for all $\vwtzptz\in\RRntz$ (vector at~$\ptz$), 
\be
\label{eqdefFtf}
%(\vwtzts(\pt)=)\quad 
\boxed{\Ftzt(\ptz).\vwtzptz := \lim_{h\rar0} {\Phitzt(\ptz{+}h\vwtzptz) - \Phitzt(\ptz) \over h}}
\eqnote (\Phitzt)_*(\vwtz)(\pt) \eqnote \vwtzs(t,\pt),
\ee
with $\pt=\Phitzt(\ptz)$. See figure~\ref{figpf}.
\findef

\mn
{\bf Marsden--Hughes notations:} %(shorten notation for~\eref{eqdefFtfdr0}) : 
$\Phi := \Phitzt$, $F := d\Phi$, $P:=\ptz$, $\vW(P):=\vwtzptz$, $p=\Phi(P)$, thus
\be
\label{eqdefFtfMH}
\boxed{F(P).\vW(P) := \lim_{h\rar0}{\Phi(P{+}h\vW(P)) - \Phi(P) \over h}}
\eqnote \Phi_*\vW(p)
\eqnote \vw_*(p).
\ee

%\vspace{-.4pt}

%\medskip\noindent
{\bf NB:} The ``deformation gradient'' $\Ftzt=d\Phitzt$ is \textbf{\textsl{not}} a ``gradient'' (its definition does \textbf{\textsl{not}} need a Euclidean dot product);
This lead to confusions when covariance-contravariance and objectivity are at stake.
It would be simpler to stick to the name ``$\Ftzt =$ the differential  of~$\Phitzt$'', but it is not the standard usage, except in thermodynamics: \Eg, the differential $dU$ of the internal energy~$U$ is \textslbf{not} called ``the gradient of~$U$'' (there is no meaningful inner dot product): It is just called ``the differential of~$U$''...

%%%%%%%%%%%%%%%%%%%%%%%%%%%%%%%%%%%%%%%%%%%%%%%%%%%%%%%%%%%%%%%%%%%%%%%%%%%%%%%%%%%

\subsubsection{Push-forward (values of $F$)}

\debdef
Let $\vwtz:\left\{\eqalign{\Omegatz & \rar \RRntz \cr \ptz & \rar \vwtzptz}\right\}$ be a vector field in~$\Omegatz$.
Its push-forward by~$\Phitzt$ is the vector field
$(\Phitzt)_*(\vwtz)$ in~$\Omegat$ defined by
\be
\label{eqdefFtfd}
(\Phitzt)_*\vwtz(\pt) = \Ftzt(\ptz).\vwtzptz \eqnote \vwtzs(t,\pt) \qwhen \pt=\Phitzt(\ptz).
%\qalsowritten \Phi_*\vW(p)= F(P).\vW(P) \eqnote \vw_*(p),
\ee
See figure~\ref{figpf}.
Marsden notation: $\Phi_*\vW(p)= F(P).\vW(P) \eqnote \vw_*(p)$ when $p=\Phitzt(P)$. 

In other words %, the push-forwarded vector field in~\eref{eqdefFtfd} is defined by %$\vwttzs \circ \Phitzt := \Ftzt.\vwtz$, \ie
\be
(\Phitzt)_*\vw_\tz := (\Ftzt.\vwtz) \circ (\Phitzt)^{-1}.
\ee
Marsden notation: $\Phi_*\vW = (F.\vW)\circ\Phi^{-1} = \vw_*$.
\findef

%%%%%%%%%%%%%%%%%%%%%%%%%%%%%%%%%%%%%%%%%%%%%%%%%%%%%%%%%%%%%%%%%%%%%%%%%%%%%%%%%%%

\subsubsection{$F$ is a two point tensors}

With~\eref{eqdefFtf}, ``the tangent map'' is
\be
\label{eqdefFtft}
\widehat\Ftzt:
\left\{\eqalign{
\Omegatz & \rar \Omegat \times \calL(\RRntz;\RRnt) \cr
\ptz & \rar \widehat\Ftzt(\ptz) = (\pt,\Ftzt(\ptz)) \qwhen  \pt=\Phitzt(\ptz).
}\right.
\ee
%so $\Ftzt(\ptz)$ is pointed at~$\pt$, \ie\ its value is considered at~$\pt$.

\debdef
\label{defMHtpt}
(Marsden--Hughes~\cite{marsden-hughes}.) %, as at \S~\ref{secbrtpt}:
The function $\widehat\Ftzt$ is called a two point tensor, referring to the points $\ptz\in\Omegatz$ (departure set)
and $\pt = \Phitzt(\ptz)\in \Omegat$ (arrival set where $\vwtzs(t,\pt)= \Ftzt(\ptz).\vwtz(\ptz)$ is drawn).
And in short $\widehat\Ftzt\eqnote \Ftzt$ is said to be a two point tensor.
\findef

\debrem
\label{remMHtpt}
The name ``two point tensor'' is a shortcut than can create confusions and errors when dealing with the transposed: $\Ftzt$ is not immediately a ``tensor'':
A tensor is a multilinear form, so gives scalar results ($\in\RR$),
while $F(P):=\Ftzt(P)\eqnote F_P\in \calL(\RRntz;\RRnt)$ gives vector results (in~$\RRnt$).
However $F_P$ can be naturally and canonically associated with the bilinear form $\tF_P\in \calL(\RRnts,\RRntz;\RR)$ defined by, 
for all $\vu_P\in\RRntz$ and $\ell_p\in\RRnts$, with $p=\Phitzt(P)$,
\be
\label{eqtFP}
\tF_P(\ell_p,\vu_P) := \ell_p.F_P.\vu_P \;(\in\RR),
\ee
see~\S~\ref{sectroflm},
and it is $\tF_P$ which defines the so-called ``two point tensor''.

But don't forget that the transposed of a linear form ($F_P$ here) is \textslbf{not} deduced from the transposed of the associated bilinear form ($\tF_P$ here). So be careful with the word ``transposed'' and its two distinct definitions.
Indeed, the transposed of a bilinear form $b\dd$ is intrinsic to~$b\dd$ (is objective),
given by $b^T(\vu,\vw)=b(\vw,\vu)$, while the transposed of a linear function~$L$ is not intrinsic to~$L$ (is subjective),
given by $(L^T.\vu,\vw)_g= (L.\vw,\vu)_h$ where $\dd_g$ and~$\dd_h$ are inner dot products chosen by human beings.
(details in~\S~\ref{secbT} and \S~\ref{seccpgdd01}).
%See~\eref{eqdefJ2g}. A
%: A bilinear form can be symmetric when the associated linear map is \textslbf{not} symmetric; 
%Such a canonical natural association $L\in\calL(A;B) \leftrightarrow \tilde F \in \calL(B^*,A;\RR)$ is widely used in differential geometry when $B{=}A$ (but here $A{=}\RRntz \ne B{=}\RRnt$); And is also widely used for computations purposes (see~\S~\ref{sectroflm}); But it also creates pure mathematical calculations which may be problematic to interpret, see  \eg\ remark~\ref{remmean}.
%So, it is better to avoid the use of~$\tilde{\Ftzt(\ptz)}$ (scalar result) and the associated tensorial product notations, see~\S~\ref{secremtensnot}: Use~$\Ftzt(\ptz)$ (vector result) even if you call it a tensor.... % since $\Ftzt(\ptz)=d\Phitzt(\ptz)$ is the differential at~$\ptz$).
\finrem

\debrem
More generally for manifolds,  %with differential geometry \cf~\S~\ref{secRRnt}: 
the differential of $\Phi:=\Phitzt$ at $P\in\Omegatz$ is
%$T_\ptau\Omegatau$ is the tangent space at $\ptau\in\Omegatau$,
%and $ \{\pt\}\times T_\pt\Omegat$ is the fiber at $\pt\in\Omegat$,
%and $T\Omegatau = \bigcup_{\ptau\in\Omegatau}(\{\ptau\}\times T_\ptau\Omegatau)$ is the tangent bundle of~$\Omegatau$. Then the covariant deformation gradient at~$\ptz$ between $\tz$ and~$t$ is
%$\Ftzt(\ptz) := d\Phitzt(\ptz)$ the differential of~$\Phitzt$ at~$\ptz$, so
$
%\label{eqnoteFgd}
F(P):=
d\Phi(P) :
\left\{\eqalign{
T_P\Omegatz & \rar T_p\Omegat \cr
\vW(P) & \rar
\vw_*(p):= d\Phi(P).\vW(P)% \eqnote F(P).\vW(P),
}\right\}
$ with $p=\Phitzt(P)$.
And the tangent map is
%: It the function defined with~\eref{eqnoteFgd} by
\be
\label{eqnoteFgd}
 T\Phi :
\left\{\eqalign{
T\Omegatz & \rar T\Omegat \cr
(P,\vW(P)) & \rar T\Phi(P,\vW(P)) := (p,d\Phi(P).\vW(P))= (p,\vw_*(p)), \qwhere p=\Phitzt(P),
}\right.
\ee
called the associated two point tensor.
\finrem

%%%%%%%%%%%%%%%%%%%%%%%%%%%%%%%%%%%%%%%%%%%%%%%%%%%%%%%%%%%%%%%%%%%%%%%%%%%%%%%%%%%

\subsubsection{Evolution: Toward the Lie derivative (in continuum mechanics)} 
\label{exapfv}

%Continuing example~\ref{exapf00}. %, see figure~\ref{figpf2}.
% $\Psi = \Phitzt$, $\UE=\Omegatz$, and $\UF = \Omegat$. Let $\bigC := \bigcup_{t\in[t_1,t_2]}(\{t\}\times\Omegat)$, cf~\eref{eqbigU}.
Consider a Eulerian vector field
$\vw : 
\left\{\eqalign{
\bigC = \bigcup_t(\{t\}\times \Omegat) & \rar \vRRn \cr
(t,p) & \rar \vw(t,p)
}\right\}$, %\cf~\eref{eqdeffspa20}, 
\eg\ a ``force field''.
%And let $\vw_t(p) := \vw(t,p)$.
Then, at~$\tz$ consider
$\vw_\tz :\left\{\eqalign{
\Omegatz & \rar \RRntz \cr
\ptz & \rar \vw_\tz(\ptz):= \vw(\tz,\ptz)
}\right\}$. 
The push-forward of~$\vw_\tz$ by~$\Phitzt$ is, \cf~\eref{eqdefFtfMH},
\be
\label{eqrapvEivei2a}
\vw_{\tz*}(t,p(t)) = \Ftzt(\ptz).\vw_\tz(\ptz) , \qwhere p(t)=\Phitz(t,\ptz).
\ee
See~figure~\ref{figpf}.
Then, without any ubiquity gift, at~$t$ at~$p(t)$ we can compare $\vw(t,p(t))$ (real value of $\vw$ at~$t$ at~$p(t)$) with $\vw_{\tz*}(t,p(t))$ (transported memory along the trajectory). Thus the rate
\be
\label{eqrapvEivei2c}
{\vw(t,p(t)) - \vw_{\tz*}(t,p(t))\over t-\tz} 
={\hbox{\footnotesize actual}(t,p(t)) - \hbox{\footnotesize memory}(t,p(t)) \over t-\tz}
\quad\hbox{is meaningful at $(t,p(t))$}
%\eqnote \calL_\vv\vw(\tz,\ptz)
\ee
(no ubiquity gift required).
%Here $\vw_{\tz*}(t,p(t)) = \Ftz(t,\ptz).\vw_\tz(\ptz)$ 
This rate gives, as $h\rar0$, the Lie derivative $\calL_\vv\vw$ (the rate of stress), and we will see at \S~\ref{secdlcv}
that $\calL_\vv\vw={D\vw\over Dt} - d\vv.\vw$
(the $d\vv$ term tells that a ``non-uniform flow'' acts on the stress).
%The difference between the actual force $\vw_t(\pt)$ at~$t$ and the result or the vector that has let itself be deformed by the flow, that is $\vwtzs(\pt)$: It 

%%%%%%%%%%%%%%%%%%%%%%%%%%%%%%%%%%%%%%%%%%%%%%%%%%%%%%%%%%%%%%%%%%%%%%%%%%%%%%%%%%%

\subsubsection{Pull-back}

Formally the pull-back is the push-forward with $(\Phitzt)^{-1}$: 
%With $d(\Phitzt)^{-1}(\pt)=(d\Phitzt(\ptz))^{-1}$ because $(\Phitzt)^{-1}(\Phitzt(\ptz))=\ptz$,

\debdef
The pull-back $(\Phitzt)^*\vw_t$ of a vector field $\vw_t$ defined on~$\Omegat$ is the vector field defined on~$\Omegatz$ by, with $\ptz=(\Phitzt)^{-1}(\pt)$, % written $P=\Phi^{-1}(p)$ (Marsden notations),
\be
\vw_t^*(\tz,\ptz)=(\Phitzt)^*\vw_t(\ptz) := (\Ftzt)^{-1}(\pt).\vw_t(\pt), \qwritten \vW^*(P) = F^{-1}(p).\vw(p).
\ee
\findef
%(We can't relive the past... So the pull-back is usually defined after the push-forward.)

%%%%%%%%%%%%%%%%%%%%%%%%%%%%%%%%%%%%%%%%%%%%%%%%%%%%%%%%%%%%%%%%%%%%%%%%%%%%%%%%%%%

\subsection{Quantification with bases}
\label{secFbase}

(Simple Cartesian framework.)
$(\va_i)$ is a Cartesian basis in~$\vRRntz$, $(\vb_i)$ is a Cartesian basis in~$\vRRnt$, $o_t$ is an origin in~$\RRn$ at~$t$, $\Phitzt \eqnote \Phi$ supposed~$C^1$, $\phi_i:\Omegatz\rar\RR$ is its components in the referential $(o_t,(\vb_i))$:
\be
\label{eqvwF30}
%\pt = 
\Phi(\ptz) = o_t+\sumin \phi_i(\ptz)\vb_i, \qie \ora{o_t \Phi(\ptz)} = \sumin \phi_i(\ptz)\vb_i.
\ee
Thus, with the classic notation $d\phi_i(\ptz).\va_j \eqnote {\pa \phi_i \over \pa X_j}(\ptz)$ since $(\va_i)$ is a Cartesian basis, and $(\vb_i)$ being a Cartesian basis,
$$
d\Phi(\ptz).\va_j 
= \sumin (d\phi_i(\ptz).\va_j)\vb_i
= \sumin {\pa \phi_i \over \pa X_j}(\ptz) \vb_i, \qthus
[d\Phi(\ptz)]_{[\va,\vb]} = [{\pa \phi_i \over \pa X_j}(\ptz)] = [F(\ptz)]_{[\va,\vb]},
$$
$[d\Phi(\ptz)]_{[\va,\vb]}= [F(\ptz)]_{[\va,\vb]}$ being the Jacobian matrix of~$\Phi$ at~$\ptz$ relative to the chosen bases.
In short:
\be
d\Phi.\va_j 
= \sumin {\pa \phi_i \over \pa X_j} \vb_i, \qthus
[d\Phi]_{[\va,\vb]} = [{\pa \phi_i \over \pa X_j}] = [F]_{[\va,\vb]} = [F_{ij}],
\ee
Thus, if $\vW\in\RRntz$ is a vector at~$\ptz$ and $\vW=\sumjn W_j\va_j$ then, by linearity of differentials,
\be
\label{eqvwF3t0}
d\Phi.\vW = F.\vW = \sumin F_{ij}W_j\vb_i, \qie
[F.\vW]_{|\vb} =  [F]_{|\va,\vb}.[\vW]_{|\va}
\ee
(more precisely: $\Ftzt(\ptz).\vW(\ptz) = \sumin F_{ij}(\ptz)W_j(\ptz)\vb_i$).

Similarly, for the second order derivative $d^2\Phi = dF$ (when $\Phi$ is~$C^2$):
With $\vU = \sumjn U_j\va_j$ and $\vW = \sumkn W_k\va_k$, and with $(\va_i)$ and $(\vb_i)$ Cartesian bases, we get
\be
\label{eqvwF0d0d}
dF(\vU,\vW) = d^2\Phi(\vU,\vW)
= \sumin d^2\phi_i(\vU,\vW) \vb_i
= \sumijkn {\pa^2 \phi_i \over \pa X_j \pa X_k} U_j W_k \vb_i 
=  \sumin \; \Bigl([\vU]^T_{|\va} .[d^2\phi_i]_{|\va}.[\vW]_{|\va}\Bigr)\; \vb_i,
\ee
$[d^2\phi_i(\ptz)]_{|\va} = [{\pa^2\phi_i \over \pa X_j\pa X_k}(\ptz)]_{j=1,...,n \atop k=1,...,n}$ being the Hessian matrix of~$\phi_i$ at~$\ptz$ relative to the basis~$(\va_i)$.

With Marsden duality notations:
\be
\eqalign{
\bullet & p=\Phi(P) = o_t+\sumin \phi^i(P)\ve_i, \quad
\FiJ(P) = {\pa \phi^i \over \pa X^J}(P) \quad (= d\phi^i(P).\vE_J), \cr
\bullet & F(P).\vW = \sumiJn \FiJ(P)\,W^J \ve_i, \quad [F] = [\FiJ] = [d\Phi], \cr
\bullet & dF(\vU,\vW)= d^2\Phi(\vU,\vW)= \sum_{i,J,K=1}^n {\pa^2 \phi^i \over \pa X^J \pa X^K}U^JW^K \ve_i
=  \sumin \; \Bigl([\vU]^T .[d^2\phi^i].[\vW]\Bigr)\; \ve_i.
}.
\ee
% where $[d^2\phi^i]_{|\vE} = [{\pa^2\phi^i \over \pa X^j\pa X^k}]_{j=1,...,n \atop k=1,...,n} \eqnote [d^2\phi^i]$.

\debrem
$J,j$ are dummy variables when used in a summation: \Eg, 
$df.\vW
= \sumjn {\pa f \over \pa X^j}W^j
= \sumJn {\pa f \over \pa X^J}W^J
= \sum_{\alpha=1}^n {\pa f \over \pa X^\alpha}W^\alpha
= {\pa f \over \pa X^1}W^1+{\pa f \over \pa X^2}W^2+...
$
(there is no uppercase for 1, 2...). And Marsden--Hughes notations (capital letters for the past) are not at all compulsory, classical notations being just as good and even preferable if you hesitate (because they are not misleading). See~\S~\ref{secann1}.
\finrem

%%%%%%%%%%%%%%%%%%%%%%%%%%%%%%%%%%%%%%%%%%%%%%%%%%%%%%%%%%%%%%%%%%%%%%%%%%%%%%%%%%%
%%%%%%%%%%%%%%%%%%%%%%%%%%%%%%%%%%%%%%%%%%%%%%%%%%%%%%%%%%%%%%%%%%%%%%%%%%%%%%%%%%%

\subsection{The unfortunate notation $d\vec x = F.d\vec X$}
\label{secnm}

%%%%%%%%%%%%%%%%%%%%%%%%%%%%%%%%%%%%%%%%%%%%%%%%%%%%%%%%%%%%%%%%%%%%%%%%%%%%%%%%%%%

\subsubsection{Issue}

\eref{eqdefFtfd}, \ie\ $\vw_*(p) := F(P).\vW(P)$, is sometimes written
\be
\label{eqdefFtfda}
%\vw_* = F.\vw,\quad\hbox{which is then written}\quad 
d\vx = F.d\vX %,  \qie d\vx = d\Phi.d\vX  
\hbox{ : ``a very unfortunate and misleading notation''}
\ee
which amounts to ``confuse a length and a speed''... And you also the phrase ``\eref{eqdefFtfda} is still true if $||d\vX||=1$''... while $d\vX$ is supposed to be small...
%NB: \eref{eqdefFtfd}-\eref{eqdefFtfdM} is indeed valid if $||\vW(P)||=1$.
%Explanations:

 % (and this notation results in misinterpretations).

\comment{
In particular, despite the ``infinitesimal notation'', $d\vX$ does not need to be ``a small vector'': In fact \eref{eqdefFtfda} is also true when $||d\vX||=1$ (!), weird isn't it? (Incompatible with usual infinitesimal notations.)

\Eg, in~$\vRRt$ with a basis $(\vE_i)$ at~$\tz$,  % at $ \ptz\in\Omegatz$,
with $\vwtz=\vE_j$ in~\eref{eqdefFtf} we get
$F.\vE_j = \lim_{h\rar0} {\Phi(\ptz+h\vE_j) - \Phi(\ptz) \over h}$ the derivative at~$\ptz$ in the direction~$\vE_j$
(derivative of the function $h \rar \Phi(\ptz+h\vE_j)$ at $h=0$);
And with a Euclidean dot product~$\dd_g$, we get ${||F.\vE_j||_g \over ||\vE_j||_g}=$ the stretch ratio at~$\ptz$ in the direction~$\vE_j$, and when $(\vE_i)$ is a Euclidean basis we get 
$\det(F.\vE_1,F.\vE_2,F.\vE_3)=J(\ptz)$ the dilatation rate at~$\ptz$ (Jacobian of~$\Phitzt$ at~$\ptz$).
And, it is not a good idea to write $\vE_j=d\vX$, \cf~\eref{eqdefFtfda}.
}

%Two explanations are detailed in the next paragraphs \S~\ref{secavo} and~\S~\ref{secado}.

%(The notation~\eref{eqdefFtfda} amounts to confuse a length $\delta\vX$ and a rate ${d\vX\over ds}$, see next~\S~\ref{secavo0}-\S~\ref{secavo}.)

%%%%%%%%%%%%%%%%%%%%%%%%%%%%%%%%%%%%%%%%%%%%%%%%%%%%%%%%%%%%%%%%%%%%%%%%%%%%%%%%%%%

\subsubsection{Where does this unfortunate notation come from?}
\label{secavo0}

The notation~\eref{eqdefFtfda} comes from the first order Taylor expansion $\Phi(Q)=\Phi(P) + d\Phitzt(P).(Q{-}P) +o(||Q{-}P||)$, where $P,Q\in\Omegatz$, \ie, with $p=\Phitzt(P)$ and $q=\Phitzt(Q)$ and $h=||Q{-}P||$,
\be
\label{eqdefFtfdr00}
q - p  = F(P).(Q{-}P) + o(h), \qwritten \delta \vx = F.\delta \vX+o(\delta\vX),
\ee
or $\ora{pq}  = F(P).\ora{PQ} + o(h)$.
So as $Q\rar P$ we get $0=0$... Quite useless, isn't it? %: % which is noted $d\vx = F.d\vX$...!?

While %, with $h=o(||Q{-}P||)$,
\be
\label{eqnotmec2}
{q - p \over h} =  F(P). {Q-P \over h} + o(1)
%{q - p \over ||Q{-}P||} =  F(P). {Q-P \over ||Q{-}P||} + o(1)
\quad \hbox{is useful:}
\ee
As $Q\rar P$ we get $\vw_*=F(P).\vW$ which relates tangent vectors, see figure~\ref{figpf}
Details:

% the relation  $\vw_t(p) = \Ftzt(P).\vW$ between tangent vectors, \ie
%However $d\vX$ and~$d\vx$ are not always described as tangent vectors.

%%%%%%%%%%%%%%%%%%%%%%%%%%%%%%%%%%%%%%%%%%%%%%%%%%%%%%%%%%%%%%%%%%%%%%%%%%%%%%%%%%%

\subsubsection{Interpretation: Vector approach}
\label{secavo}

Consider a spatial curve
$
%\label{eqnadm0}
c_\tz :
\left\{\eqalign{
[s_1,s_2] & \rar \Omegatz  \cr
s & \rar P := c_\tz(s) 
}\right\}
%\qand \vX(s) : =\ora{O_\tz c_\tz(s)} =\ora{O_\tz P}.
$
in~$\Omegatz$, \cf\ figure~\ref{figpf}.
It is deformed by $\Phitzt$ to become the spatial curve defined by
$
%\label{eqnadm2}
c_t := \Phitzt\circ c_\tz :
\left\{\eqalign{
[s_1,s_2] & \rar \Omegat  \cr
s & \rar p := c_t(s) = \Phitzt(c_\tz(s)).
}\right. 
$
 in~$\Omegat$.
Hence, relation between tangent vectors:
\be
\label{eqnotmec3}
{d c_t \over ds}(s) = d\Phitzt(c_\tz(s)).{dc_\tz \over ds}(s), , \qie \vw_*(p)=F(P).\vW(P)
\qwritten \boxed{{d\vx \over ds} = F.{d\vX \over ds}}, %\quad( \hbox{relates tangent vectors}),
\ee
But you \textsl{\textbf{can't}} simplify by~$ds$ to get $d\vx = F.d\vX$: It is absurd to confuse ``a slope ${d \vX \over ds}(s)$'' and ``a length~$\delta\vX$''.

NB: $||{d c_t \over ds}(s)||=||{d\vX \over ds}(s)||=1$ is meaningful in~\eref{eqnotmec3}:
%$||{dc_\tz \over ds}(s)||=1 = ||\vw_\tz(\ptz)||=||{d\vX \over ds}(s)||$ is meaningful:
It means that the parametrization of the curve~$c_\tz$ in~$\Omegatz$
uses a spatial parameter~$s$ such that $||c_\tz{}'(s)||=1$ for all~$s$,
% (such a parameter is called an ``intrinsic parameter''), 
\ie\ \st\ $||\vW_P||=1$ in figure~\ref{figpf}.
You \textsl{\textbf{cannot}} simplify by~$ds$: $||d\vX||=1$ is absurd together with $d\vX$ ``small''.

%%%%%%%%%%%%%%%%%%%%%%%%%%%%%%%%%%%%%%%%%%%%%%%%%%%%%%%%%%%%%%%%%%%%%%%%%%%%%%%%%%%

\subsubsection{Interpretation: Differential approach}
\label{secado}

%Differential approach (unmissable in thermodynamics): 
\eref{eqdefFtfda} is a relation between differentials... if you adopt the correct notations; Let us do it: With~\eref{eqvwF30},
%$(\va_i)$ and $(\vb_i)$ are Cartesian bases in~$\RRntz$ and~$\RRnt$, $o_t$ is a point in~$\RRn$ (a chosen origin at~$t$), and
\be
\label{eqdxiXj1}
\vx =\ora{o_t p} = \ora{o_t\Phitzt(P)} = \sumin \phi_i(P)\vb_i \eqnote \sumin x_i(P)\vb_i ,
\qwhere %\phi_i(P) \eqnote x_i(P) \qand 
\phi_i \eqnote x_i \quad(\hbox{function of $P$}).
\ee
Thus, with $(\piai)= (dX_i)$ the (covariant) dual basis of~$(\va_i)$ %(unmissable in thermodynamics), 
we get the system of $n$ equations (functions): %which can be written
\be
d\Phi=F, \qie
\left\{\eqalign{
& \textstyle d\phi_1(P) = \sumjn {\pa \phi_1\over \pa X_j}(P) \; dX_j \cr
& \vdots \cr
& \textstyle d\phi_n(P) = \sumjn {\pa \phi_n\over \pa X_j}(P) \; dX_j \cr
}\right\},  \quad\hbox{which is noted}\quad 
d\vx = F.d\vX,
\ee
\nobreak
this last notation being often misunderstood\footnote{
Spivak~\cite{spivak} chapter~4:
Classical differential geometers (and classical analysts) did not hesitate to talk
about ``infinitely small'' changes $dx^i$ of the coordinates $x^i$, just as Leibnitz had.
No one wanted to admit that this was nonsense, because true results were
obtained when these infinitely small quantities were divided into each other
(provided one did it in the right way).
%[${\pa \phi^i\over \pa X^j}\eqnote{\pa x^i\over \pa X^j}$ with duality notations].
Eventually it was realized that the closest one can come to describing an
infinitely small change is to describe a direction in which this change is supposed
to occur, \ie, a tangent vector. Since $df$ is supposed to be the infinitesimal
change of $f$ under an infinitesimal change of the point, $df$ must be a function
of this change, which means that $df$ should be a function on tangent vectors.
The $dX_i$ themselves then metamorphosed into functions, and it became clear
that they must be distinguished from the tangent vectors $\pa/\pa X_i$.
Once this realization came, it was only a matter of making new definitions,
which preserved the old notation, and waiting for everybody to catch up.
}\label{ftnoteS}: It is nothing more than $d\Phi=F$ (coordinate free notation).
% when you consider coordinates (bases and coordinates: $[d\Phi]_{|\va,\vb} = [F_{ij}]$). 

\comment{
So
\be
\label{eqdxiXj0}
\forall i=1,...,n, \quad 
d\phi_i(P) = \sumjn {\pa \phi_i\over \pa X_j}(P) \; dX_j, \qwritten
dx_i = \sumjn {\pa x_i\over \pa X_j}\,dX_j,
\ee
%for all $i=1,...,n$, 
%where ${\pa \phi_i\over \pa X_j}(P) := d\phi_i(P).\va_j$ (derivative in the $j$-th direction).
%and $[F_{ij}]=[{\pa \phi_i\over \pa X_j}]$ (Jacobian matrix).
%It means $dx_i(P).\vW = d\phi_i(P).\vW = \sumjn F_{ij}(P) \;W_j$ when $\vW=\sumjn W_j\va_j$.
which is 
}

%%%%%%%%%%%%%%%%%%%%%%%%%%%%%%%%%%%%%%%%%%%%%%%%%%%%%%%%%%%%%%%%%%%%%%%%%%%%%%%%%%%

\subsubsection{The ambiguous notation $\protect\overbigdot{d\vec x\,} = \protect\overbigdot{F}.d\vec X$}
\label{secpbcomp}

The bad notation $d\vx = F.d\vX$  gives the unfortunate and misunderstood notations
$\overbigdot{d\vx} = \overbigdot{F}.d\vX$, and then
\be
\label{eqdwdt0301m}
%\overbigdot{d\vx} = \overbigdot{F}.d\vX, \qthen 
\overbigdot{d\vx} = L.d\vx \qwhere L= \overbigdot{F}.F^{-1}. %\;\;(\mope^{\eref{eqdefL2}} d\vv).
\ee

Question: What is the meaning (and legitimate notation) of~\eref{eqdwdt0301m}? 

Answer: $\overbigdot{d\vx} = L.d\vx$ means
\be
\label{eqnadm3b}
\boxed{{D\vwtzs \over Dt} = d\vv.\vwtzs} \;\;=\hbox{evolution rate of tangent vectors along a trajectory}
\ee
see figure~\ref{figpf}.
% (since $\overbigdot{F}.F^{-1}\equalref{eqdefL2} d\vv$).
Indeed, $\vwtzs(t,p(t)) \equalref{eqrapvEivei2a} \Ftz(t,p_\tz).\vwtz(\ptz)$  gives
\be
{D\vwtzs \over Dt}(t,p(t))
= {\pa\Ftz \over \pa t}(t,\ptz).\vwtz(\ptz)
= {\pa\Ftz \over \pa t}(t,\ptz).(\Ftzt(\ptz)^{-1}.\vwtzs(t,p(t))),
%= d\vv(t,\pt).\vwtzs(t,p(t)).
\ee
\ie\ ${D\vwtzs \over Dt}(t,p(t)) = d\vv(t,p(t)).\vwtzs(t,p(t))$, \ie ~\eref{eqnadm3b}.
%And ${D\vwtzs \over Dt}(t,p(t))$ is the rate of evolution along a trajectory of the tangent vector $\vwtzs(t,p(t))$, 
In particular ${D\vwtzs \over Dt}(\tz,\ptz) = d\vv(\tz,\ptz).\vwtz(\ptz)$ is the evolution rate of tangent vectors at $\tz$ at~$\ptz$.

%%%%%%%%%%%%%%%%%%%%%%%%%%%%%%%%%%%%%%%%%%%%%%%%%%%%%%%%%%%%%%%%%%%%%%%%%%%%%%%%%%%

\subsection{Tensorial notations, warnings, remarks}
\label{secremtensnot}

%%%%%%%%%%%%%%%%%%%%%%%%%%%%%%%%%%%%%%%%%%%%%%%%%%%%%%%%%%%%%%%%%%%%%%%%%%%%%%%%%%%

%\subsubsection{Definition}

As already noted, \cf~\eref{eqtFP}, the linear map $F:=d\Phitzt(\ptz)\in\calL(\RRntz;\RRnt)$ is naturally canonically associated with the bipoint tensor $\tF \in\calL(\RRnts,\RRntz;\RR)$ defined by, for all $(\ell,\vW)\in\RRnts \times \RRntz$,
\be
\tF(\ell,\vW) := \ell.F.\vW,
\ee

Quantification of~$\tF$: basis~$(\va_i)$ with dual basis $(\pi_{ai})$ in~$\RRntz$, basis $(\vb_i)$ in~$\RRnt$:
%$d\Phi.\va_j = \sumin (d\phi_i.\va_j)\vb_i$ (where $d\phi_i.\va_j = {\pa \phi^i \over \pa x_j} = F_{ij}$): Hence
%= \sumin {\pa \phi_i \over \pa x_j}(\ptz) \vb_i, \qthus
%[d\Phi]_{[\va,\vb]} = [{\pa \phi_i \over \pa x_j}]\eqnote [F]_{[\va,\vb]}.
\be
\hbox{if}\quad F.\va_j = \sumin {\pa \phi^i \over \pa X_j} \vb_j \qthen
\label{eqvwF31} 
\tF = \sumijn {\pa \phi^i \over \pa X_j} \,\vb_i\otimes \pi_{aj}  = \sumin \vb_i\otimes d\phi_i.
\ee
And similarly
\be
\label{eqvwF0d0c}
d\tF
= \sumijkn {\pa^2 \phi^i \over \pa X_j \pa X_k}\; \vb_i \otimes (\pi_{aj} \otimes \pi_{ak})
 = \sumin \vb_i\otimes d^2\phi_i.
\ee

\noindent
{\bf Warning:}
The tensorial notation can be misleading, in particular if you use the transposed, see remark~\ref{remMHtpt}.
So, you should always use the standard notation for the linear form $F\in\calL(\RRntz;\RRnt)$ to begin with, \ie\ use
$F.\va_j = \sumjn F_{ij} \vb_i$ or $F.\vE_J=\sumijn \FiJ \ve_i$ (Marsden notations).
And only use the tensorial notations for calculations purposes at the end (after application of the proper definitions).

\debrem
In some manuscripts you find the notation $F = d\Phi \eqnote \Phi\otimes \nabla_X$.
It does not help to understand what $F$ is (it is the differential $d\Phi$),
and should not be used as far as objectivity is concerned:

$\bullet$ A differentiation is \textsl{\textbf{not}} a tensorial operation, see %\S~\ref{secmodule} and 
example~\ref{remunpeuplus}, so why use the tensor product notation $\Phi\otimes \nabla_X$, when the standard notation $d\Phi\simeq \tF=\sumin \ve_i\otimes d\phi^i$ is legitimate, explicit, objective and easy to manipulate? %, \cf~\eref{eqvwF31}?

$\bullet$ And it could be misinterpreted, since, in mechanics, $\nabla f$ is often understood to be
the vector $\sum_i {\pa f\over \pa x_i}\ve_i$ (contravariant) which needs a Euclidean dot product to be defined (which one?), 
while the differential $df$ is covariant (a differential is unmissable in thermodynamics because you can't use gradients).

$\bullet$
It gives the confusing notation $\Phi\otimes \nabla_X\otimes \nabla_X$, instead of the legitimate $d^2\Phi = \sumin \vb_i\otimes d^2\phi_i$ which is explicit, objective and easy to manipulate:
$d^2\Phi(\vU,\vW)= \sumin d^2\phi_i(\vU,\vW) \,\vb_i$.
\finrem

\debexe
Use Marsden duality notations for~\eref{eqvwF31}-\eref{eqvwF0d0c}.

\debrep
Cartesian bases, with $(dX^i)$ the (covariant) dual basis of~$(\vE_i)$: with
$\FiJ ={\pa \phi^i \over \pa X^J}$, we get
$d\Phi \eqnote \tF = \sumin \ve_i\otimes d\phi^i = \sumiJn \FiJ\,\ve_i\otimes dX^J$,
and $d^2\Phi = \sumin \ve_i\otimes d^2\phi^i
%= \sum_{i,J,K=1}^n {\pa^2 \phi^i \over \pa x^J \pa x^K}\; \ve_i \otimes (E^J \otimes E^K)
= \sum_{i,J,K=1}^n {\pa^2 \phi^i \over \pa X^J \pa X^K}\; \ve_i \otimes (dX^J \otimes dX^K)
$.
\finrep
\finexe

\comment{

%%%%%%%%%%%%%%%%%%%%%%%%%%%%%%%%%%%%%%%%%%%%%%%%%%%%%%%%%%%%%%%%%%%%%%%%%%%%%%%%%%%

\subsection{Remark: Tensorial notations}
\label{secremtensnot}

%%%%%%%%%%%%%%%%%%%%%%%%%%%%%%%%%%%%%%%%%%%%%%%%%%%%%%%%%%%%%%%%%%%%%%%%%%%%%%%%%%%

%\subsubsection{Definition}

The linear map $F:=\Ftzt(\ptz)\in\calL(\RRntz;\RRnt)$ can be canonically naturally associated with the bipoint tensor $\tF \in\calL(\RRnts,\RRntz;\RR)$ defined by, for all $(\ell,\vW)\in\RRnts \times \RRntz$,
\be
\tF(\ell,\vW) := \ell.F.\vW,
\ee
see~\S~\ref{sectroflm}.
Then, $(\pi_{ai})=(E^I)$ being the dual basis of~$(\va_i)$ (classical and duality notations),  \eref{eqvwF30} gives
$d\phi_i
= \sumjn {\pa \phi_i\over \pa x_j}\pi_{aj}
= \sumjn F_{ij}\pi_{aj}
$, thus
\be
\label{eqvwF31}
\left\{\eqalign{
\clanot: & \hbox{if }\; F.\va_j = \sumin F_{ij}\,\vb_i \qthen
\tF = \sumin \vb_i\otimes d\phi_i = \sumiJn F_{ij}\,\vb_i\otimes \pi_{aj}, \cr
\duanot: & \hbox{if }\; F.\vE_J = \sumin \FiJ\,\ve_i \qthen
\tF = \sumin \ve_i\otimes d\phi^i = \sumiJn \FiJ\,\vb_i\otimes E^j, \cr
}\right.
\ee
\comment{
And we do recover, with the contraction rule (see~\eref{eqobjcL}),
$\tF.\vW 
:= (\sumiJn F_{ij}\,\vb_i\otimes \pi_{aj}).\vW
= \sumiJn F_{ij}\,\vb_i(\pi_{aj}.\vW)
= \sumiJn F_{ij} W_j\,\vb_i = F.\vW$.
(Duality notations:  $(dX^I)$ is the dual basis of~$(\vE_I)$ and $F.\vE_J = \sumin F^i_J\,\ve_i$ and
$\tF = \sumiJn F^i_J\,\ve_i\otimes dX^J$.)
}
and
\be
\label{eqvwF0d0c}
\left\{\eqalign{
\clanot: & d^2\Phi \mathop{\simeq}^{\rm isom} d\tF \simeq \sumin \vb_i\otimes d^2\phi_i
= \sumijkn {\pa^2 \phi^i \over \pa x_j \pa x_k}\; \vb_i \otimes \pi_{aj} \otimes \pi_{ak}, \cr
\duanot: & d^2\Phi \mathop{\simeq}^{\rm isom} d\tF \simeq \sumin \ve_i\otimes d^2\phi^i
= \sum_{i,J,K=1}^n {\pa^2 \phi^i \over \pa X^J \pa X^K}\; \ve_i \otimes E^J \otimes E^K. \cr
}\right.
\ee
Of course we recover~\eref{eqvwF0d0d} since 
$d\tF(\vU,\vW)
= (\sumin \vb_i\otimes d^2\phi_i)(\vU,\vW)
= \sumin \vb_i (d^2\phi_i(\vU,\vW))
= \sumijkn ({\pa^2 \phi^i \over \pa x_j \pa x_k}U_jW_k) \vb_i
$
as well as
$d\tF(\vU,\vW)
= (\sumijkn {\pa^2 \phi^i \over \pa x_j \pa x_k}\; \vb_i \otimes \pi_{aj} \otimes \pi_{ak})(\vU,\vW)
= \sumijkn {\pa^2 \phi^i \over \pa x_j \pa x_k}\; \vb_i (\pi_{aj}.\vU)(\pi_{ak}.\vW)
= \sumijkn {\pa^2 \phi^i \over \pa x_j \pa x_k}U_jW_k \vb_i
$.
(Use duality notations if you prefer.)

\debrem
In some manuscripts you find the notation $F = d\Phi \eqnote \Phi\otimes \nabla_X$.
It does not help to understand what $F$ is (it is the differential $d\Phi$),
and should not be used as far as objectivity is concerned:

$\bullet$ A differentiation is \textsl{\textbf{not}} a tensorial operation, see %\S~\ref{secmodule} and 
remark~\ref{remunpeuplus}, so why use the tensor product notation $\Phi\otimes \nabla_X$, when the standard notation $d\Phi\simeq \tF=\sumin \ve_i\otimes d\phi^i$ is legitimate, explicit and easy to manipulate? %, \cf~\eref{eqvwF31}?

$\bullet$ It could be misinterpreted, since, in mechanics, $\nabla$ is often understood to be the gradient (a gradient needs a Euclidean dot product: Which one?) which is contravariant, 
while a differential is covariant (a differential is unmissable in thermodynamics).

$\bullet$
It gives the confusing notation $\Phi\otimes \nabla_X\otimes \nabla_X$, instead of the legitimate~\eref{eqvwF0d0c} which is explicit and easy to manipulate.
\finrem

}

%%%%%%%%%%%%%%%%%%%%%%%%%%%%%%%%%%%%%%%%%%%%%%%%%%%%%%%%%%%%%%%%%%%%%%%%%%%%%%%%%%%

\subsection{Change of coordinate system at~$t$ for~$F$}
\label{seccpgdd2}

Let $\ptz\in\Omegatz$, $\pt=\Phitzt(\ptz)\in \Omegat$,
$\vW(\ptz)\in\RRntz$, $\vw(\pt)  = \Ftzt(\ptz).\vW(\ptz) \in \RRnt$ (its push-forward), written
$\vw  = F.\vW$ for short. %(abusive misleading notation: Should be written $\vw\circ\Phi = F.\vW$).
The observer at~$\tz$ used a basis $(\va_i)$ in~$\RRntz$.
At~$t$, in~$\RRnt$, a first observer chooses a Cartesian basis $(\vbio)$,
and a second observer chooses a Cartesian basis $(\vbin)$.
Let $P=[P_{ij}]$ be the transition matrix from~$(\vbio)$ to~$(\vbin)$, \ie\ $\vbjn = \sumin P_{ij}\vbio$ for all~$j$.
The change of basis formula for vectors from~$(\vbio)$ to~$(\vbin)$ (in~$\RRnt$) gives
\be
\label{eqcbvw}
[\vw]_{|\vbn} = P^{-1}.[\vw]_{|\vbo}, \qthus
[F.\vW]_{|\vbn} = P^{-1}.[F.\vW]_{|\vbo}.
\ee
Thus
\be
[F]_{|\va,\vbn}.[\vW]_{|\va}
=  P^{-1}.[F]_{|\va,\vbo}.[\vW]_{|\va},
\ee
true for all $\vW$, thus
\be
\label{eqFtztnew}
\boxed{[F]_{|\va,\vbn} = P^{-1}.[F]_{|\va,\vbo}}.
\ee
NB: \eref{eqFtztnew} is \textslbf{not} the change of basis formula $[L]_{|\new} = P^{-1}.[L]_{|\old}.P$ for endomorphisms,
which would be nonsense since $F:=\Ftzt(\ptz):\RRntz \rar \RRnt$ is not an endomorphism; %, \cf~\eref{eqnoteFgd}: 
\eref{eqFtztnew} is just the usual change of basis formula for vectors $\vw$ in~$\RRnt$, \cf~\eref{eqcbvw}.

%%%%%%%%%%%%%%%%%%%%%%%%%%%%%%%%%%%%%%%%%%%%%%%%%%%%%%%%%%%%%%%%%%%%%%%%%%%%%%%%%%%
%%%%%%%%%%%%%%%%%%%%%%%%%%%%%%%%%%%%%%%%%%%%%%%%%%%%%%%%%%%%%%%%%%%%%%%%%%%%%%%%%%%

%\newpage

%\section{More on the deformation gradient $F$}

%We concentrate on the linear map~$\Ftzt(P) = d\Phitzt(P) : \RRntz \rar \RRnt$ and its action (push-forward) expressed in bases (quantification).

%%%%%%%%%%%%%%%%%%%%%%%%%%%%%%%%%%%%%%%%%%%%%%%%%%%%%%%%%%%%%%%%%%%%%%%%%%%%%%%%%%%

\subsection{Spatial Taylor expansion of $\Phi$ and~$F$}

$\Phitzt\eqnote \Phi$ is supposed to be~$C^2$ for all~$\tz,t$.
Let $P\in\Omegatz$, $d\Phi = F$, and $\vW\in\RRntz$ vector at~$P$. Then, in~$\Omegat$,
\be
\label{eqdefFt0}
\eqalign{
\Phi(P{+}h\vW)
%= & \Phi(P) + h\, d\Phi(P).\vW + o(h) \cr
= & \Phi(P) + h\, F(P).\vW + {h^2 \over 2}\, dF(P)(\vW,\vW) + o(h). \cr
}
\ee
\comment{
And $F(P{+}h\vW) = F(P) + h\, dF(P).\vW + o(h)$. % \;\in\calL(\RRntz;\RRnt),
And
\be
\eqalign{
\Phi(P{+}h\vW)
%= & \Phi(P) + h\, d\Phi(P).\vW + {h^2 \over 2}\, d^2\Phi(P)(\vW,\vW) + o(h^2) \cr
= & \Phi(P) + h\, F(P).\vW + {h^2 \over 2}\, dF(P)(\vW,\vW) + o(h^2). \cr
}
\ee
}

%%%%%%%%%%%%%%%%%%%%%%%%%%%%%%%%%%%%%%%%%%%%%%%%%%%%%%%%%%%%%%%%%%%%%%%%%%%%%%%%%%%

\subsection{Time Taylor expansion of $F$}

The motion $\tPhi$ is supposed to be $C^3$. $\tz\in\RR$, $\Phitz$ be the associated motion,
$p(t) = \tPhi(t,\Pobj)$ and $\ptz = \tPhi(\tz,\Pobj)$, with $\vv(t,\pt) = {\pa \tPhi \over \pa t}(t,\Pobj)$ the Eulerian velocity and $\vVtz(t,\ptz) := {\pa \Phitz \over \pa t}(t,\ptz) = \vv(t,\pt)$ the Lagrangian velocity, and $\Ftzptz(t)=\Ftz(t,\ptz) = d\Phitz(t,\ptz) \eqnote F(t)$.
We have
\be
\label{eqdlF2}
\eqalign{
{\pa\Ftz \over \pa t}(t,\ptz) 
= &{\pa (d\Phitz) \over \pa t}(t,\ptz)
= d({\pa \Phitz \over \pa t})(t,\ptz) 
%= {\pa (d\Phitz) \over \pa t}(t,\ptz) 
\cr
= & d\vVtz(t,\ptz) = d\vv(t,p(t)).F(t), \qinshort \boxed{\overbigdot F=d\vV=d\vv.F}.
}
\ee
Then ${\pa^2 \Phitz \over \pa t^2}(t,\ptz) = \vAtz(t,\ptz) = \vgamma(t,p(t))$
(Lagrangian and Eulerian accelerations),
hence
\be
\label{eqdl2wadz0a}
{\pa^2\Ftz \over \pa t^2}(t,\ptz) = d\vAtz(t,\ptz) = d\vgamma(t,p_t).F(t), \qinshort \overbbigdot F=d\vA=d\vgamma.F.
\ee

Thus, the second order time Taylor expansion of $\Ftzptz\eqnote F$ is, in the vicinity of~$t$,
\be
\label{eqdPhi2a}
\eqalign{
F(t{+}h)
= & F(t) + h\, d\vV(t) + {h^2\over 2}\, d\vA(t) + o(h^2) \cr
= & \Bigl(I + h\, d\vv + {h^2\over 2}\, d\vgamma\Bigr)(t,p(t)).F(t) + o(h^2)
\qwhen p(t)=\Phitz(t,\ptz). \cr
}
\ee
NB: They are \textslbf{three} times are involved: $t$ and $t{+}h$ as usual, and $\tz$ through $F:=\Ftzptz$, $\vV:=\vVtzptz$ and $\vA:=\vAtzptz$ (observer dependent), as for~\eref{eqttexp1}.

In particular $F(\ptz):=F^\tz_\tz(\ptz)=I$ gives, in the vicinity of~$\tz$,
\be
\label{eqdPhi2b}
F(\tz{+}h)
=  \bigl(I + h\, d\vv + {h^2\over 2}\, d\vgamma\bigr)(\tz,\ptz) + o(h^2).
\ee

\debrem
$\gamma = {\pa \vv \over \pa t} + d\vv.\vv$ is not linear in~$\vv$. Idem,
\be
\label{eqvgnlv}
\eqalign{
d\vgamma
= & d({D\vv \over Dt}) = d({\pa\vv \over \pa t} +d\vv.\vv)
= d{\pa \vv\over \pa t} + d^2\vv.\vv + d\vv.d\vv 
\quad (=  {D(d\vv) \over Dt} + d\vv.d\vv)
} % , \quad\hbox{non linéaire en~$\vv$}.
\ee 
is non linear in~$\vv$, and gives
$
\Ftzptz''(t) = (d{\pa \vv\over \pa t} + d^2\vv.\vv + d\vv.d\vv)(t,p_t).\Ftzptz(t),
$
non linear in~$\vv$.
\finrem

\debexe
Directly check that $F'=d\vv.F$ gives $F''=d\vgamma.F$.

\debrep
$F'(t)=d\vv(t,p(t)).F(t)$ gives
$F''(t) = {D(d\vv) \over Dt}(t,p(t)).F(t) + d\vv(t,p(t)).F'(t)$
with ${D(d\vv) \over Dt} = d\vgamma - d\vv.\vv$, \cf~\eref{eqvgnlv},
thus
$F''(t) = (d\vgamma-d\vv.d\vv)(t,p(t)).F(t) + d\vv(t,p(t)). d\vv(t,p(t)).F(t) = d\vgamma(t,p(t)).F(t)$. 
\finrep
\finexe

%%%%%%%%%%%%%%%%%%%%%%%%%%%%%%%%%%%%%%%%%%%%%%%%%%%%%%%%%%%%%%%%%%%%%%%%%%%%%%%%%%%
%%%%%%%%%%%%%%%%%%%%%%%%%%%%%%%%%%%%%%%%%%%%%%%%%%%%%%%%%%%%%%%%%%%%%%%%%%%%%%%%%%%

\subsection{Homogeneous and isotropic material}
\label{secmath}

\def\func{\vec{\hbox{fun}}}

%Framework: linear modelization and elastic materials.
Let $P\in\Omegatz$, let $\Ftzt(P) := d\Phitzt(P)$;
Suppose that the ``Cauchy stress vector'' $\vf_t(\pt)$
à~$t$ at $\pt = \Phitzt(P)$ only depends on $P$, \ie\ there exists a function $\func$ such that
\be
\label{eqvecf}
\vf_t(\pt) = \func(P,\Ftzt(P)).
\ee

\debdef
A material is homogeneous iff $\func$ doesn't depend on the first variable~$P$,
i.e., iff, for all $P \in \Omega_\tz$,
\be
\func(P,\Ftzt(P)) = \func(\Ftzt(P)).
\ee
(Same mechanical property at any point.)
\findef

\debdef (Isotropy.)
Consider a Euclidean dot product, the same at all time.
A material is isotropic at $P\in\Omegatz$ iff $\func$
is independent of the direction you consider,
\ie, iff, for any rotation $R_\tz(P)$ in~$\RRntz$,
\be
\func(P,\Ftzt(P) = \func(P,\Ftzt(P).R_\tz(P)).
\ee
(Mechanical property unchanged when rotating the material first.)
\findef

\debdef
A material is isotropic homogeneous iff it is isotropic and homogeneous.
\findef

%%%%%%%%%%%%%%%%%%%%%%%%%%%%%%%%%%%%%%%%%%%%%%%%%%%%%%%%%%%%%%%%%%%%%%%%%%%%%%%%%%%

\subsection{The inverse of the deformation gradient}

%%%%%%%%%%%%%%%%%%%%%%%%%%%%%%%%%%%%%%%%%%%%%%%%%%%%%%%%%%%%%%%%%%%%%%%%%%%%%%%%%%%

%\subsection{Definition of $H = F^{-1}$}

%%%%%%%%%%%%%%%%%%%%%%%%%%%%%%%%%%%%%%%%%%%%%%%%%%%%%%%%%%%%%%%%%%%%%%%%%%%%%%%%%%%

%\subsubsection{Definition of $\Htzt$}

\comment{
Let $\tz,t \in \RR$, % (fixed).
$\Phitzt :
\left\{\eqalign{
\Omegatz & \rar \Omegat \cr
P & \rar p= \Phitzt(P)
}\right\}
$.
Thus 
$(\Phitzt)^{-1} :
\left\{\eqalign{
\Omegat & \rar \Omegatz \cr
p & \rar P= (\Phitzt)^{-1}(p)
}\right\}
$.
}

$((\Phitzt)^{-1}\circ\Phitzt)(P) = P$ gives, with $p=\Phitzt(P)$,
\be
d(\Phitzt)^{-1}(p). d\Phitzt(P) = I_\tz,\qthus
d(\Phitzt)^{-1}(p) = d\Phitzt(P)^{-1} = \Ftzt(P)^{-1}, %,\qie F(P)^{-1} \eqnote F^{-1}(p).
\ee
where $\Ftzt = d\Phitzt$ is the deformation gradient. 
We have thus define the two point tensor
\be
\label{eqHtz}
%d(\Phitzt)^{-1} \eqnote 
\Htzt := (\Ftzt)^{-1} :
\left\{\eqalign{
\Omegat & \rar \calL(\RRnt;\RRntz) \cr
p & \rar %(\Ftzt)^{-1}(p) = 
\boxed{\Htzt(p) = (\Ftzt)^{-1}(p) := (\Ftzt(P))^{-1}}  \qwhen p = \Phitzt(P).
}\right.
\ee
So % (in details: for all $\vw(p)\in T_p(\Omegat)$),
\be
%(\vW(P) =)\quad  %(\Ftzt)^{-1}(p).\vw(p) = 
\Htzt(p).\vw(p) = (\Ftzt)^{-1}(p).\vw(p) := \Ftzt(P)^{-1}.\vw(p) \;\in\RRntz, \qinshort H.\vw=F^{-1}.\vw,
\ee
for all $\vw(p)\in\RRnt$ vector at~$p$. This defines, with $\pt = \Phitz(t,P)$,
\be
\label{eqHtz1}
\Htz :
\left\{\eqalign{
\bigC=\bigcup_t(\{t\} \times \Omegat) &\rar \calL(\RRnt;\RRntz) \cr
(t,\pt) & \rar \Htz(t,\pt) \eqdef \Htzt(\pt) = (\Ftz(t,P))^{-1}.
}\right.
\ee
NB: $\Htz$ looks like a Eulerian map, but isn't:
$\Htz$ depends on a initial time~$\tz$ and is a two point tensor (starts in~$\RRnt$, arrives in~$\RRntz$). We will however use the material time derivative ${D\over Dt}$ notation in this case, that is, we define, along a trajectory $t\rar p(t) = \Phitz(t,P)$,
\be
{D\Htz \over Dt}(t,p(t)) := {\pa \Htz \over \pa t}(t,p(t)) + d\Htz(t,p(t)).\vv(t,p(t)),
\qie {D\Htz \over Dt} = {\pa \Htz \over \pa t} + d\Htz.\vv,
\ee
which is the time derivative $g'(t)$ of the function $g : t \rar g(t)=\Htz(t,\Phitz(t,P))$ (\ie\ $g(t)=\Htz(t,p(t))$).

Hence, with $p(t)=\Phitz(t,P)$ and $\Htz(t,p(t)).\Ftz(t,P) = I_\tz$, written $H.F=I$, we get
\be
\label{eqdPhi3}
{DH \over Dt}.F + H.{\pa F \over \pa t} = 0, \qthus {DH \over Dt} = - H.d\vv, % = - H.{\pa F \over \pa t}.F^{-1}
\ee
since ${\pa F \over \pa t}(t,P).F^{-1}(t,p(t)) = d\vv(t,p(t))$ \cf~\eref{eqdlF2}.

\debexe
With $\vwtzs(t,p(t)) = \Ftz(t,P).\vW(P)$, \ie\ $\Htz(t,p(t)).\vwtzs(t,p(t))=\vW(P)$, when $p(t) = \Phitz(t,P)$,
prove~\eref{eqdPhi3}.

\debrep
${D\vwtzs \over Dt}(t,p(t)) =d\vv(t,p(t)).\vwtzs(t,p(t))$, \cf~\eref{eqnadm3b};
And $(\Htz.\vwtzs)(t,p(t))=\vW(P)$ gives
${D\Htz \over Dt}.\vwtzs + \Htz.{D\vwtzs \over Dt} = 0$;
Thus ${D\Htz \over Dt}.\vwtzs + \Htz.d\vv.\vwtzs = 0$,
thus ${D H \over Dt} = - H.d\vv$.
\finrep
\finexe

\debexe
Prove: $\Htzt = H^\tz_{t_1} \circ H^{t_1}_t$ and ${D\Htz \over Dt}(t,p(t)) = H^\tz_{t_1}(p_{t_1}).{ DH^{t_1} \over Dt}(t,p(t))$ for all $\tz,t_1$ with $p_{t_1}=\Phi^\tz_{t_1}(p_\tz)$.

\debrep
We have $\Phitzt(p_\tz) = \Phi^{t_1}_{t}(\Phi^\tz_{t_1}(p_\tz))$, \cf~\eref{eqcompf0b0},
hence $\Ftzt(\ptz) = F^{t_1}_t(p_{t_1}).F^\tz_{t_1}(\ptz)$,
thus  $\Ftzt(\ptz)^{-1} = F^\tz_{t_1}(\ptz)^{-1}.F^{t_1}_t(p_{t_1})^{-1}$,
\ie\ $\Htzt(\pt) = H^\tz_{t_1}(p_{t_1}). H^{t_1}_t(p(t))$,
thus, $\Htz(t,p(t)) = H^\tz_{t_1}(p_{t_1}). H^{t_1}(t,p(t))$,
thus ${D\Htz \over Dt}(t,p(t)) = H^\tz_{t_1}(p_{t_1}).{ DH^{t_1} \over Dt}(t,p(t))$.
%Hence the result.
\finrep
\finexe

\comment{
\debexe
Montrer que $(F^{-1})' = - F^{-1}.d\vv$ redonne $(F^{-1})'' = F^{-1}.(2d\vv.d\vv - d\vgamma)$.

\debrep
$(F^{-1})'' = - (F^{-1})'.d\vv - F^{-1}.{D(d\vv) \over Dt}
= F^{-1}.d\vv.d\vv -F^{-1}.(d\vgamma - d\vv.d\vv)$, \cf~\eref{eqvgnlv}.
% Ou bien $F.(F^{-1})' = - d\vv$ donne $F'.(F^{-1})' + F.(F^{-1})'' = -{D(d\vv) \over Dt}$, soit $-d\vv.F.F^{-1}.d\vv + F.(F^{-1})'' = -(d\vgamma - d\vv.d\vv)$
\finrep
\finexe
}

\comment{
\debexe
Donner le développement formel au second ordre de~$F^{-1}$,
et vérifier formellement que ce développement et~\eref{eqdPhi2a}
redonnent $F^{-1}.F=I$, au second ordre près.

\debrep
\eref{eqdPhi3} donne le développement limité formel au second ordre en temps au voisinage de~$t$ :
\be
\label{eqdPhi2b}
\eqalign{
\FtzP(t{+}h)^{-1} 
= &\FtzP(t)^{-1}.(I - h\,d\vv - {h^2 \over 2}\,(d\vgamma - 2d\vv.d\vv))(t,\pt) + o(h^2). \cr
}
\ee
C'est formel car
$(\FtzP)^{-1}(t{+}h) : \vec\RR^n_{t+h} \rar\RRntz$
et $(\FtzP)^{-1}(t) : \vec\RR^n_{t} \rar\RRntz$,
et ce calcul nécessite une structure cartésienne
permettant d'avoir $\vec\RR^n_{t+h} =\RRnt \eqnote \vRRn$.
%(Pour le calcul sur les variétés voir les dérivées de Lie).
Donc (notations allégées) :
$$
\eqalign{
\FtzP(t{+}h)^{-1}.\FtzP(t{+}h)
& = [F^{-1}.(I - h\,d\vv - {h^2 \over 2}\,(d\vgamma - 2d\vv.d\vv)) + o(h^2)]
.[\bigl(I + h\, d\vv + {h^2\over 2}\, d\vgamma\bigr).F + o(h^2)]
 \cr
&= F^{-1}.(I + h\,(-d\vv+d\vv) + {h^2\over 2}(d\vgamma - 2d\vv.d\vv -  (d\vgamma - 2d\vv.d\vv)) + o(h^2)).F = I + o(h^2).
}
$$
\finrep
\finexe
}

%\newpage
%%%%%%%%%%%%%%%%%%%%%%%%%%%%%%%%%%%%%%%%%%%%%%%%%%%%%%%%%%%%%%%%%%%%%%%%%%%%%%%%%%%
%%%%%%%%%%%%%%%%%%%%%%%%%%%%%%%%%%%%%%%%%%%%%%%%%%%%%%%%%%%%%%%%%%%%%%%%%%%%%%%%%%%

\section{Flow}
\label{secflot}

%%%%%%%%%%%%%%%%%%%%%%%%%%%%%%%%%%%%%%%%%%%%%%%%%%%%%%%%%%%%%%%%%%%%%%%%%%%%%%%%%%%

\subsection{Introduction: Motion versus flow}

\leavevmode

$\bullet$
Motion: A motion $\tPhi : (t,\Pobj) \rar \pt = \tPhi(t,\Pobj)$ locates at~$t$ a particle $\Pobj$ in the affine space~$\RRn$, \cf~\eref{eqdeftPhi0}; From which the Eulerian velocity field~$\vv$ is deduced:
$\vv(t,\pt) := {d\tPhiPobj\over dt}(t,\Pobj)$, \cf~\eref{eqdefve}.

$\bullet$ Flow: A flow starts with a Eulerian velocity field $\vv$, from which we deduce a motion
by solving the ODE (ordinary differential equation) ${d\Phi \over d t}(t) = \vv(t,\Phi(t))$.
%(a flow deals with Eulerian velocities).
%by solving the ordinary differential equation ${d\Phi \over d t}(t) = \vv(t,\Phi(t))$ (with the Cauchy--Lipschitz theorem).

%The solution $\Phi \eqnote \Phitzptz$ (``integral curve'' of~$\vv$) is then the motion $\tPhiPobj$ of a particle $\Pobj$ animated with the velocity $\vv(t,\pt)$ at~$t$ at $\pt = \tPhi(t,\Pobj)$.
% (and  with the velocity $\vv(\tz,\ptz)$ at~$\tz$ at $\ptz=\tPhi(\tz,\Pobj) = \Phitzptz(\tz)$).
%. Here $\tz$ is any time sufficiently close to~$t$ (for a solution to exists in the vicinity of~$t$): So $\tz$ is not interpreted as a ``Lagrangian time''.

%%%%%%%%%%%%%%%%%%%%%%%%%%%%%%%%%%%%%%%%%%%%%%%%%%%%%%%%%%%%%%%%%%%%%%%%%%%%%%%%%%%

\subsection{Definition}

%Let $\Omega$ be an open set in~$\RRn$. % (more generally in a differential manifold).
Let $\vv:
\left\{\eqalign{
\RR\times\RRn & \rar \vRRn \cr
(t,p) & \rar \vv(t,p) \cr
}\right\}
$ be a unstationary vector field (\eg, a Eulerian velocity field which definition domain is $\bigC=\bigcup_{t\in[t_1,t_2]} (\{t\}\times \Omegat)$). %
%The domain of definition of $\vv$ is \eg\ a set $\bigC=\bigcup_{t\in]t_1,t_2[} (\{t\} \times \Omegat)$, \cf~\eref{eqbigU}.
%, where the $\Omegat$ are open subsets in~$\RRn$.
%Et dans le cas d'une variété différentiable on remplace $\vRRn$ par le fibré tangent.
We look for maps
$\Phi : 
\left\{\eqalign{
\RR & \rar  \RRn \cr
t & \rar p=\Phi(t) 
}\right\}
$
which are locally (\ie\ in the vicinity of some~$\tz$) solutions of the ODE (ordinary differential equation)
\be
\label{eqcl0}
{d\Phi\over d t}(t) = \vv(t,\Phi(t)),\quad\hbox{also written}\quad
{dp\over d t}(t) = \vv(t,p(t)), \qor {d\vx\over d t}(t) = \vv(t,\vx(t))
\ee
where $\vx(t)=\ora{\calO p(t)}$ after a choice of an origin. Also written ${dp\over d t} = \vv(t,p)$ or ${d\vx\over d t} = \vv(t,\vx)$.

\debdef
\label{defflow}
A solution $\Phi$ of~\eref{eqcl0} is a flow of~$\vv$;
Also called an integral curve of~$\vv$ since \eref{eqcl0} also reads
$\Phi(t) = \int_{\tau=t_1}^t \vv(\tau,\Phi(\tau))\,d\tau + \Phi(t_1)$.
\findef

\debrem
Improper notation for~\eref{eqcl0}: 
\be
\label{eqabsu}
{dp\over d t}(t) \eqnote {dp(t)\over d t} \qquad (= \vv(t,p(t))) .
\ee

Question: If the notation ${dp(t)\over d t}$ is used, then  what is the meaning of ${dp(f(t))\over d t}$?

Answer: It means,
either ${dp\over d t}(f(t))$,
or ${d(p\circ f) \over dt}(t) = {dp\over d t}(f(t)){df\over dt}(t)$: Ambiguous.
So it is better to use ${dp\over d t}(t)$, and to avoid ${dp(t)\over d t}$,
unless the context is clear (no composite functions).
\finrem

\comment{
\debrem
Another improper notation for~\eref{eqcl0}: 
${dp\over d t} = \vv(t,p)$,
which does not mean ${dp\over d t}(u) = \vv(t,p(u))$ for any~$u$ (used for streamlines, \cf~\eref{eqlc1}),
but means ${dp\over d t}(t) = \vv(t,p(t))$.
\finrem
}

%%%%%%%%%%%%%%%%%%%%%%%%%%%%%%%%%%%%%%%%%%%%%%%%%%%%%%%%%%%%%%%%%%%%%%%%%%%%%%%%%%%

\subsection{Cauchy--Lipschitz theorem}

Let $(\tz,\ptz)$ be in the definition domain of~$\vv$. %, and $(\tz,\ptz)$ will be called an initial condition.
We look for $\Phi$ solution of ``the ODE with initial condition $(\tz,\ptz)$'', in some vicinity of~$\tz$, \ie\ such that
\be
\label{eqcl}
{d\Phi\over d t}(t) = \vv(t,\Phi(t)) \qand \Phi(\tz)=\ptz.
\ee
(The couple $(\tz,\ptz)$ is the initial condition, and the values $\tz$ and $\ptz$ are the initial conditions.)

%\debdef A solution of~\eref{eqcl} is called a flow of~$\vv$. \findef

\debdef
\label{defclip}
Let $t_1,t_2\in\RR$, $t_1<t_2$.
Let $\Omega$ be an open set in~$\RRn$ and $\bOmega$ its closure supposed to be a regular domain.
Let $||.||$ be a norm in~$\vRRn$.
A continuous map $\vv : [t_1,t_2]\times\bOmega \rar \vRRn$
is Lipschitzian iff it is ``space Lipschitzian, uniformly in time'', that is, iff
\be
\label{eqclip}
\exists k>0, \; \forall t\in [t_1,t_2], \; \forall p,q \in \bOmega, \; ||\vv(t,q) - \vv(t,p)||\le k ||q-p||.
\ee
So, ${||\vv_t(q) - \vv_t(p)|| \over  ||q-p|| }\le k$, for all $t$ and all $p\ne q$
(the variations of $\vv$ are bounded in space, uniformly in time).
\findef

\begin{theorem} [and definifion] (Cauchy--Lipschitz).
\label{propCL}
If $\vv : [t_1,t_2]\times\bOmega \rar \vRRn$ is Lipschitzian and $(\tz,\ptz)\in]t_1,t_2[ \times \Omega$, then there exists $\eps = \eps_{\tz,\ptz}>0$ \st~\eref{eqcl} has a unique solution
$\Phi:]\tz{-}\eps,\tz{+}\eps[\rar \RRn$, noted~$\Phitzptz$:
\be
\label{eqclb}
{d\Phitzptz\over d t}(t) = \vv(t,\Phitzptz(t)) \qand \Phitzptz(\tz)=\ptz.
\ee
Moreover, if  $\vv$ is~$C^k$ then $\Phi$ is~$C^{k+1}$.
\finthm

\debdem
See \eg\ Arnold~\cite{arnolded}, or any ODE course.
In particular
$\ds ||\vv||_\infty \eqdef \sup_{t\in]\tz{-}\eps,\tz{+}\eps[,\; p\in\Omega}||\vv(t,p)||_\RRn$
(maximum speed) exists since $\vv\in C^0$ on the compact $[t_1,t_2]\times \bOmega$),
see definition~\ref{defclip},
hence we can choose $\eps = \min(\tz{-}t_1,t_2{-}\tz,{ d(\ptz,\pa\Omega) \over ||\vv||_\infty})$
(the time needed to reach the border~$\pa\Omega$ from~$\ptz$).
\findem

We have thus defined the function, also called ``a flow'',
\be
\label{eqdeff3}
\Phi : 
\left\{\eqalign{
]t_1,t_2[\times ]t_1,t_2[ \times\, \Omegatz & \rar  \Omega \cr
(t,\tz,\ptz) & \rar p = \Phi(t,\tz,\ptz) := \Phitzptz(t) \eqnote \Phi(t;\tz,\ptz) .
}\right.
\ee
And~\eref{eqclb} reads
\be
\label{eqdeff10}
{\pa \Phi\over \pa t}(t;\tz,\ptz) = \vv(t,\Phi(t;\tz,\ptz)),\qwith \Phi(\tz;\tz,\ptz)=\ptz.
\ee
We have thus defined the function, also called ``a flow'',
\be
\label{eqdefphitz}
\Phitz :
\left\{\eqalign{
[\tz{-}\eps,\tz{+}\eps] \times \Omegatz & \rar \RRn \cr
(t,\ptz) & \rar p=\Phitz(t,\ptz) \eqdef \Phitzptz(t): \cr
}\right.
\ee
And~\eref{eqclb} reads
\be
\label{eqcl2}
{\pa \Phitz\over \pa t}(t,\ptz) = \vv(t,\Phitz(t,\ptz)),\qand \Phitz(\tz,\ptz)=\ptz.
\ee
Other definition and notation (can be ambiguous): $\Phi_{t;\tz} = \Phitzt : \Omegatz \rar  \RRn$, and \eref{eqdeff10} is written
\be
\label{eqdeff1}
{d\Phi_{t;\tz}(\ptz)\over d t} = \vv(t,\Phi_{t;\tz}(\ptz)), \qand \Phi_{\tz;\tz}(\ptz)=\ptz.
\ee

\debthm
\label{thmexfl}
Let $\vv$ be Lipschitzian, let $\tz\in]t_1,t_2[$, and let $\Omegatz$ be an open set \st\ $\Omegatz\subset\subset \Omega$ (\ie\ there exists a compact set $K \in \RRn$ \st\ $\Omegatz\subset K\subset\Omega$).
%And let $K$ be such a compact.
Then there exists $\eps>0$ \st\ a flow $\Phitz$ exists on $]\tz{-}\eps,\tz{+}\eps[\times \Omegatz$.
\finthm

\debdem
Let $d= d(K,\RRn{-}\Omega)$ (la distance of $K$ to the border of~$\Omega$.

Let $\ds ||\vv||_\infty \eqdef \sup_{t\in[t_1,t_2],p\in \bOmega}||\vv(t,p)||_\RRn$
(exists since $\vv\in C^0$ on the compact $[t_1,t_2]\times \bOmega$).

Let $\eps=\min(\tz{-}t_1 , t_2{-}\tz,{d\over||\vv||_\infty})$
(less that the minimum time to reach the border from~$K$ at maximum speed $||v||_\infty$).

Let $\ptz\in K$ and $t \in ]\tz{-}\eps,\tz{+}\eps[$. Then $\Phitzptz$ exists, \cf theorem~\ref{propCL},
and
$||\Phitzptz(t)-\Phitzptz(\tz)||_\RRn
\le [t-\tz|\,\sup_{\tau\in ]\tz{-}\eps,\tz{+}\eps[}(||(\Phitzptz)'(\tau)||_\RRn)$
(mean value theorem since, $\vv$ being~$C^0$, $\Phi$ is~$C^1$).
Thus
$||\Phitzptz(t)-\Phitzptz(\tz)||_\RRn \le  [t-\tz|\, ||v||_\infty$,
thus $\Phitzptz(t) \in \Omega$. 
Thus $\Phitzptz$ exists on~$]\tz{-}\eps,\tz{+}\eps[$, for all $\ptz\in K$.
%Thus $\Phitz$ exists on~$]\tz{-}\eps,\tz{+}\eps[\times K$.
\findem

\comment{
\debrem
To consider a flow allows to consider any local time approach: for any $t \in ]t_1,t_2[$,
for $\tau$ close to~$t$,
\be
\label{eqcl2l}
\left\{\eqalignrll{
&{\pa \Phi^t\over \pa \tau}(\tau,\pt) = \vv(\tau,\Phi^t(\tau,\pt))
&\quad (={\pa \Phi\over \pa \tau}(\tau;t,\pt) = (\Phi^t_\pt)'(\tau)), \cr
&\Phi^t(t,\pt)=\pt &\quad (=(\Phi(t;t,\pt)= \Phi^t_\pt(t)).
}\right.
\ee
This is also referred to as an ``updated Lagrangian setup''.
\finrem
}

\debrem
The definition of a flow starts with a Eulerian velocity (independent of any initial time),
and then, due to the introduction of initial conditions, leads to the Lagrangian functions~$\Phitz$, \cf~\eref{eqdefphitz}.
%But remember that an initial ODE purely deals with the Eulerian equation~\eref{eqcl0}.
Once again, Lagrangian functions are the result of Eulerian functions.
%: And in this manuscript we were careful to introduce~\S~\ref{seceul} (Eulerian) before~\S~\ref{seclag} (Lagrangian).
\finrem

%%%%%%%%%%%%%%%%%%%%%%%%%%%%%%%%%%%%%%%%%%%%%%%%%%%%%%

\subsection{Examples}

%%%%%%%%%%%%%%%%%%%%%%%%%%%%%%%%%%%%%%%%%%%%%%%%%%%%%%

\paragraph{Example 1}
%\label{secexa}
$\RR^2$ with an origin $\calO$, a Euclidean basis $(\ve_1,\ve_2)$ and $\Omega=[0,2]\times[0,1]$ (observation window).
Let $p\in\RR^2$, $\ora{\calO p}\eqnote \vx = x\ve_1+y\ve_2 \eqnote (x,y)$.
Let $t_1=-1$, $t_2=1$, $\tz\in ]t_1,t_2[$, $a,b\in\RR$, $a\ne0$, and
\be
\label{eqfigv0}
\vv(t,p)
%\eqnote \vv(t,x,y)
=\left\{\eqalign{
& %v^1(t,p) \eqnote 
v^1(t,x,y) = a y , \cr
& %v^2(t,p) \eqnote 
v^2(t,x,y) = b \sin (t{-}\tz). \cr
}\right.
\ee
%voir figure~\ref{figcisailiv}. 
($b=0$ corresponds to the stationary case = shear flow.) %$\vv$ is a $C^\infty$ vector field.
$\vx(\tz)=\pmatrix{x_0 \cr y_0}$,
$\vx(t)= \pmatrix{x(t) \cr y(t)} = \ora{\calO \Phitzptz(t)}$ and \eref{eqcl2} give
\be
\left\{\eqalign{
& {d x\over dt}(t) = v^1(t,x(t),y(t)) = a y(t), \cr
& {d y\over dt}(t) = v^2(t,x(t),y(t)) = b \sin (t{-}\tz), \cr
}\right. \qwith
\left\{\eqalign{
& x(\tz)=x_0,\cr
& y(\tz)=y_0.\cr
}\right.
\ee
Thus
\be
\label{eqexacisi10}
\vx(t)=\ora{\calO p(t)} = \ora{\calO \Phitzptz(t)}
= \pmatrix{
x(t) = x_0 + a (y_0+b) (t{-}\tz)  - ab\sin(t{-}\tz) \cr
y(t) = y_0 + b - b\cos(t{-}\tz)\hfill \cr
}.
\ee

%%%%%%%%%%%%%%%%%%%%%%%%%%%%%%%%%%%%%%%%%%%%%%%%%%%%%%

\paragraph{Example 2}
%\label{secexa2}
Similar framework. Let $\omega>0$ and consider (spin vector field)
\be
\label{eqflotdspin}
\vv(t,x,y)= \pmatrix{-\omega y \cr
\omega x
} = \omega \pmatrix{0 & -1 \cr 1 & 0} \pmatrix{x \cr y}
\eqnote \vv(x,y).
\ee 
With $\ora{\calO \ptz} = \vx_\tz =\pmatrix{x_\tz \cr y_\tz}$,
$r_\tz = \sqrt{x_\tz^2+y_\tz^2}$, and $\theta_0$ \st\
$\vx_\tz=\pmatrix{x_\tz = r_\tz\cos(\omega\tz) \cr y_\tz = r_\tz\sin(\omega\tz)}$,
the solution $\Phitzptz$ of~\eref{eqcl2} is
\be
\label{eqexaspin0}
\vx(t)=\ora{\calO p(t)} = \ora{\calO \Phitzptz(t)}
= \pmatrix{
x(t) = r_\tz\cos(\omega t) \cr %=\PhitzP^1(t)
y(t) = r_\tz\sin(\omega t) \hfill }. %=\PhitzP^2(t)
\ee
Indeed,
$
\pmatrix{{\pa x \over \pa t}(t,\vx_0) \cr {\pa y \over \pa t}(t,\vx_0)}
=\pmatrix{v^1(t,x(t,\vx_0) ,y(t,\vx_0)) \cr v^2(t,x(t,\vx_0) ,y(t,\vx_0))}
= \pmatrix{-\omega y(t,\vx_0) \cr \omega x(t,\vx_0)}
$,
thus
${\pa x \over \pa t}(t,\vx_0) =-\omega y(t,\vx_0)$
and ${\pa y \over \pa t}(t,\vx_0) = \omega x(t,\vx_0)$,
thus ${\pa^2 y \over \pa t^2}(t,\vx_0) = -\omega^2 y(t,\vx_0)$,
hence $y$; Idem for~$x$.
Here $d\vv(t,x,y)=\omega\pmatrix{0 & -1 \cr 1 & 0}
=\omega\pmatrix{\cos{\pi\over 2} & -\sin{\pi\over 2} \cr \sin{\pi\over 2} & \cos{\pi\over 2}}$
is the $\pi/2$-rotation composed with the homothety with ratio~$\omega$.

%%%%%%%%%%%%%%%%%%%%%%%%%%%%%%%%%%%%%%%%%%%%%%%%%%%%%%%%%%%%%%%%%%%%%%%%%%%%%%%%%%%

\subsection{Composition of flows}

Let $\vv$ be a vector field on $\RR \times \Omega$ and $\Phitzptz$ solution of~\eref{eqclb}.
We use the notations
\be
\pt = \Phitzt(\ptz) = \Phi_{t;\tz}(\ptz) := \Phitzptz(t)= \Phitz(t,\ptz) = \Phi(t;\tz,\ptz) = \Phi_{\tz,\ptz}(t).
\ee

%%%%%%%%%%%%%%%%%%%%%%%%%%%%%%%%%%%%%%%%%%%%%%%%%%%%%%%%%%%%%%%%%%%%%%%%%%%%%%%%%%%

\subsubsection{Law of composition of flows (determinism)}

\debprop
For all $t_0,t_1,t_2 \in \RR$, we have (determinism)
\be
\label{eqcompf01}
\Phi^{t_1}_{t_2} \circ \Phi^\tz_{t_1} = \Phi^\tz_{t_2},
\qie \Phi_{t_2;t_1}\circ \Phi_{t_1;\tz} = \Phi_{t_2;\tz}.
\ee
(``The composition of the photos gives the film'').
So, %if $p_\tz\in \Omegatz$ then
\be
\label{eqcompf0b0}
p_{t_2}=\Phi^{t_1}_{t_2}(p_{t_1}) = \Phi^\tz_{t_2}(p_\tz)
\qwhen p_{t_1}=\Phi^\tz_{t_1}(p_\tz) ,%\qand p_{t_2}=\Phi^\tz_{t_2}(p_\tz),
\ee
\ie,
\be
\label{eqcompf0b}
p_{t_2}=\Phi_{t_2;t_1}(p_{t_1}) = \Phi_{t_2;\tz}(p_\tz)
\qwhen p_{t_1}=\Phi_{t_1;\tz}(p_\tz) .%\qand p_{t_2}=\Phi_{t_2,\tz}(p_\tz).
\ee
Thus
\be
\label{eqcompf02}
d\Phi^{t_1}_{t_2}(p_{t_1}). d\Phi^\tz_{t_1}(p_\tz) = d\Phi^\tz_{t_2}(p_\tz),
\qie d\Phi_{t_2;t_1}(p_{t_1}). d\Phi_{t_1;\tz}(p_\tz) = d\Phi_{t_2;\tz}(p_\tz).
\ee
Summary with commutative diagrams:
$$
\label{diag1}
\eqalign{
\xymatrix{
       && p_{t_1}  \ar[drr]^{\ts \Phi^{t_1}_{t_2}} 
\\
p_\tz \ar[urr]^{\ts \Phi^\tz_{t_1}}  \ar[rrrr]_{\ts \Phi^\tz_{t_2}}  &&&& p_{t_2}
}
}
\qie
\eqalign{
\xymatrix{
       && p_{t_1}  \ar[drr]^{\ts \Phi_{t_2;t_1}} 
\\
p_\tz \ar[urr]^{\ts \Phi_{t_1;\tz}}  \ar[rrrr]_{\ts \Phi_{t_2;\tz}}  &&&& p_{t_2}
}
}
$$
\finprop

\debdem
Let $p_{t_1} = \Phitzptz(t_1)$.
\eref{eqcl2} gives
$$
\left\{\eqalign{
& {d\Phitzptz\over dt}(t) = \vv(t,\Phitzptz(t)), \cr
& {d\Phi^{t_1}_{p_{t_1}}\over dt}(t) = \vv(t,\Phi^{t_1}_{p_{t_1}}(t)),
}\right\}
\qwith p_{t_1} = \Phitzptz(t_1) = \Phi^{t_1}_{p_{t_1}}(t_1).
$$
Thus $\Phitzptz$ and $\Phi^{t_1}_{p_{t_1}}$ satisfy the same ODE
with the same value at~$t_1$;
Thus they are equal (uniqueness thanks to  Cauchy--Lipschitz theorem),
thus $\Phi^{t_1}_{p_{t_1}}(t)=\Phitzptz(t)$ when $p_{t_1} = \Phi^\tz_{t_1}(\ptz)$,
that is, $\Phi^{t_1}_t(p_{t_1}) = \Phitzt(\ptz)$ when $p_{t_1} = \Phi^\tz_{t_1}(\ptz)$,
which is~\eref{eqcompf01} for any $t=t_2$. Thus~\eref{eqcompf02}.
\findem

\debcor
A flow is compatible with the motion~$\tPhi$ of an object~$\Obj$:
\eref{eqPhitb} %, that is $\Phitzt \eqdef \tPhi_t\circ(\tPhi_{t_0})^{-1}$, 
gives
$\Phi^{t_1}_{t_2} \circ \Phi^\tz_{t_1}
= (\tPhi_{t_2}\circ(\tPhi_{t_1})^{-1}) \circ(\tPhi_{t_1}\circ(\tPhi_{t_0})^{-1})
= \tPhi_{t_2}\circ(\tPhi_{t_0})^{-1} = \Phi^{t_0}_{t_2}$, that is~\eref{eqcompf01}.
\fincor

%%%%%%%%%%%%%%%%%%%%%%%%%%%%%%%%%%%%%%%%%%%%%%%%%%%%%%%%%%%%%%%%%%%%%%%%%%%%%%%%%%%

\subsubsection{Stationnary case} % $\tcc_{\tau_1}\circ \tcc_{\tau_2} = \tcc_{\tau_1+\tau_2}$
\label{seccfcs}

\debdef
$\vv$ is a stationary vector field iff ${\pa \vv \over \pa t}=0$;
And then $\vv(t,p)\eqnote \vv(p)$.
And the associated flow $\Phitz$ which satisfies
\be
\label{eqvstat}
{\pa \Phitz \over \pa t}(t,\ptz)= \vv(\pt) \qwhen \pt = \Phitz(t,\ptz),
\ee
is said to be stationary.
\findef

\debprop
If $\vv$ is a stationary vector field then, for all $\tz,t_1,h$, when meaningful ($h$ small enough and $t_1$ close enough to~$\tz$),
\be
\label{eqpropcs20}
\Phi^{t_1}_{t_1+h} = \Phi^\tz_{\tz+h}, \qie \Phi_{t_1+h;t_1} = \Phi_{\tz+h;\tz},
\ee
\ie\ $\Phi^{t_1}_{t_1+h}(q) = \Phi^\tz_{\tz+h}(q)$, \ie\ $\Phi(t_1{+}h;t_1,q) = \Phi(\tz{+}h;\tz,q)$ for all $q\in\Omegatz$ (see theorem~\ref{thmexfl}).
In other words,
\be
\label{eqpropcs2}
\Phi^{\tz+h}_{t_1+h} = \Phi^\tz_{t_1}, \qie \Phi_{t_1+h;\tz+h} = \Phi_{t_1;\tz},
\ee
\ie\ $\Phi^{\tz+h}_{t_1+h}(q) = \Phi^\tz_{t_1}(q)$, \ie\ $\Phi(t_1{+}h;\tz{+}h,q) = \Phi(t_1;\tz,q)$ for all $q\in\Omegatz$.
\finprop

\debdem
Let $q\in\Omegatz$, $\alpha(h)=\Phi^\tz_{\tz{+}h}(q) = \Phi^\tz_q(\tz{+}h)$ and
$\beta(h)=\Phi^{t_1}_{t_1{+}h}(q) = \Phi^{t_1}_q(t_1{+}h)$.

Thus $\alpha'(h) = {d \Phi^\tz_q \over dt}(\tz{+}h)
= \vv(\tz{+}h,\Phi^\tz_q(\tz{+}h))
= \vv(\Phi^\tz_q(\tz{+}h))
= \vv(\alpha(h))$ (stationary flow), and
$\beta'(h) = {d \Phi{t_1}_q \over dt}(t_1{+}h)
= \vv(t_1{+}h,\Phi^{t_1}_q(t_1{+}h))
= \vv(\Phi^{t_1}_q(t_1{+}h))
= \vv(\beta(h))$ (stationary flow).

Thus $\alpha$ and $\beta$ satisfy the same ODE with the same initial condition $\alpha(0)=\beta(0)=q$.
Thus $\alpha = \beta$. Hence~\eref{eqpropcs20}.
Thus, with $h=t_1{-}\tz$, \ie\ with $t_1=\tz{+}h$ and $\tz{+}h=t_1$, we get~\eref{eqpropcs2}.
\findem

\debcor
If $\vv$ is a stationary vector field, \cf~\eref{eqvstat}, then
\be
\label{eqpf0}
d\Phitzt(\ptz).\vv(\ptz) = \vv(\pt) \qwhen \pt=\Phitzt(\ptz),
\ee
that is, if $\vv$ is stationary, then $\vv$ is transported (push-forwarded by~$\Phitzt$) along itself.
% (see the push-forward \S~\ref{secsa}).
\fincor

\debdem
\eref{eqcompf0b0}, $t_2=t_1{+}s$ and $t_1=\tz{+}s$ give
$\Phi^{\tz{+}s}_{t_1{+}s}(\Phi^\tz_{\tz{+}s}(p_\tz)) = \Phi^\tz_{t_1{+}s}(p_\tz)$,
and $\vv$ is stationary,
thus $\Phi^{\tz}_{t_1}(\Phi^\tz_{\tz{+}s}(p_\tz)) = \Phi^\tz_{t_1{+}s}(p_\tz)$, 
\ie\ $\Phi(t_1;\tz,\Phi_{\tz,\ptz}(\tz{+}s)) = \Phi_{\tz,\ptz}(t_1{+}s)$, thus ($s$ derivative)
$$
d\Phi(t_1;\tz,\Phi(\tz{+}s;\tz,\ptz)).\Phi_{\tz,\ptz}{}'(\tz{+}s)
= \Phi_{\tz,\ptz}{}'(t_1{+}s),
%d\Phi(t_1;\tz,\Phi(\tz{+}s;\tz,\ptz)).{\pa \Phi \over \pa s}(\tz{+}s;\tz,\ptz)
%= {\pa \Phi \over \pa s}(t_1{+}s;\tz,\ptz),
$$
%(see~\S~\ref{secrnna} for non ambiguous calculations),
thus $d\Phi^\tz_{t_1}(\Phi(\tz{+}s;\tz,\ptz)).\vv(\tz{+}s, \Phi_{\tz,\ptz}(\tz{+}s))
= \vv(t_1{+}s,\Phi_{\tz,\ptz}(t_1{+}s))$. Thus with $s=0$, and $\vv$ being stationary,
$d\Phi^\tz_{t_1}(\Phi(\tz;\tz,\ptz)).\vv(\Phi_{\tz,\ptz}(t_0)) = \vv(\Phi_{\tz,\ptz}(t_1))$, thus~\eref{eqpf0}.
\findem

%%%%%%%%%%%%%%%%%%%%%%%%%%%%%%%%%%%%%%%%%%%%%%%%%%%%%%%%%%%%%%%%%%%%%%%%%%%%%%%%%%%

\subsection{Velocity on the trajectory traveled in the opposite direction}

Let $\tz,\tu\in\RR$, $t_1>\tz$, and $\ptz\in \RRn$. Consider the trajectory
$
\Phitzptz :
\left\{\eqalign{
[\tz,\tu] &\rar \RRn \cr
t & \rar  p(t) = \Phitzptz(t)
}\right\}
$.
So $\ptz$ is the beginning of the trajectory, $\ptu = \Phi^\tz_\tu(\ptz)$ at the end, $\vv(t,p(t)) = {d \Phitzptz \over d t}(t)$ being the velocity.

Define the trajectory traveled in the opposite direction, \ie\ define
\be
\label{eqtssi}
\Psi^\tu_\ptu : 
\left\{\eqalign{
[\tz,\tu] &\rar \RRn \cr
u & \rar  q(u) = \Psi^\tu_\ptu(u) \eqdef \Phitzptz(\tz{+}\tu{-}u) = \Phitzptz(t) = p(t)
\qwhen t = \tz{+}\tu{-}u.
}\right.
\ee
In particular
$q(\tz) = \Psi^\tu_\ptu(\tz) = \Phitzptz(\tu) = p(\tu)$ and
$q(\tu) = \Psi^\tu_\ptu(\tu) = \Phitzptz(\tz) = p(\tz)$.

\debprop
The velocity on the trajectory traveled in the opposite direction is
the opposite of the velocity on the initial trajectory:
\be
\label{eqvi3}
{d\Psi^\tu_\ptu\over du}(u) = q'(u) = -p'(t) = - \vv(t,p(t)) \qwhen t = \tz{+}\tu{-}u,
\ee
\finprop

\debdem
$\Psi^\tu_\ptu(u) = \Phitzptz(\tz{+}\tu{-}u)$ gives
${d \Psi^\tu_\ptu \over du}(u) = -{d\Phitzptz \over dt}(\tz{+}\tu{-}u)
= - \vv(\tz{+}\tu{-}u, \Phitzptz(\tz{+}\tu{-}u))
= - \vv(t,\Phitzptz(t))$ when $t = \tz{+}\tu{-}u$.
\findem

%%%%%%%%%%%%%%%%%%%%%%%%%%%%%%%%%%%%%%%%%%%%%%%%%%%%%%%%%%%%%%%%%%%%%%%%%%%%%%%%%%%

\subsection{Variation of the flow as a function of the initial time}
\label{secvfCI}

%On utilise~$\Phi$ donnée en~\eref{eqdeff3}.

%%%%%%%%%%%%%%%%%%%%%%%%%%%%%%%%%%%%%%%%%%%%%%%%%%%%%%%%%%%%%%%%%%%%%%%%%%%%%%%%%%%

\subsubsection{Ambiguous and non ambiguous notations}
\label{secrnna}

Let $\Phi :(t,u,p)\in \RR\times\RR\times \RRn \rar \Phi(t,u,p) \in\RRn$ be a $C^1$ function.
The partial derivatives are
\be
\pa_1\Phi(t,u,p) \eqdef \lim_{h\rar0} {\Phi(t{+}h,u,p) - \Phi(t,u,p) \over h},
\ee
\be
\label{eqna0}
\pa_2\Phi(t,u,p) \eqdef \lim_{h\rar0} {\Phi(t,u{+}h,p) - \Phi(t,u,p) \over h},
\ee
and $\pa_3\Phi(t,u,p)$, defined for all $\vw\in\vRRn$ (a vector at~$p$) by,
\be
\pa_3\Phi(t,u,p).\vw \eqdef \lim_{h\rar0} {\Phi(t,u,p {+} h\vw) - \Phi(t,u,p) \over h}
\eqnote d\Phi(t,u,p).\vw,
\ee

When the name of the first variable is systematically noted $t$, then
\be
\pa_1\Phi(t,u,p) \eqnote {\pa\Phi\over \pa t}(t,u,p) \eqnote {\pa\Phi(t,u,p)\over \pa t} .
\ee
NB: This notation can be ambiguous:
What is the meaning of ${\pa\Phi\over \pa t}(t;t,p)$?
In ambiguous situations, use the notation~$\pa_1\Phi$, or (if no composed functions inside) use
${\pa\Phi(t,u,p)\over \pa t}_{|u=t}$
(so $t$ is the derivation variable, and after the calculation you take $u=t$).

When the name of the second variable is systematically noted $u$, then
\be
\pa_2\Phi(t,u,p) \eqnote {\pa\Phi\over \pa u}(t,u,p) \eqnote {\pa\Phi(t,u,p)\over \pa u} .
\ee
NB: Idem this notation can be ambiguous:
What is the meaning of ${\pa\Phi\over \pa u}(u;u,p)$? 
In ambiguous situations, use the notation~$\pa_2\Phi$, or use
${\pa\Phi(t,u,p)\over \pa u}_{|t=u}$.

When the name of the third variable is systematically a space variable noted $p$, then
\be
\pa_3\Phi(t,u,p) \eqnote d\Phi(t,u,p) \eqnote {\pa\Phi\over \pa p}(t,u,p) \eqnote {\pa\Phi(t,u,p)\over \pa p}.
\ee

%%%%%%%%%%%%%%%%%%%%%%%%%%%%%%%%%%%%%%%%%%%%%%%%%%%%%%%%%%%%%%%%%%%%%%%%%%%%%%%%%%%

\subsubsection{Variation of the flow as a function of the initial time}

The law of composition of the flows gives~\eref{eqcompf0b} gives $\Phi(t;u,\Phi(u;\tz,p_0)) = \Phi(t;\tz,p_0)$. 
Thus the derivative in~$u$ gives %with $\pa_1 \Phi(t;u,p_0) = \vv(t,\Phi(t;u,p_0))$ (definition of~$\vv$),
\be
\label{eqna}
\eqalign{
&\pa_2\Phi(t;u,\Phi(u;\tz,p_0)) + d\Phi(t;u,\Phi(u;\tz,p_0)). \pa _1\Phi(u;\tz,p_0) = 0, \cr
\qie &  \pa_2\Phi(t;u,p(u)) =- d\Phi(t;u,p(u)). \vv(u,p(u)) \qwhen p(u)=\Phi(u;\tz,p_0). \cr
}
\ee
In particular $u=\tz$ gives, for all $(t,\tz,p_0)\in \RR^2\times \Omegatz$,
\be
\label{eqna2}
( %{\pa\Phi\over \pa \tz} (t;\tz,p_0) = 
{\pa\Phi (t;\tz,p_0)\over \pa \tz}=) \quad
\pa_2\Phi(t;\tz,p_0) = - d\Phi(t;\tz,p_0). \vv(\tz,p_0).
\ee
In particular
\be
({d\Phi (t;\tz,p_0)\over d \tz}_{|t=\tz}=) \quad \pa_2\Phi(\tz;\tz,p_0) = - \vv(\tz,p_0).
\ee
%(Also abusively written ${\pa \Phi (t;u,p(u)) \over \pa u} = - d\Phi(t;u,p(u)). \vv(u,p(u))$.)

\newpage
\part{Push-forward}

\section{Push-forward}
\label{secpf}

%We tackle an essential question: 
%What does ``to transport a quantity'' mean?
The general tool to describe ``transport'' is ``push-forward by a motion'' (the ``take with you'' operator), \cf~\S~\ref{secremF0} and figure~\ref{figpf}.
The push-forward also gives the tool needed to understand the velocity addition formula:
In that case, the push-forward is the translator between observers.
The push-forward can also be used to write coordinate systems.
%This translator enables to  define and understand the ``covariant objectivity'' that enables an English observer with his foot and a French observer with his metre to work together, without causing an accident like the Mars climate orbiter crash.
As usual, we start with qualitative results (observer independent results); Then, quantitative results are deduced.

%%%%%%%%%%%%%%%%%%%%%%%%%%%%%%%%%%%%%%%%%%%%%%%%%%%%%%%%%%%%%%%%%%%%%%%%%%%%%%%%%%%

\subsection{Definition}
\label{secpf0}

$\calE$ and~$\calF$ are affine spaces,
$E$ and $F$ are the associated vector spaces equipped with norms $||.||_E$ and~$||.||_F$ with $\dim E =\dim F=n$,
$\UE$ and $\UF$ are open sets in the affine space $\calE$ and~$\calF$,
or possibly the vector spaces~$E$ and~$F$, and
\be
\label{eqPsi}
\Psi : 
\left\{\eqalign{
\UE & \rar \UF \cr
\pE & \rar \pF = \Psi(\pE) \quad \hbox{is a diffeomorphism}
}\right.
\ee
(a $C^1$ invertible map which inverse is $C^1$),  
called the push-forward, and $\Psi^{-1}$ is the pull-back (push-forward with $\Psi^{-1}$).

\begin{figure}[!h]
%\begin{center}
\qquad\qquad\qquad\qquad\includegraphics[width=0.6\textwidth]{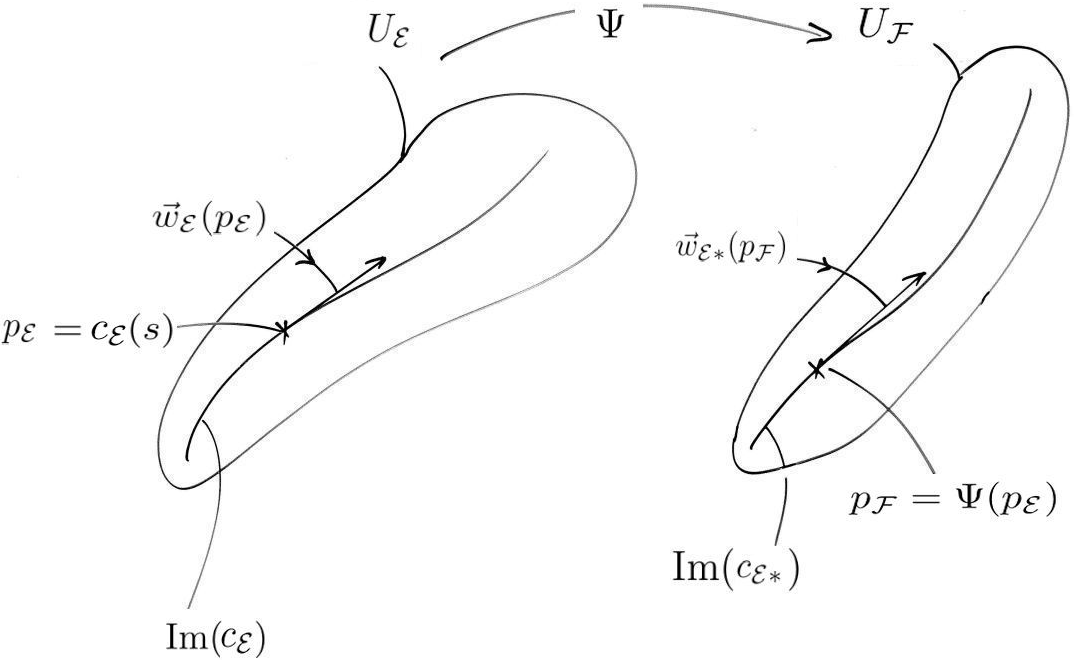}
\vspace{-10pt}
%\end{center}
\caption{$c_\calE:s\rar \pE=c_\calE(s)$ is a curve in~$\UE$.
Push-forwarded by $\Psi$ it becomes the curve $c_{\calE*}: = \Psi\circ c_\calE$ in~$\UF$.
The tangent vector at $\pE = c_\calE(s)$ is $\vw_\calE(\pE) = c_\calE{}'(s)$,
and the tangent vector at $\pF = c_\calF(s) = \Psi(c_\calE(s))$ is 
$\vw_{\calE*}(\pF) = c_\calF{}'(s) = d\Psi(\pE).\vw_\calE(\pE)$.
Other illustation: See figure~\ref{figpf}.
% (push-forward of vector fields by~$\Phitzt$).
%And if $\vw_\tz$ is a vector field in~$\Omegatz$, then $c_\tz$ is an integral curve of $\vw_\tz$, and the above steps apply to define $\vwtzs$ (the push-forward by~$\Phitzt$).
}
\label{figpf2}
\end{figure}

%\medskip
Example: $\Psi = \Phitzt : \Omegatz \rar \Omegat$, the motion that transforms $\Omegatz$ into~$\Omegat$,
\cf~\eref{eqPhit}.

Example: $\Psi:U_E \rar U_F$ a coordinate system, see~example~\ref{exarempfcb}. %In that case $\UE$ and~$\UF$ are open subset of the vector spaces $E$ and~$F$.

Example: $\Psi = \Thetat : \calRB \rar \calRA$, a change of referential at~$t$ (change of observer), see~\S~\ref{seccdr}.

%%%%%%%%%%%%%%%%%%%%%%%%%%%%%%%%%%%%%%%%%%%%%%%%%%%%%%%%%%%%%%%%%%%%%%%%%%%%%%%%%%%

\subsection{Push-forward and pull-back of points}

\debdef
\label{defpfpbp}
%With~\eref{eqPsi}.
If $\pE \in \UE$ (a point in~$\UE$) then its push-forward by~$\Psi$ is the point %$\pF=\Psi(\pE) \eqnote \Psi_*\pE \in \UF$ :
\be
\label{eqpfpbp}
\pF= 
\boxed{\Psi_*\pE := \Psi(\pE)} =\pEs \; \in \UF,
%,\quad\hbox{also written}\quad \pF = \Psi_*\pE .
\ee
see figure~\ref{figpf2}, the last notation if $\Psi$ is implicit. 
And if $\pF \in \UF$ then its pull-back by~$\Psi$ is the point %$\pE=\Psi^{-1}(\pF)\eqnote \Psi^*\pF\in\UE$:
\be
\pE = 
\boxed{\Psi^*\pF := \Psi^{-1}(\pF)} = \pFb \; \in \UE.
\ee
%the last notation if $\Psi$ is implicit.
\findef

We immediately have $\Psi^*\circ \Psi_* = I$.

\medskip
The notations $_*$ for push-forward and $^*$ for pull-back have been proposed by Spivak; Also see Abraham and Marsden~\cite{abraham-marsden} (second edition) who adopt this notation.
%; And the last notation $\Psi_*\pE$ is used when an explicit reference to~$\Psi$ is needed.

%%%%%%%%%%%%%%%%%%%%%%%%%%%%%%%%%%%%%%%%%%%%%%%%%%%%%%%%%%%%%%%%%%%%%%%%%%%%%%%%%%%

\subsection{Push-forward and pull-back of curves}

We push-forward (and pull-back) the points on a curve:

\debdef
Let
$c_\calE :
\left\{\eqalign{
]-\eps,\eps[ & \rar \UE \cr
s & \rar \pE = c_\calE(s)
}\right\}
$ be a curve in~$\UE$.
Its push-forward by~$\Psi$ is the curve
\be
\label{eqgammas}
%c_\calF := \Psi \circ c_\calE \eqnote 
\Psi_* c_\calE := \Psi \circ c_\calE : %= \curveEs  :
\left\{\eqalign{
]-\eps,\eps[ & \rar \UF \cr
s & \rar \pF %=c_\calF(s) 
= \Psi_* c_\calE(s) := \Psi(c_\calE(s)) \eqnote \curveEs(s) \quad( = \Psi(\pE)), % \qwhen \pE = c_\calE(s),
}\right.
\ee
see figure~\ref{figpf2}. ($\Psi_*c_\calE \eqnote \curveEs$ when $\Psi$ is implicit.)
This defines
\be
\Psi_* :
\left\{\eqalign{
\calF(]-\eps,\eps[;\UE) &\rar \calF(]-\eps,\eps[;\UF) \cr
c_\calE & \rar \Psi_* (c_\calE)  := \Psi \circ c_\calE \eqnote \Psi_* c_\calE = \curveEs.
}\right.
\ee
\findef

\debdef
Let
$%\label{eqgammau2}
c_\calF :
\left\{\eqalign{
]-\eps,\eps[ & \rar \UF \cr
s & \rar \pF = c_\calF(s)
}\right\}
$
is a curve in~$\UF$. Its pull-back by~$\Psi$ is
\be
\label{eqgammaub}
\Psi^* c_\calF := \Psi^{-1} \circ c_\calE %= \curveFb :
\left\{\eqalign{
]-\eps,\eps[ & \rar \UE \cr
s & \rar \pE = \Psi^* c_\calF(s) := \Psi^{-1}(c_\calF(s)) \eqnote \curveFb(s) \quad( = \Psi^{-1}(\pF)). % \qwhen \pF = c_\calF(s).
}\right.
\ee
We have thus defined
\be
\Psi^* :
\left\{\eqalign{
\calF(C^1(]-\eps,\eps[;\UF) &\rar \calF(C^1(]-\eps,\eps[;\UE) \cr
c_\calF & \rar \Psi^* (c_\calF) := \Psi^{-1} \circ c_\calF \eqnote \Psi^* c_\calF = \curveFb.
}\right.
\ee
\findef

%%%%%%%%%%%%%%%%%%%%%%%%%%%%%%%%%%%%%%%%%%%%%%%%%%%%%%%%%%%%%%%%%%%%%%%%%%%%%%%%%%%

\subsection{Push-forward and pull-back of scalar functions} 
\label{sectransf}

%%%%%%%%%%%%%%%%%%%%%%%%%%%%%%%%%%%%%%%%%%%%%%%%%%%%%%%%%%%%%%%%%%%%%%%%%%%%%%%%%%%

\subsubsection{Definitions} 

%We use the $*$-notation of Spivak (interpretation at~\S~\ref{exapff}.) Also see Abraham and Marsden~\cite{abraham-marsden} (second edition).

\debdef
\label{defpc}
Let 
$\fE :
\left\{\eqalign{
\UE & \rar \RR \cr
\pE & \rar \fE(\pE)
}\right\}
$
(scalar valued function).
Its push-forward by~$\Psi$ is the (scalar valued) function 
%$\fF=\Psi_*\fE := \fE\circ\Psi^{-1} \eqnote \fEs$ (when $\Psi$ is implicit), so
\be
\label{eqdefpff}
%\fF  \eqnote 
\Psi_*\fE := \fE\circ\Psi^{-1}: %=\fEs :
\left\{\eqalign{
\UF & \rar \RR \cr
\pF & \rar \Psi_*\fE(\pF) := \fE(\pE) \eqnote \fEs(\pF)
\qwhen \pE = \Psi^{-1}(\pF),
}\right.
\ee
(noted $\fEs$ when $\Psi$ is implicit), \ie\ $\Psi_*\fE(\Psi_*\pE) := \fE(\pE)$, or
$\fEs(\pEs):=\fE(\pE)$ when $\pEs = \Psi(\pE)$.
%In other words, $\Psi_*\fE:= \fE\circ\Psi^{-1}$.
We have thus defined
\be
\Psi_* :
\left\{\eqalign{
\calF(\UE;\RR) &\rar \calF(\UF;\RR) \cr
\fE & \rar \fF := \Psi_*(\fE) = \fE \circ \Psi^{-1} \eqnote \Psi_*\fE, % = 
}\right.
\ee
the notation $\Psi_*(\fE) = \Psi_*\fE$ since $\Psi_*$ is linear:
$((f_\calE + \lambda g_\calE)\circ \Psi^{-1})(\pF)
=(f_\calE + \lambda g_\calE)(\pE)
=f_\calE(\pE) + \lambda g_\calE(\pE)
 = (f_\calE\circ \Psi^{-1})(\pF)+\lambda (g_\calE\circ \Psi^{-1})(\pF)$
gives $\Psi_*(f_\calE + \lambda g_\calE) = \Psi_*(\fE) + \lambda \Psi_*(g_\calE)$.
\findef

\debdef
Let 
$\fF :
\left\{\eqalign{
\UF & \rar \RR \cr
\pF & \rar \fF(\pF)
}\right\}
$.
Its pull-back by~$\Psi$ is the push-forward by~$\Psi^{-1}$, \ie\ is
%$\fE=\Psi^*\fF:= \fF \circ\Psi \eqnote \fFb$ (when $\Psi$ is implicit), so
\be
\label{eqdefpff2}
\Psi^*\fF := \fF \circ \Psi :
\left\{\eqalign{
\UE & \rar \RR \cr
\pE & \rar \Psi^*\fF(\pE)  := \fF(\pF) \eqnote \fFb(\pE)
\qwhen \pF = \Psi(\pE),
%\fFb \eqdef \fF \circ \Psi
}\right.
\ee
\ie\ $\Psi^*\fF(\Psi^*\pF) := \fF(\pF)$,
\ie\ $\fFb(\pFb) := \fF(\pF)$ when $\pF=\Psi^*(\pF)$.
We have thus defined
\be
\Psi^* :
\left\{\eqalign{
\calF(\UF;\RR) &\rar \calF(\UE;\RR) \cr
\fF & \rar \Psi^*(\fF) = \fFb := \fF \circ \Psi \eqnote \Psi^*\fF.
}\right.
\ee
%the notation $\Psi^*(\fF) = \Psi^*\fF$ since $\Psi_*$ is linear.
\findef

We immediately have
$\Psi^*\circ \Psi_* = I \qand \Psi_*\circ \Psi^* = I$ (the first $I$ is the identity in~$\calF(\UE;\RR)$, the second  $I$ is the identity in~$\calF(\UF;\RR)$).

NB: We used the same notations $\Psi_*$ and $\Psi^*$ than for the push-forward and pull-backs of points: The context removes ambiguities.

%%%%%%%%%%%%%%%%%%%%%%%%%%%%%%%%%%%%%%%%%%%%%%%%%%%%%%%%%%%%%%%%%%%%%%%%%%%%%%%%%%%

\subsubsection{Interpretation: Why is it useful?} 
\label{exapff}

\Eg: Let $\tPhi:\RR\times\Obj \rar \RRn$ be a motion of an object~$\Obj$.
An observer records the temperature $\theta$ at all $t\in [\tz,T]$ and all $p=\tPhi(t,\Obj)$:
He gets $\theta : 
\left\{\eqalign{
\bigC = \bigcup_t(\{t\}\times \Omegat) & \rar \RR \cr
(t,p) & \rar \theta(t,p)
}\right\}$ a Eulerian scalar valued function,
\cf~\eref{eqdeffspa20}.
Then he chooses an initial time~$\tz$ and considers the associated motion~$\Phitz$, \cf~\eref{eqdefPhi},
and considers
$\theta_\tz :
\left\{\eqalign{
\Omegatz & \rar \RR \cr
\ptz & \rar \theta_\tz(\ptz):= \theta(\tz,\ptz)
}\right\}$
(snapshot of the temperatures at~$\tz$ in~$\Omegatz$).
The push-forward of $\theta_\tz$ by $\Phitzt$ is $(\Phitzt)_*\theta_\tz := \theta_\tz \circ (\Phitzt)^{-1}$ % \eqnote \theta_{t,\tz*} $, which 
defines the ``memory function''
\be
\label{eqdefpfft}
%\thetatzs : %= \Phitzts\theta_\tz %:= \theta_\tz \circ (\Phitzt)^{-1} :
(\Phitzt)_*\theta_\tz :
\left\{\eqalign{
\Omegat & \rar \RR \cr
\pt & \rar (\Phitzt)_*\theta_\tz(\pt) := \theta_\tz(\ptz) \qwhen \pt = \Phitzt(\ptz)
,}\right.
\ee
And he writes $(\Phitzt)_*\theta_\tz(\pt) \eqnote \thetatzs(t,\pt)$, so the memory transported is at~$t$ at~$\pt$ (along a trajectory) by
\be
\thetatzs(t,p(t)) = \theta_\tz(\ptz).
\ee
%and $\thetatzs$ defines a (virtual) unstationary vector field (not Eulerian since it depends on a~$\tz$).

%\medskip
{\bf Question:} Why do we introduce $\thetatzs$ since we have~$\theta_\tz$?

{\bf Answer:}
An observer does not have the gift of temporal and/or spatial ubiquity;
He has to do with values at the actual time~$t$ and position~$\pt$ where he is (Newton and Einstein's point of view).
So, when he was at~$\tz$ at~$\ptz$ the observer wrote the value $\theta_\tz(\ptz)$ on a piece of paper (for memory), puts the piece of paper is his pocket, then once at~$t$ at $p(t)=\Phitz(t,\ptz)$, he takes the paper out of his pocket, and renames the value he reads as
$\thetatzs(t,\pt)$ because he is now at $t$ at~$\pt$.
%so with $\thetatzs(\pt):=\theta_\tz(\ptz)$.
And, now at~$t$ at~$\pt$, he can compare the past and present value.
In particular the rate
\be
\label{eqdefpfft0}
{\theta(t,p(t)) - \thetatzs(t,p(t)) \over t-\tz} 
={\hbox{\footnotesize actual}(t,p(t)) - \hbox{\footnotesize memory}_*(t,p(t)) \over t-\tz}
%\eqnote \calL_\vv\theta(\tz,\ptz)
\ee
is physically meaningful for one observer at~$t$ at~$\pt$ (no ubiquity gift required).
For scalar value functions, we get the usual rate
${\theta(t,p(t)) - \theta(\tz,p(\tz)) \over t-\tz} $, but it isn't that simple for vector valued functions.

And the limit $t\rar\tz$ in \eref{eqdefpfft0} defines the Lie derivative for scalar valued functions.

\comment{
\medskip
For the pull-back,
for $\theta_t :
\left\{\eqalign{
\Omegat & \rar \RR \cr
\pt & \rar \theta_t(\pt):= \theta(t,\pt)
}\right\}$, we get its pull-back
\be
\thetatzb
:= \theta_t \circ \Phitzt :
\left\{\eqalign{
\Omegatz & \rar \RR \cr
\ptz & \rar \thetatzb(\ptz) := \theta_t(\pt) \qwhen \pt = \Phitzt(\ptz).
}\right.
\ee
}

%%%%%%%%%%%%%%%%%%%%%%%%%%%%%%%%%%%%%%%%%%%%%%%%%%%%%%%%%%%%%%%%%%%%%%%%%%%%%%%%%%%

\subsection{Push-forward and pull-back of vector fields}
\label{sectrans}

This is one of the most important concept for mechanical engineers.
%We need a norm $||.||_E$ in~$E$ (we need first-order Taylor expansions).

%Let $\Psi : \UE \rar \UF$ be a diffeomorphism, \cf~\eref{eqPsi}.

%%%%%%%%%%%%%%%%%%%%%%%%%%%%%%%%%%%%%%%%%%%%%%%%%%%%%%%%%%%%%%%%%%%%%%%%%%%%%%%%%%%

\subsubsection{A definition by approximation}
\label{secsa}

%\leavevmode

Elementary introduction.
Let $\pE$ and $\qE$ be points in $\UE$, %with $\qE$ close to~$\pE$.
and let $\pF = \pEs = \Psi(\pE)$ and $\qF = \qEs = \Psi(\qE)$ in $\UF$
be the push-forwards by~$\Psi$ \cf~\eref{defpfpbp}.
The first order Taylor expansion  gives
\be
%\label{eqqpdPhi}
(\Psi(\qE)-\Psi(\pE) = )\quad 
\qF-\pF = d\Psi(\pE).(\qE-\pE) + o(||\qE-\pE||_E),
\ee
thus,
\be
\label{eqqpdPhi2}
{\ora{\pF\qF}\over ||\ora{\pE\qE}||_E} = d\Psi(\pE).{\ora{\pE\qE}\over ||\ora{\pE\qE}||_E} + o(1).
\ee
And the definition of the push-forward is obtained by ``neglecting'' the $o(1)$ (limit as $\qE\rar \pE$):

\debdef
\label{eqdefpfbp}
If $\vwE(\pE) \in E$ is a vector at $\pE \in U$ then its push-forward by~$\Psi$
is the vector $\vwF(\pF) \eqnote \vwEs(\pF) \eqnote \Psi_*\vwE(\pF)\in F$
defined at $\pF = \pEs = \Psi(\pE) \in \UF$ by
\be
\label{eqqpdPhi21}
\vwF(\pF)=
\boxed{\vwEs(\pF) \eqdef  d\Psi(\pE).\vwE(\pE)} \eqnote \Psi_*\vwE(\pF).
\ee
\findef

%%%%%%%%%%%%%%%%%%%%%%%%%%%%%%%%%%%%%%%%%%%%%%%%%%%%%%%%%%%%%%%%%%%%%%%%%%%%%%%%%%%

\subsubsection{The definition of the push-forward of a vector field}

To fully grasp the definition, and to avoid making interpretation errors as in~\S~\ref{secnm} (the unfortunate notation $d\vec x = F.d\vec X$), we use the following definition of ``a vector'': It is a ``tangent vector to a curve'' (needed for surfaces and manifolds). Details:

\medskip
$\bullet$
Let $c_\calE :
\left\{\eqalign{
]-\eps,\eps[ & \rar \UE \cr
s & \rar \pE = c_\calE(s)
}\right\}$ be a $C^1$ curve  in~$\UE$.
Its tangent vector at $\pE=\curveE(s)$ is
\be
\label{eqvwU}
\vwE(\pE) := \curveEp(s) \quad ( = \lim_{h\rar0} {\curveE(s+h) - \curveE(s) \over h}),
%\qwhen \pE=\curveE(s).
\ee
see~figure~\ref{figpf2}.
This defines the function
$\vwE : 
\left\{\eqalign{
\Im(\curveE) & \rar E \cr
\pE & \rar \vwE(\pE)
}\right\}
$
called a vector field along~$\Im(\curveE) {\subset} \UE$.

\medskip
$\bullet$
The push-forward of $c_\calE$ by~$\Psi$ being the image curve $\curveEs = \Psi\circ\curveE$ (the curve transformed by~$\Psi$) \cf~\eref{eqgammas}, its tangent vector at $\pF=\curveEs(s)$ is
\be
\label{eqvwV}
\vwEs(\pF) := \curveEs{}'(s) \qthus
= d\Psi(\pE).\curveEp(s) = d\Psi(\pE).\vwE(\pE). %\eqnote \Psi_*\vw_E(\pF) \eqnote \vwEs(\pF) \;(\in F).
\ee
Thus we have defined the vector field $\vwEs$ along $\Im(\curveEs)$ called the push-forward of~$\vwE$ by~$\Psi$.

With all the integral curves of a vector field defined in~$\UE$, we get:

\debdef
\label{defrapvEivei}
The push-forward by~$\Psi$ of a $C^0$ vector field $\vwE : 
\left\{\eqalign{
\UE & \rar E \cr
\pE & \rar \vwE(\pE)
}\right\}
$ is the vector field
\be
\label{eqdefrapvEivei}
\Psi_*\vwE = \vwEs : 
\left\{\eqalign{
\UF & \rar F \cr
\pF & \rar \boxed{\Psi_*\vwE(\pF) := d\Psi(\pE).\vwE(\pE)} \eqnote \vwEs(\pF) \qwhen \pF=\Psi(\pE),
}\right.
\ee
see~figure~\ref{figpf2}. ($\Psi_*\vwE \eqnote \vwEs$ if $\Psi$ is implicit).
In other words, %with $d\Psi(\pE).\vwE(\pE) = (d\Psi.\vwE)(\pE)$,
\be
\label{eqrapvEiveia}
%(\vwF=)\quad 
%\vwEs = 
\Psi_*\vwE := (d\Psi.\vwE)\circ \Psi^{-1}.
\ee
%($\vwEs$ is defined in~$\UF$ such that its integral curve is $\curveEs = \Psi\circ c_\calE$.)
This defines the map
$
\Psi_* :
\left\{\eqalign{
C^\infty(\UE;E) & \rar C^\infty(\UF;F) \cr
\vwE & \rar \Psi_* (\vwE) := \Psi_* \vwE = \vwEs
}\right\}
$.
(We use the same notation $\Psi_*$ as in definition~\ref{defpc} for scalar valued functions: The context removes ambiguity.)
\findef

\debrem
\label{rempfLie}
Unlike scalar functions, \cf\ \S~\ref{exapff}: At~$\tz$ at~$\ptz$
you cannot just draw a vector $\vw_\tz(\ptz)$ on a piece of paper, put the paper in your pocket, then let yourself be carried by the flow $\Psi = \Phitzt$ (push-forward), then, once arrived at~$t$ at~$\pt$, take the paper out of your pocket and read it to get the push-forward: The direction and length of the vector $\vw_{\tz*}(t,\pt)$ are modified by the flow (a~vector is not just a collection of scalar components). %, unless $\Ftz(t,\ptz)=I$ for all~$t$.
%(See prop.~\ref{peqdwdt04}.)
\finrem

\debexe
%(Toward curvatures.) Avec $\pE=\curveE(s)$ et $\vcurveEp(s) \eqnote \vwE(\pE)$, montrer :
Prove:
\be
\label{eqvcou0}
\vgammaEpp(s) = d\vwE(\pE). \vwE(\pE),
\ee
and
\be
\label{eqvcou1}
d\vwEs(\pF). d\Psi(\pE) = d\Psi(\pE).d\vwE(\pE) + d^2\Psi(\pE).\vwE(\pE) ,
\ee
and
\be
\label{eqvcou2}
\eqalign{
\curveEs''(s)
= & d\vwEs(\pF). \vwEs(\pF) \quad
(=  d\Psi(\pE).\vgammaEpp(s) + d^2\Psi(\pE).\vgammaEp(s).\vgammaEp(s)). \cr
}
\ee

\debrep
$\vgammaEp(s) = \vwE(\gammaE(s))$ gives $\vgammaEpp(s) = d\vwE(\gammaE(s)).\vgammaEp(s)$,
hence~\eref{eqvcou0}.

$\vwEs(\Psi(\pE)) = d\Phi(\pE).\vwE(\pE)$ by definition of $\vwEs$, hence~\eref{eqvcou1}.

$\gammaF(s) = \Psi(\gammaE(s))$
gives $\vgammaFp(s)
= d\Psi(\gammaE(s)).\vgammaEp(s)
= d\Psi(\gammaE(s)).\vwE(\gammaE(s))
= \vwEs(\gammaF(s))
$.
Thus $\vgammaFpp(s)
= (d^2\Psi(\gammaE(s)).\vgammaEp(s)).\vgammaEp(s) + d\Psi(\gammaE(s)).\vgammaEpp(s)
= d\vwEs(\gammaF(s)).\vgammaFp(s)$, hence~\eref{eqvcou2}.
\finrep
\finexe

\comment{
\debexe
%\label{exepfLie}
Let $\vw$ be a (Eulerian) vector field (a field of forces). %, and let $\vw_t(\pt) := \vw(t,\pt)$.
The Lie derivative $\calL_\vv\vw$ of~$\vw$ along the flow which velocity field is~$\vv$ is defined by
\be
\calL_\vv\vw := {D\vw \over Dt} - d\vv.\vw \quad (= {\pa \vw \over \pa t} + d\vw.\vv - d\vv.\vw).
\ee
At $\tz$ consider $\vw_\tz(\ptz) := \vw(\tz,\ptz)$ for all $\ptz\in\Omegatz$.
And consider its push-forward at all time~$t$, that is,
\be
\vwtzs(\pt) = (\Phitzts\vw_\tz)(\pt)=d\Phitz(t,\ptz).\vw(\tz,\ptz) \eqnote \vw_*(t,\pt)
\ee
when $\pt=\Phitzt(\ptz)$, \cf~\eref{eqrapvEivei2a}.
Prove:
\be
\label{eqtparlef}
\hbox{If}, \; \forall t,\;
\vw_*(t,\pt) =\vw(t,\pt),
\qthen
%\quad \Longrightarrow \quad
\calL_\vv\vw(t,\pt) = \vec0.
\ee
Interpretation: $\vw(t,\pt) = \vwtzs(\pt)$ means that~$\vw$ has let itself be deformed by the flow
(the actual value is equal to the value transformed by the flow),
which translates as: Its Lie derivative vanishes = no resistance to deformation.
\Eg, if $\vv$ is a stationary velocity field then $\calL_\vv\vv=\vec0$.
 %Voir~\S~\ref{secdlcv} et~\ref{secLieei}.

\debrep
$p(t)=\Phitzt(\ptz)=\Phitz(t,\ptz)$ and
$\vw_*(t,p(t)) = d\Phitz(t,\ptz).\vw(\tz,\ptz)$ give (with $\Phitz$ $C^2$)
$$
\eqalign{
{D\vw_* \over Dt}(t,p(t))
= {\pa (d\Phitz) \over \pa t}(t,\ptz).\vw(\tz,\ptz)
= & d({\pa \Phitz \over \pa t})(t,\ptz).\vw(\tz,\ptz) \cr
= & (d\vv(t,\pt).d\Phitzt(\pt)).\vw(\tz,\ptz)
= d\vv(t,\pt).\vw_*(t,\pt),
}
$$
thus ${D\vw_* \over Dt}(t,\pt) - d\vv(t,\pt).\vw_*(t,\pt) = 0$, i.e.~\eref{eqtparlef}. Or see prop.~\ref{peqdwdt04}.
\finrep
\finexe

}

%%%%%%%%%%%%%%%%%%%%%%%%%%%%%%%%%%%%%%%%%%%%%%%%%%%%%%%%%%%%%%%%%%%%%%%%%%%%%%%%%%%

\subsubsection{Pull-back of a vector field}

%The pull-back is technically the push-forward by~$\Psi^{-1}$:

\debdef
If $\vwF:
\left\{\eqalign{
\UF & \rar F \cr
\pF & \rar \vwF (\pF)
}\right\}
$
is a vector field on~$\UF$, then its pull-back by~$\Psi$ is the push-forward by~$\Psi^{-1}$, \ie\ is the vector field on~$\UE$ defined by
\be
\label{eqrapvEiveib}
\Psi^*\vwF: %=\vwFb :
\left\{\eqalign{
\UE & \rar E \cr
\pE & \rar \boxed{\Psi^*\vwF(\pE) :=  d\Psi^{-1}(\pF).\vwF(\pF)} \eqnote \vwFb(\pE), \qwhen \pF = \Psi(\pE).
}\right.
\ee
In other words, %since $d\Psi(\pE)^{-1} = d\Psi^{-1}(\pF)$,
\be
\Psi^*\vwF := (d\Psi^{-1}.\vwF)\circ\Psi \eqnote \vwFb.
%,\qand \Psi^*\vwF(\pE) = \vwEb(\pE) = d\Psi(\pE)^{-1}.\vwF(\pF).
\ee
\findef

And we get
\be
\label{eqpfpb}
\Psi^*\circ \Psi_* = I \qand \Psi_*\circ \Psi^* = I.
\ee
%that is, $\Psi^*(\Psi_*\vwE)=\vwE$ or $\Psi_*\vwE = (\Psi^*)^{-1}(\vwE)$, and $\Psi_*(\Psi^*\vwE)=\vwE$ or $\Psi^*\vwE = (\Psi_*)^{-1}(\vwE)$.
Indeed, $\Psi^*(\Psi_* \vwE)(\pE)= d\Psi^{-1}(\pF).\Psi_* \vwE(\pF)
=d\Psi^{-1}(\pF).d\Psi(\pE).\vwE(\pE) = \vwE(\pE)$, for all~$\pE$. Idem for the second equality.
\findem

%%%%%%%%%%%%%%%%%%%%%%%%%%%%%%%%%%%%%%%%%%%%%%%%%%%%%%%%%%%%%%%%%%%%%%%%%%%%%%%%%%%

\subsection{Quantification with bases}
\label{secqwb}

%%%%%%%%%%%%%%%%%%%%%%%%%%%%%%%%%%%%%%%%%%%%%%%%%%%%%%%%%%%%%%%%%%%%%%%%%%%%%%%%%%%

\subsubsection{Usual result}

$(\va_i)$ is  a Cartesian basis in~$E$,
$O_\calF$ and $(\vb_i)$ are an origin in~$\calF$ and a Cartesian basis in~$F$,
$\pE\in \UE$, % let % $\pF = \Psi(\pE) = O_\calF + \ora{O_\calF\Psi(\pE)} \in \UF$ and
\be
\label{eqoraoVp}
\pF = \Psi(\pE) %= O_\calF + \ora{O_\calF\Psi(\pE)} 
= O_\calF + \sumin \psi_i(\pE)\,\vb_i , \qie
[\ora{O_\calF \pF} ]_{|\vb} %= [\ora{O_\calF\Psi(\pE)} ]_{|\vb}
= \pmatrix{ \psi_1(\pE) \cr \vdots \cr \psi_n(\pE)}.
\ee
Then, if $\vwE$ is a vector field in~$\UE$ and $\vwE=\sum_i w_j\va_i$, we get
$%\label{eqoraoVp3}
%(\vwEs(\pF)=)\; 
\Psi_*\vwE(\pF) = d\Psi(\pE).\vwE(\pE)
= \sumin (d\psi_i(\pE).\vwE(\pE)) \,\vb_i
= \sumijn w_j(\pE) (d\psi_i(\pE).\va_j) \,\vb_i
= \sumijn {\pa \psi_i \over \pa x_j}(\pE) w_j(\pE) \,\vb_i
$, so %with $d\psi_i(\pE).\va_j = {\pa \psi_i \over \pa x_j}(\pE)$ (usual notation),
\be
[\Psi_*\vwE(\pF)]_{|\vb} = [d\Psi(\pE)]_{|\va,\vb}.[\vwE(\pE)]_{|\va},
\ee
where $[d\Psi(\pE)]_{|\va,\vb} = [d\psi_i(\pE).\va_j] \eqnote [{\pa \psi_i \over \pa x_j}(\pE)]$
is the Jacobian matrix.

%%%%%%%%%%%%%%%%%%%%%%%%%%%%%%%%%%%%%%%%%%%%%%%%%%%%%%%%%%%%%%%%%%%%%%%%%%%%%%%%%%%

\subsubsection{Example: Polar coordinate system}
\label{secrempfcb}

\debexa
\label{exarempfcb}
Change of coordinate system interpreted as a push-forward:
Paradigmatic example of the polar coordinate system  (model generalized for the parametrization of any manifold). % (change of point of view ``Polar vs Cartesian'').

Parametric Cartesian vector space $\RR\times \RR \eqnote \vec\RR^2_p=\{\vq=(r,\theta)\}$,
with its canonical basis $(\va_1,\va_2)$, and $\vq = r\va_1+\theta\va_2 \eqnote (r,\theta)$, so $[\vq]_{|\va} = \pmatrix{r \cr \theta}$.
Geometric affine space~$\RR^2$ (of positions), $p\in\RR^2$, associated vector space~$\vRRd$, $O\in\RR^2$ (origin),
$\vx=\ora{Op}$, and a Euclidean basis $(\vb_1,\vb_2)$ in~$\vRRd$.
The ``polar coordinate system'' is the associated map
$
\Psi :
\left\{\eqalign{
\vec \RR_+^*\times\RR \subset \vec\RR^2_p %= (\RR \times \RR){-}\{\vec0\} 
& \rar \vRRd \cr
\vq=(r,\theta) & \rar \vx
= \Psi(\vq)=\Psi(r,\theta), \cr
}\right\}
$
defined by
\be
\label{eqPpc}
\vx=\Psi(\vq) := r \cos\theta\, \vb_1+  r \sin\theta\,\vb_2 ,
\qie
[\vx]_{|\vb} = \pmatrix{ x = r \cos\theta \cr y = r \sin\theta} .
\ee
The $i$-th coordinate line at $\vq$ in $\vec\RR^2_p$ (parametric space) is the straight line
$\vc_{\vq,i} : 
\left\{\eqalign{
\RR & \rar \vec\RR^2_p \cr
s & \rar \vc_{\vq,i}(s) =  \vq+s\va_i
}\right\}
$,
and its tangent vector at $\vc_{\vq,i}(s)$ is $\vc_{\vq,i}{}'(s)=\va_i$ for all~$s$.
This line is transformed by~$\Psi$ into the curve
$\Psi_*(c_{q,i}) = \Psi\circ \vc_{\vq,i} \eqnote c_{\vx,i}: 
\left\{\eqalign{
\RR & \rar \RR^2 \cr
s & \rar c_{\vx,i}(s) =  \Psi(\vq+s\va_i)
}\right\}
$ %which is the polar coordinate curve at $\vx=\Psi(\vq)$  
(in particular $c_{\vx,i}(0)=\vx$). So
\be
[\ora{Oc_{\vx,1}(s)}]_{|\vb}=\pmatrix{(r{+}s) \cos\theta \cr (r{+}s) \sin\theta} \quad \hbox{(straight line)}, \qand
[\ora{Oc_{\vx,2}(s)}]_{|\vb}=\pmatrix{r \cos(\theta{+}s) \cr r \sin(\theta{+}s)} \quad \hbox{(circle)}.
\ee
And the tangent vector at $c_{\vx,i}(s)$ is $c_{\vx,i}{}'(s) \eqnote \va_{i*}(\vx)$
(push-forward by~$\Psi$), so %with $\ora{Op}=\Psi(\vq)=\Psi(r,\theta)$,
\be
\eqalign{
&\va_{1*}(\vx) := \Psi_*\va_1(\vx)
= d\Psi(\vq).\va_1
= \lim_{h\rar0} {\Psi(\vq{+}h\va_1) - \Psi(\vq) \over h}
= \lim_{h\rar0} {\Psi(r{+}h,\theta) - \Psi(r,\theta) \over h}
= {\pa \Psi \over \pa r}(\vq), \cr
&\va_{2*}(\vx) := \Psi_*\va_2(\vx)
= d\Psi(\vq).\va_2 
= \lim_{h\rar0} {\Psi(\vq{+}h\va_2) - \Psi(\vq) \over h}
= \lim_{h\rar0} {\Psi(r,\theta{+}h) - \Psi(r,\theta) \over h}
= {\pa \Psi \over \pa \theta}(\vq),
 \cr
}
\ee
Thus
\be
\va_{1*}(\vx) = \cos\theta \vb_1 +  \sin\theta \vb_2 \qand 
\va_{2*}(\vx) = -r\sin\theta \vb_1 +  r\cos\theta \vb_2,
\ee
\ie
\be
\label{eqrempfcb}
[\va_{1*}(\vx)]_{|\vb} = \pmatrix{\cos\theta \cr \sin\theta} \qand 
[\va_{2*}(\vx)]_{|\vb} = \pmatrix{-r\sin\theta \cr r\cos\theta}.
\ee
The basis $(\va_{1*}(\vx),\va_{2*}(\vx))$ is called the %(holonomic) 
basis of the polar coordinate system at~$\vx$
(it is orthogonal but not orthonormal since $||\va_{2*}(\vx)||=r\ne 1$ in general);
% (from which we get the normalized basis $(\va_{1*}(\vx),{1\over r}\va_{2*}(\vx))$ ).
%The columns of the Jacobian matrix of~$\Psi$ at~$\vq$ gives the components of the $\va_{j*}(\vx)$ in the Euclidean basis~$(\vb_i)$: 
And $[d\Psi(\vq)]_{|\va,\vb}
= \pmatrix{[{\pa \Psi \over \pa r}(\vq)]_{|\vb} & [{\pa \Psi \over \pa \theta}(\vq)]_{|\vb}}
= \pmatrix{[\va_{1*}(\vx)]_{|\vb} & [\va_{2*}(\vx)]_{|\vb}}
= \pmatrix{\cos\theta & -r\sin\theta \cr \sin\theta & r\cos\theta}
= [{\pa \Psi^i \over \pa q^j}(\vq)]$ is the Jacobian matrix of~$\Psi$ at~$\vq$. % when $p=\Psi(\vq)$.

And the dual basis of the polar system basis $(\va_{1*}(\vx),\va_{2*}(\vx))$ is called $(dq_1(\vx),dq_2(\vx))$
(defined by $dq_i(\vx).\va_{j*}(\vx)=\delta_{ij}$),
so
\be
\label{eqrempfcb2}
dq_1(\vx) = \cos\theta\, dx_1 +  \sin\theta\, dx_2 \qand
dq_2(\vx) = -{1\over r}\sin\theta\, dx_1 + {1\over r} \cos\theta\, dx_2,
\ee
\ie\ $[dq_1(\vx)]_{|\vb} = \pmatrix{\cos\theta & \sin\theta}$ and
$[dq_2(\vx)]_{|\vb} = -{1\over r} \pmatrix{\sin\theta & \cos\theta}$ (row matrices) when $\vx=\Psi(\vq)$.
\finexa

\debrem
\label{rempfcb}
The components $\gamma^k_{ij}(\vx)$ of the vector $d\va_{j*}(\vx).\va_{i*}(\vx)\in\vRRd$ in the basis $(\va_{i*}(\vx))$ are the Christoffel symbols of the polar coordinate system (with duality notations as it is usually presented):
\be
d\va_{j*}(\vx).\va_{i*}(\vx) = \sumkn \gamma^k_{ij}(\vx)\va_{k*}(\vx).
\ee
At $\vx=\Psi(\vq)$, with $\va_{j*}(\vx) = d\Psi(\vq).\va_j$, \ie\ $(\va_{j*}\circ\Psi)(\vq) = {\pa\Psi \over \pa q^j}$,
we get
\be
d\va_{j*}(\vx).\va_{i*}(\vx)  = {\pa^2\Psi \over \pa q^i\pa q^j}(\vq) %= d^2\Psi(\vq)(\va_i,\va_j)
= d\va_{i*}(\vx).\va_{j*}(\vx), \qso
\gamma^k_{ij}=\gamma^k_{ji}
\ee
for all~$i,j$ (symmetry of the bottom indices as soon as $\Psi$ is~$C^2$).

Here for the polar coordinates, ${\pa \Psi \over \pa r}(\vq) = \cos\theta \vb_1+\sin\theta \vb_2$ gives
${\pa^2 \Psi \over \pa r^2}(\vq) = \vec0$, thus $\gamma^1_{11} = \gamma^2_{11} = 0$, and
${\pa^2 \Psi \over \pa \theta \pa r}(\vq) = - \sin\theta \vb_1+\cos\theta \vb_2 = {1\over r}\va_{2*}(\vx)$,
thus $\gamma^1_{12}=0=\gamma^1_{21}$ and $\gamma^2_{12} = {1\over r}=\gamma^2_{21}$.
And ${\pa \Psi \over \pa \theta}(\vq) = -r\sin\theta \vb_1+r\cos\theta \vb_2$ gives
${\pa^2 \Psi \over \pa \theta^2}(\vq) = -r\cos\theta \vb_1-r\sin\theta \vb_2 = -r\va_{1*}(\vx)$,
thus $\gamma^1_{22}=-r$ and $\gamma^2_{22} = 0$.
\finrem

\debrem
The (widely used) normalized polar coordinate basis $(\vn_1(\vx),\vn_2(\vx))=(\va_{1*}(\vx),{1\over r}\va_{2*}(\vx))$ is not holonomic, \ie\ is not the basis of a coordinate system (and its use makes higher derivation formulas complicated).
Indeed $\vn_2(\vx)={1\over r}\va_{2*}(\vx)$
gives
$d\vn_2(\vx).\vn_1(\vx) = (d({1\over r})(\vx).\vn_1(\vx))\va_{2*}(\vx) + {1\over r} d\va_{2*}(\vx).\vn_1(\vx)$,
and $\vn_1(\vx)=\va_{1*}(\vx)$ gives $d\vn_1(\vx).\vn_2(\vx)=d\va_{1*}(\vx).({1\over r}\va_{2*})$,
thus 
$d\vn_2(\vx).\vn_1(\vx)-d\vn_1(\vx).\vn_2(\vx)
=(d({1\over r})(\vx).\vn_1(\vx))\va_{2*}(\vx) \ne \vec0
$, since ${1\over r}=(x^2+y^2)^{-\demi}$ gives 
$d({1\over r})(\vx).\vn_1(\vx)
=\pmatrix{-x(x^2+y^2)^{-{3\over 2}} & -y (x^2+y^2)^{-{3\over 2}}}
.\pmatrix{\cos\theta \cr \sin\theta}
={1\over r^3}(-r\cos^2\theta - r\sin^2\theta) = {-1\over r^2}\ne0$.
\finrem

\debrem
(Pay attention to the notations.)
Let $f : \vq\in \vec\RR^2_p \rar f(\vq) \in \RR$ be $C^2$.
Call $g$ its push-forward by~$\Psi$, \ie\ $g
: \vx\in\RR^2 \rar g(\vx)=f(\vq)\in\RR$ when $\vx=\Psi(\vq)$.
So $f(\vq) = (g\circ\Psi)(\vq)$% and $df(\vq) = dg(\Psi(\vq) ).d\Psi(\vq)$
and
\be
df(\vq).\va_j = dg(\Psi(\vq)).d\Psi(\vq).\va_j = dg(\vx).\va_{j*}(\vx).
\ee
With $df(\vq).\va_j \eqnote {\pa f \over \pa q^j}(\vq)$
and $dg(\vx).\vb_j \eqnote {\pa g \over \pa x^j}(\vx)$
and $\va_{j*}(\vx)=d\Psi(\vq).\va_j = \sum_i {\pa \Psi^i \over \pa q^j}(\vq)\va_j$, we get
%(stocked in the $j$-column of~$[d\Psi(\vq)]_{|\va,\va}$, 
\be
{\pa f \over \pa q^j}(\vq) = \sum_i  {\pa g \over \pa x^i}(\vx){\pa \Psi^i \over \pa q^j}(\vq) \eqnote {\pa g \over \pa q^j}(\vx)\;\;...\;(!!)
\ee
Mind this notation!! $g$ is a function of~$\vx$, not of~$\vq$, so
$\ds {\pa g \over \pa q^i}(\vx) \mope^{\hbox{\footnotesize means}} {\pa f \over \pa q^i}(\vq)$,
\ie\ $\ds {\pa g \over \pa q^i}(\vx) \mope^{\hbox{\footnotesize means}} {\pa (g\circ\Psi) \over \pa q^i}(\vq)$...
which is $[df(\vq)] = [dg(\vx)].[d\Psi(\vq)$...

Then (with $f$ and~$\Psi$ $C^2$)
\be
\eqalign{
{\pa {\pa g \over \pa q^i} \over \pa q^j}(\vx)
\mope^{\hbox{\footnotesize means}} {\pa {\pa (g\circ\Psi) \over \pa q^i} \over \pa q^j}(\vq)
= &d(dg.\va_{i*})(\vx).d\Psi(\vq).\va_j
= d(dg.\va_{i*})(\vx).\va_{j^*}(\vx) \cr
= &d((dg(\vx).\va_{j^*}(\vx)).\va_{i*}(\vx) + dg(\vx).(d\va_{i*}(\vx).\va_j(\vx))
\eqnote {\pa^2 g \over \pa q^i \pa q^j}(\vx).
}
\ee
So
\be
{\pa^2 g \over \pa q^i \pa q^j}(\vx)
\mope^{\hbox{\footnotesize means}} d^2g(\vx)(\va_{i*}(\vx),\va_{j*}(\vx)) + \sumkn {\pa g \over \pa x^k}(\vx) \gamma^k_{ij}(\vx)\va_k(\vx),
\ee
and ${\pa^2 g \over \pa q^i \pa q^j}(\vx)$ is \textsl{\textbf{not}} reduced to $d^2g(\vx)(\va_{i*}(\vx),\va_{j*}(\vx))$ (the Christoffel symbols have appeared): First order derivatives ${\pa g \over \pa x^k}$ are still alive.
(Contrary to ${\pa^2 g \over \pa x^i \pa x^j}(\vx) = d^2g(\vx)(\vb_i,\vb_j)$ with a Cartesian basis $(\vb_i)$.)

NB: The independent variables $r$ and $\theta$ don't have the same dimension (a length and an angle):
There is no physical meaningful inner dot product in the parameter space $\vec\RR^2_p=\RR\times \RR = \{(r,\theta)\}$,
but this space is very useful... (As in thermodynamics: No meaningful inner dot product in the $(T,P)$ space.)
%NB: $\va_{i*}(p)$ is also noted $\va_{i*}(\vx)$: Here we are in the flat geometric space~$\RR^2$, and a point $p$ can be spotted thanks to a bi-point vector $\vx=\ora{Op}$.
\finrem

%%%%%%%%%%%%%%%%%%%%%%%%%%%%%%%%%%%%%%%%%%%%%%%%%%%%%%%%%%%%%%%%%%%%%%%%%%%%%%%%%%%
%%%%%%%%%%%%%%%%%%%%%%%%%%%%%%%%%%%%%%%%%%%%%%%%%%%%%%%%%%%%%%%%%%%%%%%%%%%%%%%%%%%

\section{Push-forward and pull-back of differential forms}
\label{sectransfd}

%%%%%%%%%%%%%%%%%%%%%%%%%%%%%%%%%%%%%%%%%%%%%%%%%%%%%%%%%%%%%%%%%%%%%%%%%%%%%%%%%%%

\subsection{Definition}

 Setting of~\S~\ref{secpf0}.
%$\Psi:\UE \rar \UF$ is a diffeomorphism, \cf~\eref{eqPsi}.
Consider a differential form $
\alphaE : 
\left\{\eqalign{
\UE & \rar \Es = \calL(E;\RR) \cr
\pE & \rar \alphaE(\pE) 
}\right\}$ on~$\UE$ (a field of linear forms), and a vector field
$\vwE :
\left\{\eqalign{
\UE & \rar E \cr
\pE & \rar \vwE(\pE) 
}\right\}
$.
Hence
$$
\fE = \alphaE.\vwE : 
\left\{\eqalign{
\UE & \rar \RR \cr
\pE & \rar \fE(\pE) =(\alpha.\vwE)(\pE) = \alphaE(\pE).\vwE(\pE)
}\right.
$$
is a scalar valued function (value of~$\vwE$ given by~$\alphaE$).
And~\eref{eqdefpff} gives (push-forward $\fE=\alphaE.\vwE$ by~$\Psi$)
\be
\Psi_*(\alphaE.\vwE)(\pF) = (\alphaE.\vwE)(\pE) = \alphaE(\pE).\vwE(\pE) \qwhen\pF = \Psi(\pE).
\ee
With $\vwEs(\pF) = d\Psi(\pE).\vwE(\pE)$ \cf~\eref{eqdefrapvEivei} (push-forward of $\vwE$), we get
\be
\label{eqdefpfalpha0}
\Psi_*(\alphaE.\vwE)(\pF)
%=   \alphaE(\pE).d\Psi(\pE)^{-1}.\vwEs(\pF)
=   \underbrace{\alphaE(\pE).d\Psi(\pE)^{-1}}_{\eqnote\alphaEs(\pF)}.\vwF(\pF) \qwhen\pF = \Psi(\pE)\;:
\ee
%so (objectivity of measurements hypothesis):

\debdef
\label{deftransfd}
The push-forward of a differential form $\alphaE\in\Omega^1(\UE)$ is the differential form $\in\Omega^1(\UF)$ given by
\be
\label{eqtransfd}
\Psi_*\alphaE : 
\left\{\eqalign{
\UF & \rar F^*=\calL(F;\RR) \cr
\pF & \rar \boxed{\Psi_*\alphaE(\pF) := \alphaE(\pE).d\Psi(\pE)^{-1}} \eqnote \alphaEs(\pF) \qwhen \pF=\Psi(\pE),
}\right.
\ee
the last notation when $\Psi$ is implicit.
In other words, $\Psi_*\alphaE(\pF)=\alphaE(\Psi^{-1}(\pF)).d\Psi^{-1}(\pF)$, \ie
\be
\Psi_*\alphaE := (\alphaE\circ\Psi^{-1}).d\Psi^{-1}. % \quad(= \alphaEs),
\ee
(Once again, we used the same notation $\Psi_*$ than for the push-forward of vector fields and functions: The context removes any ambiguities.)
\findef

\debrem
We cannot always see a vector field (\eg\ we can't see an internal force field): To know it we need to measure it with a well defined tool, the tool being here a differential form; And the definition~\ref{deftransfd} is a compatbility definition so that we can recover the push-forward of the vector field.
\finrem

\debdef
\label{deftransfdb}
The pull-forward of a a differential form $\alphaF\in\Omega^1(\UF)$ is the differential form
\be
\label{eqtransfdb}
\Psi^*\alphaF : 
\left\{\eqalign{
\UE & \rar \calL(E;\RR) \cr
\pE & \rar \Psi^*\alphaF(\pE) := \alphaF(\pF).d\Psi(\pE) \eqnote \alphaFb(\pE) \qwhen \pF=\Psi(\pE),
}\right.
\ee
In other words,
\be
\label{eqtransfdb2}
\Psi^*\alphaF := (\alphaF\circ\Psi).d\Psi. % \quad (=\alphaFb).
\ee
(For an alternative definition, see remark~\ref{rempb}.)
\findef

\debprop
For all $\alphaE\in\Omega^1(\UE)$ and $\alphaF\in\Omega^1(\UF)$ (differential forms),
and $\vwE\in\Gamma(\UE)$ and $\vwF\in\Gamma(\UF)$ (vector fields), we have (objectivity result) %, with $\pF=\Psi(\pE)$,
\be
\label{eqasw}
(\Psi_*\alphaE)(\pF).\vwF(\pF) = \alphaE(\pE).(\Psi_*\vwF)(\pE)  \qwhen \pF=\Psi(\pE),
\ee
\ie\ $\alphaEs(\pF).\vwF(\pF) = \alphaE(\pE).\vwFb(\pE)$. In particular with $\alphaE=df$ (exact differential form) where $f\in C^1(\UE;\RR)$,
\be
\label{eqtransfdc}
d( \Psi_* f) =\Psi_*( df ).
\ee
(This commutativity result is very particular to the case $\alpha=df$: In general
$d( \Psi_* T) \ne \Psi_*( dT )$ for a tensor of order $\ge2$, see~\eg~\eref{eqpsanocom}).
\finprop

\debdem
$\alphaEs(\pF).\vwF(\pF) = (\alphaE(\pE).d\Psi^{-1}(\pF)).\vwF(\pF) = \alphaE(\pE).(d\Psi^{-1}(\pF).\vwF(\pF)) = \alphaE(\pE).\vwF^*(\pE)$, for all $\pF=\Psi(\pE)\in\UF$.

And
$\Psi_* f(\pF) := f(\pE) = f(\Psi^{-1}(\pF))$,
thus $d(\Psi_* f)(\pF) = df(\pE).d\Psi^{-1}(\pF) = \Psi_*( df )(\pF)$.
\findem

And we have
\be
\Psi^*\circ \Psi_* = I \qand \Psi_*\circ \Psi^* = I.
\ee
Indeed
$\Psi^*(\Psi_* \alphaE)(\pE)= \Psi_* \alphaE(\pF).d\Psi(\pE)
= \alphaE(\pE).d\Psi^{-1}(\pF).d\Psi(\pE) = \alphaE(\pE)$.
Idem for $\Psi_*\circ \Psi^* = I$.

\debrem
\label{rempb}
The pull-back $\alphaF^*$ can also be defined thanks to
the natural canonical isomorphism
$
\left\{\eqalign{
\calL(E;F) & \rar \calL(\Fs;\Es) \cr
L & \rar L^*
}\right\}$
given by $L^*(\ell_F).\vu_E = \ell_F.(L.\vu_E)$ for all $(\vu_E,\ell_F)\in E\times F^*$,
and $L^*(\ell_F) = \ell_F.L$ is called the pull-back of~$\ell_F$ by~$L$.
In particular with $\ell_F = \alphaF(\pF)$ and $L = d\Psi(\pE)$ we get
$d\Psi(\pE)^*(\alphaF(\pF)) = \alphaF(\pF).d\Psi(\pE)$,
\ie~\eref{eqtransfdb}. % and~\eref{eqtransfdb2}.
%And this approach gives: $\alphaFb=\Psi^* \alphaF$ is ``covariant objective'' (independent of any observer).
%And, if $\Psi$ is invertible, then the push-forward is the pull-back considered with~$\Psi$ replaced by~$\Psi^{-1}$.
%(and here $E$ and $F$ are the tangent spaces at $\pE\in\OmegaE$ and at $\pF=\Psi(\pE)\in\OmegaF$).
\finrem

%%%%%%%%%%%%%%%%%%%%%%%%%%%%%%%%%%%%%%%%%%%%%%%%%%%%%%%%%%%%%%%%%%%%%%%%%%%%%%%%%%%

\subsection{Incompatibility: Riesz representation and push-forward}
\label{secrvdfdepf}

A push-forward is independent of any inner dot product: It is objective.

But here we introduce inner dot products $\dd_g$ in~$E$ and~$\dd_h$ in~$F$, \eg\ Euclidean dot products in~$\RRntz$ and~$\RRnt$ (observer dependent therefore subjective), because some mechanical engineers can't begin with their beloved Euclidean dot products. 
%necessary to extract from among the proposed constitutive laws of materials the ``good'' ones. But laws must first be proposed...

Let $\alphaE \in \Omega^1(\UE)$ and call $\betaF:=\Psi_*\alphaE$ its push-forward by~$\Psi$, \ie
\be
\betaF(\pF) := \alphaE(\pE).d\Psi(\pE)^{-1} \qwhen \pF=\Psi(\pE).
\ee
Then call $\vag(\pE)\in E$ and $\vbh(\pF)\in F$ the $\dd_g$ and $\dd_h$-Riesz representation vectors of~$\alphaE$ and~$\betaF$, so, for all $\vuE\in\Gamma(\UE)$ and all $\vwF\in\Gamma(\UF)$, in short,
\be
\label{eqpgrdfdepf}
\alphaE.\vuE = (\vag,\vuE)_g,\qand \betaF.\vwF = (\vbh,\vwF)_h,
\ee
which means
$\alphaE(\pE).\vuE(\pE) = (\vag(\pE),\vuE(\pE))_g$ and $\betaF(\pF).\vwF(\pF) = (\vbh(\pF),\vwF(\pF))_h
$, for all $\pE\in \UE$ and $\pF\in \UF$. %, \ie\ in short
This defines the vector fields $\vag\in\Gamma(\UE)$ and $\vbh\in\Gamma(\UF)$.
% (dependent on~$\dd_g$ and~$\dd_h$).

\debprop
$\vbh \ne \Psi_*\vag$ in general (although $\betaF=\Psi_*\alphaE$), because
% is the push-forward of~$\alphaE$,  its Riesz-representation $\vbh$ is \textslbf{not} the push-forward of $\vag$:  we have, with $\pF=\Psi(\pE)$,
%If $\betaF$ is the push-forward of~$\alphaE$ by~$\Psi$), \ie\ if $\alpha_F(\pF)=\alpha_E(\pE).d\Psi^{-1}(\pF)$ when $\pE=\Psi^{-1}(\pF)$, then
\be
\label{eqpgrdfdepf2}
\eqalign{
\vbh(\pF) 
= &  d\Psi(\pE)^{-T}.\vag(\pE) \cr
\ne & d\Psi(\pE).\vag(\pE) \;\hbox{ in general}
%\qso \vbh \ne \Psi_*\vag \hbox{ in general},
}
\ee
%where $\pE=\Psi^{-1}(\pF)$
(unless $d\Psi(\pE)^{-T}=d\Psi(\pE)$, \ie\ $d\Psi(\pE)^T.d\Psi(\pE)^{-1}=I$, as a rigid body motion).

So the Riesz representation vector of the push-forwarded linear form is not the push-forwarded representation vector of the linear form push-forwarded. % (\eg\ for the push-forward of a general motion in continuum mechanics).

%NB: 
This is not a surprise: A push-forward is independent of any inner dot product, while a Riesz representation vector depends on a chosen inner dot product (\Eg\ Euclidean foot? metre?).

So, as long as possible (not before you need to quantify), you should avoid using a Riesz representation vector, \ie\ you should use the original (the qualitative differential form) as long as possible, and delay the use of a representative (quantification with which dot product?) as late as possible. %, \cf~\eref{eqsecemp}?).
\finprop

\debdem
Recall:
The transposed relative to~$\dd_g$ and~$\dd_h$ of the linear map $d\Psi(\pE) \in\calL(E;F)$ is the linear map
$d\Psi(\pE)_{gh}^T \eqnote d\Psi(\pE)^T \in \calL(F; E)$ defined by, for all $\vuE\in E$ and $\vwF\in F$ vectors at~$\pE$ and~$\pF$, \cf~\eref{eqseccpgdd0},
\be
(d\Psi(\pE)^T.\vwF,\vuE)_g = (\vwF,d\Psi(\pE).\vuE)_h.
\ee
\eref{eqpgrdfdepf} gives, with $\pF=\Psi(\pE)$,
\be
\eqalign{
(\vag(\pE),\vuE)_g
= & \alphaE(\pE).\vuE = \bigl(\betaF(\pF).d\Psi(\pE)\bigr).\vuE
= \betaF(\pF).\bigl(d\Psi(\pE).\vuE\bigr) \cr
= &(\vbh(\pF),d\Psi(\pE).\vuE)_h = (d\Psi(\pE)^T.\vbh(\pF),\vuE)_g,
}
\ee
true for all~$\vuE$, thus $\vag(\pE) =  d\Psi(\pE)^T.\vbh(\pF)$, thus~\eref{eqpgrdfdepf2}.
\findem

%%%%%%%%%%%%%%%%%%%%%%%%%%%%%%%%%%%%%%%%%%%%%%%%%%%%%%%%%%%%%%%%%%%%%%%%%%%%%%%%%%%

\section{Push-forward and pull-back of tensors}

To lighten the presentation, we only deal with order 1 and 2 tensors.
%Pour les détails voir poly ``Tenseurs...''.
Similar approach for any tensor.

%%%%%%%%%%%%%%%%%%%%%%%%%%%%%%%%%%%%%%%%%%%%%%%%%%%%%%%%%%%%%%%%%%%%%%%%%%%%%%%%%%%

\subsection{Push-forward and pull-back of order 1 tensors}

\debprop
If $T$ is either a vector field or a differential form, then its push-forward satisfies,
for all~$\xi$ vector field or differential form (when required) in~$\UF$,
\be
\label{eqobj100}
\hbox{in short:}\quad(\Psi_* T)(\xi) = T(\Psi^*\xi), \qwritten \Psi_* T(.) = T(\Psi^*.),
\ee
\ie\ $(\Psi_* T)(\pF).\xi(\pF) = T(\pE).\Psi^*\xi(\pE)$ when $\pF=\Psi(\pE)$. Similarly:
\be
\label{eqobj100b}
\hbox{in short:}\quad(\Psi^* T)(\xi) = T(\Psi_*\xi), \qwritten \Psi^* T(.) = T(\Psi_*.),
\ee
\ie\ $(\Psi^* T)(\pE).\xi(\pE) = T(\pF).\Psi_*\xi(\pF)$ when $\pF=\Psi(\pE)$.
\finprop

\debdem
$\bullet$ Case $T = \alphaE \in \Omega^1(\UE)$ (differential form = a ${0\choose1}$ tensor),
then here $\xi = \vwF\in\Gamma(\UF)$ and we have to check: %, $(\Psi_*\alphaE)(\pF)$ and $\alphaE(\pE)$ being linear,
$(\Psi_*\alphaE)(\pF).\vwF(\pF) = \alphaE(\pE).\Psi^*\vwF(\pE)$, \ie\
$
(\alphaE(\pE).d\Psi^{-1}(\pE)).\vwF(\pF) = \alphaE(\pE).(d\Psi^{-1}(\pE).\vwF(\pF))
$: True.

$\bullet$  Case $T = \vwE \in \Gamma(\UE)$ (vector field $\simeq$ a ${1\choose 0}$ tensor), then here $\xi=\alphaF\in \Omega^1(\UF)$ we have to check:
$(\Psi_*\vwE)(\pF).\alphaF(\pF) = \vwE(\pE).\Psi^*(\alphaF)(\pE)$,
where we implicitly use to the natural canonical isomorphism
$\calJ:
\left\{\eqalign{
E &\rar \Ess \cr
\vw &\rar w \eqnote \vw
}\right\}
$ 
defined by $w(\ell) = \ell.\vw$ for all $\ell\in \Es$.
So we have to check:
$\alphaF(\pF).(\Psi_*\vwE)(\pF) = \Psi^*(\alphaF)(\pE).\vwE(\pE)$,
\ie\ 
$\alphaF(\pF).(d\Psi(\pE).\vwE(\pE)) = (\alphaF(\pF).d\Psi(\pE)^{-1}).\vwE)(\pE)$ : True.

For~\eref{eqobj100b}, use $\Psi^{-1}$ instead of~$\Psi$.
\findem

%%%%%%%%%%%%%%%%%%%%%%%%%%%%%%%%%%%%%%%%%%%%%%%%%%%%%%%%%%%%%%%%%%%%%%%%%%%%%%%%%%%

\subsection{Push-forward and pull-back of order 2 tensors}

\debdef
Let $T$ be an order 2 tensor in~$\UE$. Its push-forward by~$\Psi$ is the order 2 tensor $\Psi_* T$ in~$\UF$ defined by, for all~$\xi_1,\xi_2$ vector field or differential form (when required) in~$\UF$,
\be
\label{eqpftensg1}
\hbox{in short:} \quad \Psi_* T (\xi_1,\xi_2) :=  T(\Psi^* \xi_1,\Psi^*\xi_2) \qwritten \Psi_* T \dd :=  T(\Psi^* \cdot,\Psi^*\cdot),
\ee
\ie\
$\Psi_* T(\pF)(\xi_1(\pF),\xi_2(\pF))  :=  T(\pE)(\Psi^* \xi_1(\pE),\Psi^*\xi_2(\pE))$ when $\pF=\Psi(\pE)$.

Let $T$ be an order 2 tensor in~$\UF$. Its pull-back by~$\Psi$ is the order 2 tensor $\Psi^* T$ in~$\UE$ defined by, for all~$\xi_1,\xi_2$ vector field or differential form (when required) in~$\UE$,
\be
\label{eqdefpbPsi}
%(T^*\dd :=)\quad  
\hbox{in short:} \quad \Psi^* T (\xi_1,\xi_2) :=  T(\Psi_* \xi_1,\Psi_*\xi_2) \qwritten \Psi^* T \dd :=  T(\Psi_* \cdot,\Psi_*\cdot),
\ee
\ie,
$\Psi^* T(\pE)(\xi_1(\pE),\xi_2(\pE))  :=  T(\pF)(\Psi_*\xi_1(\pF),\Psi_*\xi_2(\pF))$ when $\pF=\Psi(\pE)$.
%(Definitions generalized to any $r$ $s$ tensors.)
\findef

%(Thus $\Psi^* (\Psi_* T) = T = \Psi_* (\Psi^* T)$.)

\debexa
If $T \in T^0_2(\UE)$ (\eg, a metric) then, for all vector fields $\vwu,\vwd$ in~$\UF$,
\be
\label{eqpftzd}
\eqalign{
T_* (\vwu,\vwd)
\mathop{=}^{\eref{eqpftensg1}} & T (\vwu{}^*,\vwd{}^*)
=  T (d\Psi^{-1}.\vwu,d\Psi^{-1}.\vwd) ,\cr
}
\ee
\ie,
$
T_*(\pF) (\vwu(\pF),\vwd(\pF))
%=  T(\pE) (\vwu{}^*(\pE),\vwd{}^*(\pE)) \cr
=  T(\pE) (d\Psi^{-1}(\pF).\vwu(\pF),d\Psi^{-1}(\pF).\vwd(\pF))
$ when $\pF=\Psi(\pE)$.

Expression with bases $(\va_i)$ in~$E$ and $(\vb_i)$ in~$F$: In short we have
$(T_*)_{ij}=T_*(\vb_i,\vb_j) = T (\vb_i{}^*,\vb_j{}^*)
= [\vb_i^*]_{|\va}^T.[T]_{|\va}.[\vb_j^*]_{|\va}
= ([\vb_i]_{|\vb}^T.[d\Psi]_{|\va,\vb}^{-T}).[T]_{|\va}.([d\Psi]_{|\va,\vb}^{-1}.[\vb_j]_{|\vb})
= ([d\Psi]_{|\va,\vb}^{-T}.[T]_{|\va}.[d\Psi]_{|\va,\vb}^{-1})_{ij}
$,
thus
\be
[ T_*]_{|\vb}
= [d\Psi]_{|\va,\vb}^{-T}.[T]_{|\va}.[d\Psi]_{|\va,\vb}^{-1},
\ee
%(In particular, this is the change of metric formula due to the push-forward.)
which means
$
[(\Psi_* T)(\pF)]_{|\vb}
= ([d\Psi(\pE)]_{|\va,\vb})^{-T}.[T(\pE)]_{|\va}.([d\Psi(\pE)]_{|\va,\vb})^{-1}$ when $\pF=\Psi(\pE)$.

Particular case of an elementary tensor $T=\alpha_1\otimes \alpha_2  \in T^0_2(\UE)$,
where  $\alpha_1,\alpha_2\in\Omega^1(\UE)$, so
$T(\vu_1,\vu_2) = (\alpha_1\otimes \alpha_2)(\vu_1,\vu_2)=(\alpha_1.\vu_1)(\alpha_2.\vu_2)$:
For all $\vwu,\vwd\in\Gamma(\UF)$,
\be
\eqalign{
(\alpha_1\otimes \alpha_2)_*(\vwu,\vwd)
\mathop{=}^{\eref{eqpftensg1}} (\alpha_1\otimes \alpha_2)(\vwu^*,\vwd^*)
= (\alpha_1.\vwu^*)(\alpha_2.\vwd^*)
%\mathop{=}^{\eref{}} (\alpha_1.d\Psi.\vwu)(\alpha_2.d\Psi.\vwd)
\mathop{=}^{\eref{eqasw}}  (\alpha_{1*}.\vwu)(\alpha_{2*}.\vwd)
%\mathop{=}^{\eref{eqdefeblf}} (\alpha_{1*} \otimes \alpha_{2*})(\vwu,\vwd)
,\cr
}
\ee
thus
\be
%T_*=
(\alpha_1\otimes \alpha_2)_* = \alpha_{1*} \otimes \alpha_{2*}.
\ee
(And any tensor is a finite sum of elementary tensors.)

And for the pull-back: For all vector fields $\vu_1,\vu_2$ in~$\UE$,
\be
\label{eqpftzdb}
\eqalign{
T^* (\vu_1,\vu_2)
\mathop{=}^{\eref{eqpftensg1}} & T (\vu_{1*},\vu_{2*})
=  T (d\Psi.\vu_1,d\Psi.\vu_2) .\cr
}
\ee
\finexa

\debexa
If $T \in T^1_1(\UE)$ %($\simeq$ a field of endomorphisms) 
then for all  vector fields $\vw\in\Gamma(\UF)$ and differential forms $\beta\in\Omega^1(\UF)$,
\be
\label{eqpftzd110}
\eqalign{
T_* (\beta,\vw)
=  T(\beta^*,\vw^*)
=  T(\beta.d\Psi,d\Psi^{-1}.\vw), \cr
}
\ee
\ie, 
$T_*(\pF) (\beta(\pF),\vw(\pF))
=  T(\pE)(\beta(\pF).d\Psi(\pE),d\Psi^{-1}(\pF).\vw(\pF))
$ when $\pF=\Psi(\pE)$.

For the elementary tensor $T=\vu\otimes \alpha  \in T^1_1(\UE)$,
made of the vector field $\vu\in\Gamma(\UE)$ and of the differential form $\alpha\in\Omega^1(\UE)$:
For all $\beta,\vw\in\Omega^1(\UF)\times\Gamma(\UF)$, in short,
\be
\eqalign{
(\vu\otimes \alpha)_*(\beta,\vw)
\mathop{=}^{\eref{eqpftensg1}} (\vu\otimes \alpha)(\beta^*,\vw^*)
= (\vu.\beta^*)(\alpha.\vw^*)
%\mathop{=}^{\eref{}} (\alpha_1.d\Psi.\vwu)(\alpha_2.d\Psi.\vwd)
\mathop{=}^{\eref{eqasw}}  (\vu_*.\beta)(\alpha_*.\vw)
= (\vu_* \otimes \alpha_*)(\beta,\vw)
,\cr
}
\ee
thus 
\be
\label{eqpftzd11e}
(\vu\otimes \alpha)_* = \vu_* \otimes \alpha_*.
\ee

Expression with bases $(\va_i)$ in~$E$ and $(\vb_i)$ in~$F$: In short we have
$(T_*)_{ij} = T_*(b^i,\vb_j)
=T(\Psi^*(b^i),\Psi^*(\vb_j))
=[\Psi^*(b^i)].[T].[\Psi^*(\vb_j)]
=[b^i].[d\Psi].[T].[d\Psi^{-1}].[\vb_j]
=([d\Psi].[T].[d\Psi^{-1}])_{ij}
$,
thus
\be
\label{eqpftzd110m}
[ T_*]_{|\vb}
= [d\Psi]_{|\va,\vb}.[T]_{|\va}.[d\Psi]_{|\va,\vb}^{-1},
\ee
which means
$
[(\Psi_* T)(\pF)]_{|\vb}
= [d\Psi(\pE)]_{|\va,\vb}.[T(\pE)]_{|\va}.[d\Psi(\pE)]_{|\va,\vb}^{-1}$ when $\pF=\Psi(\pE)$.
\finexa

%%%%%%%%%%%%%%%%%%%%%%%%%%%%%%%%%%%%%%%%%%%%%%%%%%%%%%%%%%%%%%%%%%%%%%%%%%%%%%%%%%%

\subsection{Push-forward and pull-back of endomorphisms}

We have the natural canonical isomorphism
\be
\label{eqpftzd112}
\calJ_2 :
\left\{\eqalign{
\calL(E;E) & \rar \calL(E^*,E;\RR) \cr
L & \rar T_L = \calJ_2(L) \qwhere
T_L(\alpha,\vu) := \alpha.L.\vu, \quad \forall (\alpha,\vu)\in\Es\times E.
}\right.
\ee
Thus $\Psi_* T_L(m,\vw) = T_L(\Psi^* m, \Psi^* \vw) = (\Psi^* m).L.(\Psi^* \vw)=m.d\Psi.L.d\Psi^{-1}.\vw$, thus:

\debdef
The push-forward by~$\Psi$ of a field of endomorphisms $L$ on $\UE$
is the field of endomorphisms $\Psi_* L = L_*$ on~$\UF$ defined by
\be
\label{eqpfL}
\hbox{in short:}\quad \Psi_* L = \boxed{L_* = d\Psi . L .d\Psi^{-1}},
\ee
\ie, $L_*(\pF) = d\Psi(\pE) . L(\pE) .d\Psi^{-1}(\pF)$ when $\pF=\Psi(\pE)$.
\findef

Thus with bases we get $[L_*]_{|\vb} = [d\Psi]_{|\va,\vb}.[L]_{|\va}.[d\Psi]_{|\va,\vb}^{-1}$,
``as in~\eref{eqpftzd110m}''.

%$(\Psi_* T)(\pF)(\alphaF(\pF),\vw(\pF)) = \alphaF(\pF).d\Psi(\pE) . L_T(\pE) .d\Psi(\pE)^{-1}. \vw(\pF)$.)

\debexa
Elementary field of endomorphisms $L =  (\calJ_2)^{-1}(\vu\otimes \alpha)$, where $\vu\in\Gamma(E)$ and $\alpha\in\Omega^1(E)$: So $T_L=\vu\otimes \alpha$ and $L.\vu_2 = (\alpha.\vu_2)\vu$ for all $\vu_2\in\Gamma(\UE)$).
Thus $L_*.\vw_2
= d\Psi . L .d\Psi^{-1}.\vw_2
= d\Psi . L .\vw_2{}^*
= (\alpha.\vw_2{}^*)d\Psi.\vu
= (\alpha_*.\vw_2)\vu_*
$ for all $\vw_2\in \Gamma(E)$, thus $(T_L)_* = \vu_*\otimes\alpha_*$.
\finexa

\debdef
Let $L$ be a field of endomorphisms on~$\UF$.
Its pull-back by~$\Psi$
is the field of endomorphisms $\Psi^* L = L^*$ on~$\UE$ defined by
\be
\label{eqbfL}
\hbox{in short:}\quad \Psi^* L  = \boxed{L^* = d\Psi^{-1} . L .d\Psi},
\ee
\ie,
$L^*(\pE) = d\Psi^{-1}(\pF) . L(\pF) .d\Psi(\pE)$ when $\pF = \Psi(\pE)$.
\findef

\comment{
\Eg, if $\vu$ is a vector field on~$\UE$, then its derivative $d\vu$ 
is a field of endomorphisms on~$\UE$ which push-forward $\Psi_* d\vu$ is given by
\be
(\Psi_* d\vu)(\pF) = d\Psi(\pE) . d\vu(\pE) . d\Psi^{-1}(\pF).
\ee
}

%%%%%%%%%%%%%%%%%%%%%%%%%%%%%%%%%%%%%%%%%%%%%%%%%%%%%%%%%%%%%%%%%%%%%%%%%%%%%%%%%%%

\subsection{Application to derivatives of vector fields}

$\vu\in \Gamma(\UE)$ is a $C^1$ vector field in~$\UE$), $\pE\in\UE$,
so $d\vu:\UE \rar \calL(E;E)$ (given by
$d\vu(\pE) .\vw(\pE) = \lim_{h\rar 0} {\vu(\pE+h\vw(\pE)) - \vu(\pE) \over h}$
for all $\vw\in \Gamma(\UE)$).
%(identified to a ${1\choose1}$ tensor).
Thus its push-forward: % $\Psi_*(d\vu) \eqnote (d\vu)_*$ is given by~\eref{eqpfL}:
\be
\label{eqZ_t}
 ((d\vu)_*=) \quad \Psi_*(d\vu) =  d\Psi . d\vu .d\Psi^{-1}
\ee
\ie\ $(d\vu)_*(\pF) = d\Psi(\pE) . d\vu(\pE) .d\Psi(\pE)^{-1}$ when $\pF = \Psi(\pE)$.

%%%%%%%%%%%%%%%%%%%%%%%%%%%%%%%%%%%%%%%%%%%%%%%%%%%%%%%%%%%%%%%%%%%%%%%%%%%%%%%%%%%

\subsection{$\Psi_*(d\vu)$ versus $d(\Psi_*\vu)$: No commutativity}

Here $\Psi$ is~$C^2$, $\vu \in \Gamma(\UE)$, $\pE\in\UE$, $\pF=\Psi(\pE)$, so
$%\vu_*(\pF) = 
\Psi_*\vu(\pF)
= d\Psi(\pE).\vu(\pE)
= (d\Psi(\Psi^{-1}(\pF)).(\vu(\Psi^{-1}(\pF))
$, and, for all $\vw\in \Gamma(\UF)$,
\be
d(\Psi_*\vu)(\pF).\vw(\pF)
= (d^2\Psi(\pE).(d\Psi^{-1}(\pF).\vw(\pF))).\vu(\pE)
+ d\Psi(\pE).d\vu(\pE).d\Psi^{-1}(\pF).\vw(\pF), % \; \in\calL(F;F)
\ee
with $\Psi_*(d\vu)(\pF) = d\Psi(\pE).d\vu(\pE).d\Psi^{-1}(\pF)$, thus, in short,
\be
\label{eqpsanocom}
d(\Psi_*\vu).\vw
= \Psi_*(d\vu).\vw
+ d^2\Psi(\Psi^*\vw,\vu)
\ne \Psi_*(d\vu) \quad\hbox{in general}.
\ee
So the differentiation $d$ and the push-forward ${}_*$ do not commute ($d(\Psi_*\vu) = \Psi_*(d\vu)$ iff $\Psi$ is affine).
%(It should not be a surprise: A derivation is \textslbf{not} a tensorial operation.)

%%%%%%%%%%%%%%%%%%%%%%%%%%%%%%%%%%%%%%%%%%%%%%%%%%%%%%%%%%%%%%%%%%%%%%%%%%%%%%%%%%%

\subsection{Application to derivative of differential forms}

Let $\alpha\in \Omega^1(\UE)$ (a differential form on~$\UE$).
Its derivative $d\alpha : \UE \rar \calL(E;\Es)$ is given by
$
d\alpha(\pE) .\vu(\pE)
= \lim_{h\rar 0} {\alpha(\pE+h\vu(\pE)) - \alpha(\pE) \over h} \;\in \Es
$,
for all $\vu\in \Gamma(\UE)$, \ie, for all $\vu_1,\vu_2\in \Gamma(\UE)$,
\be
(d\alpha(\pE) .\vu_1(\pE)).\vu_2(\pE)
= \lim_{h\rar 0} {\alpha(\pE+h\vu_1(\pE)).\vu_2(\pE) - (\alpha(\pE).\vu_1(\pE)).\vu_2(\pE) \over h} \quad\in \RR.
\ee
With the natural canonical isomorphism $\calL(E;\Es) \simeq \calL(E,E;\RR)$, \cf~\eref{eqtJ2f} with $\Ess \simeq E$,
we can write $d\alpha(\pE)(\vu_1(\pE)).\vu_2(\pE) = d\alpha(\pE)(\vu_1(\pE),\vu_2(\pE))$, \ie
\be
d\alpha(\vu_1).\vu_2 = d\alpha(\vu_1,\vu_2).
\ee
Thus the push-forward $\Psi_*(d\alpha) \eqnote (d\alpha)_*$ of~$d\alpha$, is given by, for all $\vwu,\vwd \in \Gamma(\UF)$,
in short,
\be
\label{eqpfda2}
(d\alpha)_*(\vwu,\vwd) = d\alpha(\vwu^*,\vwd^*), % \quad ( = d\alpha(d\Psi^{-1}.\vwu,d\Psi^{-1}.\vwd)),
\ee
\ie, with $\pF=\Psi(\pE)$, $(d\alpha)_*(\pF).\vwu(\pF)).\vwd(\pF)
= (d\alpha(\pE).d\Psi^{-1}(\pF).\vwu(\pF)).d\Psi^{-1}(\pF).\vwd(\pF)
$. %when $\pF=\Psi(\pE)$.

In particular,
$(d^2f)_*(\vwu,\vwd) = d^2f(d\Psi^{-1}.\vwu,d\Psi^{-1}.\vwd)$ ($=d^2f(\vwu^*,\vwd^*) $).

%%%%%%%%%%%%%%%%%%%%%%%%%%%%%%%%%%%%%%%%%%%%%%%%%%%%%%%%%%%%%%%%%%%%%%%%%%%%%%%%%%%

\subsection{$\Psi_*(d\alpha)$ versus $d(\Psi_*\alpha)$: No commutativity}

Here $\Psi$ is~$C^2$, $\vu \in \Gamma(\UE)$, $\pE\in\UE$ and $\pF=\Psi(\pE)$.
We have
$
\Psi_* \alpha(\pF)
= \alpha(\pE).d\Psi^{-1}(\pF)
= \alpha (\Psi^{-1}(\pF)).d\Psi^{-1}(\pF)
$, thus, for all $\vwu\in \Gamma(\UF)$,
\be
d(\psi_*\alpha)(\pF).\vwu(\pF) = (d\alpha(\pE).d\Psi^{-1}(\pF).\vwu(\pF)).d\Psi^{-1}(\pF)
+ \alpha(\pE).d^2\Psi^{-1}(\pF).\vwu(\pF) \; \in\Fs,
\ee
thus, for all $\vwu,\vwd \in \Gamma(\UF)$, in short
\be
d(\psi_*\alpha)(\vwu,\vwd)
=  d\alpha(d\Psi^{-1}.\vwu,d\Psi^{-1}.\vwd)
 + \alpha.d^2\Psi^{-1}(\vwu,\vwd) \ne  d\alpha(\vwu^*,\vwd^*)  \quad\hbox{in general}.
\ee
So the differentiation $d$ and the push-forward ${}_*$ do not commute ($d(\Psi_*\alpha) = \Psi_*(d\alpha)$ iff $\Psi$ is affine).

\comment{
%%%%%%%%%%%%%%%%%%%%%%%%%%%%%%%%%%%%%%%%%%%%%%%%%%%%%%%%%%%%%%%%%%%%%%%%%%%%%%%%%%%

\subsection{Changement de variable dans les intégrales}

Ici $E=F=\vRRn$, $\Omega_F = \Psi(\Omega_E)$ et $\Omega_F$ sont des ouverts dans~$\RRn$,
$\Psi : \Omega_E\rar\Omega_F$ est un difféomorphisme.
Et le cadre est euclidien (on a besoin des ``éléments de volume'').

Pour les fonctions $f \in C^0(\RRn;\RR)$ à valeurs scalaires, on~a :
\be
\label{eqcvipb}
\int_{p\in \Omega_F} f(p)\,d\Omega_F
= \int_{P\in \Omega_E} f(\Psi(P))\,|J_\Psi(P)|\,d\Omega_E
= \int_{P\in \Omega_E} (\Psi^*f)(P)\,\Psi^*(d\Omega_F),
\ee
où $J_\Psi(P) = \det(d\Psi(P))$ est le jacobien de~$\Psi$ en~$P$,
la dernière égalité en termes de pull-back
car $\Psi^*(d\Omega_F) = |J_\Psi(P)|\,d\Omega_E$ (pull-back d'une forme $n$-linéaire alternée).

%Pour les fonctions à valeurs vectoriels $\vf \in C^0(\RRn;\RRn)$, comme $(\Psi^*f)(P) = d\Psi^{-1}(\Psi(P)).\vf(\Psi(P))$ la formule~\eref{eqcvipb} n'est pas généralisable.
Par exemple, c'est~\eref{eqcvipb} qui est utilisée dans les formules donnant les puissances virtuelles.
%à ceci près que pour avoir la cohérence des résultats il faut exclure tout produit scalaire (euclidien on non).
Ainsi avec $f = \alpha.\vw$ où $\alpha$ est une forme différentielle et $\vw$ un champ de vecteurs,
on~a $\Psi^* f = \Psi^* \alpha.\Psi^* \vw$ par définition de~$\Psi^*\alpha$.

Par exemple si on utilise un produit scalaire euclidien $\dd_g = \dd_\RRn$ à tout~$t$,
et avec $f=(\vu,\vw)_g$ :
le pull-back de la métrique~$\dd_g$
(le champ de produits scalaires constant égal à~$\dd_g$ en tout point $p\in F$)
est donné par $(\Psi^* g)(P) = d\Psi(P).g(p).d\Psi(P)$ quand $p=\Psi(P)$.
Et on~a $(\Psi^*g)(P)(\vU(P),\vW(P))_G = g(\vu(p),\vw(p)) = (C(P).\vU(P),\vW(P))_\RR
= C^\flat(P)(\vU(P),\vW(P))$,
donc $\Psi^*g = C^\flat = $ le transformé du tenseur de Cauchy $C=F^T.F$ par Riesz, voir~\S~\ref{sectcpbm}.
Donc :
\be
\eqalign{
\int_{p\in \Omega_F} (\vu(p),\vw(p))_\RRn\,d\Omega_F
= & \int_{p\in \Omega_E} (\Psi^*g)(P)((\Psi^*\vu)(P),(\Psi^*\vw)(P))\,\Psi^*(d\Omega_F) \cr
= & \int_{P\in \Omega_E} (F(P).\vU(P),F(P).\vW(P))_\RRn\,|J_\Psi(P)|\,d\Omega_E,
}
\ee
ce qui est immédiat dans une base euclidienne, la même à~$\tz$ et à~$t$.
}

%%%%%%%%%%%%%%%%%%%%%%%%%%%%%%%%%%%%%%%%%%%%%%%%%%%%%%%%%%%%%%%%%%%%%%%%%%%%%%%%%%%
%%%%%%%%%%%%%%%%%%%%%%%%%%%%%%%%%%%%%%%%%%%%%%%%%%%%%%%%%%%%%%%%%%%%%%%%%%%%%%%%%%%

\newpage
\part{Lie derivative}

%%%%%%%%%%%%%%%%%%%%%%%%%%%%%%%%%%%%%%%%%%%%%%%%%%%%%%%%%%%%%%%%%%%%%%%%%%%%%%%%%%%
%%%%%%%%%%%%%%%%%%%%%%%%%%%%%%%%%%%%%%%%%%%%%%%%%%%%%%%%%%%%%%%%%%%%%%%%%%%%%%%%%%%

\section{Lie derivative}
\label{secdL}

%%%%%%%%%%%%%%%%%%%%%%%%%%%%%%%%%%%%%%%%%%%%%%%%%%%%%%%%%%%%%%%%%%%%%%%%%%%%%%%%%%%

\setcounter{subsection}{-1}

\subsection{Purpose and first results}

%%%%%%%%%%%%%%%%%%%%%%%%%%%%%%%%%%%%%%%%%%%%%%%%%%%%%%%%%%%%%%%%%%%%%%%%%%%%%%%%%%%

\subsubsection{Purpose?}

Cauchy's approach may be insufficient, \eg: 

\begin{enumerate}[leftmargin=10pt,parsep=3pt plus 1pt minus 1pt,itemsep=0cm,topsep=0cm] %leftmargin=*,labelindent=-1cm

\item 
- Cauchy's approach aims to compare two vectors deformed by a motion, thanks to a Euclidean dot product and the deformation gradient~$F$, with the deformation tensor~$C$ defined by $(C.\vW_1) \bcdot \vW_2 := (F.\vW_1)\bcdot (F.\vW_2)$.
It is a quantitative approach (needs a chosen Euclidean dot product: foot? metre?).

- Cauchy's approach is a first order method (dedicated to linear material): Only the first order Taylor expansion of the motion is used: Only $d\Phi=F$ is used (the ``slope''), not $d^2\Phi=dF$ (the ``curvature'') or higher derivatives. %(Also see~\S~\ref{secCG1}.)

\item
- Lie's approach aims to build qualitative ``covariant objective constitutive laws'' (some will be discredited \textbf{afterward}, because of invariance or thermodynamical requirements).

% (no a priori use of man-made tools like inner dot products),- Lie's approach is  (Cauchy's approach is concerned with isometric objectivity).

- Lie's approach ``naturally'' applies to non-linear materials thanks to second order Lie derivatives which uses the second order Taylor expansion of the motion.
%(at first order gives linearity, at second order gives curvature).

- In a non planar surface~$S$, you need the Lie derivative if you want to derive along a trajectory.
 % (and gives the expected results).

- In a Galilean Euclidean framework (quantification), the first order Lie derivatives approach give the same results than Cauchy's approach.

(Cauchy died in 1857, and Lie was born in 1842: Unfortunately Cauchy could not use the Lie derivative.)
\end{enumerate}

\comment{
: The composition of deformation directly applies
($F^{t_1}_{t_2} \circ F^{t_0}_{t_1} = F^{t_0}_{t_2}$), which is not the case for Cauchy's tensor
($C^{t_1}_{t_2} \circ C^{t_0}_{t_1} \ne C^{t_0}_{t_2}$ in general).
%  since it doesn't need to compare to vectors and their linear relative deformations
}

% The Lie derivative of a vector field gives the rate of stress applied on this vector field due to a flow: this section will explain this fact.

%%%%%%%%%%%%%%%%%%%%%%%%%%%%%%%%%%%%%%%%%%%%%%%%%%%%%%%%%%%%%%%%%%%%%%%%%%%%%%%%%%%

\subsubsection{Basic results}

The Eulerian velocity of the motion is~$\vv$. With the material derivative is ${D\eul\over Dt}:={\pa \eul \over \pa t} + d\eul.\vv$:

\begin{enumerate}[leftmargin=13pt,parsep=3pt plus 1pt minus 1pt,itemsep=0cm,topsep=0cm] %leftmargin=*,labelindent=-1cm
\item The Lie derivative $\calL_\vv f$ of a Eulerian scalar valued function~$f$ is the material derivative %, see~\eref{eqdl0}:
\be
\label{eqcallw00}
\calL_\vv f = {Df\over Dt}. % \quad (= {\pa f \over \pa t} + df.\vv),
\ee

\item The Lie derivative $\calL_\vv \vw$ of a (Eulerian) vector field~$\vw$ is more than just the material derivative ${D \vw \over D t}$:
% (a vector is not just a collection of scalar components) %, see~\eref{eqdl}:
\be
\label{eqcallw0}
\calL_\vv \vw = {D \vw \over D t}  - d\vv.\vw. % \quad (= {\pa \vw \over \pa t} + d\vw.\vv - d\vv.\vw),
\ee
$\calL_\vv \vw$ gives the rate of stress on~$\vw$ due to a flow, and in particular
the $- d\vv.\vw$ term in $\calL_\vv \vw$ tells that the spatial variations of~$\vv$ (variations of the flow) act on the evolution of the stress (anticipated).

\item \eref{eqcallw00}-\eref{eqcallw0} enable to define the Lie derivatives of tensors of any order.
\end{enumerate}

\smallskip

%Interpretation of~\eref{eqcallw0}
% (so could be useful to build constitutive laws)T; And . %, see \S~\ref{sectl1}.

 % and footnote page~\pageref{footrem}.

%%%%%%%%%%%%%%%%%%%%%%%%%%%%%%%%%%%%%%%%%%%%%%%%%%%%%%%%%%%%%%%%%%%%%%%%%%%%%%%%%%%

\subsection{Definition}
\label{secdLo}

%%%%%%%%%%%%%%%%%%%%%%%%%%%%%%%%%%%%%%%%%%%%%%%%%%%%%%%%%%%%%%%%%%%%%%%%%%%%%%%%%%%

\subsubsection{Issue (ubiquity gift)...}
\label{secdLos1}

$\tPhi$ is supposed to be regular.
$\vv(t,p(t))={\pa \tPhi \over \pa t}(t,\Pobj)$ is the Eulerian velocity at $t$ at $p(t) = \tPhi(t,\Pobj)$.
Recall: If $\eul$ is a Eulerian function then its material time derivative is %, \cf~\eref{eqdefdfdt2},
\be
\label{eqo1}
{D\eul\over Dt}(t,p(t))
=\lim_{h \rar 0} {\eul(\tph,p(\tph)) - \eul(t,p(t)) \over h}
%= {\pa \eul \over \pa t} + d\eul.\vv)(t,p(t)
.
\ee
Issue: The rate ${\eul(\tph,p(\tph)) - \eul(t,p(t)) \over h}$ raises questions:

1- The difference $\eul(\tph,p(\tph)) - \eul(t,p(t))$ requires the time and space ubiquity gift to be calculated by an observer,
since it mixes two distinct times, $t$ and $\tph$, and two distinct locations, $p(t)$ and~$\ptph$. %See remark~\ref{remhypNE}.

2- The difference $\eul(\tph,p(\tph)) - \eul(t,p(t))$ can be impossible: \Eg\
if $\eul=\vw$ is a vector field in a ``non planar surface considered on its own'' (manifold)
then $\eul(\tph,p(\tph))$ and $\eul(t,p(t))$ don't belong to the same (tangent) vector space (so the difference $\vw(\tph,p(\tph)) - \vw(t,p(t))$ is meaningless).

%%%%%%%%%%%%%%%%%%%%%%%%%%%%%%%%%%%%%%%%%%%%%%%%%%%%%%%%%%%%%%%%%%%%%%%%%%%%%%%%%%%

\subsubsection{...Toward a solution (without ubiquity gift)...}
\label{sectl1}

To compare $\eul(\tph,p(\tph))$ and $\eul(t,p(t))$ (to get the evolution of~$\eul$ along a trajectory),
you need the duration~$h$ to get from $t$ to $\tph$ 
and to move from $p(t)$ to~$\ptph$.
So, you must: % (without any gift of ubiquity):

$\bullet$ take the value $\eul(t,\pt))$ with you (for memory),

$\bullet$ move along the considered trajectory, % $\tPhi_\Pobj$, 
and doing so, the value $\eul(t,\pt)$ has possibly changed to, with $\tau = \tph$,
\be
\label{eqPhittph}
((\Phittau)_*\eul_t)(\ptau) 
\eqnote \eul_{t*}(\tau,\ptau) \quad (\hbox{push-forward});
\ee
%\Eg\ for vector fields \cf~\eref{eqdefrapvEivei} (push-forward).

$\bullet$ And now, at $(\tau,\ptau)$ where you are, you can compare the actual value $\eul(\tau,\ptau)$
with the value $\eul_{t*}(\tau,\ptau)$ you arrived with (the transported memory), % (no gift of ubiquity required):
thus the difference %, see figure~\ref{figpfl},
\be
\label{eqtauxlie0}
%\hbox{Rate} = 
\eul(\tau,\ptau) - \eul_{t*}(\tau,\ptau) 
\ee
is meaningful for a human being
since it is computed at a unique time~$\tau$ and at a unique point~$\ptau$ (no gift of ubiquity required).

%NB: In a ``non planar surface considered on its own'' (not considered as a subset of an affine space), \ie\ in a manifold, and with $\eul$ a vector field, \eref{eqtauxlie} is the only meaningful rate you can consider if you want to compare an actual value with a past value (along a trajectory). % a ``rate of evolution'').
%(this is why the Lie derivative is so important in general relativity).

\begin{figure}[!ht]
%\begin{center}
%\epsfig{figure=cisail.eps,height=6.cm} %,width=5in
%\qquad\qquad\includegraphics[width=0.7\textwidth]{Push02tv.png}
%\qquad
\qquad
\includegraphics[width=.8\textwidth]{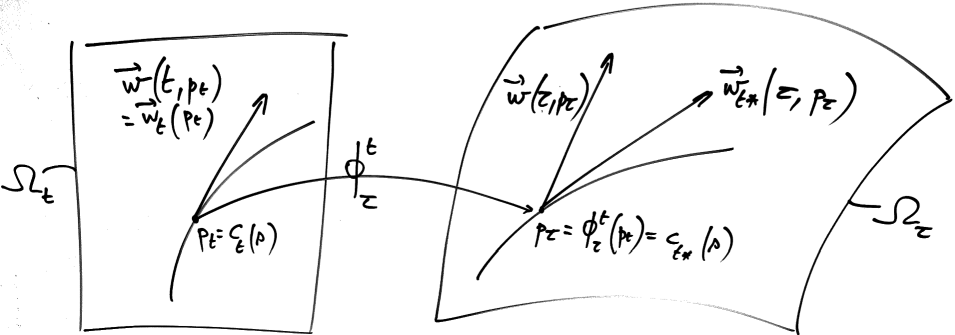}
%\qquad\qquad\includegraphics{cisail2.png}
%\epsfig{figure=cisail2.eps}
%\end{center}
\caption{To compute~\eref{eqtauxlie0} with $\eul=\vw$ a (Eulerian) vector field:
At~$t$ define the vector field $\vw_t$ in~$\Omegat$ by $\vw_t(\pt):=\vw(t,\pt)$.
The (spatial) curve $c_t:s\rar \pt=c_t(s)$ in~$\Omegat$ is an integral curve of~$\vw_t$, \ie\ satisfies $c_t{}'(s) = \vw_t(c_t(s))$. $c_t$ is transformed by $\Phittau$ into the (spatial) curve $c_\tau = \Phittau\circ c_t :
s\rar \ptau=c_\tau(s) {=} \Phittau(c_t(s))$ in~$\Omegatau$;
Hence $c_\tau{}'(s) = d\Phittau(\pt).c_{}'(s) = d\Phittau(\pt).\vw_t(\pt) \eqnote \vw_{t*}(\tau,\ptau)$ is the tangent vector at~$c_\tau$ at~$\ptau$ (push-forward).
Thus the difference $\vw(\tau,\ptau) - \vw_{t*}(\tau,\ptau)$ can be computed by a human being,
\ie\ without ubiquity gift.
%, $\vw(\tau,\ptau)$ being the value of~$\vw$ at $\tau$ at~$\ptau$.
}
\label{figpfl}
\end{figure}

%%%%%%%%%%%%%%%%%%%%%%%%%%%%%%%%%%%%%%%%%%%%%%%%%%%%%%%%%%%%%%%%%%%%%%%%%%%%%%%%%%%

\subsubsection{... The Lie derivative, first definition}
\label{sectl2}

\debdef
%Let $\vv(t,p)={\pa \tPhi \over \pa t}(t,\Pobj)$ be the Eulerian velocity at~$t$ at $p=\tPhi(t,\Pobj)$, \cf~\eref{eqdefve},let $\Phittau$
%In~$\RRn$ (affine space), 
The Lie derivative $\calL_\vv \eul$ along~$\vv$ of an Eulerian function~$\eul$
is the Eulerian function $\calL_\vv \eul$ defined by, at $t$ at $\pt=\tPhi(t,\Pobj)$,
%with $\eul_{t*}(\tph,\ptph)=((\Phittph)_*\eul_t)(\tph,\ptph)$,
\be
\label{eqdefDLrn}
\calL_\vv \eul(t,\pt)
\eqdef \lim_{h\rar 0} {\eul(\tph,\ptph) - (\Phittph)_*\eul_t(\ptph) \over h}.
%\eqdef \lim_{h\rar 0} {\eul(\tph,\ptph) - \eul_{t*}(\tph,\ptph) \over h}.
\ee
\findef

%\medskip
\noindent
{\bf Interpretation:} 
$\calL_\vv \eul $ measures the rate of change of~$\eul$ along a trajectory:

$\bullet$ $\eul(\tph,\ptph)$ is the value of~$\eul$ at~$\tph$ at~$\ptph$, see figure~\ref{figpfl}.

$\bullet$ $\eul_{t*}(\tph,\ptph)=((\Phittph)_*\eul_t)(\tph,\ptph)$ is exclusively strain related:
It is the memory transported along a flow, \ie\ the value $\eul(t,\pt)$ distorted by the flow.

\medskip
So, with $g$ defined by $g(\tau) = ((\Phi^t_\tau)_* \eul_t)(\ptau)$
(in particular $g(t) = \eul_t(\pt)$):
\be
\label{eqdefDLrn1}
\calL_\vv \eul(t,\pt) := g'(t)
= \lim_{\tau\rar t} {g(\tau) - g(t) \over \tau-t}
\mathop{=}^{\hbox{\scriptsize also written}}\quad {d ((\Phi^t_\tph)_* \eul_t)(\ptph) \over dt}_{|\tau=t}.
\ee

%%%%%%%%%%%%%%%%%%%%%%%%%%%%%%%%%%%%%%%%%%%%%%%%%%%%%%%%%%%%%%%%%%%%%%%%%%%%%%%%%%%

\subsubsection{A more general definition}

The rate in~\eref{eqdefDLrn} has to be slightly modified to be adequate in all situations: 
$\eul(\tph,\ptph) - \eul_*(\tph,\ptph)$ is computed at $(\tph,\ptph)$ which moves as $h\rar0$,
and on a ``non-planar manifold'' this is problematic.
The ``natural'' definition %, as far as mechanic and physic are concerned, 
is to arrive with the memory: %alternative definition: %(rewrite~\eref{eqdefDLrn} with $-h$ insead of~$h$):

\debdef
\label{defeqdL2}
The Lie derivative $\calL_\vv \eul$ along~$\vv$ of an Eulerian function~$\eul$
is the Eulerian function $\calL_\vv \eul$ defined by, at $t$ at $\pt=\tPhiPobj(t)$,
\be
\label{eqdefDLrn2}
\calL_\vv \eul(t,\pt)
\eqdef \lim_{h\rar 0} {\eul(t,\pt) - (\Phitmht)_*\eul_{t{-}h}(\pt) \over h}. %, \quad\hbox{rate in~$\Omegat$}.
\ee
\def\tg{\tilde g}
\Ie\ with $\tg$ defined by $\tg(\tau) = ((\Phi^\tau_t)_* \eul_\tau)(\pt)$ (in particular $\tg(t) = \eul(t,\pt)$):
\be
%\label{eqdefDL2}
\calL_\vv \eul(t,\pt) := \tg'(t)
= \lim_{\tau\rar t} {\tg(t) - \tg(\tau) \over t-\tau}
= \lim_{\tau\rar t} {\tg(\tau) - \tg(t) \over \tau-t}
\mathop{=}^{\hbox{\scriptsize also written}}\quad {d ((\Phi^\tau_t)_* \eul_\tau)(\pt) \over d\tau}_{|\tau=t}
.
\ee
%(And in~$\RRn$ the equivalence with~\eref{eqdefDLrn1} is obvious.)
\findef

Here the observer must:

$\bullet$ At~$\tmh$ at $\ptmh=\tPhiPobj(\tmh)$, take the value $\eul(\tmh,\ptmh)$,

$\bullet$ travel along the trajectory $\tPhiPobj$,

$\bullet$ once at $t$ at $\pt=\tPhiPobj(t)$, this value has become $((\Phittmh)_* \eul_\tmh)(\pt)$ (transported memory),

$\bullet$ and then the comparison with $\eul(t,\pt)$ can be done in~$\Omegat$ (no ubiquity gift required).

\debexe
Prove: \eref{eqdefDLrn} and~\eref{eqdefDLrn2} are equivalent.

\debrep
With $(\Phittph)^*.(\Phittph)_*=I$, 
%the limit of a product being the product of the limit, 
\eref{eqdefDLrn} gives
$
\calL_\vv \eul(t,\pt)
=\lim_{h\rar 0} {(\Phittph)^*\eul(t,\pt) - \eul_t(t,\pt) \over h}
= \lim_{h\rar 0} {(\Phittmh)^*\eul_\tmh)(\pt) - \eul_t(\pt) \over -h}
= \lim_{h\rar 0} {\eul_t(\pt) - ((\Phittmh)^*\eul_\tmh)(\pt)  \over h},
$
and $(\Phittmh)^* %= ((\Phittmh)^{-1})_* 
= (\Phitmht)_*$.
\finrep
\finexe

%%%%%%%%%%%%%%%%%%%%%%%%%%%%%%%%%%%%%%%%%%%%%%%%%%%%%%%%%%%%%%%%%%%%%%%%%%%%%%%%%%%

\subsubsection{Equivalent definition (differential geometry)}

\debdef
The Lie derivative of a Eulerian function $\eul$ along a flow of Eulerian velocity~$\vv$
is the Eulerian function $\calL_\vv \eul$ defined at $(t,\pt)$ by
\be
\label{eqdefDL0}
\calL_\vv \eul(t,\pt)
\eqdef \lim_{h\rar 0} {((\Phittph)^*\eul_\tph)(\pt) - \eul(t,\pt) \over h}, \quad\hbox{rate in~$\Omegat$}.
\ee
\def\hg{\hat g}
In other words, if $\hg$ is defined by $\hg(\tau) = ((\Phi^t_\tau)^* \eul_\tau)(\pt)$
(in particular $\hg(t) = \eul(t,\pt)$), then
\be
\label{eqdefDL2}
\calL_\vv \eul(t,\pt) := \hg'(t) = \lim_{\tau\rar t} {\hg(\tau) - \hg(t) \over \tau-t}
\mathop{=}^{\hbox{\scriptsize also written}}\quad {d ((\Phi^t_\tau)^* \eul_\tau)(\pt) \over d\tau}_{|\tau=t}.
\ee
\findef

\debexe
Prove: \eref{eqdefDLrn2} and~\eref{eqdefDL0} are equivalent.

\debrep
\eref{eqdefDL0} also reads
$
\calL_\vv \eul(t,\pt)
= \lim_{h\rar 0} {((\Phittmh)^*\eul_\tmh)(\pt) - \eul_t(\pt) \over -h}
$,
and $(\Phittmh)^*.(\Phitmht)_*=I$.
\finrep
\finexe

\comment{
Here the observer must:

$\bullet$ At~$\tph$ and at some $\ptph=\tPhi(\tph,\Pobj)$, take the value $\eul(\tph,\ptph)$,

$\bullet$ go back (in the past) along the trajectory $\tPhi_\Pobj$,

$\bullet$ once at $t$ at~$\pt$, this value has become $((\Phittph)^* \eul_\tph)(\pt)$,

$\bullet$ and then the comparison with $\eul(t,\pt)$ can be done (no ubiquity gift required):
}

\debrem
More precise definition, as in~\eref{eqdeffspa2}: \Eg\ with~\eref{eqdefDL0},
the Lie derivative $\tilde\calL_\vv \eul$ of a Eulerian function $\tilde\eul$ along a flow of Eulerian velocity~$\vv$ is the Eulerian function defined by, at $t$ at $\pt=\tPhi(t,\Pobj)$,
\be
\tilde\calL_\vv \eul(t,\pt) :=
%\underbrace{
((t,\pt),\calL_\vv \eul(t,\pt)
%}_{\hbox{function at $t$ at $\pt$}}
\quad (\hbox{pointed function at $(t,\pt)$}),
\ee
%\ie, $\tilde\calL_\vv \eul(t,\pt)$ is the ``pointed function $\calL_\vv \eul(t,\pt)$ at $(t,\pt)$.
And, to lighten the notation, $\tilde\calL_\vv \eul(t,\pt) \eqnote \calL_\vv \eul(t,\pt)$ (second component of $\tilde\calL_\vv \eul(t,\pt)$).
\finrem

%%%%%%%%%%%%%%%%%%%%%%%%%%%%%%%%%%%%%%%%%%%%%%%%%%%%%%%%%%%%%%%%%%%%%%%%%%%%%%%%%%%

\subsection{Lie derivative of a scalar function}

%Let $p(\tau) = \tPhi(\tau,\Pobj) = \ptau$.
Let $f$ be a $C^1$ Eulerian scalar valued function. With $(\Phitmht)_*f_\tmh(\pt) = f_\tmh(\ptmh)$, \cf~\eref{eqdefpff2}, we get
\be
\label{eqdl0}
\calL_\vv f(t,\pt) %= & \lim_{h\rar 0} {f(t,p(t)) -  f(\tmh,p(\tmh)) \over h}
\equalref{eqdefDLrn2} \lim_{h\rar 0} {f(t,\pt) - f(\tmh,p(\tmh)) \over h},
\qie
\boxed{\calL_\vv f = {Df\over Dt}} = {\pa f \over \pa t} + df.\vv .
\ee
So, for scalar functions, the Lie derivative is the material derivative.

\medskip
\noindent
{\bf Interpretation:} 
$\calL_\vv f $ measures the rate of change of~$f$ along a trajectory.

\debprop
$\calL_\vv f =0$ iff $f$ is constant along any trajectory (the real value is the memory value):
\be
\label{eqmrf0}
\calL_\vv f=0 %\;\hbox{ in }\bigC
\quad \Longleftrightarrow \quad 
\forall t,\tau\in[\tz,T],\;
%(f(\tau,p(\tau)) =)\; 
(\Phittau_*)f_t(\ptau) = f(t,p(t)) \;\;\hbox{when}\;\; \ptau=\Phittau(\pt),
\ee
\ie\ iff $f(t,p(t))=f(\tz,\ptz)$ when $p(t)=\Phitz(t,\ptz)$,
\ie\ iff $f$ let itself be carried by the flow (unchanged).
\finprop

\debdem
Let $p(t) = \tPhi(t,\Pobj)=\pt$ for all~$t$, so $p(\tau) = \tPhi(\tau,\Pobj) = \ptau = \Phittph(\pt) = \Phi^t(\tau,\pt)$.

$\Leftarrow$:
If $f_\tau = (\Phittph)_* f_t$, then $f_\tau(\ptau) = f_t(\pt)$,
thus $\lim_{\tau\rar t}{f(\tau,p(\tau)) - f(t,p(t)) \over \tau-t} = 0$, that is, ${D f \over Dt} = 0$.

$\Rightarrow$:
If ${Df \over Dt}=0$ then $f(t,p(t))$ is a constant function on the trajectory $t \rar \tPhi(t,\Pobj)$,
for any particle~$\Pobj$,
so $f(\tau,p(\tau)) = f(t,\pt)$ when $p(\tau)=\Phittph(\pt)$, that is, $f(\tau,\ptau) = (\Phittph)_*f_t(\ptau)$.
\findem

\debexe
Prove:
$
\calL_\vv (\calL_\vv f) =  {D^2 f \over D t^2}
= {\pa^2f \over \pa t^2} + 2 d({\pa f\over \pa t}).\vv +  d^2f(\vv,\vv) + df.({\pa \vv \over \pa t}+d\vv).
$

\debrep
See~\eref{eqdfddtd}.
\finrep
\finexe

\comment{
\debrem
With $\phi'(t)
= \lim_{h\rar 0} {\phi(t{+}h) - \phi(t) \over h}
= \lim_{h\rar 0} {\phi(t{-}h) - \phi(t) \over -h}
= \lim_{h\rar 0} {\phi(t) - \phi(t{-}h) \over h}$, we get
an equivalent definition of $\calL_\vv f$ with the push-forward:
\be
\label{eqdlifeq}
\calL_\vv f(t,\pt)
:=  \lim_{h\rar 0} {f_t(\pt) - ((\Phi^{t{-}h}_t)_*f_{t-h})(\pt) \over h}
\quad (=  \lim_{h\rar 0} {f(t,p(t)) - f(t{-}h,p(t{-}h)) \over h}).
\ee
\finrem
}

%%%%%%%%%%%%%%%%%%%%%%%%%%%%%%%%%%%%%%%%%%%%%%%%%%%%%%%%%%%%%%%%%%%%%%%%%%%%%%%%%%%

\subsection{Lie derivative of a vector field}
\label{secdlcv}

%%%%%%%%%%%%%%%%%%%%%%%%%%%%%%%%%%%%%%%%%%%%%%%%%%%%%%%%%%%%%%%%%%%%%%%%%%%%%%%%%%%

\subsubsection{Formula}

Let $\vw$ be a $C^1$ (Eulerian) vector field (interpreted as an ``internal force field'' in the following).

\debprop
%On~a sur~$\bigC$, quand $\Phitz$ est~$C^2$ :
\be
\label{eqdl}
\boxed{\calL_\vv \vw = {D \vw \over D t}  - d\vv.\vw} = {\pa \vw \over \pa t} + d\vw.\vv - d\vv.\vw.
\ee
So the Lie derivative is not reduced to the material derivative ${D \vw \over D t}$ (unless $d\vv=0$, \ie\ unless $\vv$ is uniform):
The spatial variations $d\vv$ of~$\vv$ influences the rate of stress: $\vv$ tries to bend~$\vw$ (which is expected).
%See remark~\ref{reml1}.
%(To compare with~\eref{eqdl0}: A vector is not reduced to a collection of scalar components.)
\finprop

\debdem
\def\vgp{{\vg\,'}}
%\Eg, with~\eref{eqdefDL0}-\eref{eqdefDL2}: % (applies in any manifold).
Let $\vg : \tau \rar \vg(\tau) = (\Phittaub \vw)(t,p(t))=d\Phittau(\pt)^{-1}.\vw(\tau,p(\tau))$ when $p(\tau) = \Phi^t(\tau,\pt)$, so \eref{eqdefDL0} reads $\calL_\vv\vw(t,\pt) = \vgp(t)$.
With $\vz(\tau):=\vw(\tau,p(\tau)) = d\Phit(\tau,\pt).\vg(\tau)$,
\be
\label{eqdld2}
\eqalign{
\vz\,'(\tau) = {D\vw \over D\tau}(\tau,p(\tau))
= &{\pa (d\Phit) \over \pa \tau}(\tau,\pt).\vg(\tau) + d\Phit(\tau,\pt).\vgp(\tau) \cr
= & (d\vv(\tau,p(\tau)).\Ft(\tau,\pt)).(\Ft(\tau,\pt)^{-1}.\vw(\tau,p(\tau))) + \Fttau(\pt).\vgp(\tau) \cr
= & d\vv(\tau,p(\tau)).\vw(\tau,p(\tau)) + \Fttau(\pt).\vgp(\tau). \cr
}
\ee
Thus ${D\vw \over Dt}(t,\pt)
= d\vv(t,\pt).\vw(t,\pt) + I.\vgp(t)$,
thus $\vgp(t) = {D\vw \over Dt}(t,\pt) - d\vv(t,\pt).\vw(t,\pt)$.
\findem

\noindent
{\bf Quantification:}
Basis $(\ve_i)$,
$\vv=\sum_i v_i \ve_i$, $\vw=\sum_i w_i \ve_i$,
$d\vv.\ve_j = \sum_{ij} v_{i|j} \ve_i$,
$d\vw.\ve_j = \sum_{ij} w_{i|j} \ve_i$; Then
\be
\calL_\vv \vw
= \sumin {\pa w_i \over \pa t} \ve_i
+ \sumijn  w_{i|j} v_j \ve_i
- \sumijn v_{i|j} w_j  \ve_i.
\ee
So (column matrix), with $[\cdot]:=[\cdot]_{|\ve}$,
\be
\boxed{[\calL_\vv\vw] = [{D \vw \over D t}] - [d\vv].[\vw]} \quad ( = [{\pa \vw \over \pa t}] + [d\vw.\vv] - [d\vv].[\vw]).
\ee
(And $[d\vw.\vv] = [d\vw].[\vv]$.)
Duality notations:
$\calL_\vv \vw = \sum_i {\pa w^i \over \pa t} \ve_i
+ \sum_{ij} w^i_{|j} v^j \ve_i
- \sum_{ij}  v^i_{|j} w^j \ve_i
$.

%%%%%%%%%%%%%%%%%%%%%%%%%%%%%%%%%%%%%%%%%%%%%%%%%%%%%%%%%%%%%%%%%%%%%%%%%%%%%%%%%%%

\subsubsection{Interpretation: Flow resistance measurement}
\label{secdlresist}

\debprop
$\Phitz$ is supposed to be a $C^2$ motion and a $C^1$ diffeomorphism in space,
and $\vw$ is a vector field.
\be
\label{eqmrf}
\calL_\vv \vw=0
\quad \Longleftrightarrow \quad 
\forall t\in[\tz,T],\;\; \vw_t = (\Phitzt)_*\vw_\tz.
\ee
\ie, ${D\vw \over Dt} = d\vv.\vw$ $\Leftrightarrow$ 
the actual vector $\vw(t,p(t))$ is equal to $\Ftzt(\ptz).\vw_\tz(\ptz) = \vwtzs(t,p(t))$ the deformed vector by the flow, see figure~\ref{figpfl}.
So: The Lie derivative $\calL_\vv\vw$ vanishes
iff $\vw$ does not resist the flow (let itself be deformed by the flow), \ie\ iff $\vw(t,\pt)=\vwtzs(t,\pt)$.
\finprop

\debdem
\def\vfp{\vf\,'}%
We have $\calL_\vv \vw={D \vw \over Dt} - d\vv.\vw$ and ${\pa \Ftz \over \pa t}(t,\ptz) = d\vv(t,p(t)).\Ftzt(\ptz)$, \cf~\eref{eqvVvv2b}.

$\Leftarrow$ (derivation):
Suppose $\vw(t,p(t)) = \Ftz(t,\ptz).\vw(\tz,\ptz)$ %($=(\Phitzt)_*\vw_\tz(t,p(t))$) 
when $p(t)=\Phitzt(\ptz)$.
Then ${D \vw \over Dt}(t,p(t))
= {\pa \Ftz \over \pa t}(t,\ptz).\vw(\tz,\ptz)
= (d\vv(t,p(t)).\Ftzt(\ptz)).(\Ftzt(\ptz)^{-1}.\vw(t,p(t)))
= d\vv(t,p(t)).\vw(t,p(t))
$,
thus ${D \vw \over Dt} - d\vv.\vw=0$. (See proposition~\ref{peqdwdt04}.)

$\Rightarrow$ (integration):
Suppose ${D\vw \over Dt} = d\vv.\vw$.
%We want to deduce $\vw(t,p(t)) = \Ftzt(\ptz).\vwtz(\ptz)$ when $p(t)=\Phitz(t,\ptz)$, \ie\ $\vwtz(\ptz)= (\Ftzt(\ptz))^{-1}.\vw(t,p(t))$ for all~$t$ when $p(t)=\Phitz(t,\ptz)$.
%\ie\ $\vw(t,\Phitz(t,\ptz)) = F^\tz_t(\ptz).\vw(\tz,\ptz)$.So, 
Let $\vf(t) = (\Ftzt(\ptz))^{-1}.\vw(t,p(t))$ (= pull-back $(\Phitzt)^*\vw(\tz,\ptz)$) when $p(t) = \Phitz(t,\ptz)$;
So $\vw(t,p(t)) = \Ftz(t,\ptz). \vf(t)$
and ${D\vw \over Dt} (t,p(t))
={\pa \Ftz \over \pa t}(t,\ptz). \vf(t) + \Ftzt(\ptz). \vfp(t)
= d\vv(t,p(t)).\Ftzt(\ptz). \vf(t) + \Ftzt(\ptz). \vfp(t)
= d\vv(t,p(t)).\vw(t,p(t)) + \Ftzt(\ptz). \vfp(t)
\mathop{=}^{hyp.} {D \vw \over Dt}(t,p(t)) + \Ftzt(\ptz). \vfp(t)
$ for all~$t$;
Thus $\Ftzt(\ptz). \vfp(t)=\vec0$, thus $\vfp(t)=\vec0$ (because $\Phitzt$ is a diffeomorphism), 
thus $\vf(t)=\vf(\tz)$,
\ie\ $\vw_t = (\Phitzt)_*\vw_\tz$, for all~$t$. %, thus $(\Ftzt(\ptz))^{-1}.\vw(t,p(t)) = \vw(\tz,\ptz)$, for all~$t$, qed.
\findem

\comment{
Interpretation with~\eref{eqdefDLrn2}: 

$\bullet$
$\vw(t,p(t))$ is the ``actual'' value of the force field $\vw$ at~$t$ at $p(t)$,

$\bullet$
while the push-forward $(\Phitzt)_*\vw_\tz(\pt)$
is the ``transported memory'' = the ``virtual'' value that $\vw$ would have had at~$t$ at~$p(t)$
if it had let itself be carried by the flow.
%(strain dependent, see~\eref{eqvwU}-\eref{eqdefrapvEivei}). %-\eref{eqvwV}

$\bullet$
}

%%%%%%%%%%%%%%%%%%%%%%%%%%%%%%%%%%%%%%%%%%%%%%%%%%%%%%%%%%%%%%%%%%%%%%%%%%%%%%%%%%%

\subsubsection{Autonomous Lie derivative and Lie bracket}

The Lie bracket of two vector fields~$\vv$ and~$\vw$ is
\be
[\vv,\vw] := d\vw.\vv - d\vv.\vw \eqnote \calL^0_\vv \vw.
\ee
And $\calL^0_\vv \vw=[\vv,\vw]$ is called the autonomous Lie derivative of $\vw$ along~$\vv$.
Thus
\be
\label{eqdl2}
\calL_\vv \vw = {\pa \vw \over \pa t} + [\vv,\vw] = {\pa \vw \over \pa t} + \calL^0_\vv \vw.
\ee
NB: $\calL^0_\vv \vw$ is used when $\vv$ et $\vw$ are stationary vector fields,
thus does not concern objectivity: A~stationary vector field in a referential
is not necessary stationary in another (moving) referential.

%%%%%%%%%%%%%%%%%%%%%%%%%%%%%%%%%%%%%%%%%%%%%%%%%%%%%%%%%%%%%%%%%%%%%%%%%%%%%%%%%%%

\subsection{Examples}
\label{secLieei}

%%%%%%%%%%%%%%%%%%%%%%%%%%%%%%%%%%%%%%%%%%%%%%%%%%%%%%%%%%%%%%%%%%%%%%%%%%%%%%%%%%%

%%%%%%%%%%%%%%%%%%%%%%%%%%%%%%%%%%%%%%%%%%%%%%%%%%%%%%

\subsubsection{Lie Derivative of a vector field along itself}

\eref{eqdl} with $\vw=\vv$ gives
$
\calL_\vv\vv={\pa\vv\over\pa t}.
$
In particular, if $\vv$ is a stationary vector field then
$\calL_\vv\vv=\vec0$ ($=[\vv,\vv]$).

%This is also a direct consequence of~\eref{eqpf0}. % and~\eref{eqdlieq}.

%%%%%%%%%%%%%%%%%%%%%%%%%%%%%%%%%%%%%%%%%%%%%%%%%%%%%%

\subsubsection{Lie derivative along a uniform flow}

Here $d\vv=0$, thus
\be
\calL_\vv \vw = {D\vw \over Dt} = {\pa\vw\over\pa t} + d\vw.\vv\quad (\hbox{when $d\vv=0$}).
\ee
Here the flow is rectilinear ($d\vv=0$): there is no curvature (of the flow)
to influence the stress on~$\vw$.

Moreover, if $\vw$ is stationary, that is ${\pa \vw\over\pa t} = 0$, then
$
%\label{eqdlfce}
\calL_{\vv} \vw = d\vw.\vv
$
= the directional derivative ${\pa \vw\over \pa \vv}$ of the vector field~$\vw$ in the direction~$\vv$.

%%%%%%%%%%%%%%%%%%%%%%%%%%%%%%%%%%%%%%%%%%%%%%%%%%%%%%

\subsubsection{Lie derivative of a uniform vector field}
\label{eqdlue}

Here $d\vw(t,p)=0$, thus
\be
\calL_\vv \vw = {\pa\vw\over\pa t} - d\vv.\vw \quad (\hbox{when $d\vw=0$}),
\ee
thus the stress on~$\vw$ is due to the space variations of~$\vv$. % (due to the curvature of the flow).
Moreover, is $\vw$ is stationary then
$
\calL_\vv \vw = - d\vv.\vw.
$

%%%%%%%%%%%%%%%%%%%%%%%%%%%%%%%%%%%%%%%%%%%%%%%%%%%%%%%%%%%%%%%%%%%%%%%%%%%%%%%%%%%

\subsubsection{Uniaxial stretch of an elastic material}
%\subsubsection{Uniaxial stretch (linear elasticity)}

$\bullet$ Strain. With $[\ora{OP}]_{|\ve} %=[\ora{O\tPhi(\tz,\Pobj)}]_{|\ve}
= [\vX]_{|\ve} = \pmatrix{X \cr Y}$,
with $\xi>0$, $t\ge \tz$, $p(t)=\Phitz(t,P)$ and $[\vx]_{|\ve} = [\ora{Op(t)}]_{|\ve}$:
%and $[\ora{Op}]_{|\ve} = [\ora{O\Phitz(t,P)}]_{|\ve} = [\vx]_{|\ve} = \pmatrix{x \cr y}$: given by
\be
[\vx]_{|\ve}
= \pmatrix{x \cr y}
%= [\ora{Op(t)}]_{|\ve} 
%= [\ora{O\Phitz(t,P)}]_{|\ve}
= \pmatrix{X \cr Y} + \xi (t{-}\tz) \pmatrix{X \cr 0} % = X+\xi X (t{-}\tz)
= \pmatrix{X (1+ \xi (t{-}\tz)) \cr Y \hfill}. % = X+\xi X (t{-}\tz)
\ee

\noindent
$\bullet$ Eulerian velocity $\vv(t,p)
= \pmatrix{\xi X\cr 0\hfill}
= \pmatrix{{\xi \over 1+ \xi (t{-}\tz)}x \cr 0\hfill}
$,
%thus 
$d\vv(t,p)= \pmatrix{ {\xi \over 1+ \xi (t{-}\tz)} & 0 \cr 0 & 0} %= d\vv(t)
$ (independent of~$p$).

\noindent
$\bullet$ Deformation gradient (independent of~$P$), with $\kappa_t = \xi (t{-}\tz)$:
\be
F_t=d\Phitzt(P)
=\pmatrix{1+ \kappa_t & 0 \cr 0 & 1} = I+\kappa_t\pmatrix{1 & 0 \cr 0 & 0}.
\ee
%In particular $||F_t.\ve_1|| = 1+ \kappa_t$ (stretch rate at~$\pt$).
Infinitesimal strain tensor, with $F_t^T=F_t$ here: 
\be
\uuepstzt(P)
= F_t - I
=\kappa_t\pmatrix{1 & 0\cr 0 & 0}=\uueps_t.
\ee
%Lagrangian velocity $\vVtzt(P) = \pmatrix{X \xi \cr 0}$.

\noindent
$\bullet$ Stress. Constitutive law = Linear isotropic elasticity: %(requires a Euclidean dot product)
\be
\uusigma_t(\pt) = \lambda \Tr(\uueps_t) I + 2\mu\uueps_t
= \kappa_t\pmatrix{\lambda{+}2\mu  & 0 \cr 0 & \lambda } = \uusigma_t.
\ee
Cauchy stress vector $\vec T$ on a surface at~$p$ with normal $\vn_t(p) = \pmatrix{n^1\cr n^2}=\vn$:
\be
\label{eqvTSS2}
\vec T_t(\pt)= \uusigma_t.\vn
= \kappa_t \pmatrix{(\lambda{+}2\mu)n_1 \cr \lambda n_2 }
= \xi (t{-}\tz) \pmatrix{(\lambda{+}2\mu)n_1 \cr \lambda n_2 }
= \vec T_t.
\ee
\comment{
Cauchy stress vector $\vec T$ on a surface around~$p$ with normal $\vn_t(p) = \ve_1 = \pmatrix{n^1=1\cr n^2=0}=\vn$:
\be
\label{eqvTSS2}
\vec T_t(\pt)= \uusigma_t.\vn
= \kappa_t \pmatrix{\lambda{+}2\mu \cr 0 }
= \xi (t{-}\tz) \pmatrix{\lambda{+}2\mu \cr 0 }
= \vec T_t.
\ee
}

\noindent
$\bullet$ Push-forwards:
$\vec T_\tz(\ptz)=0$,
thus $\Ftztzph(\ptz).\vec T_\tz(\ptz)=\vec0$.

\noindent
$\bullet$ Lie derivative:
\be
\calL_\vv \vec T(\tz,\ptz) = \lim_{t\rar\tz}{\vec T_t(\pt) -\Ftzt(\ptz).\vec T_\tz(\ptz) \over t-\tz}
=\xi \pmatrix{(\lambda{+}2\mu)n_1 \cr \lambda n_2 } \quad\hbox{(rate of stress at $(\tz,\ptz)$)}.
\ee
\comment{
(Remark: %If we suppose that this rate of stress is constant over time, then
we recover $\vec T(t,\pt)= \vec T(\tz,\ptz)+\int_\tz^t \calL_\vv \vec T(\tau,\ptau)\,d\tau
= \int_\tz^t \calL_\vv \vec T(\tz,\ptz)\,d\tau
= \mu\xi (t{-}\tz) \pmatrix{\lambda{+}2\mu \cr 0 }$, \cf~\eref{eqvTSS2}.) % (for $t$ close to~$\tz$).
}

\comment{
\noindent
$\bullet$ Push-forwards:
$F^\tz_\tph(\ptz) =  \Fttph(\pt) \circ \Ftzt(\ptz)$ gives
$\Fttph(\pt) = F^\tz_\tph(\ptz) \circ\Ftzt(\ptz)^{-1}
=\pmatrix{{1+ \xi (\tph{-}\tz) \over 1+ \xi (t{-}\tz)} & 0 \cr 0 & 1 }$, thus
$\Fttph(\pt).\vec T_t(\pt)=\pmatrix{ {1+ \xi (\tph{-}\tz) \over 1+ \xi (t{-}\tz)}\xi (\tph{-}\tz)(\lambda{+}2\mu) & 0 \cr 0 & 0}$,
thus
$\calL_\vv \vec T(t,\pt)
= \lim_{h\rar0}{\vec T_\tph(\ptph) -\Fttph(\pt).\vec T_t(\pt) \over h}
= \lim_{h\rar0}{1\over h}(\lambda{+}2\mu)\xi (\tph{-}\tz)(1-{1+ \xi (\tph{-}\tz) \over 1+ \xi (t{-}\tz)})\pmatrix{1&0\cr0&0}
= {\xi^2 \over 1+\xi}(\lambda{+}2\mu)\pmatrix{1&0\cr0&0}
$.

\noindent
$\bullet$ With push-forwards:
$\vec T_\tz(\ptz) =0$, $\Ftztzph(\pt).\vec T_\tz(\ptz)=0$,
and $\vec T_\tzph(\ptzph) =\xi h \pmatrix{(\lambda{+}2\mu) \,n^1 \cr \lambda \,n^2 }$,
thus
$\calL_\vv \vec T(t,\pt) = \lim_{h\rar0}{\vec T_\tph(\ptph) -\Fttph(\pt).\vec T_t(\pt) \over h}
=\xi \pmatrix{(\lambda{+}2\mu) \,n^1 \cr \lambda \,n^2 }$.
}

\noindent
$\bullet$ Generic computation with 
$\calL_\vv \vec T = {\pa \vec T \over \pa t} + d\vec T.\vv - d\vv.\vec T$:
\eref{eqvTSS2} gives
${\pa \vec T\over \pa t}
= \xi  \pmatrix{(\lambda{+}2\mu) \,n^1 \cr \lambda \,n^2 }
$
and 
$d\vec T = 0$
and
$d\vv_t.\vec T_t
= \pmatrix{ {\xi \over 1+ \xi (t{-}\tz)} & 0 \cr 0 & 0}
.\xi (t{-}\tz) \pmatrix{(\lambda{+}2\mu) \,n^1 \cr \lambda \,n^2 }
= {\xi^2(t{-}\tz) \over 1+ \xi (t{-}\tz)} \pmatrix{(\lambda{+}2\mu) \,n^1 \cr 0}
$.
In particular,
$d\vv(\tz,\ptz).\vec T(\tz,\ptz)
= \vec0
$. Thus
$\calL_\vv\vec T(\tz,\ptz)
= \xi  \pmatrix{(\lambda{+}2\mu) \,n^1 \cr \lambda \,n^2 }=$ rate of stress at the initial $(\tz,\ptz)$.

\comment{
\noindent
$\bullet$ Generic computation: 
$\calL_\vv \vec T = {\pa \vec T \over \pa t} + d\vec T.\vv - d\vv.\vec T$
and relative to a given surface with normal $\vn(\tz,\ptz)$ at~$\tz$ at~$\ptz$ which is deformed by the flow (strain effect)
into $\vn(t,p(t)) = \Ftz(t,\ptz).\vn_\tz(\ptz)
=\pmatrix{(1+ \xi (t{-}\tz))n^1(\tz,\ptz) \cr n^2(\tz,\ptz)}
$ when $p(t)= \Phitz(t,\ptz)=\pt$, or $\ptz = \Phitzt^{-1}(\pt)$. 
Thus ${\pa \vn\over \pa t}(t,\pt) = \pmatrix{\xi n^1(\tz,\ptz) \cr 0}
$
and $d\vn(t,\pt)
=\pmatrix{(1+ \xi (t{-}\tz))dn^1(\tz,\ptz).d\Phitzt^{-1}(\pt) \cr dn^2(\tz,\ptz).d\Phitzt^{-1}(\pt)}$.
In particular, 
${\pa \vn\over \pa t}(\tz,\ptz) 
 = \pmatrix{\xi n^1(\tz,\ptz) \cr 0}$,
and 
$d\vn(\tz,\ptz)
=\pmatrix{dn^1(\tz,\ptz) \cr dn^2(\tz,\ptz)}
$.

And
${\pa \vec T\over \pa t}(t,p)
= \xi  \pmatrix{(\lambda{+}2\mu) \,n^1(t,p) \cr \lambda \,n^2(t,p) }
+  \xi (t{-}\tz) \pmatrix{(\lambda{+}2\mu) \,{\pa n^1\over \pa t}(t,p) \cr \lambda \,{\pa n^2\over \pa t}(t,p) }
$
and 
$d\vec T(t,p) = \xi (t{-}\tz) \pmatrix{(\lambda{+}2\mu) \,dn^1(t,p) \cr \lambda \,dn^2(t,p) }$
and
$d\vv_t(p).\vec T_t(p)
= \pmatrix{ {\xi \over 1+ \xi (t{-}\tz)} & 0 \cr 0 & 0}
.\xi (t{-}\tz) \pmatrix{(\lambda{+}2\mu) \,n^1(t,p) \cr \lambda \,n^2(t,p) }
= {\xi^2(t{-}\tz) \over 1+ \xi (t{-}\tz)} \pmatrix{(\lambda{+}2\mu) \,n^1(t,p) \cr 0}
$.
In particular,
${\pa \vec T\over \pa t}(\tz,\ptz)
= \xi  \pmatrix{(\lambda{+}2\mu) \,n^1(\tz,\ptz) \cr \lambda \,n^2\tz,\ptz) }$,
and 
$d\vec T(\tz,\ptz) = 0 $,
and
$d\vv(\tz,\ptz).\vec T(\tz,\ptz)
= \vec0
$. Thus
$\calL_\vv\vec T(\tz,\ptz)
= \xi  \pmatrix{(\lambda{+}2\mu) \,n^1(\tz,\ptz) \cr \lambda \,n^2\tz,\ptz) }=$ rate of stress at $(\tz,\ptz)$.
}

%%%%%%%%%%%%%%%%%%%%%%%%%%%%%%%%%%%%%%%%%%%%%%%%%%%%%%%%%%%%%%%%%%%%%%%%%%%%%%%%%%%

\subsubsection{Simple shear of an elastic material}

Euclidean basis $(\ve_1,\ve_2)$ in~$\RR^2$, the same basis at any time.
Initial configuration $\Omegatz=[0,L_1]\otimes [0,L_2]$.
Initial position
$[\ora{OP}]_{\ve} = [\ora{O\ptz}]_{\ve} %= [\ora{O\tPhi(\tz,\Pobj)}]_{\ve}
=[\vX]_{\ve} = \pmatrix{X\cr Y}$.
Let $\xi\in\RR^*$, $\pt=\Phitzt(\ptz)$, $[\vx]_{|\ve} = [\ora{Op(t)}]_{|\ve}$, and
\be
\label{eqvTSS0}
%[\ora{O\pt}]_{\ve} %= [\ora{O\tPhi(t,\Pobj)}]_{\ve}
[\vx]_{\ve}
= \pmatrix{x=\phi^1(t,X,Y)=X \cr y=\phi^2(t,X,Y) \hfill}
= \pmatrix{X + \xi (t{-}\tz) Y \cr Y \hfill}.
\ee

\noindent
$\bullet$ Eulerian velocity $\vv_t(\pt) %= {\pa\tPhi \over \pa t}(t,\Pobj) 
= \pmatrix{\xi Y \cr 0\hfill} = \pmatrix{\xi y \cr 0\hfill}$, % at $\pt=\tPhi(t,\Pobj)$,
thus $d\vv_t(\pt)= \pmatrix{ 0 & \xi \cr 0 & 0}$.

\noindent
$\bullet$ Strain. With $\kappa_t = \xi (t{-}\tz)$, deformation gradient (independent of~$P$):
\be
\label{eqvTSS1}
d\Phitzt(P)
%=\pmatrix{{\pa \phi^1 \over \pa X}(t,X,Y) & {\pa \phi^1 \over \pa Y}(t,X,Y) \cr {\pa \phi^2 \over \pa X}(t,X,Y) & {\pa \phi^2 \over \pa Y}(t,X,Y)}(P)
=\pmatrix{1 &\kappa_t \cr 0 & 1} = \Ftzt, \qthus
\Ftzt-I=\kappa_t\pmatrix{0 & 1 \cr 0 & 0}.
\ee

\noindent
$\bullet$ Infinitesimal strain tensor: 
\be
\uuepstzt(P) = {\Ftzt(P){-}I+(\Ftzt(P){-}I)^T\over 2}
%=\demi\pmatrix{0 & \xi (t{-}\tz)\cr \xi (t{-}\tz) & 0}
={ \kappa_t \over 2}\pmatrix{0 & 1 \cr 1 & 0}
=\uueps_t
.
\ee

\noindent
$\bullet$ Stress. Constitutive law, usual linear isotropic elasticity (requires a Euclidean dot product): % And small deformation hypothesis. Thus
\be
\uusigma(t,\pt) = \lambda \Tr(\uueps_t)I + 2\mu\uueps_t
= \mu\kappa_t\pmatrix{0 & 1 \cr1 & 0} = \uusigma_t.
\ee
Cauchy stress vector $\vec T(t,\pt)$ (at~$t$ at~$\pt$) on a surface at~$p$ with normal
$\vn_t(p) %= \pmatrix{n^1_t(p)\cr n^2_t(p)} 
= \pmatrix{n^1\cr n^2} = \vn $: % (independent of $t$ and~$p$):
\be
\label{eqvTSS}
\vec T_t= \uusigma_t.\vn
%= \pmatrix{\mu\xi (t{-}\tz) \,n^2 \cr \mu\xi (t{-}\tz) \,n^1 }
= \mu\kappa_t \pmatrix{\,n^2 \cr \,n^1 }
= \mu\xi (t{-}\tz) \pmatrix{\,n^2 \cr \,n^1 }
=\vec T(t) \quad(\hbox{stress independent of~$\pt$}).
\ee

\comment{
\noindent
$\bullet$ Push-forward
\be
\vw_*=\Ftzt.\vW 
=\pmatrix{W_1 + \kappa_t W_2 \cr W_2}, 
%\quad \vw_*-\vW=\Ftzt.\vW - \vW = \pmatrix{\xi (t{-}\tz) W_2 \cr 0},
\ee
}

\noindent
$\bullet$ Lie derivative, with $\vec T_\tz=\vec0$:
\be
\calL_\vv \vec T(\tz,\ptz) = \lim_{t\rar\tz}{\vec T_t(\pt) -\Ftzt(\ptz).\vec T_\tz(\ptz) \over t-\tz}
=\mu\xi \pmatrix{\,n^2 \cr \,n^1 } \quad\hbox{(rate of stress at $(\tz,\ptz)$)}.
\ee
%If we suppose that this rate of stress is constant over time, then
\comment{
(Remark: we recover $\vec T(t,\pt)= \vec T(\tz,\ptz)+\int_\tz^t \calL_\vv \vec T(\tau,\ptau)\,d\tau
= \int_\tz^t \calL_\vv \vec T(\tz,\ptz)\,d\tau
= \mu\xi (t{-}\tz) \pmatrix{\,n^2 \cr \,n^1 }$, \cf~\eref{eqvTSS}.) % (for $t$ close to~$\tz$).
}

%\Eg, if $\vn=\pmatrix{0\cr n_2}=\pmatrix{0\cr 1}$ (horizontal surface) then $\calL_\vv\vec T(t,\pt) \simeq \mu \xi \pmatrix{ 1 \cr  0 } %+ o(t{-}\tz) $ (horizontal).

\noindent
$\bullet$ Generic computation: $\calL_\vv \vec T = {\pa \vec T \over \pa t} + d\vec T.\vv - d\vv.\vec T$.
\eref{eqvTSS} gives
${\pa \vec T\over \pa t}(t,p) = \mu\xi \pmatrix{ \,n^2 \cr  \,n^1 }$ and $d\vec T=0$.
With $d\vv_\tz.\vec T_\tz=\vec 0$.
Thus 
$\calL_\vv \vec T(\tz,\ptz) = \mu\xi \pmatrix{ \,n^2 \cr  \,n^1 }$.

%%%%%%%%%%%%%%%%%%%%%%%%%%%%%%%%%%%%%%%%%%%%%%%%%%%%%%%%%%%%%%%%%%%%%%%%%%%%%%%%%%%

\subsubsection{Shear flow}

Stationary shear field, see~\eref{eqfigv0} with $\alpha=0$ and $\tz=0$:
%pour $(x,y)\in[0,1]^2$ et pour $\lambda\in\RR$ :
\be
\label{eqexacis2}
\vv(x,y)=\left\{\eqalign{
& v^1(x,y)= \lambda y, \cr
& v^2(x,y) = 0,\cr
}\right.
\qquad
d\vv(x,y) = \pmatrix{0 & \lambda \cr 0 & 0}.
\ee
Let $\vw(t,p)=\pmatrix{0\cr b}=\vw(\tz,\ptz)$ (constant in time and uniform in space).
Then $\calL_\vv \vw = -d\vv.\vw = \pmatrix{-\lambda b \cr 0}$ measures ``the resistance to deformation due to the flow''. See figure~\ref{figcisail2},
the virtual vector $\vw_*(t,p) = d\Phi(\tz,\ptz).\vw(\tz,\ptz)$ being the vector
that would have let itself be carried by the flow (the push-forward).

\begin{figure}[!ht]
\qquad\qquad\qquad\qquad\includegraphics[width=0.4\textwidth]{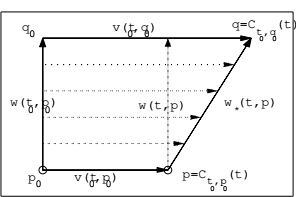}
%\qquad\qquad\includegraphics{cisail2.png}
%\epsfig{figure=cisail2.eps}
\caption{Shear flow, \cf~\eref{eqexacis2}, with $\vw$ constant and uniform. $\calL_\vv \vw$ measures the resistance to the deformation.}
\label{figcisail2}
\end{figure}

%%%%%%%%%%%%%%%%%%%%%%%%%%%%%%%%%%%%%%%%%%%%%%%%%%%%%%%%%%%%%%%%%%%%%%%%%%%%%%%%%%%

\subsubsection{Spin}

%\debexa
%\label{exaspin}
Rotating flow: Continuing~\eref{eqflotdspin}:
\be
\vv(x,y)
%= \pmatrix{-\omega y \cr\omega x}
= \omega \pmatrix{0 & -1 \cr 1 & 0} \pmatrix{x \cr y},\quad
d\vv(x,y)
%= \pmatrix{ 0 & -\omega \cr \omega & 0}
=\omega \pmatrix{0 & -1 \cr 1 & 0} = \omega\,Rot(\pi/2).
\ee
In particular $d^2\vv=0$. With $\vw=\vw_0$ constant and uniform we get
\be
\calL_\vv\vw_0 = -d\vv(p).\vw_0 = - \omega\,Rot(\pi/2). \vw_0
\qquad (\perp \; \pmatrix{a \cr b} = \vw_0).
%= \pmatrix{-\omega b \cr \omega a} = \omega\pmatrix{-b \cr a} \quad \perp \vw(p).
\ee
gives ``the force at which $\vw$ refuses to turn with the flow''.
%\finexa

%%%%%%%%%%%%%%%%%%%%%%%%%%%%%%%%%%%%%%%%%%%%%%%%%%%%%%%%%%%%%%%%%%%%%%%%%%%%%%%%%%%

%\subsubsection{Elongation}

%%%%%%%%%%%%%%%%%%%%%%%%%%%%%%%%%%%%%%%%%%%%%%%%%%%%%%%%%%%%%%%%%%%%%%%%%%%%%%%%%%%

\subsubsection{Second order Lie derivative}

\debexe
Let $\vv,\vw$ be $C^2$. Prove:
\be
\label{eqcalvv}
\eqalign{
\calL_\vv (\calL_\vv \vw)
= & {D^2 \vw \over D t^2} - 2 d\vv.{D\vw\over Dt} - {D (d\vv)\over Dt}.\vw + d\vv.d\vv.\vw , \cr
= &  {\pa ^2 \vw \over\pa t^2} + 2 d{\pa \vw \over \pa t}.\vv - 2d\vv.{\pa \vw \over \pa t}
 + d\vw.{\pa \vv \over \pa t} - d{\pa \vv \over \pa t}.\vw \cr
 & \qquad + (d^2\vw.\vv).\vv + d\vw.d\vv.\vv  - 2 d\vv.d\vw.\vv
   - (d^2\vv.\vv).\vw  + d\vv.d\vv.\vw. \cr
}
\ee

\debrep
$$
\eqalign{
\calL_\vv (\calL_\vv \vw)
= &{D (\calL_\vv \vw) \over Dt} - d\vv.(\calL_\vv \vw)
= {D({D \vw \over D t}  - d\vv.\vw)\over Dt} - d\vv.({D \vw \over D t}  - d\vv.\vw) \cr
= & {D^2 \vw \over D t^2} - {D (d\vv)\over Dt}.\vw - d\vv.{D\vw\over Dt}
 - d\vv.{D \vw \over D t} + d\vv.d\vv.\vw,  \cr
%= & {\pa^2 \vw\over \pa t^2} + 2d{\pa \vw\over \pa t}.\vv + d^2\vw(\vv,\vv)
%+ d\vw.{\pa\vv\over \pa t} + d\vw.d\vv.\vv 
%- d{\pa \vv\over \pa t}.\vw - (d^2\vv.\vw).\vv 
%- 2d\vv.{\pa \vw\over \pa t} - 2d\vv.d\vw.\vv + d\vv.d\vv.\vw
}
$$
thus~\eref{eqcalvv}$_1$, thus~\eref{eqcalvv}$_2$.
\comment{
And %\eref{eqdl} gives : 
$$
\eqalign{
\calL_\vv (\calL_\vv \vw)
= & \calL_\vv({\pa \vw \over \pa t}) + \calL_\vv(d\vw.\vv) - \calL_\vv(d\vv.\vw) \cr
= & {\pa ^2 \vw \over\pa t^2} + {\pa d\vw \over \pa t}.\vv - d\vv.{\pa \vw \over \pa t}
+ {\pa (d\vw.\vv) \over \pa t} + d(d\vw.\vv).\vv - d\vv.(d\vw.\vv)
- {\pa (d\vv.\vw) \over \pa t} - d(d\vv.\vw).\vv + d\vv.(d\vv.\vw) \cr
= & {\pa ^2 \vw \over\pa t^2} + d{\pa \vw \over \pa t}.\vv - d\vv.{\pa \vw \over \pa t}
+ {\pa d\vw \over \pa t}.\vv + d\vw.{\pa \vv \over \pa t} + (d^2\vw.\vv).\vv + d\vw.d\vv.\vv  - d\vv.d\vw.\vv \cr
& - {\pa d\vv \over \pa t}.\vw -  d\vv.{\pa \vw \over \pa t} - (d^2\vv.\vv).\vw - d\vv.d\vw.\vv + d\vv.d\vv.\vw \cr
= & {\pa ^2 \vw \over\pa t^2} + 2 d{\pa \vw \over \pa t}.\vv - 2d\vv.{\pa \vw \over \pa t}
 + d\vw.{\pa \vv \over \pa t} + (d^2\vw.\vv).\vv + d\vw.d\vv.\vv  - 2 d\vv.d\vw.\vv
 - d{\pa \vv \over \pa t}.\vw  - (d^2\vv.\vv).\vw  + d\vv.d\vv.\vw \cr
}
$$
thus~\eref{eqcalvv}$_2$.
And
%${D \vw \over D t} = {\pa \vw\over dt} + d\vw.\vv$, thus
%${D^2 \vw \over D t^2}
%= {\pa^2 \vw\over \pa t^2} + 2d{\pa \vw\over \pa t}.\vv + d^2\vw(\vv,\vv) + d\vw.({\pa \vv\over dt} + d\vv.\vv)$,
%\cf~\eref{eqdfddtd}, thus
}
\finrep
\finexe

%%%%%%%%%%%%%%%%%%%%%%%%%%%%%%%%%%%%%%%%%%%%%%%%%%%%%%%%%%%%%%%%%%%%%%%%%%%%%%%%%%%

\subsection{Lie derivative of a differential form}
\label{secdlcdf}

When the Lie derivative of a vector field~$\vw$ cannot be obtained by direct measurements, you need to use a ``measuring device'' (Germain: To know the weight of a suitcase you have to lift it: You use work).

Here we consider a measuring device which is a differential form $\alpha$. So, if $\vw$ is a vector field
then $f=\alpha.\vv$ is a scalar function, and~\eref{eqdl0} gives
$\calL_\vv(\alpha.\vw) = {D(\alpha.\vw) \over Dt}= {D\alpha \over Dt}.\vw + \alpha.{D\vw \over Dt}$, thus
\be
%\label{eqdla0}
\calL_\vv(\alpha.\vw) 
= 
\underbrace{{D\alpha \over Dt}.\vw + \alpha.d\vv.\vw}_{\rar(\calL_\vv \alpha).\vw} + 
\underbrace{\alpha.{D\vw \over Dt} - \alpha.d\vv.\vw}_{=\alpha.\calL_\vv\vw}
:
\ee

\debdef
Let $\alpha$ be a differential form.
%Let $\vv$ be the velocity field of a flow. 
The Lie derivative of~$\alpha$ along~$\vv$ is
the differential form
\be
\label{eqdla}
\boxed{\calL_\vv \alpha : = {D\alpha \over Dt}  + \alpha.d\vv} = {\pa \alpha \over \pa t} + d\alpha.\vv + \alpha.d\vv.
\ee
(An equivalent definition is given at~\eref{eqdlfd1}.) \Ie, for all vector field $\vw$,
\be
\label{eqdla2}
\calL_\vv \alpha.\vw
:= {D \alpha \over D t}.\vw  + \alpha.d\vv.\vw
\quad (= {\pa \alpha \over \pa t}.\vw + (d\alpha.\vv).\vw + \alpha.d\vv.\vw).
\ee
\findef

The definition of $\calL_\vv\alpha$, \cf~\eref{eqdla}, immediately gives the ``derivation property''
\be
\label{eqdlfd0}
\calL_\vv(\alpha.\vw) = (\calL_\vv\alpha) . \vw + \alpha.(\calL_\vv\vw) \quad(\hbox{\ie\ $\calL_\vv$ is a derivation}).
\ee

\noindent
{\bf Quantification:} Relative to a basis~$(\ve_i)$ and with $[\cdot]:=[\cdot]_{|\ve}$, % (and $[\alpha]$ is a row matrix)
\be
\label{eqdlfd0c}
\boxed{[\calL_\vv\alpha] 
= [{D \alpha \over D t}]  + [\alpha].[d\vv]} \quad\hbox{(row matrix) } 
= [{\pa \alpha \over \pa t}] + [d\alpha.\vv]  + [\alpha].[d\vv].
\ee
Thus
\be
[\calL_\vv\alpha.\vw] =[\calL_\vv\alpha].[\vw]
= [{\pa \alpha \over \pa t}].[\vw] + [d\alpha.\vv].[\vw]  + [\alpha].[d\vv].[\vw].
\ee

\debexe
Prove~\eref{eqdlfd0c} with components. And prove $[d\alpha.\vv] = [\vv]^T.[d\alpha]^T$ (row matrix),
thus $[d\alpha.\vv].[\vw] = [\vv]^T.[d\alpha]^T.[\vw]= [\vw]^T.[d\alpha].[\vv]$.

\debrep
Basis $(\ve_i)$, dual basis $(\piei)$, thus \eref{eqdla} gives
$[\calL_\vv \alpha] = [{D\alpha \over Dt}]  + [\alpha.d\vv]$.
%So we just have to prove that $[\alpha.d\vv] = [\alpha].[d\vv]$.
Let $\alpha = \sum_i \alpha_i \piei$, $\vv=\sum_i v_i \ve_i$,  $d\vv = \sum_{ij} v_{i|j} \ve_i\otimes \piej$
(tensorial writing convenient for calculations), \ie\ $[d\vv]_{|\ve}=[v_{i|j}]$, thus
$\alpha.d\vv = \sum_{ij} \alpha_i v_{i|j} \piej$,
thus $[\alpha.d\vv]_{|\pi_e} = [\alpha]_{|\pi_e}.[d\vv]_{|\ve}$ (row matrix).
And $d\alpha = \sum_{ij} \alpha_{i|j} \piei \otimes \piej$, \ie\ $[d\alpha]_{|\pi_e} = [\alpha_{i|j}]$, gives
$d\alpha.\vv = \sum_{ij} \alpha_{i|j} v_j \piei = \sum_{ij}  v_i\alpha_{j|i} \piej$,
and $[d\alpha.\vv]_{|\pi_e}$ is a row matrix ($d\alpha.\vv$ is a differential form), thus
$[d\alpha.\vv]_{|\pi_e} = [\vv]_{|\ve}^T.[d\alpha]_{|\pi_e}^T$.
(Or compute $(d\alpha.\vv).\vw=  \sum_{ij} \alpha_{i|j} v_j w_i = [\vw]_{|\ve}^T.[d\alpha]_{|\ve}.[\vv]_{|\ve}
= [\vv]_{|\ve}^T.[d\alpha]_{|\pi_e}^T.[\vw]_{|\ve}$.)
\finrep
\finexe

\debexe
Let $\alpha$ be a differential form, and let $\alpha_t(p) := \alpha(t,p)$. Prove, when $\Phitzt$ is a diffeomorphism,
\be
\label{eqmrf2}
\calL_\vv \alpha=0
\quad \Longleftrightarrow \quad 
\forall t\in[\tz,T],\;
\alpha_t = (\Phitzt)_*\alpha_\tz.
\ee
\Ie: ${D\alpha \over Dt} = -\alpha.d\vv$ $\Longleftrightarrow$ $\alpha_t(\pt)=\alpha_\tz(\ptz).\Ftzt(\ptz)^{-1}$ for all~$t$, when $\pt=\Phitzt(\ptz)$.

\debrep
%Let $p(t) = \Phitz(t,\ptz)$. 
$\Leftarrow$:
If $\alpha_t(p(t)) = \alpha_\tz(\ptz).\Ftzt(\ptz)^{-1}$,
then $\alpha(t,p(t)).\Ftz(t,\ptz) = \alpha_\tz(\ptz)$,
thus ${D\alpha \over D t}(t,\pt).\Ftzt(\ptz) + \alpha_t(\pt).{\pa\Ftz \over \pa t}(t,\ptz) = 0$,
thus ${D\alpha \over D t}(t,p(t)).\Ftzt(\ptz) + \alpha_t(\pt).d\vv(t,\pt).\Ftzt(\ptz) = 0$,
thus $\calL_\vv \alpha=0$, since $\Phitzt$ is a diffeomorphism.

$\Rightarrow$:
If $\beta(t):=(\Phitzt)_*\alpha_\tz(\ptz) = \alpha_t(p(t)).\Ftzt(\ptz)$ (pull-back at $(\tz,\ptz)$),
then $\beta(t) = \alpha(t,p(t)).\Ftz(t,\ptz)$,
thus $\beta'(t)
= {D\alpha \over D t}(t,\pt).\Ftzt(\ptz)
+ \alpha(t,\pt). d\vv(t,\pt).\Ftzt(\ptz)
=0$ (hypothesis $\calL_\vv\alpha=0$), thus $\beta(t) = \beta(\tz) = \alpha_\tz(\ptz)$.
\finrep
\finexe

\debrem
A definition equivalent to~\eref{eqdla} is, \cf~\eref{eqdefDL0},
\be
\label{eqdlfd1}
\eqalign{
\calL_\vv \alpha(t,\pt)
%\eqdef & \beta'(t) \cr
& :=  \lim_{\tau\rar t} {(\Phittau)^*\alpha_\tau(\pt) - \alpha_t(\pt) \over \tau-t}
\quad (= \lim_{\tau\rar t} {\alpha_\tau(\ptau).d\Phittau(\pt) - \alpha_t(\pt) \over \tau-t}) \cr
& \eqnote  {D(\Phittaub\alpha_\tau(\pt)) \over D\tau}_{|\tau=t}
\eqnote {D(\alpha_\tau^*(\pt)) \over D\tau}_{|\tau=t}
\quad (={D(\alpha_\tau(\ptau).d\Phittau(\pt)) \over D\tau}_{|\tau=t})
.\cr
}
\ee
Indeed, if $\beta(\tau) = (\Phittau)^*\alpha_\tau(\pt) = \alpha_\tau(\ptau).d\Phittau(\pt)$,
then $\beta'(\tau)$ and then $\tau=t$ give~\eref{eqdla}.
\finrem

\debexe
$\vv$ and $\alpha$ being $C^2$, prove:
\be
\label{eqcalva2}
\eqalign{
\calL_\vv (\calL_\vv \alpha)
= & {\pa ^2 \alpha \over\pa t^2} + 2 d{\pa \alpha \over \pa t}.\vv
+ 2{\pa \alpha \over \pa t}.d\vv + d\alpha.{\pa \vv \over \pa t} + \alpha.{\pa d\vv \over \pa t} \cr
 & + (d^2\alpha.\vv).\vv + d\alpha.(d\vv.\vv) + 2(d\alpha.\vv).d\vv
 +\alpha.(d^2\vv.\vv)  + (\alpha.d\vv).d\vv. \cr
}
\ee

\debrep
\eref{eqdla} gives
$$
\eqalign{
\calL_\vv (\calL_\vv \alpha)
= & \calL_\vv({\pa \alpha \over \pa t}) + \calL_\vv(d\alpha.\vv) + \calL_\vv(\alpha.d\vv) \cr
= & {\pa ^2 \alpha \over\pa t^2} + d{\pa \alpha \over \pa t}.\vv + {\pa \alpha \over \pa t}. d\vv
+ {\pa (d\alpha.\vv) \over \pa t} + d(d\alpha.\vv).\vv + (d\alpha.\vv).d\vv
+ {\pa (\alpha.d\vv) \over \pa t} + d(\alpha.d\vv).\vv + (\alpha.d\vv).d\vv \cr
= & {\pa ^2 \alpha \over\pa t^2} + d{\pa \alpha \over \pa t}.\vv + {\pa \alpha \over \pa t}. d\vv
+ {\pa d\alpha \over \pa t}.\vv + d\alpha.{\pa \vv \over \pa t} + (d^2\alpha.\vv).\vv
 + d\alpha.(d\vv.\vv) + (d\alpha.\vv).d\vv \cr
& + {\pa \alpha \over \pa t}.d\vv + \alpha.{\pa d\vv \over \pa t} + (d\alpha.\vv).d\vv
+\alpha.d^2\vv.\vv + (\alpha.d\vv).d\vv \cr
= & {\pa ^2 \alpha \over\pa t^2} + 2 d{\pa \alpha \over \pa t}.\vv + 2{\pa \alpha \over \pa t}. d\vv
+ d\alpha.{\pa \vv \over \pa t} + (d^2\alpha.\vv).\vv + d\alpha.(d\vv.\vv) + 2(d\alpha.\vv).d\vv
+ \alpha.{\pa d\vv \over \pa t} \cr
& + \alpha.(d^2\vv.\vv)  + (\alpha.d\vv).d\vv. \cr
}
$$
\finrep
\finexe

%%%%%%%%%%%%%%%%%%%%%%%%%%%%%%%%%%%%%%%%%%%%%%%%%%%%%%%%%%%%%%%%%%%%%%%%%%%%%%%%%%%

\subsection{Incompatibility with Riesz representation vectors}
\label{secpbplvder}

The Lie derivative has nothing to do with any inner dot product
(the Lie derivative does not compare two vectors, contrary to a Cauchy type approach).

Here we introduce a Euclidean dot product $\dd_g$ and show that the Lie derivative of a linear form~$\alpha$ is not trivially deduced from the Lie derivative of a Riesz representation vector of~$\alpha$ (which one?).
(Same issue as at~\S~\ref{secrvdfdepf}.)

Let $\alpha$ be a Eulerian differential form; Then let $\valphag(t,p)\in\vRRn$ be the $\dd_g$-Riesz representation vector of the linear form $\alpha(t,p)\in\RRns$: So, for all Eulerian vector field~$\vw$,
\be
\label{eqpbdat}
\alpha.\vw = (\valphag,\vw)_g  \qquad ( = \valphag \bcdotg \vw),
\ee
which means $\alpha(t,p).\vw(t,p) = (\valphag(t,p), \vw(t,p))_g $ at all admissible $(t,p)$.
This defines the Eulerian vector field~$\valphag$ (not intrinsic to~$\alpha$: $\valphag$ depends on the choice of~$\dd_g$, \cf~\eref{eqrtr20}).

\debprop
For all $\vv,\vw \in \vRRn$,
\be
\label{exerieszaux}
{\pa \alpha\over \pa t}.\vw = ({\pa \valphag \over \pa t} , \vw)_g,
\quad
(d\alpha.\vv).\vw = (d\valphag.\vv , \vw)_g,
\quad {D\alpha \over Dt}.\vw = ({D\valphag \over Dt} , \vw)_g.
\ee
Thus
\be
\label{eqpbvr}
\calL_\vv \alpha.\vw = (\calL_\vv\valphag , \vw)_g + (\valphag , (d\vv{+}d\vv^T).\vw)_g, \qand
\boxed{\calL_\vv \alpha.\vw \ne  (\calL_\vv\valphag, \vw)_g} \quad\hbox{in general}.
\ee
So $\calL_\vv\valphag$ is \textbf{not} the Riesz representation vector of~$\calL_\vv\alpha$ (but for solid body motions).
(Expected: A Lie derivative is covariant objective, see~\S~\ref{secodl}, and the use of an inner dot product ruins this objectivity.)
\finprop

\debdem
\def\valphagp{\vec\alpha_{gp}}%
A Euclidean dot product $g\dd$ is bilinear constant and uniform, thus: % (omitting $(t,p)$ to lighten the writings):

$\alpha.\vw = (\valphag,\vw)_g$ gives
${\pa\alpha\over\pa t}.\vw + \alpha.{\pa \vw\over \pa t}
= ({\pa \valphag\over \pa t},\vw)_g + (\valphag,{\pa \vw\over \pa t})_g$, with
$\alpha.{\pa \vw\over \pa t} = (\valphag,{\pa \vw\over \pa t})_g$, thus we are left with
${\pa \alpha\over \pa t}.\vw = ({\pa\valphag \over \pa t},\vw)_g$, for all~$\vw$.

$\alpha.\vw = (\valphag,\vw)_g$ gives
$d(\alpha.\vw).\vv = d(\valphag,\vw)_g.\vv$ for all~$\vv,\vw$,
thus
$(d\alpha.\vv).\vw + \alpha.(d\vw.\vv) = (d\valphag.\vv,\vw)_g + (\valphag,d\vw.\vv)_g$, with
$\alpha.(d\vw.\vv) = (\valphag,d\vw.\vv)_g$, thus we are left with
$(d\alpha.\vv).\vw = (d\valphag.\vv,\vw)_g$.

Thus ${D\alpha \over Dt}.\vw = ({D\valphag \over Dt} , \vw)_g$.

Thus
$(\calL_\vv\alpha).\vw
= {D\alpha\over D t}.\vw + \alpha.d\vv.\vw
= ({D\valphag \over Dt} , \vw)_g  + (\valphag,d\vv.\vw)_g
= (\calL_\vv\valphag + d\vv.\valphag , \vw)_g  + (d\vv^T_g.\valphag,\vw)_g
$.
\findem

\debrem
Chorus: a ``differential form'' (measuring instrument, covariant) should not be confused with a ``vector field'' (object to be measured, contravariant); Thus, the use of a dot product (which one?) and the Riesz representation theorem should be restricted for computational purposes, after an objective equation has been established.
%(If you use components relative to a basis, the change of index position from top $()^i$ to bottom $()_i$ is not compatible with the brutal use of the Lie derivative).
See also~remark~\ref{remisonat}.
%It has consequences for constitutive laws of fluids (usually modeled with vector fields) and for constitutive laws of solids (usually modeled with differential forms).
\finrem

%%%%%%%%%%%%%%%%%%%%%%%%%%%%%%%%%%%%%%%%%%%%%%%%%%%%%%%%%%%%%%%%%%%%%%%%%%%%%%%%%%%

\subsection{Lie derivative of a tensor}

%%%%%%%%%%%%%%%%%%%%%%%%%%%%%%%%%%%%%%%%%%%%%%%%%%%%%%%%%%%%%%%%%%%%%%%%%%%%%%%%%%%

%\subsubsection{Formula}

%A tensor gives values to vector fields and differential forms; And the Lie derivation is a derivation; Thus, with the Lie derivative of scalar functions~\eref{eqdl0}, vector fields~\eref{eqdl} and differential form~\eref{eqdla}, you can define:

The Lie derivative of any tensor of order $\ge2$ is defined thanks to
\be
\label{eqldoat}
\calL_\vv(T\otimes S) = (\calL_\vv T) \otimes S + T\otimes (\calL_\vv S) \quad(\hbox{derivation formula}).
\ee
(Or direct definition: $\calL_\vv T(\tz,\ptz) = {D((\Phitzt)^*T_t)(\ptz) \over Dt}_{|t=\tz}$). % for any tensor~$T$ at any $(\tz,\ptz)\in\bigC$.)

%%%%%%%%%%%%%%%%%%%%%%%%%%%%%%%%%%%%%%%%%%%%%%%%%%%%%%%%%%%%%%%%%%%%%%%%%%%%%%%%%%%

\subsubsection{Lie derivative of a mixed tensor}

%Voir poly ``Tenseurs...'' pour les push-forward et dérivées de Lie des tenseurs.

Let $T_m\in \Tuuo$, and $T_m$ is called a mixed tensor; Its Lie derivative, called the Jaumann derivative, is given by
\be
\label{eqJaumann}
\boxed{\calL_\vv T_m = {D T_m \over Dt} - d\vv.T_m + T_m.d\vv} = {\pa T_m\over \pa t} + dT_m.\vv - d\vv.T_m + T_m.d\vv.
\ee
Can be checked with an elementary tensor $T=\vw\otimes \alpha$: 
we have $d(\vw\otimes \alpha).\vv = (d\vw.\vv)\otimes\alpha + \vw \otimes (d\alpha.\vv)$
and $(d\vv.\vw)  \otimes \alpha = d\vv.(\vw\otimes \alpha)$, and
$\vw\otimes (\alpha.d\vv) = (\vw\otimes \alpha).d\vv$
, thus
\eref{eqldoat} gives 
$\calL_\vv(\vw\otimes \alpha) = (\calL_\vv \vw) \otimes \alpha + \vw\otimes (\calL_\vv \alpha)
$
\\
$
= {\pa \vw \over \pa t} \otimes \alpha + (d\vw.\vv) \otimes \alpha - (d\vv.\vw)  \otimes \alpha
%= {D \vw \over D t}  - d\vv.\vw}
+ \vw\otimes {\pa \alpha \over \pa t} + \vw\otimes (d\alpha.\vv) + \vw\otimes (\alpha.d\vv)
$
\\
$
= {\pa \vw\otimes \alpha \over \pa t} + d(\vw\otimes \alpha).\vv  - d\vv.(\vw\otimes \alpha) + (\vw\otimes \alpha).d\vv
$.

\mn
{\bf Quantification.} Relative to a basis $(\ve_i)$:
\be
\label{eqJaumannq}
[\calL_\vv T_m] = [{D T_m \over Dt}] - [d\vv].[T_m] + [T_m].[d\vv] 
%\quad (= [{\pa T_m\over \pa t}] + [dT_m.\vv] - [d\vv].[T_m] + [T_m].[d\vv])
\ee
(the signs $\mp$ are mixed).
``Mixed'' also refers to positions of indices (up and down with duality notations):
$T_m = \sumijn T^i{}_j \ve_i\otimes e^j$ with the dual basis~$(e^i)$, \ie\ $[T_m]_{|\ve}=[T^i{}_j]$.

\debexe
With components, prove~\eref{eqJaumannq}.

\debrep
${\pa T_m \over \pa t} = \sum_{ij}{\pa T^i{}_j \over \pa t} \ve_i\otimes e^j$, 
$dT_m = \sum_{ijk}T^i{}_{j|k} \ve_i\otimes e^j\otimes e^k$, 
$\vv = \sum_i v^i\ve_i$, $d\vv = \sum_{ij} v^i_{|j} \ve_i\otimes e^j$, thus
$dT_m.\vv = \sum_{ijk} T^i{}_{j|k} v^k\ve_i\otimes e^j$,
$d\vv.T_m = \sum_{ijk} v^i_{|k} T^k{}_j \ve_i\otimes e^j$,
$T_m.d\vv =  \sum_{ijk} T^i{}_k v^k_{|j} \ve_i\otimes e^j$.
\comment{, thus
$\calL_\vv T_m = {D T_m \over Dt} - d\vv.T_m + T_m.d\vv
= \sumijn {\pa T^i{}_j\over \pa t} \ve_i\otimes e^j
+ \sum_{ijk}  T^i{}_{j|k} v^k\ve_i\otimes e^j
- \sum_{ijk} v^i_{|k} T^k{}_j \ve_i\otimes e^j
+\sum_{ijk} T^i{}_k v^k_{|j} \ve_i\otimes e^j
$.
}
\finrep
\finexe

%%%%%%%%%%%%%%%%%%%%%%%%%%%%%%%%%%%%%%%%%%%%%%%%%%%%%%%%%%%%%%%%%%%%%%%%%%%%%%%%%%%

\subsubsection{Lie derivative of a up-tensor}

Recall: If $L\in\calL(E;F)$ (a linear map) then its adjoint $L^* \in\calL(F^*;E^*)$ is defined by, \cf~\S~\ref{secadjlm},
\be
\label{eqLs}
\forall m \in F^*,\quad \boxed{L^*.m := m.L}, \quad\hbox{\ie,}\quad
\forall m,\vu \in (F^*\times E),\quad (L^*.m).\vu= m.L.\vu.
\ee
(There is no inner dot product involved here.)
In particular, %$d\vv^*(t,\pt)\in \calL(\RRnts;\RRnts)$ is given by
$d\vv^*.m := m.d\vv$ for all $m\in\RRnts$, \ie\
$(d\vv^*.m).\vu = (m.d\vv).\vu = m.(d\vv.\vu)$ for all $m\in\RRnts$ and all $\vu\in\RRnt$.

\medskip
Let $T_u\in \Tdzo$, and $T_u$ is called a up tensor; Its Lie derivative is called the upper-convected (Maxwell) derivative or the Oldroyd derivative and is given by
\be
\label{equpmaxw}
\boxed{\calL_\vv T_u = {D T_u \over Dt} - d\vv.T_u - T_u.d\vv^*}
= {\pa T_u\over \pa t} + dT_u.\vv - d\vv.T_u - T_u.d\vv^*.
\ee
Can be checked with an elementary tensor $T=\vu\otimes \vw$ and 
$\calL_\vv(\vu\otimes \vw) = (\calL_\vv \vu) \otimes \vw + \vu\otimes (\calL_\vv \vw)$.

\mn
{\bf Quantification.} Relative to a basis $(\ve_i)$:
\be
[\calL_\vv T_u] = [{D T_u \over Dt}] - [d\vv].[T_u] - [T_u].[d\vv]^T.
\ee
``up'' also refers to positions of indices (with duality notations):
$T_u = \sumijn T^{ij} \ve_i\otimes \ve_j$ with the dual basis~$(e^i)$, \ie\ $[T_u]_{|\ve}=[T^{ij}]$.

\debexe
With components, prove~\eref{equpmaxw}.

\debrep
${\pa T_u \over \pa t} = \sum_{ij}{\pa T^{ij} \over \pa t} \ve_i\otimes \ve_j$, 
$dT_u = \sum_{ijk} T^{ij}_{|k} \ve_i\otimes \ve_j\otimes e^k$, 
$\vv = \sum_i v^i\ve_i$, $d\vv = \sum_{ij} v^i_{|j} \ve_i \otimes e^j$,
$d\vv^* = \sum_{ij} v^j_{|i} e^i \otimes \ve_j$,
thus
$dT_u.\vv = \sum_{ijk} T^{ij}_{|k} v^k\ve_i\otimes e^j$,
$d\vv.T_u = \sum_{ijk} v^i_{|k} T^{kj} \ve_i\otimes \ve_j$,
$T_u.d\vv^* = \sum_{ijk} T^{ik}v^j_{|k} e^i\otimes \ve_j$.
\finrep
\finexe

%%%%%%%%%%%%%%%%%%%%%%%%%%%%%%%%%%%%%%%%%%%%%%%%%%%%%%%%%%%%%%%%%%%%%%%%%%%%%%%%%%%

\subsubsection{Lie derivative of a down-tensor}

Let $T_d\in \Tzdo$, and $T_d$ is called a down tensor; The Lie derivative is called the lower-convected Maxwell derivative and is given by
\be
\label{eqdlt02}
\boxed{\calL_\vv T_d = {D T_d \over Dt} + T_d.d\vv + d\vv^*.T_d}
= {\pa T_d\over \pa t} + dT_d.\vv + T_d.d\vv + d\vv^*.T_d.
\ee
Can be checked with an elementary tensor $T=\ell\otimes m$ and 
$\calL_\vv(\ell\otimes m) = (\calL_\vv \ell) \otimes m + \ell\otimes (\calL_\vv m)$.

\mn
{\bf Quantification.} Relative to a basis $(\ve_i)$:
\be
\label{eqdmaxw}
[\calL_\vv T_d] = [{D T_d \over Dt}] + [T_d].[d\vv] + [d\vv]^T.[T_d].
\ee
``down'' also refers to positions of indices (with duality notations):
$T_d = \sumijn T_{ij} e^i\otimes e^j$ with the dual basis~$(e^i)$, \ie\ $[T_d]_{|\ve}=[T_{ij}]$.

\debexe
With components, prove~\eref{eqdmaxw}.

\debrep
${\pa T_d \over \pa t} = \sum_{ij}{\pa T_{ij} \over \pa t} e^i\otimes e^j$, 
$dT_d = \sum_{ijk} T_{ij|k} e^i\otimes e^j\otimes e^k$, 
$\vv = \sum_i v^i\ve_i$, $d\vv = \sum_{ij} v^i_{|j} \ve_i \otimes e^j$,
$d\vv^* = \sum_{ij} v^j_{|i} e^i \otimes \ve_j$,
thus
$dT_d.\vv = \sum_{ijk} T_{ij|k} v^k e^i\otimes e^j$,
$T_d.d\vv = \sum_{ijk} T_{ik} v^k_{|j} e^i\otimes \ve_j$,
$d\vv^*.T_d = \sum_{ijk} v^k_{|i} T_{kj} e^i\otimes \ve_j$.
\finrep
\finexe

\debexa
Let $g=\dd_g \in \Tzdo$ be a constant and uniform metric (a unique inner dot product for all $t,p$, \eg, a Euclidean dot product at all~$t$). Then ${Dg\over Dt}=0$, thus $\calL_\vv g = 0+g.d\vv + d\vv^*.g$, 
thus $[\calL_\vv g] = [g].[d\vv] + [d\vv]^T.[g]$.
\finexa

\newpage
\part{Velocity-addition formula}

\section{Change of referential and velocity-addition formula}
\label{seccdr}

\def\vvBs{\vec v_{\!B*}}
\def\vwDt{\vec w_{\!Dt}}
\def\vxA{\vec x_{\!A}}
\def\vxB{\vec x_{\!B}}
\def\vyD{\vec y_{\!D}}
\def\vxAt{\vec x_{\!At}}
\def\vxAtz{\vec x_{\!At_0}}
\def\vxBt{\vec x_{\!Bt}}
\def\vxBtz{\vec x_{\!Bt_0}}
\def\vyDt{\vec y_{\!Dt}}
\def\vyS{\vec y_{\!S}}
\def\vySi{\vec y_{\!Si}}
\def\vySh{\vec y_{\!Sh}}
\def\vySz{\vec y_{\!S0}}
\def\vyDtz{{\vec y_{\!Dt_0}}}
\def\vyStzs{{\vec y_{S\tz*}}}

\def\calRAt{{\calR_{\!At}}}
\def\calRBt{{\calR_{\!Bt}}}
\def\ObjRA{\Obj\!\calR_{\!A}}
\def\ObjRB{\Obj\!\calR_{\!\!B}}
\def\tPhiRB{{\tPhi_{\!\calR_{\!\!B}}}}
\def\vvRB{{\vv_{\!\calR_{\!\!B}}}}
\def\vvRBt{{\vv_{\!\calR_{\!\!Bt}}}}
\def\vgammaRB{{\vec\gamma_{\!\calR_{\!\!B}}}}
\def\vgammaRBt{{\vec\gamma_{\!\calR_{\!\!Bt}}}}
\def\QRB{{Q_{\!\calR_{\!\!B}}}}
\def\QBz{{Q_{\!B0}}}
\def\QBu{{Q_{\!B1}}}
\def\QBd{{Q_{\!B2}}}
\def\QBi{{Q_{\!Bi}}}
\def\QRBh{{Q_{\!\calR_{\!\!Bh}}}}
\def\QRBi{{Q_{\!\calR i}}}
\def\QRBih{{Q_{\!\calR ih}}}
\def\QRBz{{Q_{\!\OB}}}

\def\OA{{O_{\!A}}}
\def\OAt{{O_{\!At}}}
\def\PAti{{P_{\!Ati}}}
\def\PBti{{P_{\!Bti}}}
\def\OB{{O_{\!B}}}
\def\OBt{{O_{\!Bt}}}
\def\RRnC{\vec\RR^n_{\rm cart}}
\def\vomegaD{{\vec\omega_{\!D}}}
\def\tvphiA{{\vec\phi_{\!A}}}
\def\tvphiAPobj{{\vec\phi_{\!A\Pobj}}}
\def\tvphiAt{{\vec\phi_{\!At}}}
\def\tvphiAtz{{\vec\phi_{\!At_0}}}
\def\tvphiB{{\vec\phi_{\!B}}}
\def\tvphiBt{{\vec\phi_{\!Bt}}}
\def\tvphiBtz{{\vec\phi_{\!Bt_0}}}
\def\tvphiD{{\vec\phi_{\!D}}}
\def\tvphiDt{{\vec\phi_{\!Dt}}}
\def\tvphiDtz{{\vec\phi_{\!Dt_0}}}
\def\tvphiS{{\vec\phi_{\!S}}}
\def\PhiAtz{{\vec\phi^\tz_{\!A}}}
\def\PhiAtzt{{\vec\phi^\tz_{\!At}}}
\def\PhiBtz{{\vec\phi^\tz_{\!B}}}
\def\PhiBtzt{{\vec\phi^\tz_{\!Bt}}}
\def\PhiDtz{{\vec\phi^\tz_{\!D}}}
\def\PhiDtzt{{\vec\phi^\tz_{\!Dt}}}
\def\PhiDtztz{{\vec\phi^\tz_{\!D\tz}}}
\def\PhiStzt{{\vec\phi^\tz_{\!St}}}
\def\vvD{{\vec v_{\!D}}}
\def\vvS{{\vec v_{\!S}}}
\def\vgammaA{{\vec \gamma_A}}
\def\vgammaAt{{\vec \gamma_{At}}}
\def\vgammaB{{\vgamma_{B}}}
\def\vgammaBt{{\vgamma_{Bt}}}
\def\vgammaC{{\vgamma_{C}}}
\def\vgammaCt{{\vgamma_{Ct}}}
\def\vgammaD{{\vec \gamma_D}}
\def\vgammaDt{{\vgamma_{Dt}}}
\def\vgammaS{{\vec \gamma_S}}
\def\vgammaA{{\vgamma_{A}}}
\def\vgammaAt{{\vgamma_{At}}}
\def\vfA{{\vf_{A}}}
\def\vfAt{{\vf_{At}}}
\def\vfB{{\vf_{B}}}
\def\vfBt{{\vf_{Bt}}}
\def\vfC{{\vf_{C}}}
\def\vfCt{{\vf_{Ct}}}

\def\vAit{{\vec A_{it}}}
\def\vAjt{{\vec A_{jt}}}
\def\vBit{{\vec B_{it}}}
\def\vBjt{{\vec B_{jt}}}

\def\QOB{Q_{O_B}}
\def\Rtz{{R^\tz}}
\def\Rtzt{{R^\tz_t}}

\def\Mnu{\calM_{\!n\!1}}
\def\MA{\calM_{\!n\!1}\!(\!\hbox{A}\!)}
\def\MB{\calM_{\!n\!1}\!(\!\hbox{B}\!)}

\setcounter{subsection}{-1}
%%%%%%%%%%%%%%%%%%%%%%%%%%%%%%%%%%%%%%%%%%%%%%%%%%%%%%%%%%%%%%%%%%%%%%%%%%%%%%%%%%%

%%%%%%%%%%%%%%%%%%%%%%%%%%%%%%%%%%%%%%%%%%%%%%%%%%%%%%%%%%%%%%%%%%%%%%%%%%%%%%%%%%%

\subsection{Issue and result (summary)}

The velocity-addition formula is (in classical mechanics)
\be
\label{eqceloidcdv}
\vvA = \vvB + \vvD,
%(\vvA\hbox{ the absolute velocity}) = (\vvB\hbox{ the relative velocity}) + (\vvD\hbox{ the drive velocity}).
\ee
where
%, for a given particle, 
$\vvA$, $\vvB$ and $\vvD$ are the absolute, relative and drive velocity,
$\vvA$ and~$\vvD$ being velocities described by an observer~A with his referential $\calRA=(\OA,(\vA_i))$ and $\vvB$ being a velocity described by an observer~B with his referential $\calRB=(\OB,(\vB_i))$.
%Thus, if $\vvB$ and $\vvD$ are known, then $\vvA$ can be calculated.
But~\eref{eqceloidcdv} is problematic (inconsistent):

%\begin{enumerate}[leftmargin=13pt, rightmargin=0pt, itemsep=0pt, label*=\arabic*.]

%\item \eref{eqceloidcdv} can't be a vectorial equality with the usual definition of a velocity,because the velocity is the tangent vector to a trajectory in the Universe, the same for all observer (it is its quantification which is different). %, see \eg~\eref{eqdefve}.

%\item
%\eref{eqceloidcdv} can't be a simple matrix equality (quantification) because

$\bullet$ The velocities~$\vvA$ and~$\vvD$ are quantified in $\calRA$, \eg\ expressed in foot/s by the absolute observer,

$\bullet$ The velocity $\vvB$ is a quantified in~$\calRB$, \eg\ expressed in metre/s by the relative observer,
\\
%$\bullet$ 
Thus~\eref{eqceloidcdv} with $\vvB + \vvD$ tells that you add metre/s and foot/s... absurd. So:
%\\(And classical ``proofs'' can be unsatisfying because of the lack of definitions.)

%\end{enumerate}

\medskip
Question: What are we missing (and what does~\eref{eqceloidcdv} really mean)?  

Answer: We miss a functional link: The translator between A and~B. Summary:

Call $\tPhi$ the motion of a observed object~$\Obj$; 
$\tPhi$ is {\sl quantified} by A in his referential $\calRA=(\OA,(\vA_i))$ as the ``motion'' $\tvphiA=[\ora{\OA\tPhi}]_{|\vA}$, and 
is {\sl quantified} by B in his referential $\calRB=(\OB,(\vB_i))$ as the ``motion'' $\tvphiB=[\ora{\OB\tPhi}]_{|\vB}$.
% (numerical values stored in column matrices).
At~$t$, the translator $\Theta$ connects these numerical values: $\tvphiA(t,\Pobj) = \Theta(t,\tvphiB(t,\Pobj))$.
Thus
${\pa \tvphiA \over \pa t}(t,\Pobj) = {\pa \Theta \over \pa t}(t,\vxBt) + d\Theta(t,\vxBt).{\pa \tvphiB \over \pa t}(t,\Pobj)$, \ie\
$\vvA(t,\vxAt) = {\pa \Theta \over \pa t}(t,\vxBt) + d\Theta(t,\vxBt).\vvB(t,\vxBt)$
where
%with $\vxAt = \Thetat(\vxBt)$ 
$\vxAt=\tvphiA(t,\Pobj)$ and $\vxBt=\tvphiB(t,\Pobj)$.
Then call
\be
d\Thetat(\vxBt).\vvBt(\vxBt) = \vvBts(\vxAt) = \hbox{``the translated relative velocity at~$t$ from B to~A''},
\ee
thus, with ${\pa \Theta \over \pa t}(t,\vxBt) = \vvD(t,\vxAt)$ the drive velocity,
which gives $\vvA(t,\vxAt) = \vvBs(t,\vxAt) + \vvD(t,\vxAt)$:
% [vitesse d'entrainement];
so
\be
\label{eqceloidcdvs}
\vvA = \vvBs + \vvD \quad = \quad \hbox{the velocity addition formula in $\calRA$},
\ee
$$
\hbox{\ie: (Absolute velocity) = (Translated relative velocity) + (Drive velocity)}.
$$
In other words, with $\vv$ the velocity of~$\Obj$ and with $\vvRB$ the velocity of~$\calRB$ in~$\calRA$: For all $ \pt=\tPhi(t,\Pobj)$,
\be
\label{eqceloidcdvs2}
[\vv_t(\pt)]_{|\vA} = d\Thetat.[\vv_t(\pt)]_{\vB} + [\vvRBt(\pt)]_{|\vA}, %\quad\hbox{for all } \pt=\tPhi(t,\Pobj),
\ee
relation between the numerical values of the velocities stored by~A and~B.

\debexa
Translation motion of $\calRB$ in~$\calRA$, so $[\vvRBt(\pt)]_{|\vA} =[ \vvRBt]_{|\vA}$ is independent of~$\pt$;
And, \eg\ with $(\vBit) = \lambda(\vAit)$ (\eg\ $\vA_i$ in foot and $\vB_i$ in meter give $\lambda \simeq 3.28$),
$d\Thetat=\lambda I$, hence $[\vv_t(\pt)]_{|\vA} = \lambda[\vv_t(\pt)]_{\vB} + [\vvRBt]_{|\vA}$,
which is the expected relation (``sum of the velocities with the good units'').
\finexa

\debexa
Motion of the Earth around the Sun: See~\S~\ref{secsummaryC}.
\finexa

%Details:

%%%%%%%%%%%%%%%%%%%%%%%%%%%%%%%%%%%%%%%%%%%%%%%%%%%%%%%%%%%%%%%%%%%%%%%%%%%%%%%%%%%

\subsection{Referentials and ``matrix motions''}
\label{secsettinvaf}

%%%%%%%%%%%%%%%%%%%%%%%%%%%%%%%%%%%%%%%%%%%%%%%%%%%%%%%%%%%%%%%%%%%%%%%%%%%%%%%%%%%

%\subsubsection{Absolute $\calRA$ and relative $\calRB$ referentials}
\subsubsection{Absolute and relative referentials}

Classical mechanics framework: Time and space are decoupled, all the observers share the same time unit (\eg\ the second)
and live in ``our'' Universe modeled as~$\RRt$ (affine space) with its usual associated vector space~$\vRRt$. In the following, the affine space is~$\RRn$ associated to the vector space~$\vRRn$, $n\in\{1,2,3\}$.

An observer A, which we will call the absolute observer, chooses a (rigid body) object $\ObjRA$ in the Universe,
chooses one particle in $\ObjRA$, calls $\OAt$ its position at~$t$,
and chooses three more particles in~$\ObjRA$, calls $\PAti$ their positions at~$t$ (in the Universe),
such that the bi-point vectors $\vAit:=\ora{\OAt\PAti}$ make a basis in~$\vRRn$.
He has thus built his (Cartesian) referential $\calRAt=(\OAt,(\vAit))$, called the absolute referential,
and written $\calRA=(\OA,(\vA_i))$ when used by~A. % (implicitly A is fixed in~$\calRA$).
\Eg\ $\ObjRA$ is the ``Sun extended to infinity'', and at~$t$, $\OAt$ is the position of the center of the Sun in the Universe, $(\vA_{it})$ is a Euclidean basis in foot fixed relative to stars. %, and A is a virtual observer fixed in~$\calRA$.

An observer B, which we will call the relative observer, proceeds similarly: He chooses a (rigid body) object $\ObjRB$ in the Universe,
builds his Cartesian referential $\calRBt=(\OBt,(\vBit))$, called the relative referential,
written $\calRB=(\OB,(\vB_i))$ when used by~B.
\Eg\ $\ObjRB$ is the ``Earth extended to infinity'', and at~$t$, $\OBt$ is the position of the center of the Earth and $(\vB_{it})$ is a Euclidean basis in metre fixed relative to the Earth.
%And $\ObjRB$ is ``extended to infinity''.

$\Mnu$ is the vectorial space of $n*1$ matrices (column matrices).
A~and~B call $\MA$ and~$\MB$ the affine spaces of $n*1$ matrices made of the ``matrix positions''
$[\ora{\OAt\pt}]_{|\vA}$ and $[\ora{\OBt\pt}]_{|\vB}$ where $\pt$ is the position at~$t$ of a particle in the Universe.
%The associated vector space is also called~$\Mnu$.

If a function $\phi$ is given as $\phi(t,x)$, then $\phi_t(x):=\phi(t,x)$, and conversely.

%%%%%%%%%%%%%%%%%%%%%%%%%%%%%%%%%%%%%%%%%%%%%%%%%%%%%%%%%%%%%%%%%%%%%%%%%%%%%%%%%%%

\subsubsection{Motion of a material object $\Obj$}

An object $\Obj$ is considered by all observers. Its motion in the Universe is
\be
\label{eqtP}
\tPhi:
\left\{\eqalign{
[t_1,t_2]\times \Obj &\rar \RRn \cr
(t,\Pobj) &\rar \pt=\tPhi(t,\Pobj) = \hbox{position of the particle~$\Pobj$ at t in the Universe}.
}\right.
\ee
At~$t$ at $\pt=\tPhi(t,\Pobj)$, the Eulerian velocities and accelerations of~$\Pobj$ are
\be
\label{eqtP1v}
\vv(t,\pt)={\pa\tPhi\over \pa t}(t,\Pobj)  \qand \vgamma(t,\pt)={\pa^2\tPhi\over \pa^2 t}(t,\Pobj) \quad(\in\vRRn).
\ee

%%%%%%%%%%%%%%%%%%%%%%%%%%%%%%%%%%%%%%%%%%%%%%%%%%%%%%%%%%%%%%%%%%%%%%%%%%%%%%%%%%%

\subsubsection{Quantification: Absolute and relative ``motion'' of $\Obj$}

At~$t$, the position $\pt=\tPhi(t,\Pobj)$ of a particle $\Pobj\in\Obj$
is spotted by~A, resp.~B, with the bi-point vectors $\ora{\OAt\pt}$, resp. $\ora{\OBt\pt}$ in~$\vRRn$,
which components is stored by A, resp.~B, in his referentials: With
\be
\ora{\OAt\pt} = \sumin x_{\!Ati}\vAit \qand \ora{\OBt\pt} = \sumin x_{\!Bti}\vBit,
\ee
and with $(\vE_i)$ the canonical basis in~$\Mnu$, the $n*1$ matrices
% , so with the matrices $\vxAt:=[\ora{\OA\pt}]_{|\vA}\in\MA$ and $\vxBt:=[\ora{\OB\pt}]_{|\vB}\in\MB$ given by
\be
\label{eqtPhiA0}
\vxAt := [\ora{\OA\pt}]_{|\vA} = \pmatrix{x_{\!At1}\cr \vdots \cr x_{\!Atn}\cr} = \sumin x_{\!Ati}\vE_i , \qand 
\vxBt := [\ora{\OB\pt}]_{|\vB} = \pmatrix{x_{\!Bt1}\cr \vdots \cr x_{\!Btn}\cr} = \sumin x_{\!Bti}\vE_i,
\ee
are stored by A and B.
(Initial notation: $\vxAt:=[\ora{\OAt\pt}]_{|(\vAit)}$, %and $\vxBt:=[\ora{\OBt\pt}]_{|(\vBit)}$,
but here $\OAt$ and $(\vAit)$ are fixed in~$\calRA$, idem for~B.)

Mind the notations: $\pt$ is a point, $\ora{\OAt\pt}$ is a vector, $\vxAt$ is a column matrix (components).

This defines the ``absolute motion'' $\tvphiA$ and ``relative motion'' $\tvphiB$ of~$\Obj$ (matrix valued):
\be
\label{eqtPhiA}
\tvphiA:
\left\{\eqalign{
[t_1,t_2]{\times} \Obj & \rar \MA \cr
(t,\Pobj) & \rar \boxed{\tvphiA(t,\Pobj) := [\ora{\OA\tPhi(t,\Pobj)}]_{|\vA}} = \sumin x_{\!Ai}(t)\vE_i \eqnote \vxA(t)
= [\ora{\OA p(t)}]_{|\vA}
,
}\right.
\ee
\be
\label{eqtPhiB}
\tvphiB:
\left\{\eqalign{
[t_1,t_2]{\times} \Obj & \rar \MB \cr
(t,\Pobj) & \rar \boxed{\tvphiB(t,\Pobj) := [\ora{\OB\tPhi(t,\Pobj)}]_{|\vB}} = \sumin x_{\!Bi}(t)\vE_i \eqnote \vxB(t) = [\ora{\OB p(t)}]_{|\vB}
.
}\right.
\ee
And the ``absolute'' and ``relative'' velocities and accelerations of~$\Pobj$ are (matrix valued in~$\Mnu$):
\be
\label{eqvvA}
\boxed{\vvA(t,\vxAt) : = [\vv(t,\pt)]_{|\vA}} \qand \vgammaA(t,\vxAt):=[\vgamma(t,\pt)]_{|\vA},
\qwhen \vxAt:= [\ora{\OA\pt}]_{|\vA}
,
\ee
\be
\label{eqvvB}
\boxed{\vvB(t,\vxBt) : = [\vv(t,\pt)]_{|\vB}} \qand \vgammaB(t,\vxBt):=[\vgamma(t,\pt)]_{|\vB},
\qwhen \vxBt:= [\ora{\OB\pt}]_{|\vB}
.
\ee

\debexe
Prove: $\vvA(t,\vxAt) = {\pa \tvphiA \over \pa t}(t,\Pobj)$.

\debrep
$\ora{\OAt\tPhiPobj(t)} = \sum_i x_{\!Ai}(t) \vA_i(t)$ gives 
$\vv(t,\pt)= \sum_i x_{\!Ai}{}'(t) \vA_i(t) + x_{\!Ai}(t) \vA_i{}'(t)$, 
thus 
$[\vv(t,\pt)]_{|\vA} = \sum_i x_{\!Ai}{}'(t) [\vA_i(t)]_{|\vA} + x_{\!Ai}(t) [\vA_i{}'(t)]_{|\vA}$
(since $\Mnu$ is a vector space) $= \sum_i x_{\!Ai}{}'(t)\vE_i + [\vec0]
$ (the $\vA_i(t)$ are static in~$\calRA$: $[\vA_i{}'(t)]_{|\vA}=[\lim_{h\rar0} {\vA_i(t{+}h)-\vA_i(t) \over h}]_{|\vA}
=\lim_{h\rar0} {[\vA_i(t{+}h)]_{|\vA}-[\vA_i(t)]_{|\vA} \over h}
=\lim_{h\rar0} {\vE_i-\vE_i \over h}=0
$).
And $\tvphiAPobj(t) = [\ora{\OA\tPhiPobj(t)}]_{|\vA} = \sum_i x_{\!Ai}(t) [\vA_i(t)]_{|\vA}
= \sum_i x_{\!Ai}(t) \vE_i
$, thus $\tvphiAPobj{}'(t) = \sum_i x_{\!Ai}{}'(t) \vE_i$.
\finrep
\finexe

\debexe
$\vu$ is a $C^1$ vector field, $p$ is a point, $\vxA:=[\ora{\OA p}]_{|\vA}$ and
$\vu_A(\vxA):= [\vu(p)]_{|\vA}$ (matrices). % for all $p\in\Omegat$ and all $\vw\in\RRnt$,
Prove:
$d\vuA(\vxA) = [d\vu(p)]_{|\vA}$ (endomorphism in~$\Mnu$), \ie\ 
$d\vuA(\vxA).[\vw]_{|\vA} = [d\vu(p).\vw]_{|\vA}$ for all $\vw\in\vRRn$.

\debrep
The point $p{+}h\vw\in\RRn$ is referenced by A as 
$[\ora{\OA p} + h\vw]_{|\vA}=[\ora{\OA p}]_{|\vA}+h[\vw]_{|\vA}=\vxA + h[\vw]_{|\vA}$.
Thus
$d\vuA(\vxA).[\vw]_{|\vA}
= \lim_{h\rar0} {\vuA(\vxA {+} h[\vw]_{|\vA}) - \vuA(\vxA) \over h}
= \lim_{h\rar0} {[\vu(p{+}h\vw)]_{|\vA} - [\vu(p)]_{|\vA} \over h}
= \lim_{h\rar0} {[\vu(p{+}h\vw) - \vw(p)]_{|\vA} \over h}
= [\lim_{h\rar0} {\vu(p{+}h\vw) - \vw(p) \over h}]_{|\vA}
= [d\vu(p).\vw]_{|\vA}
= [d\vu(p)]_{|\vA}.[\vw]_{|\vA}
$, true for all~$\vw$. %, thus $d\vwA(\vxA) = [d\vw(p)]_{|\vA}$.
\finrep
\finexe

\debexe
Call $Q_t$ the transition matrix from~$(\vA_{it})$ to~$(\vB_{it})$ at~$t$.
Prove $\vxAt = [\ora{\OA\OBt}]_{|\vA} + Q_t.\vxBt$.

\debrep
$\vxAt=[\ora{\OA\pt}]_{|\vA} = [\ora{\OA\OBt} + \ora{\OBt\pt}]_{|\vA} = [\ora{\OA\OBt}]_{|\vA} + [\ora{\OBt\pt}]_{|\vA}$,
and the change of basis formula gives $[\ora{\OBt\pt}]_{|\vB}=Q_t^{-1}.[\ora{\OBt\pt}]_{|\vA}$.
\finrep
\finexe

%%%%%%%%%%%%%%%%%%%%%%%%%%%%%%%%%%%%%%%%%%%%%%%%%%%%%%%%%%%%%%%%%%%%%%%%%%%%%%%%%%%

\subsubsection{Motion of $\calRB$}

Particular case $\Obj=\ObjRB$:
Its motion in the Universe, also called the motion of~$\calRB$, is noted
\be
\label{eqtPRB}
\tPhiRB:
\left\{\eqalign{
[t_1,t_2]\times \ObjRB &\rar \RRn \cr
(t,\QRB) &\rar \qt=\tPhiRB(t,\QRB).
}\right.
\ee
At~$t$ at $\qt=\tPhiRB(t,\QRB)$, the Eulerian velocities and accelerations of~$\QRB$ are
\be
\label{eqtPRB1v}
\vvRB(t,\qt)={\pa\tPhiRB\over \pa t}(t,\QRB) \qand \vgammaRB(t,\qt)={\pa^2\tPhiRB\over \pa^2 t}(t,\QRB).
\ee
%(Remark: $\ObjRB$ is supposed to be ``extended to infinity'', \eg\ the solid the Earth is virtually extended to infinity.)

%%%%%%%%%%%%%%%%%%%%%%%%%%%%%%%%%%%%%%%%%%%%%%%%%%%%%%%%%%%%%%%%%%%%%%%%%%%%%%%%%%%

\subsubsection{Quantification: Drive and static ``motion'' of~$\calRB$}

The ``drive motion''~$\tvphiD$, also called the motion of~$\calRB$ in~$\calRA$, and the ``static motion''~$\tvphiS$ is the quantification of $\tPhiRB$ by~A and by~B:
\be
\label{eqtPhiD}
\tvphiD:
\left\{\eqalign{
[t_1,t_2]\times \ObjRB & \rar \MA \cr
(t,\QRB) & \rar \tvphiD(t,\QRB) := [\ora{\OA\tPhiRB(t,\QRB)}]_{|\vA} \eqnote \vyD(t) = [\ora{\OA q(t)}]_{|\vA} ,
}\right.
\ee
\be
\label{eqtPhiS}
\tvphiS:
\left\{\eqalign{
\ObjRB & \rar \MB \cr
\QRB & \rar \tvphiS(\QRB) := [\ora{\OB\tPhiRB(t,\QRB)}]_{|\vB} \eqnote \vyS = [\ora{\OB q(t)}]_{|\vB}.
}\right.
\ee
($\tvphiS$ is independent of $t$ since $\ObjRB$ is fixed in~$\calRB$.)

The drive velocity, also called the velocity of~$\calRB$ in~$\calRA$, and static velocity of~$\QRB$ are
\be
\label{eqvvD}
\vvD(t,\vyDt):= [\vvRB(t,\qt)]_{|\vA}
\qwhen \vyDt:= [\ora{\OA\qt}]_{|\vA},
\ee
\be
\vvS(t,\vyS):= [\vvRB(t,\qt)]_{|\vB} = [\vec0] \eqnote \vec0 \hbox{  (null matrix)}.
\ee
And the drive and static accelerations are
$\vgammaD(t,\vyDt)=[\vgammaRB(t,\qt)]_{|\vA}$
 and $\vgammaS(t,\vyS)=\vec0$.

\debexe
Why introduce $\tvphiS$ (static)?

\debrep
You can't confuse a particle $\QRB$ with its stored positions $\vyS$ or $\vyDt$ at~$t$. And see~\eref{eqdefThetat0}.
\finrep
\finexe

%%%%%%%%%%%%%%%%%%%%%%%%%%%%%%%%%%%%%%%%%%%%%%%%%%%%%%%%%%%%%%%%%%%%%%%%%%%%%%%%%%%

\subsection{The translator $\Theta_t$}
\label{secThetat}

%%%%%%%%%%%%%%%%%%%%%%%%%%%%%%%%%%%%%%%%%%%%%%%%%%%%%%%%%%%%%%%%%%%%%%%%%%%%%%%%%%%

\subsubsection{Definition of~$\Thetat$}

% (we take a photo of the Universe at~$t$).

\debdef
At $t$, the translator $\Thetat : \MB\rar \MA$ is defined with~\eref{eqtPhiD}-\eref{eqtPhiS} by:
\be
\label{eqdefThetat0}
\left\{\eqalign{
&[\ora{\OB q(t)}]_{|\vB} = \tvphiS(\QRB) = \vyS \hbox{ position of $\QRB$ in $\calRB$ (static), and} \cr
&[\ora{\OA q(t)}]_{|\vA} = \tvphiDt(\QRB) = \vyDt \hbox{ position of $\QRB$ at $t$ in $\calRA$ (moving)}  \cr
}\right\}
\Longrightarrow 
\vyDt = \Thetat(\vyS),
\ee
\ie\ $\Thetat$ is the ``inter-referential function at~$t$'' which translates the ``matrix position''
$\vyS=\tvphiS(\QRB)=[\ora{\OB\tPhiRB(t,\QRB)}]_{|\vB}\in\MB$ (position of~$\QRB$ as stored by~B) to the ``matrix position'' 
$\vyDt=\tvphiD(t,\QRB)=[\ora{\OA\tPhiRB(t,\QRB)}]_{|\vA}\in\MA$ (position of~$\QRB$ as stored by~A).  
So $\Thetat$ is defined by
\be
\label{eqdefPsit0}
\boxed{\tvphiDt = \Thetat \circ \tvphiS} : 
\left\{\eqalign{
\ObjRB & \rar \MA \cr
\QRB & \rar \tvphiDt(\QRB) := \Thetat (\tvphiS(\QRB)),
}\right.
\ee
\ie\ defined by
\be
\label{eqdefPsit00}
\Thetat := \tvphiDt\circ \tvphiS^{\,-1}:
\left\{\eqalign{
\MB & \rar \MA \cr
\vyS & \rar \vyDt = \Thetat(\vyS) :=\tvphiDt(\tvphiS^{\,-1}(\vyS))
}\right.
\ee
(stored position by~B to stored position by~A).
\findef

\Eg, for $\QRBz$ the particle in~$\ObjRB$ at $t$ at~$\OBt$ (chosen by B to locate its origin), \eref{eqdefPsit0} gives, with $\vec0$ the null matrix in~$\Mnu$,
\be
\label{eqdefThetatv0}
[\ora{\OA \OBt}]_{|\vA} = \Thetat(\vec0).
\ee

So, $\Thetat$ is defined such that the following diagram commutes:
\be
\label{psietdg0}
\xymatrix{
          &  \qquad \vyS = \tvphiS(\QRB) = \hbox{localization of $\QRB$ by B}  \ar[dd]^{\ts \Thetat} \\
\QRB \in \ObjRB \ar[ur]^{\ts \tvphiS\qquad}  \ar[dr]^{\ts \tvphiDt\qquad} \\
					&  \qquad \vyDt= \tvphiDt(\QRB) = \Thetat(\vyS) = \hbox{localization at $t$ of $\QRB$ by A} .  %\\ = q_{rt*}
}
\ee

\mn

%%%%%%%%%%%%%%%%%%%%%%%%%%%%%%%%%%%%%%%%%%%%%%%%%%%%%%%%%%%%%%%%%%%%%%%%%%%%%%%%%%%

\subsubsection{Translation at $t$ for the motion~$\tPhi$}

$t$ is fixed, the position $\pt=\tPhi(t,\Pobj)$ of a particle $\Pobj \in\Obj$ is also the position $\qt=\tPhiRB(t,\QRB)$ of a particle $\QRB\in\ObjRB$, so $\tvphiAt(\Pobj) = [\ora{\OA \pt}]_{|\vA} = [\ora{\OA \qt}]_{|\vA} = \tvphiDt(\QRB)$,
and $\tvphiBt(\Pobj) = [\ora{\OB \pt}]_{|\vB} = [\ora{\OB \qt}]_{|\vB} = \tvphiS(\QRB)$, thus \eref{eqdefPsit0} gives
\be
\label{eqtvpsiAB}
%\tvphiAt(\Pobj) \equalref{eqdefPsit0} \Thetat(\tvphiBt(\Pobj)), \qso 
\boxed{\tvphiAt = \Thetat \circ \tvphiBt}.
\ee

%%%%%%%%%%%%%%%%%%%%%%%%%%%%%%%%%%%%%%%%%%%%%%%%%%%%%%%%%%%%%%%%%%%%%%%%%%%%%%%%%%%

\subsection{$d\Thetat$}

%%%%%%%%%%%%%%%%%%%%%%%%%%%%%%%%%%%%%%%%%%%%%%%%%%%%%%%%%%%%%%%%%%%%%%%%%%%%%%%%%%%

\subsubsection{Push-forward}
%\label{secvft}

\def\vEits{\vec E_{it*}}
\def\vEjts{\vec E_{jt*}}
\def\vwS{{\vec w_S}}
\def\vwSt{{\vec w_{St}}}
\def\vySt{{\vec y_{St}}}
\def\vwSts{{\vec w_{St*}}}
\def\Qit{Q_{it}}
\def\Qjt{Q_{jt}}
\def\Qith{Q_{ith}}
\def\vAit{\vec A_{it}}
\def\vAjt{\vec A_{jt}}
\def\vBit{\vec B_{it}}
\def\vBjt{\vec B_{jt}}

If $\vyS\in\MB$, $\vwS\in\Mnu$ and $\vyDt = \Thetat(\vyS)$ , then 
\be
\label{eqdThetatvyS}
\vwSts(\vyDt) := d\Thetat(\vyS).\vwS  \quad (= \lim_{h\rar0} {\Thetat(\vyS+h\vwB) - \Thetat(\vyS) \over h})
\ee
is the push-forward of the matrix $\vwS\in\Mnu$ by~$\Thetat$.

%So, $\vwSts([\ora{\OA\qt}]_{|\vA}) = d\Thetat([\ora{\OB\qt}]_{|\vB}).\vwS([\ora{\OB\qt}]_{|\vB})$
%for all $\qt$ and all $\vw_t$ with $\vwS([\ora{\OB\qt}]_{|\vB}) = [\vw_t(\qt)]_{|\vB}$.
So, $\vwSts([\ora{\OA\qt}]_{|\vA}) = d\Thetat([\ora{\OB\qt}]_{|\vB}).[\vw_t(\qt)]_{|\vB}$
for all $\qt\in\RRn$ and all $\vw_t:\RRn\rar\vRRn$ (vector field).

%%%%%%%%%%%%%%%%%%%%%%%%%%%%%%%%%%%%%%%%%%%%%%%%%%%%%%%%%%%%%%%%%%%%%%%%%%%%%%%%%%%

\subsubsection{$\Thetat$ is affine in classical mechanics}

\def\vySz{\vec y_{\!S0}}
\def\vySu{\vec y_{\!S1}}
\def\vyDz{\vec y_{\!D0}}
\def\vyDu{\vec y_{\!D1}}

\def\qtDz{\qt_{\!D_0}}
\def\qtDu{\qt_{\!D_1}}
\def\qtDi{\qt_{\!Di}}
\def\qSz{q_{\!S_0}}
\def\qSu{q_{\!S_1}}
\def\qSi{q_{\!Si}}
\def\vyDtu{\vy_{Dt1}}

\debprop
\label{propThetaff}
In $\RR^3$ (in classical mechanics), $\Thetat$ is affine: 
For all $\QBz,\QBu\in\ObjRB$ and all $t,u\in\RR$, with $\qti = \tPhiRB(t,\QBi) \in\RRn$ (positions at~$t$ in our Universe),
\be
\label{eqvyDtnb0}
\Thetat([\ora{\OB \qtz}]_{|\vB} + u\,[\ora{\qtz \qtu}]_{|\vB})
= [\ora{\OA \qtz}]_{|\vA} + u\,[\ora{\qtz \qtu}]_{|\vA}, \qand [\ora{\qtz\qtu}]_{|\vA} = d\Thetat.[\ora{\qtz\qtu}]_{|\vB}
\ee
the differential $d\Thetat(\vySz)\eqnote d\Thetat$ being independent of~$\vySz$.
In particular $[\vBit]_{\vA} = d\Thetat.[\vB_i]_{|\vB}$. %, \ie\ $d\Thetat.\vE_i = \vEits$.
In other words, for all $\vySz,\vySu \in \MB$ and all $t,u\in\RR$,
\be
\label{eqThetaff}
\Thetat((1{-}u)\vySz + u\,\vySu) = (1{-}u)\Thetat(\vySz) + u\,\Thetat(\vySu), 
\qand
\Thetat(\vySu) = \Thetat(\vySz) + d\Thetat.(\vySu{-}\vySz).
\ee
\finprop

\debdem
\comment{
Consider two particles $\QBz,\QBu\in\ObjRB$ and $\qti = \tPhiRB(t,\QBi) \in\RRn$ their positions at~$t$, and
$\vySi = [\ora{\OB \qti}]_{|\vB} \in \MB$. We want:
$\Thetat((1{-}u)[\ora{\OB \qtz}]_{|\vB} + u\,[\ora{\OB \qtu}]_{|\vB})
= (1{-}u)[\ora{\OA \qtz}]_{|\vA} + u\,[\ora{\OA \qtu}]_{|\vA}$,
\ie\ $\Thetat([\ora{\OB \qtz}]_{|\vB} + u\,[\ora{\qtz \qtu}]_{|\vB})
= [\ora{\OA \qtz}]_{|\vA} + u\,[\ora{\qtz \qtu}]_{|\vA}$.
}
Consider the straight line (possible in classical mechanics in~$\RRt$)
$\qt: u  \rar \qt(u)=\qtz + u\,\ora{\qtz\qtu} \in\RRn$ (fixed in~$\calRB$),
in particular, $\qt(0)=\qtz$ and~$\qt(1)=\qtu$.
Let
$\vyS(u)=[\ora{\OB \qt(u)}]_{|\vB}$ (positions stored by~B), so
$\vyS(u)=[\ora{\OB \qtz} + u\,\ora{\qtz\qtu}]_{|\vB} 
=[(1{-}u)\ora{\OB \qtz} + u\,\ora{\OB \qtu}]_{|\vB} 
= (1{-}u)[\ora{\OB \qtz}]_{|\vB} + u\,[\ora{\OB \qtu}]_{|\vB}
= (1{-}u)\vySz + u\,\vySu$, where $\vySz=[\ora{\OB \qtz}]_{|\vB} = \vyS(0)$ and $\vySu=[\ora{\OB \qtu}]_{|\vB} = \vyS(1)$. 
Idem for~A:
%$\vyDtz=[\ora{\OA \qtz}]_{|\vA}$, $\vyDtu=[\ora{\OA \qtu}]_{|\vA}$ and 
$\vyDt(u)=[\ora{\OA \qt(u)}]_{|\vA} = (1{-}u)\vyDtz + u\vyDtu$ (positions stored by~A).
Thus 
$$
\eqalign{
(1{-}u)\Thetat(\vySz) + u\Thetat(\vySu)
\equalref{eqdefThetat0} (1{-}u)\vyDtz + u\vyDtu  = \vyDt(u) 
\equalref{eqdefThetat0} \Thetat(\vyS(u))
= \Thetat((1{-}u)\vySz + u\vySu)
%\equalref{eqdefPsit0}  (1{-}u)\vyDtz + u\vyDtu 
 ,
}
$$
thus~\eref{eqThetaff}$_1$, thus~\eref{eqvyDtnb0}$_1$.
%for all~$u$, $\vySz,\vySu$; Thus~\eref{eqThetaff}$_1$ And (derivation in~$u$)
Hence (derivation in~$u$):
$-\Thetat(\vySz) + \Thetat(\vySu) = d\Thetat(\vyS(u)).(-\vySz+\vySu)$,
true for all~$u$, thus $d\Thetat(\vyS(u))$ is independent of~$u$, $d\Thetat(\vyS(u)) = d\Thetat(\vySz)$, true for all $\vySz$, so $d\Thetat(\vySz)\eqnote d\Thetat$, thus~\eref{eqThetaff}$_2$, thus~\eref{eqvyDtnb0}$_2$.
\comment{
Hence
$\ds %[\ora{\qtz\qtu}]_{|\vA} = 
[\ora{\OA\qtu}]_{|\vA} - [\ora{\OA\qtz}]_{|\vA}
\equalref{eqdefThetat0}\Thetat([\ora{\OB\qtu}]_{|\vB}) - \Thetat([\ora{\OB\qtz}]_{|\vB})
\equalref{eqThetaff} d\Thetat.([\ora{\OB\qtu}]_{|\vB} - [\ora{\OB\qtz}]_{|\vB})
%= d\Thetat.[\ora{\qtz\qtu}]_{|\vB}
$, thus~\eref{eqThetaff}$_2$.
}
Thus $[\ora{\OBt\PBti}]_{|\vA} = d\Thetat.[\ora{\OBt\PBti}]_{|\vB}$ where $\PBti$ is \st\ $\vBit=\ora{\OBt\PBti}$, thus $[\vBit]_{\vA} = d\Thetat.[\vB_i]_{|\vB}$.
\findem

\debexe
Call $Q_t=[Q_{t,ij}]$ the transition matrix from $(\vAit)$ to~$(\vBit)$ in~$\vRRn$,
and $(\vE_i)$ the canonical basis in~$\Mnu$.
Prove
\be
\label{eqcalPo30}
[d\Thetat]_{|\vE} = Q_t, \qie d\Thetat.\vE_j = \sumin Q_{t,ij} \vE_i, \;\;\forall j.
\ee

\debrep
%$Q_t$ is defined by 
$\vBjt = \sumin Q_{t,ij} \vAit$ gives
$[\vBjt]_{|\vA} = \sumin Q_{t,ij} \vE_i$, and
$[\vBjt]_{|\vA} \equalref{eqvyDtnb0} d\Thetat.[\vBjt]_{|\vB} = d\Thetat.\vE_j$.
\finrep
\finexe

%%%%%%%%%%%%%%%%%%%%%%%%%%%%%%%%%%%%%%%%%%%%%%%%%%%%%%%%%%%%%%%%%%%%%%%%%%%%%%%%%%%

\subsection{Translated velocities}

$t$ is fixed, $\vv_t(\pt)={\pa\tPhi\over \pa t}(t,\Pobj)$ is the velocity of a particle $\Pobj\in\Obj$ at~$t$ at~$\pt$,
$\vxAt := [\ora{\OAt\pt}]_{|\vA}$, $\vxBt := [\ora{\OBt\pt}]_{|\vB}$, and $\Thetat$ affine.
% and $\vvBt(\vxBt) := [\vv_t(\pt)]_{|\vB}$, 

\debdef
The translated relative velocity and acceleration from B to~A at~$t$ at~$\pt$ are the matrices
\be
\label{eqvyDtn2}
\boxed{\vvBts(\vxAt) := d\Thetat.\vvBt(\vxBt)} \qand \boxed{\vgammaBts(\vxAt) = d\Thetat.\vgammaBt(\vxBt)}.
\ee
\Ie, $\vvBts(\vxAt)=d\Thetat.[\vv_t(\pt)]_{\vB}$ and $\vgammaBts(\vxAt)=d\Thetat.[\vgamma_t(\pt)]_{\vB}$.
% (If $\Thetat$ is not affine, then $\vvBts(\vxAt) := d\Thetat(\vxBt).\vvBt(\vxBt)$.)
\findef

\noindent
{\bf Interpretation:} %In~\eref{eqvyDtn2}, $t$ is fixed (the scene is frozen), and 
Let $\qtz$ and $\qtu$ be particles in~$\ObjRB$ \st\
$\vvBt(\vxBt) = [\ora{\qtz\qtu}]_{|\vB}$ where $\vxBt=[\ora{\OB\qtz}]_{|\vB}$
(here $\ora{\qtz\qtu}$ is a tangent vector at~$\qtz$ to the curve $\qt: u  \rar \qt(u)=\qtz + u\,\ora{\qtz\qtu}$ in the proof of~prop.~\ref{propThetaff}); Then $[\ora{\qtz\qtu}]_{|\vA} \equalref{eqvyDtnb0} d\Thetat.[\ora{\qtz\qtu}]_{|\vB}$ gives $[\ora{\qtz\qtu}]_{|\vA} = d\Thetat.\vvBt(\vxBt)=\vvBts(\vxAt)$.
Similarly for $\vgammaBts(\vxAt)$.

\comment{
\noindent
{\bf Interpretation:} In~\eref{eqvyDtn2}, $t$ is fixed (the scene is frozen), and the position $\ptz=\tPhi(t,\Pobj)$ of a particle $\Pobj \in\Obj$ is also the position $\qtz=\tPhiRB(t,\QRB)$ of a particle $\QRB\in\ObjRB$;
So $\vxBtz=[\ora{\OB \ptz}]_{|\vB}=[\ora{\OB \qtz}]_{|\vB}=\vySz$
and $\vxAtz=[\ora{\OA \ptz}]_{|\vA}=[\ora{\OA \qtz}]_{|\vA}=\vyDtz$; 
And $\vvBt(\vxBt) = \vwS(\vyS)= [\ora{\qtz\qtu}]_{|\vB}$, where $\qtu$ is position of a second particle in~$\ObjRB$ at~$t$,
thus \eref{eqdThetatvyS} gives
$d\Thetat.\vvBt(\vxBtz) = d\Thetat.\vwS(\vySz)= \vwSts(\vyDtz)= [\ora{\qtz\qtu}]_{|\vA}$
and $\vwSts(\vyDtz) \eqnote \vvBts(\vxAtz)$ (push-forward).
}

\debexe
%\label{exeEF20}
$(\vA_i)$ and $(\vB_i)$ are Euclidean basis (\eg\ in foot and metre),
$\dd_A$ and $\dd_B$ are the associated Euclidean dot products, $\lambda = ||\vB_i||_A$ (\eg\ $\simeq3.28$),
$(\vE_i)$ is the canonical basis in~$\Mnu$, and 
$\dd_\calM$ is the canonical inner dot product in~$\Mnu$.
Call $\vEits := d\Thetat.\vE_i$ and prove:
\be
\label{eqEF20}
\forall i,j,\;\; (\vEits,\vEjts)_\calM = \lambda^2 \delta_{ij}, \qand d\Thetat^T.d\Thetat = \lambda^2 I.
\ee

\debrep
$(\vBit)$ is a Euclidean basis for~B, thus is a Euclidean orthogonal basis for all observers, in particular for~A, with
$||\vBit||_A = \lambda$ for all~$i$. And $\vEits = d\Thetat.[\vBjt]_{|\vB} \equalref{eqvyDtnb0} [\vBit]_{|\vA}$.
Thus 
$(\vEits,\vEjts)_\calM 
=[\vEits]^T.[\vEjts]
= [\vBit]_{|\vA}^T.[\vBjt]_{|\vA}
= (\vBit,\vBjt)_A
= \lambda^2(\vBit,\vBjt)_B
= \lambda^2 \delta_{ij}
$, thus~\eref{eqEF20}$_1$;
Then 
$\lambda^2 \delta_{ij}
%= (\vEits,\vEjts)_\calM 
= (d\Thetat.\vE_i,d\Thetat.\vE_j)_\calM
= (d\Thetat^T.d\Thetat.\vE_i,\vE_j)_\calM
$, true for all $i,j$,
thus $d\Thetat^T.d\Thetat = \lambda^2 I$, % since $(\vE_i)$ is a basis in~$\Mnu$, 
thus~\eref{eqEF20}$_2$.
\finrep
\finexe

%%%%%%%%%%%%%%%%%%%%%%%%%%%%%%%%%%%%%%%%%%%%%%%%%%%%%%%%%%%%%%%%%%%%%%%%%%%%%%%%%%%

\subsection{Definition of~$\Theta$}

\debdef
The translator from~B to~A is the function $\Theta$
defined with~\eref{eqdefPsit0} by
\be
\label{eqdefPsit3}
\Theta :
\left\{\eqalign{
\bigcup_{t\in[t_1,t_2]} (\{t\} \times \MB) & \rar \MA \cr
(t,\vyS) & \rar \boxed{\Theta(t,\vyS) := \Thetat(\vyS)},
}\right.
\ee
\ie, for all $\QRB\in\ObjRB$ and all~$t$,
\be
\Theta(t,[\ora{\OB\tPhiRB(t,\QRB)}]_{|\vB}) = [\ora{\OA\tPhiRB(t,\QRB)}]_{|\vA},
\qie \Theta(t,\tvphiS(\QRB)) = \tvphiD(t,\QRB).
\ee
\findef

%\ie\ $\Theta(t,[\ora{\OB\tPhiRB(t,\QRB)}]_{|\vB}) = [\ora{\OA\tPhiRB(t,\QRB)}]_{|\vA}$ for all $t\in\RR$ and all $\QRB\in\ObjRB$.

\Eg, \eref{eqdefThetatv0} gives
$\Theta(t,\vec0)= [\ora{\OA \OB(t)}]_{|\vA}$.

\debrem
\label{remT1}
The translator $\Theta$ looks like a motion, but is not: A ``usual'' motion is defined by one observer and connects one particle to its position; % in the Universe; 
While $\Theta$ connects two ``matrix positions'' of one particle relative to two referentials:
$\Theta$~is an ``inter-referential'' function. 
\finrem

%%%%%%%%%%%%%%%%%%%%%%%%%%%%%%%%%%%%%%%%%%%%%%%%%%%%%%%%%%%%%%%%%%%%%%%%%%%%%%%%%%%

\subsection{The ``$\Theta$-velocity'' is the drive velocity}

\debdef
The ``$\Theta$-velocity'' and ``$\Theta$-acceleration'' are defined by (Eulerian type definition)
\be
\label{eqvvrat}
\hbox{with}\quad \vyDt = \Theta(t,\vyS), \quad
\left\{\eqalign{
&\vvTheta (t,\vyDt) := {\pa \Theta \over \pa t} (t, \vyS) \;\;(\in\Mnu), \cr
&\vgamma_\Theta (t,\vyDt)  ={\pa^2 \Theta \over \pa t^2} (t, \vyS) \;\;(\in\Mnu).
}\right.
\ee
%(They look like Eulerian functions but are not since $\Theta$ is not a usual motion.)
\findef

\debprop
\be
\label{eqiwt}
\boxed{\vvTheta = \vvD} \qand \boxed{\vgamma_\Theta = \vgamma_D},
\ee
\ie\ 
$\vvTheta(t,\vy) = \vvD(t,\vy)$ and $\vgamma_\Theta(t,\vy) = \vgamma_D(t,\vy)$ in~$\Mnu$, for all $t$ and all $\vy\in\MA$.
\finprop

\debdem
$\Theta(t,\tvphiS(\QRB)) = \tvphiD(t,\QRB)$ gives
\be
\label{eqiwtd}
{\pa \Theta \over \pa t}(t,\tvphiS(\QRB)) = {\pa \tvphiD \over \pa t}(t,\QRB), \qie
\vvTheta(t,\Theta(t,\tvphiS(\QRB))) = \vvD(t,\tvphiD(t,\QRB)),
\ee
thus $\vvTheta(t,\tvphiD(t,\QRB)) = \vvD(t,\tvphiD(t,\QRB))$, thus
$\vvTheta(t,\vy) = \vvD(t,\vy)$ for all $\vy\in\MA$.
% since $\Theta(t,\tvphiS(\QRB)) = \tvphiD(t,\QRB) \eqnote \vyDt$.
%\ie\ $\vvTheta(t,\vyDt) = \vvD(t,\vyDt)$ since $\Theta(t,\tvphiS(\QRB)) = \tvphiD(t,\QRB) \eqnote \vyDt$.
Idem with~${\pa^2 \over \pa t^2}$.
\findem

%%%%%%%%%%%%%%%%%%%%%%%%%%%%%%%%%%%%%%%%%%%%%%%%%%%%%%%%%%%%%%%%%%%%%%%%%%%%%%%%%%%

\subsection{The velocity-addition formula}

\eref{eqtvpsiAB} gives
\be
\tvphiA(t,\Pobj) = \Theta(t,\tvphiB(t,\Pobj)),
\ee
thus
\be
\label{eqvaf0}
{\pa\tvphiA \over \pa t}(t,\Pobj)
= {\pa\Theta\over \pa t}(t,\tvphiB(t,\Pobj)) 
+ d\Theta(t,\tvphiB(t,\Pobj)).{\pa\tvphiB\over \pa t}(t,\Pobj) 
.
\ee
Thus
\be
\vvA(t,\vxAt) = \vvTheta(t,\vxAt) +  d\Theta(t,\vxBt).\vvB(t,\vxBt),
%\vvAt(\vxAt) = \vvThetat(\vxAt) + d\Thetat(\vxBt).\vvBt(\vxBt) \qwhen \vxAt = \Thetat(\vxBt).
\ee
where 
$\vxBt = \tvphiB(t,\Pobj)$ and $\vxAt = \tvphiA(t,\Pobj) = \Thetat(\vxBt)$.
Thus, with $\vvTheta \equalref{eqiwt} \vvD$,
\be
\label{eqloicv20}
\boxed{\vvAt = \vvBts + \vvDt} \qwhere \vvBts(\vxAt) := d\Thetat(\vxBt).\vvBt(\vxBt),
\ee
which is the velocity-addition formula in~$\calRA$:
\be
\eqalign{
\hbox{$\vvAt$ the absolute velocity}
= &\hbox{$\vvBts$ the translated relative velocity from B to~A} \cr
& + \hbox{$\vvDt$ the drive velocity}.
}
\ee
In other words (relation between the numerical values of the velocities stored by~A and~B),
\be
[\vv_t(\pt)]_{|\vA} = d\Thetat.[\vv_t(\pt)]_{\vB} + [\vvRBt(\pt)]_{|\vA}. %\quad\hbox{for all } \pt=\tPhi(t,\Pobj),
\ee

%%%%%%%%%%%%%%%%%%%%%%%%%%%%%%%%%%%%%%%%%%%%%%%%%%%%%%%%%%%%%%%%%%%%%%%%%%%%%%%%%%%

\subsection{Coriolis acceleration, and the acceleration-addition formula}

\eref{eqvaf0} gives 
\be
\label{eqacc00}
\eqalign{
{\pa^2\tvphiA \over \pa t^2}&(t,\Pobj)
= {\pa^2\Theta\over \pa t^2}(t,\vxBt)
+ d{\pa \Theta \over \pa t}(t,\vxBt). {\pa \tvphiB \over \pa t}(t,\Pobj) \cr
& + \bigl({\pa (d\Theta) \over \pa t}(t,\vxBt)
+ d^2\Theta(t,\vxBt). {\pa \tvphiB \over \pa t}(t,\Pobj)\bigr).{\pa \tvphiB \over \pa t}(t,\Pobj)
 + d\Theta(t,\vxBt).{\pa^2\tvphiB\over \pa t^2}(t,\Pobj)
.
}
\ee
And $d^2\Thetat=0$ in our classical framework ($\Thetat$ is affine);
And ${\pa \Theta \over \pa t}(t, \vyS) = \vvThetat(\Thetat(\vyS))$ gives
${\pa (d\Theta) \over \pa t}(t, \vyS)=d({\pa \Theta \over \pa t})(t, \vyS)
= d\vvThetat(\Thetat(\vyS)).d\Thetat(\vyS)$;
And ${\pa \tvphiB \over \pa t}(t,\Pobj)=\vvBt(\vxBt)$ where $\vxBt=\tvphiB (t,\Pobj)$; Thus
\be
\label{eqacc00b}
\eqalign{
\vgamma_{At}(\vxAt)
= &
\vgamma_{\Theta t}(\vxAt) + 2 d\vvThetat(\vxAt).d\Thetat(\vxBt).\vvBt(\vxBt) + d\Thetat(\vxBt).\vgamma_{Bt}(\vxBt) \cr
= &
\vgammaDt(\vxAt) + 2 d\vvDt(\vxAt).\vvBts(\vxAt) + \vgammaBts(\vxAt). \cr
}
\ee

\debdef
At $t$, the Coriolis acceleration $\vgammaCt$ at $\vxAt$ is
\be
\label{eqdefcor}
\vgammaCt(\vxAt) = 2 d\vvDt(\vxAt).\vvBts(\vxAt) %= 2 d\vvThetat(\vxAt).d\Thetat(\vxBt).\vvBt(\vxBt)
 , \qie
\boxed{\vgamma_{Ct} = 2 d\vvDt.\vvBts}.
\ee
And the Coriolis acceleration $\vgammaC$ at $t$ at $\vxAt$ is $\vgammaC(t,\vxAt) :=  \vgammaCt(\vxAt)$.
\findef

Thus \eref{eqacc00b} gives the acceleration-addition formula in~$\calRA$:
\be
\label{eqc2}
\boxed{\vgamma_{At} =  \vgamma_{Bt*} + \vgammaDt + \vgamma_{Ct}}, \qie
\ee
\be
\eqalign{
\hbox{$\vgamma_{At}$ the absolute acceleration}
= & \hbox{$\vgamma_{Bt*}$ the translated relative acceleration from B to~A} \cr
 & + \hbox{$\vgammaDt$ the drive acceleration} %(acc\'el\'eration d'entrainement)}
+ \hbox{$\vgamma_{Ct}$ the Coriolis acceleration}.
}
\ee
In other words (relation between the numerical values of the acceletations stored by~A and~B),
\be
[\vgamma_t(\pt)]_{|\vA} = d\Thetat.[\vgamma_t(\pt)]_{\vB} + [\vgammaRBt(\pt)]_{|\vA}
+ 2[d\vvRBt]_{|\vA}.d\Thetat.[\vv_t(\pt)]_{|\vB}.
\ee

%%%%%%%%%%%%%%%%%%%%%%%%%%%%%%%%%%%%%%%%%%%%%%%%%%%%%%%%%%%%%%%%%%%%%%%%%%%%%%%%%%%

\subsection{With an initial time}

\def\OmegaRBt{\Omega_{\!R\!Bt}}
\def\OmegaRBtz{\Omega_{\!R\!B\tz}}
\def\PhiRBtz{\Phi_{\!R\!B}^\tz}
\def\PhiRBtzt{\Phi_{\!R\!Bt}^\tz}
\def\vuBtz{{\vu_{B\tz}}}
\def\vxBtzs{p_{B\tz*}}

%%%%%%%%%%%%%%%%%%%%%%%%%%%%%%%%%%%%%%%%%%%%%%%%%%%%%%%%%%%%%%%%%%%%%%%%%%%%%%%%%%%

%\subsubsection{For $\Obj$}

%(For Lagrangian formulation addicts.)	
Let $\tz,t\in\RR$.
Consider the Lagrangian associated function $\Phitzt$ with the motion $\tPhi$ of~$\Obj$:
\be
\Phitzt :
\left\{\eqalign{
\Omegatz &\rar\Omegat \cr
\ptz{=}\tPhi(\tz,\Pobj) & \rar \pt= \Phitzt(\ptz) :=\tPhi(t,\Pobj).
}\right.
\ee
And, with $\vxAt=\tvphiA(t,\Pobj)=[\ora{\OA \pt}]_{|\vA}$ and $\vxBt=\tvphiB(t,\Pobj)=[\ora{\OB \pt}]_{|\vB}$, define the ``matrix motions'' $\PhiAtzt:\MA\rar\MA$ and $\PhiBtzt:\MB \rar\MB$ by
\be
\left\{\eqalign{
&\PhiAtzt(\vxAtz) := \vxAt \quad 
(= [\ora{\OA \tPhi(t,\Pobj)}]_{|\vA} = [\ora{\OA \Phitzt(\ptz)}]_{|\vA} = \tvphiAt(\Pobj) )
,\cr
&\PhiBtzt(\vxBtz) := \vxBt \quad 
(= [\ora{\OB \tPhi(t,\Pobj)}]_{|\vB} = [\ora{\OB \Phitzt(\ptz)}]_{|\vB} = \tvphiBt(\Pobj) )
.\cr
}\right.
\ee
And $\Thetat(\vxBt) = \vxAt$, \ie\  $\Thetat(\PhiBtzt(\vxBtz)) = \PhiAtzt(\vxAtz)$
with $\vxAtz = \Thetatz(\vxBtz)$, thus
\be
\label{eqdefpsie1}
\boxed{\Thetat \circ \PhiBtzt = \PhiAtzt \circ \Thetatz} : \MB \rar \MA.
\ee
In other words, the following diagram commutes:
\be
%\label{diag1}
\xymatrix{
         & \vxBtz = \tvphiB(\tz,\Pobj) \ar@<1ex>[dd]_{\ts\Thetatz} \ar[r]_{\ts\PhiBtzt}
                      & \vxBt  = \PhiBtzt(\vxBtz) 
											\ar@<1ex>[dd]_{\ts\Thetat}
\\
\Pobj\in\Obj \ar[ur]^{\ts \tvphiBtz}  \ar[dr]^{\ts \tvphiAtz}
\\
         & \vxAtz = \tvphiA(\tz,\Pobj) = \Thetatz(\vxBtz) \ar[r]^{\;\ts\PhiAtzt}
                      & \vxAt = \PhiAtzt(\vxAtz) = \Thetat(\vxBt).
}
\ee
Thus, for any vector field $\vuBtz$ in~$\calRB$,
\be
\underbrace{d\Thetat(\vxBt)}_{\hbox{\footnotesize (translation at t)}}.
\underbrace{d\PhiBtzt(\vxBtz).\vuBtz(\vxBtz)}_{\hbox{\footnotesize (deformation from $\tz$ to t)}}
= 
\underbrace{d\PhiAtzt(\vxAtz)}_{\hbox{\footnotesize (deformation from $\tz$ to t)}}.
\underbrace{d\Thetatz(\vxBtz).\vuBtz(\vxBtz)}_{\hbox{\footnotesize (translation at $\tz$)}}.
\ee
%\ie, the translation of the transported = the transport of the translated.

\debexe
Redo the above steps with $\ObjRB$ instead of~$\Obj$.

\debrep
Consider the Lagrangian associated function $\PhiRBtzt$ with the motion $\tPhiRB$ of~$\ObjRB$:
\be
\PhiRBtzt :
\left\{\eqalign{
\OmegaRBtz=\RRn &\rar\OmegaRBt=\RRn \cr
\qtz=\tPhiRB(\tz,\QRB) & \rar \qt= \PhiRBtzt(\qtz) :=\tPhiRB(t,\QRB),
}\right\}
\ee
then define the ``matrix motions'' $\PhiDtzt:\MA\rar\MA$ and $\PhiStzt:\MB \rar\MB$ by
\be
\left\{\eqalign{
&\PhiDtzt(\vyDtz) := \vyDt \quad 
(= [\ora{\OA \tPhiRB(t,\QRB)}]_{|\vA} = [\ora{\OA \PhiRBtzt(\ptz)}]_{|\vA} = \tvphiDt(\QRB) )
,\cr
&\PhiStzt(\vyS) := \vyS \quad 
(= [\ora{\OB \tPhiRB(t,\QRB)}]_{|\vB} = [\ora{\OB \PhiRBtzt(\qtz)}]_{|\vB} = \tvphiS(\QRB) )
,\cr
}\right.
\ee
Thus $\tvphiS$ is a time-shift, which is also abusively noted $\PhiStzt=I$ (algebraic identity).
So with $\Thetat(\vyS) = \vyDt$ we get $\Thetat(\PhiDtzt(\vyS)) = \PhiDtzt(\vyDtz)$,
with $\vyDtz = \Thetatz(\vyS)$, thus
\be
\label{eqeor2b0}
\boxed{\Thetat \circ \PhiStzt = \PhiDtzt \circ\Thetatz} : \MB\rar\MA 
\ee
(also abusively written $\Thetat = \PhiDtzt \circ\Thetatz$). 
In other words, the following diagram commutes:
\be
\label{psietdgv}
\xymatrix{
           &  \vyS = \tvphiS(\QRB) \ar[dd]^{\ts \Thetatz} \ar[r]_{\ts \PhiStzt = \hbox{time shift} }
					        & \vyS = \tvphiS(\QRB) \ar@<1ex>[dd]_{\ts\Thetat}  
\\
\QRB \in \ObjRB \ar[ur]^{\ts \tvphiS\quad}  \ar[dr]^{\ts \PhiDtz\quad} 
\\
					 &\vyDtz= \tvphiDtz(\QRB) = \Thetatz(\vyS)\ar[r]^{\ts\PhiDtzt\qquad\;\;}
                      & \vyDt = \tvphiDt(\QRB) = \PhiDtzt(\vyDtz) = \Thetat(\vyS).
}
\ee
\def\vuS{\vu_{\!S}}
And \eref{eqeor2b0} gives, for any $\vyS=\tvphiS(\QRB)$ and all vector field $\vuS$ (static in~$\calRB$),
with $\vyDtz = \Thetatz(\vyS)$,
\be
\label{eqeor2bd}
\underbrace{d\Thetat(\vyS)}_{\hbox{\footnotesize (translation at t)}}.
\underbrace{d\PhiStzt(\vyS).\vuS(\vyS)}_{\hbox{\footnotesize (time shift from $\tz$ to t)}}
= 
\underbrace{d\PhiDtzt(\vyDtz)}_{\hbox{\footnotesize (Drive motion from $\tz$ to t)}} .
\underbrace{d\Thetatz(\vyS).\vuS(\vyS)}_{\hbox{\footnotesize (translation at $\tz$)}} .
\ee
\finrep
\finexe

%%%%%%%%%%%%%%%%%%%%%%%%%%%%%%%%%%%%%%%%%%%%%%%%%%%%%%%%%%%%%%%%%%%%%%%%%%%%%%%%%%%

%%%%%%%%%%%%%%%%%%%%%%%%%%%%%%%%%%%%%%%%%%%%%%%%%%%%%%%%%%%%%%%%%%%%%%%%%%%%%%%%%%%

\subsection{Drive and Coriolis forces}
\label{secfc}

%%%%%%%%%%%%%%%%%%%%%%%%%%%%%%%%%%%%%%%%%%%%%%%%%%%%%%%%%%%%%%%%%%%%%%%%%%%%%%%%%%%

\subsubsection{Fundamental principal: requires a Galilean referential}

Second Newton's law of motion (fundamental principle of dynamics):
In a Galilean referential,
% \ie\ you quantify vectors by giving their components relative to the Galilean Cartesian basis, 
the sum of the external forces~$\vf$
on an object is equal to its mass multiplied by its acceleration:
\be
\label{eqFP2}
\sum\hbox{external}\vf = m\vgamma \quad\hbox{(in a Galilean referential)}.
\ee

Question: And in a Non Galilean referential?

Answer: Then you have to add ``observer dependent forces'',
\ie\ you have to add ``apparent forces'' due to the motion of the non Galilean observer.
Indeed, the motion of an object in our Universe does not care about the observer motion (his accelerations and velocities).

%\Eg, your referential $\calRB$ is fixed on a carousel (a spinning merry-go-round):
See \eg\
\verb+https://www.youtube.com/watch?v=_36MiCUS1ro+ for a carousel (a merry-go-round),

See \eg\
\verb+https://www.youtube.com/watch?v=aeY9tY9vKgs+ for tornadoes.
%t of Atmospheric Science at the University of Illinois Urbana-Champaign.

%%%%%%%%%%%%%%%%%%%%%%%%%%%%%%%%%%%%%%%%%%%%%%%%%%%%%%%%%%%%%%%%%%%%%%%%%%%%%%%%%%%

\subsubsection{Drive + Coriolis forces = the inertial force} %, and Fundamental Principle}
\label{secicf}

\def\vfBD{{\vec f_{\!B\!D}}}
\def\vfBDt{{\vec f_{\!B\!Dt}}}
\def\vfBC{{\vec f_{\!B\!C}}}
\def\vfBCt{{\vec f_{\!B\!Ct}}}

Consider
%a motion $\tPhi$ of~$\Obj$, a particle $\Pobj\in\Obj$, $t\in\RR$, $\pt=\tPhi(t,\Pobj)$, $\vv(t,\pt)={\pa\tPhi\over \pa t}(t,\Pobj)$, $\vgamma(t,\pt)={\pa^2\tPhi\over \pa t^2}(t,\Pobj)$, and
 $\vf(t,\pt)=$ the sum of the external forces acting on~$\Pobj$ at~$t$ at $\pt=\tPhi(t,\Pobj)$.
%$\calRA$ is a Galilean referential, $\calRB$ is a non Galilean referential, $\Theta$ is the translator from B to~A, \cf~\eref{eqdefThetat}.

In a Galilean referential $\calRA$, Newton laws~\eref{eqFP2} means
\be
\label{eqNLA}
[\vf_t(\pt)]_{|\vA} = m\, [\vgamma_t(\pt)]_{|\vA}, \qwritten \boxed{\vfAt(\vxAt)= m\, \vgammaAt(\vxAt)} \;\in\Mnu,
\ee
with $\vxAt:=[\ora{\OA \pt}]_{|\vA}$, $\vfAt(\vxAt):=[\vf_t(\pt)]_{|\vA}$ and $\vgammaAt(\vxAt)=[\vgamma_t(\pt)]_{|\vA}$.
With $\vxAt = \Thetat(\vxBt)$, the acceleration addition formula gives
$\vfAt(\vxAt)= m(d\Thetat.\vgammaB(\vxBt) +  \vgammaDt(\vxAt) +  \vgammaCt(\vxAt))\in\calRA$, thus, in~$\calRB$,
\be
\label{eqNLA2}
\underbrace{d\Thetat^{-1}.\vfAt(\vxAt)}_{\vfAt{}^*(\vxBt)=\vfBt(\vxBt)}
= m\, \vgammaB(\vxBt)
+ \underbrace{ m\, d\Thetat^{-1}.\vgamma_{Dt}(\vxAt)}_{m\, \vgammaDt{}^*(\vxBt)}
+ \underbrace{ m\, d\Thetat^{-1}.\vgamma_{Ct}(\vxAt)}_{m\, \vgammaCt{}^*(\vxBt)},
\ee
and $d\Thetat^{-1}.[\vf_t(\pt)]_{|\vA}=d\Thetat^{-1}.\vfAt(\vxAt) \equalref{eqvyDtnb0} [\vf_t(\pt)]_{|\vB} \eqnote \vfBt(\vxBt)$ is the external forces as quantified by~B at~$t$, \cf~\eref{eqvyDtnb0} (with $\Thetat$ supposed to be affine).
And with the pull-back notation, \cf~\eref{eqvyDtnb0}:

%\ie\ $\vfBt(\vxBt) := \vfAt{}^*(\vxBt)$ is the pull-back of $\vfAt$ by~$\Thetat$ (translation for~B).

\debdef
At $t$ on~$\pt$, define %the following matrices:
\be
\label{eqpfdunrr2}
\eqalign{
\bullet&\hbox{The drive force }  \hfill  \vfBDt(\vxBt) := -m\, d\Thetat^{-1}.\vgamma_{Dt}(\vxAt) 
\quad (= - m\, \vgammaDt{}^*(\vxBt)). \cr
\bullet&\hbox{The Coriolis force } \hfill  \vfBCt(\vxBt) := -m\, d\Thetat^{-1}.\vgamma_{Ct}(\vxAt)
\quad (= - m\, \vgammaCt{}^*(\vxBt)). \cr
\bullet&(\hbox{The inertial, or fictitious, force}  := \vfBDt(\vxBt) + \vfBCt(\vxBt) = -m\, d\Thetat^{-1}.(\vgamma_{Dt}+\vgamma_{Ct})(\vxAt).)
}
\ee
\findef

Then \eref{eqNLA2} gives the fundamental principle quantified in~$\calRB$ (non Galilean referential):
\be
\boxed{\vfBt(\vxBt) + \vfBDt(\vxBt) + \vfBCt(\vxBt) = m\, \vgammaB(\vxBt)},
\ee
\ie, at~$t$, in~$\calRB$: The external force $+$ the Drive and Coriolis forces $=$ $m$ times the acceleration.

\comment{
Application: To compute a trajectory in~$\calRB$ of a particle $\Pobj$, we have to integrate twice
($\vgammaBt(t,\vxB(t))=$) $\vxB{}''(t)
= \vfB(t,\vxB(t)) + \vfBD(t,\vxB(t)) + \vfBC(t,\vxB(t))$
(without $\vfBD$ and $\vfBC$ we get erroneous results).
}

%%%%%%%%%%%%%%%%%%%%%%%%%%%%%%%%%%%%%%%%%%%%%%%%%%%%%%%%%%%%%%%%%%%%%%%%%%%%%%%%%%%

\subsection{Summary for ``Sun and Earth''}
\label{secsummaryC}

Illustation with a simplified (circular) motion of the Earth around the Sun. %, seen as a sphere, the motion of its center being circular around the Sun.
\comment{
Illustration: Observer~B is a french observer who uses a non Galilean referential and the metre,
and A is an English observer who uses a Galilean referential and the foot (it is A Galilean who computes the Coriolis acceleration due to the motion of the non Galilean~B).
}

%%%%%%%%%%%%%%%%%%%%%%%%%%%%%%%%%%%%%%%%%%%%%%%%%%%%%%%%%%%%%%%%%%%%%%%%%%%%%%%%%%%

\subsubsection{Coriolis forces on the Earth}

\def\QBi{{Q_{\!Bi}}}
\def\QBu{{Q_{\!B1}}}
\def\QBd{{Q_{\!B2}}}
\def\QBt{{Q_{\!B3}}}
\def\qut{{q_{t1}}}
\def\qdt{{q_{t2}}}
\def\qtt{{q_{t3}}}
\def\vButs{{\vB_{1t*}}}
\def\vBdts{{\vB_{2t*}}}
\def\vBtts{{\vB_{3t*}}}
\def\RA{R_{\!A}}
\def\RB{R_{\!B}}
\def\calRBAt{\calR_{\!B\!At}}
\def\vyDp{\vec y_{\!D}^{}\!'}
\def\vyDpp{\vec y_{\!D}^{}\!''}
\def\vyits{{\vec y_{it*}}}
\def\vzD{\vec z_{\!\phi}}

\begin{enumerate}[leftmargin=0pt, rightmargin=0pt, itemsep=0pt, label*=\arabic*.]

\item {\bf Referentials.}

\begin{enumerate}[leftmargin=0pt, rightmargin=0pt, itemsep=0pt, label*=\arabic*.]
\item Relative referential $\calRB=(\OB,(\vB_1,\vB_2,\vB_3))$ chosen by the observer~B fixed on the Earth,
where $\OBt = \tPhiRB(t,\QRBz)$ is the position of the particle $\QRBz$ at the center of the Earth,
written $\OB$ by~B (fixed for~B), and
$(\vB_{1t},\vB_{2t},\vB_{3t})$ is a Euclidean basis (\eg\ built with the metre) fixed in the Earth,
written $(\vB_1,\vB_2,\vB_3)$ by~B (fixed for~B),
with $\vB_3$ chosen to be along the rotation axis of the Earth and oriented from the south pole to the north pole;
And $\dd_B$ is the associated Euclidean dot product.
So, a fixed particle $\QRB$ in the Earth at 
longitude $\theta_\QRB\in]-\pi,\pi]$ and latitude $\phi_\QRB\in[-{\pi\over 2},{\pi\over 2}]$
is referenced by observer~B as the matrix
$
\vyS
=\tvphiS(\QRB)=[\ora{\OB \tPhiRB(t,\QRB)}]_{|\vB}
=\RB\pmatrix{
\cos(\theta_{\!\QRB}) \cos(\phi_{\!\QRB}) \cr 
\sin(\theta_{\!\QRB}) \cos(\phi_{\!\QRB}) \cr 
\sin(\phi_{\!\QRB}) \hfill}
$ where $\RB=||\ora{\OB \tPhiRB(t,\QRB)}||_B$ is the distance between $\QRBz$ and~$\QRB$
(\eg\ if $\QRB$ is on the surface of the Earth then $\RB\simeq 6371$ km).

\item Initial Galilean referential
$\calRA_0=(O_{\!A0},(\vA_1,\vA_2,\vA_3))$: $O_{\!A0}$ is at the center of the Sun and $(\vA_1,\vA_2,\vA_3)$ is a Euclidean basis (\eg\ built with the foot) fixed relative to the stars, such that $\vA_3=\mu\vB_3$ with $\mu>0$ (\eg\ $\mu=0.3048$ and $\lambda={1\over\mu}\simeq 3.28$); And $\dd_A$ is the associated Euclidean dot product.

\item Deduced absolute Galilean referential $\calRA= (\OAt,(\vA_1,\vA_2,\vA_3))$ chosen by observer~A fixed on Earth,
where $\OAt=\OBt$, written $\OA$ by~A  (fixed for~A).
%(It is quantified as the matrix motion $t \rar [\ora{O_{A0} \OB(t)}]_{|\vA}$ in~$\calRA$.)
Since it takes more that 365 days for $\QRBz$ to complete a rotation around the Sun, the motion of~$\QRBz$ will be considered to be rectilinear at constant velocity ``in a short interval of time''
sufficient for the computation of the Coriolis acceleration with ``sufficient accuracy'' (simplifies the calculations).
%Thus $\calRA=(\OA,(\vA_1,\vA_2,\vA_3))$ is a (approximated) Galilean referential.

\quad (If~A prefers to work with the initial Galilean referential $\calRA_0$, then the absolute matrix motion $\tvphiA(t,\Pobj) = [\ora{\OA\tPhi(t,\Pobj)}]_{|\vA}$ has to be replaced by $\tvphiA(t,\Pobj) = [\ora{O_{\!A0} \OB(t)}]_{|\vA} + [\ora{\OB(t)\tPhi(t,\Pobj)}]_{|\vA}$, idem for the drive motion~$\tvphiD$.)
\end{enumerate}

%%%%%%%%%%%%%%%%%%%%%%%%%%

\item {\bf Drive motion.}
\begin{enumerate}[leftmargin=0pt, rightmargin=0pt, itemsep=0pt, label*=\arabic*.]

\item The motion $t\rar \qt=\tPhiRB(t,\QRB)$ of a particle $\QRB$ in the Earth is stored by~A as the drive motion~$\tvphiD$ given by (matrix valued), with $\omega$ the angular velocity of the Earth in~$\calRA$,
\be
\label{eqqDE}
\vyD(t) = \tvphiD(t,\QRB)
=\RA(\QRB)\pmatrix{
\cos(\omega t) \cos\phi_{\!\QRB} \cr 
\sin(\omega t) \cos\phi_{\!\QRB} \cr 
\sin\phi_{\!\QRB} \hfill}
= [\ora{\OA q(t)}]_{|\vA}
=\pmatrix{y_{D1}(t) \cr y_{D2}(t) \cr y_{D3}}
,
\ee
where $\RA(\QRB)=||\ora{\QRBz\QRB}||_{|\vA}$ is the distance between $\QRBz$ and~$\QRB$ for~A (\eg\ $\RA\simeq 20902231$~foot if $\QRB$ is on the surface of the Earth).
(And $(\omega t)$ by replaced by $(\alpha_0{+}\omega (t{-}\tz))$ to be more general.)

\item Drive velocity: With $\vomegaD := \omega\vA_3$,
\be
\label{eqDvE}
\vvD(t,\vyD(t)) = \vyDp(t)
=\omega \RA\pmatrix{
-\sin(\omega t) \cos\phi_\QRB \cr % = -\omega y(t)
\cos(\omega t) \cos\phi_\QRB\hfill\cr  % =\omega x(t)
0\hfill} 
= \omega\pmatrix{- y_2(t) \cr  y_1(t)\cr0}
= \omega\pmatrix{0 & -1 &0 \cr 1 & 0 & 0 \cr 0 &0 &0}.\vyD(t)
= \vomegaD \wedge \vyD(t).
\ee

\item Drive acceleration:
\be
\label{eqdaE}
\vgammaD(t,\vyDt) = \vyDpp(t) = \vomegaD \wedge \vyDp(t) 
= \vomegaD \wedge \vvD(t,\vyDt)
= \vomegaD \wedge (\vomegaD \wedge \vyD(t))
= -\omega^2 \pmatrix{y_{D1}(t) \cr y_{D2}(t) \cr 0}
\ee
= the usual centrifugal acceleration (in a plane parallel to the equatorial plane, drawing).

\item Differential of the drive velocity (time and space independent here):
\eref{eqDvE} gives
\be
\label{eqdvD0}
d\vvD(t,\vyDt) = d\vvD =\pmatrix{0 & -\omega &0 \cr \omega & 0 & 0 \cr 0 &0 &0}= \vomegaD \wedge.
\ee
\end{enumerate}

%%%%%%%%%%%%%%%%%%%%%%%%%%

\item {\bf Translator.}
\begin{enumerate}[leftmargin=0pt, rightmargin=0pt, itemsep=0pt, label*=\arabic*.]

\item
Here $\OAt=\OBt$, thus $\Thetat(\vec0)=\vec0$ (with $[\vec0]\eqnote \vec0=$ the null matrix), \cf~\eref{eqdefThetatv0}. % by definition of~$\OA$ and~$\OB$.

\item Calculation of~$d\Thetat$.
With $\Thetat$ affine, $d\Thetat.[\vBit]_{|\vB} = [\vBit]_{|\vA}$.
Thus $\vB_3=\lambda \vA_3$ (hypothesis) and $d\Thetat.[\vB_3]_{|\vB} = [\vB_{3t}]_{|\vA}$ give
$d\Thetat.\vE_3 = \lambda\vE_3$ where $(\vE_i)$ is the canonical basis in~$\Mnu$.
Then let $\QBi\in\ObjRB$ be the Earth particle which position $q_{ti}=\tPhiRB(t,\QBi)$ makes $\vBit:=\ora{\OBt q_{ti}}$.
%, \ie\ $d\Thetat.\vE_i = \vEits$ where $(\vE_i)$ is the canonical basis in~$\Mnu$.
\comment{
The particle $\QBt$ is referenced by~B as $[\ora{\OB \qtt}]_{|\vB}=[\vB_3]_{|\vB}$, 
and $\vB_3=\lambda \vA_3$ (hypothesis), thus $[\vBit]_{|\vA}=d\Thetat.[\vBit]_{|\vB}$
gives 
$[\ora{\OA\qtt}]_{|\vA} = [\vB_3]_{|\vA}
=\lambda\pmatrix{
0 \cr 
0 \cr 
1 \hfill}
$ (hypothesis $\vA_3$ and~$\vB_3$ are aligned).
}
So, $\vB_1$ and $\vB_2$ being in the equatorial plane, \eref{eqqDE} gives
$d\Thetat.\vE_1 = d\Thetat.[\vB_1]_{|\vB} = [\vB_1]_{|\vA} = [\ora{\OA\qut}]_{|\vA}
=\lambda\pmatrix{\cos(\omega t) \cr \sin(\omega t) \cr 0}
$,
and
$d\Thetat.\vE_2 = d\Thetat.[\vB_2]_{|\vB} = [\vB_2]_{|\vA} = [\ora{\OA\qdt}]_{|\vA}
=\lambda\pmatrix{-\sin(\omega t) \cr \cos(\omega t) \cr 0}
$.
Thus $[d\Thetat]_{|\vE}=\lambda \pmatrix{
\cos(\omega t) & -\sin(\omega t) & 0 \cr
\sin(\omega t) & \cos(\omega t) & 0 \cr
0 & 0 & 1
}
=
$ the expected rotation matrix expanded by~$\lambda$ (change of unit of measurement).
%In particular A knows $||\vBit||_A = \lambda$, \ie\ knows the unit of measurement used by~B, \eg\ $\lambda={1\over 0.3048}\simeq 3.28$ if B uses the metre and A the foot.

\item Calculation of $\Thetat$ (affine): $\Thetat(\vyS)=\Thetat(\vec0) + d\Thetat.\vyS$,
so, with $\OAt=\OBt$ here,
\be
\label{eqdaE0}
\vyDt := \Thetat(\vyS)= d\Thetat.\vyS
\ee

\end{enumerate}

\item {\bf Motions of~$\Obj$.}
\begin{enumerate}[leftmargin=0pt, rightmargin=0pt, itemsep=0pt, label*=\arabic*.]

\item B quantifies the motion $\tPhi$ of~$\Obj$, \ie\ he stores the relative motion $\tvphiB$ of~$\Obj$, and the relative velocities and accelerations $\vvBt$ and~$\vgammaB$ (matrices), \cf~\eref{eqtPhiB}-\eref{eqvvB}.

\item 
Translations for~A: With $\vxAt=\Thetat(\vxBt)$,
%Then, thanks to~$\Thetat$, A~gets, at~$t$, the quantified translated velocity and acceleration $\vvBts$ and $\vgammaBts$ in~$\calRA$: 
\be
\vvBts(\vxAt) = d\Thetat(\vxBt).\vvBt(\vxBt) \qand \vgammaBts(\vxAt) = d\Thetat(\vxBt).\vgammaBt(\vxBt).
\ee

\end{enumerate}

\item {\bf Drive force}  (apparent force in~$\calRB$ due to the motion of~B):
\be
\vf_{\!B\!Dt}(\vxBt)
=-m\,d\Thetat^{-1}.\vgammaDt(\vxAt)
\equalref{eqdaE} \lambda m  \omega^2 d\Thetat^{-1}. \pmatrix{x_{A1}(t) \cr x_{A2}(t) \cr 0}
\equalref{eqdaE0} \lambda m  \omega^2\pmatrix{x_{B1}(t) \cr x_{B2}(t) \cr 0},
\ee
centrifugal force (in a ``parallel plane'' at latitude of~$\Pobj$).

\item {\bf Coriolis acceleration} (apparent acceleration due to the motion of~B):
\be
\label{eqexaCa}
\vgammaCt(\vxAt) = 2\,d\vvDt . (d\Thetat.\vvBt(\vxBt)) = 2\, d\Thetat.d\vvDt.\vvBt(\vxBt)
 %= 2\vomegaD \wedge \vvBts(\vxAt) . %,
\ee
because $d\Thetat$ %(rotation dilated) 
commutes with $d\vvDt$ (composition of ``rotations along the same south-north axis''
which reads as $e^{i\omega t}.e^{i{\pi\over 2}} = e^{i{\pi\over 2}}e^{i\omega t}= e^{i({\pi\over 2}+\omega t)}$ in the equatorial plane). 

\item {\bf Coriolis force} (apparent force due to the motion of~B):
\be
\vf_{BCt}(\vxBt)=-m\,d\Thetat^{-1}.\vgammaCt(\vxAt) = -2m\,d\vvDt .\vvBt(\vxBt) = -2m \vomega\wedge \vvBt(\vxBt).
\ee
\end{enumerate}

%\item Drive force and Coriolis force needed by~B. 
%(Drive and Coriolis acceleration have to be calculated in a Galilean referential~$\calRA$, from which we deduce the drive and Coriolis forces in the referential~$\calRB$ needed for Newton's law in a non Galilean referential, see~\eref{eqnlnonGR}.)
%ith $\vxBt=\Thetat^{-1}(\vxAt)$ for a particle $\Pobj\in\Obj$: 

%See \eg: \\ villemin.gerard.free.fr/Scienmod/Coriolis.htm, \\ planet-terre.ens-lyon.fr/article/force-de-coriolis.xml.

%%%%%%%%%%%%%%%%%%%%%%%%%%%%%%%%%%%%%%%%%%%%%%%%%%%%%%%%%%%%%%%%%%%%%%%%%%%%%%%%%%%
%%%%%%%%%%%%%%%%%%%%%%%%%%%%%%%%%%%%%%%%%%%%%%%%%%%%%%%%%%%%%%%%%%%%%%%%%%%%%%%%%%%
\newpage
\section{Objectivities}
\label{seco}

%\def\Laquant{La quantité est objective covariante ssi, pour tout $t$, on~a }
%The ``physical quantity'' is objective covariante iff, for all~$t$

%Framework of~\S~\ref{seccdr} (classical mechanics: The time scale is the same for all users).

Goal: To give an objective expression of the laws of mechanics; As Maxwell~\cite{maxwell} said:
``The formula at which we arrive must be such that a person of any nation, by substituting for the different symbols the numerical value of the quantities as measured by his own national units, would arrive at a true result''.

Generic notation: if a function $z$ is given as $z(t,x)$, then $z_t(x):=z(t,x)$, and conversely.

%Remark: The isometric objectivity is a restricted version of the covariant objectivity: Isometric objectivity only deals with rigid body motions between observers, when covariant objectivity allows any regular motions between observers.

%Remark: The covariant objectivity implies the  (restricted version when only rigid body motions between observers are considered).

%%%%%%%%%%%%%%%%%%%%%%%%%%%%%%%%%%%%%%%%%%%%%%%%%%%%%%%%%%%%%%%%%%%%%%%%%%%%%%%%%%%

\subsection{``Isometric objectivity'' and ``Frame Invariance Principle''}
\label{secoi}

This manuscript is not intended to describe ``isometric objectivity'':

``Isometric objectivity'' is the framework in which the ``principle of material frame-indifference''
(``frame invariance principle'') is settled, principle which states that
``Rigid body motions should not affect the stress constitutive law of a material''.
\Eg, Truesdell--Noll~\cite{truesdell-noll} p.~41:
\begin{center}
<< Constitutive equations must be invariant under changes of frame of reference. >>
\end{center}
Or Germain~\cite{germain2} :
\begin{center}
<< {\sc Axiom of power of internal forces.} The virtual power of the "internal forces" acting on a system S for a given virtual motion is an objective quantity; i.e., it has the same value whatever be the frame in which the motion is observed. >>
\end{center}

\noindent
{\bf NB:} Both of these affirmations are limited to ``isometric changes of frame''
(the same metric for all), as Truesdell--Noll~\cite{truesdell-noll} \hbox{page 42-43} explain:
The ``isometric objectivity'' 
concern one observer who defines his Euclidean dot product
and consider only orthonormal change of bases to validate a constitutive law.

If you want to interpret ``isometric objectivity'' in the ``covariant objectivity'' framework,
then ``isometric objectivity'' corresponds to a dictatorial management:
One observer with his Euclidean referential (\eg\ based on the English \foot),
imposes his unit of length to all other users (isometry hypothesis).
%So only changes of orthonormal bases are allowed.
%This is a ``subjective'' approach: 
%\Eg, an English observer (\foot) and an French observer (\metre) can't work together in this framework (they don't use the same metric).
(Note: The metre was not adopted by the scientific community until after 1875.)

Moreover, isometric objectivity leads to despise the difference between covariance and contravariance,
due to the uncontrolled use of the Riesz representation theorem.
%see \eg~\S~\ref{remisonat} and~\S~\ref{secEEsnonnat};
%And leads to the misunderstanding of the Lie derivative of order 2 tensors (Jaumann, Truesdell, Oldroyd, lower-Maxwell, upper-Maxwell...).

\debrem
Marsden and Hughes~\cite{marsden-hughes} p.~8 use this isometric framework to begin with.
But, pages~22 and~163, they write that a ``good modelization'' has to be ``covariant objective'' (observer independent) to begin with;
And they propose a covariant modelization for elasticity at~\S~3.3.
%, without mixing linear forms (``covariant vectors'') and vectors (``contravariant vectors'').
%(They use metrics but pay attention to the use of the Riesz representation theorem.)
\finrem

%%%%%%%%%%%%%%%%%%%%%%%%%%%%%%%%%%%%%%%%%%%%%%%%%%%%%%%%%%%%%%%%%%%%%%%%%%%%%%%%%%%

\subsection{Definition and characterization of the covariant objectivity}

%%%%%%%%%%%%%%%%%%%%%%%%%%%%%%%%%%%%%%%%%%%%%%%%%%%%%%%%%%%%%%%%%%%%%%%%%%%%%%%%%%%

\subsubsection{Framework of classical mechanics}

Framework of classical mechanics to simplify.
Consider two observers A and B and their referentials $\calRA=(\OA,(\vA_i))$ and $\calRB=(\OB,(\vB_i))$.
\Eg, $(\vA_i)$ and $(\vB_i)$ are Euclidean bases in foot and metre, $\dd_A$ and~$\dd_B$ is their associated Euclidean dot products.
And $\Theta$ is the translator, \cf~\eref{eqdefThetat0}.

Consider a regular motion~$\tPhi$ of an object~$\Obj$,  $\pt=\tPhi(t,\Pobj) \in \RRn$ the position at~$t$ of a particle in our Universe, $\Omegat = \tPhi(t,\Obj)$ the configuration at~$t$, and $\bigC = \bigcup_{t\in[a,b]}(\{t\}\times \Omegat)$ the set of configurations.
And $\vxAt:=[\ora{\OA \pt}]_{|\vA}\in\MA$ and $\vxBt:=[\ora{\OB \pt}]_{|\vB}\in\MB$ are the stored components
of~$\pt$ relative to the chosen referentials, $\MA$ and $\MB$ being the spaces of $n*1$ matrices as referred to by  $A$ and~$B$.

%%%%%%%%%%%%%%%%%%%%%%%%%%%%%%%%%%%%%%%%%%%%%%%%%%%%%%%%%%%%%%%%%%%%%%%%%%%%%%%%%%%

\subsubsection{Covariant objectivity of a scalar function}

Let $f:
\left\{\eqalign{
\bigC & \rar\RR \cr
(t,\pt) &\rar f(t,\pt) \cr
}\right\}
$ be a Eulerian scalar function (\eg, a temperature field). %Let $\vxAt:=[\ora{\OA \pt}]_{|\vA}\in\MA$ and $\vxBt:=[\ora{\OB \pt}]_{|\vB}\in\MB$.
$f$ is quantified by A and~B as the functions $\fA:
\left\{\eqalign{
\RR{\times} \MA & \rar\RR \cr
(t,\vxAt) &\rar \fA(t,\vxAt):=f(t,\pt) \cr
}\right\}
$
and
$\fB:
\left\{\eqalign{
\RR{\times} \MB & \rar\RR \cr
(t,\vxBt) &\rar \fB(t,\vxBt):=f(t,\pt) \cr
}\right\}
$.

\debdef
\label{defoc0}
$f$ is objective covariant iff, for all referentials $\calRA$ and~$\calRB$ and for all~$t$,
\be
\label{eqdefocv0}
\fAt(\vxAt) = \fBt(\vxBt) \qwhen \vxAt = \Thetat(\vxBt),
\ee
%\ie\ $\fA(t,\vxAt) = \fB(t,\vxBt)$ when $\vxAt = \Thetat(\vxBt)$,
\ie\ $\fAt = f_{Bt*}$ is  the push-forward %$\Thetats\fBt$ 
of $\fBt$ by $\Thetat$ \cf~\eref{eqdefpff}.
%(the value is independent of the observer).
%Or iff $\fBt = \Thetat^*\fAt =$ the pull-back of~$\fAt$ by~$\Thetat$, \cf~\eref{eqdefpff2}.
\findef

%%%%%%%%%%%%%%%%%%%%%%%%%%%%%%%%%%%%%%%%%%%%%%%%%%%%%%%%%%%%%%%%%%%%%%%%%%%%%%%%%%%

\subsubsection{Covariant objectivity of a vector field}
\label{secobjvec}

Let $\vw:
\left\{\eqalign{
\bigC & \rar\vRRn \cr
(t,\pt) &\rar \vw(t,\pt) \cr
}\right\}$ be a Eulerian vector field (\eg, a force field).
$\vw$ is quantified by A and~B as the functions
$\vwA:
\left\{\eqalign{
\RR{\times} \MA & \rar \MA \cr
(t,\vxAt) &\rar \vwA(t,\vxAt):=[\vw(t,\pt)]_{\vA} \cr
}\right\}
$
and
$\vwB:
\left\{\eqalign{
\RR{\times} \MB & \rar \MB \cr
(t,\vxBt) &\rar \vwB(t,\vxBt):=[\vw(t,\pt)]_{\vB} \cr
}\right\}
$.
So $\vwA(t,\vxAt) %=[\vw(t,\pt)]_{\vA}
$ and 
$\vwB(t,\vxBt) %=[\vw(t,\pt)]_{\vB}
$ are the column matrices of the components of~$\vw(t,\pt)$ in~$\calRA$ and~$\calRB$.

\debdef
\label{defoc1}
$\vw$ is objective covariant iff,
for all referentials $\calRA$ and~$\calRB$ and for all~$t$,
\be
\label{eqdefocv}
\vwAt(\vxAt) = d\Thetat(\vxBt).\vwBt(\vxBt) \qwhen \vxAt = \Thetat(\vxBt),
\ee
\ie\ $\vwAt = \vwBts$ is the push-forward %$\Thetats\vwBt$ 
of~$\vwBt$ by $\Thetat$ \cf~\eref{eqdefrapvEivei}.

%Or iff $\vwBt = \Thetat^*\vwAt=$ the pull-back of~$\vwAt$ by~$\Thetat$, \cf~\eref{eqrapvEiveib}.
\findef

\debexa
Fundamental counter-example: A Eulerian velocity field is not objective, \cf~\eref{eqloicv20},
because of the drive velocity $\vv_D \ne \vec 0$ in general.
Neither is a Eulerian acceleration field, \cf~\eref{eqc2}.
\finexa

\debexa
The field of gravitational forces (external forces) is objective covariant.
\comment{
Remark: 
Compare $\vg_{At} = m\vgammaAt$ in a Galilean referential, \cf~\eref{eqFP2},
and $\vg_{Bt} + \vf_{it} + \vfCt = m\vgammaBt $ in a non Galilean referential,
cf~\S~\ref{secfc}: $\vg$ is objective, so $\vg_{At}(\vxAt) = d\Thetat(\vxBt).\vg_{Bt}(\vxBt)$, but the acceleration $\vgamma$ is not objective.
}
\finexa
%(Un~champ de vecteurs qui est objectif n'est pas lié à un calcul dans référentiel.)

%%%%%%%%%%%%%%%%%%%%%%%%%%%%%%%%%%%%%%%%%%%%%%%%%%%%%%%%%%%%%%%%%%%%%%%%%%%%%%%%%%%

\subsubsection{Covariant objectivity of differential forms}

\def\alphaA{{\alpha_{A}}}
\def\alphaAt{{\alpha_{At}}}
\def\alphaB{{\alpha_{B}}}
\def\alphaBt{{\alpha_{Bt}}}

%A differential form acts on vector fields to give scalar values, hence the objectivity of a differential form is deduced from the two previous~\S, \ie\ is defined thanks to push-forwards (or pull-backs).

Let $\alpha:
\left\{\eqalign{
\bigC & \rar\RRns \cr
(t,\pt) &\rar \alpha(t,\pt) \cr
}\right\}$
be a Eulerian differential form (\eg\ a measuring device used to get the internal power).
$\alpha$ is quantified by A and~B as the functions
$\alphaA:
\left\{\eqalign{
\RR{\times} \MA & \rar \MA \cr
(t,\vxAt) &\rar \alphaA(t,\vxAt):=[\alpha(t,\pt)]_{\vA} \cr
}\right\}
$
and
$\alphaB:
\left\{\eqalign{
\RR{\times} \MB & \rar \MB \cr
(t,\vxBt) &\rar \alphaB(t,\vxBt):=[\alpha(t,\pt)]_{\vB} \cr
}\right\}
$.
So $\alphaA(t,\vxAt) %=[\vw(t,\pt)]_{\vA}
$ and 
$\alphaB(t,\vxBt) %=[\vw(t,\pt)]_{\vB}
$ are the row matrices of the components of~$\alpha(t,\pt)$ in~$\calRA$ and~$\calRB$.
%(row matrices of components used by A and~B).

%The observers $A$ and $B$ describe $\alpha$ as being the functions $\alphaA$ and $\alphaB$.

\debdef
\label{defoc1ell}
$\alpha$ is objective covariant iff,
for all referentials $\calRA$ and~$\calRB$ and for all~$t$,
\be
\label{eqdefocell}
\alphaAt(\vxAt) = \alphaBt(\vxBt) . d\Thetat(\vxBt)^{-1} \qwhen \vxAt = \Thetat(\vxBt).
\ee
\ie\ $\alphaAt = \alpha_{Bt*}$ is the push-forward %$\Thetats\alphaBt$ 
of $\alphaBt$ by $\Thetat$ \cf~\eref{eqtransfd}.
\findef

NB: \eref{eqdefocell} and~\eref{eqdefocv} are compatible:
If $\vw$ is an objective vector field and if $\alpha$ is an objective differential form,
then the scalar function $\alpha.\vw$ is objective:
\be
\alphaAt(\vxAt).\vwAt(\vxAt)
= \alphaBt(\vxBt).\vwBt(\vxBt)
\quad(=(\alpha(t,\pt).\vw(t,\pt)),
\ee
since
$
\alphaAt(\vxAt).\vwAt(\vxAt)
= (\alphaBt(\vxBt) . d\Thetat(\vxBt)^{-1}).(d\Thetat(\vxBt).\vwBt(\vxBt))
= \alphaBt(\vxBt).\vwBt(\vxBt)
$.
%Matrix expression.

%%%%%%%%%%%%%%%%%%%%%%%%%%%%%%%%%%%%%%%%%%%%%%%%%%%%%%%%%%%%%%%%%%%%%%%%%%%%%%%%%%%

\subsubsection{Covariant objectivity of tensors}
\label{secmfi}

A tensor acts on both vector fields and differential forms,
and its objectivity is deduced from the previous~\S.

So, let $T$ be a (Eulerian) tensor corresponding to a ``physical quantity''.
The observers $A$ and $B$ describe $T$ as being the functions
$T_A$ and $T_B$.

\debdef
$T$ is objective covariant iff,
for all referentials $\calRA$ and~$\calRB$ and for all~$t$,
\be
T_{At}(\vxAt) =  T_{Bt*}(\vxAt)
\ee
\ie\ $T_{At}$ is the push-forward of $T_{Bt}$ by $\Thetat$.

(Recall: 
$T_{Bt*}(\vxAt)(\alpha_1(\vxAt),...,\vw_1(\vxAt)) := T_{Bt}(\vxBt)(\alpha_1{}^*(\vxBt),...,\vw_1{}^*(\vxBt))$.)
\findef

\begin{example}[Non covariant objectivity of a differential $d\vw$] \rm
\label{exadv1}
Let $\vw$ be an objective vector field, seen as $\vwA$ by~A and $\vwB$ by~B;
So $\vwAt(\vxAt) \mope^{\eref{eqdefocv}} d\Thetat(\vxBt).\vwBt(\vxBt)$ when $\vxAt = \Thetat(\vxBt)$,
%\ie\ $\vwAt(\Thetat(\vxBt)) = d\Thetat(\vxBt).\vwBt(\vxBt)$.
thus
\be
d\vwAt(\vxAt).d\Thetat(\vxBt) =  d\Thetat(\vxBt).d\vwBt(\vxBt) + (d^2\Thetat(\vxBt).\vwBt(\vxBt)),
\ee
hence
\be
\label{eqdvqossi}
\eqalign{
d\vwAt(\vxAt)
= & d\Thetat(\vxBt).d\vwBt(\vxBt).d\Thetat(\vxBt)^{-1} + (d^2\Thetat(\vxBt).\vwBt(\vxBt)).d\Thetat(\vxBt)^{-1} \cr
\ne & d\Thetat(\vxBt).d\vwBt(\vxBt).d\Thetat(\vxBt)^{-1}\qwhen d^2\Thetat \ne 0.
}
\ee
Thus $d\vw$ is not covariant objective in general. %(but it is ``isometric objective'').
\def\vuuA{\vu_{A1}}\def\vudA{\vu_{A2}}\def\vuA{\vu_A}%
\def\vuuB{\vu_{B1}}\def\vudB{\vu_{B2}}\def\vuB{\vu_B}%
However in classical mechanics for ``change of Cartesian referentials'' $\Thetat$ is affine, so $d^2\Thetat=0$, and in particular $d\vw$ is objective when $\vw$~is.
And
\be
\label{eqpropod2}
\eqalign{
& (d^2\vwAt(\vxAt).d\Thetat(\vxBt)).d\Thetat(\vxBt) + d\vwAt(\vxAt).d^2\Thetat(\vxBt) \cr
& = d\Thetat(\vxBt).d^2\vwBt(\vxBt) + 2\,d^2\Thetat(\vxBt).d\vwBt(\vxBt) + d^3\Thetat(\vxBt).\vwBt(\vxBt). \cr
}
\ee
Thus $d^2\vw$ is not covariant objective in general (but if $\Thetat$ is affine then $d^2\vw$ is objective if $\vw$~is).
\comment{
when $\vw$~is.
Indeed here $d\vwAt(\vxAt)
= d\Thetat.d\vwBt(\vxBt).d\Thetat^{-1}
$
thus, with $\vuAt(\vxAt):=d\Thetat.\vuBt(\vxBt)$ and $\ell_{At}(\vxAt) = \ell_{Bt}(\vxBt).d\Theta^{-1}$ (objectivity),
we get $\ell_{At}(\vxAt).d\vwAt(\vxAt).\vuAt(\vxAt)
= (\ell_{Bt}(\vxBt).d\Theta^{-1}).(d\Theta.d\vwBt(\vxBt).d\Theta^{-1}).(d\Theta.\vuBt(\vxBt))
= \ell_{Bt}(\vxBt).d\vwBt(\vxBt).\vuBt(\vxBt)
$.
And the natural canonical isomorphism $\tJ:L\in\calL(E;F) \rar \tL\in\calL(F^*,E;\RR)$
gives $\tilde{d\vwAt(\vxAt)}(\ell_{At}(\vxAt),\vuAt(\vxAt))
= \tilde{d\vwBt(\vxBt)}(\ell_{Bt}(\vxBt),\vuBt(\vxBt))$, \ie\ $\tilde{d\vw_t}$ is isometric objective,
thus $d\vw_t$ is said to be isometric objective.
;
Indeed, with shortened notations, 
$(d^2\vwA.d\Theta).d\Theta = d\Theta.d^2\vwB$
gives (natural canonical isomorphism) $d^2\vwA(d\Theta.,d\Theta.) = d\Theta.d^2\vwB(.,.)$, thus
$(\ell_B.d\Theta).d^2\vwA(d\Theta.\vuuB,d\Theta.\vudB)
=(\ell_B.d\Theta).d\Theta.d^2\vwB(\vuuB,\vudB)
=\ell_B.d^2\vwB(\vuuB,\vudB)
$, thus (natural canonical isomorphism)
$d^2\vwA(\ell_A,\vuuA,\vudA)
= d^2\vwB(\ell_B,\vuuB,\vudB)
$.
}
\finexa

%%%%%%%%%%%%%%%%%%%%%%%%%%%%%%%%%%%%%%%%%%%%%%%%%%%%%%%%%%%%%%%%%%%%%%%%%%%%%%%%%%%

\subsection{Non objectivity of the velocities}

%%%%%%%%%%%%%%%%%%%%%%%%%%%%%%%%%%%%%%%%%%%%%%%%%%%%%%%%%%%%%%%%%%%%%%%%%%%%%%%%%%%

\subsubsection{Eulerian velocity~$\vv\,$: not covariant (and not isometric) objective}
\label{seccdve}

Velocity addition formala: With $\vvBts(\vxAt) = d\Thetat(\vxBt).\vw(\vxBt)$ when $\vxAt= \Thetat(\vxBt)$, \cf~\eref{eqloicv20},
\be
\eqalign{
\vvAt(\vxAt) = & \vvBts(\vxAt) + \vvDt(\vxAt) \cr
\ne & \vvBts(\vxAt) \qwhen \vvDt(\vxAt)\ne \vec 0,
}
\ee
thus a~Eulerian velocity field is not covariant objective (and not isometric objective).

%%%%%%%%%%%%%%%%%%%%%%%%%%%%%%%%%%%%%%%%%%%%%%%%%%%%%%%%%%%%%%%%%%%%%%%%%%%%%%%%%%%

\subsubsection{$d\vv$ is not objective}

The velocity addition formula, $(\vvAt - \vvDt)(\vxAt) = \vvBts(\vxAt) = d\Thetat(\vxBt).\vvBt(\vxBt)$ when $\vxAt  = \Thetat(\vxBt)$, gives 
\be
\label{eqdvva1}
d(\vvAt - \vvDt)(\vxAt).d\Thetat(\vxBt)
= d\Thetat(\vxBt).d\vvBt(\vxBt) + d^2\Thetat(\vxBt).\vvBt(\vxBt),
\ee
thus $d\vv$ is neither covariant objective nor isometric objective because of~$d\vvD$:
\be
\label{eqnoL2}
d\vvAt(\vxAt) = d\vvBts(\vxAt) + d\vvDt(\vxAt) + d^2\Thetat(\vxBt).\vvBt(\vxBt).d\Thetat(\vxBt)^{-1} \ne d\vvBts(\vxAt) \quad\hbox{in general}.
\ee
% then $d\vvAt = d\vvBts + d\vvDt$ and $d\vvDt\ne 0$ in general.

\debrem
Recall: ``Isometric objective'' implies

$\bullet$ The use of the same Euclidean metric in~$\calRB$ and~$\calRA$, \ie\ $\dd_A = \dd_B$,

$\bullet$ $\tPhiRB$ (motion of $\calRB$) is a solid body motion, and 

$\bullet$ $\Thetat$ is affine (so $d^2\Thetat=0$ for all~$t$).
%$\bullet$ $\PhiDtz$ is a solid body motion, \ie\ $(d\PhiDtzt)^T.d\PhiDtzt=I$ and $\Thetat$ is affine (so $d^2\Thetat=0$ for all~$t$).
\finrem

\debexe
\def\Fatzpatz{F_{\!\!A}}%
%Isometric setting: Both $(\vA_i)$ and~$(\vB_i)$ are orthonormal relative to the same Euclidean dot product.
Prove, with $Q_t$ the (orthonormal) transition matrix from~$(\vA_i)$ to~$(\vB_i)$:
\be
\label{eqnoL50}
[d\vv_t]_{|\vB} = Q_t.[d\vv_t]_{|\vA}. Q_t^{-1} + Q'(t). Q_t^{-1}, \qwritten
[L]_{|\vB} = Q.[L]_{|\vA}. Q^T + \overbigdot{Q}. Q^T .
\ee
(Used in classical mechanics courses, to prove that $d\vv$ isn't ``isometric objective'' because of $ \overbigdot{Q}. Q^T$.)

\debrep
$\tz,t\in\RR$, 
$\ptz = \tPhi(\tz,\Pobj)$, $\pt = \tPhi(t,\Pobj)=\Phitzt(\ptz)$, $\vv(t,\pt)= {\pa \tPhi\over \pa t}(t,\Pobj)$,
and $\Ftzt(\ptz)=d\Phitzt(\ptz)$.
So $\vv(t,\Phitzt(\ptz))= {\pa\Phitzptz \over \pa t}(t,\ptz)$, 
thus $d\vv(t,\pt).\Ftzptz(t) = {\pa \Ftzptz \over \pa t}(t)$.
%\ie\ $L(t,\pt) = \Ftzptz'(t).\Ftzptz(t)^{-1}$ 
And~\eref{eqFtztnew}, with $\Ftzptz \eqnote F$, gives $[F(t)]_{|\va_\tz,\vB} = Q(t).[F(t)]_{|\va_\tz,\vA}$,
thus $[F'(t)]_{|\va_\tz,\vB} = Q'(t).[F(t)]_{|\va_\tz,\vA} + Q(t).[F'(t)]_{|\va_\tz,\vA}$.
Thus $[d\vv(t,\pt)]_{|\vB} = [\Ftzptz'(t).\Ftzptz(t)]_{|\vB}
=[\Ftzptz'(t)]_{|\vB}.[\Ftzptz(t)]_{|\vB}
=(Q'(t).[F(t)]_{|\va_\tz,\vA} + Q(t).[F'(t)]_{|\va_\tz,\vA}).[F(t)]_{|\va_\tz,\vA}^{-1}.Q(t)^{-1}
=Q'(t).Q(t)^{-1} + Q(t).[F'(t)]_{|\va_\tz,\vA}.[F(t)]_{|\va_\tz,\vA}^{-1}.Q(t)^{-1}
=Q'(t).Q(t)^{-1} + Q(t).[d\vv(t,\pt)]_{|\vA}.Q(t)^{-1}
$. And \cf~\eref{eqdefL}-\eref{eqdefL2}.
\finrep
\finexe

\debexe
Prove that $d^2\vv$ is ``isometric objective'' when $\tPhiRB$ is a rigid body motion.
%Express $d^2\vvAt$ as a function of~$d^2\vvBt$ in the ``isometric objective'' framework.

\debrep
\eref{eqpropod2} with $\vvA-\vv_D$ instead of~$\vw_A$, and $\vv_B$ instead of~$\vw_B$
give, in an ``isometric objective'' framework,
\be
d^2(\vvAt - \vvDt)(\vxAt).(\vuBts,\vwBts)
=  d\Thetat(\vxBt).d^2\vvBt(\vxBt)(\vuB,\vwB).
\ee
Here $d^2\vvDt=0$ (rigid body motion), thus $d^2\vv$ is ``isometric objective''.
\finrep
\finexe

%%%%%%%%%%%%%%%%%%%%%%%%%%%%%%%%%%%%%%%%%%%%%%%%%%%%%%%%%%%%%%%%%%%%%%%%%%%%%%%%%%%

\subsubsection{$d\vv + d\vv^T$ is ``isometric objective''}

\debprop
If $\tPhiRB$ is a rigid body motion then $d\vv_{t} + d\vv_{t}^T$ is ``isometric objective''
\be
\label{eqdvvtisom}
d\vvAt+d\vvAt^T
= (d\vvBt + d\vvBt^T)_*.
\ee
(Isometric framework: The rate of deformation tensor is independent of an added added rigid motion.)
\finprop

\debdem
$Q.Q^T = I$ gives $\overbigdot{Q}. Q^T + (\overbigdot{Q}. Q^T)^T = 0$, then apply~\eref{eqnoL50}.
\findem

\debexe
\label{exeOmnonO}
Prove that $\Omega={d\vv-d\vv^T \over2}$ is not isometric objective.

\debrep
\eref{eqnoL2} gives $d\vvAt^T = d\vvBts^T + d\vvDt^T$, thus
${d\vvAt - d\vvAt^T \over 2} = {d\vvBts - d\vvBts^T \over 2} + {d\vvDt - d\vvDt^T \over 2}
\ne {d\vvBts - d\vvBts^T \over 2}$,
even if $\tPhiRB$ is a solid body motion (then ${d\vvDt - d\vvDt^T \over 2} = \vomega\wedge$ is a rotation time a dilation).
%, in which case ${d\vvDt - d\vvDt^T \over 2}={d\vvDt + d\vvDt \over 2} = d\vvDt \ne 0$ (except when $\vvDt$ is uniform, \ie, except for a change of Galilean referential).
\finrep
\finexe

%%%%%%%%%%%%%%%%%%%%%%%%%%%%%%%%%%%%%%%%%%%%%%%%%%%%%%%%%%%%%%%%%%%%%%%%%%%%%%%%%%%

\subsubsection{Lagrangian velocities}
\label{seccdvl}

The Lagrangian velocities do not define a vector field, \cf~\S~\ref{secbrtpt}.
Thus asking about the objectivity of Lagrangian velocities is meaningless.

%%%%%%%%%%%%%%%%%%%%%%%%%%%%%%%%%%%%%%%%%%%%%%%%%%%%%%%%%%%%%%%%%%%%%%%%%%%%%%%%%%%

\subsection{The Lie derivatives are covariant objective}
\label{secodl}

Framework of~\S~\ref{seccdr}. In particular we have the velocity-addition formula $\vvAt = \vvBts + \vvDt$ in~$\calRA$ where $\vvBts(\vxAt)=d\Thetat(\vxBt).\vvBt(\vxBt)$ and $\vxBt=\Thetat(\vxAt)$, \cf~\eref{eqloicv20}.

The objectivity under concern is the covariant objectivity
(no inner dot product or basis required).
The Lie derivatives are also called ``objective rates'' because they are covariant objectives.
Easy proofs.

%%%%%%%%%%%%%%%%%%%%%%%%%%%%%%%%%%%%%%%%%%%%%%%%%%%%%%%%%%%%%%%%%%%%%%%%%%%%%%%%%%%

\subsubsection{Scalar functions}

\debprop
If $f$ be a covariant objective function, \cf~\eref{eqdefocv0},
then its Lie derivative $\calL_\vv f$ is covariant objective:
\be
\label{eqdol0f1}
\calL_{\vvA} \fA = \Theta_{*}(\calL_{\vvB} \fB), 
\qie
\calL_{\vvA} \fA(t,\vxAt) = \calL_{\vvB} \fB(t,\vxBt) \qwhen \vxAt = \Thetat(\vxBt),
\ee
\ie, ${D \fA \over Dt}(t,\vxAt) = {D \fB \over Dt}(t,\vxBt)$, \ie\
$({\pa \fA \over \pa t}+d\fA.\vvA)(t,\vxAt) = ({\pa  \fB \over \pa t}+d\fB.\vvB)(t,\vxBt)$.
\finprop

\debdem
Consider the motion $t \rar p(t)=\tPhi(t\Pobj)$ of a particle~$\Pobj$,
and $\vxA(t)=[\ora{\OA p(t)}]_{|\vA}$ and $\vxB(t)=[\ora{\OB p(t)}]_{|\vB}$.
With $f$ objective, \eref{eqdefocv0} gives  $\fB(t,\vxB(t)) = \fA(t,\Theta(t,\vxB(t)))$ ($ = \fA(t,\vxA(t))$),
thus
\be
\eqalign{
{D\fB \over D t}(t,\vxB(t))
= &{\pa \fA \over \pa t}(t,\vxA(t)) + d\fAt(\vxA(t)).
(
\underbrace{ {\pa \Theta \over \pa t}(t,\vxB(t))}_{\vvDt(\vxAt)}
+ 
\underbrace{ d\Thetat(\vxB(t)).\vvBt(\vxB(t)))}_{\vvBts(\vxAt)} \cr
= &{\pa \fA \over \pa t}(t,\vxAt) + d\fAt(\vxAt).\vvAt(\vxAt)
= {D \fA \over D t}(t,\vxAt),
}
\ee
thanks to velocity addiction formula $\vvAt = \vvBts + \vvDt$. 
\findem

%%%%%%%%%%%%%%%%%%%%%%%%%%%%%%%%%%%%%%%%%%%%%%%%%%%%%%%%%%%%%%%%%%%%%%%%%%%%%%%%%%%

\subsubsection{Vector fields}

\debprop
Let $\vw$ be a covariant objective vector field, \cf~\eref{eqdefocv}.
Then its Lie derivative $\calL_\vv \vw$ is covariant objective:
\be
%\label{eqlieo}
\calL_{\vvA}\vw_A = \Theta_{*}(\calL_{\vvB}\vwB),
\ee
\ie, when $\vxAt = \Thetat(\vxBt)$,
\be
\label{eqlieo}
\calL_{\vvA}\vw_A(t,\vxAt) = d\Thetat(\vxBt).\calL_{\vvB}\vwB(t,\vxBt),
\ee
\ie,
\be
({D\vw_A \over Dt} - d\vvA.\vw_A)(t,\vxAt)
= d\Theta(t,\vxBt).({D\vwB \over Dt} - d\vvB.\vwB)(t,\vxBt),
\ee
\ie,
\be
({\pa\vw_A \over \pa t} + d\vw_A.\vvA - d\vvA.\vw_A)(t,\vxAt)
= d\Theta(t,\vxBt).({\pa\vwB \over \pa t} + d\vwB.\vvB - d\vvB.\vwB)(t,\vxBt).
\ee

But the partial, convected, material, and Lie autonomous derivatives are not covariant objective
(not even isometric objective because of the drive velocity~$\vvD$): We have
\be
\label{eqlieo1}
(d\vwAt.(\vvAt {-} \vvDt))(\vxAt)
=  (d\Thetat.(d\vwBt.\vvBt) + (d^2\Thetat.\vwBt).\vvBt)(\vxBt),
\ee
\be
\label{eqlieo4}
(d(\vvAt {-} \vvDt).\vwAt)(\vxAt)
=  (d\Thetat.(d\vvBt.\vwBt) +  (d^2\Thetat.\vvBt).\vwBt)(\vxBt),
\ee
\be
\label{eqlieo6}
(d(\vvAt {-} \vvDt).(\vvAt {-} \vvDt))(\vxAt)
=  (d\Thetat.(d\vvBt.\vvBt) +  d^2\Thetat(\vvBt,\vvBt))(\vxBt),
\ee
\be
\label{eqlieo10}
\calL^0_{(\vvAt-\vvDt)}\vwAt(\vxAt)
= d\Thetat(\vxBt).\calL^0_{\vvBt}\vwBt(\vxBt) ,
\ee
\comment{
\be
(d\vwAt.\vvAt - d\vvAt.\vwAt) (\vxAt)
= d\Thetat(\vxBt).(d\vwBt.\vvBt - d\vvBt.\vwBt)(\vxBt) 
+ (d\vwAt.\vvDt - d\vvDt.\vwAt) (\vxAt).
\ee
}
\be
\label{eqlieo2}
{\pa \vwA \over \pa t}(t,\vxAt) + \calL^0_{\vv_D}\vwAt(\vxAt)
=  d\Thetat(\vxBt).{\pa \vwB \over \pa t}(t,\vxBt),
\ee
%(En particulier~\eref{eqlieo10} $+$ \eref{eqlieo2} donnent~\eref{eqlieo}.) Et :
\be
\label{eqlieo2b}
\eqalign{
{D \vwA \over D t}(t,\vxAt) - d\vvDt.\vwAt(\vxAt)
= d\Thetat.(\vxBt).{D \vwB \over D t}(t,\vxBt) + d^2\Thetat(\vvBt,\vwBt)(\vxBt),
}
\ee
%Et :
\be
\label{eqlieo22}
\eqalign{
{\pa (\vvA{-}\vv_D)\over \pa t}(t,\vxAt) +  \calL^0_{\vv_D}(\vvA{-}\vv_D)(t,\vxAt)
= & d\Thetat(\vxBt).{\pa \vvB\over \pa t}(t,\vxBt).
}
\ee
%(Recall: $\vvAt \ne \vvBts$ in general since $\vvAt = \vvBts +\vvDt$.)
\finprop

\comment{
(Pour se convaincre que les dérivées de Lie autonomes ne sont pas objectives,
il suffit de noter que si une fonction objective est stationnaire dans un référentiel,
alors elle ne l'est pas dans un autre en mouvement par rapport au premier :
c'est $\calL$ et non $\calL^0$ qu'il faut considérer pour l'objectivité.)
}

\debdem
$\bullet$ $\vwAt(\Thetat(\vxBt)) = d\Thetat(\vxBt).\vwBt(\vxBt)$ gives
\be
\label{eqdvqossi2}
d\vwAt(\vxAt).d\Thetat(\vxBt) = d^2\Thetat(\vxBt).\vwBt(\vxBt) + d\Thetat(\vxBt).d\vwB(\vxBt),
\ee
thus, with $d\Thetat(\vxBt).\vvBt(\vxBt) = (\vvAt {-} \vvDt)(\vxAt) = \vvBts(\vxAt)$ (velocity-addition formula),
$$
d\vwAt(\vxAt). (\vvAt {-} \vvDt)(\vxAt)
= (d^2\Thetat(\vxBt).\vvBt(\vxBt)).\vwBt(\vxBt)
+ d\Thetat(\vxBt).d\vwBt(\vxBt).\vvBt(\vxBt),
$$
hence~\eref{eqlieo1}. In particular
$d\vwAt(\vxAt).\vvAt(\vxAt) \ne d\Thetat(\vxBt).(d\vwBt(\vxBt).\vvBt(\vxBt))$
(the vector field $d\vw.\vv$ is not objective).

\medskip
$\bullet$ 
$(\vvAt {-} \vvDt)(\Thetat(\vxBt))
= d\Thetat(\vxBt).\vvBt(\vxBt)$ gives
$$
d(\vvAt {-} \vvDt)(\vxAt).d\Thetat(\vxBt)
=  d^2\Thetat(\vxBt).\vvBt(\vxBt) + d\Thetat(\vxBt).d\vvBt(\vxBt),
$$
so, applied to~$\vwBt$ (resp.~$\vvBt$), we get~\eref{eqlieo4} (resp.~\eref{eqlieo6}).
Hence~\eref{eqlieo10}. % et~\eref{eqlieo40}.

\medskip
$\bullet$
If $\vxAt = \Thetat(\vxB)$, then $\vw_A(t,\Theta(t,\vxB)) = d\Theta(t,\vxB).\vwB(t,\vxB)$, so,
with ${\pa \Theta \over \pa t}(t,\vxB) = \vvThetat(\vxAt)$, we get
$$
\eqalign{
%{d \vw_A(t,q_a(t))\over dt}
{\pa \vw_A \over \pa t}(t, \vxAt) + d\vwAt(\vxAt).\vvThetat(\vxAt)
= & d{\pa \Theta \over \pa t}(t,\vxB).\vwBt(\vxB) +  d\Thetat(\vxB).{\pa \vwB \over \pa t}(t,\vxB) \cr
= & (d\vvThetat(\vxAt).d\Thetat(\vxB)).\vwBt(\vxB) +  d\Thetat(\vxB).{\pa \vwB \over \pa t}(t,\vxB),
}
$$
Thus~\eref{eqlieo2} since $\vvTheta = \vv_D$;
Then \eref{eqlieo1} gives~\eref{eqlieo2b}.

\medskip
$\bullet$
$\vv_{B*}(t,\Theta(t,\vxB)) = d\Theta(t,\vxB).\vvB(t,\vxB)$ gives
$$
\eqalign{
{\pa \vv_{B*} \over \pa t}(t,\vxAt) + d\vv_{B*}(\vxAt).\vvTheta(t,\vxAt)
= &
\underbrace{{\pa d\Theta \over \pa t}(t,\vxB)}_{d\vvThetat(\vxAt).d\Thetat(\vxB)}
.\vvBt(\vxB) 
+ d\Theta(t,\vxB).{\pa \vvB\over \pa t}(t,\vxB,) \cr
%= & d\vvThetat(\vxAt).d\Thetat(\vxBt).\vvBt(\vxBt) + d\Theta(t,\vxBt).{\pa \vvB\over \pa t}(t,\vxBt).
}
$$
since ${\pa d\Theta \over \pa t}(t,\vxB)=d({\pa \Theta \over \pa t})(t,\vxB)$ and
${\pa \Theta \over \pa t}(t,\vxB) = \vv_\Theta(t,\vxAt)=\vv_\Thetat(\Thetat(\vxB))$;
hence~\eref{eqlieo22}.
\findem

%%%%%%%%%%%%%%%%%%%%%%%%%%%%%%%%%%%%%%%%%%%%%%%%%%%%%%%%%%%%%%%%%%%%%%%%%%%%%%%%%%%

\subsubsection{Tensors}

\debprop
It $T$ is a covariant objective tensor, then its Lie derivatives are covariant objectives:
\be
\calL_{\vvA}T_A = \Theta_{*}(\calL_{\vvB}T_B) . %\qquad (= (\calL_{\vvB}\alpha_r).d\Theta_{t}^{-1}) .
\ee
\finprop

\debdem
Corollary of~\eref{eqdol0f1} and~\eref{eqlieo}
to get $\calL_\vv(\alpha.\vw) = (\calL_\vv\alpha).\vw  + \alpha.(\calL_\vv\vw)$;
Then use $\calL_\vv(t_1\otimes t_2) = (\calL_\vv t_1)\otimes t_2 + t_1\otimes (\calL_\vv t_2)$.
\findem

%%%%%%%%%%%%%%%%%%%%%%%%%%%%%%%%%%%%%%%%%%%%%%%%%%%%%%

\subsection{Taylor expansions and ubiquity gift}

%%%%%%%%%%%%%%%%%%%%%%%%%%%%%%%%%%%%%%%%%%%%%%%%%%%%%%

\subsubsection{In $\RRn$ with ubiquity}

Generic formula:
\be
f(t) = f(\tz) + (t{-}\tz)\,f'(\tz) + {(t{-}\tz)^2\over 2}\,f'(\tz)^2 + o((t{-}\tz)^2).
\ee
In particular $f(t) = \vw(t,p(t))$ gives
\be
\label{eqdl2cla00}
\eqalign{
\vw(t,p(t))
= & \vw(\tz,p(\tz)) + (t{-}\tz)\,{D\vw\over D t}(\tz,p(\tz))
+ {(t{-}\tz)^2 \over 2}\,{D\vw\over D t}(\tz,p(\tz))^2
+ o((t{-}\tz)^2). \cr
}
\ee
%Donc uniquement à l'aide des valeurs ponctuelles en $(\tz,p(\tz))$ on estime $\vw(t,p(t))$ (ici au second ordre).

\medskip
\noindent
{\bf Problem : }
$\vw(t,p(t))$ is a vecteur at~$t$ at~$p(t)$
while $\vw(\tz,p(\tz))$ is a vecteur at~$\tz$ at~$p(\tz)$,
so \eref{eqdl2cla00} cannot be written
\be
\label{eqdl2cla00b}
\vw(t,p(t))
- \bigl(\vw(\tz,p(\tz)) + (t{-}\tz)\,{D\vw\over D t}(\tz,p(\tz))
+ {(t{-}\tz)^2 \over 2}\,{D\vw\over D t}(\tz,p(\tz))^2\bigr)
= o((t{-}\tz)^2),
\ee
since the left-hand side supposes the ubiquity gift.

\Eg\ in a non-planar manifold (\eg\ a surface in~$\RRt$ considered on its own),
$\vw(t,\pt) \in \TptOmegat = $ the linear tangent space at $p(t)=\pt$,
whereas $\vw(\tz,\ptz) \in \TptzOmegatz = $ the linear tangent space at $p(\tz)=\ptz$,
and the tangent spaces $\TptOmegat$ and $\TptzOmegatz$ are distinct at two distinct points in general;
Thus the left-hand side of~\eref{eqdl2cla00b} is meaningless.

In $\RRt$ our affine space (our Universe), $\TptOmegat$ and $\TptzOmegatz$ are identified with~$\vRRt$,
and~\eref{eqdl2cla00} is well defined, and very useful!
%Néanmoins, elles posent problème au niveau de l'interprétation : elles imposent à l'observateur d'avoir le don d'ubiquité (temporel et spatial), comme décrit au~\S~\ref{sedipapf}.

%%%%%%%%%%%%%%%%%%%%%%%%%%%%%%%%%%%%%%%%%%%%%%%%%%%%%%

\subsubsection{General case}

By definition, \cf~\eref{eqdefDL0}, with $p(t)=\Phitz(t,\ptz)=\Phitzt(\ptz)$,
\be
\label{eqdlpb1}
\calL_\vv\vw(\tz,\ptz)
= {d\Phitzt(\ptz)^{-1}.\vw(t,p(t)) - \vw(\tz,\ptz) \over t-\tz} + o(1).
%= {((\Phitzt)^* \vw_t)(\tz,\ptz) - \vw(\tz,\ptz) \over t{-}\tz} + o(1).
\ee
%(To compare with ${D\vw\over D t}(\tz,p(\tz)) = {\vw(t,p(t)) - \vw(\tz,p(\tz)) \over t-\tz} + o(1)$, \cf~\eref{eqdl2cla00}.)
Thus,
\be
d\Phitzt(\ptz)^{-1}.\vw(t,p(t))
=  \vw(\tz,\ptz)
+ (t{-}\tz)\, \calL_\vv\vw(\tz,\ptz) + o(t{-}\tz).
\ee
Hence we get the first order Taylor expansion without ubiquity gift:
\be
\label{eqdlpb20}
\vw(t,p(t))
= d\Phitzt(\ptz). \Bigl(\vw + (t{-}\tz)\, \calL_\vv\vw\Bigr)(\tz,\ptz) + o(t{-}\tz),
%= d\Phitzt(\ptz). \Bigl(\vw(\tz,\ptz) + (t{-}\tz)\, \calL_\vv\vw(\tz,\ptz) + o(t{-}\tz)\Bigr),
\ee
both side of the equality being in $\TptOmegat$ (meaningful in any manifold).

%So, at first order, $\vw(t,p(t))$ is given by the push-forward of $\vw + (t{-}\tz)\, \calL_\vv\vw$ by~$\Phitzt$: At $(\tz,\ptz)$, we take the value $\Bigl(\vw + (t{-}\tz)\calL_\vv\vw \Bigr)(\tz,\ptz)$ which we transport along the motion~$\Phitz$ until we reach $(t,p(t))$. % (along the trajectory $\Phitzptz$).

\debprop
In~$\RRn$, with the gift of ubiquity, % (so here $\RRntz=\RRnt=\vRRn$), 
\eref{eqdlpb20} gives~\eref{eqdl2cla00}. % at first order.
\finprop

\debdem
%$h=t{-}\tz$ and \eref{eqdPhi2b} give 
%$d\Phitz(\tz{+}h,\ptz)) \equalref{eqdPhi2b} I_\tz + h\, d\vv(\tz,\ptz) + o(h)$, thus 
$d\Phitz(\tz{+}h,\ptz)).\vw(\tz,\ptz) \equalref{eqdPhi2b} \vw(\tz,\ptz) + h\, d\vv(\tz,\ptz).\vw(\tz,\ptz) + o(h)$, % for any $\vWptz$,
% $\Ftzptz(\tz{+}h) = I_\tz + h\, d\vv(\tz,\ptz) + o(h)$, thus $\Ftzptz(\tz{+}h).\vWptz = \vWptz + h\, d\vv(\tz,\ptz).\vWptz + o(h)$ for any $\vWptz$.
thus %\eref{eqdlpb20} gives:
$$
\eqalign{
\vw(t,\pt)
\equalref{eqdlpb20} & (I + h\, d\vv(\tz,\ptz) + o(h)).
\Bigl((\vw + h\,{D\vw\over Dt} - h\,d\vv.\vw)(\tz,\ptz) + o(h)\Bigr) \cr
%= & \vw(\tz,\ptz) + h\,(d\vv(\tz,\ptz).\vw(\tz,\ptz) + \calL_\vv\vw(\tz,\ptz)) + o(h) \cr
= & (\vw + h\,d\vv.\vw + h\,{D\vw\over Dt} - h\,d\vv.\vw)(\tz,\ptz) + o(h)
=  (\vw + h\,{D\vw\over Dt})(\tz,\ptz) + o(h), \cr
}
$$
which is~\eref{eqdl2cla00}.
\findem

\debprop
In $\RRn$, at second order,
\be
\label{eqvqFvg5}
\vw(t,p(t)) = d\Phitzt(\ptz).
\bigl((\vw + h\calL_\vv\vw + {h^2 \over 2}\calL_\vv(\calL_\vv\vw) )(\tz,\ptz) + o(h^2)\bigr).
\ee
So if the values  $\vw(\tz,\ptz)$, $\calL_\vv\vw(\tz,\ptz)$ and $\calL_\vv(\calL_\vv\vw)(\tz,\ptz)$
are known, then $\vw(t,p(t))$ is estimated at second order thanks to the push-forward of
$(\vw + h\calL_\vv\vw + {h^2 \over 2}\calL_\vv(\calL_\vv\vw) )(\tz,\ptz)$ by $\Phitzt$.
\finprop

\debdem
\def\vgp{{\vg\,'}}\def\vgpp{{\vg\,''}}%
Let $d\Phitz(t,\ptz) = \Ftzptz(t)$. % and $h=t{-}\tz$.
Let $\vg(t) = d\Phitz(t,\ptz)^{-1}.\vw(t,p(t))$ when $p(t) = \Phitz(t,\ptz)$.
So $\calL_\vv\vw(\tz,\ptz) = \vgp(\tz)$, \cf~\eref{eqdefDL2}.
And
${D\vw \over Du}(u,p(u))
= d\vv(u,p(u)).\vw(u,p(u)) + \Ftu(\pt).\vgp(u)$, \cf~\eref{eqdld2}.
Thus
${D^2\vw \over Du^2}(u,p(u))
= {D(d\vv) \over Du}(u,p(u)).\vw(u,p(u)) + d\vv(u,p(u)).{D\vw \over Du}(u,p(u))
+ d\vv(u,p(u).\Ftu(\pt).\vgp(u)+ \Ftu(\pt).\vgpp(u)$.
Thus
${D^2\vw \over Dt^2}(t,p(t))
= ({D(d\vv) \over Dt}.\vw + d\vv.{D\vw \over Dt} + d\vv.\calL_\vv\vw)(t,p(t)) + \vgpp(t)$.
With \eref{eqcalvv} we get
$\vgpp(t) = \calL_\vv(\calL_\vv\vw)(t,p(t))$, \cf~\eref{eqcalvv}.

Alternate proof (calculation): \eref{eqdPhi2a} gives
$\Ftzptz(t) = I_\tz + h\, d\vv(\tz,\ptz) + {h^2\over 2}\, d\vgamma(\tz,\ptz) + o(h^2)$.
Thus, omitting the reference to $(\tz,\ptz)$ to lighten the writing,
\be
\eqalign{
&d\Phitzt(\ptz).(\vw + h\calL_\vv\vw + {h^2 \over 2}\calL_\vv\calL_\vv\vw  + o(h^2)) \cr
& = 
\Bigl(I + h\, d\vv + {h^2 \over 2}\,d({D\vv \over Dt}) + o(h^2)\Bigr)
.\Bigl(\vw + h\calL_\vv\vw + {h^2 \over 2}\calL_\vv\calL_\vv\vw + o(h^2) \Bigr) \cr
}
\ee

The~$h^0$ term is $I.\vw = \vw$.
The~$h$ term is $\calL_\vv\vw + d\vv.\vw = {D\vw\over D t}$.
The~$h^2$ term is the sum of
$$
\eqalign{
\bullet & \demi \calL_\vv\calL_\vv\vw
= \demi ({D^2 \vw \over Dt^2} - 2\,d\vv. {D\vw \over Dt}
- {D (d\vv) \over Dt}.\vw + d\vv.d\vv.\vw), \;\hbox{ \cf \eref{eqcalvv}}, \cr
\bullet & d\vv.\calL_\vv\vw
=  d\vv.{D\vw\over D t} - d\vv.d\vv.\vw
=  \demi (2d\vv.{D\vw\over D t} - 2d\vv.d\vv.\vw), \cr
\bullet & \demi d({D \vv \over D t}).\vw = \demi({D (d\vv) \over Dt}.\vw + d\vv.d\vv.\vw),
\;\hbox{ \cf \eref{eqdfddtd2}} .\cr
}
$$
And the sum gives ${D^2\vw\over D t^2}$.
\findem

\comment{

\def\odd{\mathop{\;\raise-1.55pt\hbox{\large$0$}\mkern-9.5mu \twodots\;}}
\def\otd{\mathop{\;\raise-1.65pt\hbox{\Large$0$}\mkern-10.2mu \treedots}\;}
\def\uukappa{{\underline{\underline{\kappa}}}}
\def\uusigma{{\underline{\underline{\sigma}}}}
\def\uutau{{\underline{\underline{\tau}}}}
\catcode`@=11
\DeclareRobustCommand{\twodots}{\t@urdots}
\def\t@urdots{\mbox{\kern1\p@\vbox to 1ex{\hbox{.}\vss\vss\hbox{.}}}}
\DeclareRobustCommand{\treedots}{\th@urdots}
\def\th@urdots{\mbox{\kern1\p@\vbox to 1.3ex{\hbox{.}\vss \hbox{.}\vss\hbox{.}}}}
\catcode`@=12

%%%%%%%%%%%%%%%%%%%%%%%%%%%%%%%%%%%%%%%%%%%%%%%%%%%%%%%%%%%%%%%%%%%%%%%%%%%%%%%%%%%

\subsection{Dérivée de Lie et principe des puissances virtuelles}

%%%%%%%%%%%%%%%%%%%%%%%%%%%%%%%%%%%%%%%%%%%%%%%%%%%%%%%%%%%%%%%%%%%%%%%%%%%%%%%%%%%

\subsubsection{Double contraction tensorielle}

Notons $\odd$ la double contraction tensorielle (valeur réelle indépendante du choix de la base) :
si $\uukappa\in\Tuuo$ et $\uusigma\in\Tuuo$ sont deux tenseurs
(identifiés à des endomorphismes, cf.~\eref{eqtJ1}), alors :
\be
\uukappa \odd \uusigma \eqdef \Tr(\uukappa . \uusigma),
\ee
trace du tenseur $\uukappa.\uusigma\in\Tuuo$
(identifié à l'endomorphisme $\uukappa \circ \uusigma$).

Donc, si $(\ve_i)$ est une base de base duale~$(e^i)$,
si $\uukappa = \sumijn \kappa^i_j\,\ve_i\otimes e^j$ et
$\uusigma = \sumijn \sigma^i_j\,\ve_i\otimes e^j$, cf.~\eref{eqrcL},
alors $\uukappa.\uusigma = \sumijkn \kappa^i_k \sigma^k_j \ve_i\otimes e^j$, et :
\be
\label{eqdefodd}
\uukappa \odd \uusigma = \sumijn \kappa^i_j\sigma^j_i,
\ee
valeur réelle indépendante de la base choisie (trace d'un endomorphisme).

Ainsi, si $\vw\in\Tuzo$, si $\alpha\in\Tzuo$ et si $\uukappa\in\Tuuo$ alors :
\be
\label{eqdefodd2}
\alpha.\uukappa.\vw = (\vw\otimes \alpha) \odd \uukappa.
\ee
(Dans une base quelconque,
$\sumijn \alpha_i \kappa^i_j w^j
= (\sum_{j,i=1}^n w^j\alpha_i \ve_j\otimes e^i)\odd (\sumijn \kappa^i_j\,\ve_i\otimes e^j)$.)

%%%%%%%%%%%%%%%%%%%%%%%%%%%%%%%%%%%%%%%%%%%%%%%%%%%%%%%%%%%%%%%%%%%%%%%%%%%%%%%%%%%

\subsubsection{Double contraction matricielle}

Si $M=[M^i_j]$ et $N=[N^i_j]$ sont deux matrices $n*n$, on définit
la double contraction matricielle par (produit terme à terme) :
\be
M:N \eqdef \sumijn M^i_j N^i_j.
\ee
Donc :

1- si $\uukappa\in\Tuuo$ et $\uusigma\in\Tuuo$ sont deux endomorphismes, et

2- si, cadre classique, on dispose d'une structure euclidienne où
donc on s'est donné une unité de mesure, une base euclidienne $(\ve_i)$ associée,
et le produit scalaire euclidien~$\dd_g$ associé,

3- alors avec $[\uukappa]_{|\ve} = [\kappa^i_j]$ et $[\uusigma]_{|\ve} = [\sigma^i_j]$ donnent :
\be
[\uukappa]_{|\ve} : [\uusigma]_{|\ve}
\eqdef \sumijn \kappa^i_j\sigma^i_j
\eqnote \uukappa : \uusigma.
\ee
N.B. : ce produit matriciel terme à terme ne vérifie pas la convention d'Einstein :
il n'est objectif covariant à cause de l'utilisation d'un produit scalaire euclidien (de qui ?),
contrairement à $\uukappa \odd \uusigma$, cf.~\eref{eqdefodd}.
En particulier la dernière notation $\uukappa : \uusigma$ est abusive,
i.e. n'est pas intrinsèque à~$\uukappa$ et à~$\uusigma$
 (n'est pas objective : dépend de l'unité de mesure).

%%%%%%%%%%%%%%%%%%%%%%%%%%%%%%%%%%%%%%%%%%%%%%%%%%%%%%%%%%%%%%%%%%%%%%%%%%%%%%%%%%%

\subsubsection{Dérivée de Lie et principe des puissances virtuelles linéaire}
\label{secppv}

On s'intéresse ici à une forme locale de la puissance interne, à l'aide d'une densité de puissance,
dont l'intégration sur un ``volume'' donne une puissance mesurable. %Ici $\RRn = \RR^3$.

Soit un mouvement~$\tPhi$ de~$\Obj$, cf.~\eref{eqdeftPhi}, objet constitué d'un matériau donné.
On note $\vv$ le champ des vitesses eulériennes, cf.~\eref{eqdefve}.

On suppose que
$n$ champs de vecteurs $\vw_i$ caractérisent le matériau,
par exemple $n$ champs de vecteurs tangents à $n$ courbes dans le matériau, cf.~\S~\ref{secsa}.

Et on suppose que la densité de puissance interne s'exprime, à~$t$ en un point $\pt \in \Omegat$,
à l'aide des $n$ dérivées de Lie $\calL_\vv \vw_i(t,\pt)$
et de $n$ formes différentielles~$\alpha^i$ (instruments de mesure)
indépendantes pour donner :
\be
\label{eqppv0}
\calP(\vv) = \sumjn \alpha^j.\calL_\vv \vw_j
= \sumjn \alpha^j.{\pa \vw_j\over \pa t} + \sumjn \alpha^j.d\vw_j.\vv - \sumjn \alpha^j.d\vv.\vw_j,
\ee
cf.~\eref{eqdl}.
(Si $(\ve_i)$ est une base de base duale $(e^i)$,
on peut par exemple essayer de prendre $\alpha^i=e^i = $ les $n$ instruments de mesure
des efforts dus à la déformation.)

Plaçons-nous un cadre galiléen (cadre classique)
dans lequel $\calP(\vv) = 0$ quand $\vv=0$ et $d\vv=0$.
%$\tPhi$ est un mouvement de translation stationnaire et uniforme, i.e. t.q. ${\pa \vv \over \pa t}=0$ et $d\vv=0$.
On obtient $\sumjn \alpha^j.{\pa \vw_j\over \pa t} = 0$ quand $\vv=0$,
$\sumjn \alpha^j.d\vw_j.\vv = 0$ pour tout $\vv \in \vRRn$ quand $d\vv=0$,
donc $\sumjn \alpha^j.d\vw_j=0$, 
et il reste :
\be
\label{eqexas3}
\calP(\vv) 
= -\sumjn\alpha^j.d\vv.\vw_j
= -\uutau  \odd d\vv \qo \uutau  = -\sumjn \vw_j\otimes \alpha^j,
\ee
où on~a utilisé~\eref{eqdefodd2}.
On retrouve ainsi un résultat similaire à la formulation classique (cadre galiléen) de la puissance interne, avec ici un résultat indépendant de la base choisie (contraction tensorielle).

Exemple : cas $\alpha^j = e^j$ pour tout~$j$ : on~a :
\be
\label{eqexas3e}
\uutau = -\sumjn \vw_j \otimes e^j,\qe
[\uutau]_{|\ve}= - \pmatrix{[\vw_1]_{|\ve} & ... & [\vw_n]_{|\ve}},
\ee
et $[\uutau]_{|\ve}$ stocke dans sa $j$-ème colonne les
composantes des $\vw_j = \sumin \tau^i_j\ve_i$.
Et avec $\vv = \sumin v^i\ve_i$, et donc
$d\vv = \sumin \ve_i\otimes dv^i$,
on obtient 
$\uutau.d\vv = \sumjn \vw_j \otimes dv^j$ et
$\uutau  \odd d\vv = \Tr(\uutau.d\vv)
= \sumjn \Tr(\vw_j \otimes dv^j)
= \sumjn dv^j.\vw_j$, donc :
\be
\calP(\vv) 
= -\sumijn \tau^i_j{\pa v^j \over \pa x^i},
\ee
résultat indépendant de la base cartésienne choisie.

Ajoutons une structure euclidienne (cas classique), et notons $\dd_g$ le produit scalaire euclidien choisi.
Pour un endomorphisme $\uukappa \in \Tuuo$ on dispose de son
$\dd_g$-transposé $\uukappa_g^T \in \Tuuo$ défini par
$(\uukappa_g^T.\vu,\vw)_g = (\vu,\uukappa.\vw)_g$ pour tout $\vu,\vw\in\vRRn$.
Et, dans la base euclidienne $(\ve_i)$ choisie,
notant $\uukappa = \sumijn \kappa^i_j\,\ve_i\otimes e^j$ et
$\uukappa_g^T = \sumijn (\kappa_g^T)^i_j\,\ve_i\otimes e^j$,
on~a $(\kappa_g^T)^i_j = \kappa^j_i$ pour tout~$i,j$.
Ainsi, dans le cadre galiléen euclidien choisit, \eref{eqexas3} s'écrit :
\be
\calP(\vv) = - \uutau_g^T : d\vv , %= - \sumjn \tau^i_j{\pa v^i \over \pa x^j}.
\ee
et on retrouve le ``tenseur de Cauchy'' $\uusigma = \uutau_g^T$.
Cas particulier $\uutau$ symétrique : on~a $\uusigma = \uutau_g^T = \uutau$, et donc :
%$\uutau : d\vv = \uutau : d\vv^T_g$ et :
\be
%\label{eqexas5}
\calP(\vv)
= -\uusigma  : d\vv 
= -\uusigma  : {d\vv + d\vv_g^T \over 2},
\ee
et on~a retrouvé ``la forme usuelle'' de la puissance virtuelle des efforts intérieurs dans le cas linéaire.
% (après choix d'une unité de mesure et d'un cadre euclidien associé, et avec $\uutau$ tenseur ${1\choose1}$ symétrique relativement à~$\dd_g$).

%%%%%%%%%%%%%%%%%%%%%%%%%%%%%%%%%%%%%%%%%%%%%%%%%%%%%%%%%%%%%%%%%%%%%%%%%%%%%%%%%%%

\subsubsection{Commentaire}

%La formulation~\eref{eqppv0} est fondamentalement différente de la formulation usuelle~\eref{eqexas5} (quand est $\uutau$ symétrique).

La formulation~\eref{eqppv0} est objective covariante
car elle utilise les dérivées de Lie objectives covariantes
(quand les $\vw_j$ sont objectifs covariants),
dérivées qui permettent de mesurer les efforts exercés sur un matériau par un flot,
voir~\S~\ref{secLieei}.
Et~\eref{eqexas3} est l'expression a posteriori (représentation dans un base)
issue d'une formulation objective covariante (qui fait apparaître a posteriori
la quantité non objective covariante~$d\vv$).
Cette démarche est applicable en relativité.

Cela diffère de la formulation de la puissance interne classique construite dans un cadre euclidien
permettant de définir et utiliser le tenseur de Cauchy
(basé sur la comparaison euclidienne de deux vecteurs), cf.~\eref{eqdefCt} et~\eref{eqdefC}.
Voir également le~\S~\ref{secFIP} (``Frame Invariance Principle'')
issu par exemple de Truesdell et Noll~\cite{truesdell-noll} ou de Germain~\cite{germain2}
qui supposent ``l'objectivité isométrique''.
%La quantification des efforts à l'aide de la dérivée de Lie ne nécessite pas l'utilisation d'un produit scalaire, et permet de travailler dans un cadre objectif covariant.

%%%%%%%%%%%%%%%%%%%%%%%%%%%%%%%%%%%%%%%%%%%%%%%%%%%%%%%%%%%%%%%%%%%%%%%%%%%%%%%%%%%

\subsubsection{Tenseur de Cauchy : remarque...}
\label{secremliecau}

On peut introduire le tenseur de Cauchy classique $\uusigma$ sans faire appel,
dans un premier temps, à un produit scalaire euclidien.
Il suffit pour cela :

1- d'appliquer la remarque de Germain :
``pour connaître le poids d'une valise il faut la soulever''
(il ne suffit pas de la regarder : il faut faire un travail).

2- D'utiliser l'approche thermodynamique du travail :
la quantité de travail élémentaire est une forme différentielle
$\delta W \eqnote T^\flat \in \Tzuo$ (non exacte en général) qui,
le long d'un chemin régulier $\gamma : s\in[0,\eps] \rar x= \gamma(s)$, donne
la quantité de travail
$W(\gamma) = \int_\gamma T^\flat = \int_{s=0}^\eps T^\flat(\gamma(s)).\vgammap(s)\,ds
\simeq \eps T^\flat(\gamma(0))\vgammap(0)$ (approximation de Riemann à gauche),
soit $W(\gamma) = \eps T^\flat(p).\vu(p)$ où $p=\gamma(0)$ et $\vu(p) = \vgammap(0)$.
Et $T^\flat(p)$ est notre ``densité de travail''.

3- Puis, on introduit un produit scalaire euclidien~$\dd_g$
qui permet de disposer d'un vecteur normal unitaire $\vn(p)$ en un point $p$ d'une surface
dans le matériau, et sa forme duale $n^\flat(p)$, cf.~\eref{eqedf}.

4-
Et on suppose (Cauchy) que $T^\flat(p)$ est une fonction linéaire en~$\vn(p)$,
i.e. qu'il existe un endomorphisme $\uutau(p) : \vRRn \rar\vRRn$
t.q. $T^\flat(p)=n^\flat(p).\uutau(p)$.
(Dans une base $\dd_g$-euclidienne, si $\uutau = \sumijn \tau^i_j \ve_i \otimes e^j$
et $n^\flat = \sumin n_i e^i$
alors $T^\flat=n^\flat.\uutau = \sumijn n_i\tau^i_j e^j$.)

5-
Donc :
\be
T^\flat.\vu = n^\flat.\uutau.\vu
= (\vn,\uutau.\vu)_g
= (\uusigma.\vn,\vu)_g, \qo \uusigma = \uutau^T_g,
\ee
où on retrouve le tenseur de Cauchy usuel $\uusigma$,
avec $\vec t$ le vecteur de Cauchy de $\dd_g$ représentation de~$T^\flat$,
qui vérifie donc $\vec t = \uusigma.\vn$.

6- Avec~\eref{eqexas3e}
on~a $T^\flat = n^\flat.\uutau = -\sumjn (n^\flat.\vw_j)\, e^j$,
donc $\vec t = -\sumjn (\vn,\vw_j)_g\, \ve_j$,
et donc $\uusigma = \ve_j\otimes w_j^\flat$,
où la forme linéaire $w_j^\flat$ est associée par Riesz
au vecteur~$\vw_j$, soit $w_j^\flat.\vn = (\vn,\vw_j)_g$ pour tout~$\vn$.
%et donc $T^\flat.\vu  = -\sumjn (n^\flat.\vw_j)\, (e^j.\vu)$.
%= -\sumjn (n^\flat \otimes e^j)(\vw_j,\vu)$.

N.B. : cela laisse entrevoir la possibilité qu'un matériau
puisse être modélisé par $n$ formes linéaires $w_j^\flat$
dont l'action sur un champ de vecteurs vitesses (de nature vectorielle)
donne les réels $w_j^\flat.\vv$, réels indépendants de la base choisie.
Et non par $n$ champs de vecteurs $\vw_i$ (formant~$\uutau$).
%Voir preprint.

%%%%%%%%%%%%%%%%%%%%%%%%%%%%%%%%%%%%%%%%%%%%%%%%%%%%%%%%%%%%%%%%%%%%%%%%%%%%%%%%%%%

\subsubsection{... et dérivée de Lie et principe des puissances virtuelles linéaire bis}

La remarque précédente, cf.~\S~\ref{secremliecau}, suggère de proposer :

On suppose que
$n$ champs de formes linéaires $\beta^i$ caractérisent le matériau.

Et on suppose que la densité de puissance interne s'exprime, à~$t$ en un point $\pt \in \Omegat$,
à l'aide des $n$ dérivées de Lie $\calL_\vv \beta^i(t,\pt)$
et de $n$ champs de vecteurs~$\vu_i$ indépendants pour donner :
\be
\calP(\vv) = \sumjn \calL_\vv \beta^j.\vu_j
= \sumjn {\pa \beta^j\over \pa t}.\vu_j
+ \sumjn (d\beta^j.\vv).\vu_j
+ \sumjn \beta^j.d\vv.\vu_j,
\ee
cf.~\eref{eqdla2}.
(Si $(\ve_i)$ est une base,
on peut par exemple essayer de prendre $\vu_i=\ve_i$.)

Plaçons-nous un cadre galiléen (cadre classique)
dans lequel $\calP(\vv) = 0$ quand $\vv=0$ et $d\vv=0$.
On obtient $\sumjn {\pa \beta^j\over \pa t}.\vu_j$ quand $\vv=0$,
$ \sumjn (d\beta^j.\vv).\vu_j$ pour tout $\vv \in \vRRn$ quand $d\vv=0$,
et il reste :
\be
\calP(\vv) 
= \sumjn \beta^j.d\vv.\vu_j
= -\uutau  \odd d\vv \qo \uutau  = -\sumjn \vu_j\otimes \beta^j,
\ee
où on~a utilisé~\eref{eqdefodd2}.
On retrouve ainsi un résultat similaire à la formulation classique (cadre galiléen) de la puissance interne, avec ici un résultat indépendant de la base choisie (contraction tensorielle).

%%%%%%%%%%%%%%%%%%%%%%%%%%%%%%%%%%%%%%%%%%%%%%%%%%%%%%%%%%%%%%%%%%%%%%%%%%%%%%%%%%%

\subsubsection{Commentaire : fluides vs solides}

Un fluide de Stokes au repos ne stocke pas d'énergie.
Et son mouvement (son flot) est caractérisé pas les vecteurs vitesses (tangents aux trajectoires).

Il peut alors sembler naturel de mesurer la puissance sous la forme~\eref{eqppv0}.

En revanche un solide élastique stocke de l'énergie en position statique

%%%%%%%%%%%%%%%%%%%%%%%%%%%%%%%%%%%%%%%%%%%%%%%%%%%%%%%%%%%%%%%%%%%%%%%%%%%%%%%%%%%

\subsection{Perspectives non linéaires}

%%%%%%%%%%%%%%%%%%%%%%%%%%%%%%%%%%%%%%%%%%%%%%%%%%%%%%%%%%%%%%%%%%%%%%%%%%%%%%%%%%%

\subsubsection{Perspectives : principe des puissances virtuelles non linéaires}

\def\vwzj{{\vw_{0j}}}
\def\vwuj{{\vw_{1j}}}
\def\vwzuj{{\vw_{01j}}}
\def\vwuzj{{\vw_{10j}}}
\def\vwuuj{{\vw_{11j}}}
\def\vwzdj{{\vw_{02j}}}
\def\vwdzj{{\vw_{20j}}}
\def\vwudj{{\vw_{12j}}}
\def\vwduj{{\vw_{21j}}}
\def\vwddj{{\vw_{22j}}}
\def\alphazj{{\alpha_{0j}}}
\def\alphauj{{\alpha_{1j}}}
\def\alphauuj{{\alpha_{11j}}}

On complète~\eref{eqppv0} pour, par exemple, obtenir la densité de puissance linéaire générale :
\be
\label{eqppv1l}
\calP_{1\ell}(\vv)
= \sumjn \alpha^j.\vwzj 
+  \alpha^j.\calL_\vv \vwzuj +  \calL_\vv\alpha^j.\vwuzj,
\ee
les $\vw_{\cdot j}$ étant supposés objectifs covariants.

\debexa
Cadre galiléen avec par exemple $\alpha^j = e^j$ :
l'hypothèse $\calP(\vv)=0$ quand $d\vv=0$ donne à la place de~\eref{eqexas3} :
\be
\label{eqexas1l}
\calP_{1\ell}(\vv)
%= -e^j.d\vv.\vwzuj + e^j.d\vv.\vwuzj
= -\uutau  \odd d\vv \qo \uutau = -\sumjn e^j.d\vv.(\vwzuj-\vwuzj),
\ee
et on retrouve~\eref{eqexas3} avec ici $2n$ champs de vecteurs $\vw_{\cdot j}$.
\finexa

On peut compléter~\eref{eqppv1l} pour obtenir, par exemple, une puissance virtuelle (non linéaire)
avec des dérivées premières de Lie, de type :
\be
\label{eqppv1nl}
\calP_{1n\ell}(\vv)
= \calP_{1\ell}(\vv)
+ \sumjn \calL_\vv \alpha^j.\calL_\vv \vwuuj,
\ee
méthode qui pourrait être baptisée ``méthode de premier gradient non linéaire''.

\debexa
Cadre galiléen avec par exemple $\alpha^j = e^j$ :
l'hypothèse $\calP(\vv)=0$ quand $d\vv=0$ donne à la place de~\eref{eqexas1l},
et avec 
$\calL_\vv e^j.\calL_\vv \vwuuj = (e^i.d\vv).({\pa \vwuuj \over \pa t} + d\vwuuj.\vv - d\vv.\vwuuj)$ :
\be
\label{eqexas1nl}
\eqalign{
\calP(\vv)
= & \uutau \odd d\vv + \sumjn ({\pa \vwuuj \over \pa t}\otimes e^j) \odd d\vv
- (\vwuuj\otimes e^j) \odd (d\vv.d\vv) + e^j.d\vv.d\vwuuj.\vv \cr
= & (\uutau + {\pa \uukappa \over \pa t}) \odd d\vv
+ \uukappa \odd (d\vv.d\vv) + d\uukappa \otd (\vv \otimes d\vv),
\qo
\left\{\eqalign{
& \uutau = -\sumjn (\vwzuj-\vwuzj)\otimes e^j, \cr
& \uukappa  = \sumjn \vwuuj \otimes e^j,
}\right. \cr
}
\ee
%où $\uutau = -\sumjn (\vwzuj-\vwuzj)\otimes e^j$ et $\uukappa  = \sumjn \vwuuj \otimes e^j$,
et où on~a utilisé la triple contraction tensorielle définie par :
\be
S \odd T = \sumijkn S^i{}_{jk} T^{kj}{}_i
\qq 
\left\{\eqalign{
& S = \sumijkn S^i{}_{jk} \ve_i \otimes e^j \otimes e^k \in \Tudo, \cr
& T = \sumijn T^{ij}{}_k \ve_i \otimes \ve_j \otimes e^k \in \Tduo.
}\right.
\ee
En effet, avec $\vv= \sumjn v^j \ve_j$, on~a $d\vv = \sumjkn {\pa v^j \over \pa x^k}\ve_j \otimes e^k$,
et donc $e^j.d\vv = \sumkn {\pa v^j \over \pa x^k} e^k$ ($ =d v^j$),
et avec $\vwuuj = \sumin w^i_j \ve_i$, on~a
$d\vwuuj = \sumikn {\pa w^i_j \over \pa x^k}\ve_i \otimes e^k$, et donc
$d\vwuuj.\vv = \sumikn {\pa w^i_j \over \pa x^k} v^k\ve_i$. Et donc :
\be
\sumjn e^j.d\vv.d\vwuuj.\vv = \sumijkn {\pa w^i_j \over \pa x^k} v^k {\pa v^j \over \pa x^i}
=d\uukappa \otd (\vv \otimes d\vv),
\ee
Et dans~\eref{eqexas1nl} les deux derniers termes sont non linéaires en~$\vv$
(similaire à un méthode de premier ordre ``non linéarisé'').
\finexa

\debexa
Suite.
Par exemple avec $\uukappa = \uutau$
(par exemple en prenant $\vwuzj=0$ et $\vwuuj = \vwzuj$,
le matériau étant alors caractérisé  par $n$ champs de vecteurs~$\vwzuj$),
\eref{eqexas1nl} donne :
\be
\label{eqexas1nlt}
\calP(\vv)
=  (\uutau + {\pa \uutau \over \pa t}) \odd d\vv
+ \uutau \odd (d\vv.d\vv) + d\uutau \otd (\vv \otimes d\vv),
\ee
formulation qui peut être un candidat à un modélisation simplifiée de turbulence au premier ordre non linéaire (dans un cadre galiléen).
\finexa

\debexa
Suite. Par exemple, \eref{eqexas1nlt} s'écrivant :
\be
\label{eqexas1nlv}
\calP(\vv)
=  \uutau \odd d\vv + {\pa \uukappa \over \pa t} \odd d\vv
+ d\uukappa \otd (\vv \otimes d\vv)+ \uukappa \odd (d\vv.d\vv) ,
\ee
on peut penser à une modélisation de matériau visco-élastique (dans un cadre galiléen),
basée sur une formulation objective,
avec $\uutau$ pour la partie fluide complétée par $\uukappa$ pour le caractère ``visco-élastique''.
Cette formulation d'origine objective covariante
diffère notoirement de formulations classiques de type ``matériau viscoélastiques de Maxwell''.
\finexa

Et on compléter~\eref{eqppv1nl} pour obtenir, par exemple, une puissance virtuelle au second ordre (non linéaire) de type :
\be
\label{eqppv2nd}
\calP_2(\vv)
= \calP_{1n\ell}(\vv) + \alpha^j.\calL_\vv\calL_\vv \vwzdj
+ \calL_\vv\alpha^j.\calL_\vv\calL_\vv \vwudj
+ \calL_\vv\calL_\vv\alpha^j.\vwdzj
+ \calL_\vv\calL_\vv\alpha^j.\calL_\vv \vwduj
+ \calL_\vv\calL_\vv\alpha^j.\calL_\vv\calL_\vv \vwddj
,
\ee
méthode qui pourrait être baptisée ``méthode de second gradient non linéaire''.
(Simplifiable dans le cadre galiléen.)

Et on peut généraliser à tout ordre supérieur.

On peut ainsi obtenir, de manière objective, des modélisations de turbulence,
de viscoélasticité... 

\medskip
N.B. : ces méthodes, basées sur les dérivées objectives de Lie,
ne font pas appel à un produit scalaire euclidien donné a priori,
en particulier ne font pas appel aux dérivées secondes de Cauchy
qui d'ailleurs sont non linéaires, cf.~\eref{eqdPhi302} (et qui représenteraient quoi ?).
Noter que les méthodes classiques ``de second gradient'',
proposées par exemple par Germain~\cite{germain2}, sont linéaires,
ce qui peut sembler étonnant (pour des méthodes de second ordre).

\comment{
3- Pour une formulation du principe des puissances virtuelles, faite à un instant~$t$, les dérivées de Lie
semblent appropriées (point de départ : le principe des travaux virtuels).

4- Pour le principe des puissances virtuelles où on considère les champs des vitesses~$\vv$ stationnaires,
on~a, cf.~\eref{eqpropcalv20} et~\eref{eqccvw} :
\be
\calL_\vv\vw
=  {\pa\vw\over\pa t} + d\vw.\vv - d\vv.\vw,
\ee
\be
\calL_\vv (\calL_\vv\vw)
=  {\pa^2 \vw \over \pa t^2} + 2d{\pa\vw\over\pa t}.\vv - 2d\vv.{\pa\vw\over\pa t} 
 +  d^2\vw(\vv,\vv) + d\vw.d\vv.\vv - d^2\vv(\vw,\vv) - 2d\vv.d\vw.\vv  + d\vv.d\vv.\vw.
\ee
}

}

\newpage
\appendix
\part{Appendix}

\def\bL{\beta_{\!L}}
\def\bLs{\beta_{\!(\!L^{\!*}\!)}}
\def\bLT{(\beta_{\!L})^T}

In this appendix, we tried to give standard results useful in mechanics, results that are scattered in the existing literature, and sometimes difficult to find except in math books (differential geometry).

The definitions, notations and results are detailed, so that no ambiguity is possible
(some notations can be nightmarish when not understood, or misused, or come like a bull in a china-shop).
All the results presented apply to solids, fluids, thermodynamics, general relativity, electromagnetism, quantum mechanics, chemistry... (the same math applies to all... even applies to mechanical engineers...).
%, despite the introduction of certain (apparently opposed notations) dedicated to some particular field of physics. 
% (usually due to traditions). % or a lack of basic mathematical knowledge).

%%%%%%%%%%%%%%%%%%%%%%%%%%%%%%%%%%%%%%%%%%%%%%%%%%%%%%%%%%%%%%%%%%%%%%%%%%%%%%%%%%%
%%%%%%%%%%%%%%%%%%%%%%%%%%%%%%%%%%%%%%%%%%%%%%%%%%%%%%%%%%%%%%%%%%%%%%%%%%%%%%%%%%%

\section{Classical and duality notations}
\label{secann1}

%%%%%%%%%%%%%%%%%%%%%%%%%%%%%%%%%%%%%%%%%%%%%%%%%%%%%%%%%%%%%%%%%%%%%%%%%%%%%%%%%%%

\subsection{Contravariant vector and basis}
\label{secbased}

%%%%%%%%%%%%%%%%%%%%%%%%%%%%%%%%%%%%%%%%%%%%%%%%%%%%%%%%%%%%%%%%%%%%%%%%%%%%%%%%%%%

\subsubsection{Contravariant vector}

Let $(E,+,.) \eqnote E$ be a real vector space (= a linear space on the field~$\RR$).

\debdef
An element $\vx\in E$ is called a vector, or a ``contravariant vector''.
\findef

A vector is a vector... 
So why this name contravariant? 
Historical answer: Because of the change of basis formula
$[\vx]_{|\new} = P^{-1}.[\vx]_{|\old}$, see~\eref{eqdefP1}, which uses~$P^{-1}$.

So, what is a covariant vector? Answer: From the vector space~$E$, you can build the vector space (an overlay) $\calL(E;\RR)\eqnote E^*=$ the space of linear forms on~$E$ (a linear form is a measuring instrument that gives values to vectors).
%; Such a space does not exist without the introduction at first of a vector space~$E$.
Then an element $\ell\in E^*$ will be called a covariant vector, because of the change of basis formula $[\ell]_{|\new} = [\ell]_{|\old}.P$. See~\S~\ref{secfcb} for details.

%%%%%%%%%%%%%%%%%%%%%%%%%%%%%%%%%%%%%%%%%%%%%%%%%%%%%%%%%%%%%%%%%%%%%%%%%%%%%%%%%%%

\subsubsection{Basis}

Definitions:
$\bullet$ 
$n$ vectors $\ve_1,...,\ve_n\in E$ are linearly independent iff,
for all $\lambda,...,\lambda_n\in\RR$, the equality $\sumin \lambda_i \ve_i=\vec0$ implies
$\lambda_i=0$ for all~$i=1,...,n$.

$\bullet$ $n$ vectors $\ve_1,...,\ve_n\in E$ span $E$ iff, for all $\vx\in E$, $\exists \lambda_1,...,\lambda_n\in\RR$
such that $\vx=\sumin \lambda_i\ve_i$.

$\bullet$ A basis in~$E$ is a set $\{\ve_1,...,\ve_n\}  \subset E$
made of $n$ linearly independent vectors which span~$E$, in which case the dimension of~$E$ is~$n$.

%%%%%%%%%%%%%%%%%%%%%%%%%%%%%%%%%%%%%%%%%%%%%%%%%%%%%%%%%%%%%%%%%%%%%%%%%%%%%%%%%%%

\subsubsection{Canonical basis}

Consider the field~$\RR$ of reals and the Cartesian product $\vRRn=\RR\times ... \times \RR$, $n$ times.
The canonical basis is
\be
\label{eqdefbc}
\ve_1=(1,0,...,0), \; ...,\; \ve_n=(0,...,0,1),
\ee
with $0=$ the addition identity element used $n{-}1$ times, and $1=$ the multiplication identity element used once.

\debrem
\label{remnocanb}
The 3-D geometric space we live in has no canonical basis: What would the identity element~$1$ mean?
1~metre? 1~foot? And there is no ``intrinsic'' preferred direction to define~$\ve_1$.
So the Cartesian product $\vRRn=\RR\times ... \times \RR$ and its canonical basis form an abstract mathematical model.
%(The vector space $\vRRt$ is however isomorphic to the mathematical Cartesian product $\RR\times \RR\times \RR$).
\comment{
\Eg, the Cartesian space $\vE=\{(T,P)\} = \RR\times \RR$ of temperature and pressure (thermodynamic)
has no ``natural'' canonical basis in practice: 
In practice, $\ve_1=(1,0)$ is defined with $1=$  a unit of temperature (\eg\ Celsius),
and $\ve_2=(0,1)$ with $1=$ a unit of pressure (\eg\ Pascal)
(the chosen unit of measurement are observer dependent).
\Eg, the geometric space $\RR^3$ (our usual space) has no ``natural'' canonical basis:
a basis $(\ve_1,\ve_2,\ve_3)$ used for measurements need a unit of measurement and directions
(observer dependent).
}%
\finrem

%%%%%%%%%%%%%%%%%%%%%%%%%%%%%%%%%%%%%%%%%%%%%%%%%%%%%%%%%%%%%%%%%%%%%%%%%%%%%%%%%%%

\subsubsection{Cartesian basis}

(Ren\'e Descartes 1596-1650.)
Let $n=1,2,3$, let $\RRn$ be the usual affine space (space of points), and let $\vRRn = (\vRRn,+,.)$ be the usual real vector space of bipoint vectors with its usual algebraic operations.

Let $p \in\RRn$, and let $(\ve_i(p))$ be a basis at~$p$.

A Cartesian basis in~$\vRRn$ is a basis independent of~$p$ (the same at all~$p$), and then  $(\ve_i(p))\eqnote (\ve_i)$.

Example of a non Cartesian basis:
The polar basis, see example~\ref{exarempfcb} (polar coordinate system).

And a Euclidean basis is a particular Cartesian basis described in~\S~\ref{secbe}.

More generally, a Cartesian basis refers to $E^n=E\times...\times E$ ($n$-times) where $E$ is a dimension~1 vector space.

%NB: In a non-planar surface (manifold), a Cartesian basis does not exist since the direction of a tangent vectors $\ve_i(p)$ changes with~$p$ (\eg, in the curved space-time in which we live, see the general relativity model.)
%A Cartesian basis depends on an observer.

%%%%%%%%%%%%%%%%%%%%%%%%%%%%%%%%%%%%%%%%%%%%%%%%%%%%%%%%%%%%%%%%%%%%%%%%%%%%%%%%%%%

\subsection{Representation of a vector relative to a basis}

We give:

$\bullet$ the classical notation (non ambiguous), \eg\ used by~Arnold~\cite{arnold} and Germain~\cite{germain}, and 

$\bullet$ the duality notation (can be ambiguous because of misuses), \eg\ used by Marsden and Hughes~\cite{marsden-hughes}.
\\
Both classical and duality notation are equally good, but if you have any doubt, use the classical notations.

\debdef
Let $\vx\in E$. Let $(\ve_i)$ be a basis in~$E$.
The components of~$\vx$ relative to the basis~$(\ve_i)$ are the $n$ real numbers $x_1,...,x_n$ (classical notation) also named $x^1,...,x^n$ (duality notation) such that
\be
\label{eqxi0}
\vx
=\underclas{ x_1\ve_1+...+x_n\ve_n}
=\underdual{ x^1\ve_1+...+x^n\ve_n}
, \qie [\vx]_{|\ve} 
= \underclas{\pmatrix{x_1\cr\vdots\cr x_n}}
= \underdual{\pmatrix{x^1\cr\vdots\cr x^n}},
\ee
$[\vx]_{|\ve}$ being the column matrix representing~$\vx$ relative to the basis~$(\ve_i)$.
(Of course $x_i=x^i$ for all~$i$.)
And the column matrix $[\vx]_{|\ve}$ is simply named $[\vx]$ if one chosen basis is imposed to all.
With the sum sign:
\be
\vx =\underclas{\sumin x_i\ve_i} = \underdual{\sumin x^i\ve_i} \quad (= \sumJn x_J\ve_J %= \sumKn x^K\ve_K 
= \sum_{\alpha=1}^n x^\alpha\ve_\alpha).
\ee
(The index in a summation is a dummy index, even if you do not write the sum sign~$\sum$ as can be done with Enstein's convention:
$\vx =\sumjn x^j\ve_j \eqnote x^j\ve_j =  x^i\ve_i = x^J\ve_J = x^\alpha\ve_\alpha$.)
\findef

%The use of the notation $[\vx]_{|\ve}$ avoids dealing explicitly with the dual or classical notation choice.

\debexa
In $\vRRd$ with $\vx= 3\ve_1+4\ve_2 = \sum_{i=1}^2 x_i\ve_i = \sum_{i=1}^2 x^i\ve_i$: We have $x_1{=}x^1{=}3$ and $x_2{=}x^2{=}4$. And 
$[\vx]_{|\ve} 
= 3[\ve_1]_{|\ve} + 4[\ve_2]_{|\ve}
=\sum_{i=1}^2 x_i[\ve_i]_{|\ve}
=\sum_{i=1}^2 x^i[\ve_i]_{|\ve}
$.
In particular, with $\delta^i_j=\delta_{ij}
:=\left\{\eqalign{=&1\hbox{ if } i{=}j \cr =&0\hbox{ if } i{\ne}j }\right\}
$
the Kronecker symbols,
\be
\label{eqxi0b}
%\forall j\in [1,n]_\NN, \quad
\ve_j
=\underclas{ \sumin \delta_{ij} \ve_i}
=\underdual{\sumin \delta^i_j\ve_i},\qie
[\ve_1]_{|\ve} = \pmatrix{1\cr0\cr\vdots\cr 0},\; ...,\;
[\ve_n]_{|\ve} = \pmatrix{0\cr\vdots\cr0\cr 1} ,
\ee
that is, %relative to the basis~$(\ve_i)$,
the components of~$\ve_j$ are 
$\delta_{ij}$ with classical notations, and
$\delta^i_j$ with duality notations. %, for $i=1,...,n$.
And the matrices $[\ve_j]_{|\ve}$ mimic the use of theoretical Cartesian space $\vRRn = \RR\times ... \times \RR$ and its canonical basis.
%, but see remark~\ref{remnocanb}, and $\delta_{ij} = \delta^i_j$ is the $i$-th component of~$\ve_j$ relative to the basis~$(\ve_i)$: $\ve_j = \delta_{1j}\ve_1+...+\delta_{nj}\ve_n = \delta^1_j\ve_1+...+\delta^n_j\ve_n$.
\finexa

\debrem
The column matrix $[\vx]_{|\ve}$ is also called a ``column vector''. NB: A ``column vector'' is not a vector, but just a matrix (a collection of real numbers). See the change of basis formula~\eref{eqdefP1} where the same vector is represented by two ``column vectors'' (two column matrices).
\finrem

%(We follow Einstein's convention given in~\S~\ref{secEC}.)

%%%%%%%%%%%%%%%%%%%%%%%%%%%%%%%%%%%%%%%%%%%%%%%%%%%%%%%%%%%%%%%%%%%%%%%%%%%%%%%%%%%

\subsection{Dual basis}

Recall: Let $E$ and $F$ be vector spaces and $(\calF(E;F),+,.) \eqnote \calF(E;F)$ be the usual real vector space of functions
% $f:E\rar F$
with the internal addition $(f,g) \rar f+g$
defined by $(f+g)(x) := f(x) + g(x)$
and the external multiplication $(\lambda,f) \rar \lambda .f$ defined by
$(\lambda .f)(x) := \lambda (f(x))$, for all $f,g\in\calF(E;F)$, $x\in E$, $\lambda\in\RR$.
And $\lambda.f\eqnote \lambda f$ for all $f\in\calF(E;F)$ and $\lambda\in\RR$.

%%%%%%%%%%%%%%%%%%%%%%%%%%%%%%%%%%%%%%%%%%%%%%%%%%%%%%%%%%%%%%%%%%%%%%%%%%%%%%%%%%%

\subsubsection{Linear forms = ``Covariant vectors''}

%Particular case $F=\RR$ in the definition~\ref{defal}:

\debdef
\label{defal2}
The set $E^* := \calL(E;\RR)$ of linear scalar valued functions is called the dual of~$E$:
\be
\label{eqdefal2}
E^* := \calL(E;\RR) = \hbox{the dual of $E$}.
\ee
And a linear scalar valued function $\ell\in E^*$ is called a linear form.

More precisely, $E^*$ as defined in~\eref{eqdefal2} is the algebraic dual of~$E$;
To define the topological dual usually needed with $L^2$ functions in mechanics, $E$ needs to be a Banach space
(a vector space equipped with a norm with which $E$ is complete),
and $E^*$ is then the set of continuous linear forms.
(If $E$ is finite dimensional then
any linear form is continuous relative to any norm since all norms are equivalent in finite dimension.)
\findef

$E^*$ is a vector space: sub-space of~$(\calF(R;\RR),+,.)$ (trivial).

\mn
%\medskip
\noindent
{\bf Interpretation:} It answers the question: What does a function $E\rar\RR$ do?
Answer: Like any function, it gives values to vectors: $\ell(\vu)=$ the value of $\vu$ through~$\ell$.
That is, a $\ell\in E^*$ is a measuring tool for vectors: If $\vu\in E$ then $\ell(\vu)=$ real value given by~$\ell$.

\mn
{\bf Notation}: If $\ell \in E^*$ then
\be
\label{eqellu}
\forall \vu\in E, \quad \ell(\vu) \eqnote \ell.\vu, % \quad\hbox{also named}\quad \la\ell,\vu\ra_{E^*,E}.
\ee
also written $\la\ell,\vu\ra_{E^*,E}$ where $\la . , .\ra_{E^*,E}$ is the duality bracket: The dot in $\ell.\vu$ is ``the distributivity dot'' since linearity $\ell(\vu+\lambda\vv) = \ell(\vu) + \lambda \ell(\vv)$ = distributivity
$\ell.(\vu+\lambda\vv) = \ell.\vu + \lambda \ell.\vv$.

NB: The dot in $\ell.\vu$ is \textsl{\textbf{not}} an inner dot product (since $\ell\notin E$ while $\vu\in E$).

\debdef
A linear form $\ell$ in $E^*$ is also called a ``covariant vector'';
Co-variant refers to:

1- The action of a function on a vector, \cf~\eref{eqellu} (co-variant calculation), and

2- The change of coordinate formula $[\ell]_{\new} = [\ell]_{|\old}.P$, see~\eref{eqdefP1} (covariant formula).
\findef

\noindent
{\bf NB:} $E^*$ being a vector space, an element $\ell\in E^*$ is indeed a vector.
But $E^*$ has no existence if~$E$ has not been specified first since
$E^*:=\calL(E;\RR)$. And $\ell\in E^*$ can't be confused with a vector $\vu\in E$ since there is no natural canonical isomorphism between $E$ and~$E^*$ (no ``intrinsic representation''), see~\S~\ref{secEEsnonnat}.
%In the following, a ``covariant vector'' $\ell\in E^*$ will be called a linear form. % (rather than a vector). 

\debrem
\label{remMisner}
Misner--Thorne--Wheeler~\cite{misner-thorne-wheeler}, box 2.1, insist:
%\begin{center}
``Without it [the distinction between covariance and contravariance],
one cannot know whether a vector is meant or the very different object that is a linear form.''
%\end{center}
%The confusion between a linear function (covariant) and a vector (contravariant), together with the use of the Riesz-representation theorem, often leads to consider that any phenomenon (its modelization) is linear... Which is a rather restrictive point of view...
\finrem

\comment{
\mn
{\bf Vocabulary.}
1- Algebra: $E^*$ is the ``algebraic dual'' of~$E$. 
2- Functional analysis: In infinite dimension, $E^*$ is also named $E'$ and is the set of scalar valued linear functions which are continuous relative to given norms, and is named the ``topological dual''.
%In finite dimension, all norms are equivalent, a linear function is continuous, so $E'=E^*$ is both the algebraic and topological dual of~$E$.
}

%%%%%%%%%%%%%%%%%%%%%%%%%%%%%%%%%%%%%%%%%%%%%%%%%%%%%%%%%%%%%%%%%%%%%%%%%%%%%%%%%%%

\subsubsection{Covariant dual basis (= the functions that give the components of a vector)}

Notation:
If $\vu_1,...,\vu_k$ are vectors in~$E$, then $\Vect\{\vu_1,...,\vu_k\}:=$ the vector space spanned by $\vu_1,...,\vu_k$.

Let $E$ be a finite dimensional vector space, and let $(\ve_i)_{i=1,...,n}$ be a basis in~$E$

\debdef
Let $i\in[1,n]_\NN$.
The scalar projection on~$\Vect\{\ve_i\}$ parallel to $\Vect\{\ve_1,...,\ve_{i-1},\ve_{i+1},...,\ve_n\}$ is the linear form named $\pi_{ei}\in E^*$ with the classical notation, named $e^i\in E^*$ with the duality notation, defined by, for all $i,j$,
\be
\label{eqdefbd}
\left\{\eqalignrll{
& \clasnot:
&\quad \pi_{ei}(\ve_j) = \delta_{ij},\qie \pi_{ei}.\ve_j = \delta_{ij}, \cr
& \dualnot:
&\quad  e^i(\ve_j) = \delta^i_j,\qie e^i.\ve_j = \delta^i_j. \cr
}\right.
\ee
\findef

Thus, $\pi_{ei}=e^i$ being linear, if $\vx \eqclas\sumin x_i\ve_i \eqdual \sumin x^i\ve_i$ (classical or duality notations),
then \eref{eqdefbd} gives
\be
\pi_{ei}.\vx \eqclas x_i \quad=\quad x^i \eqdual e^i.\vx,
%)\quad  \pi_{ei}.\vx = x_i = e^i.\vx = x^i.
\ee
\ie\ $\pi_{ei} = e^i$ gives
the $i$-th component of a~vector~$\vx$ relative to the basis~$(\ve_i)$, see figure~\ref{figbnon}.

\begin{figure}[!h]
\qquad\qquad\qquad\qquad\qquad\qquad\qquad\qquad\qquad\includegraphics[width=0.2\textwidth]{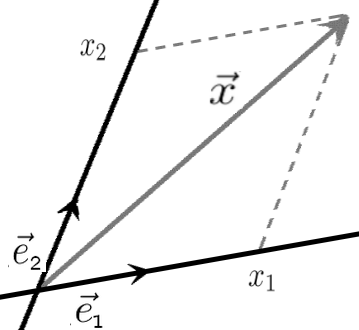}
\caption{
Parallel projections: $\pi_{e1}(\vx)=x_1$ and $\pi_{e2}(\vx)=x_2$ (dual not.: $e^1(\vx)=x^1$ and $e^2(\vx)=x^2$).
}\label{figbnon}
\end{figure}

NB: 
The dual basis $(\pi_{ei})$ is intrinsic to~$(\ve_i)$; And
there can't be any notion of orthogonality in~$E$ here since we can't use a  inner dot product:
The functions $\pi_{ei}=e^i$ and vectors $\vx$ do not belong to a same vector space.
%The dot notation $\pi_{ei}.\vx=e^i.\vx$ only means $\pi_{ei}(\vx)=e^i(\vx)$, the function $\pi_{ei}=e^i$ being linear.

\debprop {\bf and definition of the dual basis.}
\label{propbd}
$(\pi_{ei})_{i=1,...,n} = (e^i)_{i=1,...,n}$ is a basis in~$E^*$, called the (covariant) dual basis of the basis~$(\ve_i)$.
\finprop

\debdem
If $\sumin \lambda_i\pi_{ei}=0$, then $0=(\sumin \lambda_i\pi_{ei})(\ve_j)=\sumin \lambda_i \pi_{ei}(\ve_j)=\sumin \lambda_i \delta_{ij} = \lambda_j$ for all~$j$,
thus $(\pi_{ei})_{i=1,...,n}$ is a family of $n$ independent vectors in~$E^*$.
Then let $\ell\in E^*$ and $m=\sum_i (\ell.\ve_i)\pi_{ei}$. Thus $m\in E^*$ (since $E^*$ is a vector space),
and $m(\ve_j) = \sum_i (\ell.\ve_i)(\pi_{ei}.\ve_j) = \sum_i (\ell.\ve_i)\delta_{ij} =(\ell.\ve_j)$,
thus $m=\ell$, thus $\ell = \sum_i (\ell.\ve_i)\pi_{ei}$, thus $\Vect\{(\pi_{ei})_{i=1,...,n}\}$ span~$E^*$;
Thus $(\pi_{ei})_{i=1,...,n}$ is a basis in~$E^*$; Thus $\dim E^*=n$. %\cf~prop.~\ref{propLab}, thus $(\pi_{ei})_{i=1,...,n}$ is a basis in~$E^*$.
(Use duality notations if you prefer.)
\findem

\debexa
\label{exa11b}
Following example~\ref{exa11}.
The size of a child is represented on a wall by a bipoint vector~$\vu$.
And English observer chooses the foot as unit of length, represented by a vertical bipoint vector which he names~$\ve$.
And then defines the linear form $\pi_e : \vec\RR \rar \RR$ by $\pi_e.\ve=1$.
Thus $\pi_e$ is a measuring instrument, which gives
$s=\pi_e.\vu=$ the size of the child in foot, \ie\ $\vu = s\ve$.
\finexa

\debexe
\label{exebd}
Let $(\va_i)$ and $(\vb_i)$ be bases and let $(\pi_{ai})$ and $(\pi_{bi})$ be the dual bases.
Let $\lambda\ne0$. Prove:
\be
\label{eqexebd}
\hbox{If }\; \forall i=1,...,n,\; \vb_i=\lambda \va_i,
\qthen \forall i=1,...,n,\; \pi_{bi} = {1\over \lambda}\,\pi_{ai}.
\ee
(With duality notations, $b^i = {1\over \lambda}\,a^i$.)
%And recover $\pi_{bi}(\vb_j)=\pi_{ai}(\va_j)=\delta_{ij}=\delta^i_j =b^i(\vb_j)=a^i(\va_j)=\delta^i_j$. 

\debrep
$\pi_{bi}.\vb_j = \delta_{ij} = \pi_{ai}.\va_j
= \pi_{ai}.{\vb_j\over \lambda}= {1\over \lambda}\,\pi_{ai}.\vb_j$ for all~$j$ (since $\pi_{ai}$ is linear),
thus $\pi_{bi} = {1\over \lambda}\,\pi_{ai}$, true for all~$i$. %, \ie~\eref{eqexebd}.
%And then $b^i(\vb_j)=({1\over \lambda}\,a^i)(\lambda \va_j)={1\over \lambda}\lambda\,a^i(\va_j) = a^i(\va_j)$.
\finrep
\finexe

%%%%%%%%%%%%%%%%%%%%%%%%%%%%%%%%%%%%%%%%%%%%%%%%%%%%%%%%%%%%%%%%%%%%%%%%%%%%%%%%%%%

\subsubsection{Example: aeronautical units}

\debexa
\label{exaavion1}
International aeronautical units:
Horizontal length = nautical mile (NM),
altitude = English foot (ft).
% (It is NOT the metre, and a pilot knows that: Fundamental at the time of landing!).
Application: An air traffic controller chooses the point $O=$ the position of its control tower,
and a plane $p$ is located thanks to the bipoint vector $\vx = \ora{\calO p}$.
And the traffic controller chooses
$\ve_1=$ the vector of length $1$~NM oriented South (first runway),
$\ve_2=$ the vector of length $1$~NM oriented Southwest (second runway),
$\ve_3=$ the vertical vector of length $1$~ft.
Thus his referential is $\calR =(\calO ,(\ve_1,\ve_2, \ve_3))$,
and his dual basis $(\pi_{e1},\pi_{e2},\pi_{e3})$ is defined by $\pi_{ei}(\ve_j)=\delta_{ij}$ for all~$i,j$, \cf~\eref{eqdefbd}. He writes $\vx = \sumin x_i\ve_i \in \vRRn$, so that
$x_1=\pi_{e1}(\vx)=$ the distance to the south in~NM,
$x_2=\pi_{e2}(\vx)=$  the distance to the southwest in~NM,
$x_3=\pi_{e3}(\vx)=$  the altitude in~ft.
%(If you prefer, with duality notations, $\pi_{ei}{=}e^i$ and $\vx = \sumin x^i\ve_i$ and $x^i=e^i.\vx$.)

Here the basis $(\ve_i)$ is not a Euclidean basis.
This non Euclidean basis $(\ve_i)$ is however vital if you take a plane.
(A Euclidean basis is not essential to life...). See next remark~\ref{remMCOC}.
\finexa

\debrem
\label{remMCOC}
The \metre\ is the international unit for NASA that launched the Mars Climate Orbiter probe,
and the \foot\ is the international vertical unit for aviation;
And for the Mars Climate Orbiter landing procedure,
NASA (uses the metre) asked Lockheed Martin (uses the foot) to do the computation.
%Not the \foot\ used by Boeing (aviation) for landing computation.
Result? The Mars Climate Orbiter space probe burned in the Martian atmosphere
because of $\lambda \sim3$ times too high a speed during the landing procedure:
One metre is $\lambda\sim3$ times one foot, and someone forgot it...
Although NASA and Lockheed Martin used a Euclidean dot product... But not the same (one based on a metre, and one based on the foot). Objectivity and covariance can be useful... %, see~\S~\ref{secbe}.
\finrem

%%%%%%%%%%%%%%%%%%%%%%%%%%%%%%%%%%%%%%%%%%%%%%%%%%%%%%%%%%%%%%%%%%%%%%%%%%%%%%%%%%%

\subsubsection{Matrix representation of a linear form}

Let $\ell\in E^*$, let $(\ve_i)$ be a basis: The components of~$\ell$ are the $n$ reals % $\ell_i=\ell(\ve_i)=\ell.\ve_i$, \cf\ definition~\ref{defpropLab0}:
\be
\label{eqxi2}
\ell_i:=\ell(\ve_i) = \ell.\ve_i,\qand [\ell]_{|\ve} = \pmatrix{\ell_1 & ... & \ell_n}
\ee
is the row matrix of~$\ell$, called the matrix of~$\ell$ relative to~$(\ve_i)$. Thus, if $\vx\in E$
and $\vx \eqclas \sumin x_i\ve_i \eqdual \sumin x^i\ve_i$, then
\be
\label{eqxi3}
\ell.\vx
\eqclas \sumin \ell_i x_i
\eqdual \sumin \ell_i x^i  = [\ell]_{|\ve}.[\vx]_{|\ve}
\ee
with usual matrix computation rules (a $1*n$ matrix times a $n*1$ matrix).

In particular for the dual basis $(\pi_{ei}) = (e^i)$ (classical and duality notations),
\be
\label{eqdefbd1}
[\pi_{ej}]_{|\ve}=[e^j]_{|\ve} =(0\;\; ... \;\;0 \;\; \underbrace{1}_{\makebox[1cm]{\footnotesize\rm $j$th position}} \;\; 0 \;\; ... \;\; 0) \quad(\hbox{= row matrix } = [\ve_j]_{|\ve}^T).   
\ee
Thus we have, with classical and duality notations,
\be
\label{eqxi1}
\ell
\eqclas\sumin \ell_i\,\pi_{ei}
\eqdual\sumin \ell_i\,e^i.
\ee

\debrem
\label{remconve}
Relative to a basis,
a vector is represented by a column matrix, \cf~\eref{eqxi0},
and a linear form by a row matrix, \cf~\eref{eqxi2}. This enables:

$\bullet$ The use of matrix calculation to compute
$\ell.\vx=[\ell]_{|\ve}.[\vx]_{|\ve}$, \cf~\eref{eqxi3}, not to be confused with an inner dot product calculation
$\vx\bcdot\vy$ relative to an inner dot product in~$E$ for $\vx,\vy\in E$.% (``product of two column matrix'' which depends on the choice of some inner dot product),

$\bullet$ Not to confuse the ``nature of objects'': Relative to a basis,
a (contravariant) vector is a mathematical object represented by a column matrix,
while a linear form (covariant vector) is a mathematical object represented by a row matrix.
\Cf\ remark~\ref{remMisner}.
\finrem

\comment{
\debrem
In elementary courses, a classical notation often used is, with $\vx=\sumin x_i\ve_i$,
\be
\ell.\vx = a_1 x_1+...+ a_n x_n.
\ee
And $a_i\eqnamed \ell_i$ in~\eref{eqxi1}. (And the graph of a linear form $\ell \in E^*$ is a hyperplane)
\finrem
}

\comment{
\debrem
%\label{exa112}
{\bf Fundamental:} 
We always have $e^i(\ve_j) = \delta^i_j$, \cf~\eref{eqdefbd}, and there is no use of any inner dot product to define~$e^i$.
%, and Einstein convention is satisfied for all basis, \eg, $a^1(\va_1) = b^1(\vb_1)$ ($ = 1$), see~\S~\ref{secEconv1}.

Whereas if $\dd_g$ is an inner dot product, then $(\va_i,\va_j)_g \ne (\vb_1,\vb_1)_g$ in general.
\Eg\ with $\dd_a = \lambda^2\dd_b$ and $\lambda\ne1$.
\Eg, $(\va_i)$ is the Euclidean basis made by an English observer using the foot
and $\dd_a$ is the associated Euclidean dot product,
and $(\vb_i)$ is a Euclidean basis made by a French observer using the metre
and $\dd_b$ is the associated Euclidean dot product;
Then with~$\dd_g=\dd_a$ we get $||\va_i||^2_g=(\va_1,\va_1)_g = 1$ since $\va_1$ is 1 foot long,
when $||\vb_i||^2_g=(\vb_1,\vb_1)_g >10$ since $\vb_1$ is 1 metre long and $1$~metre $\simeq 3.28$~foot.
Here $(\vb_1,\vb_1)_g\ne (\va_1,\va_1)_g$ although $a^1(\va_1) = b^1(\vb_1)$ ($ = 1$).
\finrem
}

%%%%%%%%%%%%%%%%%%%%%%%%%%%%%%%%%%%%%%%%%%%%%%%%%%%%%%%%%%%%%%%%%%%%%%%%%%%%%%%%%%%

\subsubsection{Example: Thermodynamic}
\label{secthermo}

Consider the Cartesian space $\vRRd=\{(T,P)\in\RR\times \RR\}= \{$(temperature,pressure)$\}$.
There is no meaningful inner dot product in this $\vRRd$: What would $\sqrt{T^2{+}P^2}$ mean (Pythagoras: Can you add Kelvin degrees and kg/(m·$s^2$)? Thus, in thermodynamics, the (covariant) dual bases are the main ingredient for calculations.

\Eg, in the Cartesian product $\vRRd=\RR\times \RR$ consider the basis $(\vE_1{=}(1,0),\vE_2{=}(0,1))$ (after a choice of temperature and pressure units); Let $\vX\in \vRRd$, $\vX= T\vE_1+P\vE_2 \eqnote (T,P)$, and let $(\pi_{E1},\pi_{E2})=(E^1,E^2)\eqnote(dT,dP)$ be the (covariant) dual basis.
The first principle of thermodynamics tells that the density $\alpha$ of internal energy is an exact differential form: $\exists U\in C^1(\vRRd;\RR)$ \st\ $\alpha=dU$. So, at any $\vX_0=(T_0, P_0)$,
\be
\label{eqthermo1}
\alpha(\vX_0) = dU(\vX_0) ={\pa U \over\pa T}(\vX_0)\,dT+ {\pa U \over\pa P}(\vX_0)\,dP
\qand [dU(\vX_0)]_{|\vE} = \pmatrix{{\pa U \over\pa T}(\vX_0) & {\pa U \over\pa P}(\vX_0)}
\ee
(row matrix).%
\comment{
And the variation rate at $\vX_0=(T_0, P_0)$ in the direction $\vV=\delta T\vE_1+\delta P\vE_2 = (\delta T,\delta P)$ in $\vRRd$ is
\be
\label{eqthermo1c}
dU(\vX_0).\vV ={\pa U \over\pa T}(T_0, P_0)\,\delta T+ {\pa U \over\pa P}(T_0, P_0)\,\delta P.
\ee
}
And we have the first order Taylor expansion in the vicinity of $\vX_0$,
% \ie, with $\delta X=h\Delta \vX$ and in the vicinity of $h{=}0$):
\be
\eqalign{
U(\vX_0+\delta\vX)
= & U(\vX_0) + dU(\vX_0).\delta \vX + o(\delta \vX ) \cr
= & U(T_0,P_0) + \delta T{\pa U \over\pa T}(T_0,P_0) + \delta P{\pa U \over\pa T}(T_0,P_0) + o((\delta T,\delta P)).
}
\ee

\noindent
{\bf Matrix computation:} Column matrices for vectors, row matrices for linear forms:
\be
[\vE_1]_{|\vE} = \pmatrix{1 \cr 0},
\quad [\vE_2]_{|\vE} = \pmatrix{0 \cr 1},
\quad [\vX_0]_{|\vE} = \pmatrix{T_0 \cr P_0},
\quad [\delta\vX]_{|\vE}
= \pmatrix{\delta T \cr \delta P},
\ee
\be
[E^1]_{|\vE} = [dT]_{|\vE} 
= \pmatrix{1 & 0},\quad
[E^2]_{|\vE} = [dP]_{|\vE}
= \pmatrix{0 & 1} ,\quad
[dU]_{|\vE} = \pmatrix{{\pa U \over\pa T} & {\pa U \over\pa P}}
% \quad\hbox{(row matrices)},
\ee
give
\be
\eqalign{
dU(\vX_0).\delta\vX
= & \pmatrix{{\pa U \over\pa T}(\vX_0) & {\pa U \over\pa P}(\vX_0)}.\pmatrix{\delta T \cr \delta P}
= {\pa U \over\pa T}(\vX_0)\delta T + {\pa U \over\pa P}(\vX_0)\delta P.
}
\ee
This is a ``covariant calculation'' (in particular no inner dot product has been used).

\comment{
\debrem
The first principle of thermodynamics states that an ``internal state energy''~$U$ must exist. And because the mechanical work $W$ is not a state function (it depends on the trajectory which connects two points), we the heat $Q$ is defined to be $Q := U-W$.
\finrem
}

%%%%%%%%%%%%%%%%%%%%%%%%%%%%%%%%%%%%%%%%%%%%%%%%%%%%%%%%%%%%%%%%%%%%%%%%%%%%%%%%%%%

\subsection{Einstein convention}
\label{secEconv}

%%%%%%%%%%%%%%%%%%%%%%%%%%%%%%%%%%%%%%%%%%%%%%%%%%%%%%%%%%%%%%%%%%%%%%%%%%%%%%%%%%%

\subsubsection{Definition}
\label{secEconv1}

%When, for some mathematical object (qualitative approach), you work with its components (quantification) after having chosen some bases:

When you work with components (after a choice of a basis), the goal is to visually differentiate a linear form from a vector (to visually differentiate covariance from contravariance).
Framework: a finite dimension vector space $E$, $\dim E=n$, and duality notations.

%\goodbreak
\mn
{\bf Einstein Convention:}
%\smallskip
\begin{enumerate}[leftmargin=5pt ,parsep=3pt plus 1pt minus 1pt,itemsep=0cm,topsep=0cm]

\item 
A basis in~$E$ (contravariant) is written with bottom indices: \Eg, $(\ve_i)$ is a basis in~$E$.

\item 
A vector $\vx\in E$ (contravariant) has its components relative to~$(\ve_i)$ (quantification) written with top indices: $\vx = \sumin x^i \ve_i$,
and is represented by the column matrix $[\vx]_{|\ve} =\pmatrix{x^1\cr\vdots\cr x^n}$.
(Classical notations: $\vx = \sumin x_i \ve_i$, and column matrix of $x_i$.)

\item 
A basis in~$E^*=\calL(E;\RR)$ (covariant) is written with top indices: \Eg, $(e^i)\in \Es^n$ is the dual basis of the basis~$(\ve_i)$. (Classical notations: $(\pi_{ei})$.)

\item 
A linear form $\ell \in E^*$ (covariant) has its components relative to~$(e^i)$ (quantification) written with bottom indices: $\ell = \sumin \ell_i e^i$,
and its matrix representation is the row matrix $[\ell]_{|\ve} = \pmatrix{\ell_1 & ... & \ell_n}$.

\item 
You can also omit the sum sign~$\sum$ when there are repeated indices at a different position; \Eg\ 
$\sumin x^i\ve_i \eqnote x^i\ve_i $, and 
$\sumin \Lij \ve_i \eqnote L^i_j\ve_i$. In fact, before computers and word processors, to print $\sumin$ was not easy. But with \LaTeX\ this is no more a problem, so in this manuscript the sum sign~$\sum$ is not omitted (and some confusions are avoided).
%: To insist on the difference between covariance and contravariance, see remark~\ref{remMisner}.
%(And the sum sign $\int$ is not omitted either...)

\end{enumerate}

\debrem
Einstein's convention is not mandatory. \Eg\ Arnold doesn't use it when he doesn't need~it, or when it makes reading difficult, or when it induces misunderstandings. In classical mechanics, Einstein's convention may induce more confusion than understandings, and may be misused... so it is better \textslbf{not} to use it:
Golden rule: Use classical notations when in doubt.
\finrem

%%%%%%%%%%%%%%%%%%%%%%%%%%%%%%%%%%%%%%%%%%%%%%%%%%%%%%%%%%%%%%%%%%%%%%%%%%%%%%%%%%%

\subsubsection{Do not mistake yourself}

%\begin{enumerate}[leftmargin=* ,parsep=3pt plus 1pt minus 1pt,itemsep=0cm,topsep=0cm]
\begin{enumerate}[leftmargin=0pt ,parsep=3pt plus 1pt minus 1pt,itemsep=0cm,topsep=0cm]

\item Einstein's convention is just meant not to confuse a linear function with a vector.

\item It only deals with quantification relative to a basis.
%: The same mathematical object has multiple components representations and notations, see \eg\ the classical or duality notations.

\item Classical notations are as good as duality notations, even you are told that classical notations cannot detect obvious errors in component manipulations... But duality notations can be misused in classical mechanics (\cf\ the paradigmatic example of the vectorial dual basis, correctly treated at~\S~\ref{secvectdb}); And thus add confusion to the confusion.
%So keep using the classical notations at first.

\item The convention does not admit shortcuts; \Eg\ with a metric: $g(\vu,\vv) = \sumijn g_{ij}u^iv^j$ shows the observer dependence on a choice of a basis thanks to the $g_{ij}$; And even if $g_{ij}=\delta_{ij}$ you can\textslbf{not} write $g(\vu,\vv) = \sumijn u^iv^j$: You have to write $g(\vu,\vv) = \sumijn g_{ij}u^iv^j$: Unmissable in physics because you need to see the metric and bases in use.

\comment{
\item Einstein's convention can be satisfied while the results are meaningless.
Trivial example: $\sumin \ell_i w^i$ when $\ell=\sumin \ell_i e^i\in E^*$ is a linear form on~$E$
and $\vw=\sumin w^i \vb_i\in F$ is some vector with $F\ne E$, as can be the case
with the deformation gradient (absurd results which are taken for granted...). 
}

\item Golden rule: Return to classical notations if in doubt.
(Einstein's convention can add confusions, untruths, misinterpretations, absurdities, misuses...)
\end{enumerate}

\comment{

\medskip
Remark: In general relativity, even if $g_{ij} = \delta_{ij}$ at some point, we always write $(\vx,\vy)_g = \sumijn x^i g_{ij} y^j$, for the simple reason that: If we have $g_{ij}=\delta_{ij}$ at some point, we can also have $g_{ij}\ne \delta_{ij}$ at some other point, even if $\dd_g$ is some usual Riemannian metric, because the basis $(\ve_j)$ is most of the time a coordinate basis which thus is not $\dd_g$-orthonormal.
When in classical mechanics, with a {\sl a priori} given inner dot product, we often write $(\vx,\vy)_g = \sumin x^iy^i$ when $[g]=I$, the result being meaningful, also written $(\vx,\vy)_g = \sumin x_iy_i$ with classical notations.

}

%Thus the point 2- above is to be forgotten.

%%%%%%%%%%%%%%%%%%%%%%%%%%%%%%%%%%%%%%%%%%%%%%%%%%%%%%%%%%%%%%%%%%%%%%%%%%%%%%%%%%%

\subsection{Change of basis formulas}
\label{secfcb}

$E$ being a finite dimension vector space, $\dim E = n$, let $(\veio)$ and $(\vein)$ be two bases in~$E$,
and let $(\pi_{\old,i})$ and $(\pi_{\new,i})$ be the dual bases in $E^*$, written $(\eio)$ and $(\ein)$ with duality notations. % (and Einstein's convention).
%and let $(\pi_{\old,i}) = (\eio)$ and $(\pi_{\new,i}) = (\ein)$ be the dual bases in $E^*$ (classical and duality notations).

%%%%%%%%%%%%%%%%%%%%%%%%%%%%%%%%%%%%%%%%%%%%%%%%%%%%%%%%%%%%%%%%%%%%%%%%%%%%%%%%%%%

\subsubsection{Change of basis endomorphism and transition matrix}

\debdef
The change of basis endomorphism $\calP\in\calL(E;E)$ from $(\veio)$ to~$(\vein)$ is 
the endomorphism (= the linear map $E\rar E$) defined by, for all $j\in[1,n]_\NN$,
\be
\label{eqdefP0}
\boxed{\calP.\vejo = \vejn}.
\ee
\findef

\debdef
The transition matrix from $(\veio)$ to~$(\vein)$ is the matrix
$P:= [\calP]_{|\veo}=[P_{ij}]$ of the endomorphism~$\calP$ relative to the basis~$(\veio)$,
\ie\ defined by, for all~$j$,
\be
\label{eqvaon}
%\forall j,\;\; 
\vejn = \calP.\vejo  = \sumin P_{ij}\,\veio, \qie
[\vejn]_{|\veo} = P.[\vejo]_{|\veo}
= \pmatrix{\ds P_{1j} \cr \vdots \cr \ds P_{nj}},
\ee
\ie, $[\vejn]_{|\veo}$ is the $j$-th column of $P= [\calP]_{|\veo}$.
Duality notations: $\vejn=\sumin \Pij\,\veio$ and $P:= [\calP]_{|\veo}=[\Pij]$.
\findef

Apart from the classical and notations, you may find other ``component type'' notations:
\be
\vejn = \sumin P_{ij}\,\veio = \sumin (P_j)_i\,\veio = \sumin \Pij\,\veio = \sumin (P_j)^i\,\veio,
\ee
\ie\ $P_{ij} = (P_j)_i = \Pij = (P_j)^i$ are four notations for the $i$-th component of~$\ve_j$,
\ie
\be
\label{eqaltnotvejn}
[\vejn]_{|\veo}
= \pmatrix{\ds P_{1j} \cr \vdots \cr \ds P_{nj}}
= \pmatrix{\ds (P_j)_1 \cr \vdots \cr \ds (P_j)_n}
= \pmatrix{\ds P^1{}_j \cr \vdots \cr \ds P^n{}_j}
= \pmatrix{\ds (P_j)^1 \cr \vdots \cr \ds (P_j)^n}
 \quad\hbox{(= the $j$-th column of $P$)}.
\ee

%%%%%%%%%%%%%%%%%%%%%%%%%%%%%%%%%%%%%%%%%%%%%%%%%%%%%%%%%%%%%%%%%%%%%%%%%%%%%%%%%%%

\subsubsection{Inverse of the transition matrix}

The inverse endomorphism $\calQ := \calP^{-1} \in\calL(E;E)$, \cf~\eref{eqdefP0}, is given by, for all $j\in[1,n]_\NN$,
\be
\label{eqdefQ00}
\vejo = \calQ.\vejn \quad ( = \calP^{-1}.\vejn),
\ee
\ie\ $\calQ$ is change of basis endomorphism from $(\vein)$ to~$(\veio)$.
And $\ds Q:=[\calQ]_{|\ven} = [Q_{ij}] $ is the transition matrix from $(\vein)$ to~$(\veio)$: %for all $j\in[1,n]_\NN$,
\be
\label{eqdefQ0}
\vejo %= \calQ.\vejn 
= \sumin Q_{ij}\vein  ,\quad 
[\vejo]_{|\ven} = \pmatrix{
Q_{1j}\cr \vdots \cr Q_{nj}}.
\ee
\ie\ $\calQ$ is change of basis endomorphism from $(\vein)$ to~$(\veio)$.

Use other notation if you prefer: $Q_{ij}=(Q_j)_i = \Qij = (Q_j)^i$

\debprop
\label{propdefP0i0}
\be
\label{eqdefP0i0}
Q=P^{-1}.
\ee
\finprop

\debdem
$\vejn = \calP.\vejo
= \sumin P_{ij} \veio
= \sumin P_{ij} (\sumkn Q_{ki} \vekn)
= \sumkn (\sumin Q_{ki} P_{ij}) \vekn
= \sumkn (Q.P)_{kj} \vekn
$ for all~$j$,
thus $(Q.P)_{kj} = \delta_{kj}$ for all~$j,k$. Hence $Q.P=I$, \ie~\eref{eqdefP0i0}.
\findem

\debexe
Prove $
\left\{\eqalign{
&[\calP]_{|\veo} = [\calP]_{|\ven} = P, \cr
&[\calQ]_{|\ven} = [\calQ]_{|\veo} = Q, \cr
}\right\}$, 
\ie
\be
\label{eqdefP0i}
\left\{\eqalign{
& \textstyle\calP.\vejn = \sumijn P_{ij} \vein \quad ( = \sumijn \Pij \vein = \sumijn (P_j)^i \vein), \cr
& \textstyle\calQ.\vejo = \sumijn Q_{ij} \veio \quad ( = \sumijn \Qij \veio= \sumijn (Q_j)^i\veio). \cr
}\right.
\ee

\debrep
$Z = [Z_{ij}] = [\calP]_{|\ven}$ means $\calP.\vejn = \sum_i Z_{ij} \vein$, \ie\
$\vejn 
= \calQ.(\sumin Z_{ij} \vein)
= \sumin Z_{ij} \calQ.\vein
= \sumin Z_{ij} (\sumkn Q_{ki} \vekn)
= \sumkn(\sumin Q_{ki} Z_{ij})\vekn
= \sumkn (Q.Z)_{kj}\vekn$ for all~$j$,
thus $(Q.Z)_{kj} =  \delta_{kj}$ for all~$j,k$, thus $Q.Z=I$, thus $Z=P$.
Idem for~$Q$, thus~\eref{eqdefP0i}.
\comment{
And $Z = [Z^i{}_j] = [\calQ]_{|\veo}$ means $\calQ.\vejo = \sum_i Z^i{}_j \veio$, \ie\
$\vejo = \calP.\sum_i Z^i{}_j \veio
= \sum_i Z^i{}_j (\sum_k P^k{}_i \veko)
= \sum_k(\sum_i P^k{}_iZ^i{}_j)\veko
= \sum_k (P.Z)^k{}_j\veko$,
thus $(P.Z)^k{}_j =  \delta^k_j$,
true for all~$j$, thus $P.Z=I$, thus $Z=Q$.
Idem for~$P$, thus~\eref{eqdefP0i}.

And $Z = [Z^i{}_j] = [\calP]_{|\ven}$ means $\calP.\vejn = \sum_i Z^i{}_j \vein$, \ie\
$\vejn = \calQ.\sum_i Z^i{}_j \vein
= \sum_i Z^i{}_j (\sum_k Q^k{}_i \vekn)
= \sum_k(\sum_i Q^k{}_iZ^i{}_j)\vekn
= \sum_k (Q.Z)^k{}_j\vekn$,
thus $(Q.Z)^k{}_j =  \delta^k_j$,
true for all~$j$, thus $Q.Z=I$, thus $Z=P$.
Idem for~$Q$. %, thus~\eref{eqdefP0i}.
}
\finrep
\finexe

\debrem
$P^T \ne P^{-1}$ in general.
\Eg, $(\veio) = (\va_i)$ is a foot-built Euclidean basis,
and $(\vein) = (\vb_i) $ is a metre-built Euclidean basis,
and $\vb_i = \lambda \va_i$ for all~$i$ (the basis are ``aligned''), so $P=\lambda I$; Thus $P^T= \lambda I$ and $P^{-1} = {1\over \lambda} I \ne P^T$,
since $\lambda ={1\over 0.3048}\ne 1$.
Thus it is essential not to confuse $P^T$ and~$P^{-1}$ (not to confuse covariance with contravariance), %if an English observer works with a French observer,
\cf\ \eg\ the Mars Climate Orbiter crash (remark~\ref{remMCOC}).
\finrem

%%%%%%%%%%%%%%%%%%%%%%%%%%%%%%%%%%%%%%%%%%%%%%%%%%%%%%%%%%%%%%%%%%%%%%%%%%%%%%%%%%%

\subsubsection{Change of dual basis}

\def\piin{\pi_{\new,i}}
\def\pijn{\pi_{\new,j}}
\def\piio{\pi_{\old,i}}
\def\pijo{\pi_{\old,j}}

\debprop
$(\piin)$ and $(\piio)$ being the dual bases of~$(\vein)$ and $(\veio)$, for all $i\in[1,n]_\NN$,
\be
\label{eqdefP2}
\piin=\sumjn Q_{ij} \pi_{\old,i},   \qand
[\piin]_{|\veo} = \pmatrix{ Q_{i1} & ... & Q_{in}}\quad\hbox{($i$-th row of $Q$)},
\ee
to compare with~\eref{eqvaon} (matrices of linear forms are row matrices).

Duality notations: $\ein = \sumjn \Qij \ejo$ and $[\ein]_{|\veo} = \pmatrix{ Q^i{}_1 & ... & Q^i{}_n}$.
\finprop

\debdem
$\piin(\veko) 
\mathop{=}^{\eref{eqdefQ0}} \piin(\sum_j Q_{jk}\vejn )
= \sum_j Q_{jk}\,\piin(\vejn) 
= \sum_j Q_{jk}\,\delta_{ij}
= Q_{ik}$,
and
$\sum_j Q_{ij} \pijo(\veko)
= \sum_j Q_{ij} \delta_{jk}
= Q_{ik}
$, 
true for all $i,k$,
thus $\piin =\sum_j Q_{ij}$,
\ie~\eref{eqdefP2}
\findem

\comment{
\debdem
On the one hand
$\piin. \veko 
\mathop{=}^{\eref{eqdefQ0}} \piin.(\sum_j Q^j{}_k\vejn )
= \sum_j Q^j{}_k\,(\piin.\vejn) 
= \sum_j Q^j{}_k\,\delta^i_j = Q^i{}_k
$,
and on the other hand
$\piin. \veko
=(\sum_j \Qij \pijo).\veko 
= \sum_j \Qij (\pijo.\veko)
= \sum_j \Qij \delta^j_k 
= Q^i{}_k
$, 
true for all $i,k$,
hence~\eref{eqdefP2}. Similarly we get~\eref{eqdefP2b}.
\comment{
On the one hand
$\ein. \veko 
\mathop{=}^{\eref{eqdefQ0}} \ein.(\sum_j Q^j{}_k\vejn )
= \sum_j Q^j{}_k\,(\ein.\vejn) 
= \sum_j Q^j{}_k\,\delta^i_j = Q^i{}_k
$,
and on the other hand
$(\sum_j \Qij \ejo).\veko 
= \sum_j \Qij (\ejo.\veko)
= \sum_j \Qij \delta^j_k 
= Q^i{}_k
$, 
true for all $i,k$,
hence~\eref{eqdefP2}. Similarly we get~\eref{eqdefP2b}.
}
\findem
}

%%%%%%%%%%%%%%%%%%%%%%%%%%%%%%%%%%%%%%%%%%%%%%%%%%%%%%%%%%%%%%%%%%%%%%%%%%%%%%%%%%%

\subsubsection{Change of coordinate system for vectors and linear forms}

\debprop
Let $\vx\in E$ and $\ell\in \Es$. Then
\be
\label{eqdefP1}
\eqalign{
& \bullet \; [\vx]_{|\ven} = P^{-1}.[\vx]_{|\veo} \quad \hbox{(contravariance formula for vectors: between column matrices)}, \cr
& \bullet \; [\ell]_{|\ven} = [\ell]_{|\veo}.P \qquad \hbox{(covariance formula for linear forms: between row matrices)}.
}
\ee
%(The contravariance formula deals with column matrices, the covariance formula deals with row matrices.)
And the scalar value $\ell.\vx$ %(dimensionless result of the measure of~$\vx$ by~$\ell$) 
is computed indifferently with one or the other basis (objective result):
\be
\label{eqdefP1b}
\ell.\vx = [\ell]_{|\veo}.[\vx]_{|\veo} = [\ell]_{|\ven}.[\vx]_{|\ven}.
\ee
\finprop

\debdem
Let $\vx = \sum_j x_j\vejo = \sum_i y_i\vein$.
%, \ie\$[\vx]_{|\veo} = \pmatrix{x_1 \cr \vdots \cr x_n}$ and $[\vx]_{|\ven} = \pmatrix{y_1 \cr \vdots \cr y_n}$.
We have $\vx = \sum_j x_j\vejo = \sum_j x_j (\sumin Q_{ij}\vein) = \sum_{ij} Q_{ij} x_j\vein$,
thus $y_i= \sum_{j} Q_{ij} x_j$ for all~$i$, thus~\eref{eqdefP1}$_1$.

And 
%$[\ell]_{|\veo} = \pmatrix{\ell_1 & ... & \ell_n}$, and $[\ell]_{|\ven} = \pmatrix{m_1 & ... & m_n}$, we have
$\ell 
=\sum_j m_j \pijn
= \sum_i \ell_i \piio
\mathop{=}^{\eref{eqdefP2}} \sum_{ij} \ell_i P_{ij} \pijn $
gives $m_j = \sum_i\ell_i P_{ij}$ for all~$j$, thus~\eref{eqdefP1}$_2$.
(Use duality notations if you prefer.)

Thus
%$[\ell]_{|\veo}.[\vx]_{|\veo} = \ell.\vx = [\ell]_{|\ven}.[\vx]_{|\ven}$, thus~\eref{eqdefP1b} (or \eref{eqdefP1} gives
$[\ell]_{|\ven}.[\vx]_{|\ven}
= ([\ell]_{|\veo}.P).(P^{-1}.[\vx]_{|\veo} )
%= [\ell]_{|\veo}.(P.P^{-1}).[\vx]_{|\veo} 
= [\ell]_{|\veo}.[\vx]_{|\veo}$,
hence~\eref{eqdefP1b}.
\findem

\def\xin{{x^i_\new}}
\def\xjn{{x^j_\new}}
\def\xio{{x^i_\old}}
\def\xjo{{x^j_\old}}

\noindent
{\bf Notation:}
\eref{eqdefP1} and $\vx = \sum_j x_j\vejo = \sum_i y_i\vein$ give $y_i = \sumjn Q_{ij} x_j$,
which means: $y_i$ is the function defined by $y_i(x_1,...,x_n) = \sumjn  Q_{ij} x_j$,
thus $Q_{ij} = {\pa y_i \over \pa x_j}(x_1,...,x_n)$; Similarly with $P_{ij}$; Which is written
\be
\label{eqnotePij}
Q_{ij} = {\pa y_i \over \pa x_j},\qand P_{ij} = {\pa x_i \over \pa y_j}.
\ee
(Use duality notations if you prefer, \eg\ $\Qij = {\pa y^i \over \pa x^j}$.)

\comment{
\noindent
{\bf Notation:}
\eref{eqdefP1}  gives $\xin = \sumjn  \Qij \xjo$ and $\xio = \sumjn  \Pij \xjn$, and thus the notation
\be
\label{eqnotePij}
\Qij = {\pa \xin \over \pa \xjo},\qand \Pij = {\pa \xio \over \pa \xjn}.
\ee
Full notation: $\xin$ is the function defined by $\xin(x^1_\old,...,x^n_\old) = \sumjn  \Qij \xjo$,
thus $\Qij = {\pa \xin \over \pa \xjo}(x^1_\old,...,x^n_\old)$,
idem $\xio(x^1_\new,...,x^n_\new) = \sumjn  \Pij \xjn$
gives $\Pij = {\pa \xio \over \pa \xjn}(x^1_\new,...,x^n_\new)$,
}

\comment{

%%%%%%%%%%%%%%%%%%%%%%%%%%%%%%%%%%%%%%%%%%%%%%%%%%%%%%%%%%%%%%%%%%%%%%%%%%%%%%%%%%%

\subsection{Tensorial product, multilinear forms, contraction}

%%%%%%%%%%%%%%%%%%%%%%%%%%%%%%%%%%%%%%%%%%%%%%%%%%%%%%%%%%%%%%%%%%%%%%%%%%%%%%%%%%%

\subsubsection{Tensorial product}

\debdef
The tensorial product of $n$ functions $f_i:A_i\rar \RR$, $i=1,...,n$, is the function
$f_1\otimes...\otimes f_n : A_1 \times...\times A_n \rar \RR$ (scalar valued) defined by
\be
\label{eqdefpt}
%\forall (\vx_1,...,\vx_n) \in A_1\times...\times A_n,\quad
(f_1\otimes... \otimes f_n)(\vx_1,...,\vx_n) = f_1(\vx_1)...f_n(\vx_n),
\ee
and is called a function with separated variables.
\findef

\Eg, $n{=}2$, $A_1{=}\RR$, $A_2{=}\RR^*$, $f_1(x)=\cos x$, $f_2(x)={1\over x}$ give $(f\otimes g)(x,y)={\cos(x) \over y}$.

%%%%%%%%%%%%%%%%%%%%%%%%%%%%%%%%%%%%%%%%%%%%%%%%%%%%%%%%%%%%%%%%%%%%%%%%%%%%%%%%%%%

\subsubsection{Multilinear forms}

\debdef
$A_1,...,A_n$ being $n$ vector spaces, a $n$-multilinear form on $A_1,...,A_n$ is a function $M : A_1 \times ... \times A_n \rar \RR$ (scalar valued) \st,
for all $i=1,...,n$, all $\vx_i,\vy_i\in A_i$ and all $\lambda\in\RR$,
\be
M(...,\vx_i+\lambda \vy_i,...) = M(...,\vx_i,...) + \lambda\; M(...,\vy_i,...),
\ee
the other variables being unchanged.
And $\calL(A_1,...,A_n ; \RR)$ be the set of $n$-multilinear forms on the Cartesian product $A_1\times ...\times A_n$.
(It is a vector space, sub-space of $\calF(A_1 \times...\times A_n;\RR)$: trivial).
\findef

%%%%%%%%%%%%%%%%%%%%%%%%%%%%%%%%%%%%%%%%%%%%%%%%%%%%%%%%%%%%%%%%%%%%%%%%%%%%%%%%%%%

\subsubsection{Multilinear forms}

\debdef
The tensorial product of $n$ functions $f_i:A_i\rar \RR$, $i=1,...,n$, is the function
$f_1\otimes...\otimes f_n : A_1 \times...\times A_n \rar \RR$ defined by
\be
\label{eqdefpt}
%\forall (\vx_1,...,\vx_n) \in A_1\times...\times A_n,\quad
(f_1\otimes... \otimes f_n)(\vx_1,...,\vx_n) = f_1(\vx_1)...f_n(\vx_n),
\ee
and is called a function with separated variables.
\findef

\Eg, $n{=}2$, $A_1{=}\RR$, $A_2{=}\RR^*$, $f_1(x)=\cos x$, $f_2(x)={1\over x}$ give $(f\otimes g)(x,y)={\cos(x) \over y}$.

%%%%%%%%%%%%%%%%%%%%%%%%%%%%%%%%%%%%%%%%%%%%%%%%%%%%%%%%%%%%%%%%%%%%%%%%%%%%%%%%%%%

\subsubsection{... and contraction}

The notion of can lead to confusions, so pay attention...

\debdef
The contraction of the tensorial product $f_1\otimes... \otimes f_n : A_1 \times...\times A_n \rar \RR$
with a $\vx_n\in A_n$ is defined by
\be
(f_1\otimes... \otimes f_n)(\vx_n) := f_n(\vx_n) f_1\otimes... \otimes f_{n-1} : A_1 \times...\times A_{n-1} \rar \RR.
\ee
More generally, the contraction of the tensorial product $f_1\otimes... \otimes f_n :  : A_1 \times...\times A_n \rar \RR$
with a $\vx_i\in A_i$ is defined by
\be
(f_1\otimes... \otimes f_n)(\vx_i) := f_i(\vx_i) f_1\otimes...f_{i-1}\otimes f_{i+1}... \otimes f_{n-1} .
\ee
\findef

\debdef
The contraction of the tensorial product $f_1\otimes... \otimes f_n : A_1 \times...\times A_n \rar \RR$
with a $\vx_n\in A_n$ is defined by
\be
(f_1\otimes... \otimes f_n)(\vx_n) := f_n(\vx_n) f_1\otimes... \otimes f_{n-1} : A_1 \times...\times A_{n-1} \rar \RR.
\ee
More generally, the contraction of the tensorial product $f_1\otimes... \otimes f_n :  : A_1 \times...\times A_n \rar \RR$
with a $\vx_i\in A_i$ is defined by
\be
(f_1\otimes... \otimes f_n)(\vx_i) := f_i(\vx_i) f_1\otimes...f_{i-1}\otimes f_{i+1}... \otimes f_{n-1} .
\ee
\findef

\debexa
If $\ell\in E^*$ then its contraction with a $\vx\in E$ is the real $\ell.\vx$.

If $\ell\in E^*$ and $u\in\Ess$, then its contraction with a $\vx\in E$ is the real
\finexa
}

%%%%%%%%%%%%%%%%%%%%%%%%%%%%%%%%%%%%%%%%%%%%%%%%%%%%%%%%%%%%%%%%%%%%%%%%%%%%%%%%%%%

\subsection{Bidual basis (and contravariance)}

\debdef
The dual of~$E^*$ is $\Ess:= (E^*)^* = \calL(E^*;\RR)$ and is named the bidual of~$E$.

$\Ess$ is also called the space of contravariant vectors = the space of directional derivatives.
\findef

\debdef
Let $(\ve_i)$ be a basis in~$E$,
let $(\pi_{ei})$ be its dual basis (basis in~$\Es$).
The dual basis $(\pa_i)$ of $(\pi_{ei})$ is called the bidual basis of~$(\ve_i)$.
(Duality notations: $(\pi_{ei})=(e^i)$.)

(The notation $\pa_i$ refers to the derivation in the direction~$\ve_i$:
$\pa_i(df(\vx)) = df(\vx).\ve_i={\pa f \over \pa x_i}(\vx)$, see~\S~\ref{seccontree1}.)
\findef

Thus, the linear form $\pa_i \in \Ess =\calL(E^*;\RR)$ are characterized by, for all~$j$,
\be
\label{eqpa}
\pa_i.\pi_{ej} = \delta_{ij} = \pi_{ej}.\ve_i, \qso
\ell = \sumin \ell_i \pi_{ei}  \qiff \ell_i = \pa_i.\ell  \;\; (= \ell.\ve_i ).
\ee
Indeed, $\pa_i(\ell) = \pa_i( \sumjn \ell_j \pi_{ej}) = \sumjn \ell_j \pa_i(\pi_{ej}) = \sumjn \ell_j \delta_{ij} = \ell_i$. (Duality notation: $\pa_i.e^j = \delta^j_i= e^j.\ve_i$ and $\ell=\sumin \ell_i e^i$.)

\mn
{\bf Remark:}
With the natural canonical isomorphism
$
\calJ : 
\left\{\eqalign{
E & \rar E^{**} = \calL(\Es;\RR)\cr
\vu & \rar \calJ(\vu), \;\;\hbox{where}\;\; \calJ(\vu).\ell := \ell.\vu,\;\; \forall \ell\in \Es
}\right\}
$
see~\eref{eqisomcan0},
we can identify $\vu$ and $\calJ(\vu)$ (observer independent identification), thus $\pa_i=\calJ(\ve_i)\eqnote \ve_i$, and \eref{eqpa} reads (usual notation in differential geometry)
$\ve_i.\pi_{ej} = \delta_{ij}$ and $\ell_i = \ve_i.\ell$.

%%%%%%%%%%%%%%%%%%%%%%%%%%%%%%%%%%%%%%%%%%%%%%%%%%%%%%%%%%%%%%%%%%%%%%%%%%%%%%%%%%%

\subsection{Bilinear forms}
\label{secfb}

%%%%%%%%%%%%%%%%%%%%%%%%%%%%%%%%%%%%%%%%%%%%%%%%%%%%%%%%%%%%%%%%%%%%%%%%%%%%%%%%%%%

\subsubsection{Definition}

Let $E$ and $F$ be vector spaces.

\debdef
$\bullet$ A bilinear form is a $2$-multilinear form $\beta\dd : 
\left\{\eqalign{
E \times F &\rar  \RR \cr
(\vu,\vw)   &\rar \beta(\vu,\vw)
}\right\}$. 
\\
So, $\beta(\vu_1+\lambda\vu_2,\vw) = \beta(\vu_1,\vw) + \lambda \beta(\vu_2,\vw)$ and $\beta(\vu,\vw_1+\lambda\vw_2) = \beta(\vu,\vw_1) + \lambda \beta(\vu,\vw_2)$ for all $\vu,\vu_1,\vu_2 \in E$, $\vw,\vw_1,\vw_2 \in F$, $\lambda\in\RR$.

$\bullet$  $\calL(E,F;\RR)$ is the set of bilinear forms $E \times F \rar  \RR$.

$\bullet$ If $(\ell,m)\in E^* \times F^*$, then the bilinear form $\ell\otimes m \in \calL(E,F;\RR)$ is defined by
\be
\label{eqdefelemb}
(\ell\otimes m)(\vu,\vw) = \ell(\vu)m(\vw) \quad (=(\ell.\vu)(m.\vw))
\ee
for all $(\vu,\vw)\in E\times F$, and is called an elementary bilinear form.
\findef

%%%%%%%%%%%%%%%%%%%%%%%%%%%%%%%%%%%%%%%%%%%%%%%%%%%%%%%%%%%%%%%%%%%%%%%%%%%%%%%%%%%

\subsubsection{The transposed of a bilinear form}
\label{secbT}

(Warning: Not to be confused with the subjective definition of a transposed of a linear map which requires inner dot products to be defined, see \eg~\eref{eqseccpgdd0e}.)

\debdef
If $\beta \in \calL(E,F;\RR)$ then its transposed is the bilinear form $\beta^T \in \calL(F,E;\RR)$ defined by,
for all $(\vw,\vu)\in F\times E$,
\be
\label{eqbT}
\beta^T(\vw,\vu)=\beta(\vu,\vw).
\ee
(This definition is observer independent: no basis or inner dot product is required in this definition.)
\findef

%%%%%%%%%%%%%%%%%%%%%%%%%%%%%%%%%%%%%%%%%%%%%%%%%%%%%%%%%%%%%%%%%%%%%%%%%%%%%%%%%%%

\subsubsection{Symmetric and definite positive bilinear forms}

\debdef
Here $F=E$ (no choice), and  $\beta\in \calL(E,E;\RR)$.

$\bullet$ $\beta$ is semi-positive, iff for all $\vu\in E$,
\be
\beta(\vu,\vu)\ge0.
\ee

$\bullet$ $\beta$ is  definite positive, iff for all $\vu\ne\vec0$,
\be
\beta(\vu,\vu)>0.
\ee

$\bullet$ $\beta$ is symmetric iff $\beta^T=\beta$, \ie, for all $\vu,\vv\in E$,
\be
\beta(\vu,\vv)=\beta(\vv,\vu).
\ee
\comment{
$\bullet$ $\beta\in \calL(E,E;\RR)$ is definite positive iff $\beta(\vu,\vu) >0$ for all $\vu\ne \vec0$.

$\bullet$ $\beta\in \calL(E,E;\RR)$ is an inner dot product iff $\beta$ is symmetric definite positive. In which case the associated norm is the function $||.||_\beta : E\rar \RR_+$ is defined by $||\vu||_\beta := \sqrt{\beta(\vu,\vu)}$.

$\bullet$ $\beta\in \calL(E,E;\RR)$ is a semi-inner dot product iff $\beta$ is symmetric and $\beta(\vu,\vu) \ge 0$ for all $\vu\in E$.
In which case the associated semi-norm is the function $||.||_\beta : E\rar \RR_+$ is defined by $||\vu||_\beta := \sqrt{\beta(\vu,\vu)}$.
}
\findef

%%%%%%%%%%%%%%%%%%%%%%%%%%%%%%%%%%%%%%%%%%%%%%%%%%%%%%%%%%%%%%%%%%%%%%%%%%%%%%%%%%%

\subsubsection{Inner dot product, and metric}

\debdef
$\bullet$ An ``inner dot product'' (or ``scalar inner dot product'',
or ``inner scalar product'', or ``inner product'') 
in a vector space~$E$ is a bilinear form $\beta\eqnote g\eqnote g\dd\in \calL(E,E;\RR)$ which is symmetric and definite positive.
%(The name $g$ for a symmetric definite positive bilinear form in~$E$ is often chosen in physics.)
And then (for inner dot products)
\be
g\dd \eqnote \dd_g \eqnote \cdot \bcdotg \cdot, \qie
g(\vu,\vw) = (\vu,\vw)_g \eqnote \vu \bcdotg \vw,\;\;\forall \vu,\vw\in E.
\ee

$\bullet$ Then two vectors $\vu,\vw\in E$ are $\dd_g$-orthogonal iff $(\vu,\vw)_g=0$.

%for all $\vu,\vw\in E$.
$\bullet$ And the associated norm with~$\dd_g$ is the function $||.||_g : E \rar \RR_+$ given by, for all $\vu\in E$,
\be
\label{eqnaps}
||\vu||_g = \sqrt{(\vu,\vu)_g}.
\ee
(To prove that it is a norm, use the Cauchy--Schwarz inequality~\eref{eqCS}.)

$\bullet$ An ``semi-inner dot product'' $\dd_g$ (or ``semi-scalar inner dot product'') 
in a vector space~$E$ is a bilinear form $\beta\eqnote g\dd\in \calL(E,E;\RR)$ which is symmetric and semi-positive.
And the associated semi-norm is given by~\eref{eqnaps}.
\findef

\debprop
(Cauchy--Schwarz inequality.) $\dd_g$ being an inner dot product in~$E$,
\be
\label{eqCS}
\forall \vu,\vw \in E,\quad |(\vu,\vw)_g| \le ||\vu||_g||\vw||_g.
\ee
And $|(\vu,\vw)_g| = ||\vu||_g||\vw||_g$ iff $\vu$ and $\vw$ are parallel.
\finprop

\debdem
Let $p(\lambda) = ||\vu{+}\lambda \vw||_g^2=(\vu{+}\lambda \vw ,\vu{+}\lambda \vw)_g$,
so  $p(\lambda)= a \lambda^2 + b \lambda + c$ where $a=||\vw||_g^2$, $b=2(\vu,\vw)_g$ and $c=||\vu||^2_g$.
With $p(\lambda)\ge0$ (since$\dd_g$ is positive), we get $b^2-4ac\ge0$, thus~\eref{eqCS},
and $p(\lambda)=0$ iff $\vu{+}\lambda \vw=0$.
\findem

\debdef (Metric.)
With $\RRn$ our usual affine geometric space, $n=1$, $2$ or~$3$, and $\vRRn=$ the usual associated vector space made of bipoint vectors.
Let $\Omega\subset \RRn$ be open in~$\RRn$.
A metric in~$\Omega$ is a $C^\infty$ function
$g:
\left\{\eqalign{
\Omega & \rar \calL(\vRRn,\vRRn;\RR) \cr
p & \rar g(p) \eqnote g_p
}\right\}
$
such that $g_p$ is an inner dot product in~$\vRRn$ at each $p\in\Omega$.

Particular Case: % of continuum mechanics: 
When the $g_p$ is independent of~$p$ (general case in continuum mechanics),
a metric is simply called a inner dot product (\eg\ a Euclidean metric is called a Euclidean dot product).

(In a differentiable manifold~$\Omega$, a metric is a $C^\infty$ ${0\choose2}$ tensor $g$ \st\ $g(p)$ is an inner dot product at each $p\in\Omega$.
A Riemannian metric is a metric \st\ $g(p)$ is a Euclidean dot product at each $p\in\Omega$.)
\findef

%%%%%%%%%%%%%%%%%%%%%%%%%%%%%%%%%%%%%%%%%%%%%%%%%%%%%%%%%%%%%%%%%%%%%%%%%%%%%%%%%%%

\subsubsection{Quantification: Matrice $[\beta_{ij}]$ and tensorial representation}

$\dim E=n$, $\dim F=m$, $\beta\in \calL(E,F;\RR)$,
$(\va_i)$ is a basis in~$E$ which dual basis is $(\piai)$, $(\vb_i)$ is a basis in~$F$ which dual basis is $(\pibi)$.
(With duality notations, $(\piai)=(a^i)$ and $(\pibi)=(b^i)$.)

\debdef
The components of $\beta\in \calL(E,F;\RR)$ relative to the bases~$(\va_i)$ and $(\vb_i)$ are the $nm$ reals
\be
\label{eqbdd}
\beta_{ij} := \beta(\va_i,\vb_j), \qand [\beta]_{|\va,\vb} = [\beta_{ij}]_{i=1,...,n \atop j=1,...,m}
\ee
is the matrix of~$\beta$ relative to the bases $(\va_i)$ and~$(\vb_i)$, simply written $[\beta_{ij}]$ if the bases are implicit. 

And if $F=E$ and $(\vb_i)=(\va_i)$ then $[\beta]_{|\va,\vb} \eqnote [\beta]_{|\va}$.
\findef

\debprop
\label{propbilb}
A bilinear form $\beta\in\calL(E,F;\RR)$ is known as soon as the $nm$ scalars $\beta_{ij}= \beta(\va_i,\vb_j)$ are known, and, %with usual matrix computation rules, 
for all $(\vu,\vw)\in E \times F$, 
\be
\label{eqdmg}
\beta(\vu,\vw) = [\vu]_{|\va}{}^T.[\beta]_{|\va,\vb}.[\vw]_{|\vb} ,\qwritten
\boxed{\beta(\vu,\vw) = [\vu]^T.[\beta].[\vw]},
\ee
so
\be
\label{eqbilint}
\beta = \sumin\sumjm \beta_{ij} \pi_{ai} \otimes \pi_{bj}, % \quad \hbox{(tensorial representation)}.
\ee
and a basis in $\calL(E,F;\RR)$ is made of the $nm$ functions $\pi_{ai} \otimes \pi_{bj}$, and $\dim\calL(E,F;\RR) = nm$.
(Duality notations: $\beta = \sumin\sumjm \beta_{ij} a^i \otimes b^j$.)
%with shortened notation when the bases in use are implicit.
\finprop

\debdem
$\beta$ being bilinear, $\vu=\sumin u_i\va_i$ and $\vw=\sumjn w_j\vb_j$ give
$\beta(\vu,\vw)=\sumijn u_i w_j \beta(\va_i,\vb_j)=\sumijn u_i \beta_{ij} w_j
=([\vu]_{|\va})^T.[\beta]_{|\va,\vb}.[\vw]_{|\vb}$, 
thus~\eref{eqdmg}. In particular, if the $\beta_{ij}$ are known,
then $b$ is known.
And $(\piai\otimes \pibj)(\va_k,\vb_\ell) \equalref{eqdefelemb} (\piai.\va_k)(\pibj.\vb_\ell) = \delta_{ik}\delta_{j\ell}$
(all the elements of the matrix $[\piai\otimes \pibj]_{|\va,\vb}$ are zero except the element at the intersection of row~$i$ and column~$j$ which is equal to~$1$).
And $(\piai\otimes \pibj)(\vu,\vw) \equalref{eqdefelemb} (\piai.\vu)(\pibj.\vw) = u_iw_j$, thus
$\beta(\vu,\vw) = \sumijn \beta_{ij} u_i w_j = \sumijn \beta_{ij}(\piai\otimes \pibj)(\vu,\vw)$, thus
$\beta := \sumijn \beta_{ij}(\piai\otimes \pibj)$, thus the $\piai\otimes \pibj$ span $\calL(E,F;\RR)$.
And $\sum_{ij}\lambda_{ij}(\piai\otimes \pibj)=0$ implies
$0=(\sum_{ij}\lambda_{ij}(\piai\otimes \pibj))(\va_k,\vb_\ell)
=\sum_{ij}\lambda_{ij}(\piai\otimes \pibj)(\va_k,\vb_\ell)=\lambda_{k\ell}=0$ for all $k,\ell$;
Thus the $\piai\otimes \pibj$ are independent.
Thus $(\piai\otimes \pibj)$ is a basis in~$\calL(E,F;\RR)$ and $\dim(\calL(E,F;\RR))=nm$.
(Duality notations: $\beta(\vu,\vw)=\sumijn \beta_{ij}u^iw^j$ and $\beta := \sumijn \beta_{ij}\,a^i\otimes b^j$.)
\findem

\debexa
$\dim E = \dim F =2$.
$[\beta]_{|\va,\vb} = \pmatrix{1 & 2 \cr 0 & 3}$
means $\beta(\va_1,\vb_1) = \beta_{11} = 1$, $\beta(\va_1,\vb_2) = \beta_{12} = 2$, $\beta(\va_2,\vb_1) = \beta_{21} = 0$, $\beta(\va_2,\vb_2) = \beta_{22} = 3$.
And $\beta_{12} = [\va_1]_{|\va}^T.[\beta]_{|\va,\vb}.[\vb_2]_{|\vb}
=\pmatrix{1 & 0}.\pmatrix{1 & 2 \cr 0 & 3}.\pmatrix{0\cr 1} = 2$.
\finexa

\debexe
Let $\beta\in\calL(E,E;\RR)$, let $(\va_i)$ and $(\vb_i)$ be two bases in~$A$, and let $\lambda\in\RR^*$. Prove:
\be
\label{eqlabidb}
\hbox{if, } \;\forall i\in[1,n]_\NN,\;\; \vb_i=\lambda\va_i, \qthen
[\beta]_{|\vb} = \lambda^2[\beta]_{|\va}.
\ee
(A change of unit, \eg\ from foot to metre, has a ``big'' influence on the matrix.)

\debrep
$\vb_i=\lambda\va_i$ give
$\beta(\vb_i,\vb_j) = \beta(\lambda\va_i,\lambda\va_j) = \lambda^2 \beta(\va_i,\va_j)$ (bilinearity),
thus $[\beta]_{|\vb} = \lambda^2[\beta]_{|\va}$.
\finrep
\finexe

\debexe
Prove
\be
\label{eqbetaT}
[\beta^T]_{\vb,\va} = ([\beta]_{\va,\vb})^T, \qwritten [\beta^T]=[\beta]^T.
\ee

\debrep
Let$[\beta]_{\va,\vb} = [\beta_{ij}]_{i=1,...,n \atop j=1,...,m}$ and
$[\beta^T]_{\vb,\va} = [\gamma_{ij}]_{i=1,...,m \atop j=1,...,n}$.
We have $\gamma_{ij} = \beta^T(\vb_i,\va_j) = \beta(\va_j,\vb_i)=\beta_{ji}$, qed.
\finrep
\finexe

%%%%%%%%%%%%%%%%%%%%%%%%%%%%%%%%%%%%%%%%%%%%%%%%%%%%%%%%%%%%%%%%%%%%%%%%%%%%%%%%%%%

\subsection{Linear maps}

%%%%%%%%%%%%%%%%%%%%%%%%%%%%%%%%%%%%%%%%%%%%%%%%%%%%%%%%%%%%%%%%%%%%%%%%%%%%%%%%%%%

\subsubsection{Definition}

Let $E$ and $F$ be vector spaces.

\debdef
\label{defal}
$\bullet$ A function $L : E \rar  F$ is linear iff
$L(\vu_1+\lambda\vu_2) = L(\vu_1)+ \lambda L(\vu_2)$ for all $\vu_1,\vu_2 \in E$ and all $\lambda\in\RR$
(distributivity type relation).
And (distributivity notation):
\be
\label{eqellpv}
L(\vu) \eqnote L.\vu,\qso L(\vu_1+\lambda\vu_2) = L.(\vu_1+\lambda\vu_2) = L.\vu_1+ \lambda L.\vu_2.
\ee
NB: This dot notation $L.\vu$ is a linearity notation (distributivity type notation); 
It is an ``outer'' dot product between a (linear) function and a vector;
It is \textsl{\textbf{not}} an ``inner'' dot product since $L$ and $\vu$ don't belong to a same space.
It is \textsl{\textbf{not}} a matrix product since no basis has been introduced yet (no quantification has been done yet).

$\bullet$  $\calL(E;F)$ is the set of linear maps $E\rar F$ (vector space, subspace of $(\calF(E;F),+,.)$).

$\bullet$ If $F=E$ then a linear map $L\in\calL(E;E)$ is called an endomorphism in~$E$.

(If $F=\RR$ then a linear map $E\rar \RR$ is called a linear form, and $E^* := \calL(E;\RR)$ is the dual of~$E$.)
\findef

\def\GL{G\!L}
\def\calMn{{\cal M}_{\!n}}

\mn
{\bf Vocabulary:}
Let $L_i(E;E)$ be the space of linear invertible linear maps.
If $E$ is a finite dimension vector space, $\dim E=n$,
then, in algebra, the set $(L_i(E;E),\circ)$ of linear maps equipped with the composition rule is named $\GL_n(E)$ = ``the linear group'' (it is indeed a group, easy check). And the ``linear group'' of $n*n$ invertible matrices is 
$\GL_n(\calMn):=(L_i(\calMn;\calMn),.)$, \ie\ $L_i(\calMn;\calMn)$ with the matrix product rule.

\debexe
(Math exercise.)
Let $E=(E,||.||_E)$ and $F=(F,||.||_F)$ be Banach spaces, and let $L_{ic}(E;F)$ be the space of invertible linear continuous maps $E\rar F$, with its usual norm $||L||= \sup_{||\vx||_E=1}||L.\vx||_F$.
Let 
$Z : 
\left\{\eqalign{
L_{ic}(E;F) &\rar  L_{ic}(E;F) \cr
L & \rar L^{-1}
}\right\}$. 
Prove $dZ(L).M = -L^{-1} \circ M \circ L^{-1}$, for all $M\in L_{ic}(E;F)$.
(Recall: In finite dimension, a linear map is always continuous.)

\debrep
Consider $\lim_{h\rar0}{Z(L+hM) - Z(L) \over h}
=\lim_{h\rar0}{(L+hM)^{-1} - L^{-1} \over h} $ ($\eqnote dZ(L).M$ %for all $M\in L_i(E;F)$ 
if the limit exists).
With $N=L^{-1}.M$ we have 
$L+hM
%= L(I+hL^{-1}.M)
= L(I+hN)
$, and $(I+hN)$ is invertible as soon as $||hN|| <1$, \ie\ $h<{1\over ||N||}={1\over ||L^{-1}.M||}$,
its inverse being $I-hN+h^2N-...$ (Neumann serie);
Thus $I+hN=I-hN+o(h)$,
and $(L+hM)^{-1} = (I+hN)^{-1}.L^{-1} = (I-hN + o(h)).L^{-1} = L^{-1}-hN.L^{-1} + o(h) $.
Thus
$ {(L+hM)^{-1} - L^{-1} \over h}
%= {(I+hM.L^{-1})^{-1}L^{-1} - L^{-1} \over h}
={L^{-1} - hN.L^{-1} + o(h) - L^{-1} \over h}
=-N.L^{-1} + o(1) \mrar_{h\rar0} -N.L^{-1}$.
\finrep
\finexe

%%%%%%%%%%%%%%%%%%%%%%%%%%%%%%%%%%%%%%%%%%%%%%%%%%%%%%%%%%%%%%%%%%%%%%%%%%%%%%%%%%%

\subsubsection{Quantification: Matrices $[L_{ij}]=[\Lij]$} %Classical or duality notation: 

%$E$ and $F$ are finite dimensional vector spaces, 
$\dim E=n$, $\dim F=m$, $L\in \calL(E;F)$, 
$(\va_i)$ is a basis in~$E$ which dual basis is $(\piai)$, $(\vb_i)$ is a basis in~$F$ which dual basis is $(\pibi)$.
(With duality notations, $(\piai)=(a^i)$ and $(\pibi)=(b^i)$.)

\debdef
\label{defpropLab0}
The components of a linear map $L\in \calL(E;F)$ relative to the bases~$(\va_i)$ and $(\vb_i)$ are the $nm$ reals named $L_{ij}$ (classical notation) $=\Lij$ (duality notation), 
which are the components of the vectors $L.\va_j$ relative to the basis~$(\vb_i)$. That is:
\be
\label{eqpropLab0}
\left\{\eqalignrll{
& \clasnot:
&\; L.\va_j = \sumim L_{ij}\vb_i, \cr
& \dualnot
&\; L.\va_j = \sumim \Lij\vb_i, \cr
}\right\},
\qie
[L.\va_j]_{|\vb}
\eqclas \pmatrix{L_{1j}\cr\vdots\cr L_{mj}}
\eqdual\pmatrix{L^1{}_j\cr\vdots\cr L^m{}_j}.
\ee
%is the $j$-th column of~$[L]_{|\va,\vb}$. That is, the $L_{ij}=\Lij$ for $i=1,...,n$ are the components of the vectors $L.\va_j$ relative to the basis~$(\vb_i)$.
%(Of course, $L_{ij}=\Lij$ for all~$i,j$: This is just a choice of notation for the components of~$L$.)
And
\be
\label{eqpropLab0m}
[L]_{|\va,\vb}
\eqclas[L_{ij}]_{i=1,...,m \atop j=1,...,n}
\eqdual[\Lij]_{i=1,...,m \atop j=1,...,n}
\ee
is the matrix of~$L$ relative to the bases $(\va_i)$ and~$(\vb_i)$
(so $[L.\va_j]_{|\vb}$ is the $j$-th column of~$[L]_{|\va,\vb}$).

Particular case: If $E=F$ (so $L$ is an endomorphism) and if $(\vb_i)=(\va_i)$ then $[L]_{|\va,\va} \eqnote [L]_{|\va}$.
\findef

\debexa
$n=m=2$.
$[L]_{|\va,\vb} = \pmatrix{1 & 2 \cr 0 & 3}$
means $L.\va_1 = \vb_1$ and $L.\va_2 = 2\vb_1+3\vb_2$ (column reading).
Here $L_{11} {=} 1$, $L_{12} {=} 2$, $L_{21} {=} 0$, $L_{22} {=} 3$
(duality notations: $L^1{}_1 {=} 1$, $L^1{}_2 {=} 2$, $L^2{}_1 {=} 0$, $L^2{}_2 {=} 3$).
\finexa

And $L$ being linear, for all $\vu\in E$, $\vu = \sumjn u_j \va_j = \sumjn u^j \va_j$, we get, thanks to linearity,
\be
\label{eqpropLab20}
L.\vu 
\eqclas \sumim\sumjn L_{ij} u_j \vb_i
\eqdual \sumim\sumjn \Lij u^j \vb_i,
\qie \boxed{[L.\vu]_{|\vb} = [L]_{|\va,\vb}.[\vu]_{|\va}} .
\ee
Shortened notation: $[L.\vu] = [L].[\vu]$ when the bases are implicit. 

\comment{
with the usual rules of matrix multiplication,
that is, 
\be
\label{eqpropLab20b}
L.\vu 
= \underbrace{\sumim\sumjn L_{ij} u_j \vb_i}_\clanot
= \underbrace{\sumim\sumjn \Lij u^j \vb_i}_\duanot.
\ee
where $\vu = \sumjn u_j \va_j = \sumjn u^j \va_j$
(linearity gives
$L.\vu = \sumjn u_j L.\va_j = \sumjn u^j L.\va_j$).
}

\debprop
\label{propLab}
A linear map $L\in\calL(E;F)$ is known as soon as the $n$ vectors $L.\va_j$ are known, $j\in[1,n]_\NN$. % for all~$j$.
And  the linear maps $\calL_{ij}\in\calL(E;F)$ defined by $\calL_{ij}.\va_\ell=\delta_{j\ell }\vb_i$ (all the elements of the matrix $[\calL_{ij}]_{|\va,\vb}$ vanish except the element at the intersection of row~$i$ and column~$j$ which is equal to~$1$), for $i,\ell=1,...,n$ and $j=1,...,m$, constitute a basis $\in\calL(E;F)$.
(Duality notations: $\calL_{ij}\eqnote \calL_i{}^j$, and $\calL_i{}^j.\va_\ell=\delta^j_\ell \vb_i$.)
 %With 
%Marsden notations: $\calL_{ij}=\calL_i{}^j$ and $\calL_i{}^j.\va_\ell=\delta_\ell^j\vb_i$.
%Tensorial notations: $\calL_{ij}=\vb_i \otimes \pi_{aj} = \vb_i \otimes a^j = \calL_i{}^j$.
So, $\dim(\calL(E;F)) = nm$.
\finprop

\debdem
$\vu\in E$ and $\vu = \sum_k u_j\va_j$ give $L.\vu = \sum_j u_j L.\va_j$, since $L$ is linear.
Thus $L$ is known iff the $n$ vectors $L.\va_j$ are known for all $j=1,...,n$;
And $L.\va_k=\sum_i L_{ik} \vb_i$ together with 
$\sum_{ij} L_{ij}\calL_{ij}.\va_k
= \sum_{ij} L_{ij}\delta_{jk }\vb_i
= \sum_{i} L_{ik}\vb_i
$, for all~$k$, thus $L= \sum_{ij} L_{ij}\calL_{ij}$, thus the $\calL_{ij}$ span $\calL(E;F)$.
And $\sumim\sumjn \lambda_{ij}\calL_{ij}=0$ implies, for all $\ell$,
$\vec0
=\sumim\sumjn \lambda_{ij}\calL_{ij}.\va_\ell
=\sumim\sumjn \lambda_{ij}\delta_{j\ell }\vb_i
=\sumim \lambda_{i\ell}\vb_i
$, thus $ \lambda_{i\ell}=0$, for all $i$ and~$\ell$.
Thus the $\calL_{ij}$ are independent.
Thus $(\calL_{ij})_{i=1,...,n\atop j=1,...,m}$ is a basis in $\calL(E;F)$.
%Use duality notations if you prefer: $L.\va_j = \sumin \Lij\vb_i$ and $\calL_{ij}\eqnote\calL_i{}^j$, so $\calL=\sumijn \Lij \calL_i{}^j$, indeed $(\sumijn \Lij \calL_i{}^j).\va_\ell = \sumijn \Lij \delta_\ell{}^j\vb_i = \sumin L^i{}_\ell\vb_i = L.\va_\ell$.
%(if you insist on using tensorial notations then $\calL_{ij} \simeq \vb_i \otimes \pi_{aj}$ where $(\pi_{ai})$ is the dual basis of~$(\va_i)$, or $\calL_i{}^j \simeq \vb_i \otimes a^j$ with duality notations.)
\findem

\debexe
If $L\in\calL(E;E)$ (endomorphism), if $(\va_i)$ is a basis in~$E$,  prove:
\be
\label{eqlabid0}
\hbox{if }\;\lambda\in\RR^*\: \hbox{ and }\; \vb_i=\lambda\va_i\:\; \forall i\in[1,n]_\NN, \qthen
%L = \sumijn \Lij \va_i \otimes a^i = \sumijn \Lij \vb_i \otimes b^i, \qie 
[L]_{|\vb} = [L]_{|\va},
\ee
\ie, a change of unit has not influence on the matrix of an endomorphism.
Check with the change of basis formulas.
{\bf NB:} To compare with~\eref{eqlabidb}: Covariance and contravariance should \textsl{\textbf{not}} be confused.

\debrep
Let $L.\va_j = \sumin L_{aij} \va_i$
and $L.\vb_j = \sumin L_{bij} \vb_i$. Then
$\sumin L_{bij} \vb_i
= L.\vb_j 
= L.(\lambda \va_j) 
= \lambda L.\va_j 
= \lambda \sumin L_{aij} \va_i
= \lambda \sumin L_{aij} {\vb_i \over\lambda} 
= \sumin L_{aij} \vb_i$, thus $L_{bij}=L_{aij}$.

Change of basis formula: $[L]_{|\vb} = P^{-1}.[L]_{|\va}.P$ with $P=\lambda I$ here.
\finrep
\finexe

\comment{
\debrem
\label{remLijt}
With the use of tensorial notation and the duality notations, a basis in $\calL(E;F)$ is $\vb_i \otimes a^j$; Thus $L$ also reads $\sum_{ij} \Lij  \vb_i \otimes a^j$ for calculation purposes, see~\S~\ref{sectroflm}.
\finrem
}

\comment{
\debexa
If $M = [M^i{}_j]$ is a matrix, then its transposed matrix is $M^T = [Z^i{}_j]$ where $Z^i{}_j = M^j{}_i$ for all $i,j$
\finexa
}

%%%%%%%%%%%%%%%%%%%%%%%%%%%%%%%%%%%%%%%%%%%%%%%%%%%%%%%%%%%%%%%%%%%%%%%%%%%%%%%%%%%

\subsubsection{Trace of an endomorphism}

Let $E$ be a vector space, $\dim E=n$. Let $\vu\in E$ and $\ell\in\Es$ and call $\calL_{\vw,\ell} \in \calL(E;E)$ the endomorphism, called an elementary endomorphism, defined by
\be
\label{eqwol}
\calL_{\vw,\ell}.\vu :=\vw  (\ell.\vu) = (\ell.\vu)\vw.
\ee

\debdef
The trace of the endomorphism $L_{\vw,\ell}$ is the real
\be
\label{eqdefTr0}
\Tr(\calL_{\vw,\ell}) := \ell.\vw.
\ee
And the trace operator is the linear map 
$\Tr:
\left\{\eqalign{
\calL(E;E) & \rar \RR \cr
L & \rar \Tr(L)
}\right\}
$ defined on elementary endomorphisms $\ell\otimes\vw$
by~\eref{eqdefTr0}.
\findef

\debprop
\label{propTrib}
Let $L\in\calL(E;E)$. The real $\Tr(L)$ is objective (is intrinsic to~$L$), \ie\ is independent of any basis in~$E$.
And (quantification) if $(\ve_i)$ is a basis and $L.\ve_j = \sumin L_{ij} \ve_i$ for all~$j$, then
\be
\label{eqdefTr}
\Tr(L) = \sumin L_{ii} \;(\in \RR), %  \quad (= \sumin e^i.L.\ve_i).
\ee
\ie, $\Tr(L)$ is the trace of the matrix~$[L]_{|\ve}$.
(Duality notations $L.\ve_j = \sumin \Lij \ve_i$ and $\Tr(L) = \sumin L^i{}_i$.)
\finprop

\debdem
%Let $\vw\in E$ and $\ell\in\Es$; 
$\Tr(\calL_{\vw,\ell}) := \ell.\vw$ is a real that can be considered by any observer,
and which value is the same for all observers, \cf~\eref{eqdefP1b}: It is objective.

Let $L\in\calL(E;E)$.
Let $(\va_i)$ be a basis and $(\piai)$ be its (covariant) dual basis,
and $L.\va_j = \sum_i L_{ij}\va_i$, \ie\ $[L]_{[\va}=[L_{ij}]$. 
And we have $
(\sum_{ik}  L_{ik} \calL_{\va_i,\piak}).\va_j
=\sum_{ik}  L_{ik} \calL_{\va_i,\piak}.\va_j
\equalref{eqwol}\sum_{ik}  L_{ik} \va_i(\piak.\va_j)
=\sum_{ik}  L_{ik} \va_i \delta_{kj}
=\sum_{i}  L_{ij} \va_i
$, thus $L=\sum_{ij}  L_{ij} \calL_{\va_i,\piaj}$ (sum of elementary endomorphisms), thus, $\Tr$ being linear
$\Tr(L)= \sum_{ij} L_{ij} \Tr \calL_{\va_i,\piaj} = \sum_{ij} L_{ij}\delta_{ji} = \sum_i L_{ii}$,
thus~\eref{eqdefTr}.
\findem

\debexe
Check with the change of basis formula that $\Tr(L)$ is an invariant (the same value for all observers).

\debrep
Let $(\va_i)$ and $(\vb_i)$ be two bases, $P=[P_{ij}]$ be the transition matrix from $(\va_i)$ to~$(\vb_i)$, $Q = P^{-1}$, $[L]_{[\va}=[(L_a)_{ij}]$, $[L]_{[\vb}=[(L_b)_{ij}]$.
We have $[L]_{[\vb} \equalref{eqLijoldnew2} P^{-1}.[L]_{[\va}.P$,
\ie\ 
$(L_b)_{ij} = \sum_{k\ell}Q_{ik} (L_a)_{k\ell} P_{\ell j}$,
thus 
$\sum_i (L_b)_{ii}
= \sum_{ik\ell} Q_{ik} (L_a)_{k\ell} P_{\ell i}
%= \sum_{ik\ell} P_{\ell i} Q_{ik} (L_a)_{k\ell} 
= \sum_{k\ell} (P.Q)_{\ell k} (L_a)_{k\ell} 
= \sum_{k\ell} \delta_{\ell k} (L_a)_{k\ell} 
= \sum_k  (L_a)_{kk} 
$, qed. %quod erat demonstrandum.
\finrep
\finexe

\comment{
\debexe
Let $L\in \calL(E;E)$, $(\va_i)$ and $(\vb_i)$ be bases in~$E$,
and let %$L=\sum_{i,j=1}^n L^i_j\,\vb_i\otimes a^j\in \calL(E;E)$, that is, 
$L.\va_j = \sumin L^i_j\vb_i$, so here $L$ is not treated as an endomorphism (not quantified with only one basis).
Prove: if $P$ is the change of basis from $(\va_i)$ to~$(\vb_i)$, then
$\Tr L = \sum_j(PL)^j_j$.

\debrep
$L.\va_j=\sum_{ij} L^i_j\,(\sum_k P^k_i\va_k)
=\sum_{kj}(\sum_i L^i_j P^k_i)\,\va_k
=\sum_{kj} (PL)^k_j\,\va_k$, so $[L]_{|\va} =[(PL)^i_j]$.
\finrep
\finexe
}

\comment{
\debrem
The Kronecker operator $\uudelta$ is the ${1\choose1}$ uniform tensor characterized by
$\uudelta(\vu\otimes\ell) = \ell.\vu$ for all $(\vu,\ell)\in E {\times} \Es$.
This definition is qualitative (independent of any observer)
and gives back proposition~\ref{propTrib}:
Indeed the natural canonical isomorphism $\tcalJ$, \cf~\eref{eqdefcalJ}, gives $\tL = \sumijn (L_a)^i_j \va_i \otimes a^j = \sumijn (L_b)^i_j \vb \otimes b^j$, thus $\uudelta(\tL) = \sumijn (L_a)^i_j \uudelta(\va_i \otimes a^j)
= \sumijn (L_a)^i_j\delta^j_i = \sumin (L_a)_{ii}$,
as well as $\uudelta(\tL) = \sumin (L_b)_{ii}$.
So $\Tr(L) = \uudelta(\tL)$ thanks~$\tcalJ$.
And the trace operator~$\Tr$ is identified with the Kronecker operator~$\uudelta$.
\finrem
}

Alternative definition with one-one tensors: see~\S~\ref{seckroten}.

%%%%%%%%%%%%%%%%%%%%%%%%%%%%%%%%%%%%%%%%%%%%%%%%%%%%%%%%%%%%%%%%%%%%%%%%%%%%%%%%%%%

\subsection{Transposed matrix}

The definition can be found in any elementary books, \eg, Strang~\cite{strang}: If $M = [M_{ij}]_{i=1,...,m \atop j=1,...,n}$ is an $m*n$ matrix then its transposed is the $n*m$ matrix
$M^T=[(M^T)_{ij}]_{i=1,...,n \atop j=1,...,m}$ defined by 
\be
\label{eqTMat}
(M^T)_{ij} := M_{ji}
\ee
(exchange rows and columns).
\Eg, $M=\pmatrix{1 & 2 \cr 3 & 4 }$ gives $M^T=\pmatrix{1 & 3 \cr 2 & 4 }$, and $(M^T)_{12} {=} M_{21} {=} 3$.

And $M$ is symmetric iff $M^T=M$ (this requires $m{=}n$).

And $M.N = [\sum_k M_{ik}N_{kj}]_{i \atop j}$ gives
$(M.N)^T
= [\sum_k M_{jk}N_{ki}]_{i \atop j} 
= [\sum_k (N^T)_{ik}(M^T)_{kj}]_{i \atop j} 
=N^T.M^T$.

\debexe
Prove: If $M$ is an $n*n$ invertible matrix then $M^T$ is invertible and $(M^T)^{-1}=(M^{-1})^T$ ($\eqnote M^{-T}$); And if moreover $M$ is symmetric, then $M^{-1}$ is symmetric.

\debrep
$M.M^{-1}=I$ gives $(M^{-1})^T.M^T=I^T=I$, thus $M^T$ is invertible with $(M^T)^{-1}=(M^{-1})^T$.
Moreover if $M=M^T$ then $M^{-1}=(M^{-1})^T$.
\finrep
\finexe

%%%%%%%%%%%%%%%%%%%%%%%%%%%%%%%%%%%%%%%%%%%%%%%%%%%%%%%%%%%%%%%%%%%%%%%%%%%%%%%%%%%

\subsection{A transposed endomorphism: depends on a chosen inner dot product}
\label{sectend}

Not to be confused with the transposed of a matrix, \cf~\eref{eqTMat}. And
not to be confused with the transposed of a bilinear form (observer independent), \cf~\eref{eqbT};
In particular, a transposed of a linear map depends on the observer who use it (depends on the choice of an inner dot product).

%%%%%%%%%%%%%%%%%%%%%%%%%%%%%%%%%%%%%%%%%%%%%%%%%%%%%%%%%%%%%%%%%%%%%%%%%%%%%%%%%%%

\subsubsection{Definition (requires an inner dot product: Not objective)}

Let $E$ be a finite dimensional vector space equipped with an inner dot product $g\dd=\dd_g$.
%(If~$E$ is infinite dimensional we have to consider continuous endomorphims.)

\debdef
The transpose of an endomorphism $L\in\calL(E;E)$ relative to~$\dd_g$
is the endomorphism $L^T_g \in \calL(E;E)$ defined by
\be
\label{eqseccpgdd0e}
\forall \vx,\vy\in E, \quad %(L^T_g(\vy),\vx)_g = (\vy,L(\vx))_g, \qwritten 
(L^T_g.\vy,\vx)_g = (\vy,L.\vx)_g, \qie (L^T_g.\vy) \bcdotg\vx = \vy \bcdotg (L.\vx).
\ee
(It depends on~$\dd_g$, see~\eref{eqaltltb01c}.)
If $\dd_g$ is an imposed Euclidean dot product (isometric framework) then $L^T_g \eqnote L^T$, thus $(L^T.\vy,\vx)_g = (\vy,L.\vx)_g$,
\ie\ $(L^T.\vy) \bcdot \vx = \vy \bcdot (L.\vx)$.
%the dot notation $L^T_g(\vy)\eqnote L^T_g.\vy$ since $L^T_g$ is linear.
% (it is not a matrix product: No basis has been introduced yet).
\comment{
This defines the map
$(.)^T_g : 
\left\{\eqalign{
\calL(E;E) & \rar \calL(E;E) \cr
L & \rar  L^T_g
}\right\}
$.
}
\findef

%We immediately get $(L^T_g)^T_g = L$.

\debexe (Math exercise.)
The existence and uniqueness of $L_g^T$ is \eg\ proved with a basis in~$E$ when $E$ is finite dimensional, see next~\S~\ref{secLTqc}.
More general proof: Prove: If $(E,\dd_g)$ is an infinite dimensional Hilbert space and if $L\in\calL(E;E)$ is continuous, then $L_g^T$ exists, is unique, and is continuous (apply the Riesz representation theorem~\ref{thmRiesz}).

\debrep
\def\ellyg{\ell_{\vy g}}\def\vellyg{\vell_{\vy g}}%
Let $\vy\in E$, then let
$\ellyg: \vx \in E\rar \ellyg(\vx) := (\vy,L.\vx)_g \in \RR$. $\ellyg$ is linear (trivial since $L$~is linear and $\dd_g$ is bilinear) and continuous:
$|\ellyg.\vx|\le ||\vy||_g ||L.\vx||_g \le ||\vy||_g||L||\,||\vx||_g$ gives $||\ellyg||_\Es \le ||L||\,||\vy||_g <\infty$; 
Let $\vellyg\in E$ be the $\dd_g$-Riesz representation of $\ellyg \in E^*$: $\ellyg.\vx =(\vellyg,\vx)_g$ for all~$\vx$, with $||\vellyg||_g = ||\ellyg||_\Es$;
We have thus defined $L^T_g : \vy\in E \rar L^T_g(\vy):=\vellyg\in E$;
with $(L^T_g(\vy),\vx)_g =(\vellyg,\vx)_g = \ellyg.\vx = (\vy,L.\vx)_g$, thus $L^T_g$ is linear
(since $\dd_g$ is bilinear) and continuous: $||L^T_g.\vy||_g = ||\vellyg||_g = ||\ellyg||_\Es \le ||L||\,||\vy||_g$ gives $||L^T_g|| \le ||L||_{\calL(E;E)}< \infty$.
\finrep
\finexe

\debrem
Recall: The transposed~$\beta^T$ of a bilinear form~$\beta$ is objective, \cf~\eref{eqbT}: We don't need any tool like an inner dot product to define~$\beta^T$.
Not to be confused with: The transposed $L^T_g \eqnote L^T$ of a linear map~$L$ is subjective: It depends on a choice of an inner dot products~$\dd_g$ by an observer. 
In particular it is \textsl{\textbf{dangerous}} to represent a linear map in a basis with its ``bilinear tensorial representation'' when dealing with the transposed:
$L\in\calL(E;F)$ is naturally canonically represented by the bilinear form $\bL \in \calL(F^*,E;\RR)$, and thus
$\bLT \in \calL(E,F^*;\RR)$; And
\be
L.\va_j=\sumin \Lij \vb_i 
\;\hbox{ gives }\; \bL = \sumijn \Lij \vb_i \otimes a^j,
\;\hbox{ thus }\; \bLT \mope^{\eref{eqbT}} \sumijn \Lji a^i \otimes \vb_j,
\ee
while $L^T\in \calL(F;E)$ is naturally canonically represented by the bilinear form 
$\beta_{(L^T)} \in \calL(E^*,F;\RR)$, and
\be
\label{eqbTne}
L^T.\vb_j = \sumin (L^T)^i{}_j \va_i
\;\hbox{ gives }\; b_{(L^T)} = \sumijn (L^T)^i{}_j \va_i \otimes b^j,
\;\hbox{ thus }\; \boxed{\beta_{(L^T)} \ne \bLT}
\ee
for two reasons: 1- $\va_i \otimes b^j \ne a^i \otimes \vb_j$, and 2- 
$L^T:=L^T_{gh}$ depends on chosen inner dot products $\dd_g$ and~$\dd_h$ by observers in~$E$ and~$F$, see~\eref{eqaltltb3}: $(L^T)^i{}_j \mope^{\eref{eqaltltb3}} \sumkln ([g]^{-1})_{ik} L^\ell{}_k \, h_{\ell j}$;
While $\bLT$ is independent of any inner dot products.
(In fact $\bLT \in \calL(F^*,E;\RR)$ is the tensorial representation of the adjoint $L^*\in\calL(F^*;E^*)$ of~$L$: With $L^*.b^j = \sumin (L^*)_i{}^j a^i$ we get $\tilde{(L^*)} = \sumijn (L^*)_i{}^j a^i \otimes \vb_j=\bLT$,
see~\eref{defadjLqd}.)

So in continuum mechanics it is strongly advised \textslbf{not} to use the tensorial notation for linear maps
when dealing with transposed (\eg\ when using $F^T$ the transposed of the deformation gradient).
\finrem

%%%%%%%%%%%%%%%%%%%%%%%%%%%%%%%%%%%%%%%%%%%%%%%%%%%%%%%%%%%%%%%%%%%%%%%%%%%%%%%%%%%

\subsubsection{Quantification with bases}
\label{secLTqc}

Let $(\ve_i)$ be a basis in~$E$, let $g_{ij} := g(\ve_i,\ve_j)$, so $[g]_{|\ve}:=[g_{ij}] \eqnote [g]$, and let (classical notation)
\be
\label{eqLTqc}
L.\ve_j = \sumin L_{ij} \ve_i, \quad 
L_g^T.\ve_j = \sumin (L_g^T)_{ij}, \qie
[L]_{|\ve} = [L_{ij}] \eqnote [L], \quad [L_g^T]_{|\ve} = [(L_g^T)_{ij}] \eqnote [L_g^T].
\ee
\eref{eqseccpgdd0e} gives
$\ds [\vx]^T.[g].[L^T_g.\vy]
= [L.\vx]^T.[g].[\vy]
%= [\vx]^T.[L]^T.[g].[\vy]
$ for all $\vx,\vy$, 
thus
\be
\label{eqaltltbb00}
[g].[L^T_g] = [L]^T.[g], \qie \sumkn g_{ik} (L_g^T)_{kj} = \sumkn L_{ki} \,g_{kj}
\ee
%(full notation: $[g]_{|\ve}.[L^T_g]_{|\ve} = ([L]_{|\ve})^T.[g]_{|\ve}$),
\ie,
\be
\label{eqaltltb01c}
\boxed{[L^T_g] = [g]^{-1}.[L]^T.[g]}, \qie 
(L_g^T)_{ij} = \sumkln ([g]^{-1})_{ik} L_{\ell k} g_{\ell j}.
\ee
To compare with~\eref{eqbTne}.
%Full notation: $[L^T_g]_{|\ve} =[g]_{|\ve}^{-1}.([L]_{|\ve})^T.[g]_{|\ve}$.
If and only if $(\ve_i)$ is $\dd_g$-orthonormal then $[g]=[\delta_{ij}]$ and $(L_g^T)_{ij} = L_{ji}$.
With duality notations,
$
L.\ve_j = \sumin L^i{}_j \ve_i, \; L_g^T.\ve_j  = \sumin (L_g^T)^i{}_j, \; [L]_{|\ve} = [L^i{}_j], \; [L_g^T]_{|\ve}  = [(L_g^T)^i{}_j],
$
and
\be
\label{eqaltltb0d}
\sumkn g_{ik} (L_g^T)^k{}_j= \sumkn L^k{}_i \,g_{kj}, \qie
(L_g^T)^i{}_j = \sumkln ([g]^{-1})_{ik} L^\ell{}_k \, g_{\ell j}. % \qwhen [g^{ij}]:=([g]_{\ve})^{-1} .
\ee

\debrem
\label{remgsij1}
%Warning: The Einstein convention is not satisfied since we used an inner dot product (even if $g_{ij}=\delta_{ij}$ which gives $(L_g^T)^i{}_j = L^j{}_i$).
The last equation~\eref{eqaltltb0d}$_2$ is also written
\be
\label{eqaltltb0d2}
(L_g^T)^i{}_j = \sumkln g^{ik} L^\ell{}_k \, g_{\ell j} \qwhen ([g]_{\ve})^{-1} = [g_{ij}]^{-1} \eqnote [g^{ij}].
\ee
Don't be fooled by the notation $g^{ij}$, defined by $[g^{ij}]:=[g_{ij}]^{-1}$.
(It is also the short notation for $(g^\sharp)^{ij}$, see~\eref{eqgsharpij4}.)
%: It is  \textbf{\textsl{not}} a (covariant) duality notation here;
%It is just the inverse matrix of the matrix~$[g_{ij}]$.
%\Eg, if $[g]_{\ve}=[g_{ij}]=\pmatrix{1 &0\cr0&2}$ then $[g^{ij}] %([g]_{\ve})^{-1}  := [g_{ij}]^{-1}=\pmatrix{1 &0\cr0&\demi}$.
Use classical notations to avoid misuses and misinterpretations.
%For the notation $(g^\sharp)^{ij} \eqnote g^{ij}$, go to~\S~\ref{remcarpet3}.
\finrem

\debrem
A bilinear form $\beta\in\calL(E,E;\RR)$ satisfies $[\beta^T] = [\beta]^T$.
A linear endomorphism $L\in\calL(E;E)$ satisfies $[L^T_g] = [g]^{-1}.[L]^T.[g] \ne [L]^T$ in general
(\eg\ take $[L]=\pmatrix{0&1\cr 1&0}$ and $[g]=\pmatrix{1&0\cr 0&2}$).
So do not confuse a bilinear on~$E$ (objective) with a linear endomorphism on~$E$ (subjective).
\finrem

\debexe
\label{exesymend}
In $\vRRd$, let $(\ve_1,\ve_2)$ be a basis.
Let $L\in\calL(\vRRd;\vRRd)$ be defined by $[L]_{|\ve} = \pmatrix{0 & 1 \cr 1 & 0}$.
Find two inner dot products $\dd_g$ and~$\dd_h$ in~$\vRRd$ such that $L^T_g \ne L^T_h$ (a transposed endomorphism is not unique, is not intrinsic to~$L$, since it depends on a choice of an inner dot product by an observer).

\debrep
Calculations with~\eref{eqaltltbb00}:

Choose $\dd_g$ given by $[g]_{|\ve}= \pmatrix{1 & 0 \cr 0 & 1} = [I]$. Thus
$[L_g^T]_{|\ve} = [I].[L]_{|\ve}.[I] = \pmatrix{0 & 1 \cr 1 & 0}$; %, \ie, $L_g^T.\ve_1=\ve_2$ and $L_g^T.\ve_2=\ve_1$.
So $L_g^T= L$.

Choose $\dd_h$ given by $[h]_{|\ve}= \pmatrix{1 & 0 \cr 0 & 2}$. Thus
$[L_h^T]_{|\ve} = [h]_{|\ve}^{-1}.[L]_{|\ve}.[h]_{|\ve}
%= \pmatrix{1 & 0 \cr 0 & \demi}.\pmatrix{0 & 1 \cr 1 & 0}.\pmatrix{1 & 0 \cr 0 & 2}
= \pmatrix{0 & 2 \cr \demi & 0}
$; % \ie, $L_h^T.\ve_1=\demi\ve_2$ and $L_h^T.\ve_2=2\ve_1$. 
So $L_h^T \ne L$.

Thus $L_h^T \ne L_g^T$, \eg, $\ve_2 = L_g^T.\ve_1 \ne L_h^T.\ve_1=\demi\ve_2$.
\finrep
\finexe

\debexe
Prove: If $L$ is invertible then $L_g^T$ is invertible, and $(L_g^T)^{-1} = (L^{-1})_g^T$ (written $L_g^{-T}$).

\debrep
Suppose: $\exists \vy\in E$, $\vy \ne \vec 0$, \st\ $L_g^T.\vy=0$.
%Suppose that $L_g^T$ is not invertible: Then $\exists \vy\in E$, $\vy \ne \vec 0$, \st\ $L_g^T.\vy=0$; Then, 
$L$ being invertible, $\exists!\vx\in E$ \st\ $L.\vx=\vy$, with $\vx\ne\vec0$ since $\vy\ne\vec0$
and $L$ is linear; And $L_g^T.\vy=0$ gives $L_g^T.L.\vx=0$, thus $(L_g^T.L.\vx,\vx)_g=0$, thus $||L.\vx||_g^2=0$,
thus $L.\vx=0$, thus $\vx=0$ since $L$ is linear bijective; Absurd. Thus $\Ker(L_g^T)=\{\vec0\}$, thus $L_g^T$ is invertible since it is an endomorphism.
%\eref{eqaltltb0} gives: the matrix $[L_g^T]_{|\ve}$ is invertible since $[g]_{|\ve}$ and $[L]_{|\ve}$ are, thus $L_g^T$ is invertible.
And
$\ds (L_g^T.(L^{-1})^T_g.\vx,\vy)_g
\mathop{=}^{\eref{eqseccpgdd0e}} ((L^{-1})^T_g.\vx,L.\vy)_g
\mathop{=}^{\eref{eqseccpgdd0e}} (\vx,(L^{-1}).L.\vy)_g
= (\vx,\vy)_g
%= (I.\vx,\vy)_g
= (L_g^T.(L_g^T)^{-1}.\vx,\vy)_g$, 
true $\forall \vx,\vy$, thus $L_g^T.(L^{-1})^T_g = L_g^T.(L_g^T)^{-1}$, thus $(L^{-1})^T_g = (L_g^T)^{-1}$
since $L_g^T$ is invertible.
\comment{
Or use matrices relative to a basis if you prefer:
$\det ([g]^{-1}.[L]^T.[g]) 
= \det ([g]^{-1})\det ([L]^T)\det ([g])
= \det ([g])^{-1}\det ([L])\det ([g])
\ne0$, so $[g]^{-1}.[L]^T.[g]$ is invertible and
$[L^T_g]^{-1} = [g].([L]^T)^{-1}.[g]^{-1}$.
}
\finrep
\finexe

\debexe
\label{exesameTe}
Special case of proportional inner dot products $\dd_a$ and~$\dd_b$:
$\exists\lambda>0$ \st\ $\dd_a = \lambda^2\dd_b$.
Prove: $L_{a}^T = L_{b}^T$:
Two proportional inner dot products give the same transposed endomorphism.

\debrep
$(L^T_b.\vy,\vx)_b
= (\vy,L.\vx)_b
= \lambda^2 (\vy,L.\vx)_a 
= \lambda^2 (L^T_a.\vy,\vx)_a
= (L^T_a.\vy,\vx)_b
$, for all $\vx,\vy$, so $L^T_b=L^T_a$.
\finrep
\finexe

 %See remark~\eref{remsymend}.

%%%%%%%%%%%%%%%%%%%%%%%%%%%%%%%%%%%%%%%%%%%%%%%%%%%%%%%%%%%%%%%%%%%%%%%%%%%%%%%%%%%

\subsubsection{Symmetric endomorphism}

\debdef
An endomorphism $L \in \calL(E;E)$ is $\dd_g$-symmetric iff
$L^T_g = L$:
\be
\hbox{$L$ $\dd_g$-symmetric}
\quad\Longleftrightarrow\quad
L^T_g = L
\quad\Longleftrightarrow\quad
(L.\vx,\vy)_g = (\vx,L.\vy)_g, \quad \forall\vx,\vy\in E.
\ee 
\findef

\debrem
\label{remsymend}
The symmetric character of an endomorphism~$L$ is not intrinsic to the endomorphism:
It depends on~$\dd_g$;
See exercise~\ref{exesymend} where
$L$ is $\dd_g$-symmetric while it is \textsl{\textbf{not}} $\dd_h$-symmetric.
\finrem

\comment{
Quantification: \eref{eqaltltb01c} gives
\be
%[g]_{|\ve}.[L]_{|\ve} = [L]_{|\ve}^T.[g]_{|\ve},\qie 
[L^T]_{|\ve} = [g]_{|\ve}.[L]_{|\ve}.[g]_{|\ve}^{-1}.
\ee
In particular, if $(\ve_i)$ is a $\dd_g$-orthonormal basis then $L$ is $\dd_g$-symmetric iff $[L^T]_{|\ve} = [L]_{|\ve}$.
}

%%%%%%%%%%%%%%%%%%%%%%%%%%%%%%%%%%%%%%%%%%%%%%%%%%%%%%%%%%%%%%%%%%%%%%%%%%%%%%%%%%%

\subsubsection{The general flat $^\flat$ notation for an endomorphism: Relative to a $\dd_g$}
\label{secgenflat}

Let $\dd_g$ be an inner dot product in a vector space~$E$, and let $L\in\calL(E;E)$ (a $C^0$ endomorphism).

\debdef
The bilinear form $L^\flat_g \in\calL(E,E;\RR)$ which is $\dd_g$-associated to the endomorphism $L\in\calL(E;E)$ is defined by, for all $\vu,\vw\in E$,
\be
\label{eqLRiesz}
L^\flat_g(\vu,\vw) := (\vu,L.\vw)_g.
\ee
($L^\flat_g$ depends on a choice of a~$\dd_g$.)
We have thus defined the $\dd_g$-dependent operator: 
\be
(.)^\flat_g=\calJ_g(.) : 
\left\{\eqalign{
\calL(E;E) & \rar \calL(E,E;\RR) \cr
L & \rar \calJ_g(L) := L^\flat_g ,
}\right.
\ee
If $\dd_g$ is imposed, then $L^\flat_g\eqnote L^\flat$.
\findef

(The bilinearity of~$L^\flat_g$ is trivial since $L$ is linear and $\dd_g$ is bilinear, and the bilinear form $L^\flat_g$ continuous as soon as $L$ and~$\dd_g$ are since
$|L^\flat_g(\vu,\vv)| \le ||g||\,||L.\vu||\,||\vv|| \le (||g||\,||L||)\,||\vu||\,||\vv||$.)

\debprop
With the natural canonical isomorphism $L\in\calL(E;E) \simeq T_L\in \calL(E^*,E;\RR)$ given by
$T_L(\ell,\vw)=\ell.L.\vw$, and with $T_L \eqnote L$, the function $(.)^\flat_g$ is the change of contravariance to covariance mapping given by
\be
\label{eqLRiesz2}
L^\flat_g = g.L.
\ee
\finprop

\debdem
Recall: The contraction of an elementary ${0\choose2}$ tensor $\ell_1\otimes \ell_2$ with an elementary ${1\choose1}$ tensor $\vv\otimes \ell_3$ is the ${0\choose2}$ tensor 
$(\ell_1\otimes \ell_2).(\vv\otimes \ell_3) := (\ell_2.\vv)\ell_1\otimes \ell_3$. And the contraction on any tensors is the bilinear map defined on elementaty tensors.
So, with a basis $(\ve_i)$ in~$E$ and its dual basis $(e^i)$ in~$\Es$,
if $g=\sum_{ij} g_{ij} e^i\otimes e^j$ and $L = \sum_{ij} \Lij \ve_i\otimes e^j$ then
$g.L = \sum_{ijk} g_{ik}L^k{}_j \ve_i\otimes e^j$.
Thus $(g.L)(\vu,\vw) 
= \sum_{ij} u^i w^j (g.L)(\ve_i,\ve_j)
= \sum_{ijk} u^i w^j g_{ik}L^k{}_j
= \sum_{ik} u^i g_{ik}(L.\vw)^k 
= \sum_{ik} u^i g_{ik}(L.\vw)^k 
= g(\vu,L.\vw) = L^\flat_g(\vu,\vw)
$.
\findem

\mn
{\bf Quantification:}
Let $(\ve_i)$ be a basis in~$E$, and, with duality notations motivated by the flat notation
``$i$~top changed into $i$~bottom'' in the components $\Lij$ of~$L$, let $g_{ij}:=g(\ve_i,\ve_j)$, $L.\ve_j = \sumin \Lij \ve_i$ and $L^\flat_{g,ij} = L^\flat_g(\ve_i,\ve_j)$,
\ie\ with tensorial notations for calculations
\be
g=\sum_{ij} g_{ij} e^i\otimes e^j, \quad
L = \sum_{ij} \Lij \ve_i\otimes e^j, \quad 
L^\flat_g = \sum_{ij} L^\flat_{g,ij} e^i\otimes e^j.
\ee
So $[g]_{|\ve}=[g_{ij}]$, $[L]_{|\ve}=[\Lij]$ and $[L^\flat_g]_{|\ve} = [L^\flat_{g,ij}]$.
Then~\eref{eqLRiesz2} gives (or see next exercise)
\be
\label{eqLRiesze}
\boxed{[L^\flat_g] = [g].[L]} .
\ee

\debexe
Prove~\eref{eqLRiesze} with components.

\debrep
With~\eref{eqLRiesz} we get
$\ds L^\flat_{g,ij}
=L^\flat_g(\ve_i,\ve_j) 
= (\ve_i,L.\ve_j)_g
= (\ve_i,\sum_k L^k{}_j \ve_k)_g
= \sum_k L^k{}_j g_{ik}
= ([g].[L])_{ij}$.
\finrep
\finexe

%NB: Warning: $[L]$ can be symmetric while $[L^\flat_g]$ is not: Take $[L]=\pmatrix{0&1\cr1&0}$ and $[g]=\pmatrix{1&0\cr0&2}$.

\debrem
%$L^\flat$: obtained from the Riesz representation theorem (change of variance).
%\label{secLflat2}\def\vR{{\vec R}}
A change of variance, here from the ${1\choose1}$ type tensor~$L$ to the ${0\choose2}$ tensor~$L^\flat_g$,
is necessarily observer dependent:
There is no natural canonical isomorphism between a vector space $E$ and its dual~$E^*$, see~\S~\ref{secEEsnonnat}.
Details: Here fix $\vw$ and write $\ell_{g,\vw}(\vu) = (\vu,L.\vw)_g$ ($ = L^\flat_g(\vu,\vw)$);
Thus $\ell_{g,\vw}\in E^*$ is the  $\dd_g$-representation function (linear form) of the vector $L.\vw$,
\ie\ $\ell_{g,\vw} = \vR_g(L.\vw)$ where $\vR_g$ is the $\dd_g$-Riesz-representation operator (the change of variance operator, see~\eref{eqJg}.
\finrem

\comment{
Since $L\in\calL(E;E) \simeq T_L\in \calL(E^*,E;\RR)$, with the natural canonical isomorphism given by
$T_L(\ell,\vw)=\ell.L.\vw$, we get $T_L(\ell_{g,\vu},\vw) = \ell_{g,\vu}.L.\vw = (\vR_g(\ell_{g,\vu}),L.\vw)_g$
\be
L^\flat_g = (g.T_L)^T
\ee
Indeed, the function $L^\flat_{g,\vu} : \vw\in E \rar L^\flat_{g,\vu}(\vw)$ is a linear form, \cf~\eref{eqLRiesz},
thus can be represented by the $\dd_g$-Riesz representation vector $\vell_{g,\vu}\in E$
given by $L^\flat_{g,\vu}(\vw) = (\vell_{g,\vu},\vw)_g$,
thus $\vell_{g,\vw} = L^T_g.\vw$

\cf~\eref{eqRieszbase}
$[\vell_g] = [g]^{-1}.[\ell]^T$.

$L^\flat_g(\vu,\vw) = g(L.\vu,\vw)$ and

To get from $L$ to $L^\flat_g$ we can use the Riesz representation theorem, which gives the change of variance from ${1\choose1}$ for~$L$ to~${0\choose 2}$ for~$L^\flat$: The vector part (contravariant) is transformed into a linear form (covariant). In particular such a change is necessarily observer dependent (there is no natural canonical isomorphism between $E$ and~$E^*$). 
With $E^*=\calL(E;\RR)$ the set of continuous linear forms on~$E$, for any $\ell\in E^*$, call
\be
\vell_g = \vR_g(\ell)=\hbox{ the $\dd_g$-Riesz representation vector of $\ell$ see~\eref{eqJg}},
\ee
\ie\ $\vell_g\in E$ is defined by $(\vell_g,\vw)_g \mope^{\eref{eqrtr}}\ell.\vw$ for all $\vw\in E$
(change of variance from $\ell$ to~$\vell$); Then define $L^\flat_g$ by, for all $(\vu,\ell)\in E\times E^*$,
\be
(L^\flat_g(\vu,\vR_g(\ell))=) \quad 
L^\flat_g(\vu,\vell_g) := \ell.(L.\vu) .
\ee
Thus from the ${1\choose 1}$ tensor $L$, we obtained the ${0\choose 2}$  observer dependent tensor~$L^\flat_g$.
}

\comment{
%%%%%%%%%%%%%%%%%%%%%%%%%%%%%%%%%%%%%%%%%%%%%%%%%%%%%%%%%%%%%%%%%%%%%%%%%%%%%%%%%%%

\subsubsection{The pulled-back metric $g^*$ and $C^\flat_{G} = g^*$}

\def\vwsuP{{\vw^*_{1P}}}
\def\vwsdP{{\vw^*_{2P}}}
\def\vwsiP{{\vw^*_{iP}}}

(Mathematical remark, not essential to mechanics since $g^*$ despises~$\dd_G$.)

Let $g$ be a ${0\choose2}$ tensor in~$\Omegat$ (\eg, a metric in~$\Omegat$).
Its pull-back $g^*$ by the motion $\Phi:=\Phitzt$ is the ${0\choose2}$~tensor in~$\Omegatz$ defined by~\eref{eqpftzdb}: with $p=\Phitzt(P)$,
\be
\label{eqpbg1}
g_P^*(\vWuP,\vWdP) := g_p(F_P.\vWuP,F_P.\vWdP), \quad\hbox{in short:} \quad
\boxed{g^*(\vW_1,\vW_2) := (F.\vW_1,F.\vW_2)_g}  , %g(F.\vW_1,F.\vW_2)
\ee
for all vectors $\vWuP,\vWdP\in \RRntz$ at $P$ in~$\Omegatz$.
In other words, $g^*$ is defined thanks to the push-forward of vector fields, in short,
\be
\boxed{g^*(\vW_1,\vW_2) := g(\vw_1,\vw_2)} \qwhen \vw_i=F.\vW_i \hbox{ (push-forward)},
\ee
\ie,
In particular, if $g$ is a metric in $\Omegat$ then $g^*$~is a metric in~$\Omegatz$ (for a regular motion).

The sole definition of $g^*$ does not enable to obtain $C$ or~$C^\flat$: We also need an inner dot product $\dd_G$ in~$\RRntz$, \cf~\eref{eqdefCt} (definition of the Cauchy strain tensor made to measure relative vector deformations).
So, with an inner dot product $\dd_G$ in~$\RRntz$,
%Then, by definition~\eref{eqpbg1} of~$g^*$, and  
we get, for all $\vWu,\vWd\in\RRntz$, %\eref{eqpbg2} gives
\be
\label{eqpbg2c}
g^*(\vW_1,\vW_2) = (\CGg.\vWu,\vWd)_G = C^\flat_{Gg}(\vW_1,\vW_2) ,  \qthus 
\boxed{g^* = C^\flat_{Gg}} \qwritten g^* = C^\flat.
\ee
{\bf NB: } $C^\flat$ depends on $\dd_G$, cf~\eref{eqpbg2b} (bilinear form associated to the linear map~$C$), while the pull-back $g^*$ does \textbf{\textit{not}} depend on~$\dd_G$, \cf~\eref{eqpbg1}.
%In fact,  $g^\flat$ (more generally a pull-back) is not defined with the transposed $F^T=F^T_{Gg}$ of~$F$, \cf~\eref{eqpbg1}.

}

%%%%%%%%%%%%%%%%%%%%%%%%%%%%%%%%%%%%%%%%%%%%%%%%%%%%%%%%%%%%%%%%%%%%%%%%%%%%%%%%%%%

\subsection{A transposed of a linear map: depends on chosen inner dot products}
\label{seccpgdd0}

This paragraph is needed to define the transposed of the deformation gradient.

Not to be confused with the transposed of a matrix, \cf~\eref{eqTMat}. And
not to be confused with the objective transposed of a bilinear form, \cf~\eref{eqbT};
\Eg, a transposed of a linear map is \textslbf{not} objective.

%%%%%%%%%%%%%%%%%%%%%%%%%%%%%%%%%%%%%%%%%%%%%%%%%%%%%%%%%%%%%%%%%%%%%%%%%%%%%%%%%%%

\subsubsection{Definition (subjective)} %: Needs two inner dot products
\label{seccpgdd01}

$(E,\dd_g)$ and $(F,\dd_h)$ are Hilbert spaces, and $L\in\calL(E;F)$
(which is supposed to be continuous if $E$ and $F$ are infinite dimensional).
\Eg, $E=\RRntz$, $F=\RRnt$, $L=d\Phitzt(P)\in\calL(\RRntz;\RRnt)=$ the deformation gradient, \cf~\eref{eqdefFtf},
$\dd_g$ is the foot built Euclidean dot product chosen by the observer who made the measurements at~$\tz$,
$\dd_h$ is the metre built Euclidean dot product chosen by the observer who makes the measurements at~$t$.
%(The case $E=F$ and $\dd_g=\dd_h$ was treated in~\S~\ref{sectend}.)

\debdef
The transposed of $L\in\calL(E;F)$ relative to $\dd_g$ and~$\dd_h$
is the linear map $L^T_{gh} \in \calL(F;E)$ defined by, for all $(\vx,\vy)\in E\times F$,
\be
\label{eqseccpgdd0}
%(L^T_{gh}(\vy),\vx)_g = (\vy,L(\vx))_h,\qwritten
(L^T_{gh}.\vy,\vx)_g = (\vy,L.\vx)_h,
\ee
where we used the dot notation $L^T_{gh}(\vy)\eqnote L^T_{gh}.\vy$ since $L^T_{gh}$ is linear.
This defines the map
\be
(.)^T_{gh} : 
\left\{\eqalign{
\calL(E;F) & \rar \calL(F;E) \cr
L & \rar (.)^T_{gh}(L) := L^T_{gh}
}\right.
\ee
\findef

NB: So a linear map has an infinite number of transposed (it depends on inner dot products).

Notation: If $\dd_g$ and $\dd_h$ are imposed then $L^T_{gh} \eqnote L^T$.

And if $F=E$ and $\dd_h=\dd_g$ then $L^T_{gh} = L^T_g$, see~\S~\ref{sectend}. 
%(And we immediately get $(L^T_{gh})^T_{hg} = L$.)

\comment{
\debexe
\eref{eqseccpgdd0} gives a characterization of~$L^T_{gh}$. Prove that $L^T_{gh}$ does exist.

\debrep
See next paragraph: A linear map is defined by its values on one basis (matrix~$[(L^T_{gh})_{ij}]$).
\finrep
\finexe
}

%%%%%%%%%%%%%%%%%%%%%%%%%%%%%%%%%%%%%%%%%%%%%%%%%%%%%%%%%%%%%%%%%%%%%%%%%%%%%%%%%%%

\subsubsection{Quantification with bases}

Let $(\va_i)$ and $(\vb_i)$ be bases in~$E$ and~$F$,
let $g_{ij}:=g(\va_i,\va_j)$, $h_{ij}:=h(\vb_i,\vb_j)$, $[g]_{|\va} = [g_{ij}]$, $[h]_{|\vb} = [h_{ij}]$, and let (classical notation)
\be
\eqalign{
&L.\va_j = \sumim L_{ij} \vb_i, \qie
[L]_{|\va,\vb} = [L_{ij}] \eqnote [L], \cr
&L_{gh}^T.\vb_j = \sumin (L_{gh}^T)_{ij} \va_i, \qie
[L_{gh}^T]_{|\vb,\va} = [(L_{gh}^T)_{ij}] \eqnote [L_{gh}^T].
}
\ee
\eref{eqseccpgdd0} gives
$[\vx]_{|\va}^T.[g]_{|\va}.[L_{gh}^T.\vy]_{|\vy}
= ([L.\vx]_{|\vb})^T.[h]_{|\vb}.[\vy]_{|\vb}
%= [\vx]_{|\va}^T.([L]_{|\va,\vb})^T.[h]_{|\vb}.[\vy]_{|\vb}
$
for all $\vx,\vy$, thus, 
$[g]_{|\va}.[L_{gh}^T]_{|\vb,\va} = ([L]_{|\va,\vb})^T.[h]_{|\vb}$ and
$[L_{gh}^T]_{|\vb,\va} = [g]_{|\va}^{-1}.([L]_{|\va,\vb})^T.[h]_{|\vb}$.
Shortened notation:
\be
\label{eqaltltb}
[g].[L^T] = [L]^T.[h], \qie \sumkn g_{ik} (L_{gh}^T)_{kj} = \sumkm L_{ki} \,h_{kj},
\ee
\ie\
\be
\label{eqaltltb2}
\boxed{[L^T] = [g]^{-1}.[L]^T.[h]}, \qie
(L_{gh}^T)_{ij} = \sumkn\sumlm ([g]^{-1})_{ik} L_{\ell k} h_{\ell j}.
\ee
With duality notations,
$
L.\ve_j = \sumin L^i{}_j \ve_i, \; [L]_{|\ve} = [L^i{}_j], \; L_{gh}^T.\ve_j  = \sumin (L_{gh}^T)^i{}_j, \; [L_{gh}^T]_{|\ve}  = [(L_{gh}^T)^i{}_j],
$
and
\be
\label{eqaltltb3}
\sumkn g_{ik} (L_{gh}^T)^k{}_j= \sumkn L^k{}_i \,h_{kj}, \qie
(L_{gh}^T)^i{}_j = \sumkln ([g]^{-1})_{ik} L^\ell{}_k \, h_{\ell j}
\quad(\eqnote \sumkln (g^{ik} L^\ell{}_k \, h_{\ell j}).
\ee
(Be careful with the notation $([g]^{-1})_{ik}\eqnote g^{ij}$, see remark~\ref{remgsij1}.)

\debexe
Prove: If $L$ is invertible then $(L_{gh}^T)^{-1} = (L^{-1})^T_{hg}$.

\debrep
$(L_{gh}^T.(L^{-1})^T_{hg}.\vx,\vy)_g
= ((L^{-1})^T_{hg}.\vx,L.\vy)_h
= (\vx,L^{-1}.L.\vy)_g
= (\vx,\vy)_g
= (L_{gh}^T.(L_{gh}^T)^{-1}.\vx,\vy)_g$, 
true $\forall \vx,\vy$.
\finrep
\finexe

%%%%%%%%%%%%%%%%%%%%%%%%%%%%%%%%%%%%%%%%%%%%%%%%%%%%%%%%%%%%%%%%%%%%%%%%%%%%%%%%%%%

\subsubsection{Deformation gradient symmetric: Absurd}

The symmetry of a linear map $L\in\calL(E;F)$ is a nonsense if $E \ne F$.

\Eg: The gradient of deformation $\Ftzt(\ptz) = d\Phitzt(\ptz)\eqnote F \in  \calL(\RRntz;\RRnt)$ cannot be symmetric since $F^T \in \calL(\RRnt;\RRntz)$.
% (no ubiquity gift admissible;, full notation$\Ftzt(\ptz) \in  \calL(\TptzOmegatz;\TptOmegat)$).
%(Full notation: $\Ftzt(\ptz) = d\Phitzt(\ptz) \in  \calL((\RRntz,\dd_G);(\RRnt,\dd_g))$ cannot be symmetric since $\Ftzt(\ptz)^T \in \calL((\RRnt,\dd_g);(\RRntz,\dd_G))$.)
Idem for the first Piola--Kirchhoff tensor~$\PKtzt$, which motivates
the introduction of the symmetric second Piola--Kirchhoff tensor 
$\SKtzt$, %(full notation $\in  \calL(\TptzOmegatz;\TptzOmegatz)$), 
%$\in \calL(\TptzOmegatz;\TptzOmegatz)$, 
see Marsden--Hughes~\cite{marsden-hughes} or \S~\ref{secPK2}.

%%%%%%%%%%%%%%%%%%%%%%%%%%%%%%%%%%%%%%%%%%%%%%%%%%%%%%%%%%%%%%%%%%%%%%%%%%%%%%%%%%%

\subsubsection{Isometry}

%Let $(E,\dd_g)$ and $(F,\dd_h)$ be Hilbert spaces, with $\dim E = \dim F =n$.

\debdef
\label{defoi}
A linear map $L\in\calL(E;F)$ is an isometry relative to $\dd_g$ and~$\dd_h$ iff
\be
\label{eqdefoi}
\forall \vx,\vy\in E, \quad (L.\vx,L.\vy)_h = (\vx,\vy)_g,\qie L^T_{gh} \circ L = I_E \hbox{ (identity in $E$)}.
\ee
%N.B. : la définition d'isométrie n'est pas intrinsèque (à~$L$) : elle dépend du choix des produits scalaires. Voir exercice~\ref{exet0}.
\findef
%\label{defoi2}
In particular, an endomorphism $L\in\calL(E;E)$ is a $\dd_g$-isometry iff
\be
\label{eqdefoib}
\forall \vx,\vy\in E, \quad (L.\vx,L.\vy)_g = (\vx,\vy)_g, \qie L^T_g \circ L = I_E.
\ee
%Thus, if $(\ve_i)$ is a $\dd_g$-orthonormal basis, then $(L.\ve_i)$ is a $\dd_g$-orthonormal basis.

Thus, if $L\in\calL(E;F)$ is an isometry and $(\ve_i)$ is a $\dd_g$-orthonormal basis, then $(L.\ve_i)$ is a $\dd_h$-orthonormal basis, since $(L.\ve_i,L.\ve_j)_h = (\ve_i,\ve_j)_g = \delta_{ij}$ for all $i,j$.

\debexe
Let $\vf:E\rar F$.
Prove:
\be
\label{eqfvfisom}
%\hbox{if:}\quad\forall \vx,\vy\in E,\;\; (\vf(\vx),\vf(\vy))_h=(\vx,\vy)_g \hbox{ (isomtery)},
\hbox{if \ $\vf$ is an isometry}
\;\;\hbox{then}\;\; \hbox{$\vf$ is linear}.
\ee

\debrep
Let $(\ve_i)$ be a $\dd_g$-orthonormal basis; Thus $(\vf(\ve_i))$ is a $\dd_h$-orthonormal basis (since $\vf$ is an isometry).
Thus, if $\vx=\sumin x_i\ve_i$ then
$\ds \vf(\vx)
\mope^{b.o.n.}\sumin (\vf(\vx),\vf(\ve_i))_h\vf(\ve_i)
\mope^{hyp.}\sumin (\vx,\ve_i)_g\vf(\ve_i)
\mope^{b.o.n.}\sumin x_i\vf(\ve_i)
$,
thus $\ds \vf(\vx+\lambda\vy)
=\sumin (x_i+\lambda y_i)\vf(\ve_i) 
=\sumin x_i\vf(\ve_i) + \lambda \sumin  y_i\vf(\ve_i) 
= \vf(\vx)+\lambda\vf(\vy)$,
thus $\vf$ is linear.
\finrep
\finexe

\debexe
$\RRn$ is an affine space, $\vRRn$ is the usual associated vector space, and $\dd_g$ is an inner dot product in~$\vRRn$. Definition: A distance-preserving function $f : p\in\RRn \rar f(p) \in \RRn$ is a function \st
\be
\label{eqfvfisom2}
||\ora{f(p)f(q)}||_g = ||\ora{pq}||_g, \quad \forall p,q\in\RRn.
\ee
Prove: If $f$ is a distance-preserving function, then $f$ is affine. %, and $df$ is an isometry.
%Then let $L=df$ (linear) and prove $(L.\vx,L.\vy)_g=(\vx,\vy)_g$ for all $\vx,\vy\in\vRRn$.

\debrep
Let $O\in \RRn$ (an origin)
and $\vf:\vx=\ora{Op}\in\vRRn \rar \vf(\vx):=\ora{f(O)f(p)}$
(vectorial associated function).
Let $\vx=\ora{Op}$ and $\vy=\ora{Oq}$.
Then the remarkable identity
$\ds 2(\vf(\vx),\vf(\vy))_g
= ||\vf(\vx)||^2_g + ||\vf(\vy)||^2_g - ||\vf(\vx){-}\vf(\vy)||^2_g$ gives
$\ds 2(\vf(\vx),\vf(\vy))_g
= ||\vf(\vx)||^2_g + ||\vf(\vy)||^2_g - ||\ora{f(q)f(p)}||^2_g
= ||\vf(\vx)||^2_g + ||\vf(\vy)||^2_g - ||\ora{qp}||^2_g
= ||\vx||^2_g + ||\vy||^2_g - ||\vx-\vy||^2_g
=  2(\vx,\vy)_g
$, thus $\vf$ is an isometry, thus $\vf$ is linear \cf~\eref{eqfvfisom},
thus $f$ is affine since 
$f(p)=f(O)+\vf(\ora{Op})$.
\finrep
\finexe

\comment{
$\ds 2(\vf(\vx),\vf(\vy))_g
= ||\ora{f(O)f(p)}||^2_g + ||\ora{f(O)f(q)}||^2_g - ||\ora{f(q)f(p)}||^2_g
= ||\ora{Op}||^2_g + ||\ora{Oq}||^2_g - ||\ora{qp}||^2_g
= ||\vx||^2_g + ||\vy||^2_g - ||\vx-\vy||^2_g
=  2(\vx,\vy)_g
$,
}

%%%%%%%%%%%%%%%%%%%%%%%%%%%%%%%%%%%%%%%%%%%%%%%%%%%%%%%%%%%%%%%%%%%%%%%%%%%%%%%%%%%

\subsection{The adjoint of a linear map (objective)}
\label{secadjlm}

(For mathematicians; May produce misunderstandings, misuses, problematic mechanical interpretations.)

No inner dot product is required here: A linear map $L$ has only one adjoint~$L^*$ (intrinsic to~$L$);
While $L$ has many transposed $L^T=L^T_{gh}$ which depend on inner dot products.

%%%%%%%%%%%%%%%%%%%%%%%%%%%%%%%%%%%%%%%%%%%%%%%%%%%%%%%%%%%%%%%%%%%%%%%%%%%%%%%%%%%

\subsubsection{Definition}

$E$ and $F$ are vector spaces, and $E^*=\calL(E;\RR)$ and $F^*=\calL(F;\RR)$ are the dual spaces (made of linear continuous forms). (If $E$ and $F$ are finite dimensional, the continuity is always satisfied.)

%(In infinite dimension, consider two Banach spaces $(E,||.||_E)$ and $(F,||.||_F)$, and linear continuous maps).

\debdef
\label{defadjL}
Let $L\in\calL(E;F)$ (linear and continuous); Its adjoint is the linear map $L^* \in \calL(F^*;E^*)$ canonically defined by
\be
\label{eqdefadjL0}
L^* :
\left\{\eqalign{
F^* & \rar E^* \cr
m & \rar L^*(m) := m \circ L, 
}\right.
\ee
\ie, for all $(\vx,m)\in E\times F^*$,
\be
\label{eqdefadjL0b}
(L^*(m))(\vx) := m(L(\vx)) .
\ee
(The adjoint $L^*$ cannot be confused with a transposed $L^T$ which requires inner dot products, \cf~\eref{eqseccpgdd0}.)
\findef

The linearity of~$L^*$ is trivial, thus, together with the linearity of $m$ and~$L$,  we can use the dot notation:
\be
\label{eqdefadjL}
L^*.m := m.L, \qand (L^*.m).\vx := m.L.\vx.
\ee
And $||L^*.m||_\Es = ||m.L||_\Es \le ||m||_\Fs||L||_{\calL(E;F)}$ gives $||L^*||_{\calL(F^*;E^*)} \le ||L||_{\calL(E;F)}<\infty$, thus $L^*$ is continuous (when $L$~is).

\comment{
\debrem
If $E$ and $F$ are reflexive Banach spaces, \ie\ if $(E^*)^*=E$ (it is always the case if $E$ is a Hilbert space), then the supposed continuity of~$L$ gives $L^*$ is continuous and $||L^*||_{L(F^*;E^*)}=||L||_{L(E;F)}$; Indeed,
$||L^*(m)||_\Es = ||m.L||_\Es \le ||L||\;||m||_\Fs$ gives $||L^*|| \le ||L||$,
and $L^{**}=L$ gives $||L|| \le ||L^*||$, thus $\LL||)||L^*||$.
\finrem
\debrem
If $E$ and $F$ are finite dimensional, then the continuity requirements are full filled.
\comment{
If $E$ and $F$ are infinite dimensional, then the continuity requirement enables to prove the existence of~$L^*$, thanks to the Riesz representation theorem. Indeed, for a given $y\in F$, define $a_y:\vx\in E \rar a_\vy(\vx) := (L.\vx,\vy)_F$; Then $a_\vy$ is trivially linear, and $|a_\vy(\vx)|\le ||L||\,||\vy||_F||\vx||_E$ gives $||a_\vy||\le ||L||\,||\vy||_F <\infty$ since $L$ is continuous; Then the Riesz representation theorem tells that $a_\vy\in E^*$ can be represented by a vector $\vell_\vy \eqnote L^*(\vy)\in E$ where $(L^*(\vy),\vx)_g=(L.\vx,\vy)_F$. Thus $L^*:F$
}
\finrem
}

%%%%%%%%%%%%%%%%%%%%%%%%%%%%%%%%%%%%%%%%%%%%%%%%%%%%%%%%%%%%%%%%%%%%%%%%%%%%%%%%%%%

\subsubsection{Quantification}

$E$ and $F$ are finite dimensional, $\dim E=n$, $\dim F=m$, and $(\va_i)$ and $(\vb_i)$ are bases in $E$ and~$F$.
Let $[L]_{|\va,\vb} \eqnote [L]$, $[L^*]_{|b,a} \eqnote [L^*]$, $[m]_{|b} \eqnote [m]$ and $[\vx]_{|\va} \eqnote [\vx]$ be the matrices relative to the chosen bases:
\eref{eqdefadjL} gives 
$([L^*].[m].[\vx] = [m].[L].[\vx]$ for all $\vx\in E$ and $m\in\Fs$, thus, for all $m\in F^*$ (recall that $[m]$ is a line matrix),
thus $[L^*].[m]^T = ([L]^T.[m]^T$,
thus
\be
\label{defadjLq2}
\boxed{[L^*] = [L]^T} \quad(\hbox{transposed matrix}).
\ee
(Full notation: $[L^*]_{|b,a} = ([L]_{|\va,\vb})^T$.) 

Details:
With the dual bases $(\pi_{ai})$ and $(\pi_{bi})$,
with $L.\va_j=\sumim L_{ij} \vb_i$, \ie\
$[L]_{|\va,\vb} = [L_{ij}]_{i=1,...,m \hfill\atop j=1,...,n}$,
and with
$L^*.\pi_{bj} = \sumin (L^*)_{ij} \pi_{ai}$, \ie\
$[L^*]_{|b,a} = [(L^*)_{ij}]_{i=1,...,n \hfill\atop j=1,...,m}$,
\eref{eqdefadjL} gives, for all $(i,j)\in[1,n]_\NN \times [1,m]_\NN$,
\be
\label{defadjLq}
(L^*.\pi_{bj}).\va_i =\pi_{bj}.(L.\va_i), \qthus \boxed{ (L^*)_{ij} = L_{ji} } \qand [L^*]=[L]^T.
\ee
Duality notations (warning: can be misused): $L.\va_j=\sumim \Lij \vb_i$, \ie\
$[L]_{|\va,\vb} = [\Lij]_{i=1,...,m \hfill\atop j=1,...,n}$, and
$L^*.b^j = \sumin (L^*)_i\,^j a^i$,  \ie\
$[L^*]_{|b,a} = [((L^*)_i\,^j]_{i=1,...,n \hfill\atop j=1,...,m}$, thus, for all $(i,j)\in[1,n]_\NN \times [1,m]_\NN$,
\be
\label{defadjLqd}
(L^*.b^j).\va_i =b^j.(L.\va_i), \qthus (L^*)_i\,^j = L^j{}_i \qand [L^*]=[L]^T.
\ee
(Recall: Use classical notations if in doubt, or, preferably, don't use duality notations here.)

\debrem
Reminder: The transposed $b^T$ of a bilinear~$b$ form is intrinsic to~$b$, and the adjoint $L^*$ of a linear map~$L$ is intrinsic to~$L$; But a transposed $L^T$ of a linear form $L$ is \textslbf{not} intrinsic to the linear form (it depends on chosen inner dot products): Watch out for the (unfortunate) vocabulary ``transpose''!
\finrem

\comment{
Indeed $m=\sumjn m_j\pi_{bj}$ gives
$L^*.m
%=L^*.(\sumkn m_k\pi_{bk})
=\sumjn m_j L^*.\pi_{bj}
=\sumijn m_j (L^*)_{ij} \pi_{ai}
=\sumin (\sumjn L_{ji} m_j) \pi_{ai}
$.
}

%%%%%%%%%%%%%%%%%%%%%%%%%%%%%%%%%%%%%%%%%%%%%%%%%%%%%%%%%%%%%%%%%%%%%%%%%%%%%%%%%%%

\subsubsection{Relation with the transposed when inner dot products are introduced}

\def\vmh{{\vec m_h}}

let $L\in\calL(E;F)$. We need inner dot products $\dd_g$ and $\dd_h$ in~$E$ and~$F$ to define $L^T = L^T_{gh}$. To have a functional relation between $L^*$ and~$L^T_{gh}$, we use
the $\dd_g$-Riesz representation mapping
$\vR_g :
\left\{\eqalign{
E^* & \rar E \cr
\ell & \rar \vR_g(\ell)=\vell_g
}\right\}$,
where $\ell.\vx = (\vell_g,\vx)_g$ for all $\vx\in E$, see~\eref{eqJg}; idem with~$F$.

Let $L\in\calL(E;F)$ (continuous). 
For all $\vx\in E$ and all $m\in\Fs$ we have
\be
(L^*.m).\vx \mathop{=}^{\eref{eqdefadjL0b}} m.(L.\vx),\qthus
(\vR_g(L^*.m),\vx)_g =  (\vR_h(m),L.\vx)_h, 
\ee
thus $((\vR_g \circ L^*).m),\vx)_g =  ((L^T_{gh}\circ \vR_h).m,L.\vx)_g$.
Thus $\vR_g \circ L^* = L^T_{gh}\circ \vR_h$, \ie
\be
\label{defadjLqR}
\boxed{L^T_{gh} = \vR_g \circ L^* \circ (\vR_h)^{-1}} \qie	
\eqalign{
E\;\mathop{\longleftarrow}^{\ds\;\; L^T_{gh} } \;& F
\cr
\noalign{\vskip-4pt}
\vR_g \uparrow \qquad\quad &\;\;\uparrow \vR_h
\cr
\noalign{\vskip-2pt}
E^*\;\mathop{\longleftarrow}_{\ds\; L^*} \;& F^*
\cr
}\;\hbox{ is a commutative diagram}.
\ee

\debexe
From~\eref{defadjLqR}, 
recover~\eref{eqaltltb}, \ie\ $[L_{gh}^T] = [g]^{-1}.[L]^T.[h]$.
%recover $[L_{gh}^T]_{\vb,\va} = [g]_{|\va}^{-1}.([L]_{|\va,\vb})^T.[h]_{|\vb}$, \cf~\eref{eqaltltb}.

\debrep
$[L^T_{gh}] 
\equalref{defadjLqR} [\vR_g].[L^*].[\vR_h]^{-1}
\equalref{eqRieszbase}[g]^{-1}.[L]^T.[h]
$.
\finrep
\finexe

%%%%%%%%%%%%%%%%%%%%%%%%%%%%%%%%%%%%%%%%%%%%%%%%%%%%%%%%%%%%%%%%%%%%%%%%%%%%%%%%%%%

\subsection{Tensorial representation of a linear map}
\label{sectroflm}

%%%%%%%%%%%%%%%%%%%%%%%%%%%%%%%%%%%%%%%%%%%%%%%%%%%%%%%%%%%%%%%%%%%%%%%%%%%%%%%%%%%

\subsubsection{A tensorial representation}

Consider the natural canonical isomorphism (between linear maps $E\rar F$ and bilinear forms $F^*\times E \rar \RR$)
\be
\label{eqdefcalJ}
\tcalJ : 
\left\{\eqalign{
\calL(E;F) & \rar \calL(F^*,E;\RR) \cr
L & \rar \bL=\tcalJ(L)
}\right\}
\qwhere \bL(m,\vu) := m.(L.\vu), \quad \forall (m,\vu)\in F^* \times E,
\ee
see~\S~\ref{secal}. And $\bL$ is also named~$L$ for calculations purposes, see~\eref{eqLaitJt2}.

(NB: It can be dangerous to substitute $L$ with $\bL$, see \eg~\S~\ref{secbT}.)

\medskip\noindent
{\bf Quantification:} 
Let $(\va_i)_{i=1,...,n}$ be a basis in~$E$, 
$(\vb_i)_{i=1,...,m}$ be a basis in~$F$ which dual basis is $(\pibi)$,
$L\in\calL(E;F)$. Then
\be
\label{eqLaitJ0}
\beta_L(\pibi,\va_i) = \pibi.L.\va_i.
\ee
Thus, if
\be
\label{eqLaitJ}
L.\va_j = \sumim L_{ij} \vb_i \qthen  \bL = \sumim\sumjn L_{ij} \vb_i \otimes \pi_{aj}
\ee
Indeed, 
$(\sum_{ij} L_{ij} \vb_i \otimes \pi_{aj})(\pibk,\va_\ell)
=\sum_{ij} L_{ij} (\vb_i \otimes \pi_{aj})(\pibk,\va_\ell)
=\sum_{ij} L_{ij} (\vb_i.\pibk)(\pi_{aj}.\va_\ell)
=\sum_{ij} L_{ij} (\vb_i.\pibk)(\pi_{aj}.\va_\ell)
=\sum_{ij} L_{ij} \delta_{ki}\delta_{j\ell}
= L_{k\ell} = \pibk.L.\va_\ell
$,
so~\eref{eqLaitJ0} gives~\eref{eqLaitJ}.

Duality notations:
$L.\va_j = \sumim \Lij \vb_i$ and $\bL = \sumim\sumjn \Lij \vb_i \otimes a^j$.

\mn
{\bf Contraction rule.} If you write $L = \sumim\sumjn L_{ij} \vb_i \otimes \pi_{aj}$ ($\simeq\beta_L$),
then the vector $L.\vu\in F$ is computed thanks to the ``contraction rule'':
\be
\label{eqLaitJt2}
L.\vu = (\sumim\sumjn L_{ij} \vb_i \otimes \underbrace{\pi_{aj}).\vu}_{\makebox[.1cm]{\footnotesize\rm contraction}}
:= \sumim\sumjn L_{ij} \vb_i (\pi_{aj}.\vu)
= \sumim\sumjn L_{ij} u_j\vb_i.
%\mathop{=}^{\eref{eqpropLab20}} L.\vu.
\ee
(With duality notations:
$\ds L.\vu= 
(\sumim\sumjn L^i{}_j \vb_i \otimes \underbrace{a^j).\vu}_{\makebox[.1cm]{\footnotesize\rm contraction}}
= \sumim\sumjn L^i{}_j \vb_i (a^j.\vu)
= \sumim\sumjn L^i{}_j u^j\vb_i
%\mathop{=}^{\eref{eqpropLab20}} L.\vu
$.)

\debrem
\label{remlinrbi}
Warning: The bilinear form $\bL$ should not be confused with the linear map~$L$: The domain of definition of~$\bL$ is $F^*\times E$, and $\bL$ acts on the two objects $\ell$ (linear form) and $\vu$ (vector) to get a \textbf{\textsl{scalar}} result; While the domain of definition of~$L$ is $E$, and $L$ acts one object~$\vu$ to get a \textbf{\textsl{vector}} result. However, you can use the tensorial notation for~$L$... only to calculate $L.\vu$ with~\eref{eqLaitJt2}.
%Paradigmatic example: The change of basis endomorphism~$\calP$, see~\eref{eqdefP0}: It is not represented with tensorial notations except eventually for computational purposes.
%Moreover, look at~\S~\ref{secbT}.
\finrem

%%%%%%%%%%%%%%%%%%%%%%%%%%%%%%%%%%%%%%%%%%%%%%%%%%%%%%%%%%%%%%%%%%%%%%%%%%%%%%%%%%%

\subsubsection{Warning: Confusion between transposed and adjoint}

The transposed $L^T\in\calL(F;E)$ of a linear map $L\in\calL(E;F)$ needs inner dot products to be defined, cf~\eref{eqseccpgdd0}: It is  \textslbf{not} intrinsic to~$L$, \textslbf{not} objective ; While the transposed $b^T\in\calL(B,A;\RR)$ of a bilinear form $b\in\calL(A,B;\RR)$ is intrinsic to~$L$ (it does not need inner dot products to be defined).

So if you represent a linear map $L\in\calL(E;F)$ by its tensorial representation $\bL\in\calL(F^*,E;\RR)$, \cf~\eref{eqLaitJ},
then 

1- you know the transposed $\bLT$ (given by $\bLT(\vw,\vu) = \bL(\vu,\vw)$),

2- but you \textsl{\textbf{cannot}} deduce the transposed $L^T\in\calL(F;E)$ from $\bLT$ (\ie, to start with $\bLT$ is misleading): You need to choose inner dot products, and then use the formula $(L^T.\vy,\vx)_g = (\vy,L.\vx)_h$ where $L^T:=L^T_{gh}$ to get $[L^T] = [g]^{-1}.[L]^T.[h]$ (and $[L^T]\ne[L]^T$ in general).

3- In particular: If $L \in \calL(E;E)$ is symmetric (relative to the chosen inner dot products), then $\bL\in\calL(E^*,E;\RR)$ is never symmetric because $E^*\ne E\,$!
(Recall: there is no natural canonical isomorphism between $E$ and~$E^*$.)

\comment{
\debrem
We have $L\in\calL(E;F) \simeq \bL \in \calL(F^*,E;\RR)$ (natural canonical isomorphism)
where $\bL(\pi_{b_i},\va_j) := \pi_{bi}.(L.\va_j) % = L_{ij}
$,
idem $L^*\in\calL(F^*;E^*) \simeq \bLs \in \calL(E,F^*;\RR)$
where $\bLs(\va_i,\pi_{bj}) := \va_i.(L^*.\pi_{bj}) = (L^*.\pi_{bj}).\va_i % =(L^*)_{ij}
$;
Thus \eref{eqdefadjL}$_2$ gives $\bLs(\va_i,\pi_{bj}) = \bL(\pi_{b_j},\va_i)$, \ie\ $(\bLs)_{ij} = (\bL)_{ji}$, \ie\ $[\bLs] = [\bL]^T$ which is nothing but $[L^*] = [L]^T$, relation between matrices; Which is not connected with $[L^T]$ unless you choose inner dot products and make calculations that give
$[L^T] = [g]^{-1}.[L]^T.[h]$, \cf~\eref{eqaltltb2}. 
%(Use Marsden notations if you prefer: $\bLs(\va_i,b^j) = \bL(b^j,\va_i)$.)
\finrem
}

%%%%%%%%%%%%%%%%%%%%%%%%%%%%%%%%%%%%%%%%%%%%%%%%%%%%%%%%%%%%%%%%%%%%%%%%%%%%%%%%%%%

\subsection{Change of basis formulas for bilinear forms and linear maps}

%%%%%%%%%%%%%%%%%%%%%%%%%%%%%%%%%%%%%%%%%%%%%%%%%%%%%%%%%%%%%%%%%%%%%%%%%%%%%%%%%%%

\subsubsection{Notations for transitions matrices for bilinear forms and linear maps}
\label{secannfcb}

\def\PA{{P_{\!\!A}}}
\def\PB{{P_{\!\!B}}}
\def\QA{{Q_{\!\!A}}}
\def\QB{{Q_{\!\!B}}}
\def\calPA{\calP_{\!\!A}}
\def\calPB{\calP_{\!\!B}}

Let $A$ and~$B$ be finite dimension vector spaces, $\dim A = n$, $\dim B=m$.
(\Eg\ application to the change of basis formula for the deformation gradient $A{=}\RRntz \rar B{=}\RRnt$.)

Let $(\vaio)$ and $(\vain)$ be two bases in~$A$, and $(\vbio)$ and $(\vbin)$ be two bases in~$B$.
Let $\calPA$ and $\calPB$ be the change of basis endomorphisms from old to new bases,
and $\PA:=[\calPA]_{|\vao}=[\PA_{ij}]$ and $\PB:=[\calPB]_{|\vbo}=[\PB_{ij}]$ be the associated transition matrices, and $\QA=\PA^{-1}$ and $\QB=\PB^{-1}$:
\be
\label{eqsecannfcb}
\eqalign{
&\vajn = \calPA.\vaio = \sumijn \PA_{ij} \vaio, \quad
\pi_{a\new,j} = \sumin \QA_{ij} \pi_{a\old,i},\cr
&\vbjn = \calPB.\vbio = \sumijm \PB_{ij} \vbio, \quad
\pi_{b\new,j} = \sumijn \QB_{ij} \pi_{b\old,i}.
}
\ee
Duality notations: $\vajn = \sumin \PA^i{}_j \vaio$ and $\ain = \sumjn \QA^i{}_j \ajo$
and $\vbjn = \sumin \PB^i{}_j \vbio$ and $\bin = \sumjn \QB^i{}_j \bjo$.

%%%%%%%%%%%%%%%%%%%%%%%%%%%%%%%%%%%%%%%%%%%%%%%%%%%%%%%%%%%%%%%%%%%%%%%%%%%%%%%%%%%

\subsubsection{Change of coordinate system for bilinear forms $\in\calL(A,B;\RR)$}

Let $g\in\calL(A,B;\RR)$, and, for all $(i,j)\in [1,n]_\NN\times [1,m]_\NN$,
\be
g(\vaio,\vbjo)= M_{ij}, \quad g(\vain,\vbjn)= N_{ij}, \qie
\left\{\eqalign{
&[g]_{|\old s} = M = [M_{ij}]_{i=1,...,n \atop j=1,...,m}, \cr
&[g]_{|\new s} = N = [N_{ij}]_{i=1,...,n \atop j=1,...,m}.
}\right.
\ee

\debprop
Change of basis formula:
\be
\label{eqsecannfcb20}
[g]_{|\new s} = \PA^T.[g]_{|\old s}.\PB, \qie
N = \PA^T.M.\PB.
\ee
In particular, if $A=B$ and $(\vaio) = (\vbio)$ and $(\vain) = (\vbin)$,
then $\PA = \PB \eqnote P$, and
\be
\label{eqsecannfcb2}
\boxed{[g]_{|\new} = P^T.[g]_{\old}.P}, \qie
N = P^T.M.P.
\ee
\finprop

\debdem
$N_{ij}
=g(\vain,\vbjn)
= \sum_{k\ell} \PA^k{}_i \PB^\ell{}_j g(\vako,\vblo)
= \sum_{k\ell} \PA^k{}_i M_{k\ell} \PB^\ell{}_j
= \sum_{k\ell} (\PA^T)^i{}_k M_{k\ell} \PB^\ell{}_j
%=(\PA^T.N.\PB)_{ij}
$.
\findem

\debexe
Prove (objective result):
\be
\label{eqsecannfcb3}
g(\vu,\vw) = [\vu]_{|\van}^T.[g]_{|\new s}.[\vw]_{|\vbn} = [\vu]_{|\vao}^T.[g]_{|\old s}.[\vw]_{|\vbo}.
\ee

\debrep
$[\vu]_{|\van}^T.[g]_{|\new s}.[\vw]_{|\vbn}
= (\PA^{-1}.[\vu]_{|\vao})^T.(\PA^T.[g]_{|\old s}.\PB).(\PB^{-1}.[\vw]_{|\vbo})
$.
\finrep
\finexe

%%%%%%%%%%%%%%%%%%%%%%%%%%%%%%%%%%%%%%%%%%%%%%%%%%%%%%%%%%%%%%%%%%%%%%%%%%%%%%%%%%%

\subsubsection{Change of coordinate system for bilinear forms $\in\calL(A^*,B^*;\RR)$}

Let $z\in\calL(A^*,B^*;\RR)$, and, for all $(i,j)\in [1,n]_\NN\times [1,m]_\NN$,
\be
z(\aio,\bjo)= M^{ij}, \quad z(\ain,\bjn)= N^{ij}, \qie
\left\{\eqalign{
&[z]_{|\old s} = M = [M^{ij}]_{i=1,...,n \atop j=1,...,m}, \cr
&[z]_{|\new s} = N = [N^{ij}]_{i=1,...,n \atop j=1,...,m}.
}\right.
\ee

\debprop
Change of basis formula:
\be
[z]_{|\new s} = \PA^{-T}.[z]_{|\old s}.\PB^{-1}, \qie
N = \PA^{-T}.M.\PB^{-1}.
\ee
In particular, if $A=B$ and $(\vaio) = (\vbio)$ and $(\vain) = (\vbin)$,
then $\PA = \PB \eqnote P$, and
\be
\boxed{[z]_{|\new} = P^{-T}.[z]_{\old}.P^{-1}}, \qie
N = P^{-T}.M.P^{-1}.
\ee
\finprop

\debdem
$N_{ij}=z(\ain,\bjn)
= \sum_{k\ell} \QA^k{}_i \QB^\ell_j z(\ako,\blo)
= \sum_{k\ell} \QA^k{}_i M^{k\ell} \QB^\ell{}_j
= \sum_{k\ell} (\QA^T)^i{}_k M^{k\ell} \QB^\ell{}_j
%=(\QA^T.N.\QB)_{ij}
$.
\findem

%%%%%%%%%%%%%%%%%%%%%%%%%%%%%%%%%%%%%%%%%%%%%%%%%%%%%%%%%%%%%%%%%%%%%%%%%%%%%%%%%%%

\subsubsection{Change of coordinate system for bilinear forms $\in\calL(B^*,A;\RR)$}

(Toward linear maps $L \in \calL(A;B) \simeq \calL(B^*,A;\RR)$ thanks to the natural canonical isomorphism.)

Let $T\in\calL(B^*,A;\RR)$, and, for all $(i,j)\in [1,n]_\NN\times [1,m]_\NN$,
%and $T = \sum_{ij} M^i{}_j \vbio \otimes \ajo = \sum_{ij} N^i{}_j \vbin \otimes \ajn$, \ie
\be
T(\bio,\vajo)= M^i{}_j, \quad T(\bin,\vajn)= N^i{}_j, \qie
\left\{\eqalign{
&[T]_{|\old s} = M = [M^i{}_j]_{i=1,...,n \atop j=1,...,m}, \cr
&[T]_{|\new s} = N = [N^i{}_j]_{i=1,...,n \atop j=1,...,m}.
}\right.
\ee

\debprop
Change of basis formula:
\be
[T]_{|\new s} = \PB^{-1}.[T]_{|\old s}.\PA, \qie
N = \QA.M.\PB.
\ee
In particular, if $A=B$ and $(\vaio) = (\vbio)$ and $(\vain) = (\vbin)$,
then $\PA = \PB \eqnote P$, and
\be
\boxed{[T]_{|\new} = P^{-1}.[T]_{\old}.P}, \qie
N = P^{-1}.M.P, \qie N^i{}_j = \sumkln Q^i{}_k M^k{}_\ell P^\ell{}_j.
\ee
%and $N_{ij} = \sum_{k\ell} Q_{ik} M_{k\ell} P_{\ell k}$ with classical notations.
\finprop

\debdem
$N^i{}_j = T(\bin,\vajn)
= \sum_{k\ell} \QB^i{}_k\PA^\ell{}_j T(\bio,\vajo)
= \sum_{k\ell} \QB^i{}_k M^i{}_j\PA^\ell{}_j
$
\findem

%%%%%%%%%%%%%%%%%%%%%%%%%%%%%%%%%%%%%%%%%%%%%%%%%%%%%%%%%%%%%%%%%%%%%%%%%%%%%%%%%%%

\subsubsection{Change of coordinate system for tri-linear forms $\in\calL(A^*,A,A;\RR)$}

(Toward $d^2\vu$: For a vector field $\vu \in \Gamma(U) \simeq \Tuzu$, $\vu(p) \in \vRRn$, 
its differential satisfies $d\vu(p) \in \calL(\vRRn;\vRRn) \simeq \calL(\RRns,\vRRn;\RR)$, 
and $d^2\vu(p) \in \calL(\vRRn; \calL(\vRRn;\vRRn)) \simeq \calL(\RRns,\vRRn,\vRRn;\RR)$, see~\S~\ref{secasod}.)
%So here we take $B=A$, thus $B^*=A^*$.)

Consider a tri-linear form $T\in\calL(A^*,A,A;\RR) %\eqnote \calL^1_2
$, and
%let, for all $i,j,k = 1,...,n$,
\be
%T=\sumijkn M^i_{jk} \vaio\otimes \ajo\otimes \ako=\sumijkn N^i_{jk} \vain\otimes \ajn\otimes \akn
M^i_{jk}=T(\aio,\vajo,\vako),\quad N^i_{jk}=T(\ain,\vajn,\vakn), \qie
[T]_{|\vao} = [M^i_{jk}], \quad[T]_{|\van} = [N^i_{jk}].
\ee
Then
\be
\label{eqcvTijk}
N^i_{jk} = \sum_{\lambda,\mu,\nu=1}^n Q^i_\lambda P^\mu_j P^\nu_k M^\lambda_{\mu\nu}.
\ee
Indeed
$\sum_{\lambda\mu\nu} M^\lambda_{\mu\nu} \va_{old,\lambda}\otimes a_\old^\mu\otimes a_\old^\nu
=\sum_{\lambda\mu\nu ijk} M^\lambda_{\mu\nu} Q^i_\lambda P^\mu_j P^\nu_k \vain\otimes \ajn\otimes \akn
$.
\findem

%%%%%%%%%%%%%%%%%%%%%%%%%%%%%%%%%%%%%%%%%%%%%%%%%%%%%%%%%%%%%%%%%%%%%%%%%%%%%%%%%%%

\subsubsection{Change of coordinate system for linear maps $\in\calL(A;B)$}

Notation of~\S~\ref{secannfcb}.
Let $L\in\calL(A;B)$ be a linear map, and let, for all $j=1,...,n$,
\be
\left\{\eqalign{
&  L.\vajo = \sumim M_{ij} \vbio = \sumim M^i{}_j \vbio \qie [L]_{|old s} = M=[M_{ij}]=[M^i{}_j]_{i=1,...,m \atop j=1,...,n} , \cr
&
L.\vajn = \sumim N_{ij} \vbin = \sumim N^i{}_j \vbin \qie [L]_{|new s} = N=[N_{ij}]=[N^i{}_j]_{i=1,...,m \atop j=1,...,n} , \cr
}\right.
\ee
with classical and duality notations.
%The change of basis formula for vectors $\ds [L.\vu]_{|\vbn} \equalref{eqdefP1} \PB^{-1}.[L.\vu]_{|\vbo} $ gives:

\debprop
Change of bases formula:
\be
\label{eqLijoldnew2}
[L]_{|\new s} = \PB^{-1}.[L]_{|\old s}.\PA, \qie 
N = \PB^{-1}.M.\PA.
\ee
In particular, if $A=B$, if $(\vaio) = (\vbio)$, $(\vain) = (\vbin)$, then
$\PA = \PB \eqnote P$ and
\be
\label{eqLijoldnew3}
\boxed{[L]_{|\new} = P^{-1}.[L]_{|\old}.P}, \qie
N = P^{-1}.M.P, \qie N_{ij} = \sum_{k,\ell=1}^n Q_{ik}M_{k\ell}P_{\ell j},
\ee
with $Q=P^{-1}$, and with duality notations $N^i{}_j = \sum_{k\ell} Q^i{}_k M^k{}_\ell P^\ell{}_j$.
\finprop

\debdem
$L.\vajn
= \sum_i N^i{}_j \vbin
= \sum_{ik} N^i{}_j \PB^k{}_i\vbko
= \sum_k (\PB.N)^k{}_j\vbko
$ and
$L.\vajn
=L.(\sum_i \PA^i{}_j\vaio)
=\sum_i \PA^i{}_j \sum_k M^k{}_i\vbko
=\sum_k (M.\PA)^k{}_j\vbko
$, for all~$j$, %thus $(\PB.N)^k{}_j = (M.\PA)^k{}_j$ for all $k,j$, 
thus $\PB.N = M.\PA$. %, \ie~\eref{eqLijoldnew2}.
\findem

\debexe
Prove:
\be
\ell.L.\vu = [\ell]_{|\vbn}.[L]_{|\new s}.[\vu]_{|\van} = [\ell]_{|\vbo}.[L]_{|\old s}.[\vu]_{|\vao} \quad(\hbox{objective result}).
\ee

\debrep
$[\ell]_{|\vbn}.[L]_{|\new s}.[\vu]_{|\van}
= ([\ell]_{|\vbo}.\PB).(\PB^{-1}.[L]_{|\old s}.\PA).(\PA^{-1}.[\vu]_{|\vao})
$.
\finrep
\finexe

\debrem
Bilinear forms in~$\calL(A,A;\RR)$ and endomorphisms in~$\calL(A;A)$
behave differently: The formulas~\eref{eqsecannfcb2} and~\eref{eqLijoldnew3} should not be confused
since $P^{-1} \ne P^T$ in general.
\Eg, if an English observer uses a Euclidean (old) basis $(\va_i) = (\vaio)$ in foot,
if a French observer uses a Euclidean (new) basis $(\vb_i)= (\vain)$ in metre,
and if (simple case) $\vb_i=\lambda \va_i$ for all~$i$ (change of unit),
then
\be
[L]_{|\new} =  [L]_{|\old},
\qwhile [g]_{|\new} = \underbrace{\lambda^2}_{>10}\; [g]_{|\old}. % \;\hbox{ with }\; \lambda>10.
\ee
Quite different results! \Ie\ $P^{-1}.[L]_{|\old}.P \ne P^T.[L]_{|\old}.P$ for a general change of basis. % (when $P^T\ne P^{-1}$).
%( if $P^T=P^{-1}$ which is the very restricted case of change of orthonormal bases). 
See the Mars Climate Orbiter crash, remark~\ref{remMCOC}, where someone forgot that 1 foot $\ne$ 1 metre.
%General case: Compare~\eref{eqsecannfcb20} and~\eref{eqLijoldnew2}.
\finrem

%%%%%%%%%%%%%%%%%%%%%%%%%%%%%%%%%%%%%%%%%%%%%%%%%%%%%%%%%%%%%%%%%%%%%%%%%%%%%%%%%%%
%%%%%%%%%%%%%%%%%%%%%%%%%%%%%%%%%%%%%%%%%%%%%%%%%%%%%%%%%%%%%%%%%%%%%%%%%%%%%%%%%%%

%\part{Referential, Euclidean Framework, Tensors, Differentials}
%\part{Referential, Euclidean Framework}

\section{Euclidean Frameworks}
\label{secrefE}

Time and space are decoupled (classical mechanics).
$\RRn$ is the geometric affine space, $n=1,2,3$,
and $\vRRn$ is the associated vector space made of ``bi-point vectors''.
% which will be equipped with measuring instruments: Euclidean dot products to get angles and lengths.

%%%%%%%%%%%%%%%%%%%%%%%%%%%%%%%%%%%%%%%%%%%%%%%%%%%%%%%%%%%%%%%%%%%%%%%%%%%%%%%%%%%

\subsection{Euclidean basis}
\label{secbe}

{\bf Manufacturing of a Euclidean basis}.

An observer chooses a unit of measure (\foot, \metre, a unit of length used by Euclid, the diameter a of pipe...)
and makes a ``unit rod'' of length~$1$ in this unit.

Postulate: The length of the rod does not depend on its direction in space.

\medskip
$\bullet$ Space dimension $n = 1$:
This rod models a vector~$\ve_1$ which makes a basis $(\ve_1)$ called the Euclidean basis relative
to the chosen unit of measure.
%And the dual basis is made of the linear form $e^1 : \vec\RR \rar \RR$ given by $e^1(\ve_1)=1$.
%Thus a vector $\vx$ modeling some rod, has the length $x^1 := e^1(\vx)$ in chosen unit of measure.
%Moreover, once a unit of measure is chosen (by an observer), a vector in~$\vec\RR$ is written without an arrow, \eg\ $\ve_1$ is written~$e_1$, and $\vx$ is written~$x$.

\medskip
$\bullet$ Space dimension $n \ge 2$:

- The observers makes three rods of length 3, 4 and 5,
and makes a triangle $(A,B,C)$ with $A$, $B$ and $C$ are the vertices and $A$ not on the side on length~$5$.

- Pythagoras: $3^2 + 4^2 = 5^2$ gives: The triangle $(A,B,C)$ is said to have a right angle at~$A$.

- Two vectors $\vu$ and $\vw$ in~$\vRRn$ are orthogonal iff the triangle $(A,B,C)$
can be positioned such that $\vec{AB}$ and $\vec{AC}$ are parallel to~$\vu$ and~$\vw$.

- A basis $(\ve_i)_{i=1,...,n}$ is Euclidean relative to the chosen unit of measurement
iff the~$\ve_i$ are two to two orthogonal and their length is~$1$ (relative to the chosen unit).
%Et on dit que $(\ve_1,...,\ve_n)$ est une base orthonormée euclidienne (une b.o.n. euclidienne) relativement à l'unité de mesure choisie.

\debexa
\label{exa0}
An English observer defines a Euclidean basis $(\va_i)$ using the \foot.
A French observer defines a Euclidean basis $(\vb_i)$ using the \metre. We have
\be
\label{eqexa0}
1\,\foot =  \mu\,\metre, \quad \mu = 0.3048,\qand
1\,\metre = \lambda\,\foot,\quad \lambda = {1\over \mu} \simeq 3.28.
\ee
($\mu = 0,3048$ is the official length in metre for the English~\foot.)
\Eg, the bases are ``aligned'' iff, for all~$i$,
\be
\label{eqexa0a}
%\va_i = \mu \vb_i \qand 
\vb_i = \lambda \va_i \quad (\hbox{change of measurement unit}),
\ee
thus the transition matrix from $(\va_i)$ to~$(\vb_i)$ is $P=\lambda I $, thus $P^T=P$, $P^{-1}={1\over\lambda} I$ and $P^T.P=\lambda^2 I$.
\finexa

\debrem
\label{rema}
The bases used in practice are not all Euclidean.
See example~\ref{exaavion1}, especially if you fly.
%Or see \S~\ref{seccontree}.
\finrem

%%%%%%%%%%%%%%%%%%%%%%%%%%%%%%%%%%%%%%%%%%%%%%%%%%%%%%%%%%%%%%%%%%%%%%%%%%%%%%%%%%%

\subsection{Euclidean dot product}
\label{secbepe}

\debdef
\label{defeee0}
An observer who has built) his Euclidean basis $(\ve_i)$, \cf~\S~\ref{secbe}.
The associated Euclidean dot product is the bilinear form $g\dd =\dd_g\in\calL(\vRRn,\vRRn;\RR)$ defined by
\be
\label{eqeee0}
(g_{ij}=) \quad g(\ve_i,\ve_j)=\delta_{ij},\quad \forall i,j ,\qie [g]_{|\ve} = [\delta_{ij}] = I.
\ee
%\cf~prop.~\ref{propbilb}.
In other words,
\be
\label{eqeee}
\dd_g := \sumin \pi_{ei}\otimes \pi_{ei} = \sumin e^i\otimes e^i,
\ee
with classical and duality notations, $(\pi_{ei})= (e^i)$ being the dual basis of~$(\ve_i)$.
And if you want to use the Einstein convention you have to write $\dd_g := \sumijn \delta_{ij} e^i\otimes e^j$: You cannot avoid writing $\delta_{ij}=g_{ij}$.
\findef

Thus, for all $\vx,\vy\in\vRRn$, with $\vx= \sumin x_i\ve_i$ and $\vy= \sumin y_i\ve_i$ (classical notations),
\be
\label{eqvxvye}
(\vx,\vy)_g = \sumin x_i y_i  = [\vx]_{|\ve}^T.[\vy]_{|\ve}.
\ee
With duality notations, $\vx= \sumin x^i\ve_i$, $\vy= \sumin y^i\ve_i$ and $(\vx,\vy)_g = \sumin x^i y^i$; 
And if you want to use the Einstein convention then write $(\vx,\vy)_g := \sumijn  \delta_{ij} x^i y^j$: You cannot avoid writing~$\delta_{ij}$.

\debdef
The associated norm is $||.||_g := \sqrt{\dd_g}$, and the length of a vector~$\vx$ relative to the chosen Euclidean unit of measurement is $||\vx||_g := \sqrt{(\vx,\vx)_g}$.
\findef

Thus with the Euclidean basis $(\ve_i)$ (used to build~$\dd_g$), if $\vx=\sumin x_i\ve_i$, then $||\vx||_g = \sqrt{\sumin x_i^2}$ is the length of~$\vx$ relative to the chosen Euclidean unit of measure (Pythagoras).
(With duality notations $||\vx||_g = \sqrt{\sumin(x^i)^2}$, and if you want to use the Einstein convention: $||\vx||_g = \sqrt{\sumijn \delta_{ij} x^i x^j}$.)

%\medskip Recall: Two vectors $\vx,\vy$ are $\dd_g$-orthogonal iff $(\vx,\vy)_g = 0$.

\debdef
The angle $\theta(\vx,\vy)$ between two vectors $\vx,\vy\in\vRRn-\{\vec0\}$ is defined by
\be
\label{eqangle}
\cos(\theta(\vx,\vy)) = ({\vx\over ||\vx||_g},{\vy\over ||\vy||_g})_g . %\quad (=\cos(\theta(\vy,\vx))).
\ee
(With a calculator, this formula gives $\theta(\vx,\vy)= \arccos(({\vx\over ||\vx||_g},{\vy\over ||\vy||_g})_g)$
a value in~$[0,\pi]$.)
\findef

%\debdef A basis $(\vb_i)$ such that  $(\vb_i,\vb_j)_g = \delta_{ij}$, for all $i,j=1,...,n$, is called a $\dd_g$-orthonormal basis. \findef

\comment{
\debprop
If $(\vb_i)$ is another $\dd_g$-orthonormal bases, then it is also a Euclidean basis for the same unit of measurement.
And if $P$ is the transition matrix from $(\ve_i)$ to $(\vb_i)$, then
\be
P^T.P=I,\qand
\dd_g = \sumin e^i \otimes e^i = \sumin b^i \otimes b^i \quad\hbox{(with two $\dd_g$-orthonormal bases)}.
\ee
%And $(\vb_i)$ is also a Euclidean basis for the chosen unit of measure (used to define~$(\ve_i)$): In particular, $||3\vb_i + 4 \vb_j||_g^2= 25$ for all $i\ne j$ (Pythagoras).
\finprop

\debdem
With $\vb_j = \sumin P^i_j \va_i$ we get
$\delta_{ij} = (\vb_i,\vb_j)_g
= \sumkln P^k_i P^\ell_j (\va_k,\va_\ell)_g
= \sumkln P^k_i P^\ell_j \delta_{k\ell}
=\sumkn P^k_i P^k_j
=\sumkn (P^T)^i_k P^k_j = (P^T.P)^i_j
$ for all $i,j$, thus $P^T.P = [\delta_{ij}] = I$.

And for the dual bases we have $a^i = \sumjn P^i_j b^j$, \cf~\eref{eqdefP2b}.
Thus $\sumin a^i \otimes a^i = \sumikln P^i_k P^i_\ell b^k \otimes b^\ell
= \sumkln \delta_{k\ell} b^k \otimes b^\ell
= \sumkn b^k \otimes b^k
$, that is, $[g]_{|\va} = [g]_{|\vb} = I$.
\findem

\debprop
If $(\va_i)$ and $(\vb_i)$ are two $\dd_g$-orthonormal bases, 
and if $P$ is the transition matrix from $(\va_i)$ to $(\vb_i)$, then
\be
P^T.P=I,\qand
\dd_g = \sumin a^i \otimes a^i = \sumin b^i \otimes b^i \quad\hbox{(with two $\dd_g$-orthonormal bases)}.
\ee
%And $(\vb_i)$ is also a Euclidean basis for the chosen unit of measure (used to define~$(\ve_i)$): In particular, $||3\vb_i + 4 \vb_j||_g^2= 25$ for all $i\ne j$ (Pythagoras).
\finprop

\debdem
With $\vb_j = \sumin P^i_j \va_i$ we get
$\delta_{ij} = (\vb_i,\vb_j)_g
= \sumkln P^k_i P^\ell_j (\va_k,\va_\ell)_g
= \sumkln P^k_i P^\ell_j \delta_{k\ell}
=\sumkn P^k_i P^k_j
=\sumkn (P^T)^i_k P^k_j = (P^T.P)^i_j
$ for all $i,j$, thus $P^T.P = [\delta_{ij}] = I$.

And for the dual bases we have $a^i = \sumjn P^i_j b^j$, \cf~\eref{eqdefP2b}.
Thus $\sumin a^i \otimes a^i = \sumikln P^i_k P^i_\ell b^k \otimes b^\ell
= \sumkln \delta_{k\ell} b^k \otimes b^\ell
= \sumkn b^k \otimes b^k
$, that is, $[g]_{|\va} = [g]_{|\vb} = I$.
\findem
}

%%%%%%%%%%%%%%%%%%%%%%%%%%%%%%%%%%%%%%%%%%%%%%%%%%%%%%%%%%%%%%%%%%%%%%%%%%%%%%%%%%%

\subsection{Change of Euclidean basis}

Let $(\va_i)$ (\eg\ English observer basis built with the foot) and $(\vb_i)$ (\eg\ French observer basis built with the metre) be Euclidean bases in~$\vRRn$, and let $\dd_g$ and $\dd_h$ be the associated Euclidean dot products.

%%%%%%%%%%%%%%%%%%%%%%%%%%%%%%%%%%%%%%%%%%%%%%%%%%%%%%%%%%%%%%%%%%%%%%%%%%%%%%%%%%%

\subsubsection{Two Euclidean dot products are proportional}

\debprop
\label{propbes}
If $\lambda = ||\vb_1||_g$, then $||\vb_i||_g=\lambda$ for all $i=1,...,n$ (change of unit) and
\be
\label{eqpropbes}
%\hbox{if}\quad \lambda = ||\vb_1||_g \qthen
\dd_g = \lambda^2 \dd_h, \qand ||.||_g = \lambda ||.||_h.
\ee
%(Useful if \eg\ you want to use the results of Newton, Descartes...)
%that is, $\sumin a^i\otimes a^i = \lambda^2\sumin b^i\otimes b^i$.
\finprop

\debdem
By definition of a Euclidean basis, the length of the rod that enabled to define $(\vb_i)$ is independent of~$i$, \cf~\S~\ref{secbe}, thus $||\vb_i||_g=||\vb_1||_g$ for all~$i$, and here $||\vb_i||_g \eqnote\lambda$.
Thus $%(\vb_i,\vb_i)_g = 
||\vb_i||_g^2
= \lambda^2 %= \lambda^2 (\vb_i,\vb_i)_h
= \lambda^2 ||\vb_i||_h^2
$ for all~$i$, since $%(\vb_i,\vb_i)_h = 
||\vb_i||^2_h=1$.
And if $i\ne j$ then  $(\vb_i,\vb_j)_g=0=(\vb_i,\vb_j)_h$ since $\vb_i$ and $\vb_j$ form a right angle (Pythagoras), cf.~\eref{eqeee}.
Hence $(\vb_i,\vb_j)_g=\lambda^2(\vb_i,\vb_j)_h$ for all $i,j$, thus~\eref{eqpropbes}.
\findem

\debexa
\label{secemp}
Continuation of example~\ref{exa0}: % avec~\eref{eqexa0}.
$\dd_a = \sumin a^i \otimes a^i$ is the English Euclidean dot product (\foot),
and
$\dd_b = \sumin b^i \otimes b^i$ is the French Euclidean dot product (\metre).
\eref{eqpropbes} and~\eref{eqexa0} give:
\be
\label{eqsecemp}
\dd_a = \lambda^2\dd_b \qand ||.||_a = \lambda ||.||_b,
\qwith  \lambda \simeq 3.28 \qand \lambda^2 \simeq 10.76 .
\ee
In particular, if $\vw$ is \st\ $||\vw||_b=1$ (its length is $1$ \metre),
then $||\vw||_a=\lambda$ (its length is $\lambda \simeq 3.28$ \foot).
\finexa

\comment{
%%%%%%%%%%%%%%%%%%%%%%%%%%%%%%%%%%%%%%%%%%%%%%%%%%%%%%%%%%%%%%%%%%%%%%%%%%%%%%%%%%%

\subsubsection{Change of Euclidean basis}

Let $\lambda>0$ \st\ $\lambda = ||\vb_1||_g$, and let $\calP : \vRRn\rar\vRRn$ be the change of basis endomorphism
from $(\va_i)$ to~$(\vb_i)$, that is,
\be
\label{eqclaPfa0}
\vb_j = \calP.\va_j,\quad \forall j=1,...,n .
\ee

\debprop
$\calP$ is composed of a Euclidean isometry $\Lis$ and of the dilation $\lambda I$ (with scale factor~$\lambda$):
\be
\label{eqclaPfa}
\calP = \lambda \Lis \qwhere  \Lis^T.\Lis=I, \qthus
\calP^T.\calP = \lambda^2 I \qand \calP^{-1} = {1\over \lambda^2}\calP^T.
\ee
\finprop

\debdem
For all $i,j$,
$(\va_i,\va_j)_g = \delta_{ij} = (\vb_i,\vb_j)_h
= (\calP.\va_i,\calP.\va_j)_h
=(\calP_{gh}^T.\calP.\va_i,\va_j)_g
= {1\over\lambda^2}(\calP_{gh}^T.\calP.\va_i,\va_j)_g
= ({1\over \lambda}\calP.\va_i,{1\over \lambda}\calP.\va_j)_h %= (\va_i,\va_j)_g
$, thus ${1\over \lambda}\calP \eqnamed \Lis$ is an isometry. %, \cf~prop.~\ref{propoib0},2.
And~\eref{eqclaPfa0} gives~\eref{eqclaPfa}.
\findem
}

%%%%%%%%%%%%%%%%%%%%%%%%%%%%%%%%%%%%%%%%%%%%%%%%%%%%%%%%%%%%%%%%%%%%%%%%%%%%%%%%%%%

\subsubsection{Counterexample : non existence of a Euclidean dot product}
\label{seccontree}

\quad
1- Thermodynamic: Let $T$ be the temperature and $P$ the pressure,
%They do not share the same dimension,
and consider the Cartesian vector space $\{(T,P)\} = \{\hbox{(temperature,pressure)}\} = \RR\times\RR $.
There is no associated Euclidean dot product: An associated norm would give
$||(T,P)|| = \sqrt{T^2 + P^2}\in\RR$ which is meaningless (incompatible dimensions). See~\S~\ref{secthermo}.

\medskip
2- Polar coordinate system $\vq = (r,\theta)\in \RR\times\RR$: There is no Euclidean norm $\sqrt{r^2+\theta^2}$ for~$\vq$ that is physically meaningful (incompatible dimensions), see example~\ref{exarempfcb}.

%%%%%%%%%%%%%%%%%%%%%%%%%%%%%%%%%%%%%%%%%%%%%%%%%%%%%%%%%%%%%%%%%%%%%%%%%%%%%%%%%%%

\subsection{Euclidean transposed of the deformation gradient}

Let $n\in\{1,2,3\}$ and consider a linear map $L\in\calL(\RRntz;\RRnt)$ (\eg, $L=\Ftzt(P)$).

Let $\dd_G$ be a Euclidean dot product in~$\RRntz$ (used in the past by someone), and let
$\dd_g$ and~$\dd_h$ be Euclidean dot products in~$\RRnt$ (the actual space where the results are obtained
by two observers, \eg, $\dd_g$ built with a foot and $\dd_h$ built with a metre).
Let $L^T_{Gg}$ and~$L^T_{Gh}$ be the transposed of $L$ relative to the dot products,
that is, $L^T_{Gg}$ and $L^T_{Gh}$ in $\calL(\RRnt;\RRntz)$ are characterized by,
for all $(\vX,\vy)\in\RRntz\times \RRnt$, \cf~\eref{eqseccpgdd0},
\be
\label{eqgagbpropG}
(L^T_{Gg}.\vy,\vX)_G = (L.\vX,\vy)_g \qand (L^T_{Gh}.\vy,\vX)_G = (L.\vX,\vy)_h.
\ee

\debcor
\label{propteG}
\be
\label{eqteG}
\hbox{if} \quad \dd_g = \lambda^2 \dd_h \qthen
L_{Gg}^T = \lambda^2 L_{Gh}^T . %\quad \in \calL(\RRnt;\RRntz)
\ee
%(a vector type relation, as in~\eref{eqFtztnew} or~\eref{eqfcvvr3}).
NB: Do not forget $\lambda^2$, \cf~remark~\ref{remMCOC} (Mars Climate Orbiter crash).
\fincor

\debdem
$\ds
(L_{Gg}^T.\vy,\vX)_G
\mathop{=}^{\eref{eqgagbpropG}}  (L.\vX,\vy)_g
\mathop{=}^{\eref{eqteG}_1} \lambda^2 (L.\vX,\vy)_h
\mathop{=}^{\eref{eqgagbpropG}} \lambda^2 (L^T_{Gh}.\vy,\vX)_G
$ for all $\vX\in\RRntz$ and all $\vy\in\RRnt$, thus $L_{Gg}^T.\vy = \lambda^2 L_{Gh}^T.\vy$ for all $\vy\in\RRnt$,
thus \eref{eqteG}$_2$.
\findem

%%%%%%%%%%%%%%%%%%%%%%%%%%%%%%%%%%%%%%%%%%%%%%%%%%%%%%%%%%%%%%%%%%%%%%%%%%%%%%%%%%%

\subsection{The Euclidean transposed for endomorphisms}

Let $n\in\{1,2,3\}$ and consider an endomorphism $L\in \calL(\RRnt;\RRnt)$
(\eg\ $L=d\vv_t(p) \in \calL(\RRnt;\RRnt)$ the differential of the Eulerian velocity).
Let $\dd_g$ and~$\dd_h$ be dot products in~$\vRRn$.
Let $L^T_g$ and~$L^T_h$ be the transposed of~$L$ relative to~$\dd_g$ and~$\dd_h$,
that is, $L^T_g$ and $L^T_h$ in $\calL(\RRnt;\RRnt)$ are the endomorphisms defined by, for all $\vx,\vy \in \RRnt$, \cf~\eref{eqseccpgdd0e},
\be
\label{eqgagbprop0}
(L^T_g.\vy,\vx)_g = (L.\vx,\vy)_g, \qand (L^T_h.\vy,\vx)_h = (L.\vx,\vy)_h.
\ee

\debcor
\label{propte}
\be
\label{eqte}
\hbox{if} \quad \dd_g = \lambda^2 \dd_h \qthen
L_g^T = L_h^T \eqnote L^T \quad \in \calL(\RRnt;\RRnt)
\ee
(an endomorphism type relation):
Thus we can speak of ``the Euclidean transposed of an endomorphism''.
\fincor

\debdem
$\ds
(L_g^T.\vy,\vx)_g
\mathop{=}^{\eref{eqgagbprop0}}  (L.\vx,\vy)_g
\mathop{=}^{hyp} \lambda^2 (L.\vx,\vy)_h
\mathop{=}^{\eref{eqgagbprop0}} \lambda^2 (L^T_h.\vy,\vx)_h
\mathop{=}^{hyp} (L^T_h.\vy,\vx)_g
$ for all $\vx,\vy\in\vRRn$, thus $L_g^T.\vy = L_h^T.\vy$ for all $\vy\in\vRRn$.
\findem

%Et réciproquement, si $L_g^T.L=I$ alors $L$ vérifie immédiatement $(L.\vx,L.\vy)_g$ pour tout~$\vx$, par définition de la transposée, cf.~\eref{eqgagbprop0}.

%%%%%%%%%%%%%%%%%%%%%%%%%%%%%%%%%%%%%%%%%%%%%%%%%%%%%%%%%%%%%%%%%%%%%%%%%%%%%%%%%%%
\comment{
\subsection{Euclidean isometries}
\label{sececeucis}

Let $\dd_g$ be a Euclidean dot product in~$\vRRn$.
An endomorphism $L\in \calL(\vRRn;\vRRn)$ is a $\dd_g$-isometry iff
\be
\label{eqdefoib0}
\forall \vx,\vy\in\vRRn,\quad (L.\vx,L.\vy)_g = (\vx,\vy)_g, \qie L_g^T.L=I,\qie L^{-1}=L_g^T
\ee
with~\eref{eqte}.
In particular, $||L.\vx||_g = ||\vx||_g$ for all~$\vx$ (the norm is unchanged),
and a $\dd_g$-orthonormal basis is transformed into a $\dd_g$-orthonormal basis. (See~\eref{eqdefoib}.)

\debprop
1- If $L$ is a $\dd_g$-isometry and if $(\ve_i)$ is a $\dd_g$-orthonormal basis,
then $(L.\ve_i)$ is a $\dd_g$-orthonormal basis.

2- If $\dd_g$ and $\dd_h$ are two Euclidean dot product in~$\vRRn$,
if $L$ is a $\dd_g$-isometry, then $L$ is also a $\dd_h$-isometry, simply called a Euclidean isometry.
And in this case, with $L.\ve_j = \sumin \Lij \ve_i$, we have:
\be
%\label{eqdefoib02}
L^T_g\eqnamed L^T, \qand L^T.L=I, \qand [(L^T)^i{}_j] = [\Lij]^T = [L^j{}_i] \quad\hbox{(Euclidean framework)}.
\ee
\finprop

\debdem
1- \eref{eqdefoib0} gives $(L.\ve_i,L.\ve_j)_g = (\ve_i,\ve_j)_g=\delta_{ij}$.

2- Let $\lambda>0$ \st\ $\dd_g= \lambda^2\dd_h$, \cf~\eref{eqpropbes}.
With \eref{eqdefoib0} we get
$\lambda^2(L.\vx,L.\vy)_h =(L.\vx,L.\vy)_g = (\vx,\vy)_g = \lambda^2(\vx,\vy)_h$,
true for all $\vx,\vy$.
\findem

\debexe
Let $L$ be a Euclidean isometry, let $(\ve_i)$ be an $\dd_g$-orthonormal basis.
Prove that the ``column vectors'' (meaning?) of $[L]_{|\ve}$ make an orthonormal basis.
%Give the isometries in~$\RR^2$.
%In~$\vRRt$ prove that an isometry cannot be a rotation.

\debrep
The implicit meaning is: the $j$-th ``column vector'' is the $j$-th column of~$[L]_{|\ve}$
which gives the components of the vector $\vc_j := L.\ve_j = \sumijn L^i_j \ve_i$.
And the exercise is to prove that $(\vc_j)$ is an orthonormal basis.

\eref{eqdefoib02} gives %$L^T.L=I$ we have 
$[L^T]_{|\ve}{}^T.[L]_{|\ve}=[I]$,
that is, $[\vc_i]_{|\ve}^T.[\vc_j]_{|\ve} = \delta_{ij}$ for all $i,j$,
thus $(\vc_i,\vc_j)_g = \delta_{ij}$ for all $i,j$.
\comment{
Thus in~$\RR^2$, there exists $\theta\in\RR$ \st\
$[L]_{|\ve} = \pmatrix{\cos\theta & \sin\theta \cr - \sin\theta & \cos\theta}$
(rotation) or $[L]_{|\ve} = \pmatrix{\cos\theta & \sin\theta \cr \sin\theta & -\cos\theta}$
(symmetry relative to the line at angle~${\theta\over 2}$), since the column must be orthogonal and normed.
In~$\vRRt$, we can use~\eref{eqom} to get $\Omega.\vomega_e = 0$, so $\Omega$ is not bijective.
Or, if preferred, a rotation $R$ is antisymmetric ($R^T = -R$), and $R$ being a real matrix
its eigenvalues are either pure complex that are 2 by 2 conjugate, or vanish,
thus in~$\RRt$ there is at least one vanishing eigenvalue, thus $L^T.L\ne I$.
}
\finrep
\finexe

}

%%%%%%%%%%%%%%%%%%%%%%%%%%%%%%%%%%%%%%%%%%%%%%%%%%%%%%%%%%%%%%%%%%%%%%%%%%%%%%%%%%%

\subsection{Unit normal vector, unit normal form}
\label{secnf}

\def\suminmu{\sum_{i=1}^{n-1}}

The results in this~\S\ are not objective:
We need a Euclidean dot product (need a unit of length: Foot? Meter?) to get a unit (Euclidean) normal vector.

%%%%%%%%%%%%%%%%%%%%%%%%%%%%%%%%%%%%%%%%%%%%%%%%%%%%%%%%%%%%%%%%%%%%%%%%%%%%%%%%%%%

\subsubsection{Framework}

\def\vbeta{{\vec\beta}}

$\dd_g$ is a Euclidean dot product (needed to define Euclidean orthonormality) %(defined by $g(\ve_i,\ve_j) := \delta_{ij}$ for all $i,j$),
and, for all $\vu,\vw\in\vRRn$,
\be
(\vu,\vw)_g \eqnote \vu \bcdotg \vw
\ee
(or $\eqnote \vu \bcdot \vw$ when one chosen Euclidean dot product is imposed to all).

$\Omega$ is a %oriented %(has an interior and an exterior) 
regular open bounded set in~$\RRn$, $n=2$ or~$3$, and $\Gamma := \pa\Omega$ is its regular surface (dimension $n{-}1$). If $p\in\Gamma$ then
$T_p\Gamma$ is the tangent plane at $p$ to~$\Gamma$,
and a basis $(\vbeta_1(p),...,\vbeta_{n-1}(p))$ in~$T_p\Gamma$ is known (usually obtained thanks to a coordinate system describing~$\Gamma$). 
And, to lighten the writings, $(\vbeta_1(p),...,\vbeta_{n-1}(p))$ is written $(\vbeta_1,...,\vbeta_{n-1})$.

%%%%%%%%%%%%%%%%%%%%%%%%%%%%%%%%%%%%%%%%%%%%%%%%%%%%%%%%%%%%%%%%%%%%%%%%%%%%%%%%%%%

\subsubsection{Unit normal vector}
\label{secvnu}

Call $\vn_g(p)$ the unit outward normal vector at $p\in\Gamma$ at~$T_p\Gamma$ relative to~$\dd_g$; %with shortened notations, 
So $\vn_g(p) \bcdotg \vbeta_i(p) = 0$ for all $i=1,...,n{-}1$, and $||\vn_g(p)||_g=1$, \ie\ $\vn_g$ is defined on~$\Gamma$ by (up to its sign)
\be
\label{eqed0}
\forall i=1,...,n{-}1,\;\; \vbeta_i \bcdotg \vn_g = 0,\qand \vn_g \bcdotg \vn_g=1 \quad(= ||\vn_g||_g^2),
\ee
\ie, at any $p\in\Gamma$, $\vn_g(p)$ is orthogonal to the hyperplane $\Vect\{\vbeta_1(p),...,\vbeta_{n-1}(p)\}$ and $\vn_g(p)$ is unitary.
%Also written $\vbeta_i \bcdot \vn = 0$ and $\vn \bcdotg \vn=1 = ||\vn||^2$ if the choice of~$\dd_g$ is imposed.
%(Full notation: $\vn_g(p) \bcdotg \vbeta_i(p) = 0$ and $||\vn_g(p)||_g=1$.)
So $(\vbeta_1(p),...,\vbeta_{n-1}(p),\vn_g(p))$ is a basis at $p$ in~$\vRRn$, written in short $(\vbeta_1,...,\vbeta_{n-1},\vn_g)$.
Drawing. 

Thus, for all $\vw\in\vRRn$, if $\vw = \suminmu w_i \vbeta_i + w_n \vn_g$ (classical notations) then %\eref{eqed0} gives
\be
\label{eqnflat10}
w_n = \vw \bcdotg \vn_g =\hbox{ the normal component of~$\vw$ at~$p$ at~$\Gamma$}.
\ee
($w_n$ depends on~$\dd_g$.)
(Duality notations: $\vw = \suminmu w^i \vbeta_i + w^n \vn_g$ and $w^n = \vw \bcdotg \vn_g$.)
%(Full notation: $\vw(p) =  \suminmu w^i(p) \vbeta_i(p) + w^n(p) \vn_g(p)$ gives $w^n(p) = (\vw(p),\vn_g(p))_g$.)

\debexe
Let $(\va_i)$ be a basis in~$\vRRn$, $\vbeta_j = \sumin B_{ij}\va_i$ for $j=1,...,n{-}1$, and $\vn_g = \sumin n_i\va_i$,
and $g_{ij} = g(\va_i,\va_j)$ for all $i,j$.
What equations satisfy the~$n_j$? And particular case $(\va_i)$ is $\dd_g$-orthonormal?

\debrep
\eref{eqed0} gives $[\vbeta_i]_{|\va}^T.[g]_{|\va}.[\vn_g]_{|\va}=0$ for $i=1,...,n{-}1$ (so $n{-}1$ equations), with $[\vn_g]_{|\va}^T.[g]_{|\va}.[\vn_g]_{|\va}=1$ (so 1 equation),
and $\vn_g$ is obtained up to its sign.

%$\sumjkn B_{ki} g_{kj} n_j = 0$ for $j=1,...,n{-}1$, \ie\ with $B=[Bij]_{i=1,...,n \atop j=1,...,n-1}$, $B^T.[g].[\vn]=0$ (so $n{-}1$ equations); And $\sumijn n_i g_{ij} n_j=1$, \ie\ $[\vn]^T.[g].[\vn]=1$.

If $(\va_i)$ is $\dd_g$-orthonormal, then  $\sumjn B_{ij}n_j = 0$ for $j=1,...,n{-}1$, with $\sumin n_i^2=1$.
\finrep
\finexe

\debexe
\label{exeippabn}
Let $(\va_i)$ be a Euclidean basis in~\foot, $(\vb_i)$ a Euclidean basis in~\metre,
$\dd_a$ and $\dd_b$ the associated Euclidean dot products, so $\dd_a = \lambda^2 \dd_b$ with $\lambda\simeq 3.28$, \cf~\eref{eqpropbes}.
Let $\vn_a(p)$ and $\vn_b(p)$ be the corresponding unit outward normal vectors, \cf~\eref{eqed0}.
1- Prove (up to the sign):
\be
\label{eqippabn}
\vn_b = \lambda \vn_a, \qand  (\vw,\vn_a)_a = \lambda (\vw,\vn_b)_b \quad \forall\vw\in\vRRn
\ee
%($\vn_b$ is $\lambda$ times larger than~$\vn_a$ as expected).
2- Then let $\vn_a = \sumim n_{ai} \va_i$ and $\vn_b = \sumim n_{bi} \vb_i$; Prove:
\be
\hbox{If}, \; \forall i=1,...,n,\;\vb_i=\lambda \va_i \qthen \forall i=1,...,n,\; n_{ai} = n_{bi}.
%,\qie [\vn_a]_{|\va} = [\vn_b]_{|\vb}.
\ee
%(Duality notations: $\vn_a = \sumim n_a^i \va_i$, $\vn_b = \sumim n_b^i \vb_i$ and $n_a^i = n_b^i$.)
So the vectors $\vn_a$ and~$\vn_b$ are different ($\lambda>1$), and their respective components are equal... relative to different bases! And of course $1 = ||\vn_a||_a^2 = \sumin (n_{ai})^2 = \sumin (n_{bi})^2 = ||\vn_b||_b^2=1$.

\debrep
$\vn_a(p) \parallel \vn_b(p)$, since the vectors are Euclidean and orthogonal to~$T_p\Gamma$ \cf~\eref{eqed0}.
%$(\vn_a(p),\vb_i(p))_a = 0 =(\vn_b(p),\vb_i(p))_b$, \cf~\eref{eqed0}.
And $||.||_a=\lambda||.||_b$ \cf~\eref{eqsecemp},
thus $||\vn_b||_b = 1 =||\vn_a||_a = \lambda||\vn_a||_b = ||\lambda\vn_a||_b$,
so $\vn_b = \pm\lambda \vn_a$.
And they both are outward vectors, so $\vn_b = +\lambda \vn_a$.
Thus $(\vw,\vn_a)_a = \lambda^2(\vw,\vn_a)_b
= \lambda^2(\vw,{\vn_b \over \lambda})_b = \lambda (\vw,\vn_b)_b$.

And if $\vb_i=\lambda \va_i$ \eref{eqippabn} gives
$\sumin n^i_b \vb_i = \lambda \sumin n^i_a \va_i
= \sumin n^i_a (\lambda \va_i) = \sumin n^i_a \vb_i$, then $n_a^i = n_b^i$.
\finrep
\finexe

%%%%%%%%%%%%%%%%%%%%%%%%%%%%%%%%%%%%%%%%%%%%%%%%%%%%%%%%%%%%%%%%%%%%%%%%%%%%%%%%%%%

\subsubsection{Unit normal form $n^\flat$ associated to~$\vn$}
\label{secunf}

\def\bb{\beta}

(For mathematicians; May produce misunderstandings and lack of mechanical interpretations; Don't forget: $n^\flat$ is obtained after $\vn$ has been defined.)

%To define the unit normal form $n^\flat_g$, a Euclidean dot product~$\dd_g$ and the normal unit vector $\vn_g$ must first be defined.

At $p\in\Gamma$, once you have computed~$\vn_g(p)$, you can define the associated unit normal form $n^\flat_g(p)\in\RRns$: It is the linear form defined by
$n^\flat_g(p).\vw := \vn_g(p) \bcdotg \vw$ for all $\vw\in\vRRn$, \ie\ on~$\Gamma$, for all $\vw\in\vRRn$,
%, for all $\vw\in\vRRn$, in short,
\be
\label{eqedflat}
n^\flat_g.\vw := \vn_g \bcdotg \vw
\ee
($\eqnote \vn \bcdot \vw$ if one chosen Euclidean dot product is imposed to all).
Thus $[n^\flat_g].[\vw] = [\vn_g]^T.[g].[\vw]$.
%\ie, $n^\flat(p).\vw := (\vn_g(p),\vw)_g$,
%\Ie, $n^\flat_g$ is defined by, for all $i=1,...,n$, \be \label{eqedflat2} n^\flat_g.\ve_i  = \vn_g \bcdotg \ve_i. \ee

Quantification: Let $(\ve_i)$ be a basis in~$\vRRn$; Then \eref{eqedflat} gives
$[n^\flat_g]_{|\ve}.[\vw]_{|\ve} = [\vn_g]_{|\ve}^T.[g]_{|\ve}.[\vw]_{|\ve}$ simply written $[n^\flat_g].[\vw] = [\vn_g]^T.[g].[\vw]$ if the basis $(\ve_i)$ is imposed.

So, with duality notations to justify the ${}^\flat$ notation, with
$(e^i)$ the dual basis of~$(\ve_i)$, let
\be
\vn_g = \sumin n_g^i\ve_i \qand n_g^\flat  = \sumin n_{gi}e^i.
\ee
\ie\ $n_g^i$ and $n_{gi}$ are the components of~$\vn_g$ and $n_g^\flat$ relative to the basis~$(\ve_i)$ and~$(e^i)$.
Since \eref{eqedflat} gives $n^\flat_g.\ve_i := \vn_g \bcdotg \ve_i$ for all~$i$, we get, for all~$i$,
\be
\label{eqnflat2}
n_{ig} = \sumjn g_{ij} n_g^j %\quad(\hbox{Euclidean framework}).
\ee

Particular case $(\ve_i)$ is a $\dd_g$-Euclidean basis, then $n_{ig} = n_g^i$.
%; The positions of $i$ is not compatible with the Einstein convention: You have lost the track of the chosen~$\dd_g$ (unfortunately usual in continuum mechanics).

Classical notations: $\vn_g = \sumin (\vn_g)_i\ve_i$, dual basis $(\pi_{ei})$, $n_g^\flat  = \sumin (n_g^\flat)_i\pi_{ei}$, $(n_g^\flat)_i = \sumjn g_{ij} (\vn_g)_j$.
%Particular case: $(\ve_i)$ is a $\dd_g$-Euclidean basis:  $(n_g^\flat)_i = (\vn_g)_i \eqnote n_i$.

NB: In physics don't forget to write the $g_{ij}$ in~\eref{eqnflat2} even if $g_{ij} = \delta_{ij}$, since you need to see the chosen metric and basis (and verify the Einstein convention), although \eref{eqnflat2} is simply written $n_{i} = \sumjn g_{ij} n^j$...

%%%%%%%%%%%%%%%%%%%%%%%%%%%%%%%%%%%%%%%%%%%%%%%%%%%%%%%%%%%%%%%%%%%%%%%%%%%%%%%%%%%

\subsection{Integration by parts (Green--Gauss--Ostrogradsky)}

\def\pinOmega{{p\in\Omega}}
\def\pinGamma{{p\in\Gamma}}

Let $\Omega$ be a regular bounded open set in~$\RRn$ and $\Gamma=\pa\Omega$ its frontier,
let $\phi \in C^1(\overline\Omega;\RR)$, let $(\ve_i)$ be a Euclidean basis and $\dd_g$ ites associated Euclidean dot product, let ${\pa \phi \over \pa x_i}(p) := d\phi(p).\ve_i$ (usual notation),
let $\vn_g(p)=\vn(p) = \sumin n_i(p)\ve_i$ (classical notations) be the unit outward normal at $p\in\Gamma$.
Then, for $ i=1,...,n$,
\be
\label{eqipp00}
\int_\pinOmega {\pa \phi \over \pa x_i}(p)\,d\Omega = \int_\pinGamma \phi(p) n_i(p)\,d\Gamma, 
\qinshort \int_\Omega {\pa \phi \over \pa x_i}\,d\Omega = \int_\Gamma \phi n_i\,d\Gamma.
\ee
Thus, for any $v\in C^1(\overline\Omega;\RR)$, with $\phi v$ instead of~$\phi$ in~\eref{eqipp00}, we get the integration by parts formula (Green formula):
\be
%\label{eqippv1}
\int_\Omega {\pa \phi \over \pa x_i}v\,d\Omega
=- \int_\Omega \phi \,{\pa v \over \pa x_i}\,d\Omega
+ \int_\Gamma \phi v n_i\,d\Gamma.
\ee
Thus, for any $\vv\in C^1(\overline\Omega;\vRRn)$ (vector field), with $\vv(p)= \sumin v_i(p)\ve_i)$ we get
\be
\label{eqippv1}
\int_\Omega {\pa \phi \over \pa x_i}v_i\,d\Omega
=- \int_\Omega \phi \,{\pa v_i \over \pa x_i}\,d\Omega
+ \int_\Gamma \phi v_i n_i\,d\Gamma.
\ee
%(Duality notations: $\int_\Omega {\pa \phi \over \pa x^i}v^i\,d\Omega = \int_\Gamma \phi n^i\,d\Gamma$.)
Thus, with the gradient vector $\vgrad\phi(p) = \sumin {\pa \phi \over \pa x_i} \ve_i$ and with $\dvg\vv = \sumin {\pa v_i \over \pa x_i}$, we get the Gauss--Ostrogradsky formula:
\be
\label{eqipp}
\int_\Omega \vgrad\phi \bcdot \vv\,d\Omega
=- \int_\Omega \phi \,\dvg\vv\,d\Omega
+ \int_\Gamma \phi \vv \bcdot \vn\,d\Gamma.
\ee
(And $\int_\Gamma \phi \vv \bcdot \vn\,d\Gamma$ gives the flux through~$\Gamma$.)

\debexe
Use the differential $d\phi$ instead of the gradient $\vgrad\phi$ (which is the $\dd_g$-Riesz representation vector of~$d\phi$) to express~\eref{eqippv1}. Is the use of $n^\flat$ useful in that case?

\debrep
$\int_\Omega d\phi . \vv\,d\Omega
=- \int_\Omega \phi \,\dvg\vv\,d\Omega
+ \int_\Gamma \phi \vv \bcdot \vn\,d\Gamma$.
Since $n^\flat$ depends on~$\vn$ (definition), there is no reason that justifies the use of~$n^\flat$ (unless you want to introduce useless notations here).
\comment{
\Eg\ in physics which needs to see the metric in use, if you use a basis $(\ve_i)$ and its dual basis~$(e^i)$ (duality notations),
%and use the standard notations $d\phi= \sumin {\pa \phi\over pa x^i}\ve_i$, $\vv= \sumin v^i\ve_i$ and $\vn=\sumin n^i\ve_i$,
then $\int_\Gamma \phi \vv \bcdot \vn\,d\Gamma = \sumijn \int_\Gamma \phi v^i g_{ij} n^j\,d\Gamma$
is \textsl{\textbf{not}} written $=\sumin \int_\Gamma \phi v^i n^i\,d\Gamma$.
In other words, the flux $\int_\Gamma \phi \vv \bcdot \vn\,d\Gamma$ through~$\Gamma$ depends on the chosen Euclidean dot product (foot? metre?), and writng $n^\flat.\vv$ instead $\vv \bcdot \vn$ just hides the fundamental dependence on the inner dot product.%
}
\comment{
~\eref{eqippv1} reads
$\sumin \int_\Omega {\pa \phi\over pa x^i} \,v^i\,d\Omega
=- \sumin \int_\Omega \phi \,{\pa v^i\over \pa x^i}\,d\Omega
+ \sumijn \int_\Gamma \phi v^i g_{ij} n^j\,d\Gamma
$.
}
\finrep
\finexe

\comment{
\debexe
Continuation of exercise~\ref{exeippabn}.
Prove that the volumes satisfy $\det_{|\va} = \lambda^3 \det_{|\vb}$,
and that the area satisfy $\nu_a = \lambda^2\nu_b$.
Check that~\eref{eqippv2} is compatible with the change of unit of measurement.

\debrep
In~$\RR^3$ (similar calculations in~$\RR^2$).
The observer~$A$ computes
$(\int_\Omega d\phi.\vv\,d\Omega)_A = (\int_\Gamma \phi (\vv,\vn_a)_a\, d\Gamma)_A$
(unit given by $\va_i$).
And the observer~$B$ computes
$(\int_\Omega d\phi.\vv\,d\Omega)_B = (\int_\Gamma \phi (\vv,\vn_b)_b\, d\Gamma)_B$.
Let us prove that
$(\int_\Omega d\phi.\vv\,d\Omega)_A = \lambda^3 (\int_\Omega d\phi.\vv\,d\Omega)_B$,
as well as
$(\int_\Gamma \phi (\vv,\vn_a)_a\, d\Gamma)_A = \lambda^3 (\int_\Gamma \phi (\vv,\vn_b)_b\, d\Gamma)_B$,
which is the desired result.

Volumes: $\det_{|\va}(\vb_1,\vb_2,\vb_3) = \det_{|\va}(\calP.\va_1,\calP.\va_2,\calP.\va_3)
= \det_{|\va}(\calP)\det_{|\va}(\va_1,\va_2,\va_3) = \det_{|\va}(\calP) = \lambda^3$, \cf~\eref{eqdetLv},
thus $\det_{|\va}(\vb_1,\vb_2,\vb_3) = \lambda^3.1 =  \lambda^3\det_{|\vb}(\vb_1,\vb_2,\vb_3)$,
thus $\det_{|\va} = \lambda^3.1 = \lambda^3\det_{|\vb}$.

Areas: $\nu_a(\vu,\vw) = \det_{|\va}(\vu,\vw,\vn_a)
= \lambda^3 \det_{|\vb}(\vu,\vw,\vn_a)
= \lambda^3 \det_{|\vb}(\vu,\vw,{\vn_b\over \lambda})
= \lambda^2 \det_{|\vb}(\vu,\vw,\vn_b) = \lambda^2 \nu_b(\vu,\vw)$,
thus $\nu_a = \lambda^2\nu_b$.

And $d\phi.\vv = (\sum_i {\pa \phi\over \pa x^i}e^i).(\sum_j v^j\ve_j)$
is independent of the Euclidean basis, \cf~\eref{eqdefP1b}.
Thus $(\int_\Omega d\phi.\vv\,d\Omega)_A = \lambda^3(\int_\Omega d\phi.\vv\,d\Omega)_B$.
And 
$(\int_\Gamma \phi (\vv,\vn_a)_a\, d\Gamma)_A = \lambda^2(\int_\Gamma \phi (\vv,\vn_a)_a\, d\Gamma)_B
\mathop{=}^{\eref{eqippabn}} \lambda^2(\int_\Gamma \phi \lambda(\vv,\vn_b)_b\, d\Gamma)_B
= \lambda^3(\int_\Gamma \phi (\vv,\vn_b)_b\, d\Gamma)_B$.
\finrep
\finexe
}

%%%%%%%%%%%%%%%%%%%%%%%%%%%%%%%%%%%%%%%%%%%%%%%%%%%%%%%%%%%%%%%%%%%%%%%%%%%%%%%%%%%

\section{Rate of deformation tensor and spin tensor}
\label{secdecdv}

Let $\tPhi : [t_1,t_2] \times \Obj \rar \RRn$ be a regular motion, \cf~\eref{eqdeftPhi0},
and let $\vv : \bigC \rar\vRRn$ be the Eulerian velocity field, \cf~\eref{eqdefve},
that is, $\vv(t,p) = {\pa \Phi \over \pa t}(t,\Pobj)$ when $p = \tPhi(t,\Pobj)$.
Its differential $d\vv$ is given in~\eref{eqspdv}.

At~$t$, an observer chooses a unit of measurement (foot, metre...)
and builds the associated Euclidean dot product~$\dd_g$ in~$\RRnt$, \cf~\S~\ref{secbepe}.
(We loose the objective point of view here).
And the same $\dd_g$ is used at all~$t$.
%And an Euclidean basis~$(\ve_i)$ (a $\dd_g$-orthonormal basis) may be used, see~def.~\ref{defEab}.

%%%%%%%%%%%%%%%%%%%%%%%%%%%%%%%%%%%%%%%%%%%%%%%%%%%%%%%%%%%%%%%%%%%%%%%%%%%%%%%%%%%

\subsection{The symmetric and antisymmetric parts of $d\vv$}

With the imposed chosen Euclidean dot product~$\dd_g$ in~$\RRnt$,
we can consider the transposed endomorphism $d\vv_t(p)^T_g\eqnote d\vv_t(p)^T \in \calL(\RRnt;\RRnt)$, which is defined by, for all $\vwu,\vwd\in\RRnt$ vectors at~$p$,
\be
\label{eqLT}
(d\vv_t(p)^T.\vwu, \vwd)_g = (\vwu,d\vv_t(p).\vwd)_g
\ee
\cf~\S~\ref{seccpgdd0}.
We have thus defined
\be
d\vv_t^T : 
\left\{\eqalign{
\Omegat & \rar \calL(\RRnt;\RRnt) \cr
p & \rar d\vv_t^T(p) := d\vv_t(p)^T \cr
}\right.
\ee
Other usual notations (definitions):
$d\vv_t(p)^T \eqnote d\vv(t,p)^T \eqnote d\vv^T(t,p)$.

\debdef
The (Eulerian) rate of deformation tensor, or stretching tensor, is the $\dd_g$-symmetric part of~$d\vv$:
\be
\label{eqtenstauxd}
\calD={d\vv + d\vv^T\over 2},
\quad\hbox{\ie,}\quad\forall (t,p)\in\bigcup_{t\in\RR}(\{t\}\times \Omegat),\quad
\calD(t,p)={d\vv(t,p) + d\vv(t,p)^T\over 2} .
\ee
The (Eulerian) spin tensor is the $\dd_g$-antisymmetric part of~$d\vv$:
\be
\label{eqOmegarot}
\Omega={d\vv - d\vv^T\over 2},
\quad\hbox{\ie,}\quad\forall (t,p)\in\bigcup_{t\in\RR}(\{t\}\times \Omegat),\quad
\Omega(t,p)={d\vv(t,p) - d\vv(t,p)^T\over 2} . %,\;\forall (t,p)\in\calC.
\ee
(So $d\vv= \calD+\Omega$ with $\calD$ the rate of deformation tensor and $\Omega=\vomega \wedge $ a rotation times a dilation, see the following.)
\findef

NB: The same notation is used for the set of points $\Omega_t=\Phitzt(\Omegatz) \subset\RRn$
and for the spin tensor $\Omega_t={d\vv_t - d\vv_t^T\over 2}$: The context removes ambiguities.

%%%%%%%%%%%%%%%%%%%%%%%%%%%%%%%%%%%%%%%%%%%%%%%%%%%%%%%%%%%%%%%%%%%%%%%%%%%%%%%%%%%

\subsection{Quantification with a basis}

With a basis $(\ve_i)$ in~$\RRnt$, \eref{eqLT}  gives
\be
\label{eqOmegarot21}
[g]_{|\ve}.[d\vv^T]_{|\ve} = [d\vv]_{|\ve}^T.[g]_{|\ve}, \qand [d\vv^T]_{|\ve} = [g]_{|\ve}^{-1}.[d\vv]_{|\ve}^T.[g]_{|\ve}.
\ee
%\Eg, with a Cartesian basis~$\ve_i)$, if $[g]_{|\ve}=[g_{ij}]$, if $[d\vv]_{|\ve}=[{\pa v_i \over \pa x_j}]$, and if $[d\vv]_{|\ve}^T=[T_{ij}]$, then $\sumkn g_{ik}T_{kj} = \sumkn {\pa v_i \over \pa x_k}g_{kj}$ for all $i,j$.
In particular, if $(\ve_i)$ is a $\dd_g$-orthonormal basis, then 
%$T_{ij}={\pa v_j \over \pa x_i}$ for all $i,j$, \ie\ 
$[d\vv^T]_{|\ve} = [d\vv]_{|\ve}^T$ (orthonormal basis case).
Thus for the endomorphisms $\calD$ and~$\Omega$,
and with the above Euclidean framework and its Euclidean orthonormal basis, we have
$\calD.\ve_j = \sumin \calD_{ij} \ve_i$ and $\Omega.\ve_j = \sumin \Omega_{ij} \ve_i$ with
$\calD_{ij}=\demi({\pa v_i\over\pa x_j}+{\pa v_j\over\pa x_i})$ and
$\Omega_{ij} =\demi({\pa v_i\over\pa x_j}-{\pa v_j\over\pa x_i})$, that is,
\be
\label{eqOmegarot2}
[\calD]_{|\ve}={[d\vv]_{|\ve}+[d\vv]_{|\ve}^T \over 2} \qand
[\Omega]_{|\ve}={[d\vv]_{|\ve}-[d\vv]_{|\ve}^T \over 2}
\quad \hbox{(Euclidean framework)}.
\ee
Duality notations:
$\calD.\ve_j = \sumin \calD^i_j \ve_i$, $\calD^i_j=\demi({\pa v^i\over\pa x^j}+{\pa v^j\over\pa x^i})$
and
$\Omega.\ve_j = \sumin \Omega^i{}_j \ve_i$, $\Omega^i{}_j=\demi({\pa v^i\over\pa x^j}-{\pa v^j\over\pa x^i})$,
so with $\calD^i_j=\calD^j_i$ and $\Omega^i{}_j=-\Omega^j{}_i$.

%(With classical notations: $[L]_{|\ve}=[d\vv]_{|\ve} = [L_{ij}]=[{\pa v_i \over \pa x_j}]_{i=1,...,n \atop j=1,...,n}$.)

%%%%%%%%%%%%%%%%%%%%%%%%%%%%%%%%%%%%%%%%%%%%%%%%%%%%%%%%%%%%%%%%%%%%%%%%%%%%%%%%%%%
%%%%%%%%%%%%%%%%%%%%%%%%%%%%%%%%%%%%%%%%%%%%%%%%%%%%%%%%%%%%%%%%%%%%%%%%%%%%%%%%%%%

\section{Interpretation of the rate of deformation tensor}

We are interested in the evolution of the deformation gradient $F(t):=\Ftzptz(t)$ along the trajectory of a particle~$\Pobj$ which was at $\ptz$ at~$\tz$. So:

\comment{
Let $\tPhi : [t_1,t_2] \times \Obj \rar \RRn$ be a regular motion, \cf~\eref{eqdeftPhi0},
$\tz\in[t_1,t_2]$, % (any fixed time, not considered as an initial time here), 
$\Omegatz = \tPhi(\tz,\Obj)$,
$\Pobj\in \Obj$,
$\ptz = \tPhi(\tz,\Pobj)$;
Let $\Phitz : (t,\ptz)\in [t_1,t_2]\times \rar p(t)=\Phitz(t,\ptz)=\tPhi(t,\Pobj) \in \RRn$ (trajectory of~$\Pobj$),
let
$\Phitzptz:=\Phitz(t,\ptz)$,
and let $F(t) =\Ftzptz(t):=d\Phitz(t,\ptz) = \Ftz(t,\ptz)$.
}

Let $\vA=\va(\tz,\ptz)$ and $\vB=\vb(\tz,\ptz)$ be vectors at~$\tz$ at~$\ptz$ in~$\Omegatz$,
and consider their push-forwards by the flow~$\Phitzt$ (the transported vectors), \ie\ the vectors at $t$ at~ $p(t)=\Phitzptz(t)$ given by
\be
\va(t,p(t)):= F(t).\vA \qand
\vb(t,p(t)):= F(t).\vB.
\ee
see \eref{eqdefFtfd} and~figure~\ref{figpf}.
They define the %(Eulerian type scalar valued) 
function
\be
(\va,\vb)_g:
\left\{\eqalign{
\bigC & \rar \RR \cr
(t,\pt) &\rar (\va,\vb)_g(t,\pt) := (\va(t,\pt),\vb(t,\pt))_g . %\quad (= (F(t).\vA,F(t).\vB)_g).
}\right.
\ee
%that is, $(\va,\vb)_g(t,p(t)) := (F(t)^{-1}.\va(t,p(t)),F(t)^{-1}.\vb(t,p(t)))_g$.
%Thus, with $\ptz$ \st\ $\pt=\Phitzptz(\ptz)$,  we have $(\va,\vb)_g(t,\Phitzptz(t)) = (F(t).\vA,F(t).\vB)_g)$.

\begin{prop}%[and definition]
\label{propint}
The rate of deformation tensor $\calD = {d\vv+d\vv^T\over 2}$ gives (half) the evolution rate between two vectors deformed by the flow, that is, along trajectories,
\be
\label{eqDint}
{D(\va,\vb)_g\over D t} =2(\calD.\va,\vb)_g.
\ee
\finprop

\debdem
$f(t) := (\va(t,p(t)),\vb(t,p(t)))_g = (F(t).\vA,F(t).\vB)_g$ %($=(\va,\vb)_g(t,p(t))$ when $p(t)=\Phitzt(\ptz)$), 
gives
\be
\label{eqDint2}
f'(t)
= (F'(t).\vA,F(t).\vB)_g + (F(t).\vA,F'(t).\vB)_g.
\ee
And $F'(t) = d\vv(t,p(t)).F(t)$, \cf~\eref{eqvVvv2b}. Thus, with $\va(t,p(t))=F(t).\vA$ and $\vb(t,p(t))=F(t).\vB$,
\be
\label{eqpropint3}
\eqalign{
f'(t)
= & (d\vv(t,p(t)).F(t).\vA,F(t).\vB)_g + (F(t).\vA,d\vv(t,p(t)).F(t).\vB)_g\cr
= & (d\vv(t,p(t)).\va(t,p(t)),\vb(t,p(t)))_g + (\va(t,p(t)),d\vv(t,p(t)).\vb(t,p(t)))_g\cr
= & ((d\vv(t,p(t))+d\vv(t,p(t))^T).\va(t,p(t)),\vb(t,p(t)))_g,
% = (2\calD(t,p(t)).\va(t,p(t)),\vb(t,p(t)))_g,
}
\ee
\ie~\eref{eqDint}, since $f(t)=(\va,\vb)_g(t,p(t))$ gives $f'(t)={D(\va,\vb)_g\over D t}(t,p(t))$.
\findem

\comment{
Let $\tPhi$ be a motion and $\vv(t,p(t)) = {\pa \tPhi \over \pa t}(t,\Pobj)$ when $p(t)= \tPhi(t,\Pobj)$.
Let $t \in [t_1,t_2]$, and let $\Phi^t$ be the associated motion
defined by $\Phi^t(\tau,\pt) = \tPhi(\tau,\Pobj)$ when $\pt = \tPhi(t,\Pobj)$,
% pour $\tau$ dans un voisinage de~$t$,
\cf~\eref{eqdefPhi}.
Let $\ptau = \Phi^t(\tau,\pt) = \Phittau(\pt)$, and let $\Ftpt(\tau) = d\Phi^t_\pt(\tau)$.
}%
%

%%%%%%%%%%%%%%%%%%%%%%%%%%%%%%%%%%%%%%%%%%%%%%%%%%%%%%%%%%%%%%%%%%%%%%%%%%%%%%%%%%%
%%%%%%%%%%%%%%%%%%%%%%%%%%%%%%%%%%%%%%%%%%%%%%%%%%%%%%%%%%%%%%%%%%%%%%%%%%%%%%%%%%%

\section{Rigid body motions and the spin tensor}

Choose a Euclidean dot product~$\dd_g$ (required to characterize a rigid body motion).

Result: A rigid body motion is a motion whose Eulerian velocity satisfies $d\vv + d\vv^T = 0$, \ie, $\calD=0$
(Eulerian approach independent of any initial time~$\tz$ chosen by some observer).

But the usual classical introduction to rigid body motion relies on some initial time~$\tz$ (Lagrangian approach). So, to begin with, let us do it with the Lagrangian approach.
Recall: T the first order Taylor expansion of~$\Phitzt$ in the vicinity of a $\ptz\in\Omegatz$ is
\be
\label{eqfoe}
\Phitzt(\qtz) = \Phitzt(\ptz) + \Ftzt(\ptz).\ora{\ptz\qtz} + o(\ora{\ptz\qtz}).
\ee

\comment{
$\tPhi : [t_1,t_2] \times \Obj \rar \RRn$ is a regular motion, %\cf~\eref{eqdeftPhi0},
$\Omegat=\tPhi(t,\Obj)$,
$\bigC= \bigcup_{t\in[t_1,t_2]}(\{t\}\times\Omega_t)$, %cf~\eref{eqbigU},
and $\vv : \bigC \rar\vRRn$ is the Eulerian velocity field, %\cf~\eref{eqdefve},
\ie\ $\vv(t,\pt) = {\pa \tPhi \over \pa t}(t,\Pobj)$ when $\pt = \tPhi(t,\Pobj)$.
Choose a $\tz\in[t_1,t_2]$, let $\Phitz$ be the motion associated to~$\tPhi$ relative to~$\tz$, 
%\cf~\eref{eqdefPhi} and $\Phitzt$ the motion associated to~$\tPhi$ relative to~$\tz$ and~$t$, \cf~\eref{eqPhit},
and let $\vVtz:[t_1,t_2]\times \Omegatz\rar\RRnt$ be the Lagrangian velocity field, %\cf~\eref{eqdefVl00},
\ie\ $\vVtz(t,\ptz) = {\pa \Phitz \over \pa t}(t,\ptz) = \vv(t,\pt)$ when $\pt=\Phitz(t,\ptz)$.
Let $\ptz\in\Omegatz$, let $\Phitzt(\ptz):=\Phitz(t,\ptz)$, let $\Ftzt(\ptz):=d\Phitzt(\ptz)\in\calL(\RRntz;\RRnt)$:
Taylor expansion: For all $\qtz\in\Omegatz$ (in a vicinity of~$\ptz$),
%(Recall: If you don't like the notation~$\RRnt$, then use the non ambiguous notation $\TptOmegat$, \cf~\S~\ref{secRRnt}.)
}

%%%%%%%%%%%%%%%%%%%%%%%%%%%%%%%%%%%%%%%%%%%%%%%%%%%%%%%%%%%%%%%%%%%%%%%%%%%%%%%%%%%

\subsection{Affine motions and rigid body motions}
\label{secaffm}

%%%%%%%%%%%%%%%%%%%%%%%%%%%%%%%%%%%%%%%%%%%%%%%%%%%%%%%%%%%%%%%%%%%%%%%%%%%%%%%%%%%

\subsubsection{Affine motions}

\debdef
\label{defmaff}
$\Phitz$ is an affine motion (understood ``affine motion in space'') iff $\Phitzt$ is an ``affine motion'',
\ie\ iff %$\Omegatz$ is connected, $\Phitz$ is $C^\infty$ in time, 
$\Phitzt$ is a $C^1$ diffeomorphism (in space), and
\eref{eqfoe} reads, for all $\ptz,\qtz\in\Omegatz$ and all $t\in[t_1,t_2]$,
\be
\label{eqhmh2}
\Phitzt(\qtz) = \Phitzt(\ptz) + \Ftzt(\ptz).\ora{\ptz\qtz}.
\ee
Marsden--Hughes notations: $\Phi(Q)=\Phi(P) + F(P).\ora{PQ}$.
\findef

\debprop{\bf and definition.}
\label{propaffm}
If $\Phitz$ is an affine motion, then $\Ftzt(\ptz)$ is independent of~$\ptz$, \ie,
for all $t\in]t_1,t_2[$ and all $\ptz \in \Omegatz$ and all $\qtz \in \Omegatz$, % (in an open vicinity of~$\ptz$),
\be
\label{eqhmh21}
\Ftzt(\ptz)=\Ftzt(\qtz)  \eqnote \Ftzt.
\ee
And then $d\Ftzt(\ptz)=0$, \ie\ $d^2\Phitzt(\ptz)=0$. %(no curvature).
And for all $t\in]t_1,t_2[$, $\Phi^t$ is an affine motion: % (updated Lagrangian description):
For all $\tau\in]t_1,t_2[$ and all $\pt,\qt \in \Omegat$, % (in an open vicinity of~$\pt$),
\be
\label{eqhmh21t}
\Phittau(\qt) = \Phittau(\pt) + \Fttau.\ora{\pt\qt}.
\ee
And $\tPhi$ is said to be an affine motion (understood ``affine motion in space'').
\finprop

\debdem
$\qtz=\ptz+\ora{\ptz\qtz}$ gives
$\Phitzt(\qtz)=\Phitzt(\ptz+\ora{\ptz\qtz}) = \Phitzt(\ptz) + d\Phitzt(\ptz).\ora{\ptz\qtz}$,
and, similarly,
$\Phitzt(\ptz)=\Phitzt(\qtz+\ora{\qtz\ptz}) = \Phitzt(\qtz) + d\Phitzt(\qtz).\ora{\qtz\ptz}$.
Thus (addition)
$\Phitzt(\qtz)+\Phitzt(\ptz)=\Phitzt(\ptz)+\Phitzt(\qtz) + (d\Phitzt(\ptz) - d\Phitzt(\qtz)).\ora{\ptz\qtz}$,
thus $(d\Phitzt(\ptz) - d\Phitzt(\qtz)).\ora{\ptz\qtz} = 0$, true for all $\ptz,\qtz$,
thus $d\Phitzt(\ptz) - d\Phitzt(\qtz)=0$, \ie~\eref{eqhmh21}.

Thus $d^2\Phitzt(\ptz).\vu_\tz
= \lim_{h\rar0} { d\Phitzt(\ptz+h\vu_\tz) - d\Phitzt(\ptz) \over h}
= \lim_{h\rar0} { d\Phitzt - d\Phitzt \over h}
= 0$
for all $\ptz$ and all $\vu_\tz$, thus $d^2\Phitzt(\ptz)=0$ for all $\ptz$, thus $d^2\Phitzt=0$.

And~\eref{eqcompf01} gives
$(\Phittau \circ\Phitzt)(\ptz) = \Phi^\tz_\tau(\ptz)$,
thus, with $\pt = \Phitzt(\ptz)$, we get $d\Phittau(\pt).d\Phitzt(\ptz) = d\Phi^\tz_\tau(\ptz)$,
thus $d\Phittau(\pt) = d\Phi^\tz_\tau(\ptz).d\Phitzt(\ptz)^{-1}$,
and~\eref{eqhmh2} gives %$d\Phittau(\pt).d\Phitzt = d\Phi^\tz_\tau$, thus,
\be
\label{eqhmh21t2}
d\Phittau(\pt) = d\Phi^\tz_\tau.d\Phitzt^{-1} \eqnote d\Phittau \quad(\hbox{independent of~$\pt$}),
\ee
thus~\eref{eqhmh21t}.
\findem

\debcor
%\label{cordvea}
If $\tPhi$ is affine then, $\vv_t$ is affine for all~$t$, and $\vVtzt$ is affine for all~$\tz,t$, \ie,
for all $\pt\in\Omegat$ we have $d\vv_t(\pt) = d\vv_t$ (independent of~$\pt$), and for all $\ptz\in\Omegatz$ we have $d\vVtzt(\ptz)\eqnote d\vVtzt$ (independent of~$\ptz$): For all $\qt\in\Omegat$ and all $\qtz\in\Omegatz$,
\be
\label{eqvedh0}
\left\{\eqalign{
\bullet \; &
\vv_t(\qt)=\vv_t(\pt)+ d\vv_t.\ora{\pt\qt}, \cr
\bullet \; &
\vVtzt(\qtz)=\vVtzt(\ptz)+d\vVtzt.\ora{\ptz\qtz}.\cr
}\right. 
\ee
\fincor

\debdem
\eref{eqhmh2} gives $\Phitz(t,\qtz)=\Phitz(t,\ptz)+ \Ftz(t).\ora{\ptz \qtz}$,
and the derivation in time gives~\eref{eqvedh0}$_2$,
then~\eref{eqvedh0}$_1$ thanks to $\pt=\Phitzt(\ptz)$, $\qt=\Phitzt(\qtz)$ and $\ora{\ptz\qtz} = (\Ftzt)^{-1}.\ora{\pt\qt}$, \cf~\eref{eqhmh2}.
\findem

\debexa
In $\RR^2$, with a basis $(\vE_1,\vE_2)$ in $\RRntz$ and a basis $(\ve_1,\ve_2) \in\RRnt$, % and an origin $o_t\in\RRn$,
then $\Ftzt$ given by $[\Ftzt]_{|\vE,\ve} = \pmatrix{1+t&2t^2\cr3t^3&e^t}$
derives from the affine motion
$[\ora{\Phitzt(\ptz)\Phitzt(\qtz)}]_{|\ve} = \pmatrix{1+t&2t^2\cr3t^3&e^t}.[\ora{\ptz\qtz}]_{|\vE}$.
\finexa

%%%%%%%%%%%%%%%%%%%%%%%%%%%%%%%%%%%%%%%%%%%%%%%%%%%%%%%%%%%%%%%%%%%%%%%%%%%%%%%%%%%

\subsubsection{Rigid body motion}

A Euclidean dot product~$\dd_g$ in~$\RRnt$ is chosen, the same at all time~$t$. Let $\Phi := \Phitzt$ and $F:=\Ftzt$.
Recall: If $P\in\Omegatz$ and $p=\Phi(P)$ ($\in\Omegat$) then the transposed of the linear map $F(P)\in\calL(\RRntz;\RRnt)$ relative to~$\dd_g$ is the linear map $F^T(p):=F(P)^T\in\calL(\RRnt;\RRntz)$ defined by
\be
F^T(p) :=F(P)^T :
\left\{\eqalign{
\RRnt & \rar \RRntz \cr
\vw_p & \rar F^T(p).\vw_p\qst (F^T(p).\vw_p,\vU_P)_g = (\vw_p,F(P).\vU_P)_g,\;\; \forall \vU_P\in\RRntz. \cr
}\right.
\ee
We have thus defined the function $F^T:\Omegat  \rar \calL(\RRnt;\RRntz)$. 

Particular case: For an affine motion, since $F$ is independent of~$P$, we get $F^T$ is independent of~$p$.
\comment{
: For all $(\vU_P,\vw_p)\in\RRntz\times \RRnt$,
\be
(F^T.\vw_p,\vU_P)_g = (\vw_p,F.\vU_P)_g,\quad\hbox{(affine motion)}.
\ee
}

\debdef
\label{defms00}
A rigid body motion is an affine motion $\tPhi$ such that, for all $\tz,t\in \RR$, $P\in\Omegatz$, $\vU_P,\vW_P\in\RRntz$, and with $p=\Phitzt(P)$,
\be
\label{eqdefms00}
(F.\vU_P,F.\vW_P)_g = (\vU_P,\vW_P)_g, \qie (F^T.F.\vU_P , \vW_P)_g = (\vU_P,\vW_P)_g, \qie \boxed{F^T.F=I}.
\ee
(Angles and lengths are unchanged.)
%Full notation: $(\Ftzt.\vU_P,\Ftzt.\vW_P)_g = (\vU_P,\vW_P)_g$ and $(\Ftzt)^T.\Ftzt = I$.
\findef

In other words, with the Cauchy strain tensor $C\in \calL(\RRntz;\RRntz)$ defined by $C=F^T.F$,
the motion is rigid iff it is affine and
\be
\boxed{C=I}, \qie \boxed{F^{-1} = F^T}.
\ee
%(Full notation: See~\eref{eqdefCf}.)

\begin{prop}
\label{propdefms02}
If $\Phitz$ is a rigid body motion, if $(\vA_i)$ is a $\dd_g$-Euclidean basis in~$\RRntz$,
if $P\in\Omegatz$, if $t\in[\tz,T]$ and $p=\Phitzt(P)$,
and if $\va_i(t,p)=\Ftz(t,P).\vA_i$ for all~$i$,
then $\va_i(t,p) \eqnote \va_{i,t}$ is independent of~$p$, and $(\va_{i,t})$ is a $\dd_g$-Euclidean basis with the same orientation than~$(\vA_i)$ for all~$t$.
\finprop

\debdem
$\Phitzt$ is affine, thus, for all $t,P$, $\Ftzt(P)=\Ftzt$ (independent of~$P$),
thus $\va_{i,t}(p)=\Ftzt.\vA_i \in \RRnt$ is independent of~$p$, this at all~$t$. Let $t$ be fixed and $\va_{i,t} \eqnote \va_i$ ($=F.\vA_j$).
We get
$(\va_i,\va_j)_g
= (F.\vA_i,F.\vA_j)_g
= (F^T.F.\vA_i,\vA_j)_g
= (I.\vA_i,\vA_j)_g
= (\vA_i,\vA_j)_g
= \delta_{ij}$
for all $i,j$, thus $(\va_i)$ is $\dd_g$-orthonormal basis.
And
$\det(\va_1,...,\va_n)
= \det(F.\vA_1,...,F.\vA_n)
= \det(F)\det(\vA_1,...,\vA_n)
= \det(F)
$ since $(\vA_i)$ is a $\dd_g$- orthonormal basis.
And, $\Phitzt$ being a diffeomorphism,
$t\rar\det(\Ftzt) %=\det(\Ftz(t))
$ is continuous, does not vanish, moreover with $\det(F^\tz_\tz)=\det(I)=1>0$; Thus $\det(\Ftzt)>0$ for all~$t$,
hence $\det(\va_1,...,\va_n)>0$: 
The bases have the same orientation.
\findem

\debexa
In~$\RR^2$, a rigid body motion is given by
$\Ftzt=\pmatrix{\cos(\theta (t)) & -\sin(\theta (t)) \cr\sin(\theta (t)) &\cos(\theta (t))}$
with $\theta$ a regular function \st\ $\theta(\tz)=0$.
\finexa

\debexe
%\label{propvaip}
Let $\tPhi$ be a rigid body motion. Prove
\be
\label{eqFtpT}
(F^T)'(t) = (F'(t))^T, \qand \hbox{$F^T.F'$ is antisymmetric}.
\ee
%which means that the endomorphism $F^T(t).F'(t) \in\calL(\vRRntz;\vRRntz)$ is antisymmetric at all time~$t$.

\debrep
Let $t\in\RR$, $p(t)=\Phitzt(P)$, $\vU,\vW\in\RRntz$ and $\vw(t,p(t))=F(t).\vW$. 
And recall that the function $F^T:t\rar F^T(t)$ is defined (as usual) by $F^T(t):=(F(t))^T$.
We have
$(F(t)^T.\vw(t,p(t)),\vU)_g = (\vw(t,p(t)),F(t).\vU)_g$. Thus
$((F^T)'(t).\vw(t,p(t)) + F^T(t).{D\vw\over Dt}(t,p(t)),\vU)_g
= ({D\vw\over Dt}(t,p(t)),F(t).\vU)_g + (\vw(t,p(t)),F'(t).\vU)_g $, which simplifies into
$((F^T)'(t).\vw(t,p(t)) ,\vU)_g
= (\vw(t,p(t)),F'(t).\vU)_g = ((F'(t))^T.\vw(t,p(t)),\vU)_g$, thus $(F^T)'(t) = (F'(t))^T$.

And \eref{eqdefms00} reads $F^T(t).F(t)=I_\tz$, thus
$(F^T)'(t).F(t) + F^T(t).F'(t)=0$, thus
$(F')^T(t).F(t) + F^T(t).F'(t)=0$, thus $F^T.F'$ is antisymmetric.
\finrep
\finexe

%%%%%%%%%%%%%%%%%%%%%%%%%%%%%%%%%%%%%%%%%%%%%%%%%%%%%%%%%%%%%%%%%%%%%%%%%%%%%%%%%%%

\subsubsection{Alternative definition of a rigid body motion: $d\vv+d\vv^T=0$}
\label{secsbdmv}

The stretching tensor $\calD_t= {d\vv_t + d\vv_t^T \over 2}$ and the spin tensor $\Omegat= {d\vv_t - d\vv_t^T \over 2}$ have been defined in~\eref{eqtenstauxd}-\eref{eqOmegarot}.

\debprop
\label{propvaip}
If $\tPhi$ is a rigid body motion, \cf~\eref{eqdefms00}, then the endomorphism $d\vv_t \in\calL(\vRRnt;\vRRnt)$ is antisymmetric at all~$t$:
\be
\label{eqdesa}
%{d\vv_t + d\vv_t^T \over 2} = 0, \qie
d\vv_t = \Omega_t , %\quad ( = {d\vv_t - d\vv_t^T \over 2}),
\qie %(d\vv_t + d\vv_t^T=)\quad  
\calD_t = 0.
\ee
Conversely, if $d\vv_t + d\vv_t^T = 0$ at all~$t$, then $\tPhi$ is a rigid body motion (here no initial time is required).

So the relation << $d\vv_t + d\vv_t^T = 0$ for all~$t$ >> gives an equivalent definition to the definition~\ref{defms00}.
%the characterization, but here no initial time is required: A rigid body motion does not depend on the choice of an initial time by an observer.
\finprop

\debdem
Let  $F(t):=\Ftzptz(t)$ and $F^T(t):=F(t)^T$ and $V(t):=\vVtzptz(t) = (\Phitzptz)'(t) = \vv(t,\pt)$ (the Lagrangian and Eulerian velocities). \eref{eqdefms00} gives
$\ds(F.F^T)'(t)=0
= F'(t).F(t)^T + F(t).(F^T)'(t)
\mathop{=}^{\eref{eqFtpT}} F'(t).F(t)^T + (F'(t).F(t)^T)^T
= dV(t).F(t)^{-1}+(dV(t).F(t)^{-1})^T
\mathop{=}^{\eref{eqvVvv}} d\vv(t,\pt) + d\vv(t,\pt)^T
$. Thus~\eref{eqdesa}.

%And $d\vv_t(\pt).\Ftzt(\ptz) % = d\vVtzt(\ptz)  = {\pa \Ftz \over \pa t}(t,\ptz)$, with $d\vv + d\vv^T = 0$ and $\Ftzt$ affine, gives ${d \Ftz \over d t}(t).\Ftzt^{-1}$ antisymmetric, thus ${d \Ftz \over d t}(t).\Ftzt^T$ antisymmetric (rigid motion: $F^{-1}=F^T$).

Conversely, suppose $d\vv + d\vv^T = 0$. Then \eref{eqDint} gives ${D (\va,\vb)_g\over Dt}=0$,
thus $(\va,\vb)_g(t,\pt)=(\va,\vb)_g(\tz,\ptz)$ for all $t,\tz$ and all $\ptz=\Phitzt(\pt)$,
\ie\ $(\Ftzt(\ptz).\vA,\Ftzt(\ptz).\vB)_g = (\vA,\vB)_g$ for all $t,\tz$, all $\ptz$ and all $\vA,\vB\in\RRntz$:
Thus $\tPhi$ is a rigid body motion, cf~\eref{eqdefms00}.
\findem

\comment{
\debrem
Reminder: Notations (usual definitions) used above and in the sequel.
Let $L:t\in\RR \rar L(t) \in \calL(A;B)$ is a time dependent linear map.

If $L(t)$ is invertible for all~$t$, then $L(t)^{-1} \in \calL(B;A)$ define the function
$L^{-1} : t\in\RR \rar L^{-1}(t):=L(t)^{-1} \in \calL(B;A)$.

If $\dd_A$ and $\dd_B$ are inner dot products in~$A$ and~$B$, then $L(t)^T \in \calL(B;A)$ define the function
$L^T :t\in\RR \rar L^T(t) := L(t)^T \in \calL(B;A)$.
(Recall that $L(t)^T$ is defined relatively to $\dd_A$ and $\dd_B$ by $(L(t)^T.\vb,\va)_A=(L(t).\va,\vb)_B$
for all $(\va,\vb)\in A\times B$. Thus$(L^T(t).\vb,\va)_A=(L(t).\va,\vb)_B$ for all $(\va,\vb)\in A\times B$.)

Then, if $L$ is derivable, then $L^T$ is derivable and $(L^T)'(t) = L'(t)^T$.
Indeed, $(L^T(t).\vb,\va)_A = (L(t).\va,\vb)_B$ gives
$((L^T)'(t).\vb,\va)_A = (L'(t).\va,\vb)_B = (L'(t)^T.\vb,\va)_A$,
true for all $\va,\vb$, thus $(L^T)'(t) = L'(t)^T$.
\finrem
}

%%%%%%%%%%%%%%%%%%%%%%%%%%%%%%%%%%%%%%%%%%%%%%%%%%%%%%%%%%%%%%%%%%%%%%%%%%%%%%%%%%%

\subsection{Representation of the spin tensor $\Omega$: vectors, and pseudo-vectors}
\label{secvrO}

We are dealing here with concepts that are sometimes misunderstood.
Framework: $\RRn=\RR^3$.
%, $(\ve_1,\ve_2,\ve_3)$  is a Euclidean basis, and $\dd_g$ is the associated Euclidean dot product, \cf~\S~\ref{secbe}.
%There is often some confusion between a vector and a matrix named a ``pseudo-vector'' or a ``column vector''. A description of these two different concepts for the spin tensor $\Omega = \demi(d\vv- d\vv^T)$ is given. % in this section. %, with $\RRn = \RR^3$.

%%%%%%%%%%%%%%%%%%%%%%%%%%%%%%%%%%%%%%%%%%%%%%%%%%%%%%%%%%%%%%%%%%%%%%%%%%%%%%%%%%%

\subsubsection{Reminder}

\leavevmode

$\bullet$ %Let $(\ve_1,\ve_2,\ve_3)$ be a Euclidean basis.
The determinant $\det_{|\ve}$ associated with a basis~$(\ve_i)$ in~$\RR^3$ is the alternating multilinear form defined by
$\det_{|\ve}(\ve_1,\ve_2,\ve_3)=1$; The algebraic volume (or signed volume) limited by three vectors $\vu_1,\vu_2,\vu_3$
is $\det_{|\ve}(\vu_1,\vu_2,\vu_3)$; And the (positive) volume is $|\det_{|\ve}(\vu_1,\vu_2,\vu_3)|$, see~\S~\ref{secdetend}.

$\bullet$ Let $A$ and $B$ be two observers (\eg\ $A$=English and $B$=French),
let $(\va_i)$ be a Euclidean basis chosen by~$A$ (\eg\ based on the foot),
let $(\vb_i)$ be a Euclidean basis chosen by~$B$ (\eg\ based on the metre), see~\S~\ref{secbe}.
Let $\lambda=||\vb_1||_a>0$ (change of unit of length coefficient). %, hence $\dd_a = \lambda^2\dd_b$, \cf~\eref{eqpropbes}.
The relation between the determinants %(dependent on the measuring unit and orientation)
 is:
\be
\label{eqdetlambda}
\det_{|\va} = \pm \lambda^3\,\det_{|\vb} \qwith
\left\{\eqalign{
&+\hbox{ if }\det_{|\va}(\vb_1,\vb_2,\vb_3)>0
\quad\hbox{(\ie\ if the bases have the same orientation)},
 \cr
&- \hbox{ if }\det_{|\va}(\vb_1,\vb_2,\vb_3)<0
\quad\hbox{(\ie\ if the bases have opposite orientation)}.
\cr
}\right.
\ee
\comment{
\be
\label{eqdetlambda}
\det_{|\va} = 
\left\{\eqalign{
&+\lambda^3\,\det_{|\vb} \quad\hbox{if }\det_{|\va}(\vb_1,\vb_2,\vb_3)>0
\quad\hbox{(\ie\ if the bases have the same orientation)},
 \cr
&-\lambda^3\,\det_{|\vb} \quad\hbox{if }\det_{|\va}(\vb_1,\vb_2,\vb_3)<0
\quad\hbox{(\ie\ if the bases have opposite orientation)}.
\cr
}\right.
\ee
}
In particular, if $A$ and $B$ use the same unit of length
(or if $A$ uses two $\dd_g$-Euclidean basis $(\va_i)$ and~$(\vb_i)$), then $\lambda=1$ and $\det_{|\va} = \pm \det_{|\vb}$.
\comment{
So, if $\vu_1,\vu_2,\vu_3$ define a parallelepiped, then its algebraic volume $\det_{|\va}(\vu_1,\vu_2,\vu_3)$ relative to the unit of measure of~$A$ is equal to $\pm\lambda^3$ times the algebraic volume $\det_{|\vb}(\vu_1,\vu_2,\vu_3)$, the sign depending on the relative orientation of the bases.
}

$\bullet$
With an imposed Euclidean dot product~$\dd_g$: An endomorphism~$L$ is $\dd_g$-antisymmetric iff
\be
\label{eqLant}
\forall \vu,\vv, \;\; (L.\vu,\vv)_g + (\vu,L.\vv)_g=0, \qie
L^T=-L. % \quad(\hbox{Euclidean framework}).
\ee
%That is,  $(L.\vu,\vv)_g + (\vu,L.\vv)_g=0$ for all~$\vu,\vv$.
%(See~\eref{eqte}.)

%%%%%%%%%%%%%%%%%%%%%%%%%%%%%%%%%%%%%%%%%%%%%%%%%%%%%%%%%%%%%%%%%%%%%%%%%%%%%%%%%%%

\subsubsection{Definition of the vector product (cross product)}

Let $(\ve_i)$ be a $\dd_g$-orthonormal basis,
let $\vu,\vv\in\vRRt$, and let $\ell_{\ve,\vu,\vv} \in \calL(\vRRt,\RR)$ be the linear form defined by
\be
\ell_{\ve,\vu,\vv} : 
\left\{\eqalign{
\vRRt & \rar \RR \cr
\vz & \rar \ell_{\ve,\vu,\vv}(\vz) \eqdef \det_{|\ve}(\vu,\vv,\vz)
}\right.
\ee
(the algebraic volume of the parallelepiped limited by $\vu,\vv,\vz$ in the Euclidean chosen unit).

\debdef
The vector product, or cross product, $\vu\wedge_e \vv$ of two vectors $\vu$ and~$\vv$
is the $\dd_g$-Riesz representation vector of~$\ell_{\ve,\vu,\vv}$,
that is,  $\vu\wedge_e \vv \in \vRRt$ is characterized by
$\ell_{\ve,\vu,\vv}(\vz) = (\vu\wedge_e \vv,\vz)_g$ for all $\vz\in\vRRt$, \cf~\eref{eqrtr}, \ie
\be
\label{eqw1}
\forall \vz\in\vRRt,\quad \boxed{(\vu\wedge_e \vv,\vz)_g = \det_{|\ve}(\vu,\vv,\vz)}.
\ee
NB: $\vu\wedge_e \vv$ depends on~$\dd_g$ since we need a $\dd_g$-Euclidean basis~$(\ve_i)$ (and depends on the orientation of~$(\ve_i)$.
\findef

We have thus defined the bilinear cross product operator %, relative to the chosen Euclidean basis
\be
\label{eqw1b}
\wedge_e :
\left\{ \eqalign{
\vRRt \times \vRRt & \rar \vRRt \cr
(\vu,\vv) & \rar \wedge_e(\vu,\vv) \eqdef \vu\wedge_e \vv.
}\right.
\ee
(The bilinearity is trivial thanks to the multilinearity of the determinant.)
%and $\wedge_e$ est a bilinear antisymmetric (since $\det_{|\ve}$ is alternated).
If one Euclidean basis is imposed by one observer to all the other observers, then
$\vu\wedge_e \vv$ is written $\vu \wedge \vv$ (non objective).

%%%%%%%%%%%%%%%%%%%%%%%%%%%%%%%%%%%%%%%%%%%%%%%%%%%%%%%%%%%%%%%%%%%%%%%%%%%%%%%%%%%

\subsubsection{Calculation of the vector product}

$\vu = \sum_{i=1}^3 u_i\ve_i$, $\vv = \sum_{i=1}^3 v_i\ve_i$ and~\eref{eqw1} give
\be
(\vu\wedge_e \vv,\ve_1)_g
= \det_{|\ve}(\vu,\vv,\ve_1)
%= \det_{|\ve}(\ve_1,\vu,\vv)
= \det\pmatrix{u_1 & v_1 & 1 \cr u_2 & v_2 & 0 \cr u_3 & v_3 & 0}
= \det\pmatrix{u_2 & v_2 \cr u_3 & v_3}
=u_2 v_3 - u_3 v_2.
\ee
%thanks to the multi-linearity of~$\det_{|\ve}$.
Similar calculation for $(\vu\wedge_e \vv,\ve_2)_e$ and $(\vu\wedge_e \vv,\ve_3)_e$, thus
\be
\label{eqpv}
\vu\wedge_e \vv = \sum_{i=1}^3 
(u_{i+1} v_{i+2} - u_{i+2} v_{i+1})\,\ve_i, \qie
[\vu\wedge_e \vv]_{|\ve} = \pmatrix{
u_2 v_3 - u_3 v_2 \cr
u_3 v_1 - u_1 v_3 \cr
u_1 v_2 - u_2 v_1 \cr
}.
\ee
with the generic notation $w_4:=w_1$ and $w_5=w_2$.
(In particular $\ve_i\wedge_e \ve_{i+1} = \ve_{i+2}$.)

\debprop
1- $\vu\wedge_e \vv = -\vv\wedge_e \vu$.

2- $\vu\parallel \vv$ iff $\vu\wedge_e \vv = 0$.

3- If $\vu$ and $\vv$ are independent then
$\vu\wedge_e \vv$ is  orthogonal to the linear space $\Vect\{\vu,\vv\}$ % = \Ker\ell_{\ve,\vu,\vv}$
generated by~$\vu$ and~$\vv$.
% So $\vn_e = {(\vu\wedge \vv)_e \over ||(\vu\wedge \vv)_e||_e}$ est un vecteur $\dd_e$-orthogonal unitaire au plan engendré par~$\vu$ et~$\vv$.

4- $\vu\wedge_e \vv$ depends on the unit of measurement and on the orientation of~$(\ve_i)$:
If $\dd_a$ and $\dd_b$ are two Euclidean dot products, let $\lambda>0$ such that $\dd_a = \lambda^2\dd_b$, and then
\be
\label{eqpv2}
\vu\wedge_a \vv = \pm\lambda \vu\wedge_b \vv.
\ee
\finprop

\debdem
1- $\det_{|\ve}(\vu,\vv,\vz) = - \det_{|\ve}(\vv,\vu,\vz)$ (since $\det_{|\ve}$ is alternated).

2- If  $\vu\parallel \vv$ then
$ \det_{|\ve}(\vu,\vv,\vz)=0=(\vu\wedge_e \vv,\vz)_e$, so $\vu\wedge_e \vv \perp_g\vz$, for all $\vz$.
 And if $\vu\wedge_e \vv = 0$ then \eref{eqpv} gives $\vu\parallel \vv$.

3- If $\vz\in \Vect\{\vu,\vv\}$
then $\det_{|\ve}(\vu,\vv,\vz) = 0 = (\vu\wedge_e \vv,\vz)_g$ thus $\vu\wedge_e \vv \perp_g \vz$.

4- $\ds (\vu\wedge_a \vv,\vz)_a
\mathop{=}^{\eref{eqw1}} \det_{|\va}(\vu,\vv,\vz) 
\mathop{=}^{\eref{eqdetlambda}} \pm \lambda^3 \det_{|\vb}(\vu,\vv,\vz) 
\mathop{=}^{\eref{eqw1}} \pm \lambda^3 (\vu\wedge_b \vv,\vz)_b
= \pm \lambda^3{1\over \lambda^2}  (\vu\wedge_b \vv,\vz)_a$,
true for all $\vz$, thus~\eref{eqpv2}.
\findem

\debexe
Prove that $\vu\wedge_e \vv$ is a contravariant vector. % thanks to the change of basis contravariant formula.

\debrep
It is a vector (Riesz representation vector) in~$\vRRt$, so it is contravariant;
Or calculation: It satisfies the contravariance change of basis formula, see~\eref{eqpcbR}.
\finrep
\finexe

%%%%%%%%%%%%%%%%%%%%%%%%%%%%%%%%%%%%%%%%%%%%%%%%%%%%%%%%%%%%%%%%%%%%%%%%%%%%%%%%%%%

\subsubsection{Antisymmetric endomorphism represented by a vector}

\debprop
\label{propdefvomega}
Let $(\ve_i)$ be a chosen $\dd_g$-Euclidean basis.
If an endomorphism $\Omega \in \calL(\vRRt ;\vRRt)$ is $\dd_g$-antisymmetric then there exists a unique vector $\vomega_e\in\vRRt$ \st, for all $\vy,\vz \in\vRRt$,
\be
\label{eqom0}
(\Omega.\vy,\vz)_g = \det_{|\ve}(\vomega_e,\vy,\vz),
\ee
\ie, there exists a unique vector $\vomega_e\in\vRRt$ \st, for all $\vy,\vz \in\vRRt$,
\be
\label{eqdefvomega}
\boxed{\Omega.\vy = \vomega_e\wedge_e\vy},
\ee
And
\be
\label{eqom}
[\Omega]_{|\ve}=\pmatrix{0& -c & b \cr c & 0 & -a \cr -b & a & 0} \qiff
[\vomega_e]_{|\ve} = \pmatrix{a\cr b\cr c}.
\ee
In particular $\Omega.\vomega_e=\vec0$ ($= \vomega_e\wedge_e\vomega_e$),
\ie\ $\vomega_e$ is an eigenvector associated with the eigenvalue~$0$.
\finprop

\debdem
$\Omega$ is antisymmetric, thus $[\Omega]_{|\ve}$ is given as in~\eref{eqom}.
In particular $[\Omega.\ve_1]_{|\ve} = [\Omega]_{|\ve}.[\ve_1]_{|\ve} = \pmatrix{0 \cr c \cr -b }$.
Calculation of the components of~$\vomega_e$ if it exists: 
Let $\vomega = \omega_1\ve_1 +  \omega_2\ve_2 +  \omega_3\ve_3$;
%Then $\Omega.\vy = \vomega\wedge_e\vy$ for all~$\vy$ gives $[\Omega.\ve_1]_{|\ve} = [\Omega]_{|\ve}.[\ve_1]_{|\ve} = \pmatrix{0 \cr c \cr -b }$ and 
thus $[\vomega\wedge \ve_1]_{|\ve} = \pmatrix{0 \cr \omega_3 \cr -\omega_2}$, \cf~\eref{eqpv},
thus $\omega_3=c$ and $\omega_2=b$; Idem with $\ve_2$ so that $\omega_1=a$.
Thus if it exists $\vomega$ is unique. And $\vomega_e$ given in~\eref{eqom} satisfies~\eref{eqdefvomega}: It exists.
\findem

\debprop
\label{propvom}
Let $\dd_a$ and $\dd_b$ be two Euclidean dot products (\eg\ in foot and metre),
let $(\va_i)$ and $(\vb_i)$ be Euclidean associated bases,
let $||\vb_1||_a = \lambda$ (change of unit coefficient), so $\dd_a= \lambda_2\dd_b$).
Suppose $[\Omega]_{|\va}=\pmatrix{0& -c & b \cr c & 0 & -a \cr -b & a & 0}$, thus
$[\vomega_a]_{|\va} = \pmatrix{a\cr b\cr c}$, \cf~\eref{eqom}. Then
(change of representation vector for~$\Omega$):
\be
%\label{eqvomab1}
\eqalign{
\bullet\;&\hbox{If $(\vb_i)$ and $(\va_i)$ have the same orientation, then}\quad \vomega_b = \lambda\vomega_a, \cr
\bullet\;&\hbox{If $(\vb_i)$ and $(\va_i)$ have opposite orientation, then}\quad \vomega_b = -\lambda\vomega_a, \cr
}
\ee
\Eg, if $\vb_i = \lambda \va_i$ for all~$i$ (change of unit, same orientation) then
$\vomega_b = \lambda\vomega_a$, and 
if $\vb_1 = -\lambda \va_1$, $\vb_2 = \lambda \va_2$, $\vb_3 = \lambda \va_3$ (change of unit, opposite orientation) then $\vomega_b = -\lambda\vomega_a$.

\smallskip
%\noindent
{\bf NB:} 
%1- A~representation vector $\vomega$ of~$\Omega$ is \textsl{\textbf{not}} unique since it depends on the length and orientation of a chosen Euclidean basis. 2- 
The formula $\vomega_b = \pm\lambda\vomega_a$ is a change of vector formula,
\textsl{\textbf{not}} a change of basis formula. 
\finprop

\debdem
Apply~\eref{eqpv2}.
\comment{, or:

$\Omega = -c \; \va_1 \otimes a^2 + b\; \va_1 \otimes a^3 - a\; \va_2 \otimes a^3$ + antisym.

1- If $\vb_i = \lambda \va_i$ for all~$i$ then $b^i = {1\over \lambda} a^i$,
thus $\Omega = -c \; \vb_1 \otimes b^2 + b\; \vb_1 \otimes b^3 - a\; \vb_2 \otimes b^3$ + antisym.
Thus $\vomega_b = a\; \vb_1 + b\;\vb_2 + c\; \vb_3
= \lambda(a\; \va_1 + b\;\va_2 + c\; \va_3) = \lambda\vomega_a$.

2- If $\vb_1 = -\lambda \va_1$, $\vb_2 = \lambda \va_2$, $\vb_3 = \lambda \va_3$
then $b^1 = -{1\over \lambda} a^1$, $b^2 = {1\over \lambda} a^2$, $b^3 = {1\over \lambda} a^3$;
Thus $\Omega = c \; \vb_1 \otimes b^2 - b\; \vb_1 \otimes b^3 - a\; \vb_2 \otimes b^3$ + antisym.
Thus $\vomega_b = a\; \vb_1 - b\;\vb_2 - c\; \vb_3
= \lambda(-a\; \va_1 - b\;\va_2 - c\; \va_3) = -\lambda\vomega_b$.

3- 
$(\Omega.\vv,\vz)_b = \det_{|\vb}(\vomega_b,\vv,\vz)
= \pm\lambda^3 \det_{|\va}(\vomega_b,\vv,\vz)$
and $(\Omega.\vv,\vz)_b  = \lambda^2(\Omega.\vv,\vz)_a
=\lambda^2 \det_{|\va}(\vomega_a,\vv,\vz)$.
Thus
$\pm\lambda \det_{|\va}(\vomega_b,\vv,\vz) = \det_{|\va}(\vomega_a,\vv,\vz)$,
thus
$\pm\lambda \det([\vomega_b]_{|\va},[\vv]_{|\va},[\vz]_{|\va})
=\det([\vomega_a]_{|\va},[\vv]_{|\va},[\vz]_{|\va})$,
true for all $\vv,\vz$; Thus
$\pm\lambda [\vomega_b]_{|\va} = [\vomega_a]_{|\va}$,
thus $\lambda\vomega_b = \pm\vomega_a$, with $+$ for same orientation,
and with $-$ otherwise, \cf~\eref{eqdetlambda}.
}
\findem

\noindent
{\bf Interpretation of~$\vomega_e$:}
Suppose $[\Omega]_{|\ve}=\alpha\pmatrix{0& -1 & 0 \cr 1 & 0 & 0 \cr 0 & 0 & 0}$.
So $\Omega$ is the rotation with angle~${\pi\over 2}$ in the horizontal plane
composed with the dilation with ratio~$\alpha$.
And $[\vomega_e]_{|\ve} = \alpha\pmatrix{0\cr 0 \cr 1} = \alpha\,\ve_3$ is orthogonal to the horizontal plane and gives the rotation axis and the dilation coefficient.

\debexe
\label{eqinterpo}
Let $\Omega$ \st\
$[\Omega]_{|\ve}=\pmatrix{0& -c & b \cr c & 0 & -a \cr -b & a & 0}$
(see~\eref{eqom}).
Find a direct orthonormal basis~$(\vb_i)$ (relative to~$(\ve_i)$) \st\
$[\Omega]_{|\vb} = \sqrt{a^2{+}b^2{+}c^2}\pmatrix{0& -1 & 0 \cr 1 & 0 & 0 \cr 0 & 0 & 0}$.

\debrep
Let $\vb_3 = {\vomega_e \over ||\vomega_e||_e}$, that is,
$[\vb_3]_{|\ve} = {1 \over \sqrt{a^2{+}b^2{+}c^2}}\pmatrix{a\cr b\cr c}$.
Then choose $\vb_1 \perp \vb_3$, \eg\
$[\vb_1]_{|\ve} = {1 \over \sqrt{a^2{+}b^2}}\pmatrix{-b\cr a\cr 0}$.
Then choose $\vb_2 = \vb_3 \wedge_e \vb_1$,
that is, $[\vb_2]_{|\ve} = {1 \over \sqrt{a^2{+}b^2}}{1 \over \sqrt{a^2{+}b^2{+}c^2}}
\pmatrix{-ac\cr -bc\cr a^2+b^2}$.
Thus $(\vb_i)$ is a direct orthonormal basis, and the transition matrix from $(\ve_i)$ to~$(\vb_i)$
is $P=\pmatrix{[\vb_1]_{|\ve} & [\vb_2]_{|\ve} & [\vb_3]_{|\ve}}$.
And $[\Omega]_{|\vb} = P^{-1}.[\Omega]_{|\ve}.P$ (change of basis formula),
with $P^{-1} = P^T$ (change of orthonormal basis),
thus $[\Omega]_{|\vb} = P^T.[\Omega]_{|\ve}.P$

%that is, $P.[\Omega]_{|\vb} = [\Omega]_{|\ve}.P$.
With 
% $[\Omega]_{|\ve}.P = \pmatrix{([\Omega]_{|\ve}.[\vb_1]_{|\ve}) & ([\Omega]_{|\ve}.[\vb_2]_{|\ve}) & ([\Omega]_{|\ve}.[\vb_3]_{|\ve})}$, and
$[\Omega]_{|\ve}.[\vb_1]_{|\ve}
= {1 \over \sqrt{b^2{+}c^2}}\pmatrix{0& -c & b \cr c & 0 & -a \cr -b & a & 0}.\pmatrix{-b\cr a\cr 0}
={1 \over \sqrt{b^2{+}c^2}}\pmatrix{-ac \cr -bc \cr a^2+b^2}
=\sqrt{a^2{+}b^2{+}c^2}[\vb_2]_{|\ve}$ (expected),
$[\Omega]_{|\ve}.[\vb_2]_{|\ve}
= {1 \over \sqrt{b^2{+}c^2}}{1 \over \sqrt{a^2{+}b^2{+}c^2}}
\pmatrix{0& -c & b \cr c & 0 & -a \cr -b & a & 0}.\pmatrix{-ac\cr -bc\cr a^2+b^2}
= {1 \over \sqrt{b^2{+}c^2}}{1 \over \sqrt{a^2{+}b^2{+}c^2}}
\pmatrix{bc^2 + b(a^2+b^2) \cr -ac^2-a(a^2+b^2) \cr abc-abc}
=-\sqrt{a^2{+}b^2{+}c^2}[\vb_1]_{|\ve}$ (expected),
and
$[\Omega]_{|\ve}.[\vb_3]_{|\ve} = [\vec0]$ (expected since $\vb_3\parallel\vomega_e$).
Thus $[\Omega]_{|\ve}.P = \sqrt{a^2{+}b^2{+}c^2} \pmatrix{[\vb_2]_{|\ve} & -[\vb_1]_{|\ve} & [\vec0]_{|\ve}}$.
And $(P^T.[\Omega]_{|\ve}.P)_{ij} = [\vb_i]_{|\ve}^T.[\Omega]_{|\ve}.[\vb_j]_{|\ve}$
gives the result.
\comment{
Let $A = \sqrt{a^2{+}b^2{+}c^2}\pmatrix{0& -1 & 0 \cr 1 & 0 & 0 \cr 0 & 0 & 0}$;
Then $P.A = \sqrt{a^2{+}b^2{+}c^2}\pmatrix{(P.[\ve_2]_{|\ve}) & (P.[-\ve_1]_{|\ve}) & [\vec0]_{|\ve}}
= \sqrt{a^2{+}b^2{+}c^2}\pmatrix{[\vb_2]_{|\ve} & [-\vb_1]_{|\ve} & [\vec0]_{|\ve}}$,
Thus $P.A = [\Omega]_{|\ve}.P$,
thus $[\Omega]_{|\vb} = A$.
}
\finrep
\finexe

\comment{
\debrem
Let $(dx_i)$ be the (covariant) dual basis of~$(\ve_i)$.
Then with tensorial notation %(thanks to the natural canonical usual isomorphism)
\eref{eqom} reads
\be
\Omega =
- a (dx_2 \wedge dx_3)
- b (dx_3 \wedge dx_1)
- c (dx_1 \wedge dx_2)
 \qand
\vomega_e = a \ve_1 + b \ve_2 + c\ve_3.
\ee
where $dx_i \wedge dx_{i+1} := dx_i \otimes dx_{i+1} - dx_{i+1} \otimes dx_i$ (antisymmetric bilinear form).
\comment{
$\Omega = -c(dx_1 \otimes dx_2 - dx_2 \otimes dx_1)
+ b (dx_1 \otimes dx_3 - dx_3 \otimes dx_1)
-a (dx_2 \otimes dx_3 - dx_3 \otimes dx_2)$.
Let $dx_1 \wedge dx_2 \eqdef dx_1 \otimes dx_2 - dx_2 \otimes dx_1$ (usual notation),
idem for $dx_2 \wedge dx_3$ and $dx_3 \wedge dx_1$.
Then \eref{eqom} reads with tensorial notation:
\be
\Omega =
- a (dx_2 \wedge dx_3)
- b (dx_3 \wedge dx_1)
- c (dx_1 \wedge dx_2)
 \qand
\vomega_e = a \ve_1 + b \ve_2 + c\ve_3.
\ee
}
\comment{
NB: if $\Omega = {d\vv - d\vv^T \over 2}$, \cf~\eref{eqOmegarot},
that is, if $a= {\pa v^3 \over \pa y} - {\pa v^2 \over \pa z}$,
$b= {\pa v^1 \over \pa z} - {\pa v^3 \over \pa x}$,
$c= {\pa v^2 \over \pa x} - {\pa v^1 \over \pa y}$,
%(obtained by circular permutation),
then $\vomega$ is not (isometric) objective, see the velocity addition formula and exercise~\ref{exeOmnonO}.
}
\finrem
}

\comment{
\debexe
Prove, with the change of basis formulas, that $\vomega_e$ is contravariant  (change of basis contravariant formula).

\debrep
$\Omega_e$ is a vector in~$\vRRt$, so it is also called a contravariant vector. Change of basis:
Let $\calP_a$ be the change of basis endomorphism from a basis~$(\ve_i)$ to a basis~$(\va_i)$.
Let $\calP_b$ be the change of basis endomorphism from a basis~$(\ve_i)$ to a basis~$(\vb_i)$.
Thus $\vb_j=\calP_b.\calP_a^{-1}.\va_j$ for all~$j$, thus, $\calP = \calP_b.\calP_a^{-1}$
is the  change of basis endomorphism from a basis~$(\va_i)$ to a basis~$(\vb_i)$.

%that is, $\va_j=\calP_a.\ve_j$ for all~$j$, and $[\vx]_{|\ve} = [\calP_a]_{|\ve}[\vx]_{|\va}$ (contravariance).
We have, $\det_{|\ve}(\vomega_e,\vy,\vz)
=\det([\calP_a]_{|\ve})\det_{|\va}(\vomega_e,\vy,\vz)
=\det([\calP_a]_{|\ve})\det([\vomega_e]_{|\va},[\vy]_{|\va},[\vz]_{|\va})
$, idem
$=\det([\calP_b]_{|\ve})\det([\vomega_e]_{|\vb},[\vy]_{|\vb},[\vz]_{|\vb})$.
Thus
$\det([\vomega_e]_{|\va},[\vy]_{|\va},[\vz]_{|\va})
= \det([\calP_a]_{|\ve})^{-1}\det([\calP_b]_{|\ve})\det([\vomega_e]_{|\vb},[\vy]_{|\vb},[\vz]_{|\vb})
= \det([\calP]_{|\ve})\det([\vomega_e]_{|\vb},[\vy]_{|\vb},[\vz]_{|\vb})
= \det([\calP]_{|\ve}.[\vomega_e]_{|\vb},[\calP]_{|\ve}.[\vy]_{|\vb},[\calP]_{|\ve}.[\vz]_{|\vb})
= \det([\calP]_{|\ve}.[\vomega_e]_{|\vb},[\vy]_{|\va},[\vz]_{|\va})
$, for all $\vy,\vz$, thus $[\vomega_e]_{|\va} = [\calP]_{|\ve}.[\vomega_e]_{|\vb}$,
\ie\ $[\vomega_e]_{|\vb} = [\calP]_{|\ve}^{-1}.[\vomega_e]_{|\va}$ as expected.
\finrep
\finexe
}

\comment{
\debexe
Let $\dd_a$ and $\dd_b$ be two Euclidean dot products.
Recall: If $\Omega$ is an endomorphism then $\Omega^T_a=\Omega^T_b~\eqnamed\Omega^T=$ the Euclidean-transposed endomorphism, see~\eref{eqte}. 

Suppose: $\Omega$ is Euclidean-antisymmetric, \ie, $\Omega^T=-\Omega$ in our Euclidean framework.
Let $(\va_i)$ and $(\vb_i)$ be Euclidean associated bases relative to~$\dd_a$ and~$\dd_b$.
Let $\vomega_a$ and~$\vomega_b$ defined by $\Omega.\vy = \vomega_a\wedge_a\vy = \vomega_b\wedge_b\vy$
for all $\vy\in\vRRt$, \cf~\eref{eqdefvomega}.
Gives the change of vector relation between $\vomega_a$ and~$\vomega_b$.

\debrep
We have $(\Omega.\vy,\vz)_a = \det_{|\va}(\vomega_a,\vy,\vz)$ for all $\vy,\vz$, \cf~\eref{eqdefvomega}.
Let $\lambda$ \st\ $\dd_a = \lambda^2 \dd_b$.
Thus, with~\eref{eqdetlambda},
$\lambda^2(\Omega.\vy,\vz)_b = \pm\lambda^3\det_{|\vb}(\vomega_a,\vy,\vz)$ for all $\vy,\vz$,
Then let $\vomega_b$ be the vector representation of~$\Omega$ relative to~$\dd_b$,
that is, $(\Omega.\vy,\vz)_b = \det_{|\vb}(\vomega_b,\vy,\vz)$ for all $\vy,\vz$,
thus  $\lambda^2(\Omega.\vy,\vz)_b = \lambda^2\det_{|\vb}(\vomega_b,\vy,\vz)$ for all $\vy,\vz$,
thus $\pm\lambda^3\det_{|\vb}(\vomega_a,\vy,\vz) = \lambda^2\det_{|\vb}(\vomega_b,\vy,\vz)$ for all $\vy,\vz$,
thus $\pm\lambda\vomega_a = \vomega_b$.
\finrep
\finexe
}

%%%%%%%%%%%%%%%%%%%%%%%%%%%%%%%%%%%%%%%%%%%%%%%%%%%%%%%%%%%%%%%%%%%%%%%%%%%%%%%%%%%

\subsubsection{Curl}

\def\vcurl{\vec{\rm curl}}

\debdef
%Exclusively in~$\vRRt$.
If $\vv$ is a $C^1$ vector field,
if $(\ve_i)$ is a Euclidean basis in~$\vRRt$,
and if $\vv = \sum_{i=1}^3 v^i\ve_i$,
then the curl (or rotational) of $\vv$ relative to~$(\ve_i)$ is the vector field 
$\vcurl_{e}\vv = \vrot_{e}\vv$ given by
\be
\label{eqdefrot}
\vcurl_{e}\vv = \sum_{i=1}^3 ({\pa v_{i+2}\over\pa x_{i+1}}-{\pa v_{i+1}\over\pa x_{i+2}})\,\ve_i, \qie
[\vcurl_{e}\vv]_{|\ve} = \pmatrix{
{\pa v_3\over\pa x_2}-{\pa v_2\over\pa x_3}\cr
{\pa v_1\over\pa x_3}-{\pa v_3\over\pa x_1}\cr
{\pa v_2\over\pa x_1}-{\pa v_1\over\pa x_2}\cr
}.
\ee
\findef

\debprop
Let $\Omega(t,\pt) = {d\vv(t,\pt) - d\vv(t,\pt)^T \over 2}$,
and let $\vomega_e(t,\pt)$ be the associated vector relative to the Euclidean basis~$(\ve_i)$, \cf~\eref{eqdefvomega}.
Then
\be
\label{eqvomegarot}
\vomega_e=\demi\vcurl_e\vv.
\ee
%\ie\ , at $t$ at~$\pt$, $\vrot_e\vv(t,\pt) = 2\vomega_e(t,\pt)$.
\finprop

\debdem
\eref{eqOmegarot2} gives
$[\Omega]_{|\ve}=\demi\pmatrix{
0 & {\pa v_1\over\pa x_2}-{\pa v_2\over\pa x_1}
       & {\pa v_1\over\pa x_3}-{\pa v_3\over\pa x_1} \cr
\cdot & 0 & {\pa v_2\over\pa x_3}-{\pa v_3\over\pa x_2} \cr
\cdot & \cdot & 0\cr
}$, with $[\Omega]_{|\ve}$ antisymmetric.
Thus~\eref{eqom}, \eref{eqpv} and~\eref{eqdefrot} gives~\eref{eqvomegarot}.
\findem

%%%%%%%%%%%%%%%%%%%%%%%%%%%%%%%%%%%%%%%%%%%%%%%%%%%%%%%%%%%%%%%%%%%%%%%%%%%%%%%%%%%

\subsection{Pseudo-cross product, and pseudo-vector}
\label{secpv}

Framework: $\Mtu$ the space of $3*1$ matrices, so we leave the vector framework to enter the matrix world.

%%%%%%%%%%%%%%%%%%%%%%%%%%%%%%%%%%%%%%%%%%%%%%%%%%%%%%%%%%%%%%%%%%%%%%%%%%%%%%%%%%%

\subsubsection{Definition}

\debdef
A column matrice is also called a pseudo-vector, or a column vector.
\findef

\debdef
$\ds\pmatrix{x_1 \cr x_2 \cr x_3}\eqnote [\vx]$
and $\ds\pmatrix{y_1 \cr y_2 \cr y_3}\eqnote [\vy]$
being two matrices in~$\Mtu$, % = collection of real numbers.
their pseudo-cross product is
\be
\label{eqppv}
\pmatrix{x_1 \cr x_2 \cr x_3} \cwedge \pmatrix{y_1 \cr y_2 \cr y_3}
\eqdef \pmatrix{ x_2y_3 - x_3y_2 \cr x_3y_1 - x_1y_3 \cr x_1y_2 - x_2y_1}
\eqnote [\vx]\cwedge [\vy].
\ee
Thus the pseudo-cross product of two pseudo-vectors is a pseudo-vector (is a matrix).
\findef

%%%%%%%%%%%%%%%%%%%%%%%%%%%%%%%%%%%%%%%%%%%%%%%%%%%%%%%%%%%%%%%%%%%%%%%%%%%%%%%%%%%

\subsubsection{Antisymmetric matrix represented by a pseudo-vector}

\debdef
Let $A=[A_{ij}]= \pmatrix{0& -c & b \cr c & 0 & -a \cr -b & a & 0}$ be an antisymmetric matrix ($A_{ji} = -A_{ij}$ for all $i,j$).
The pseudo-vecteur $\cwr$ associated to~$A$ is the column matrix
$
\cwr := \pmatrix{a\cr b\cr c}.
$
So,
\be
\label{eqdefvomega2}
\boxed{A. [\vy] = \cwr\, \cwedge [\vy]},\qie
A. \pmatrix{y_1 \cr y_2 \cr y_3} = \cwr\, \cwedge \pmatrix{y_1 \cr y_2 \cr y_3},\quad
\hbox{for all matrix }
[\vy]=\pmatrix{y_1 \cr y_2 \cr y_3}.
\ee
\findef

%%%%%%%%%%%%%%%%%%%%%%%%%%%%%%%%%%%%%%%%%%%%%%%%%%%%%%%%%%%%%%%%%%%%%%%%%%%%%%%%%%%

\subsubsection{Antisymmetric endomorphism and its pseudo-vectors representations}

Let $\RRt$ be our usual affine space, $\dd_g$ be a Euclidean dot product, and $(\ve_i)$ be a $\dd_g$-Euclidean associated basis.
Let $\Omega$ be an antisymmetric endomorphism relative to~$\dd_g$, so $\Omega^T = - \Omega$, \cf~\eref{eqLant}.
Thus $[\Omega]_{|\ve}$ is an antisymmetric matrix. Call $\cwr$ the associated pseudo-vector, \ie, \cf~\eref{eqdefvomega2}, for all $\vy \in\vRRt$,
\be
\label{eqdefvomega3}
[\Omega]_{|\ve}.[\vy]_{|\ve} = \cwr\cwedge [\vy]_{|\ve} .
\ee
This formula is widely used in mechanics, and unfortunately sometimes noted $\Omega.\vy = \vomega\wedge \vy$ (!):

\medskip
%\noindent
{\bf Be careful:} \eref{eqdefvomega3} is \textbf{\textsl{not}} a vectorial formula;
This is just a formula for matrix calculations which gives false result if a change of basis is considered;
%(the usual change of basis formulas don't apply).
\Eg, with $(\va_1,\va_2,\va_3)$ be a $\dd_g$-Euclidean basis, and
$(\vb_1,\vb_2,\vb_3) = (-\va_1,\va_2,\va_3)$. So $(\vb_i)$ is also a $\dd_g$-Euclidean basis, but with a different orientation.

1- Vector approach:
Let $P$ be the transition matrix from $(\va_i)$ to~$(\vb_i)$,
so $P=\pmatrix{-1 & 0 & 0 \cr 0 & 1 & 0 \cr 0 & 0 & 1}$.
Let $[\Omega]_{|\va} = \pmatrix{0& -c & b \cr c & 0 & -a \cr -b & a & 0}$.
Thus, $\Omega$ being an endomorphism, the change of basis formula gives
\be
[\Omega]_{|\vb} = P^{-1}.[\Omega]_{|\va}.P
= \pmatrix{-1 & 0 & 0 \cr 0 & 1 & 0 \cr 0 & 0 & 1}
.\pmatrix{0& -c & b \cr c & 0 & -a \cr -b & a & 0}
.\pmatrix{-1 & 0 & 0 \cr 0 & 1 & 0 \cr 0 & 0 & 1}
= \pmatrix{0& c & -b \cr -c & 0 & -a \cr b & a & 0}.
\ee
Thus the vectors $\vomega_a$ and $\vomega_b$ are given by~\eref{eqom}:
\be
\label{eqpseu1}
[\vomega_a]_{|\va}=\pmatrix{a \cr b \cr c}, \quad
[\vomega_b]_{|\vb}=\pmatrix{a \cr -b \cr -c},
\qie
\left\{\eqalign{
& \vomega_a = a \va_1 + b \va_2 + c \va_3, \cr
& \vomega_b = a \vb_1 - b \vb_2 - c \vb_3,
}\right\}
\qthus \boxed{\vomega_b = -\vomega_a}.
\ee
%since $(\vb_1,\vb_2,\vb_3) = (-\va_1,\va_2,\va_3)$. (Calculation already done in the proof of prop.~\ref{propvom}: Change of vector).

2- Matrix approach \eref{eqdefvomega2} gives 
$[\Omega]_{|\va}. [\vy] = \cwr_a\, \cwedge [\vy]$ and
$[\Omega]_{|\vb}. [\vy] = \cwr_b\, \cwedge [\vy]$,
with
\be
\cwr_a = \pmatrix{a \cr b \cr c} \qand
\cwr_b = \pmatrix{a \cr -b \cr -c}, \qso \boxed{\cwr_a \ne -\cwr_b}.
\ee
And $\cwr$ does not represent a single vector either, since it does not satisfy the vector change
of basis formula $\cwr_b \ne P^{-1}.\cwr_a$. Thus $\cwr$ is not a vector (is not tensorial): It is just a matrix
(called a ``pseudo-vector'').

%%%%%%%%%%%%%%%%%%%%%%%%%%%%%%%%%%%%%%%%%%%%%%%%%%%%%%%%%%%%%%%%%%%%%%%%%%%%%%%%%%%

\subsection{Examples}

%%%%%%%%%%%%%%%%%%%%%%%%%%%%%%%%%%%%%%%%%%%%%%%%%%%%%%%%%%%%%%%%%%%%%%%%%%%%%%%%%%%

\subsubsection{Rectilinear motion}

Let $\tPhi :[t_1,t_2]\times \Obj\rar\RRn$ be a $C^1$ motion.
Let $\tz\in]t_1,t_2[$ and $\Pobj \in \Obj$.

\comment{
Then :
\be
\tPhiPobj(t) = \tPhiPobj(\tz) + (t{-}\tz)\,\tPhiPobj{}'(\tz) + o(t{-}\tz).
\ee
}

\debdef
The motion of~$\Pobj$ is rectilinear iff, for all $\tz,t\in[t_1,t_2]$,
\be
{\tPhiPobj(t) - \tPhiPobj(\tz) \over t{-}\tz} \parallel \tPhiPobj{}'(\tz).
\ee
And the motion is rectilinear uniform iff, for all $\tz,t\in[t_1,t_2]$,
\be
\tPhiPobj(t) = \tPhiPobj(\tz) + (t{-}\tz)\,\tPhiPobj{}'(\tz) ,
\qie
p(t) = p(\tz) +  (t{-}\tz)\,\vVtz(\tz,p(\tz))
\ee
when $p(t) = \tPhi(t,\Pobj)$,
that is, the trajectory is traveled at constant velocity.
\findef

%%%%%%%%%%%%%%%%%%%%%%%%%%%%%%%%%%%%%%%%%%%%%%%%%%%%%%%%%%%%%%%%%%%%%%%%%%%%%%%%%%%

\subsubsection{Circular motion}

Let $(\vE_1,\vE_2)$ be a Euclidean basis.
Let $\tz\in[t_1,t_2]$.
A motion $\Phitz$ is a circular motion iff
\be
\ora{\calO \PhitzP(t)} = x(t) \vE_1 + y(t) \vE_2,\quad
[\ora{\calO \PhitzP(t)}]_{|\vE}
=\pmatrix{x(t)=a+R\cos (\theta(t)) \cr 
y(t)=b+R\sin(\theta(t))},
%\eqnote \pmatrix{a \cr b} + [\vphi^\tz_P(t)]_{|\vE},
\ee
for some $R>0$ (called the radius), some $a,b\in\RR$,
and some function $\theta : \RR\rar \RR$.
And $\pmatrix{a \cr b}=\calO_C\in\RR^2$ is the center of the circle
and $\theta(t)$ is the angle at~$t$.
And the particle $\Pobj$ (\st\ $\tPhi(\tz,\Pobj)=P$) stays on the circle with center $\calO_C$ and radius $R$.

The circular motion is uniforme iff, for all~$t$, $\theta''(t)=0$,
that is, $\exists\omega_0\in\RR$, $\forall t\in[t_1,t_2]$,  $\theta(t) = \omega_0 t$.

\medskip
Notation:
\be
\vphi^\tz_P(t) = R\cos (\theta(t) \vE_1 + R\sin(\theta(t)) \vE_2,
\qso  [\vphi^\tz_P(t)]_{|\vE} = \pmatrix{R\cos (\theta(t)) \cr 
R\sin(\theta(t))}.
\ee
Thus the Lagrangian velocity of a circular motion is
\be
\vVtz_P(t) = (\Phitzt)'(t) = (\vphi^\tz_P)'(t),\qso
[\vVtz_P(t)]_{|\vE}= R\theta'(t)\pmatrix{-\sin(\theta(t)) \cr \cos(\theta(t))},
\ee
and $\vVtz_P(t)$ is orthogonal to $\vphi^\tz_P(t)$ (the radius vector).
And the Lagrangian acceleration is
\be
\vGammatz_P(t)=R\theta''(t)\pmatrix{-\sin(\theta(t)) \cr \cos(\theta(t))}
   + R(\theta'(t))^2\pmatrix{-\cos(\theta(t)) \cr -\sin(\theta(t))}.
\ee

Consider
\be
\ve_r(t)={\vphi^\tz_P(t)\over||\vphi^\tz_P(t)||}=\pmatrix{\cos (\theta(t)) \cr\sin(\theta(t))},
\qand
\ve_\theta(t)=\pmatrix{-\sin (\theta(t)) \cr\cos(\theta(t))},
\ee
thus $(\ve_r(t),\ve_\theta(t))$ is an orthonormal basis.
%  and $(p(t),(\ve_r(t),\ve_\theta(t)))$ is the Frénet frame relative to the motion.
Then :
\be
\label{eqvpdr3}
\vV^\tz_P(t) = R\theta'(t) \, \ve_\theta(t),
\ee
and :
\be
\label{eqvAP2}
\vGammatz_P(t) = -R(\theta'(t))^2 \, \ve_r(t) + R\theta''(t) \, \ve_\theta(t).
\ee
%En particulier, pour un mouvement circulaire uniforme $\vA(t) = - \omega^2(t)R \, \ve_r(t)$ (centripète).

\medskip
\noindent
\Eg, in~$\RR^3$ and a motion in he ``horizontal'' plane given by $(\ve_1,\ve_2)$, the vertical line being given by~$\vE_3$.
%and the vectors $\vE_1=(1,0,0)$ and $\vE_2=(0,1,0)$ are seen as $\vE_1=(1,0)$ and $\vE_2=(0,1)$ in the horizontal plane. And let $\vE_3=(0,0,1)$.
Here
\be
\label{eqpdr3tilde}
\vV^\tz_P(t)=\vomega(t)\wedge\vphi^\tz_P(t),\qwhere
    \vomega(t)=\omega(t)\ve_3 \qand \omega(t)=\theta'(t).
\ee
%Ainsi le vecteur $\vomega$ donne l'axe de rotation (ici ``vertical'') et $\omega(t)$ la vitesse de rotation (la dimension de~$||\vomega(t)||$ est celle d'un (temps)$^{-1}$).
And
\be
\vGammatz(t)= {d\vomega\over dt}(t)\wedge\vphi^\tz_P(t) + \vomega(t)\wedge\vV^\tz_P(t)
\qquad        (  = R{d\omega\over dt}(t) \, \ve_\theta(t) - \omega^2(t)R \, \ve_r(t)).
\ee

%%%%%%%%%%%%%%%%%%%%%%%%%%%%%%%%%%%%%%%%%%%%%%%%%%%%%%%%%%%%%%%%%%%%%%%%%%%%%%%%%%%

\subsubsection{Motion of a planet (centripetal acceleration)}

Illustration: $\Obj$ is \eg\ a planet from the solar system.

Let $(\ve_1,\ve_2,\ve_3)$ be a Euclidean basis (\eg\ fixed relative to stars an $(\ve_1,\ve_2)$
define the ecliptic plane), $\dd_g$ be the Euclidean associated dot product,
$||.||$ the Euclidean associated norm, $\calO$ an origin in~$\RR^3$ (\eg\ the center of the Sun),
and $\calR = (\calO,(\ve_i))$.

Consider a motion $\tPhi$ of~$\Obj$ in~$\calR$, \cf~\eref{eqdeftPhi0}.
Let $\tz\in[t_1,t_1]$, and consider $\Phitz \eqnote \Phi$ or $\vphitz\eqnote\vphi$, \cf~\eref{eqdefPhi}-\eref{eqdefPhiv}.

\debdef
\label{defca}
The motion of a particle $\Pobj$ is a centripetal acceleration motion iff
the particle is not static and, at all time, its acceleration vector $\vA(t)$
points to a fixed point~$F$ (focus).
\findef

We will take the focus $F$ as the origin of the referential, that is, $\calO:=F$.

\medskip

Thus, for all $t\in[t_1,t_2]$, $\ora{\calO  \Phi_P(t)} \parallel \vA_P(t)$, that is,
\be
\label{eqmaaccc}
\ora{\calO  \Phi_P(t)} \wedge \vA_P(t)=\vec0.
\ee

\debrem
A rectilinear motion is a centripetal acceleration motion,
but such a motion is usually excluded in the definition~\ref{defca}.
\finrem

\debexa
The motion of a planet from the solar system is a centripetal acceleration motion:
An elliptical motion of focus the center of the Sun.
\finexa

\debexa
The second Newton's law of motion $\sum \vf = m \vgamma$ (Galilean referential) gives:
If $\sum \vf$ is, at all time, directed to a unique point~$F$,
then the motion is a centripetal acceleration motion.
\finexa

Let $\Phi$ be a centripetal acceleration motion,
let $\calO$ be the focus, and let $\vphi_P(t) := \ora{\calO  \Phi_P(t)}$.
So the Lagrangian velocity and acceleration are
\be
\vV_P(t)={d\Phi_P\over dt}(t) = {d\vphi_P\over dt}(t), \qand \vA_P(t)={d^2\Phi_P\over dt^2}(t)  = {d^2\vphi_P\over dt^2}(t),
\ee
and $\vphi_P(t) \wedge \vA_P(t) =\vec0$, \cf~\eref{eqmaaccc}.

\debdef
The areolar velocity at~$t$ is the vector
\be
\label{eqdefvaero}
\vZ(t)=\demi \vphi_P(t)\wedge\vV_P(t).
\ee
\findef

\debprop
\label{propca}
If $\Phi$ is a centripetal acceleration motion, then the areolar velocity is contant,
that is, ${d\vZ\over dt}(t)=\vec0$ pour tout~$t$, so
\be
\label{eqvaeroc}
\vZ(t)=\vZ(\tz) ,\quad\forall t.
\ee
That is, the position vectors sweep equal areas in equal times.
And $\vZ(t_0) = \vec0$ iff $\Phi$ is a rectilinear motion.

If $\vZ(t_0)\ne\vec0$ then :

- $\vphi_P(t)$ and $\vV_P(t)$ are orthogonal to~$\vZ(t_0)$ at all time~$t$,

- The motion of the particle~$\Pobj$ takes place in the affine plane orthogonal to~$\vZ(t_0)$ passing through~$O$.

- $\vV_P(t)$ never vanishes.
\finprop

\debdem
\eref{eqdefvaero} and~\eref{eqmaaccc} give
$2{d\vZ\over dt}(t)={d\vphi_P\over dt}(t)\wedge\vV_P(t)+ \vphi(t)\wedge {d\vV_P\over dt}(t)
=\vV_P(t)\wedge\vV_P(t) + \vphi(t)\wedge\vA_P(t)=\vec0+\vec0$. Thus $\vZ$ is constant, $\vZ(t)=\vZ(t_0)$ for all~$t$.
Then, if $\vZ(\tz)\ne \vec0$ then $\vZ(t)\ne \vec0$ pour tout~$t$, and

$\bullet$ $\vZ(t) = \demi \vphi_P(t)\wedge\vV_P(t)$ gives that $\vphi_P(t)$ et $\vV_P(t)$ are orthogonal to~$\vZ(t_0)$ for all~$t$, thus $\vA_P(t)$ is orthogonal to~$\vZ(t_0)$, \cf~\eref{eqmaaccc}.

$\bullet$ The Taylor expansion reads
$\vphi_P(t) = \vphi_P(t_0)+ \vV_P(t_0)(t{-}t_0) + \int_{\tau=t_0}^t \vA_P(\tau)(t{-}\tau)^2\,d\tau$, 
with $\vV_P(t_0)$ and $\vA_P(\tau)$  $\perp \vZ(t_0)$ for all~$\tau$,
thus $\vphi_P(t) - \vphi_P(t_0)\perp \vZ(t_0)$ for all~$\tau$,
that is $\ora{\calO p(t)} - \ora{\calO P} = \ora{Pp(t)}\perp \vZ(t_0)$ for all~$\tau$,
Thus $p(t)$ belongs to the affine plane containing~$P$ orthogonal to~$\vZ(t_0)$, for all~$t$.
And $\ora{\calO P} = \vphi_P(t_0) \perp \vZ(t_0)$, thus $O$ belong to the same plane.

$\bullet$ $\vZ(t)=\vZ(t_0)\ne \vec0$ implies $\vV_P(t)\ne \vec 0$ for all~$t$,
and~\eref{eqdefvaero} gives: $(\vphi_P(t),\vV_P(t),\vZ(t_0))$ is a positively-oriented basis.
Since $\vphi_P$ and $\vV$ are continuous and do not vanish, since $\vZ(t_0)\ne \vec0$,
we get: $\Pobj$ ``turns around $\vZ(t_0)$'' and keeps its direction.

If $\vZ(t) = \vec0$ then $\vphi_P(t) \parallel \vV_P(t)$ for all~$t$, \cf~\eref{eqdefvaero},
so $\vV_P(t) = f(t)\vphi_P(t)$ where $f$ is some scalar function.
And $\vV_P(t) = \vphi_P{}'(t)$ gives $\vphi_P{}'(t) = f(t)\vphi_P(t)$,
thus $\vphi_P(t) = \vphi_P(t_0)e^{F(t)}$ where $F$ is a primitive of~$f$ \st\ $F(\tz)=0$, thus
$\vphi_P(t) \parallel \vphi_P(t_0)$, so
$\ora{\calO \Phi_P(t)} \parallel \ora{\calO \Phi_P(\tz)}$, for all~$t$: The motion is rectilinear.
\findem

\noindent
{\bf Interpretation.} (Non rectilinear motion.)
\comment{
La particule $\Pobj$ tournant autour du point~$O$
dans le plan $(\calO ,\Vect\{\vZ(t_0)\}^\perp)$,
considérons ``l'aire balayée par~$\Pobj$'' pendant un intervalle de temps $[t_0,t_1]$ :
par définition c'est l'aire limité par les segments de droite $[O,P]$ et $[0,p(t)]$
et la courbe $\Phi_P([t_0,t_1])$, faire un dessin.
}
The area swept by $\vphi_P(t)$ is, at first order,
the area of the triangle whose sides are
$\vphi_P(t)$ and $\vphi_P(t+\tau)$ (``anglular sector'').
So, with $\tau$ close to~$0$, let
\be
\vS_t(\tau) = \demi \vphi_P(t) \wedge \vphi_P(t+\tau), \qand
S_t(\tau) = ||\vS_t(\tau)||,
\ee
the vectorial an scalar area.
With $\vphi_P(t{+}\tau) = \vphi_P(t) + \vV_P(t)\tau + o(\tau)$ we get
\be
\vS_t(\tau) = \demi \vphi_P(t) \wedge (\vV_P(t)\tau + o(\tau)),
\ee
Since $\vS_t(0)=0$ we get
${\vS_t(\tau) - \vS(0) \over \tau} = \demi \vphi_P(t) \wedge \vV_P(t) + o(1)$, then
\be
{d\vS_t \over d\tau}(0) = \demi \vphi_P(t) \wedge \vV_P(t) = \vZ(t) = \vZ(t_0),
\ee
thanks to~\eref{eqvaeroc}, thus
%${d\vS_t \over d\tau}(0) = {d\vS_{t_0} \over d\tau}(0)$ pour tout $t\in[t_0,T]$, donc :
\be
{d\vS_t \over d\tau}(0) = {d\vS_{t_0} \over d\tau}(0),\qquad \forall t\in[t_0,T],
\ee
that is, the rate of variation of $\vS_t$ is constant.
And with $||\vS_t(\Delta\tau)||^2 = (\vS_t(\Delta\tau),\vS_t(\Delta\tau))$ we get
\be
{d ||\vS_t||^2 \over d\tau}(\Delta\tau) = 2 ({d\vS_t \over d\tau}(\Delta\tau),\vS_t(\Delta\tau)),
\ee
so, since $\vS_t(0)=0$,
\be
{d ||\vS_t||^2 \over d\tau}(0) = 0.
\ee
Therefore the function $t\rar ||\vS_t(0)||^2 = S_t(0)^2$ is constant,
thus $t\rar S_t(0)$ est constant, and
${d S_t \over d\tau}(0)$ is constant.

\debexe
Give a parametrization of the swept area, and redo the calculations.

\debrep
Let
\be
\label{eqvermc}
  r(t)=||\vphi_P(t)||,\quad \theta(t)=\widehat{p(t)OP} \quad \hbox{(angle)},
\ee
then
\be
\vphi_P(t) = \pmatrix{r(t) \cos(\theta(t)) \cr r(t) \sin(\theta(t)) \cr 0}.
\ee
Thus
\be
\vV_P(t)= \pmatrix{
r'(t) \cos(\theta(t) - r(t))\theta'(t)\sin(\theta(t)) \cr
r'(t) \sin(\theta(t) + r(t))\theta'(t)\cos(\theta(t)) \cr 0}.
\ee
With~\eref{eqdefvaero} we get
\be
\label{eqvaeroc2}
\vZ(t)= \demi\pmatrix{0 \cr 0 \cr r^2(t)\theta'(t)},\qwith
r^2(t)\theta'(t) =  r^2(t_0)\theta'(t_0)\quad\hbox{(constant)},
\ee
\cf~\eref{eqvaeroc}.
A parametrization of the swept area is then
\be
\vcalA :
\left\{\eqalign{
[0,1]\times [t_0,T] & \rar  \RR^3 \cr
(\rho,t) & \rar \vcalA(\rho,t) \cr
}\right\}, \qquad
\vcalA(\rho,t) = \pmatrix{\rho\, r(t) \cos(\theta(t)) \cr \rho\, r(t) \sin(\theta(t)) \cr 0}.
\ee
Therefore, the tangent associated vectors are
\be
{\pa \vcalA \over \pa \rho}(\rho,t)
= \pmatrix{r(t) \cos(\theta(t)) \cr r(t) \sin(\theta(t)) \cr 0},\qquad
{\pa \vcalA \over \pa t}(\rho,t)
= \pmatrix{\rho r'(t) \cos(\theta(t) - \rho r(t))\theta'(t)\sin(\theta(t))
 \cr \rho r'(t) \sin(\theta(t) + \rho r(t))\theta'(t)\cos(\theta(t)) \cr 0},
\ee
hence the vectorial and scalare element areas are
\be
d\vec\sigma = ({\pa \vcalA \over \pa \rho}\wedge {\pa \vcalA \over \pa t}) d\rho dt=
\pmatrix{0\cr0\cr
\rho r^2\theta'\, d\rho dt
},\qquad
d\sigma = 
\rho r^2\theta'\, d\rho d\theta.
\ee
Therefore the area between $t_0$ and $t$ is
\be
\calA(t) = \calA(t_0) + \int_{\rho=0}^1 \int_{\tau=t_0}^{t} \rho r^2(\tau)\theta'(\tau)\, d\rho d\tau
= \demi  \int_{\tau=t_0}^{t} r(\tau)^2\theta'(\tau)\,d\tau.
\ee
Hence
\be
\label{eqloidesaires}
\calA'(t) = r(t)^2\theta'(t) = r(t_0)^2\theta'(t_0) \quad (= \hbox{constant} = ||\vZ(\tz)||),
\ee
\cf~\eref{eqvaeroc2}.
\finrep
\finexe

\debexe
Prove the Binet formulas (non rectilinear central motion):
\be
   V_P(t)^2=Z_0^2\Bigl({1\over r^2} + ({d{1\over r}\over d\theta}\, )^2\Bigr)(t),\qquad
   \vGamma_P(t)= -{Z_0^2\over r^2}
       \Bigl({1\over r} +{d^2{1\over r}\over d\theta^2}\Bigr)(t)\,\ve_r(t),
\ee
for the energy and the acceleration.

\debrep
Proposition~\ref{propca} tells that $\Phi$ is a planar motion.
With~\eref{eqvermc} and $\ve_r(t)=\pmatrix{\cos(\theta(t)) \cr \sin(\theta(t))}$
we have $\vphi(t)=r(t)\ve_r(t)$ (in the plane).
Let $\ve_\theta(t) = \pmatrix{-\sin(\theta(t)) \cr \cos(\theta(t))} $, thus
$$
  \vV(t)={dr\over dt}(t)\ve_r(t) + r(t){d\ve_r\over dt}(t)=r'(t)\ve_r(t) + r(t)\theta'(t)\ve_\theta(t).
$$
And $\ve_r(t) \perp \ve_\theta(t)$ gives
$$
  V^2(t)=(r'(t))^2 + (r(t)\theta'(t))^2.
$$
Since $\theta'(t)\ne0$ for all~$t$ (non rectilinear central motion)
Let $s(\theta(t))=r(t)$.
Let us suppose that $\theta$ is~$C^1$, thus $\theta'>0$ or $\theta'<0$,
and $\theta : t\rar\theta(t)$ defines a change of variable. And
$$
   r'(t)= s'(\theta(t))\theta'(t). 
$$
And~\eref{eqloidesaires} and $\theta'(t)={Z_0\over r^2(t)}$ give % $\theta(t)\eqnote \theta$ :
$$
  V^2(t(\theta)) = (s'(\theta))^2{Z_0^2\over r^4(t)} + r^2(t){Z_0^2\over r^4(t)}
     = Z_0^2( {(s'(\theta))^2\over s^4(\theta)}+ {1\over s^2(\theta)} )
  =Z_0^2[\Bigl( {d{1\over s}\over d\theta}(\theta)\Bigr)^2+ {1\over s^2(\theta)} ].
$$
Thus $r(t)=s(\theta)$ and ${dr\over d\theta}\eqdef{ds\over d\theta}$
give the first Binet formula. Then
$$
  \vGamma(t)=r''(t)\ve_r(t)+r'(t){d\ve_r\over dt}(t)+(r'(t)\theta'(t)+r(t)\theta''(t))\ve_\theta(t)
             +r(t)\theta'(t){ d\ve_\theta\over dt}(t),
$$
with $\ds {d\ve_r\over dt} \parallel \ve_\theta$,
and $\ds {d\ve_\theta\over dt}(t)=-\theta'(t)\ve_r(t)$,
and  $\ve_\theta \perp \vGamma$ (central motion), we get
$$
  \vGamma(t)=(r''(t) - r(t)(\theta'(t))^2)\ve_r(t).
$$
And
$$
  r'(t)=s'(\theta)\theta'(t)=s'(\theta){Z_0\over r^2(t)} = Z_0{s'(\theta)\over s^2(\theta)}
 = -Z_0 {d {1\over s}\over d\theta}(\theta),
$$
thus
$$
  r''(t)=-Z_0{d^2{1\over s}\over d\theta^2}(\theta)\,\theta'(t)
       =-{Z_0^2\over r^2(t)}{d^2{1\over s}\over d\theta^2}(\theta),
$$
which is the second Binet formula.
\finrep
\finexe

%%%%%%%%%%%%%%%%%%%%%%%%%%%%%%%%%%%%%%%%%%%%%%%%%%%%%%%%%%%%%%%%%%%%%%%%%%%%%%%%%%%

\section{Riesz representation theorem}
\label{secRRT}

\def\DP{{\cal G}}

%%%%%%%%%%%%%%%%%%%%%%%%%%%%%%%%%%%%%%%%%%%%%%%%%%%%%%%%%%%%%%%%%%%%%%%%%%%%%%%%%%%

\subsection{The Riesz representation theorem}
\label{ssecRRT}

Framework: $(E,\dd_g)$ is Hilbert space, \ie\ $E$ is a vector space equipped with an inner dot product~$\dd_g$ such that,
with the associated norm defined by$||\vv||_g:=\sqrt{(\vv,\vv)_g}$, $(E,||.||_g)$ is a complete space. And $\Es = \calL(E;\RR)$ is the space of the linear and continuous forms on~$E$ (the space of linear ``measuring tools'') equipped with its norm $\ds ||\ell||_\Es := \sup_{||\vx||_g=1}|\ell.\vx| <\infty$.

$\bullet$ We have the easy statement:
\be
\label{eqrtrr}
\forall \vv\in E \hbox{ (vector),}\;\; 
\exists ! v_g \in E^* \hbox{ (linear continuous form)}\qst  v_g.\vx = (\vv,\vx)_g,\;\;\forall \vx \in E,
\ee
and $||v_g||_\Es = ||\vv||_g$.
(Usual notation in finite dimension: $v_g.\vx = \vv \bcdotg \vx$, or simply $v.\vx = \vv \bcdot \vx$ if a chosen~$\dd_g$ is imposed to all observers.)

Indeed: Define $v_g:E\rar\RR$ by $v_g(\vx) = (\vv,\vx)_g$ for all $\vx \in E$; The definition domain of~$v_g$ is~$E$ and $v_g$ is trivially linear; And %(to prove the continuity) 
the Cauchy--Schwarz inequality gives
$|v_g(\vx)| = |(\vv,\vx)_g| \le  ||\vv||_g\,||\vx||_g$ for all $\vx\in E$, thus $||v_g||_\Es \le ||\vv||_g<\infty$, thus $v_g$ is continuous;
And $|v_g(\vv)| = |(\vv,\vv)_g| = ||\vv||_g\,||\vv||_g$, thus $||v_g||_\Es \ge ||\vv||_g$, thus $||v_g||_\Es = ||\vv||_g$.

\medskip
$\bullet$ The Riesz representation theorem concerns the converse: If you choose an inner dot product~$\dd_g$ in~$E$ (\eg\ English of French), then you can represent a ``measuring instrument'' $\ell\in E^*$ by a vector $\vell_g\in E$:

\begin{theorem}[Riesz representation theorem, and definition]
\label{thmRiesz}
$(E,\dd_g)$ being a Hilbert space,
\be
\label{eqrtr}
\forall \ell\in\Es \hbox{ (linear continuous form),} \;\;
\exists ! \vell_g \in E \hbox{ (vector)}\qst \ell.\vx = (\vell_g,\vx)_g,\;\;\forall \vx \in E,
\ee
and $||\vell_g||_g = ||\ell||_\Es$.
And $\vell_g$ is called the $\dd_g$-Riesz representation vector of~$\ell$ (depends on~$g$).

(Usual notation in finite dimension: $v_g.\vx = \vv \bcdotg \vx$, or simply $v.\vx = \vv \bcdot \vx$ if a chosen~$\dd_g$ is imposed to all observers.) 
\finthm

\debdem
Easy in finite dimension: With a basis~$(\ve_i)$, if $[\ell]_{|\ve}=\pmatrix{\ell_1 & \dots & \ell_n}$ (row matrix since $\ell$ is a linear form) then \eref{eqrtr} gives $[\ell]_{|\ve}.[\vx]_{|\ve} = [\vell_g]_{\ve}^T.[g]_{|\ve}.[\vx]_{|\ve}$, thus $[\vell_g]_{\ve} = [g]_{|\ve}^{-1}.[\ell]_{|\ve}^T$ (column matrix), thus $\vell_g$. 

General case (\eg\ with $E=\Ldo$ and the finite element method):
If $\ell=0$ then $\vell_g=\vec0$ (trivial).
Suppose $\ell\ne0$: Thus $\Ker\ell = \ell^{-1}(\{0\})\ne\{\vec0\}$ (the kernel).
% $\ell$ being a linear form, co$\dim(\Ker\ell)=1$ and $\dim(\Ker\ell)^\perp=1$.
If $\dim E=1$, it is trivial (exercise). Suppose $\dim E\ge 2$.
Since $\ell$ is continuous, its kernel $\Ker\ell = \ell^{-1}(\{0\})$ is closed
in~$E$.
Thus, if $\vx\in E$, then its  $\dd_g$-orthogonal projection $\vx_0\in\Ker\ell$ on~$\Ker\ell$ exists, is unique, and is given by:
$
%\vx_0\in\Ker\ell\quad\hbox{and},\quad 
\forall \vy_0\in\Ker\ell,\;  (\vx - \vx_0 , \vy_0)_g = 0.
$
(So $\vx - \vx_0\perp_g\Ker\ell$.)
Choose a $\vx\notin \Ker\ell$ (possible since $\ell\ne0$), and let $\vn := {\vx - \vx_0 \over ||\vx - \vx_0||_g}$;
So $\vn$ is a $\dd_g$-orthonormal vector to~$\Ker\ell$, and $(\Ker\ell)^\perp = \Vect\{\vn\}$
since $\dim(\Ker\ell)^\perp=1$
(in finite dimension \cf\ the Dimension Formula which states that the dimension of the domain of a linear map is the sum of the dimension of its range and the dimension of its kernel, and in infinite dimension see next exercise~\ref{exeRe}).
And $E = \Ker \ell \oplus (\Ker\ell)^\perp$ since both vector spaces are closed (an orthogonal is always closed in a Hilbert space).
Thus if $\vx\in E$ then $\vx = \vx_0+\lambda \vn \in\Ker \ell \oplus (\Ker\ell)^\perp$;
%, where $\vx_0 \in \Ker\ell$ and $\lambda\in\RR$;
Thus $(\vx,\vn)_g = \lambda$ and $\ell(\vx) = 0+\lambda\ell(\vn) = (\vx,\vn)_g \ell(\vn) = (\vx,\ell(\vn)\vn)_g $ (bilinearity of~$\dd_g$);
Thus $\vell_g := \ell(\vn)\vn$ satisfies~\eref{eqrtr}.
And if $\vell_{g1}$ and $\vell_{g2}$ satisfy~\eref{eqrtr}
then $(\vell_{g1}-\vell_{g2},\vx)_g = 0$ for all $\vx\in E$, thus $\vell_{g1}-\vell_{g2}=0$.
Thus $\vell_g$ is unique.

And %$||\ell||_\Es=\sup_{||\vx||_g=1}|\ell(\vx)|$ and 
the Cauchy--Schwarz theorem give $||\ell||_\Es:=\sup_{||\vx||_g=1}|\ell(\vx)| =\sup_{||\vx||_g=1}|(\vell_g,\vx)_g| = ||\vell_g||_g $.
% (Remark: $\vell_g$ depends on~$\dd_g$, as well as $||\vell_g||_g=||\ell||_\Es$: the norm $||\ell||_\Es:=\sup_{\vx\in E}|\ell({\vx\over ||\vx||_g})|$ depends on the norm $||.||_g$ in the Hilbert space $(E,\dd_g)$.)

$\vR_g$ is an isomorphism between Banach spaces:
linearity since 
$(\vR_g(\ell+\lambda m),\vx)_g 
= (\ell+\lambda m).\vx 
= \ell.\vx + \lambda m.\vx 
= (\vR_g(\ell),\vx)_g + \lambda (\vR_g(m),\vx)_g
= (\vR_g(\ell) + \lambda \vR_g(m),\vx)_g$ for all~$\vx$ gives $\vR_g(\ell+\lambda m) = \vR_g(\ell) + \lambda \vR_g(m)$,
bijectivity thanks to~\eref{eqrtrr} and~\eref{eqrtr}, and the norm is kept since $||\vell_g||_g = ||\ell||_\Es$.
\findem

\debexe
\label{exeRe}
Prove: If $\ell\in E^*{-}\{0\}$ then
$\dim(\Ker\ell)^\perp=1$ ($=\dim(\Im(\ell))=\dim\RR$).

\debrep
Consider the restriction
$\ell_{|\Ker\ell^\perp} :
\left\{\eqalign{
(\Ker\ell)^\perp & \rar \RR \cr
 \vx & \rar \ell_{|\Ker\ell^\perp}.\vx=\ell.\vx
}\right\}
$.
It is linear (since $\ell$~is), and thus one to one: Indeed it is onto since $\ell\ne0$, and it is one to one since 
if %$\vx\in \Ker\ell^\perp$ is \st\ 
$\ell_{|\Ker\ell^\perp}(\vx)=0=\ell(\vx)$ then $\vx\in(\Ker\ell)^\perp \bigcap \Ker\ell = \{\vec0\}$, thus $\vx=0$.
% $\Ker(\ell_{|\Ker\ell^\perp})=\{\vec0\}$ ().
Thus $\dim (\Ker\ell)^\perp \le \dim(\Im(\ell)) = 1$: Indeed, if $\vz_1,\vz_2\in(\Ker\ell)^\perp{-}\{\vec0\}$ then $\ell_{|\Ker\ell^\perp}(\vz_1)\in\RR$ and $\ell_{|\Ker\ell^\perp}(\vz_2) \in \RR$, thus
$\exists\lambda\in\RR$ \st\
$\ell_{|\Ker\ell^\perp}(\vz_2)=\lambda\ell_{|\Ker\ell^\perp}(\vz_1)$,
thus $\ell_{|\Ker\ell^\perp}(\vz_2-\lambda\vz_1)=0$, thus $\vz_2-\lambda\vz_1=\vec0$
since $\ell_{|\Ker\ell^\perp}$ is one to one.
And $\vn\in(\Ker\ell)^\perp$ gives $\dim (\Ker\ell)^\perp\ge 1$ (above proof). Thus $\dim (\Ker\ell)^\perp=1 = \Vect\{\vn\}$.
\finrep
\finexe

%%%%%%%%%%%%%%%%%%%%%%%%%%%%%%%%%%%%%%%%%%%%%%%%%%%%%%%%%%%%%%%%%%%%%%%%%%%%%%%%%%%

\subsection{The Riesz representation operator}

The Riesz representation theorem~\ref{thmRiesz} gives the $\dd_g$-Riesz representation operator %(depends on~$\dd_g$) by
\be
\label{eqJg}
\vR_g: 
\left\{\eqalign{
\Es & \rar E \cr
\ell & \rar \vR_g(\ell) := \vell_g, \qie
(\vR_g(\ell),\vv)_g = \ell.\vv,\;\;\forall \vv \in E.
}\right.
\ee
So $\vR_g$ transforms a <<~covariant~$\ell$~>> into a <<~contravariant~$\vell_g$~>> thanks to the tool~$\dd_g$. %, see~\eref{eqRieszbased} for components).

%And $\vR_g^{-1}:$ <<~contravariant~$\vv$~>> $\ds\mathop{\lrar}^{\hbox{\footnotesize transformed}}_{\hbox{\footnotesize with~$g\dd$}}$ <<~covariant~$v_g$~>>, \cf~\eref{eqrtrr}.

\mn
NB (fundamental): $\vR_g$ is a isomorphism between the Banach spaces $(E,||.||_g)$ and $(E^*,||.||_\Es)$,
but $\vR_g$ is \textslbf{not} canonical since it requires a man made tool (an inner dot product chosen by some observer) to be defined.
(An isomorphism $E\leftrightarrow E^*$ can \textslbf{never} be canonical, see~\S~\ref{secEEsnonnat}.)

\medskip
%The vector $\vell_g \eqnote \vell(g) \eqnote \vR_g(\ell)\in E$ is the $\dd_g$-Riesz representation vector of~$\ell$  (depends on~$g$).
And with $\DP$ the set of inner dot products in~$E$, we have thus defined the Riesz representation mapping
\be
\label{eqJg0}
\vR :
\left\{\eqalign{
\DP\times\Es & \rar E \cr
(g,\ell) & \rar \vR(g,\ell):=\vell_g =\vR_g(\ell)=\vell(g).  \cr
}\right.
\ee
So $\vR$ has two inputs: A choice $\dd_g$ by an observer for the first slot, a linear form for the second slot. %, the output  being the $\dd_g$ contravariant representation vector $\vell_g$ of the covariant~$\ell$.

\comment{
And, for a given $\ell\in E^*$ considered by all observers, the relation between observers is given by
\be
\vR_\ell:=\vR(\cdot,\ell) :
\left\{\eqalign{
\DP & \rar E \cr
g & \rar \vR_\ell(g):=\vR(g,\ell)=\vell(g).  \cr
}\right.
\ee
%associates an observer (the one who chooses~$\dd_g$) with $\vell(g)=\vell_g$.
}

%%%%%%%%%%%%%%%%%%%%%%%%%%%%%%%%%%%%%%%%%%%%%%%%%%%%%%%%%%%%%%%%%%%%%%%%%%%%%%%%%%%

\subsection{Quantification with a basis}

Here $E$ is finite dimensional, $\dim E=n$, $\ell\in\Es$ (a linear form), $\dd_g$ is an inner dot product,
$(\ve_i)$ is a basis, $(\pi_{ei})$ is the dual basis (classical notations). Let
\be
g_{ij}=g(\ve_i,\ve_j),\quad \ell = \sumin \ell_i \pi_{ei}, \quad \vell_g = \sumin (\vell_g)_i\ve_i,
\quad \vR_g.\pi_{ej} = \sumin R_{ij}\ve_i,
\ee
\ie\ $[g]_{|\ve}=[g_{ij}]$, $[\ell]_{|\pi_e}=\pmatrix{\ell_1&...&\ell_n}$ (row matrix), 
$[\vell_g]_{|\ve}=\pmatrix{ (\vell_g)_1 \cr \vdots \cr (\vell_g)_n}$ (column matrix),
$[\vR_g]_{\pi_e,\ve}=[R_{ij}]$.

\comment{
, $g_{ij}=g(\ve_i,\ve_j)$, $[g]_{|\ve}=[g_{ij}]$.
Let $\ell\in E^*$ and $\vell_g=\vR_g(\ell)$ be its $\dd_g$-Riesz representation vector.
Let $(\pi_{ei})$ be the dual basis of $(\ve_i)$ (classical notations)
let $\ell = \sumin \ell_i \pi_{ei}$, so $[\ell]_{|\pi_e}=\pmatrix{\ell_1&...&\ell_n}$ (row matrix),
let $\vell_g = \sumin (\vell_g)_i\ve_i$, \ie\ $[\vell_g]_{|\ve}=\pmatrix{ (\vell_g)_1 \cr \vdots \cr (\vell_g)_n}$ (column matrix), let $\vR_g.\pi_{ej} = \sumin R_{ij}\ve_i$, so $[\vR_g.\pi_{ej}]_{|\ve} = \pmatrix{R_{1j} \cr \vdots \cr R_{nj}}$ (column matrix), so $[\vR_g]_{\pi_e,\ve}=[R_{ij}]$.
}

(Duality notations: $\ell = \sumin \ell_i e^i$, $\vell_g = \sumin \ell_g^i\ve_i$,
$\vR_g.e^j = \sumin R^{ij}\ve_i$, $[\vR_g]_{e,\ve}=[R^{ij}]$.)

\debprop
%$\vell_g\in\vRRn$ (the $\dd_g$-Riesz representation vector of $\ell\in\RRns$) and $\vR_g$ (the $\dd_g$-Riesz representation operator) satisfy
%If $\ell\in\Es$ then its $\dd_g$-Riesz representation vector $\vell_g \in E$ satisfies
\be
\label{eqRieszbase}
\boxed{[\vell_g] = [g]^{-1}.[\ell]^T} \qand \boxed{[\vR_g] = [g]^{-1}}, \qie
(\vell_g)_i = \sumjn ([g]^{-1})_{ij}(\ell)_j = \sumjn (\vR_g)_{ij}(\ell)_j.
\ee
Full matrix notation: 
$[\vell_g]_{|\ve} = ([g]_{|\ve})^{-1}.([\ell]_{|\pi_e})^T$,
and 
$[\vR_g]_{|\pi_e,\ve}= ([g]_{|\ve})^{-1}$.

Duality notation to see the change of variance induced by~$\dd_g$ (bottom index for~$\ell$, top index for~$\vell_g$):
\be
\label{eqRieszbased}
\ell_g^i = \sumjn R^{ij}\ell_j,
\ee
\ie\ $\ell_g^i = \sumjn g^{ij}\ell_j$ when $([g]^{-1})_{ij}\eqnote [g^{ij}]$.

(In particular, if $(\ve_i)$ is a $\dd_g$-orthonormal basis, then $[\vR_g]=[g]^{-1}=I$ and $\ell_g^i=\ell_i$.)
\finprop

\debdem
\eref{eqrtr} gives
$
[\ell]_{|\ve}.[\vx]_{|\ve} = [\vell_g]_{|\ve}^T.[g]_{|\ve}.[\vx]_{|\ve}
$  for all~$\vx$, 
thus $[\ell]_{|\ve} = [\vell_g]_{|\ve}^T.[g]_{|\ve}$, thus
$[g]_{|\ve}.[\vell_g]_{|\ve} = [\ell]_{|\ve}^T$ (since $[g]_{|\ve}=[g]_{|\ve}^T$), thus 
$[\vell_g] = [g]^{-1}.[\ell]^T$.

And $\vR_g.\ell \mope^{\eref{eqJg}} \vell_g$ gives
$\sumjn (\ell)_j \vR_g.\pi_j = \sumin (\vell_g)_i\ve_i$, thus
$\sumijn (\ell)_j R_{ij} \ve_i = \sumin (\vell_g)_i\ve_i$, thus
$\sumjn  R_{ij}(\ell)_j = (\vell_g)_i$ for all~$i$, thus $[\vR_g].[\ell]^T = [\vell_g]$. Thus $[\vR_g] = [g]^{-1}$.
\findem

\debrem
\def\ellsharp{{\ell^\sharp}}
If a chosen inner dot product $\dd_g$ is imposed (\eg\ Euclidean foot based)
and if duality notations are used,
then a usual notation for $\vell_g$ is $\ell^\sharp$,
since %$\ell = \sumin \ell_i e^i$ with a bottom index for $\ell_i$, and
$\vell_g = \vR_g(\ell) = \sumin \ell^i \ve_i$ with a top index for $\ell^i$: the index $i$ has been raised through~$\vR_g$.
Then~\eref{eqrtr} and~\eref{eqRieszbase} read (isometric framework)
\be
\label{eqsharpn}
\ell.\vx = \ellsharp \bcdot \vx
\qand [\ellsharp]_{|\ve} = [g]_{|\ve}^{-1}.[\ell]_{|\ve}^T.
\ee
We won't use this notation (we deal with objectivity).
\finrem

%%%%%%%%%%%%%%%%%%%%%%%%%%%%%%%%%%%%%%%%%%%%%%%%%%%%%%%%%%%%%%%%%%%%%%%%%%%%%%%%%%%

\subsection{Change of Riesz representation vector, and Euclidean case}
\label{secfcvr}

For one linear form $\ell \in E^*$, two observers with their inner dot products $\dd_g$ and $\dd_h$
get two Riesz representation vectors $\vell_g = \vR_g(\ell)$ and $\vell_h = \vR_h(\ell)$ given by, \cf~\eref{eqrtr}:
\be
\label{eqrtr2}
\forall \vx \in E,\quad (\vell_g,\vx)_g  =\ell.\vx = (\vell_h,\vx)_h.
\ee

\debprop
%\label{corfcvvr}
For any basis $(\ve_i)$ in~$E$, we have the change of representation vector formula:
\be
\label{eqfcvvrb}
[h]_{|\ve}.[\vell_h]_{|\ve} = [g]_{|\ve}.[\vell_g]_{|\ve}, \qie
[\vell_h]_{|\ve} = [h]_{|\ve}^{-1}.[g]_{|\ve}.[\vell_g]_{|\ve}.
\ee
In particular (for the Euclidean case), with $\lambda>0$:
\be
\label{eqfcvvr3}
\hbox{If }\; %\exists\lambda>0\;\hbox{ \st\ }\;
\dd_g = \lambda^2 \dd_h \qthen \vell_h = \lambda^2 \vell_g.
\ee
Conversely, if $\vell_h = \lambda^2 \vell_g$ for all linear forms $\ell\in E^*$, then $\dd_g = \lambda^2 \dd_h$.

So, a linear form $\ell$ can\textslbf{not} be identified with a Riesz representation vector (which one: $\vell_g$? $\vell_h$?);
In~other words, a Riesz representation vector $\vR_g(\ell)$ is not objective, is not intrinsic to a linear form~$\ell$.

NB: \eref{eqfcvvrb}-\eref{eqfcvvr3} is a ``change of vector formula'' (one linear form gives two vectors relative to two inner dot products); It is not a ``change of basis formula'' (for one vector and its two sets of components).
% (one vector gives two sets of components relative to two different bases).
% In~\eref{eqfcvvr3} no basis is used.
\finprop

\debdem
\eref{eqrtr2}
%that is $(\vell_a,\vx)_a = (\vell_b,\vx)_b$ for all $\vx \in E$,
gives $[\vx]_{|\ve}^T.[g]_{|\ve}.[\vell_g]_{|\ve} = [\vx]_{|\ve}^T.[h]_{|\ve}.[\vell_h]_{|\ve}$
for all~$\vx$, hence $[g]_{|\ve}.[\vell_g]_{|\ve} = [h]_{|\ve}.[\vell_h]_{|\ve}$, \ie~\eref{eqfcvvrb}.

In particular $\lambda^2\dd_h = \dd_g$ give 
$\lambda^2(\vell_g,\vx)_h = (\vell_g,\vx)_g \mope^{\eref{eqrtr2}} (\vell_h,\vx)_h$ for all~$\vx$, hence $\lambda^2 \vell_g=\vell_h$.
%(or use \eref{eqfcvvrb} with $[g]_{|\ve}=\lambda^2[h]_{|\ve}$ here.)

Converse: $\lambda^2 \vell_g = \vell_h$ for all~$\ell$ gives
$\ds
\lambda^2(\vell_g,\vx)_h
=(\vell_h,\vx)_h 
\mope^{\eref{eqrtr2}} (\vell_g,\vx)_g 
$, for all~$\vx$ and for all~$\ell$, thus for all~$\vell_g$
thanks to the isomorphism $\vR_g:E^*\rar E$, \cf~\eref{eqJg}, thus 
$\lambda^2\dd_h = \dd_g$.
\findem

\debexa
\label{exavellbl}
If $\dd_g$ and $\dd_h$ are the Euclidean dot products made with the foot and the metre then, with~\eref{eqrtr2},
\be
\label{eqrtr20}
\dd_g = \lambda^2 \dd_h \quad\Longrightarrow \quad \vell_h = \lambda^2 \vell_g, \qwith \lambda^2> 10:
\ee
$\vell_g$ (English) and $\vell_h$ (French) are quite different! A Riesz representation vector is subjective, and certainly not ``canonical'' (a word that you may find in books where... nothing is defined...).

\Eg, aviation: 
If you do want to use a Riesz representation vector to represent a $\ell\in\RRns$,
it is vital to know which Euclidean dot product is in use, see also remark~\ref{remMCOC} (Mars Climate Orbiter Crash).
Recall: The foot is the international unit of altitude for aviation.
\finexa

\debexa
%Continuing example~\ref{secemp},
If $f \in C^1(\RRn;\RR)$ and $p\in\RRn$, the differential of $f$ at $p$ is the linear form $df(p)\in\RRns$ defined by, for all $\vw\in\vRRn$, 
\be
df(p).\vw := \lim_{h\rar0}{f(p+h\vw) - f(p) \over h}\quad\hbox{(definition independent of any inner dot product)},
\ee
see~\eref{eqdefdPsi}. 
If you can choose an inner dot product~$\dd_g$ then the gradient $\vgrad_g f(p)$ is the $\dd_g$-Riesz representation vector of~$df(p)$:
\be
\vgrad_g f(p) := \vR_g(df(p)), \qie
df(p).\vw=\vgrad_g f(p) \bcdotg \vw,\;\;\forall \vw\in\vRRn.
\ee
%(when $\dd_g$ exits: This is not the case in thermodynamics). 
And~\eref{eqrtr20} gives
\be
\label{eqsecemp2}
\vgrad_h f(p) = \lambda^2\vgrad_g f(p) \qwith \lambda^2>10 \;\hbox{ (English vs French)}:
\ee
The gradient is very dependent on the observer (the gradient is subjective, the differential is objective).
\finexa

\debrem The ``gradient'' is observer dependent; We already had this observer dependence for the usual derivative in the 1-D case $f:x\in\RR\rar f(x)\in\RR$; Question: What does $f'(x)=3$ mean?

Answer. 
11- For one observer, it means $f'(x)=\lim_{h\rar 0}{f(x+h)-f(x)\over h}$... but... where in the departure space this observer has chosen a basis vector $\va$ of length~$1$ for him (\eg\ length 1~foot) which he calls~$1$; So, with no abusive notations, his derivative $f'(x)$ is in fact
$f_a'(x) = \lim_{h\rar 0}{f(x+h\va)-f(x)\over h}$.

12- For some other observer,  it means $f'(x)=\lim_{h\rar 0}{f(x+h)-f(x)\over h}$... but... where in the departure space this observer has chosen a basis vector $\vb$ of length~$1$ for him (\eg\ length 1~metre) which he calls~$1$; So, with no abusive notations, his derivative $f'(x)$ is in fact
$f_b'(x) = \lim_{h\rar 0}{f(x+h\vb)-f(x)\over h}$.

13- Both observer use the same formula $f'(x)=\lim_{h\rar 0}{f(x+h)-f(x)\over h}$ but get different results! Indeed, if $\vb = \lambda \va$, then 
$\ds
= \lim_{h\rar 0}{f(x+h\vb)-f(x)\over h}
= \lim_{h\rar 0}{f(x+h\lambda\va)-f(x)\over h}
= \lambda \lim_{h\rar 0}{f(x+(h\lambda)\va)-f(x)\over (h\lambda)}
=%\mope^{(k=\lambda h)} 
\lambda\lim_{k\rar 0}{f(x+k\va)-f(x)\over k}
$ thus 
\be
\label{eqfpab}
f_b'(x)= \lambda f_a'(x), \qwith \lambda \simeq 3.28
\ee
with foot and metre... Quite different results! 
(In fact $f'(x)={\hbox{opposite side} \over \hbox{adjacent side}}$ depends on the length of the adjacent side...)
\finrem

\debrem
We insist on the subjectivity of the gradient:

20- The differential of $f$ at a point~$x$ along a vector $\vw\in\vec\RR$ is
$df(x).\vw= \lim_{h\rar0}{f(x+h\vw) - f(x) \over h}$ and is objective: The observers all use this same formula.

21- An observer chooses a Euclidean dot product $\dd_g$ (\eg\ based on the foot), then represent $df(x)$ by its $\dd_g$-Riesz representation vector $\vR_g(df(x)) \eqnote \vgrad_g f(x)$ called the gradient of~$f$ at~$x$ relative to~$\dd_g$.

22- Another observer chooses a Euclidean dot product $\dd_h$ (\eg\ based on the metre), then represent $df(x)$ by its $\dd_h$-Riesz representation vector $\vR_h(df(x)) \eqnote \vgrad_h f(x)$ called the gradient of~$f$ at~$x$ relative to~$\dd_h$.

23- Both observer use the same formula $df(x).\vw= \lim_{h\rar0}{f(x+h\vw) - f(x) \over h}$ to get
a different result: $\vgrad_h f = \lambda^2\vgrad_g f$, because they use different measuring tools
(one based on the foot, the other on the metre).

24- Recall: The gradient depends on a choice of a Euclidean unit.
\finrem

\debexe
\def\grad{{\rm grad}}
\def\vRR{\vec\RR}
In~\eref{eqfpab} we have $f_b'(x)= \lambda f_a'(x)$. And the 1-D gradient gives
$\grad_b f(x)= \lambda^2 \grad_a f(x)$. Why?

\debrep
To define a gradient $\grad_a f$ we need a Euclidean dots products~$\dd_a$ built from a basis $(\va)$ in~$\vRR$, while to define $f_a'$ we need a unit of length. Details:
$(\va)$ and $(\vb)$ are two bases in~$\vRR$ with $\vb=\lambda \va$, thus $\dd_a=\lambda^2\dd_b$
(since $1=(\va,\va)_a=(\vb,\vb)_b=(\lambda\va,\lambda\va)_b = \lambda^2(\va,\va)_b$ gives
$(\va,\va)_a = \lambda^2(\va,\va)_b$ and $(\va)$ is a basis).
And we have $f_b'(x) \mope^{\eref{eqfpab}} \lambda f_a'(x)$,
\ie\ $df(x).\vb = \lambda df(x).\va$,
thus $(\grad f_b(x),\vb)_b = \lambda (\grad f_a(x),\va)_a$,
thus $(\grad f_b(x),\vb)_b
= \lambda\lambda^2  (\grad f_a(x),{\vb\over \lambda})_b
= (\lambda^2 \grad f_a(x),\vb)_b
$,
thus $ \grad f_b(x) = \lambda^2 \grad f_a(x)$.
\finrep
\finexe

\debexe
1- Prove that $\dd_g = \lambda^2 \dd_h$ gives $||\vell_h||_g = \lambda ||\vell_h||_h $.
2- Does it contradict the Riesz representation theorem which gives $||\ell||_\RRns = ||\vell_g||_\RRn$?

\debrep
1- $\vell_h \mope^{\eref{eqfcvvr3}} \lambda^2 \vell_g$ gives
$||\vell_h||_h = \lambda^2 ||\vell_g||_h= \lambda ||\vell_g||_g$ since $||.||_h=\lambda||.||_g$.

2- No, since $||\ell||_\RRns := \sup_{||\vx||_\RRn=1}|\ell.\vx|$ depends on the norm $||.||_\RRn$ chosen;
Here $||.||_\RRn$ is either $||.||_g$ or~$||.||_h$. Thus if you write $||\ell||_\RRns \eqnote ||\ell||_{g*}$
if you use the norme~$||.||_g$, then
$||\ell||_{h*}
= \sup_{\vv\in\vRRn} {|\ell.\vv| \over ||\vv||_h}
= \sup_{\vv\in\vRRn} {|\ell.\vv| \over {1\over \lambda}||\vv||_g}
= \lambda \sup_{\vv\in\vRRn} {|\ell.\vv| \over ||\vv||_g}
=\lambda||\ell||_{g*}$.
\finrep
\finexe

%%%%%%%%%%%%%%%%%%%%%%%%%%%%%%%%%%%%%%%%%%%%%%%%%%%%%%%%%%%%%%%%%%%%%%%%%%%%%%%%%%%

\subsection{A Riesz representation vector is contravariant}
\label{secpcbR}

$\vell_g$ is a vector in~$E$, \cf~\eref{eqrtr}, so it is contravariant. To be convinced:

\debexe
Check: % that the contravariance change of basis formula
\be
\label{eqpcbR}
[\vell_g]_{|\new} = P^{-1}.[\vell_g]_{|\old} \quad\hbox{(contravariance formula)}.
\ee
%is indeed satisfied.

\debrep
Consider two bases $(\veio)$ and~$(\vein)$ in~$E$.
With the change of basis formulas $[\vx]_{|\new} = P^{-1}.[\vx]_{|\old}$ and $[g]_{|\new} = P^T.[g]_{|\old}.P$, \eref{eqrtr} gives (with~\eref{eqsecannfcb3}), for all~$\vx$,
\be
\eqalignrll{
[\vx]_{|\old}^T.[g]_{|\old}.[\vell_g]_{|\old}
= \ell.\vx 
= & [\vx]_{|\new}^T.[g]_{|\new}.[\vell_g]_{|\new}  \cr
= & ([\vx]_{|\old}^T.P^{-T}).(P^T.[g]_{|\old}.P) . [\vell_g]_{|\new}
=  [\vx]_{|\old}^T.[g]_{|\old}. (P . [\vell_g]_{|\new}),
}
\ee
thus $[\vell_g]_{|\old} = P . [\vell_g]_{|\new}$ since $[g]$ is invertible (an inner dot product is positive definite), thus~\eref{eqpcbR}.
\finrep
\finexe

%%%%%%%%%%%%%%%%%%%%%%%%%%%%%%%%%%%%%%%%%%%%%%%%%%%%%%%%%%%%%%%%%%%%%%%%%%%%%%%%%%%

\debrem
\label{remisonat}

\comment{
A linear form $\ell\in E^*$ (a ``measuring instrument'') is covariant while its Riesz representation vector $\vell_g\in E$ is contravariant (and $\dd_g$-dependent). In fact, you cannot confuse the contravariant change of basis formula $[\vx]_{|\ven} = P^{-1}.[\vx]_{|\veo}$ with the covariant change of basis formula $[\ell]_{|\ven} = [\ell]_{|\veo}.P$, \cf~\eref{eqdefP1}. 
}

$\bullet$ Dont forget: A representation vector $\vell_g$ is not intrinsic to the linear form~$\ell$ because it depends on a~$\dd_g$ (depends on a observer: foot? metre?)).
Reminder: there is no natural canonical isomorphism between $E$ and~$E^*$, \ie\ it is impossible to identify a linear form with a vector, see~\S~\ref{secEEsnonnat}.
\comment{
 (it is impossible to confuse covariance with contravariance, and recall that you cannot confuse the contravariant change of basis formula $[\vx]_{|\ven} = P^{-1}.[\vx]_{|\veo}$ with the covariant change of basis formula $[\ell]_{|\ven} = [\ell]_{|\veo}.P$, \cf~\eref{eqdefP1}).
}

\noindent $\bullet$ $\vell_g$ is incompatible with the use of push-forwards,
\cf~\S~\ref{secrvdfdepf}.

\noindent $\bullet$ $\vell_g$ is incompatible with the use of Lie derivatives, \cf~\eref{eqpbvr}.
\comment{
\noindent $\bullet$ Problem with the usual divergence $\dvg\uusigma$ of an order two tensor~$\sigma$ used by mechanical engineers (their divergence is not objective), see~\S~\ref{secdivcla}.
By the way, we can define the objective divergence $\tdvg\uutau$ of a ${1\choose1}$ tensor
$\uutau\in \Tuuu$, \cf~\eref{eqdefdvi3}.
(And see example~\ref{exadefdvi3b}.)
}
\finrem

%%%%%%%%%%%%%%%%%%%%%%%%%%%%%%%%%%%%%%%%%%%%%%%%%%%%%%%%%%%%%%%%%%%%%%%%%%%%%%%%%%%

\subsection{What is a vector versus a $\dd_g$-vector?}
\label{secRfc}

\leavevmode

1- Originally, a vector was a bipoint vector $\vv = \ora{AB}$ in $\vRRt$ used to represent of a ``material object''. %See Maxwell~\cite{maxwell}.
\Eg\ the height of a child is represented on a wall by a vertical bipoint vector $\vx$ starting from the ground up to a pencil line.
The vector $\vx$ is objective: Any observer uses this same vector to get the height of the child... 
and then use ``their subjective unit'' (foot, metre...) to give a value. %This bi-point vector is objective (the same for any observer).

\medskip
2- Then (mid 19th century), the concept of vector space was introduced:
It is a quadruplet $(E,+,K,.)$ where $+$ is an inner law, $(E,+)$ is a group, $K$ is a field, $.$ is a external law on~$E$ (called a scalar multiplication) compatible with~$+$ (see any math book).

And then the concept of scalar inner dot product (in a vector space) was introduced.
%And then an element of a vector space is called a vector, but such a vector is not a ``unique object that can be seen by any observer'' (the material bi-point in~1-). 
%\Eg, a $n*m$ matrix $M$ is an element of the vector space of matrices~$\calM_{nm}$, so it is a vector $M\in\calM_{nm}$.

\medskip
3- We can then get non ``material'' vectors (``subjectively built vectors'').
\Eg: start with our usual vector space~$\vRRn$ of bi-point vectors, then
consider its dual $(\RRns,+,\RR,.) \eqnote \RRns$. %: It is a vector space.
Then, for a given $\ell\in\RRns$ (a given measuring device), consider two observers:
An English observer with his foot built Euclidean dot product~$\dd_g$, and a French observer with with his metre built Euclidean dot product~$\dd_h$. These observers build their own artificial Riesz representation vectors $\vell_g=\vR_g(\ell)\in\vRRn$ and $\vell_h=\vR_h(\ell)$, cf~\eref{eqrtr20}; They remark that $\vell_g\ne\vell_h$: 
%Their Riesz built vectors $\vell_g$ and $\vell_h$ are \textslbf{not} ``material vectors: 
These constructions are very subjective.
%They are just mathematically built vectors that should only be used by the concerned observers, not by the other observers! (Very subjective construction.)
%Here the Riesz mapping $\vR$, \cf~\eref{eqJg0}, gives the tool equipped with two buttons: $R(a,.):\ell \rar \vell_a$ for the English man, and $R(b,.):\ell \rar \vell_b$ for the French man.

\medskip
4- Then, with differential geometry, a vector $\vv$ has been redefined: It is a ``tangent vector'', which means that there exists a $C^1$ curve $c:s\in[a,b]\rar c(s)\in E$
such that $\vv$ is defined at a $p=c(s)\in\Im(c)$ by 
$\vv(p): =\vcp(s)$ (so a vector is part of a vector field, here defined along the range of~$c$). %=\lim_{h\rar0}{c(s{+}h)-c(s) \over h}=\lim_{h\rar0}{\ora{c(s)c(s{+}h)} \over h}$. 
(This definition of a tangent vector is applicable to ``tangent vectors to a surface''
\ie\ tangent vectors to a manifold, see~\eg~\S~\ref{secdLos1},2-.) Then it is shown that $\vv$ is equivalent to ${\pa \over \pa \vv}=$ the directional derivative in the direction~$\vv$ (natural canonical isomorphism between $E$ and~$\Ess$).  For other equivalent definitions of vectors, see Abraham--Marsden~\cite{abraham-marsden}.

%%%%%%%%%%%%%%%%%%%%%%%%%%%%%%%%%%%%%%%%%%%%%%%%%%%%%%%%%%%%%%%%%%%%%%%%%%%%%%%%%%%

\subsection{The ``$\dd_g$-dual vectorial bases'' of one basis (and warnings)}
\label{secvectdb}

Framework: $E$ is a finite dimensional vector space, $\dim E=n$ (\eg\ $E=\vRRt$).
An observer chooses an inner dot product $\dd_g$
(\eg, in~$\vRRt$, a foot-built Euclidean dot product).
Hence the results will be subjective. %, or a metre-built Euclidean dot product).
And $(\ve_i)$ is some basis in~$E$.

\comment{
NB: A vectorial dual basis is made of (contravariant) vectors which depend on an inner dot product: it is not intrinsic to the initial basis (not unique: it depends on~$\dd_g$). Not to be confused with the (covariant) dual basis (made of linear forms) which is intrinsic to the initial basis (unique).
}

%%%%%%%%%%%%%%%%%%%%%%%%%%%%%%%%%%%%%%%%%%%%%%%%%%%%%%%%%%%%%%%%%%%%%%%%%%%%%%%%%%%

\subsubsection{A basis and its many associated ``dual vectorial basis''}

\def\veig{{\ve_{ig}}} \def\vejg{{\ve_{jg}}}
\def\veia{{\ve_{ia}}} \def\veja{{\ve_{ja}}}
\def\veua{{\ve_{1a}}} \def\veda{{\ve_{2a}}}
\def\veib{{\ve_{ib}}} \def\vejb{{\ve_{jb}}}
\def\veub{{\ve_{1b}}} \def\vedb{{\ve_{2b}}}

\debdef
\label{defddgdb}
The $\dd_g$-dual vectorial basis (or $\dd_g$-vectorial dual basis, or $\dd_g$-dual basis) 
of the basis $(\ve_i)$ is the basis $(\veig)$ in~$E$ defined by
\be
\label{eqdefbdv}
\forall j=1,...,n, \quad (\veig,\ve_j)_g = \delta_{ij}, \qie \boxed{\veig \bcdotg \ve_j=\delta_{ij}}.
\ee
NB: A vectorial dual basis is not unique: It depends on the chosen inner dot product, see \eg~\eref{eqexeddgdb}.
\findef

NB: Pay attention to the notations: $\veig$ is a contravariant vector ($\veig\in E$), so, even if you use the Einstein convention, the index $_i$ in~$\veig$ must be a bottom index.

\medskip 
Let $(\pi_{ei})$ be the (covariant) dual basis of the basis~$(\ve_i)$,
\ie\ the $\pi_{ei}\in E^*$ are the objective (the same for all observers) linear forms defined by $\pi_{ei}.\ve_j= \delta_{ij}$ for all~$j$, \cf~\eref{eqdefbd}.

\begin{definition}[Equivalent definition.]\rm

The $\dd_g$-dual vectorial basis of the basis $(\ve_i)$
is the basis $(\veig)$ in~$E$ made of the $\dd_g$-Riesz representative vectors of the~$\pi_{ei}$, \ie
\be
\label{eqveig2}
\boxed{\veig := \vR_g(\pi_{ei})} , \qie
\veig \bcdotg \vv = \pi_{ei}.\vv,\;\; \forall \vv\in E.
% \quad (\hbox{\ie}\;\; (\veig,\ve_j)_g = \pi_{ei}.\vv,\;\; \forall \vv\in E).
\ee
where $\vR_g$ is the $\dd_g$-Riesz operator, see~\eref{eqJg}.

With duality notations, $(e^i)$ is the dual basis and %the $\veig$ are defined by
$\veig := \vR_g(e^i)$, \ie\ $(\veig,\vv)_g = e^i.\vv$ for all $\vv\in E$ where here the position of the index~$i$
is bottom on the left and up on the right, since $\vR_g$ changes a covariant vector (a linear form) into a contravariant vector.
\findef

%With duality notations, $\pi_{ei}\eqnote e^i$, and $\boxed{\veig:=\vR_g(e^i)}$, see~\eref{eqJg}.

\debexe
\label{execheckveig}
Prove that the vectors $\veig$ satisfy the contravariant change of basis formula
\be
\label{eqveigc}
[\veig]_{|\new} = P^{-1}.[\veig]_{|\old}
\ee
(the $\vejg$ are indeed ``contravariant vectors'').

\debrep
$\bullet$ First answer: $\veig$ is a vector in~$E$, thus it is contravariant.

$\bullet$ Second answer: Apply~\eref{eqpcbR} since $\veig$ is a Riesz-representation vector.

$\bullet$ Third answer = direct computation: % to prove that $\veig$ satisfies the change of basis formula for vectors:
Consider two bases $(\va_i)$ and $(\vb_i)$, and the transition matrix $P$ from $(\va_i)$ to~$(\vb_i)$, \ie,
$\vb_j = \sumin P_{ij} \va_i$ for all~$j$.
\eref{eqdefbdv}
and the change of basis formulas for the vectors $\ve_i$ and the bilinear form~$\dd_g$ give
$[\ve_j]_{|\va}^T.[g]_{|\va}.[\veig]_{|\va}=(\veig,\ve_j)_g = [\ve_j]_{|\vb}^T.[g]_{|\vb}.[\veig]_{|\vb}
= (P^{-1}.[\ve_j]_{|\va})^T.(P^T.[g]_{|\va}.P).[\veig]_{|\va} 
=[\ve_j]_{|\va}^T.[g]_{|\va}.P.[\veig]_{|\va}
$ for all~$i,j$, thus $[\veig]_{|\va} = P.[\veig]_{|\vb}$, thus~\eref{eqveigc}.
\finrep
\finexe

\debexe
Consider two inner dot products $\dd_a$ and~$\dd_b$ (\eg, a foot-built and a metre-built Euclidean dot product),
and a basis $(\ve_i)$ in~$E$. Call $(\veia)$ and~$(\veib)$ the $\dd_a$ and $\dd_b$-dual vectorial bases.
%and $\lambda >0$, $\lambda \ne 1$. 
Prove:
\be
\label{eqexeddgdb}
\dd_a=\lambda^2\dd_b \quad\Longrightarrow\quad 
\ve_{ib} = \lambda^2 \ve_{ia}, \quad \forall i.
\ee
%thus one basis $(\ve_i)$ has two different vectorial dual bases.
\Eg, $\lambda^2>10$ with foot and metre built Euclidean bases: $\ve_{ib}$ is very different from $\ve_{ia}\,$!
A dual vectorial basis highly depends on an observer: A vectorial dual basis is \textslbf{not} intrinsic to~$(\ve_i)$
(\textslbf{not} objective).
% While the dual basis $(e^i)=(\pi_{ei})$ is unique)

\debrep
\eref{eqdefbdv} gives
$(\veib,\ve_j)_b = \delta_{ij} = (\veia,\ve_j)_a = \lambda^2(\veia,\ve_j)_b$,
thus $(\veib-\lambda^2\veia,\ve_j)_b = \delta_{ij}$, for all $i,j$.
\finrep
\finexe

\debexa
If $(\ve_i)$ is a $\dd_g$-orthonormal basis we trivially get $\veig=\ve_i$ for all~$i$, \ie, $(\veig)=(\ve_i)$.This particular case is not compatible with joint work by an English (foot) and French (metre) observer.
%See next~\S~\ref{rembad}.
\finexa

%%%%%%%%%%%%%%%%%%%%%%%%%%%%%%%%%%%%%%%%%%%%%%%%%%%%%%%%%%%%%%%%%%%%%%%%%%%%%%%%%%%

\subsubsection{Components of $\vejg$ in the basis~$(\ve_i)$}
\label{seccofdb}

\debprop
The components of~$\vejg$ in the basis~$(\ve_i)$ are given by, for any $j\in[1,n]_\NN$,
\be
\label{eqveigb2}
\boxed{[\vejg]_{|\ve} = [\vR_g]_{|\ve}.[\ve_j]_{|\ve}} = ([g]_{|\ve})^{-1}.[\ve_j]_{|\ve} =\; \hbox{the $j$-th column of~$([g]_{|\ve})^{-1}$},
\ee
\ie\ the $i$-th component of~$\vejg$ is $([g]_{|\ve}^{-1})_{ij}$.

Thus the matrix of~$g\dd$ in the basis $(\veig)$ is the inverse of the matrix of~$g\dd$ in the basis $(\ve_i)$:
\be
\label{eqveigb2b2}
%([g(\veig,\vejg)]_{i=1,...,n \atop j=1,...,n}=)\quad 
([g(\veig,\vejg)]=)\quad 
[g]_{|(\veig)} = [g]_{|(\ve_i)}{}^{-1} 
\quad (=([g(\ve_i,\ve_j)])^{-1}).
%\quad (=([g(\ve_i,\ve_j)]_{i=1,...,n \atop j=1,...,n})^{-1}).
%[g(\veig,\vejg)]_{i=1,...,n \atop j=1,...,n} = ([g(\ve_i,\ve_j)]_{i=1,...,n \atop j=1,...,n})^{-1}.
\ee
%[g(\veig,\vejg)] = ([g(\ve_i,\ve_j)])^{-1}, \qie 
\finprop

\debdem
First proof of~\eref{eqveigb2} (straight forward calculation): \eref{eqdefbdv} gives, for all $i,j$,
%By definition of the dual basis, 
%$\delta_{ij} = \pi_{ej}.\ve_i = [\pi_{ej}]_{\ve}.[\ve_i]_{|\ve} = [\ve_j]_{|\ve}^T.[\ve_i]_{|\ve}$, thus 
\be
\label{eqveigPg001}
[\ve_j]_{|\ve}^T.[g]_{|\ve}.[\veig]_{|\ve} = \delta_{ij} = [\ve_j]_{|\ve}^T.[\ve_i]_{|\ve},
\qthus [g]_{|\ve}.[\veig]_{|\ve}=[\ve_i]_{|\ve}.
\ee
Second proof of~\eref{eqveigb2}: Apply~\eref{eqRieszbase} (generic Riesz representation result) to get~\eref{eqveigb2}.

Thus, $[g]_{|\ve}$ being symmetric we have $[g]_{|\ve}{}^{-1}$ symmetric, and $g(\veig,\vejg)
=[\veig]_{|\ve}^T.[g]_{|\ve}.[\vejg]_{|\ve}
= [\ve_i]_{|\ve}^T.[g]_{|\ve}{}^{-1}.[g]_{|\ve}.[g]_{|\ve}{}^{-1}.[\ve_j]_{|\ve}
= [\ve_i]_{|\ve}^T.[g]_{|\ve}{}^{-1}.[\ve_j]_{|\ve}
= ([g]_{|\ve}{}^{-1})_{ij}
$, thus~\eref{eqveigb2b2}.
\findem

\debexa
$\vRRd$, $[g]_{|\ve}=\pmatrix{1 & 0 \cr 0 & 2}$, thus $[g]_{|\ve}^{-1}=\pmatrix{1 & 0 \cr 0 & \demi}$.
Thus $\ve_{1g}=\ve_1$, $\ve_{2g}=\demi\ve_2$. %, $g^{22}=\demi$.
\finexa

\debrem
\label{remwarnR}
Warning: 
When $([g]_{|\ve}^{-1})_{ij} \eqnote g^{ij}$ then \eref{eqveigb2} reads
\be
\label{eqveigb2c}
\vejg = \sumin g^{ij} \ve_i,
\ee
where the Einstein convention is \textslbf{not} satisfied:
The Einstein convention is satisfied with 
\be
\label{eqveigb2c2}
\vejg = \sumin (\vejg)^i \ve_i \eqnote \sumin (P_j)^i \ve_i %\qthus (P_j)^i \mope^{\eref{eqveigb2}} ([g]_{|\ve}^{-1})_{ij}, % \quad (\hbox{also noted } g^{ij}),
\ee
(the components of vectors have up indices),
%$\veig=\sumkn (P_i)^k\ve_k$, 
and this can be verified with $(\veig,\ve_j)_g \mope^{\eref{eqdefbdv}} \delta_{ij}$ which gives $\sumkln (P_i)^kg_{kj} = \delta_{ij}$. 
%This can also be verified with $(\vejg,\ve_i)_g \mope^{\eref{eqveig2}} e^j.\ve_i$ which gives $\sumkn (P_j)^k g_{ki} = \delta^j_i$
And in~\eref{eqveigb2c}
the scalars $g^{ij}$ is just another name for $(P_j)^i$, nothing more (nothing to do with the Einstein convention).

We insist:In other words: $M=[g]_{|\ve}=[M_{ij}]$ is a matrix, and its inverse is the matrix $ M^{-1}=[M_{ij}]^{-1}$:
A matrix is just a collection of scalars (has nothing to do with the Einstein convention), and its inverse is also a collection of scalars, and you do not change this fact
by calling $M^{-1} \eqnote [M^{ij}]$ (the use of up indices is irrelevant for matrices). See remark~\ref{remgsij1}.

And because $(P_j)^i$ equals $([g]_{|\ve}^{-1})_{ij} \eqnote g^{ij}$, some people rename $\vejg$ as $\ve^{\,j}$... to get $\ve^{\,j}=\sumin g^{ij}\ve_i$...
But doing so they \textslbf{despise} Einstein's convention, despite eventual claims: They confuse covariance and contravariance... and add confusion to the confusion.

NB: Recall: If in trouble with a notation which comes as a surprise (the notation $g^{ij}$ here), use classical notations: Then no misuse of Einstein's convention and no possible misinterpretation.
In particular here $\vejg$ is a (contravariant) vector. 
\finrem

%%%%%%%%%%%%%%%%%%%%%%%%%%%%%%%%%%%%%%%%%%%%%%%%%%%%%%%%%%%%%%%%%%%%%%%%%%%%%%%%%%%

\subsubsection{Multiple admissible notations for the components of $\vejg$}

Let $\calP \in\calL(E;E)$ be the change of basis endomorphism from $(\ve_i)$ to~$(\veig)$: defined by
$\calP.\ve_j = \vejg$.
And let $P=[\calP]_{|\ve}$ (the associated transition matrix).
It gives multiple admissible (non confusing) notations for the components  of $\vejg$ relative to the basis~$(\ve_i)$:
\be
\label{eqveigPb}
\vejg = \calP.\ve_j
= \underclas{
\sumjn P_{ij}\ve_i
= 
\sumjn (P_j)_i\ve_i
} 
= \underdual{
\sumjn (P_j)^i\ve_i
= 
\sumjn \Pij \ve_i
}
,
\ee
\ie\ the $i$-th component of the vector $\vejg$ has the names $P_{ij}=(P_j)_i=(P_j)^i=\Pij$,
\ie\ $P=[\calP]_{|\ve} = [P_{ij}]=[(P_j)_i]=[(P_j)^i]=[\Pij]$ (four different notations for the same matrix), \ie
\be
\forall j,\quad
[\vejg]_{|\ve} = [\calP]_{|\ve}.[\ve_j]_{|\ve}
= \pmatrix{P_{1j} \cr \vdots \cr P_{nj}} = \pmatrix{(P_j)_1 \cr \vdots \cr (P_j)_n}
= \pmatrix{P^1{}_j \cr \vdots \cr P^n{}_j} = \pmatrix{(P_j)^1 \cr \vdots \cr (P_j)^n}
\ee
= the $j$-th column of~$[\calP]_{|\ve}$.
You can choose any notation, depending on your current need or mood...

%%%%%%%%%%%%%%%%%%%%%%%%%%%%%%%%%%%%%%%%%%%%%%%%%%%%%%%%%%%%%%%%%%%%%%%%%%%%%%%%%%%

\subsubsection{(Huge) differences between ``the (covariant) dual basis'' and ``a dual vectorial basis''}
\label{secvdb}

\begin{enumerate}
\item
A basis $(\ve_i)$ has an infinite number of vectorial dual bases $(\veig)$,
as many as the number of inner dot products~$\dd_g$ (as many as observers), see~\eref{eqveigb2}.

While a basis $(\ve_i)$ has a unique intrinsic (covariant) dual basis $\ds(\pi_{ei})\eqnote(e^i)$, \cf~\eref{eqdefbd}: Two observers who consider the same basis $(\ve_i)$ have the same (covariant) dual basis.

\item $\pi_{ei}=e^i$ is covariant, while $\ve_i$ and $\veig$ are contravariant.
%We insist: $(\veig)$ is linked to~$(\ve_i)$ by an endomorphism $\calP\in\calL(E;E)$, \cf~\eref{eqveigPb}.
And there is no transition matrix between $(\ve_i)$ and $(\pi_{ei})=(e^i)$,
since $\ve_i \in E$ and $\pi_{ei}=e^i \in E^*$ %and the (linear) function $\pi_{ei}=e^i \in E^*$ 
don't live in the same vector space.
% \Cf~remark~\ref{remcarpet} and remark~\ref{remcarpet2} (to sweep dust under the rug).

\item  If you fly, it is vital to use the dual basis $(\pi_{ei})=(e^i)$: It is possibly fatal if you confuse foot and metre at takeoff and at landing (if~you survived takeoff) because of the choice of different Euclidean dot product $\dd_g$ or~$\dd_h$;
See \eg\ the Mars Climate Orbiter crash, remark~\ref{remMCOC}. Einstein's convention can help... only if it is really followed.

\comment{
\item The push-forward of a vector field~$\vv$, \cf~\eref{eqdefrapvEivei2}, is different from the push-forward of a one-form~$\alpha$, \cf~\eref{eqtransfd}. Thus vectors and linear forms (contravariance and covariance) should not be confused.

\item The Lie derivative of a vector field~$\vv$, \cf~\eref{eqdl}, is different from the Lie derivative of a one-form~$\alpha$, \cf~\eref{eqdla}. Thus vectors and linear forms (contravariance and covariance) should not be confused.

\item The trace of an endomorphism is intrinsic to the endomorphism (or to a ${1\choose1}$ uniform tensor): It is an objective notion (the same scalar for all observers), \cf~\eref{eqdefTr}. While the trace
of a ${0\choose2}$ or ${2\choose0}$ uniform tensor is not defined since the result would be observer dependent.
}
\end{enumerate}

%The $\dd_g$ dependence is also given by~\eref{eqveig2}.

%%%%%%%%%%%%%%%%%%%%%%%%%%%%%%%%%%%%%%%%%%%%%%%%%%%%%%%%%%%%%%%%%%%%%%%%%%%%%%%%%%%

\subsubsection{About the notation $g^{ij}$ = shorthand notation for $(g^\sharp)^{ij}$}
\label{remcarpet3}

%Let $\dd_g$ be an inner dot product in~$E$;
%$\ell\in E^*$ and $\vell_g$ is its $\dd_g$-Riesz representation vector: $\ell.\vw = (\vell_g,\vw)_g %= \vell_g \bcdotg  \vw$ for all $\vw\in E$. %, \cf~\eref{eqrtr}.
 
\debdef
The Riesz associated inner dot product $g^\sharp \in \calL(E^*,E^*;\RR)$
is the bilinear form defined by, for all $\ell,m\in E^*$, 
\be
\label{eqgsharpij0}
g^\sharp(\ell,m) := g(\vell_g,\vec m_g), \qie (\ell,m)_{g^\sharp} := (\vell_g,\vec m_g)_g.
\ee
where $\vell_g=\vR_g(\ell)$ and $\vec m_g=\vR_g(m)$.
\findef

Thus $g^\sharp\dd \eqnote \dd_{g^\sharp}$ is indeed an inner dot product in~$E^*$: trivial check.

\mn
{\bf Quantification: }
$(\ve_i)$ is a basis in~$E$ and $(e^i)$ is its dual basis  (duality notations).
\eref{eqgsharpij0} gives:
\be
\label{eqgsharpij2}
(g^\sharp)^{ij} := 
g^\sharp(e^i,e^j) = g(\veig,\vejg),
 \qthus [g^\sharp]_{|e} =\boxed{[(g^\sharp)^{ij} ] = [g_{ij}]^{-1}} = [g]_{|\ve}^{-1}, 
\ee
\cf~\eref{eqveigb2b2}.
And
\be
\label{eqgsharpij4}
%g^\sharp(e^i,e^j) \eqnote (g^\sharp)^{ij}\; \mathop{=}^{\rm shorthand}_{\rm notation} \; g^{ij}, \qthus 
\boxed{[(g^\sharp)^{ij}]\; \mathop{=}^{\rm shorthand}_{\rm notation} \; [g^{ij}]}. % = [g_{ij}]^{-1}}.
\ee
Classical notations: %$(\pi_{ei})$ is the dual basis, and 
$[g^\sharp]_{|e} =[(g^\sharp)_{ij}]
=[g^\sharp(\pi_{ei},\pi_{ej})]
=[g(\veig,\vejg)] =  [g_{ij}]^{-1}  = ([g]_{|\ve})^{-1}$.

\debexe
How do we compute $g^\sharp(\ell,m)$ with matrix computations?

\debrep
$\ell = \sumin \ell_i e^i$ and $m=\sumjn m_j e^j$ give
%and $(g^\sharp)^{ij} = g^\sharp(e^i,e^j)$ give
$g^\sharp(\ell,m) = \sumijn \ell_i m_j g^\sharp(e^i,e^j) = \sumijn \ell_i (g^\sharp)^{ij} m_j
 = [\ell]_{|\ve}.[g^\sharp]_{|\ve}.[m]_{|\ve}^T = [\ell]_{|\ve}.[g]_{|\ve}^{-1}.[m]_{|\ve}^T$
(a linear form is represented by a row matrix,).
\finrep
\finexe

\debexe
\eref{eqgsharpij0} tells that the ${2\choose0}$ tensor $g^\sharp\in\calL(E^*,E^*;\RR)$ was created from the ${0 \choose 2}$ tensor $g=\dd_g\in\calL(E,E;\RR)$ using twice the $\dd_g$-Riesz representation theorem.

1- Show that if you use the $\dd_g$-Riesz representation theorem just once you get the
${1\choose1}$ tensor $g^\natural\in\calL(E^*,E;\RR)\simeq \calL(E;E)$ given by
\be
g^\natural = I.
\ee
% that you create from $g=\dd_g$ when using the $\dd_g$-Riesz representation theorem just once?

2- Reciprocal: What is the ${0\choose2}$ tensor $g^\flat\in\calL(E,E;\RR)$ that you create from the identity $I\in\calL(E;E)$ when using the $\dd_g$-Riesz representation theorem once?

3- Summary: $\tilde I = g^\natural$ gives $(\tilde I)^\flat = g^\flat=g$ and $(\tilde I)^\sharp = g^\sharp$

\debrep
1- $g^\natural\in \calL(E^*,E;\RR)$ is defined by $g^\natural(\ell,\vw) = (\vell_g,\vw)_g$
for all $(\ell,\vw)\in E^*\times E$, where $\vell_g$ is the $\dd_g$-Riesz representation vector of~$\ell$. %, so $\ell.\vu = (\vell_g,\vu)_g$ for all $\vu\in E$.
Thus $g^\natural(\ell,\vw) = \ell.\vw = \ell.I.\vw$, for all $(\ell,\vw)\in E^*\times E$, hence
$g^\natural\in \calL(E^*,E;\RR)$ is naturally canonically associated with the identity $I \in \calL(E;E)$.

2- The identity operator $I\in\calL(E;E)$ (observer independent) is naturally canonically associated with the ${1\choose1}$ tensor $\tilde I \in \calL(E^*,E;\RR)$ defined by $\tilde I (\ell,\vw)=\ell.I.\vw = \ell.\vw$ for all $(\ell,\vw)\in E^*\times E$, thus $\tilde I=g^\natural$.
%So, with the $\dd_g$-Riesz representations : $\tilde I = g^\natural$ and $(\tilde I)^\flat = g$ and $(\tilde I)^\sharp = g^\sharp$.
\finrep
\finexe

\comment{
%%%%%%%%%%%%%%%%%%%%%%%%%%%%%%%%%%%%%%%%%%%%%%%%%%%%%%%%%%%%%%%%%%%%%%%%%%%%%%%%%%%

\subsubsection{The bra $\la . |$ and the ket $|.\ra$}

Dirac has introduced the notation
\be
\vw \eqnote | w \ra= \hbox{for a vector}, \qand
\ell \eqnote \la \ell | = \hbox{for a linear form},
\ee
the vector space $E=\{| w \ra\}$ being the space of ``states''.
This emphasizes the difference between a (linear) function $\la\ell|$
and a vector $|\vw\ra$. And
\be
\ell(\vw) \eqnote  \la \ell | w \ra  %\quad (= \la \ell |(| w \ra)=\ell.\vw).
\ee
gives the bra-ket notation (the bra $\la \ell |$ for linear forms, and the ket $| w \ra$ for vectors).

And more generally, if $L\in\calL(E;E)$ then $\ell(L(\vw)) \eqnote  \la \ell  | L| w \ra = \ell.L.\vw$.

\comment{
And, \eg, $| w \ra\la \ell | = \vw \otimes \ell$ is the endomorphism $\vRRn \rar \vRRn$
defined by $| w \ra\la \ell |.|\vu\ra = \la \ell |\vu\ra| w \ra$,
that is $(\vw \otimes \ell).\vu = \vw(\ell.\vu)$ (contractions rule),
that is, $(\vw \otimes \ell).\vu = (\ell.\vu)\vw$, for all $\vu\in\vRRn$.
}

\comment{
Et les lois de changement de base sont données par~\eref{eqdefP1} :
\be
\label{eqdefP1D}
\eqalign{
& \bullet \; [| w \ra]_{|\van} = P^{-1}.[| w \ra]_{|\vao} \quad \hbox{(formule de contravariance)}, \cr
& \bullet \; [\la \ell |]_{|\van} = [\la \ell |]_{|\vao}.P \qquad \hbox{(formule de covariance)}.
}
\ee
}
}

%%%%%%%%%%%%%%%%%%%%%%%%%%%%%%%%%%%%%%%%%%%%%%%%%%%%%%%%%%%%%%%%%%%%%%%%%%%%%%%%%%%
%%%%%%%%%%%%%%%%%%%%%%%%%%%%%%%%%%%%%%%%%%%%%%%%%%%%%%%%%%%%%%%%%%%%%%%%%%%%%%%%%%%

\section{Cauchy--Green deformation tensor $C = F^T.F$}

%The construction of the Cauchy--Green deformation tensor $C=F^T.F$ is detailed, starting with the definition of the transposed~$F^T$ (tensorial definition often forgotten, leading to errors and confusions).
% (also needed for the infinitesimal strain tensor $\uueps={F+F^T\over 2} -I$, the isometric objectivity and frame invariance principle, ...).

Framework: 
$\tPhi : 
\left\{\eqalign{
\RR \times \Obj & \rar \RRn \cr
(t,\Pobj) & \rar \tPhi(t,\Pobj) \cr
}\right\}$ is a motion of~$\Obj$,
$\Omegatau= \tPhi(\tau,\Pobj)$ is the configuration of~$\Obj$ at any~$\tau$,
$\tz$ and $t$ are fixed,
%$\tz,t\in\RR$ are fixed, 
$\Phi:=\Phitzt: 
\left\{\eqalign{
\Omegatz &\rar \Omegat \cr
P &\rar p=\Phi(P) \cr
}\right\}$ is the associated motion between $\tz$ and~$t$,
%so $p$ is the position at~$t$ of the particle which was at~$\tz$ at $P$. %, \cf~\eref{eqPhit}.
and $F(P) :=d\Phi(P) :
\left\{\eqalign{
\vRRntz &\rar \RRnt \cr
\vW &\rar \vw=F(P).\vW := \lim_{h\rar0}{\Phi(P{+}h\vW) - \Phi(P) \over h}\cr
}\right\}$
is the deformation gradient at~$P$ between $\tz$ and~$t$,
\cf~\eref{eqdefFtfMH}.

%%%%%%%%%%%%%%%%%%%%%%%%%%%%%%%%%%%%%%%%%%%%%%%%%%%%%%%%%%%%%%%%%%%%%%%%%%%%%%%%%%

\setcounter{subsection}{-1}
\subsection{Goal}
\label{secCG1}

{\bf Construction of~$C$ (summary of Cauchy's approach):}
%The stress may be deduced from the Lie derivative of {\bf one} vector. % to the Lie derivative.  But the Lie derivative did not exist during Cauchy's lifetime, and Cauchy proposed to get the stress from {\bf two} vectors:

1- At~$\tz$, consider two vectors $\vW_1$ and~$\vW_2$, % and consider $(\vW_1,\vW_2)_g$,

2- at~$t$, they are distorted by the motion and become the vectors $F.\vW_1$ and $F.\vW_2$;

3- Then choose a Euclidean dot product $\dd_g\eqnote \cdot \bcdot \cdot$, the same at all~$t$ (to simplify); %and write $(\vW_1,\vW_2)_g=\vWu \bcdot \vWd$,

4- Then, by definition of the transposed, $(F.\vW_1)\bcdot (F.\vW_2) = (F^T. F.\vW_1) \bcdot \vW_2$: You have got the Cauchy strain tensor $C:=F^T.F$; 

5- Then %$(C.\vW_1)\bcdot \vW_2 - \vWu \bcdot \vWd = ((C{-}I).\vW_1) \bcdot \vW_2$
$(F.\vW_1)\bcdot (F.\vW_2) - \vWu \bcdot \vWd = ((C{-}I).\vW_1) \bcdot \vW_2$
gives a measure of the deformation with~$\vW_2$ as a reference, measure that is used to build first order constitutive laws for the stress (Cauchy).
%Now you can compare $(C.\vW_1)\bcdot \vW_2$ with $\vWu \bcdot \vWd$ (comparison with~$\vW_2$ as a reference): $(C.\vW_1)\bcdot \vW_2 - \vWu \bcdot \vWd = ((C{-}I).\vW_1) \bcdot \vW_2$ enables to build stress laws.

%%%%%%%%%%%%%%%%%%%%%%%%%%%%%%%%%%%%%%%%%%%%%%%%%%%%%%%%%%%%%%%%%%%%%%%%%%%%%%%%%%%

\subsection{Transposed $F^T$: Inner dot products required}
\label{secFT}

\def\pp{p}
\def\vutz{{\vu_\tz}}
\def\vutzs{{\vu_{\tz*}}}
\def\vcptz{\vc_\tz{}'}
\def\vcpt{\vc_t{}'}

We first give the functional definition of~$F^T$ (qualitative);
Then we get the usual matrix representation of~$F^T$ relative to observers (quantification).

%%%%%%%%%%%%%%%%%%%%%%%%%%%%%%%%%%%%%%%%%%%%%%%%%%%%%%%%%%%%%%%%%%%%%%%%%%%%%%%%%%%

\subsubsection{Definition of the function $F^T$}
\label{secFTdef}

\def\FTGg{{F^T_{Gg}}}
\def\FTGgp{{F^T_{Gg,p}}}
\def\FTGgpt{{F^T_{Gg,\pt}}}
\def\FtztTGg{{(\Ftzt)^T_{Gg}}}
\def\FtztTGgpt{{(\Ftzt)^T_{Gg,\pt}}}

%Here the qualitative point of view is adopted: 

At~$\tz$, a past observer chose an inner dot product $\dd_G$ in~$\RRntz$,
and at~$t$ the present observer  chooses an inner dot product $\dd_g$ in~$\RRnt$.
%Let $P\in\Omegatz$ and $p=\Phi(P)\in\Omegat$ and $F(P)=d\Phi(P)$ (deformation gradient at~$P$).

By definition,
the transposed of the linear map $F(P)\in\calL(\RRntz;\RRnt)$ relative to~$\dd_G$ and~$\dd_g$ is the linear map $F(P)^T_{Gg}\in\calL(\RRnt;\RRntz)$ defined by,
for all $\vU_P\in\RRntz$ (vector at~$P$) and $\vw_p\in \RRnt$ (vector at~$p$),
\be
\label{eqFT00}
(F(P)^T_{Gg}.\vw_p,\vU_P)_G = (F(P).\vU_P,\vw_p)_g, \qinshort \boxed{(F^T_{Gg}. \vw,\vU)_G = (F.\vU,\vw)_g},
\ee
see~\eref{eqseccpgdd0}.
This defines $F^T_{Gg}(p) := F(P)^T_{Gg}$ when $p=\Phi(P)$:
\be
\label{eqFT0}
F^T_{Gg} : \left\{\eqalign{
\Omegat &\rar \calL(\RRnt;\RRntz) \cr
p & \rar \boxed{F^T_{Gg}(p) := F(P)^T_{Gg}}
% \qwhen p=\Phi(P).
}\right\},
\quad\hbox{so in short}\quad  \boxed{(F^T. \vw) \bcdotG \vU = \vw \bcdotg (F.\vU)},
\ee
without forgetting that $F^T := F^T_{Gg}$ depends on~$\dd_G$ and~$\dd_g$.

\debexe
1. In $F^T.\vz.\vW = \vz.F.\vW = F.\vW.\vz=\vW .F^T.\vz$, which dots are inner dot products?

2. What does $F.\vW_1.F.\vW_2= \vW_1.F^T.F.\vW_2$ mean?

\debrep 
1. No choice:
$(\vW,\vz)\in\RRntz \times \RRnt$, so
$(F^T.\vz) \bcdotG \vW = \vz \bcdotg (F.\vW) = (F.\vW) \bcdotg \vz = \vW \bcdotG (F^T.\vz)$.
%(\ie\ $(F^T.\vz,\vW)_G = (\vz,F.\vW)_g = (F.\vW,\vz)_g = (\vW,F^T.\vz)_G$).

2. No choice: %Meaning (there is no choice):
$\vW_1,\vW_2\in\RRntz$, so
$(F.\vW_1) \bcdotg (F.\vW_2) = \vW_1 \bcdotG (F^T.(F.\vW_2))$. %, \ie\ $\ds (F.\vW_1 , F.\vW_2)_g = (\vW_1,F^T.F.\vW_2)_G$.
\finrep
\finexe

\debrem
More generally, on a surface~$\Omega$ (a manifold), \eref{eqFT00} is defined for all $(\vU_P,\vw_p)\in T_P\Omegatz \times T_p\Omegat$,
where $T_{\!\ptau}\Omegatau$ is the tangent space at~$\Omegatau$ at~$\ptau$.
\finrem

%%%%%%%%%%%%%%%%%%%%%%%%%%%%%%%%%%%%%%%%%%%%%%%%%%%%%%%%%%%%%%%%%%%%%%%%%%%%%%%%%%%

\subsubsection{Quantification with bases (matrix representation)}
\label{secFqwb}

Classical notations: $(\va_i)$ is a basis in~$\RRntz$, %and $(\piai)$ is its (covariant) dual basis (basis in~$\RRntzs$), 
and $(\vb_i)$ is a basis in~$\RRnt$. % and $(\pibi)$ is its dual basis (basis in~$\RRnts$).
Marsden--Hughes duality notations: $(\vE_I)$ is a basis in~$\RRntz$ and $(\ve_i)$ is a basis in~$\RRnt$.
And the reference to the points $P$ and~$p$ is omitted to lighten the writings (use the full notation of~\S~\ref{secFTdef} if in doubt).

\medskip
Let $[G]:=[(\va_i,\va_j)_G]$, $[g]:=[(\vb_i,\vb_j)_g]$,
$[F]_{|\va,\vb}=[F_{ij}] \eqnote [F]$, 
$[F^T]_{|\vb,\va} =[(F^T)_{ij}] \eqnote [F^T]$.

\medskip
\eref{eqFT00} gives $[\vU]^T.[G].[F^T.\vw] = [F.\vU]^T.[G].[\vw]$,
thus $[\vU]^T.[G].[F^T].[\vw] = [\vU]^T.[F]^T.[g].[\vw]$, for all $\vU,\vw$, thus
\be
\label{eqr2}
[G].[F^T] = [F]^T.[g],
\qie \boxed{[F^T] = [G]^{-1}.[F]^T.[g]}.
\ee
%(Full notation: $[G]_{|\va}.[F^T]_{|\vb,\va} = [F]_{|\va,\vb}{}^T.[g]_{\vb}$, \ie\  $[F^T]_{|\vb,\va} = [G]_{|\va}{}^{-1}.([F]_{|\va,\vb})^T.[g]_{\vb}$.

\debrem
If $(\va_i)$ and $(\vb_i)$ are $\dd_G$ and~$\dd_g$-orthonormal bases, then $[C] = [F]^T.[F]$.
But recall: If you need to work with a coordinate system, then the bases in use are the coordinate system bases which are \textslbf{not} orthonormal in general, \ie\ $[G]^{-1}\ne I$ and $[g]^{-1}\ne I$ in general. 
\finrem

\debexe
Use classical notation, then Marsden duality notations, to express~\eref{eqr2} with components.

\debrep
Classical notations: %$(\va_i)$ is a basis in~$\RRntz$ and $(\vb_i)$ is a basis in~$\RRnt$. Let
\be
\label{eqr20}
\eqalign{
& G_{ij} = G(\va_i,\va_j), \quad g_{ij} = g(\vb_i,\vb_j), \qie
[G]_{\va}=[G_{ij}], \quad [g]_{|\vb}=[g_{ij}], \cr
\hbox{and }\; 
& F.\va_j = \sumin F_{ij} \vb_i, \quad F^T.\vb_j = \sumin (F^T)_{ij} \va_i, \qie
[F]_{|\va,\vb} = [F_{ij}], \quad [F^T]_{|\vb,\va} = [(F^T)_{ij}].
}
\ee
Then $ (F^T.\vb_j,\va_i)_G \mope^{\eref{eqFT00}} (\vb_j,F.\va_i)_g$ gives 
$ (\sumkn (F^T)_{kj}\va_k,\va_i)_G = (\vb_j,\sumkn F_{ki}\vb_k)_g$, thus
$\sumkn (F^T)_{kj} (\va_k,\va_i)_G = \sumkn F_{ki}(\vb_j,\vb_k)_g$ with $F_{ki}=([F]^T)_{ik}$, thus
\be
\label{eqr2d}
\sumkn G_{ik}(F^T)_{kj} %= \sumkn F_{ki} g_{kj}
= \sumkn ([F]^T)_{ik} g_{kj}, \qie
(F^T)_{ij} = \sum_{k,\ell=1}^n ([G]^{-1})_{ik} F_{\ell k} g_{\ell j} 
%= \sum_{k,\ell=1}^n ([G]^{-1})_{ik} F_{\ell k} g_{\ell j}
,
\ee
for all~$i,j$, thus~\eref{eqr2}.
%And if $(\va_i)$ and $(\vb_i)$ are $\dd_G$ and $\dd_g$-orthonormal bases, then $(F^T)_{ij} = F_{ji}$.

Marsden  notations:
$G_{IJ} = G(\vE_I,\vE_j)$, $g_{ij} = g(\ve_i,\ve_j)$, $
F.\vE_J = \sumin \FiJ \ve_i$, $
F^T.\ve_j = \sumIn \FTIj \vE_I$, thus

$\ds
\sum_{K=1}^n G_{IK}(F^T)^K_{\;\;\;j} = \sumkn F^k_{\;\;I} g_{kj},
\qie
\FTIj = \sum_{K,k=1}^n G^{IK} F^k_{\;K} g_{kj}  \qwhere [G^{IJ}] := [G_{IJ}]^{-1}.
$
%And if $(\vE_I)$ and $(\ve_i)$ are $\dd_G$ and $\dd_g$-orthonormal bases, then $\FTIj = F^j{}_{\!I}$.
\finrep
\finexe

\comment{
\debrem
In physics we need to see the metrics and their components $g_{ij}$ and~$G_{ij}$ with respect to chosen bases (usually coordinate system bases at~$t$ and at~$\tz$): Even if $g_{ij}$ or~$G_{ij}$ $=\delta_{ij}$: We write them.
\finrem
}

\comment{
%Let $F:=\Ftzt(\ptz)$ and $F^T : = (\Ftzt)^T(\pt)$.
%With Marsden and Hughes duality notations: $(\va_i)=(\vE_I)$, $(\vb_i)=(\ve_i)$, $\ptz= P$, and $\pt= p$. 

\eref{eqFT0} immediately give $[\vU]_{|\va}^T.[G]_{|\va}.[F^T.\vw]_{|\va} = [F.\vU]_{|\vb}.[g]_{|\vb}.[\vw]_{|\vb}$,
thus $[\vU]_{|\va}^T.[G]_{|\va}.[F^T]_{|\vb,\va}.[\vw]_{|\vb} = [\vU]_{|\va}^T.[F]_{|\vw,\vb}^T.[g]_{|\vb}.[\vw]_{|\vb}$,
true for all $\vU\in\RRntz$ and $\vw\in\RRnt$, thus $[G]_{|\va}.[F^T]_{|\vb,\va} = [F]_{|\vw,\vb}^T.[g]_{|\vb}$,
thus with implicit basis,
\be
\label{eqr2}
[G].[F^T] = [F]^T.[g],
\qie \boxed{[F^T] = [G]^{-1}.[F]^T.[g]}.
\ee
In particular, if the bases $(\va_i)$ and $(\vb_i)$ are $\dd_G$-orthonormal and $\dd_g$-orthonormal, then $[F^T] = [F]^T$.
Full notations:
\be
\label{eqr21}
[G]_{|\va}.[\FtztTGg(\pt)]_{|\vb,\va} = ([F(\ptz)]_{|\va,\vb})^T.[g]_{\vb}, \qie 
[\FtztTGg(\pt)]_{|\vb,\va} = [G]_{|\va}^{-1}.([F(\ptz)]_{|\va,\vb})^T.[g]_{\vb}.
\ee

Details in terms of components:
Let $[G]_{\va}=[G_{ij}] \eqnote [G]$ and $[g]_{|\vb}=[g_{ij}] \eqnote [g]$, \ie, for all $i,j$,
\be
G(\va_i,\va_j)=G_{ij} \qand g_{ij} = g(\vb_i,\vb_j),
\ee
Let $[F]_{|\va,\vb} = [F_{ij}]\eqnote [F]$ and $[F^T]_{|\vb,\va} = [(F^T)_{ij}]\eqnote [F^T]$, \ie
\be
\label{eqr20}
F.\va_j = \sumin F_{ij} \vb_i, \qand
F^T.\vb_j = \sumin (F^T)_{ij} \va_i.
\ee
Then \eref{eqr2} reads:
\be
\label{eqr2d}
\sumkn G_{ik}(F^T)_{kj} = \sumkn F_{ki} g_{kj}, \qie
\boxed{(F^T)_{ij} = \sum_{k,\ell=1}^n ([G]^{-1})_{ik} F_{\ell k} g_{\ell j}}.
\ee
In particular, if $(\va_i)$ and $(\vb_i)$ are $\dd_G$ and $\dd_g$-orthonormal bases
then $(F^T)_{ij}=F_{ji}$.

Marsden notations: $G(\vE_I,\vE_j)\eqnote G_{IJ}$, \ie\ $[G]=[G_{IJ}]$,
and $F.\vE_J = \sumin \FiJ \ve_i$, \ie\ $[F] = [\FiJ]$,
and $F^T.\ve_j = \sumIn \FTIj \vE_I$, \ie\ $[F^T] = [\FTIj]$; And
\be
\sum_{K=1}^n G_{IK}(F^T)^K_{\;\;\;j} = \sumkn F^k_{\;\;I} g_{kj},
\qie
\FTIj = \sum_{K,k=1}^n G^{IK} F^k_{\;K} g_{kj} \qwhen [G^{IJ}] := [G]^{-1} .
\ee
%Warning: Be careful with the notation $G^{IJ}$, \cf\ remark~\ref{remwarnR}.
In particular, if $(\vE_I)$ and $(\ve_i)$ are $\dd_G$ and $\dd_g$-orthonormal bases,
then $\FTIj=F^j{}_I$.
}

%%%%%%%%%%%%%%%%%%%%%%%%%%%%%%%%%%%%%%%%%%%%%%%%%%%%%%%%%%%%%%%%%%%%%%%%%%%%%%%%%%%

\subsubsection{Remark: $F^*$}
\label{secremFs}

(For mathematicians: $F^*$ doesn't seem to be very useful in mechanics, apart from making simple things difficult, and playing games with components and duality notations...).

\debdef
The adjoint of the linear map $F\in \calL(\RRntz;\RRnt)$ (acting on vectors) is the linear map
$F^* \in \calL(\RRnts;\RRntzs)$ (acting on functions) canonically defined by,
for all  $m\in \RRnts$,
\be
\label{eqFstar}
F^*(m) := m \circ F, \qwritten F^*.m = m.F \;\;(\in\RRntzs).
\ee
\findef

So, for all $(m,\vW)\in \RRnts \times \RRntz$,
\be
\label{eqfstar}
%\forall (m,\vW)\in \RRnts \times \RRntz,\quad
(F^*.m).\vW = m.F.\vW \;\;(\in\RR).
\ee

\noindent
{\bf Quantification} (matrix representation): 
$(\pi_{ai})$ and $(\pi_{bi})$ are the covariant dual bases of $(\va_i)$ and $(\vb_i)$.
Let $(F^*)_{ij}$ be the components of~$F^*$ relative to these dual bases:
\be
F^*.\pi_{bj} = \sumIn (F^*)_{ij} \pi_{ai},
\qie %[F]_{|\va,\vb}=[F_{ij}]\;\;\hbox{and}\;\;
[F^*]_{|\pi_b,\pi_a} = [(F^*)_{ij}].
\ee
\eref{eqfstar} gives $(F^*.\pi_{bj}).\va_i = \pi_{bj}.F.\va_i$, thus %$(F^*)_{ij} = F_{ji}$ for all~$i,j$, \ie
\be
\label{eqfstar2}
\forall i,j,\; \boxed{(F^*)_{ij} = F_{ji}}, \qie [F^*]_{|\pi_b,\pi_a} = ([F]_{|\va,\vb})^T, \qinshort [F^*] = [F]^T.
\ee
%(Or \cf~\eref{defadjLq}.)
Marsden duality notations: %$F.\vE_J = \sumin \FiJ \ve_i$ and
$F^*.e^j = \sumIn (F^*)_I{}^j E^I$ gives $(F^*)_I{}^j = F^j{}_I$ for all $I,j$.

\mn
{\bf Interpretation of~$F^*$.} As usual in classical mechanics, we use Euclidean dot products, here $\dd_G$ in~$\RRntz$ and~$\dd_g$ in~$\RRnt$. Then we use 
the $\dd_G$-Riesz representation vector $\vR_G(F^*.m)\in\RRntz$ of~$F^*.m\in\RRntzs$, and the $\dd_g$-Riesz representation vector $\vR_g(m)\in\RRnt$ of~$m\in\RRnts$, so, for all $m\in\RRnts$ and $\vW\in\RRntz$,
\be
(F^*.m).\vW = \vR_G(F^*.m) \bcdotG \vW, \qand m.(F.\vW) = \vR_g(m) \bcdotg F.\vW = (F^T.\vR_g(m)) \bcdotG \vW.
\ee
Thus
\eref{eqfstar} gives 
$\vR_G(F^*.m) = F^T.\vR_g(m)$, thus 
\be
\vR_G.F^* = F^T.\vR_g, \qie F^* = \vR_G{}^{-1}.F^T.\vR_g.
\ee
%Quantification with the above bases: We have $[\vR_G]=[G]^{-1}$ and $[\vR_g]=[g]^{-1}$, \cf~\eref{eqRieszbase}, thus $[F^*]_{|\pi_b,\pi_a} = ([F]_{|\vb,\va})^T$, written in short $[F^*] = [F]^T$.

\debrem
The definition of~$F^*$ is intrinsic to~$F$ (objective), while the definition of~$F^T$ is~\textsl{\textbf{not}} intrinsic to~$F$ (\textsl{\textbf{not}} objective) since it needs inner dot products (observer choices) to be defined. 
%However $F^T$ is easy to interpret with~$C=F^T.F$ (\cf~\S~\ref{secCG1}), but interpretation for $F^*$? 
%Is $F^*$ useful in mechanics, apart from math games with components and duality notations with $[(F^*)_I{}^j] = [F^j{}_I]$, \cf~\eref{eqfstar2}?
\finrem

%%%%%%%%%%%%%%%%%%%%%%%%%%%%%%%%%%%%%%%%%%%%%%%%%%%%%%%%%%%%%%%%%%%%%%%%%%%%%%%%%%%

\subsection{Cauchy--Green deformation tensor~$C$}
\label{secC}

%%%%%%%%%%%%%%%%%%%%%%%%%%%%%%%%%%%%%%%%%%%%%%%%%%%%%%%%%%%%%%%%%%%%%%%%%%%%%%%%%%%

\def\CIJ{{C^I{}_{\!\!J}}}
\def\vWu{{\vW_1}}
\def\vWd{{\vW_2}}
\def\vwu{{\vw_1}}
\def\vwd{{\vw_2}}
%\def\vwtzs{{\vw^\tz_*}}

%%%%%%%%%%%%%%%%%%%%%%%%%%%%%%%%%%%%%%%%%%%%%%%%%%%%%%%%%%%%%%%%%%%%%%%%%%%%%%%%%%%

\subsubsection{Definition of $C$}

\def\CGg{{C_{Gg}}}
\def\CGgP{{C_{Gg,P}}}
\def\CtztGg{{(C^\tz_{t,Gg}}}

Consider vectors $\vW_{i} \in \RRntz$ at~$P$, $i=1,2$,
and their push forwards $\vw_i$ toward $p=\Phi(P)$, \ie %, in short,% (the push-forwards at~$p$)
\be
\label{eqvuvvC}
\vw_{i}  = F.\vW_{i}, %,\qinshort \vw_i=F.\vW_i.
\ee
short notation for $\vw_{i}(p)  = F(P).\vW_{i}(P)$.
With the chosen inner dot products $\dd_G$ in~$\RRntz$ and $\dd_g$ in~$\RRnt$, we get
$(\vwu(p),\vwd(p))_g
= (F(P).\vWu(P),F(P).\vWd(P))_g
\equalref{eqFT0} (F^T_{Gg}(p). F(P).\vWu(P),\vWd(P))_G$ when $p=\Phi(P)$, written in short:
\be
\label{eqdefCt}
(\vwu,\vwd)_g
=  (F.\vWu,F.\vWd)_g 
=  (\underbrace{F^T. F}_{C}.\vWu,\vWd)_G,
\ee

\debdef
The (right) Cauchy--Green deformation tensor at $P\in\Omegatz$ relative to $\dd_G$ and~$\dd_g$, is the endomorphism 
$\CGg(P) \in\calL(\RRntz;\RRntz)$ defined by
\be
\label{eqdefC}
\CGg(P) := F^T_{Gg}(p) \circ F(P) ,\qinshort \boxed{C := F^T. F}. % \quad (= F^T \circ F).
\ee
\findef

So
\be
C=F^T \circ F : \vW \mathop{\lrar}^{F} F(\vW)  \mathop{\lrar}^{F^T} F^T(F(\vW)) = C(\vW),
\ee
with $F$ and $F^T$ linear, thus $C$ is linear and $C(\vW)$ is written $C.\vW = F^T.F.\vW$. 
And \eref{eqdefCt} tells that $C$ is characterized by, for all $\vWu,\vWd\in\RRntz$,
\be
\label{eqdefCt2}
%(\vwu,\vwd)_g =  (C.\vWu,\vWd)_G = (F.\vWu,F.\vWd)_g , \qie 
\vwu \bcdotg \vwd = \boxed{(C.\vWu) \bcdotG \vWd = (F.\vWu) \bcdotg (F.\vWd)}.
\ee
Moreover, $\dd_g$ being symmetric (inner dot product), $C$ is a $\dd_G$-symmetric endomorphism in~$\RRntz$,
\ie, for all $\vWu,\vWd\in\RRntz$,
\be
\label{eqdefCt2s}
(C.\vWu,\vWd)_G = (\vWu,C.\vWd)_G, \qie (C.\vWu) \bcdotG \vWd = \vWu \bcdotG (C.\vWd),
%(C.\vWu) \bcdotG \vWd = \vWu \bcdotG (C.\vWd) , %(C.\vWu,\vWd)_G = (\vWu,C.\vWd)_G,
\ee
since $(F^T.F.\vWu,\vWd)_G = (F.\vWu,F.\vWd)_g = (\vWu,F^T.F.\vWd)_G$.

%%%%%%%%%%%%%%%%%%%%%%%%%%%%%%%%%%%%%%%%%%%%%%%%%%%%%%%%%%%%%%%%%%%%%%%%%%%%%%%%%%%

\subsubsection{Quantification}

\eref{eqdefC} gives $[C]=[F^T].[F]$, with $[F^T] \mope^{\eref{eqr2}} [G]^{-1}.[F]^T.[g]$, thus
\be
\label{eqr30}
\boxed{[C] = [G]^{-1}.[F]^T.[g].[F]}, % \quad (= [F^T].[F]),
\ee
short notation for 
$[C_{Gg}]_{|\va} = [G]_{|\va}^{-1}.([F]_{|\va,\vb})^T.[g]_{|\vb}.[F]_{|\va,\vb}
%= [F_{Gg}^T]_{|\vb,\va}.[F]_{|\va,\vb}
$.

\comment{
\debrem
If $(\va_i)$ and $(\vb_i)$ are $\dd_G$ and~$\dd_g$-orthonormal bases, then $[C] = [F]^T.[F]$.
But recall: If you need to work with a coordinate system, then the bases in use are the coordinate system bases which are \textslbf{not} orthonormal in general, \ie\ $[G]^{-1}\ne I$ and $[g]^{-1}\ne I$ in general. 
\finrem
}

\debexe
Use classical notation, then duality notations, to express~\eref{eqr30} with components.

\debrep
Classical notations:
%With a basis $(\va_i)$ in~$\RRntz$ and a basis $(\vb_i)$ in~$\RRnt$, $P\in\Omegatz$ and $p=\Phi(P)$, with $F=\Ftzt(P)$ and $C=\Ctzt(P)$, with $F.\va_j=\sumin F_{ij}\vb_i$ and 
\be
F.\va_j =  \sumin F_{ij} \vb_i \qand C.\va_j =  \sumin C_{ij} \va_i, 
\qie [F]_{|\va,\vb} = [F_{ij}] \qand [C]_{|\va} = [C_{ij}] .
\ee
\eref{eqdefCt2}-\eref{eqdefCt2s} give
%$(C.\va_i,\va_j)_G = (F.\va_i,F.\va_j)_g$, 
$(\va_i,C.\va_j)_G = (F.\va_i,F.\va_j)_g$, 
so 
$(\va_i,\sum_k C_{kj} \va_k)_G = (\sum_k F_{ki}\vb_k,\sum_\ell F_{\ell j}\vb_\ell)_g$,
 thus
$\sum_k C_{kj} (\va_i,\va_k)_G
= \sum_{k\ell} F_{ki} (\vb_k,\vb_\ell)_g F_{\ell j}
%= \sum_{k\ell} (F^T)_{ik} g_{k\ell} F_{\ell j}
$,
\ie %, for all $i,j$,
\be
\label{eqr3}
\sumkn  G_{ik} C_{kj}
= \sumkln F_{ki}\, g_{k\ell} F_{\ell j}
= \sumkln ([F]^T)_{ik}\, g_{k\ell} F_{\ell j}
, \qso
\boxed{[G].[C] = [F]^T.[g].[F]},
\ee
so
$C_{ij} 
= \sum_{k,\ell,m=1}^n ([G]^{-1})_{im} F_{km}\, g_{k\ell} F_{\ell j}
= \sum_{k,\ell,m=1}^n ([G]^{-1})_{im} ([F]^T)_{mk}\, g_{k\ell} F_{\ell j}
$. %, thus~\eref{eqr30}}.
%For orthonormal bases $C_{ij} = \sum_{k=1}^n F_{ki} F_{k j}$.
Duality notations: % with standard usual $\CIJ$ and the flat $C_{IJ}$ (the index $I$ goes from top to bottom):
\be
\label{eqr3m}
\eqalign{
&F.\vE_J =  \sumin \FiJ \ve_i \qand C.\vE_J =  \sumIn \CIJ \vE_I, 
\qie [F]_{|\vE,\ve} = [\FiJ] \qand [C]_{|\vE} = [\CIJ] ,\; \hbox{ and}\cr
\;
&\sumKn G_{IK} C^K{}_{\!\!J} = \sum_{k,\ell=1}^n F^k{}_{\!I}\, g_{k \ell }F^\ell{}_{\!\!J}, 
\qand \CIJ = \sum_{k,\ell,M=1}^n G^{IM} F^k{}_{\!\!M}\, g_{k \ell}F^\ell{}_{\!\!J}
\qwhen [G^{IJ}] := [G_{IJ}]^{-1}.
}
\ee
%For orthonormal bases $\CIJ = \sum_{k=1}^n F^k{}_{\!I} F^k{}_{\!J}$.
\finrep
\finexe

%Indeed, $(C.\vWu,\vWd)_G = (F.\vWu,F.\vWd)_g = (F^T_p. F.\vWu,\vWd)_G$ and $(C.\vWu,\vWd)_G = (F.\vWu,F.\vWd)_g = (F.\vWd,F.\vWu)_g = (F^T.F.\vWd,\vWu)_G = (C.\vWd,\vWu)_G$.

\debexe
$\dd_G$ is a Euclidean dot product in foot,
$\dd_g$ is a Euclidean dot product in metre, so 
$\dd_g = \mu^2 \dd_G$ with $\mu\simeq 0.3048$; And $(\va_i)=(\vb_i)$ is a $\dd_G$-orthonormal basis. Prove
\be
\label{eqCmu}
[C]=\mu^2 [F]^T.[F].
\ee

\debrep
$[C]_{|\va} \equalref{eqr30} [G]_{|\va}^{-1}.[F]_{|\va,\va}^T.[g]_{|\va}.[F]_{|\va,\va}$ gives
$[C]_{|\va}=I. [F]_{|\va,\va}^T.\mu^2 I.[F]_{|\va,\va}$. Shorten notation = \eref{eqCmu}.
\finrep
\finexe

%%%%%%%%%%%%%%%%%%%%%%%%%%%%%%%%%%%%%%%%%%%%%%%%%%%%%%%%%%%%%%%%%%%%%%%%%%%%%%%%%%%

\subsection{Time Taylor expansion of~$C$}

Here we use a unique inner dot product $\dd_G=\dd_g$ at all time
(to compare results in the vicinity of~$\tz$). Moreover we use an orthonormal basis (to lighten the notations),
thus, in short, $[C]=[F]^T.[F]$.

$P$ is fixed,
$\Ctzt(P) \eqnote C(t)$, and $[C(t)] = [F(t)]^T.[F(t)]$ (since $[G]=[g]=I$ here),
and $\vVtzt(P)\eqnote \vV(t)$ and $\vAtzt(P)\eqnote\vA(t)$ are the Lagrangian velocities and accelerations,
and $\vv(t,p)$ and $\vgamma(t,p)$ are the Eulerian velocities and accelerations at $t$ at $p=\Phitzt(t,P)$.

%(Second order  $C(t{+}h)=  C(t) +h\,C'(t)+  {h^2\over 2} C''(t)  + o(h^2)$).)

With Lagrangian variables (used to define~$C$): 
$F(t{+}h) = F(t) + h\, d\vV(t) + {h^2\over 2}\, d\vA(t) + o(h^2)$ gives
\be
\label{eqdPhi33}
\eqalign{
[C(t{+}h)]
= & [F(t{+}h)]^T.[F(t{+}h)] %\quad\hbox{(matrix meaning $[C(t{+}h)] = [F(t{+}h)]^T.[F(t{+}h)]$)}
\cr
= & [F^T + h\, d\vV^T + {h^2\over 2}\, d\vA^T + o(h^2)](t)
[F + h\, d\vV + {h^2\over 2}\, d\vA + o(h^2)](t) \cr
= &  [C(t) + h\,
\underbrace{([F^T].[d\vV] + [d\vV]^T.[F])(t)}_{=[(\CtzP)'(t)] \eqnote [C'(t)]}
 + {h^2\over 2}\,
\underbrace{([F]^T.[d\vA]+ 2[d\vV]^T.[d\vV] + [d\vA]^T.[F])(t)}_{=[(\CtzP)''(t)] \eqnote [C''(t)]}
)(t) + o(h^2).  \cr
}
\ee
(As usual with Lagrangian variables, we have three times involved: $\tz$, $t$ and~$t{+}h)$.) In particular
\be
[\CtzP(\tz{+}h)] = I+ ([d\vV] + [d\vV]^T)(\tz) + {h^2\over 2}\,([d\vA]+ 2[d\vV]^T.[d\vV] + [d\vA]^T)(\tz) + o(h^2). 
\ee
Abusively written $\CtzP(\tz{+}h) = I+ (d\vV + d\vV^T)(\tz) + {h^2\over 2}\,(d\vA+ 2d\vV^T.d\vV + d\vA^T)(\tz) + o(h^2)$, but don't forget it is a matrix meaning.

With Eulerian variables: With
$p(t)=\Phitz(t,P)$, we have
$d\vVtz(t,P) = d\vv(t,p(t)).F(t)$ and $d\vAtz(t,P) = d\vgamma(t,p(t)).F(t)$, thus
writing $d\vv:=d\vv(t,p(t))$ and $d\vgamma:=d\vgamma(t,p(t))$ (for short),
\be
\label{eqCpe}
\eqalign{
\CtzP(t{+}h)
= & \CtzP(t) + h\,(F^T(t).(d\vv + d\vv^T)(t,p(t)).F(t))  \cr
%\underbrace{(F^T.(d\vv + d\vv^T).F)(t)}_{=(\CtzP)' \eqnote C'(t)}
&\quad  + {h^2\over 2}\,(F^T(t).(d\vgamma + 2d\vv^T.d\vv + d\vgamma^T)(t,p(t)).F(t))
%\underbrace{(F^T.(d\vgamma + 2d\vv^T.d\vv + d\vgamma^T).F)(t)}_{=(\CtzP)''(t) \eqnote C''(t)}
 + o(h^2).
}
\ee
abusive notation of $[\CtzP(t{+}h)] = ...$ (matrices relative to a basis).

\debrem
\label{remsquarem}
$F'' = d\vA$ is easy to interpret, but 
$C''=F^T.d\vA+ 2d\vV^T.d\vV + d\vA^T.F = (F^T.d\vA+ d\vV^T.d\vV) + (F^T.d\vA+ d\vV^T.d\vV)^T
%=(F^T.F''+ (F')^T.F) + (F^T.F''+  (F')^T.F)^T
$
is not that easy to interpret (and in not linear in~$\vV$). 

We already had a problem with the composition of flows: The formula
$F^\tz_{t_2} = F^{t_1}_{t_2}.F^\tz_{t_1}$ is simple (determinism), but the formula 
$C^\tz_{t_2} = (F^\tz_{t_2})^T.F^\tz_{t_2}
= (F^\tz_{t_1})^T.(F^{t_1}_{t_2})^T.F^{t_1}_{t_2}.F^\tz_{t_1}
= (F^\tz_{t_1})^T.C^{t_1}_{t_2}.F^\tz_{t_1}
$ is ``not that simple'' ($\ne C^{t_1}_{t_2}.C^\tz_{t_1}$).
(Indeed, to consider $C$ instead of $F$ amounts to consider the ``motion squared'', \cf\
$(C.\vW,\vW)_g = ||F.\vW||_g^2$.)

Since $C'(\tz) =  d\vV(\tz) + d\vV(\tz)^T$ this may have little consequences for linear approximation near~$\tz$, but ultimately not small consequences for second-order approximations (and large deformations) if $C''$ is used to make constitutive laws. The consideration of Lie derivatives may be an interesting alternative.
\finrem

%%%%%%%%%%%%%%%%%%%%%%%%%%%%%%%%%%%%%%%%%%%%%%%%%%%%%%%%%%%%%%%%%%%%%%%%%%%%%%%%%%%

\subsection{Remark: $C^\flat$}

For mathematicians: May produce errors, misuses, covariance-contravariance confusion, see next \S~\ref{secCflatr}.
For the general ${}^\flat$ notation see~\S~\ref{secgenflat}.

%%%%%%%%%%%%%%%%%%%%%%%%%%%%%%%%%%%%%%%%%%%%%%%%%%%%%%%%%%%%%%%%%%%%%%%%%%%%%%%%%%%

\subsubsection{Definition of $C^\flat$...}

\debdef
At $P\in\Omegatz$, the bilinear form $C^\flat_{Gg}(P) \eqnote C^\flat \in\calL(\RRntz,\RRntz;\RR)$ associated with the linear map $C_{Gg}(P) \eqnote C\in \calL(\RRntz;\RRntz)$ is defined by, for all $\vWu,\vWd\in\RRntz$ vectors at~$P$,
% $C^\flat(\vWu,\vWd) := (C.\vWu,\vWd)_G$ , \ie\ in short
\be
\label{eqCfla2}
C^\flat(\vWu,\vWd) := (\vWu,C.\vWd)_G \quad (= (F.\vWu,F.\vWd)_g).
\ee
\findef

%\medskip
Then $C^\flat$ is a bilinear symmetric form (trivial) and is a metric in~$\RRntz$ when $\Ftzt\eqnote F$ is a diffeomorphism (usual hypothesis),
but not a Euclidean one (it is iff $C=I$ \ie\ for rigid body motions).

\medskip
\noindent
{\bf Quantification:} \eref{eqCfla2} gives $[\vWd]^T.[C^\flat].[\vWu] = [\vWd]^T.[G].[C].[\vWu]$ for all $\vWu,\vWd$ since $C^\flat$ and $\dd_G$ are symmetric, thus
\be
\label{eqCfM0}
[C^\flat] = [G].[C] \quad (=[F]^T.[g].[F]) .
\ee

\debexe
Use duality notations to express~\eref{eqCfM0} with components,
and explain the flat $^\flat$ notation.

\debrep
%Marsden duality notations:
\eref{eqCfla2} gives $C^\flat(\vE_J,\vE_I) := (C.\vE_J,\vE_I)_G$.
Thus with %$[C]=[\CIJ]$ and $[C^\flat] = [C_{IJ}]$ %\ie\ 
$C^\flat(\vE_I,\vE_J) = C_{IJ}$ and $C.\vE_J = \sum_I \CIJ \vE_I$
we get
$C_{JI} = \sum_K C^K{}_{\!J}(\vE_K,\vE_I)_G = \sum_K C^K{}_{\!J} G_{KI}$;
And $C^\flat$ and  $\dd_G$ are symmetric, thus 
%$C^\flat(\vE_J,\vE_I)=C^\flat(\vE_I,\vE_J) = C_{IJ}$ and $(\vE_K,\vE_I)_G = (\vE_I,\vE_K)_G = G_{IK}$, thus
\be
\label{eqCfM1}
C_{IJ} = \sum_K  G_{IK} C^K{}_{\!J}, \qie [C^\flat]_{|\vE} = [G]_{|\vE}.[C]_{|\vE} .
\ee
The flat notation $C^\flat$ is due to:
The top index~$I$ in~$\CIJ$ has been transformed into a bottom index in $C_{IJ}$ in~$C^\flat$, which characterizes a change of variance because of the use of an inner dot product.
%(Orthonormal basis : $[C^\flat] = [C]$.)

And \eref{eqCfM0} also gives 
$C_{IJ} = (F.\vE_I,F.\vE_J)_g = \sum_{k\ell} F^k{}_I F^\ell{}_J (\vb_k,\vb_\ell)_g$, thus
\be
C_{IJ} 
= \sum_{k\ell} F^k{}_{\!I} g_{k\ell} F^\ell{}_{\!J}
= \sum_{k\ell} (F^T)^I{}_{\!k} g_{k\ell} F^\ell{}_{\!J}
, \qie [C^\flat]_{|\vE} = ([F]_{|\vE,\ve})^T.[g]_{|\ve}.[F]_{|\vE,\ve}.
\ee
\finrep
\finexe

%%%%%%%%%%%%%%%%%%%%%%%%%%%%%%%%%%%%%%%%%%%%%%%%%%%%%%%%%%%%%%%%%%%%%%%%%%%%%%%%%%%

\subsubsection{... and remarks about $C^\flat$... and Jaumann}
\label{secCflatr}

$C^\flat$ can also be defined only with~$\dd_g$ by, for all $\vWu,\vWd\in\RRntz$,
\be
\label{eqCfla}
C^\flat_g(\vWu,\vWd) := (F.\vWu,F.\vWd)_g,
\ee
\ie, $C^\flat_g:=g^* \eqnote C^\flat$. So we can also say that $C^\flat_g$ is the pull-back of the metric~$\dd_g$ by $\Phi$, see~\eref{eqpftzdb}.

$\bullet$ However $C^\flat=C^\flat_g$ is useless in itself: $C^\flat$ is \textslbf{not} a Euclidean dot product (it is a metric defined at each~$P$ by $C^\flat_g(P)(\vWu,\vWd) := (F(P).\vWu,F(P).\vWd)_g$ for all $\vWu,\vWd\in\RRntz$ vectors at~$P$).
In fact, $C^\flat$ is only useful to characterize a deformation if the value $C^\flat(\vWu,\vWd)$ can be compared with the initial value $(\vWu,\vWd)_G$, \ie\ if a Euclidean dot product $\dd_G$ was introduced in~$\RRntz$: This is why $C^\flat$ is classically defined from~$C$, \cf~\eref{eqCfla2}.
%(it is not $C$ that is defined from~$C^\flat_g$ by $C=(C^\flat_g)^\sharp$).

%Recall: The purpose is to compare vectors relative to the initial configuration, so~$\dd_G$ is needed.
%(At~$t$, the Finger strain tensor can be used.)

$\bullet$~You may want to use the infinitesimal strain tensor $\uueps = {F+F^T \over 2} - I$, or 
the Green--Lagrange deformation tensor $E=\demi(C-I)$, obtained from $F^T:=F_{Gg}^T$ (essential).

%$\bullet$~The case $C=I=F^T.F$ gives you rigid body motions (here $F^{-1}=F^T\in\calL(\RRnt;\RRntz)$).

$\bullet$~There is no objective ``trace'' for a ${0\choose2}$ tensor like~$C^\flat$,
while $\Tr(C)$ is objective  since $C$ is an endomorphism ($\simeq$ a ${1\choose 1}$ tensor).

$\bullet$~The Lie derivatives of a second order tensor depends on the type of the tensor, and the Lie derivative of the ${1\choose 1}$ tensor like~$C$ gives the Jaumann derivative, which is usually preferred
to the Lie derivative of the ${0\choose 2}$ tensor like~$C^\flat$ which is the lower convected Lie derivative, see~remark~\ref{remDerlieCf}. 

So the introduction and use of~$C^\flat$ in mechanics mostly complicate things unnecessarily, and interferes with basic understandings like the distinction between covariance and contravariance.

\debrem
\label{remDerlieCf}
\def\tC{\tilde C}%
Interpretation issue (with Jaumann).

$2\calD = d\vv+d\vv^T$ gives $2{D \calD \over Dt}
= {D (d\vv) \over Dt} + {D (d\vv)^T \over Dt} %= d\vgamma - d\vv.d\vv + d\vgamma^T - d\vv^T.d\vv^T
= d\vgamma + d\vgamma^T - d\vv.d\vv - d\vv^T.d\vv^T$, thus, with~\eref{eqdPhi33} and keeping in mind the matrix meaning,
\be
\label{eqCpp2}
\eqalign{
C''(t)
= & F(t)^T.(2{D \calD \over Dt} + d\vv.d\vv + d\vv^T.d\vv^T + 2d\vv^T.d\vv)(t,p(t)).F(t) \cr
= & 2F(t)^T.({D \calD \over Dt} + \calD.d\vv + d\vv^T.\calD)(t,p(t)).F(t). \cr
}
\ee
The ${D \calD \over Dt} + \calD.d\vv + d\vv^T.\calD$ term looks like a lower-convected Lie derivative,
but with $d\vv^T$ instead of~$d\vv^*$, \cf~\eref{eqdlt02};
So you may find \eref{eqCpp2} written as $C''=2F^T.\calL_\vv \calD.F$. 
But you get disappointing results when using the the lower convected Lie derivative (Jaumann is usually preferred).
In fact, it is $\calL_\vv \calD^\flat$ (lower convected Lie derivative) that should be used, 
where $\calD_g^\flat := {d\vv^\flat_g+(d\vv^\flat_g)^T \over 2}$, to get $(C^\flat)'' = 2F^T.\calL_\vv \calD_g^\flat.F$.
\finrem
%$\bullet$ $d\vv^T$ is not (covariant) objective unlike $d\vv$ (and~$d\vv^*$), and $\calD$ is not (covariant) objective.

%(although the way the Jaumann derivative is obtained has little to do with the definition of a Lie derivative, see~footnote~\ref{footrem} page~\pageref{footrem}).
%In general the use of so-called Lie derivatives of order 2 tensors are disappointing as far as constitutive laws are concerned (when the Cauchy strain tensor is used).

\comment{
\debrem
May be the use of the (covariant objective) Lie derivatives of vectors fields $\vu$, or of differential forms~$\alpha$, could be preferred to characterize and build constitutive laws or materials, instead of the Lie derivatives of order two tensors~$T$ which are measuring tools used to give values $T(\alpha,\vu)\in \RR$ to $\vu$'s and~$\alpha$'s;
Illustration: see~\S~\ref{secelaoa} or, more precisely, %\verb+
https://www.isima.fr/leborgne/IsimathMeca/PpvObj.pdf. %+.
%Donc pour l'obtention de~\eref{eqCpp2} la structure euclidienne (non objective covariante) est essentielle.
\finrem
}

%%%%%%%%%%%%%%%%%%%%%%%%%%%%%%%%%%%%%%%%%%%%%%%%%%%%%%%%%%%%%%%%%%%%%%%%%%%%%%%%%%%

\subsection{Stretch ratio and deformed angle}

Here $\dd_g=\dd_G$, \ie\ at~$\tz$ and~$t$ we use the same Euclidean dot product, to be able to compare the lengths relative to the same unit of measurement.
(If $\dd_g\ne\dd_G$ then use $\dd_g = \mu^2\dd_G$.)

%%%%%%%%%%%%%%%%%%%%%%%%%%%%%%%%%%%%%%%%%%%%%%%%%%%%%%%%%%%%%%%%%%%%%%%%%%%%%%%%%%%

\subsubsection{Stretch ratio}

The stretch ratio at $P\in\RRntz$ between $\tz$ and~$t$ for a $\vW_P \in\RRntz$ is defined by
\be
\label{eqstrech}
\lambda(\vW_P)
:
= {||\vw_p||_G \over ||\vW_P||_G}
= {||F_P.\vW_P||_G \over ||\vW_P||_G}
\quad (= ||F_P.({\vW_P \over ||\vW_P||_G})||_G)
\ee
where $\vw_p = F_P.\vW_P$ is the deformed vector by the motion at $p=\Phi(P)$.
\Ie, in short 
\be
\forall \vW\in\RRntz \;\;\hbox{\st}\;\; ||\vW||=1,\quad
\lambda(\vW) := ||F.\vW||.  %= \sqrt{(C.\vW) \bcdot \vW} = \sqrt{C^\flat(\vW,\vW)}.
\ee
%(So you also have $\lambda(\vW) = \sqrt{(C.\vW) \bcdot \vW} = \sqrt{C^\flat(\vW,\vW)}$ with the implicit use~$\dd_G$.)
(You may find:
$\lambda(d\vX) = ||F.d\vX|| $ %= \sqrt{(C.d\vX) \bcdot d\vX}$
with $d\vX$ a unit vector(!); This notation should be avoided, see~\S~\ref{secnm}.)
%(amounts to confuse a position and a speed).

\comment{
\debexe
\def\valphap{{\vec\alpha\,'}}
\def\vbetap{{\vec\beta\,'}}
Consider a curve $\alpha : s\in[a,b] \rar \alpha(s)\in\Omegatz$ (drawn in~$\Omegatz$).
Prove: The length of the transported curve $\beta = \Phitzt\circ \alpha : [a,b] \rar \Omegat$ (drawn in~$\Omegat$)
is $|\beta|=\int_a^b \sqrt{(C(\alpha(s)).\valphap(s),\valphap(s))_G}\,ds$.

\debrep
%$\vbetap(s)=d\Phitzt(\alpha(s)).\valphap(s)$ (definition of the push-forward of vectors, \cf~\eref{eqdefrapvEivei2}).
By definition of the length, the length of~$\beta$ is $L_t := \int_a^b ||\vbetap(s)||_g\,ds$.
Here $\vbetap(s) = \Ftzt(\alpha(s)).\valphap(s)$. Thus
$||\vbetap(s)||_g^2=(\vbetap(s),\vbetap(s))_g
=(\Ftzt(\alpha(s)).\valphap(s),\Ftzt(\alpha(s)).\valphap(s))_g
=(\Ctzt(\alpha(s)).\valphap(s),\valphap(s))_G
$.
\finrep
\finexe
}

\comment{
\debdef
L'allongement unitaire est $\delta(\vW_P) = \lambda(\vW_P) - 1$ (est négatif pour une contraction).
\findef
}

%%%%%%%%%%%%%%%%%%%%%%%%%%%%%%%%%%%%%%%%%%%%%%%%%%%%%%%%%%%%%%%%%%%%%%%%%%%%%%%%%%%

\subsubsection{Deformed angle} %Glissement

Recall: The angle $\theta_\tz = \widehat{(\vWu,\vWd)}$
formed by two vectors $\vWu$ and~$\vWd$ in $\vRRntz{-}\{\vec0\}$ at $P\in\Omegatz$ is given by
$\cos(\theta_\tz)
=  {\vWu\over ||\vWu||_G} \bcdot {\vWd\over ||\vWd||_G}
$
($=  ({\vWu\over ||\vWu||_G},{\vWd\over ||\vWd||_G})_G$).
%=  {(\vWuP,\vWdP)_g \over ||\vWuP||_g||\vWdP||_g}$.

With the deformed vectors $\vw_{i} = F.\vW_{i}$ at $p = \Phitzt(P)$, the deformed angle is $\theta_t$ defined by
\be
\label{eqcofaf}
\cos(\theta_t) := \widehat{(\vwu,\vwd)}
=  {\vwu \over ||\vwu||} \bcdot {\vwd \over ||\vwd||}
=  {F.\vWu \over ||F.\vWu||}\bcdot {F.\vWd \over ||F.\vWd||}
\quad (= {(C.\vWu)\bcdot \vWd \over ||\vwu||\,||\vwd||})
%= {C^\flat(\vWu,\vWd) \over ||\vwu||\,||\vwd||})
.
\ee 

\comment{
\be
\cos(\theta_t) 
=  ({\vwu \over ||\vwu||_G},{\vwd \over ||\vwd||_G})_g
=  ({F.\vWu\over ||F.\vWu||_G},{F.\vWd \over ||F.\vWd||_G})_g
,
\ee
\ie
}

\comment{
Details with components:
That is, if $C.\va_j =  \sumin C_{ij} \va_i$, then
\be
\label{eqCdb0}
C_{ij} = \sumkn (F^T)_{ik}F_{kj}
=\sum_{k,\ell, m=1}^n ([G]^{-1})_{ik} F_{\ell k}\, g_{\ell m} F_{mj}.
\ee
}

%$\Phitz : \Omegatz \rar \Omegat$ is supposed to be a diffeomorphism. Let $\tz, t\in \RR$, $P\in\Omegatz$, and (shorten notation) $F:= \Ftzt(P)=d\Phitzt(P) \in\calL(\RRntz;\RRnt)$ (linear bijective since $\Phitz$ is a diffeomorphism).

%%%%%%%%%%%%%%%%%%%%%%%%%%%%%%%%%%%%%%%%%%%%%%%%%%%%%%%%%%%%%%%%%%%%%%%%%%%%%%%%%%%

\subsection{Decompositions of~$C$}

%%%%%%%%%%%%%%%%%%%%%%%%%%%%%%%%%%%%%%%%%%%%%%%%%%%%%%%%%%%%%%%%%%%%%%%%%%%%%%%%%%%

\subsubsection{Spherical and deviatoric tensors}

\debdef
The deformation spheric tensor is
\be
C_{sph}={1\over n}\Tr(C)\;I,
\ee
with $\Tr(C)= $ the trace of the endomorphism~$C$ (there is no ``trace'' for the ${0\choose2}$ tensor~$C^\flat$).
\findef

\debdef
The deviatoric tensor is
\be
C_{dev} = C -  C_{sph}.
\ee
\findef

(So $\Tr(C_{dev})=0$ , and $C = C_{sph} + C_{dev}$.)

%%%%%%%%%%%%%%%%%%%%%%%%%%%%%%%%%%%%%%%%%%%%%%%%%%%%%%%%%%%%%%%%%%%%%%%%%%%%%%%%%%%

\subsubsection{Rigid motion}

The deformation is rigid iff, for all $\tz,t$,
\be
\label{eqCri}
(\Ftzt)^T.\Ftzt = I,\qie \Ctzt = I,
\qwritten
C=I= F^T.F .
\ee
Thus, after a rigid body motion, lengths and angles are left unchanged.

%%%%%%%%%%%%%%%%%%%%%%%%%%%%%%%%%%%%%%%%%%%%%%%%%%%%%%%%%%%%%%%%%%%%%%%%%%%%%%%%%%%

\subsubsection{Diagonalization of~$C$}

\debprop
$C=F^T.F$ being symmetric positive,
$C$ is diagonalizable, its eigenvalues are positive, and $\vRRntz$ has an orthonormal basis made of eigenvectors of~$C$.
\finprop

\debdem
$(C(P).\vWu,\vWd)_G
= (F(P).\vWu,F(P).\vWd)_g
= (\vWu,C(P).\vWd)_G
$, thus $C$ is $\dd_G$-symmetric.

$(C.\vWu,\vWu)_G = (F.\vWu,F.\vWu)_g = ||F.\vWu||_g^2 >0$ when $\vWu\ne\vec0$,
since $F$ invertible ($\Phitzt$ is supposed to be a diffeomorphism).
Thus $C$ est $\dd_G$-symmetric definite positive real endomorphism.
% Therefore, $C$ being a real endomorphism, $C$ is diagonalizable with positive eigenvalues.
\findem

\debdef
Let  $\lambda_i$ be the eigenvalues of~$C$.
Then the $\sqrt{\lambda_i}$ are called the principal stretches.
And the associated eigenvectors give the principal directions.
\findef

%%%%%%%%%%%%%%%%%%%%%%%%%%%%%%%%%%%%%%%%%%%%%%%%%%%%%%%%%%%%%%%%%%%%%%%%%%%%%%%%%%%

\subsubsection{Mohr circle}

This \S~deals with general properties of $3*3$ symmetric positive endomorphism, like~$\Ctzt(P)$.

Consider $\vRRt$ with a Euclidean dot product $\dd_{\RRt}$ and a $\dd_\RRt$-orthonormal basis~$(\va_i)$.

Let $\calM : \vRRt\rar\vRRt$ be a symmetric positive endomorphism.
Thus $\calM$ is diagonalizable in a $\dd_\RRt$-orthonormal basis $(\ve_1,\ve_2,\ve_3)$, that is,
$\exists \lambda_1,\lambda_2,\lambda_3 \in \RR$, $\exists \ve_1,\ve_2,\ve_3\in\vRRt$ \st
\be
\calM.\ve_i = \lambda_i\ve_i \qand (\ve_i,\ve_j)_\RRt = \delta_{ij}, \qso
[\calM]_{|\ve} = {\rm diag}(\lambda_1,\lambda_2,\lambda_3)=\pmatrix{\lambda_1&0&0\cr0&\lambda_2&0\cr0&0&\lambda_3}.
\ee
%So $[\calM]_{|\ve} = {\rm diag}(\lambda_1,\lambda_2,\lambda_3)=\pmatrix{\lambda_1&0&0\cr0&\lambda_2&0\cr0&0&\lambda_3}$. %(The matrice $[M^i_j]:=[\calM]_{|\va}$ is not supposed to be diagonal).
And the orthonormal basis $(\ve_1,\ve_2,\ve_3)$ is ordered \st\ $\lambda_1 \ge \lambda_2 \ge \lambda_3$ ($> 0$).

Let $S$ be the unit sphere in~$\RR^3$, that is the set $\{(x,y,z) : x^2+y^2+z^2=1\}$.
Its image $\calM(S)$ by~$\calM$ is the ellipsoid
$\{(x,y,z) : {x^2\over \lambda_1^2}+{y^2\over \lambda_2^2}+{z^2\over \lambda_3^2}=1\}$.
Then consider $\vn=\sum_i n_i\ve_i$ \st\ $||\vn||_\RRt=1$:
\be
\label{eqmohr1}
[\vn]_{|\ve}=\pmatrix{n_1\cr n_2\cr n_3}\qwith   n_1^2+n_2^2+n_3^2=1.
\ee
Thus its image $\vA=\calM.\vn \in \calM(S)$ satisfies
\be
\label{eqdefvA}
\vA=\calM.\vn,\quad [\vA]_{|\ve} = \pmatrix{\lambda_1 n_1\cr\lambda_2 n_2\cr\lambda_3 n_3\cr}.
\ee
Then define
\be
\label{eqdefvA2}
A_n = (\vA,\vn)_{\RR^3}, % \eqnote \vA\cdot \vn, 
\quad \vA_\perp=\vA-A_n\vn,\quad A_\perp \eqdef ||\vA_\perp||.
\ee
So $\vA=A_n \vn+ \vA_\perp\;\in\; \Vect\{\vn\}\otimes \Vect\{\vn\}^\perp$.
(Remark: $\vA_\perp$ is not orthonormal to the ellipsoid $\calM(S)$, but is orthonormal to the initial sphere~$S$.)

\medskip
\noindent
{\bf Mohr Circle purpose:} To find a relation:
\be
A_\perp = f(A_n),
\ee
relation between ``the normal force $A_n$'' (to the initial sphere) and the ``tangent force$ A_\perp$'' (to the initial sphere).

\eref{eqmohr1}, \eref{eqdefvA} and $A_n=(\calM.\vn,\vn)_{\RR^3}$ give
\be
\label{eqmohr2}
\left\{\eqalign{
& n_1^2+n_2^2+n_3^2=1, \cr
& \lambda_1n_1^2+\lambda_2n_2^2+\lambda_3n_3^2 = A_n \cr
& \lambda_1^2n_1^2+\lambda_2^2n_2^2+\lambda_3^2n_3^2 = ||\vA||^2=A_n^2+A_\perp^2.
}\right.
\ee
\comment{
That is,
$
\left\{\eqalign{
& n_1^2+n_2^2+n_3^2=1, \cr
& \lambda_1n_1^2+\lambda_2n_2^2+\lambda_3n_3^2 = A_n, \cr
& \lambda_1^2n_1^2+\lambda_2^2n_2^2+\lambda_3^2n_3^2 = A_n^2+A_\perp^2.\cr
}\right\}
$
}
This is linear system with the unknowns $n_1^2,n_2^2,n_3^2$. The solution is
\be
\left\{\eqalign{
 & n_1^2= {A_\perp^2 + (A_n-\lambda_2)(A_n-\lambda_3)\over (\lambda_1-\lambda_2)(\lambda_1-\lambda_3)}, \cr
 & n_2^2= {A_\perp^2 + (A_n-\lambda_3)(A_n-\lambda_1)\over (\lambda_2-\lambda_3)(\lambda_2-\lambda_1)}, \cr
 & n_3^2= {A_\perp^2 + (A_n-\lambda_1)(A_n-\lambda_2)\over (\lambda_3-\lambda_1)(\lambda_3-\lambda_2)}. \cr
}\right.
\ee
The $n_i^2$ being non negative, and with $\lambda_1 > \lambda_2>\lambda_3\ge0$, we get
\be
\label{eqmohr4}
\left\{\eqalign{
 & A_\perp^2 + (A_n-\lambda_2)(A_n-\lambda_3)\ge0, \cr
 & A_\perp^2 + (A_n-\lambda_3)(A_n-\lambda_1)\le0, \cr
 &A_\perp^2 + (A_n-\lambda_1)(A_n-\lambda_2)\ge0. \cr
}\right.
\ee
Then let $x=A_n$ and $y=A_\perp$, and consider, for some $a,b\in\RR$, the equation
$$
y^2+(x-a)(x-b)=0, \qso  (x-{a{+}b\over2})^2+y^2={(a{-}b)^2\over4}.
$$
This is the equation of a circle centered at $({a{+}b\over2},0)$ with radius ${|a{-}b|\over2}$.

Thus~\eref{eqmohr4}$_2$ tells that $A_n$ and $A_\perp$ are inside the circle
centered at $({\lambda_1+\lambda_3\over2},0)$ with radius ${\lambda_1-\lambda_3\over2}$,
and~\eref{eqmohr4}$_{1,3}$ tell that $A_n$ and $A_\perp$ are outside the other circles
(adjacent and included in the first, drawing).

\debexe
What happens if $\lambda_1=\lambda_2=\lambda_3>0$?

\debrep
Then
$
\left\{\eqalign{
& n_1^2+n_2^2+n_3^2=1, \cr
& n_1^2 + n_2^2 + n_3^2 = {A_n\over\lambda_1}, \cr
& n_1^2+n_2^2+n_3^2 = {A_n^2+A_\perp^2\over\lambda_1^2}.\cr
}\right\}
$
Thus $A_n=\lambda_1$ and $A_n^2+A_\perp^2 = \lambda_1^2$, thus $A_\perp=0$.
Here $C=\lambda_1 I$, and we deal with a dilation: $A_\perp=0$.
\finrep
\finexe

\debexe
What happens if $\lambda_1=\lambda_2>\lambda_3>0$?

\debrep
Then
$
\left\{\eqalign{
& n_1^2+n_2^2+n_3^2=1, \cr
& \lambda_1 (1 - n_3^2) + \lambda_3 n_3^2 = A_n, \cr
& \lambda_1^2 (1 - n_3^2) + \lambda_3^2 n_3^2 = A_n^2+A_\perp^2.\cr
}\right\}
$
Thus $A_n = \lambda_1 - (\lambda_1-\lambda_3) n_3^2 \in [\lambda_3,\lambda_1]$,
and $A_\perp= \pm (\lambda_1^2 - (\lambda_1^2-\lambda_3^2) n_3^2 - A_n^2)^\demi$,
with $A_n^2+A_\perp^2$ a point on the circle with radius $\lambda_1^2 (1 - n_3^2) + \lambda_3^2 n_3^2$.
\finrep
\finexe

%%%%%%%%%%%%%%%%%%%%%%%%%%%%%%%%%%%%%%%%%%%%%%%%%%%%%%%%%%%%%%%%%%%%%%%%%%%%%%%%%%%

\subsection{Green--Lagrange deformation tensor $E$}
\label{secGL}

\eref{eqdefCt} gives
$(\vwu,\vwd)_g = (F.\vWu,F.\vWd)_g %= (F_P^T. F_P.\vWuP,\vWdP)_G 
= (C.\vW,\vW)_G$ at $p=\Phi(P)$, thus
%\eref{eqdefCt} gives
\be
\label{eqGSV1}
(\vwu,\vwd)_g - (\vWu,\vWd)_G
=  ((C-I).\vWu,\vWd)_G  .
\ee

\debdef
The Green--Lagrange tensor (or Green--Saint Venant tensor) at~$P$ relative to $\tz$ and~$t$
is the endomorphism $E^\tz_t(P) \in \calL(\RRntz ; \RRntz)$ defined by
\be
\label{eqdefGSV}
\Etzt(P) \eqdef {\Ctzt(P)-I_\tz \over 2 } ,\qinshort \boxed{E= {C-I\over 2}} \quad (= {F^T.F - I\over 2}).
\ee
(In particular $E=0$ for rigid body motions.)
And
$\Etzt :\Omegatz \rar\calL(\RRntz;\RRntz)$ is the Green--Lagrange tensor relative to $\tz$ and~$t$.
\findef

%thus $2E$ quantifies the angle of vectors after their transport by the flow (vectors initially at right angle). 
The $\demi$ because $(C.,.)=(F.,F.)$ corresponds to the ``motion squared'', see the following linearization.

\medskip
And we get the time Taylor expansion of $\EtzP(t) = \demi (\CtzP(t) - I_\tz)$ with $p(t)=\PhitzP(t)$
and~\eref{eqCpe}:
\be
\eqalign{
\EtzP(t{+}h)
= & \FtzP(t)^T.\bigl(h\,{d\vv + d\vv^T \over 2}
+  {h^2\over 2}\,({d\vgamma + d\vgamma^T \over 2} + d\vv^T. d\vv)\bigr)(t,p(t)).\FtzP(t)  + o(h^2) \cr
= & \FtzP(t)^T.\bigl(h\,\calD
+ h^2\,(({D \calD \over Dt} + \calD.d\vv + d\vv^T.\calD)(t,p(t))).\FtzP(t)  + o(h^2). \cr
}
\ee
%See \S~\ref{remDerlieCf}.

%%%%%%%%%%%%%%%%%%%%%%%%%%%%%%%%%%%%%%%%%%%%%%%%%%%%%%%%%%%%%%%%%%%%%%%%%%%%%%%%%%%

\subsection{Small deformations (linearization): The infinitesimal strain tensor $\protect\uueps$}
\label{secsdl}

%%%%%%%%%%%%%%%%%%%%%%%%%%%%%%%%%%%%%%%%%%%%%%%%%%%%%%%%%%%%%%%%%%%%%%%%%%%%%%%%%%%

\subsubsection{Landau notations big-$O$ and little-$o$}

Reminder. Let $f,g:\RR\rar\RR$ and $x_0\in\RR$.
\be
\hbox{$f=O(g)$ near $x_0$} \quad\Longleftrightarrow\quad
\exists C>0,\; \exists \eta>0,\; \forall x \;\st\ |x-x_0|<\eta,\; |f(x)|<C|g(x)|.
\ee
and $f$ is said to be ``comparable with~$g$'' near~$x_0$.

If $|g|>0$ then the conclusion reads ${|f(x)|\over |g(x)|} < C$;
And $f=O(x^n)$ near $x{=}0$ iff ${|f(x)| \over |x^n|}<C$ near $x{=}0$.

\medskip
\noindent
And
\be
\hbox{$f=o(g)$ near $x_0$} \quad\Longleftrightarrow\quad
\forall \eps>0, \; \exists \eta>0,\; \forall x \;\st\ |x-x_0|<\eta,\; |f(x)|<\eps|g(x).
\ee
and $f$ is said to be ``negligible compared with $g$ near~$x_0$''.

If $|g|>0$ then the conclusion reads ${|f(x)|\over |g(x)|} \mrar_{x\rar x_0}0$.
And $f=o(x^n)$ near $x{=}0$ iff ${|f(x)|\over |x^n|} \mrar_{x\rar 0}0$.

\comment{
Consider a motion
$\tPhi:
\left\{\eqalign{
[\tz,T] \times \Obj &\rar \RRn \cr
(t,\Pobj) &\rar p(t)= \tPhi(t,\Pobj) \cr
}\right\}
$
of a material object~$\Obj$, and $\Omegat = \tPhi(t,\Obj)$ for all~$t$.
Consider the associated motion
$\Phitz:
\left\{\eqalign{
[\tz,T] \times \Omegatz &\rar \RRn \cr
(t,P) &\rar p(t)= \Phitz(t,P) \cr
}\right\}
$
where $\Phitz(t,P):=\tPhi(t,\Pobj)$ when $P=\tPhi(\tz,\Pobj)$, supposed to be regular. And consider the deformation gradient
$\Ftz:
\left\{\eqalign{
[\tz,T] \times \Omegatz &\rar \calL(\RRntz;\RRnt) \cr
(t,P) &\rar \Ftz(t,P) = d\Phitz(t,P) \cr
}\right\}
$. 
}

%%%%%%%%%%%%%%%%%%%%%%%%%%%%%%%%%%%%%%%%%%%%%%%%%%%%%%%%%%%%%%%%%%%%%%%%%%%%%%%%%%%

\subsubsection{Definition of the infinitesimal strain tensor $\protect\uueps$}
\label{seceps}

The motion is supposed to be~$C^2$.
Along a trajectory, with $\FtzP(\tz)=I$ we have, near~$\tz$,
%$\FtzP(t{+}h) = \FtzP(t) + h\,(\Ftzt)'(t)+ o(h) = \FtzP(t) + O(h)$. In particular 
\be
\label{equuepsm00}
\FtzP(\tz{+}h) = I + O(h),
\ee
thus $\FtzP(\tz{+}h).\vW = \vW + O(h)$ for all $\vW\in\RRntz$, \ie, near~$\tz$,
\be
\label{equuepsm01}
||\vw - \vW|| = O(h) \qwhen \vw=\FtzP(\tz{+}h).\vW.
\ee
This supposes the use of a unique inner dot product $\dd_G=\dd_g$ at all time, and~\eref{equuepsm01} means
$||\vw - \vW||_g = O(h)$ near~$\tz$.

%And the small displacement displacement hypothesis is applicable when $h$ is small enough.

%\medskip So, a unique Euclidean dot product $\dd_G=\dd_g$ is given in~$\vRRntz$ and in~$\vRRnt$.

%Moreover the inner dot product $\dd_g$ is supposed Euclidean (in practice), and a $\dd_g$-Euclidean basis $(\ve_i)$ is chosen, the same at all~$t$. Then:

%Thus $[F^T]=[F]^T$, \cf~\eref{eqr2}, more precisely $[(\Ftzt)^T(p)]_{|\ve} = [\Ftzt(P)]_{|\ve}$ when $p=\Phitzt(P)$.

\debdef
If $\dd_g$ is an inner dot product, the same at all time, and if $(\ve_i)$ is a $\dd_g$-orthonormal basis, the same at all time, then the infinitesimal strain tensor at~$P$ is the matrix defined by
\be
[\uueps(P)]_{|\ve} = {[F(P)]_{|\ve} + [F(P)]_{|\ve}^T \over 2} - [I],
\ee
abusively written in short,
\be
\label{equuepsm}
\uueps \eqdef {F + F^T\over 2} - I \quad\hbox{(matrix meaning)}.
\ee
(And more precisely, at $P\in\Omegatz$ and between $\tz$ and~$t$,
$[\uuepstzt(P)]_{|\ve} = {[\Ftzt(P)]_{|\ve} + [\Ftzt(P)]_{|\ve}^T \over 2} - [I]$.)
\findef

So %, when $\vW\in\RRntz$, the quantity 
$\uueps.\vW = {F.\vW + F^T.\vW\over 2} - .\vW$ 
means
$[\uueps]_{|\ve}.[\vW]_{|\ve}= {[F]_{|\ve}.[\vW]_{|\ve} + [F]_{|\ve}^T.[\vW]_{|\ve}\over 2} - [\vW]_{|\ve}$.

\debrem
\label{remuuepsnaf}
$\uueps$ in~\eref{equuepsm} can\textslbf{not} be a tensor (cannot be a function) since
$\Ftzt(P) : \vRRntz \rar \vRRnt$ 
and $\Ftzt(P)^T %= (\Ftzt)^T(p) 
: \vRRnt \rar \vRRntz$ and $I_\tz : \vRRntz \rar \vRRntz$ don't have the same definition domain.

So $\uueps$ is not a function, is not a tensor: It is a matrix...  But is called ``the infinitesimal strain tensor''... 
\finrem

\debprop
The Green--Lagrange tensor $E={F^T.F - I \over 2}\in\calL(\RRntz;\RRntz)$ satisfies near~$\tz$:
%, and relative to the chosen $\dd_g$-Euclidean basis,
\be
\label{equuepsm2}
E = \uueps + o(t{-}\tz) 
\quad (={F + F^T \over 2} - I  + o(t{-}\tz)) \quad\hbox{(matrix meaning)},
\ee
which means $[E] = [\uueps] + o(t{-}\tz)$.
Thus, ``for small deformations'' we write $E \simeq \uueps$, \ie\ $E \simeq {F + F^T \over 2} - I$.

\comment{
(Full notation: $[\Etzt(P)]_{|\ve}
= [\uuepstzt(P)]_{|\ve} + o(t{-}\tz)$, with
$[\uuepstzt(P)]_{|\ve}= {[\Ftzt(P)]_{|\ve} + [\Ftzt(P)]_{|\ve}^T \over 2} - [I]$.)
}
Interpretation: \eref{equuepsm2} is a linearization of~$E$, since we keep the linear part of
the ``quadratic'' $E = \demi(F^T.F- I)$ %=\demi(C-I)
%$E.\vW=\demi(C-I).\vW = \demi(F^T.F.\vW- \vW)$
given by $(E.\vW,\vU)_g= \demi \bigl((F.\vW,F.\vU)_g - (\vW,\vU)_g\bigr)$ for all $\vU,\vW\in\RRntz$ (``motion squared'' \cf\ the $(F\cdot,F\cdot)_g$ term).

\finprop

\debdem
A $\dd_g$-orthonormal basis being chosen, $[F^T] \mope^{\eref{eqr2}} [F]^T$, thus 
$[C] %=[F^T.F]=[F^T].[F]
= [F]^T.[F]$, thus %\eref{eqdefGSV} gives % (matrix meaning with a chosen )
\be
\label{eqdefap}
2[E] = [C] - [I] = [F]^T.[F] - [I] = ([F]^T-[I).([F]-[I) + [F]^T + [F] - 2[I].
\ee
Then, near~$\tz$ and with $h=t{-}\tz$, \eref{equuepsm00} gives
$([F]^T-[I]).([F]-I]) = O(h) O(h) = O(h^2)$, thus
$2[E] = [F]^T + [F] - 2[I] + O(h)$, 
thus~\eref{equuepsm2}.
\findem

%%%%%%%%%%%%%%%%%%%%%%%%%%%%%%%%%%%%%%%%%%%%%%%%%%%%%%%%%%%%%%%%%%%%%%%%%%%%%%%%%%%
%%%%%%%%%%%%%%%%%%%%%%%%%%%%%%%%%%%%%%%%%%%%%%%%%%%%%%%%%%%%%%%%%%%%%%%%%%%%%%%%%%%

\section{Finger tensor $F.F^T$ (left Cauchy--Green tensor)}

Finger's approach is consistent with the foundations of relativity (Galileo classical relativity or Einstein general relativity):
We can only do measurements at the current time~$t$, and we can refer to the past.

There is a lot of misunderstandings, as was the case for the Cauchy--Green deformation tensor~$C$, due to the lack of precise definitions:
Definition domain? Value domain? Points at stake ($p$ or $P$)? Euclidean dot product (English? French?)?
Covariance? Contravariance?...

%, see hypothesis \S~\ref{sechypNE}.
%but we cannot refer to the future.

%%%%%%%%%%%%%%%%%%%%%%%%%%%%%%%%%%%%%%%%%%%%%%%%%%%%%%%%%%%%%%%%%%%%%%%%%%%%%%%%%%%

\subsection{Definition}

Let $\tPhi$ be motion, $\tz\in\RR$, $\Phitz$ the associated motion, $P\in\Omegatz$,
$t\in\RR$, and $\Ftzt(P) := d\Phitzt(P) \in \calL(\vRRntz;\vRRnt)$. % le gradient de déformation (covariant).
And let $\dd_G$ and $\dd_g$ be Euclidean dot products in~$\vRRntz$ and~$\vRRnt$.

\debdef
The Finger tensor $\uubtzt(\pt)$, or left Cauchy--Green deformation tensor,
at~$t$ at~$\pt$ relative to~$\tz$ is the endomorphism $\in\calL(\vRRnt;\vRRnt)$ defined by, with $P = \Phitzt^{-1}(\pt)$,
\be
\label{eqbtzt}
\uubtzt(\pt) \eqdef \Ftzt(P).(\Ftzt)_{Gg}^T(\pt) \quad\hbox{written in short}\quad \boxed{b=F.F^T} ,
\ee
\ie\ is defined by
$
(\uubtzt(\pt).\vwu,\vwd)_g
=  (\Ftzt(P)^T.\vwu,\Ftzt(P)^T.\vwd)_G 
= ((\Ftzt)^T(\pt).\vwu,(\Ftzt)^T(\pt).\vwd)_G,
$
for all $\vwu,\vwd$ vectors at $\pt \in\Omegat$, written in short
\be
(\uub.\vw_{1},\vw_{2})_g =  (F^T.\vw_{1},F^T.\vw_{2})_G.
\ee
(To compare with $C=F^T.F$ and $(C.\vW_{1},\vW_{2})_G =  (F.\vW_{1},F.\vW_{2})_g$.)
\findef

And the Finger tensor relative to~$\tz$ is
\be
\uubtz :
\left\{\eqalign{
\bigC = \bigcup_t(\{t\} \times \Omegat) & \rar \calL(\vRRnt;\vRRnt) \cr
(t,\pt) & \rar \uubtz(t,\pt) \eqdef \uubtzt(\pt).
}\right.
\ee
NB: $\uubtz$ looks like a Eulerian function, but isn't, since it depends on a~$\tz$.

Other definition found:
\be
\Btzt := \uubtzt \circ (\Phitzt)^{-1}, \qie
\Btzt(P) :=\uubtzt(\pt) = \Ftzt(P).\Ftzt(P)^T, \qwritten B=F.F^T.
\ee
Pay attention: $\Btzt(P)\in \calL(\vRRnt;\vRRnt)$ is an endomorphism at~$t$ at~$\pt$, not at~$\tz$ at~$P$:
\Eg, $\Btzt(P).\vw_t(\pt) = \uubtzt(\pt).\vw_t(\pt)$ is meaningful, while $\Btzt(P).\vW_\tz(P)$ is absurd.
%et $\Btzt(P).\vw_t(\pt) = \Ftzt(P).\Ftzt(P)^T.\vw_t(\pt)$ pour tout $\vw_t(\pt)$ vecteur en $\pt\in\Omegat$.

\debrem
\label{rempfC}
The push-forward by $\Phi:=\Phitzt$  of the Cauchy--Green deformation tensor $C = F^T.F$ is
$\Phi_*(C)=F.C.F^{-1} = F.F^T = \uub$, \cf~\eref{eqpfL}: It is the Finger tensor.
So the endomorphism $C$ in~$\RRntz$ is the pull-back of the endomorphism $\uub$ in~$\RRnt$.
(However a push-forward and a pull-back don't depend on any inner dot product while the transposed $F^T$ does...).
%The Finger tensor, or more precisely its flat representation~$b^\flat$, can also be interpreted as being the metric $\dd_g$ at~$t$ (with the help of the Riesz representation theorem), since $C^\flat$ is the pull-back of the metric~$\dd_g$.
\finrem

%%%%%%%%%%%%%%%%%%%%%%%%%%%%%%%%%%%%%%%%%%%%%%%%%%%%%%%%%%%%%%%%%%%%%%%%%%%%%%%%%%%

\subsection{$\protect\uub^{-1}$}

With pull-backs (towards the virtual power principle at~$t$).
With $\pt=\Phitzt(P)$ and $\vW_i(P) = (\Ftzt(P))^{-1}.\vw_i(\pt)$:
\be
(\vW_1,\vW_2)_G  = (F^{-1}.\vw_1,F^{-1}.\vw_2)_G = (F^{-T}.F^{-1}.\vw_1,\vw_2)_g
= (\uub^{-1}.\vw_1,\vw_2)_g.
\ee
So $\uub^{-1}:=(\uubtzt)^{-1}$ is useful:
\be
\label{equbbmu0}
(\uubtzt)^{-1} :
\left\{\eqalign{
\Omegat &\rar \calL(\RRnt;\RRnt) \cr
\pt & \rar (\uubtzt)^{-1}(\pt) = \Ftzt(P)^{-T}.\Ftzt(P)^{-1} = \Htzt(\pt)^T.\Htzt(\pt)
}\right. 
\ee
with $\pt= \Phitzt(P)$ and $\Htzt(\pt) = (\Ftzt(P))^{-1}$ \cf~\eref{eqHtz}. Thus we can define
\be
\label{equbbmu1}
(\uubtz)^{-1} :
\left\{\eqalign{
\bigcup_t(\{t\}\times \Omegat) &\rar \calL(\RRnt;\RRnt) \cr
(t,\pt) & \rar (\uubtz)^{-1}(t,\pt) \eqdef (\uubtzt)^{-1}(\pt).
}\right. 
\ee
Remark: $(\uubtz)^{-1}$ looks like a Eulerian function, but isn't, since it depends on~$\tz$.

In short:
\be
\label{equbbmu}
\uub^{-1} = H^T.H %= F^{-T}.F^{-1}
, \quad\hbox{to compare with}\quad C=F^T.F,
\ee
and with $\vw=F.\vW$,
\be
\uub^{-1}.\vw %= F^{-T}.\vW
= H^T.\vW,
\quad\hbox{to compare with}\quad C.\vW = F^T.\vw,
\ee
and with $\vW_i= F^{-1}.\vw_i$, \ie\ $\vw_i= F.\vW_i$,
\be
\label{equubmu}
(\uub^{-1}.\vw_1,\vw_2)_g = (\vW_1,\vW_2)_G,
\quad\hbox{to compare with}\quad(C.\vW_1,\vW_2)_G = (\vw_1,\vw_2)_g.
\ee

\debrem
$\pt=\Phitzt(P)$ and $b(\pt) = F(P).F(P)^T$ and $C(P)=F(P)^T.F(P)$ give
\be
\uub(\pt).F(P) = F(P).C(P), %\qie \uub(\pt) = F(P).C(P).F^{-1}(\pt),
\ee
written $\uub = F.C.F^{-1}$. Thus $\uub^{-1} = F.C^{-1}.F^{-1}$, so
\be
\Phitztb\uub^{-1} = F^{-1}.\uub^{-1}.F = F^{-1}.F^{-T} = (F^T.F)^{-1} = C^{-1},
\ee
\ie\ the pull-back of~$\uub^{-1}$ is $C^{-1}$, \ie\
$\uub^{-1}$ is the push-forward of~$C^{-1}$.
%However see remark~\ref{rempfC}.
\finrem

%%%%%%%%%%%%%%%%%%%%%%%%%%%%%%%%%%%%%%%%%%%%%%%%%%%%%%%%%%%%%%%%%%%%%%%%%%%%%%%%%%%

\subsection{Time derivatives of $\protect \uub^{-1}$}

With~\eref{equbbmu1} let $(\uubtz)^{-1} \eqnote \uub^{-1} = H^T.H$.
Thus, along a trajectory, and with~\eref{eqdPhi3}, we get
\be
\label{eqHtz3}
\eqalign{
{D\uub^{-1} \over Dt}
= & {D H^T \over Dt}.H + H^T.{D H \over Dt} = - d\vv^T.H^T.H - H^T.H.d\vv \cr
= &  - \uub^{-1}.d\vv - d\vv^T.\uub^{-1}.
}
\ee
\comment{
So,
\be
\label{eqHtz4}
\eqalign{
{D^2\uub^{-1} \over Dt^2}
= & - {D\uub^{-1} \over Dt}. d\vv - \uub^{-1}.{D(d\vv) \over Dt}+ (idem)^T \cr
= & (\uub^{-1}.d\vv + d\vv^T.\uub^{-1}) . d\vv - \uub^{-1}.(d\vgamma - d\vv.d\vv)
+ d\vv^T.(\uub^{-1}.d\vv + d\vv^T.\uub^{-1}) - (d\vgamma^T - d\vv^T.d\vv^T).\uub^{-1} \cr
= &\uub^{-1}.(2d\vv.d\vv - d\vgamma) + 2 d\vv^T.\uub^{-1}.d\vv + (2d\vv^T.d\vv^T - d\vgamma^T).\uub^{-1} \cr
= &\uub^{-1}.(d\vv.d\vv - {D(d\vv) \over Dt}) + 2 d\vv^T.\uub^{-1}.d\vv + (d\vv^T.d\vv^T - {D(d\vv)^T \over Dt}).\uub^{-1}.
}
\ee
}

\debexe
Prove~\eref{eqHtz3} with~\eref{equubmu}.

\debrep
\eref{equubmu} gives
${D\over Dt}(\uub^{-1}.\vw_1,\vw_2)_g
=0
= ({D\uub^{-1}\over Dt}.\vw_1,\vw_2)_g + (\uub^{-1}.{D\vw_1\over Dt},\vw_2)_g 
+ (\uub^{-1}.\vw_1,{D\vw_2\over Dt})_g
$,
and $\vw_i(t,p(t)) = \Ftz(t,P).\vW\tz(P)$ gives ${D\vw_i \over Dt} = d\vv.\vw_i$,
thus
$({D\uub^{-1}\over Dt}.\vw_1,\vw_2)_g + (\uub^{-1}.d\vv.\vw_1,\vw_2)_g 
+ (\uub^{-1}.\vw_1,d\vv.\vw_2)_g
=0
$,
thus~\eref{eqHtz3}.
\finrep
\finexe

\debexe
Prove~\eref{eqHtz3} with $F^T.\uub^{-1}.F=I_\tz$.

\debrep
$\uub^{-1} = (F.F^T)^{-1} = F^{-T} . F^{-1}$ gives $F^T.\uub^{-1}.F=I_\tz$,
thus $(F^T)'.\uub^{-1}.F
+F^T.{D\uub^{-1} \over Dt}.F
+F^T.\uub^{-1}.F' = 0$,
thus
$F^T.d\vv^T.\uub^{-1}.F
+F^T.{D\uub^{-1} \over Dt}.F
+F^T.\uub^{-1}.d\vv.F = 0$,
thus~\eref{eqHtz3}.
\finrep
\finexe

\comment{
\debrem
Une fois encore, l'utilisation a priori du produit scalaire euclidien ne semble pas satisfaisant :
\eref{eqHtz3} ne donne pas
${D\uub^{-1} \over Dt}$ égal à $- (d\vv + d\vv^T)$ comme espérer (taux de déformation mesuré à~$t$).
Voir~\S~\ref{secPK}.
Pour se passer du produit scalaire euclidien (et du transposé),
on peut exprimer les taux de déformations à l'aide des dérivées de Lie,
pour, a posteriori, utiliser un produit scalaire euclidien et retrouver les résultats classiques
dans le cadre des approximations linéaires.
\finrem
}

%%%%%%%%%%%%%%%%%%%%%%%%%%%%%%%%%%%%%%%%%%%%%%%%%%%%%%%%%%%%%%%%%%%%%%%%%%%%%%%%%%%
%%%%%%%%%%%%%%%%%%%%%%%%%%%%%%%%%%%%%%%%%%%%%%%%%%%%%%%%%%%%%%%%%%%%%%%%%%%%%%%%%%%

\subsection{Euler--Almansi tensor $\protect\uua$}
\label{seceulal}

%\debrem
Euler--Almansi approach is consistent with the foundations of relativity (Galileo relativity or Einstein general relativity): %, see hypothesis \S~\ref{sechypNE}:
We can only do measurements at the current time~$t$, and we can refer to the past.

%\finrem

At~$t$ in~$\Omegat$, %, we use the pull-back
consider the Finger tensor $\uub=F.F^T$
and its inverse $\uub^{-1} = F^{-T}.F^T= H^T.H$ \cf~\eref{equbbmu}.

\debdef
Euler--Almansi tenor at $\pt\in\Omegat$ is the endomorphism $\uuatzt(\pt) \in \calL(\vRRnt;\vRRnt)$ defined by
\be
\label{eqEA1}
\uuatzt(\pt) = \demi(I_t-\uubtzt(\pt)^{-1})= \demi(I_t- H(\pt)^T.H(\pt)),
\ee
written
\be
\label{eqEA1b}
\uua = \demi(I-\uub^{-1}) = \demi(I- H^T.H),
\ee
to compare with the Green--Lagrange tensor $E = \demi(C-I) = \demi(F^T.F-I) \in \calL(\vRRntz;\vRRntz)$.
\findef

Remark: $\uuatz$ looks like a Eulerian function, but isn't, since it depends on~$\tz$.

\medskip
\eref{equubmu} gives ($\vw_i = F.\vW_i$)
\be
\label{eqEA2}
(\vw_1,\vw_2)_g - (\vW_1,\vW_2)_G  = 2(\uua.\vw_1,\vw_2)_g,
\ee
to compare with $(\vw_1,\vw_2)_g - (\vW_1,\vW_2)_G =2(E.\vW_1,\vW_2)_G$.
(This also gives  $(\uua.\vw_1,\vw_2)_g = (E.\vW_1,\vW_2)_G$.)
And~\eref{eqEA1b} gives
\be
\label{eqEA0}
F^T.\uua.F=E, \qie \uua = F^{-T}.E.F^{-1},
\ee
standing for $\Ftzt(P)^T.\uuatzt(p).\Ftzt(P) = \Etzt(P)$ when $p = \Phitzt(P)$.

\debrem
$\uuatzt$ is not the push-forward of~$\Etzt$ by~$\Phitzt$ (the push-forward is $F.E.F^{-1}$).
\comment{, which is not a surprise:
A push-forward is independent of any inner dot product, whereas the transposed $F^T$ does depend on an inner dot product.
Once again, an imposed Euclidean structure does not allow to go ``objectively''
from a configuration at~$\tz$ to a configuration at~$t$.}
\finrem

%%%%%%%%%%%%%%%%%%%%%%%%%%%%%%%%%%%%%%%%%%%%%%%%%%%%%%%%%%%%%%%%%%%%%%%%%%%%%%%%%%%

\subsection{Time Taylor expansion for $\protect \uua$}

\eref{eqHtz3} gives % and~\eref{eqHtz4}
\be
\label{eqHtz5}
{D\uua \over Dt} = {\uub^{-1}.d\vv + d\vv^T.\uub^{-1}\over 2}.
\ee
\comment{
and with $\uub^{-1} = F^{-T}.F^{-1}$ we get
\be
F^T.{D\uua \over Dt}.F
= {F^{-1}.d\vv.F + F^T.d\vv^T.F^{-T}\over 2}
= {\Phitztb(d\vv) + \Phitztb(d\vv)^T\over 2},
\ee
où $\Phitztb(d\vv)$ est le pull-back de~$d\vv_t$
(on entrevoit une possible formulation objective des puissances à~$t$ ramenées sur une configuration initiale à l'aide de~$\Phitztb(d\vv)$).

Then
\be
{D^2\uua \over Dt^2}
= {\uub^{-1}.( d\vgamma- 2d\vv.d\vv) + (d\vgamma^T - 2d\vv^T.d\vv^T).\uub^{-1} - 2 d\vv^T.\uub^{-1}.d\vv \over 2}.
\ee
\comment{
donc :
\be
\eqalign{
F^T.{D^2\uua \over Dt^2}.F
= & {F^{-1}.(d\vgamma- 2d\vv.d\vv).F + F^T.(d\vgamma^T - 2d\vv^T.d\vv^T).F^{-T}
 - 2 F^T.d\vv^T.F^{-T}.F^{-1}.d\vv.F \over 2} \cr
= & {\Phitztb(d\vgamma- 2d\vv.d\vv) + \Phitztb(d\vgamma- 2d\vv.d\vv)^T \over 2}
 -  \Phitztb(d\vv)^T.\Phitztb(d\vv). \cr
}
\ee
}
}

%%%%%%%%%%%%%%%%%%%%%%%%%%%%%%%%%%%%%%%%%%%%%%%%%%%%%%%%%%%%%%%%%%%%%%%%%%%%%%%%%%%

\subsection{Almansi modified Infinitesimal strain tensor $\protect\tuueps$}
\label{secea}

We are at~$t$ (present time) and remember the past: We prefer a definition of a infinitesimal strain tensor $\tuueps$ from the Euler--Almansi tensor~$\uua$,
instead of~$\uueps$ from Euler--Lagrange tensor~$E$, \cf~\S~\ref{seceps}.

Same Euclidean framework as in~\S~\ref{seceps}, and matrix meaning again.

We have $I-\uub^{-1} = I-H^T.H =-(I-H^T).(I-H) + 2I - H^T - H$ where $H$ stands for $\Htzt(\pt)$.
Thus, for small displacement we get
$I-\uub^{-1} = 2I - H^T - H + O(h)$,
so
\be
%\uua \simeq {I - (H+H^T) \over 2}.
\uua(t,p(t)) = \tilde\uueps(t,p(t))+ O(h) \qwhere \tilde\uueps:= I - {H+H^T\over 2}.
\ee

And, with $t=\tz+h$ we have $\Ftz(t,P) = I + (t{-}\tz)\,d\vv(t,P) + o(t{-}\tz)$, \cf~\eref{eqdPhi2b},
thus we have $\Htz(t,p(t))=\Ftz(t,P)^{-1} = I - (t{-}\tz)\,d\vv(t,P) + o(t{-}\tz)$ when $p(t)= \Phitz(t,P)$.
Thus
\be
\Ftz(t,P)-I = I-\Htz(t,p(t)) + O(t{-}\tz).
\ee
Therefore, for small displacements ($|t-\tz|<<1$):
\be
\label{eqeapd}
\uua(t,p(t)) \simeq \tilde\uueps(t,p(t)) \simeq \uuepstz(t,P) \quad\hbox{(matrix meaning)}.
\ee

\comment{
\debexa
An elastic solid could thus be modeled as:
\be
\label{eqsm200a}
\tilde\uusigma^\tz_t(\pt) 
= \tilde\lambda \Tr(\tilde\uua(t,\pt))I_t + 2\tilde\mu\,\tilde\uua(t,\pt),
\ee
%written $\tilde\uusigma =\tilde\lambda \Tr(\uuatzt)I + 2\tilde\mu\,\uuatzt$,
with $\tilde\lambda,\tilde\mu\in\RR$,
to compare with $\uusigmatzt(P) 
= \lambda \Tr(\Etzt(P))I_\tz + 2\mu\,\Etzt(P)$ with $\lambda,\mu\in\RR$.
And for small deformations and with matrix computations, $[\uua]$ can be approximated by~$[\tilde\uueps]$, just as $[E]$ can be approximated by~$[\uueps]$.
\finexa
}

%%%%%%%%%%%%%%%%%%%%%%%%%%%%%%%%%%%%%%%%%%%%%%%%%%%%%%%%%%%%%%%%%%%%%%%%%%%%%%%%%%%
%%%%%%%%%%%%%%%%%%%%%%%%%%%%%%%%%%%%%%%%%%%%%%%%%%%%%%%%%%%%%%%%%%%%%%%%%%%%%%%%%%%

\section{Polar decomposition, elasticity and objectivity}
\label{secelasobj}

\def\diag{{\rm diag}}
\def\Rtzt{{R^\tz_t}}
\def\tSigma{\tilde\Sigma}
\def\Utzt{{U^\tz_t}}
\def\tuusigma{{\tilde\uusigma}}
\def\tuuepstzt{{\tilde\uuepstzt}}

%%%%%%%%%%%%%%%%%%%%%%%%%%%%%%%%%%%%%%%%%%%%%%%%%%%%%%%%%%%%%%%%%%%%%%%%%%%%%%%%%%%

\subsection{Polar decompositions of~$F$ (``isometric objectivity'')}

The motion is supposed regular, $\tz,t\in\RR$, $\ptz\in\Omegatz$, $F:=\Ftzt(\ptz)$ ($= d\Phitzt(\ptz)$), $\dd_G$ and $\dd_g$ are Euclidean dot products in~$\RRntz$ and~$\RRnt$, and $C = F^T.F \in\calL(\RRntz;\RRntz)$.
Here the covariant objectivity is abandoned due to the need for inner dot products.
% because we need~$C$, given by $(C.\vW_1,\vW_2)_G : = (F.\vWu,F.\vWd)_g = (\vW_1,C.\vW_2)_G$.
%, needed to be able to define the transposed $(\Ftzt)^T_{Gg}(\pt) \eqnamed F^T$ at $\pt=\Phitzt(\ptz)$ relative to~$\dd_G$ and~$\dd_g$, \cf~\eref{eqFT0}.

%%%%%%%%%%%%%%%%%%%%%%%%%%%%%%%%%%%%%%%%%%%%%%%%%%%%%%%%%%%%%%%%%%%%%%%%%%%%%%%%%%%

\subsubsection{$F=R.U$ (right polar decomposition)}
\label{secrpd}

%Let $\ds C = F^T.F \in\calL(\RRntz;\RRntz)$ (the right Cauchy--Green deformation tensor at~$\ptz$ relative to~$\dd_G$ and~$\dd_g$).
The endomorphism $C$ being $\dd_G$-symmetric definite positive (the motion is supposed to be regular), 
$\exists \alpha_1,...,\alpha_n\in\RR_+^*$ (the eigenvalues), $\exists \vc_1,...,\vc_n\in\RRntz$ (associated eigenvectors), such that, for all $i=1,...,n$,
\be
\label{eqCeigval}
C.\vc_i = \alpha_i \vc_i \qand \hbox{$(\vc_i)$ is a $\dd_G$-orthonormal basis in~$\RRntz$}.
\ee
So, if $(\va_i)$ is a $\dd_G$-Euclidean basis then $D = P^{-1}.[C]_{\va}.P$,
where $D=\diag(\alpha_1,...,\alpha_n) = [C]_{\vc}$ and $P^{-1} = P^T$, $P=[P_{ij}]$ being the transition matrix from $(\va_i)$ to~$(\vc_i)$, \ie\ defined by $\vc_j=\sum_i P_{ij}\va_i$ for all~$j$.

%(Full notation: $\CtztGg(\ptz).\vc_j(\ptz) = \alpha_j(\ptz) \vc_j(\ptz)$ and $(\vc_i(\ptz),\vc_j(\ptz))_G=\delta_{ij}$ for all $\ptz$ and $i,j$.)

Then, define the endomorphism $U \in\calL(\RRntz;\RRntz)$, called the right stretch tensor, by, for all $i=1,...,n$,
\be
\label{eqUeigval}
U.\vc_i = \sqrt{\alpha_i}\, \vc_i, \qand U \eqnote \sqrt C,
\ee
the $\sqrt{\alpha_i}$ being called the principal stretches.
%(Full notation: $\Utzt(\ptz)_{Gg} = \sqrt{\Ctzt(\ptz)_{Gg}}=$ the right stretch tensor at~$\ptz$ between $\tz$ and~$t$ relative to~$\dd_G$ and~$\dd_g$.)
Then, define the linear map $R\in\calL(\RRntz;\RRnt)$, called the rotation map, by 
\be
\label{eqFeRU}
R := F \circ U^{-1} \eqnote F.U^{-1} ,
\ee
%(full notation: $\Rtzt(\ptz)_{Gg}=\Ftzt(\ptz) \circ (\Utzt(\ptz)_{Gg})^{-1}$,
so that
\be
\label{eqFeRU2}
\boxed{F=R\circ U} \eqnote R.U, \quad\hbox{called the right polar decomposition of~$F$}.
\ee

\def\hR{\widehat{R}}
\def\hR{R_0}
\def\bR{\widehat{R}}
\def\bU{\widehat{U}}
\def\Utz{U^\tz}
\def\UtztGg{U^\tz_{t,Gg}}
\def\RtztGg{R^\tz_{t,Gg}}
\def\StztGg{S^\tz_{t,Gg}}
\def\hRtztGg{(\hR)^\tz_{t,Gg}}

\debprop
We have:
\be
\label{eqFeRUu0}
C=U\circ U \eqnote U^2,\quad \hbox{$U$ is symmetric definite positive}, \quad R^{-1}=R^T . %\in\calL(\RRnt;\RRntz).
\ee
%where $R:(\RRntz,\dd_G) \rar (\RRnt,\dd_g)$ and $(R^T.\vz_t,\vw_\tz)_G = (R.\vw_\tz,\vz_t)_g$ for all $\vwtz\in\RRntz$ and all $\vz_t\in\RRnt$ (definition of the transposed $R^T$ relative to~$\dd_G$ and~$\dd_g$).
And 
\be
\label{eqFeRUu}
\hbox{the right polar decomposition}\quad F=R\circ U\quad\hbox{is unique}:
\ee
If $F=\bR\circ \bU$ where $\bU\in\calL(\RRntz;\RRntz)$ is symmetric definite positive and $\bR\in\calL(\RRntz;\RRnt)$ satisfies $\bR^{-1} = \bR^T$,
then $\bU=U$ and $\bR=R$.
\finprop

\debdem
%(From Marsden and Hughes.)
\def\vztz{{\vz_\tz}}
\eref{eqUeigval} yields $(U \circ U).\vc_j = \lambda \vc_j = C.\vc_j$ for all~$j$, \cf~\eref{eqCeigval}, thus $U\circ U = C$ ;
Then $(U^T.\vc_i,\vc_j)_G = (\vc_i,U.\vc_j)_G =(\vc_i, \sqrt{\alpha_j}\vc_j)_G= \sqrt{\alpha_j}(\vc_i,\vc_j)_G = \sqrt{\alpha_j} \delta_{ij} = \sqrt{\alpha_i} \delta_{ij} = \sqrt{\alpha_i}(\vc_i, \vc_j)_G =(\sqrt{\alpha_i}\vc_i, \vc_j)_G =(U.\vc_i,\vc_j)_G$ for all $i,j$, thus $U^T=U$ (symmetry).

Then
$R^T \circ R = U^{-T} \circ F^T \circ F \circ U^{-1}
= U^{-T} \circ C \circ U^{-1} = U^{-1} \circ (U \circ U) \circ U^{-1} = I_\tz$ identity in~$\RRntz$.
(Details:
$(R^T.R.\vW,\vZ)_G
= (R.\vW,R.\vZ)_g
= (F.U^{-1}.\vW,F.U^{-1}.\vZ)_g
= (F^T.F.U^{-1}.\vW,U^{-1}.\vZ)_G
= (U^2.U^{-1}.\vW,U^{-1}.\vZ)_G
= (U^{-1}.U.\vW,\vZ)_G
= (\vW,\vZ)_G
$.)
Thus $R^{-1}=R^T\in\calL(\RRnt;\RRntz)$, thus $R \circ R^T=R \circ R^{-1} = I_t$ identity in~$\RRnt$.

%(Matrix computation, $[R^{-1}]_{|\ve,\vE} = [R^T]_{|\ve,\vE} = ([G]_{|\vE})^{-1}.([R]_{|\vE,\ve})^T.[g]_{|\ve}$.)

And $F = \bR \circ \bU = R \circ U$ gives
$F^T \circ F = \bU^T \circ \bR^T \circ \bR \circ \bU = U^T \circ \bR^T \circ \bR \circ U$, thus
$F^T \circ F = \bU^T \circ \bU$, with $F^T \circ F = U^T \circ U$, thus $\bU \circ \bU =U \circ U = \sqrt C$, thus 
$\bU=U$ (uniqueness of the positive square root eigenvalues). Hence $\bR=R$.
\findem

%%%%%%%%%%%%%%%%%%%%%%%%%%%%%%%%%%%%%%%%%%%%%%%%%%%%%%%%%%%%%%%%%%%%%%%%%%%%%%%%%%%

\subsubsection{$F=S.R_0.U$ (shifted right polar decomposition for covariant objectivity)}

In fact we need to be more specific if the gift of ubiquity is prohibited:
Since we work with the affine space~$\RRn$, consider the Marsden's shifter
\be
S:=\Stzt(\ptz):
\left\{\eqalign{
\TptzOmegatz \eqnote \RRntz & \rar \TptOmegat \eqnote \RRnt \cr
\vw_{\tz,\ptz} & \rar (S.\vw_{\tz,\ptz})(t,\pt) = \vw_{\tz,\ptz}\qwhere \pt=\Phitzt(\ptz). \cr
}\right.
\ee
%The shifter $S$ will be considered between the Hilbert spaces $(\TptzOmegatz,\dd_G)$ and $(\TptOmegat,\dd_g)$:
NB: 1- $S$ looks like the algebraic identity if you have time and space ubiquity gift (otherwise it is not),

2- $S$ is not a topological identity since it changes the norms in the general case: You consider $||\vw_{\tz,\ptz}||_G$ at~$\tz$ and $||S.\vw_{\tz,\ptz}||_g=||\vw_{\tz,\ptz}||_g$ at~$t$.

\medskip
Then, let $\hR\in\calL(\TptzOmegatz;\TptzOmegatz) \eqnote \calL(\RRntz;\RRntz)$ be the endomorphism defined by, in short,
\be
\label{eqhR}
\hR := S^{-1}\circ R \eqnote  S^{-1}. R, \qso \boxed{R=S . \hR} \quad (=S \circ \hR) .
\ee
Full notations: $\hRtztGg(\ptz) := (\Stzt)^{-1}(\RtztGg(\ptz))$. 
\comment{
Thus
\be
\hR :
\left\{\eqalign{
\RRntz & \rar \RRntz \cr
\vwptz & \rar \hR.\vwptz := S^{-1} . R.\vwptz
}\right\}, \quad\hbox{endomorphism in~$\RRntz$}.
\ee
}

\debprop
The endomorphism $\hR= S^{-1}\circ R$ is a rotation operator in~$(\RRntz,\dd_G)$: % (endomorphism in~$\RRntz$):
\be
%\label{eqFeRUS0}
\hR^{-1}=\hR^T \quad \hbox{in }  (\RRntz,\dd_G).
\ee
And
\be
\label{eqFeRUS}
F= S\circ \hR\circ U. %, \qwith  \hR\circ U \in\calL(\RRntz;\RRntz).
\ee
Interpretation: $F$ is composed of: The pure deformation~$U$ (endomorphism in~$\RRntz$), the rotation $\hR$ (endomorphism in~$\RRntz$), and the shift operator $S:\RRntz\rar\RRnt$ (from past to present time and position).
\finprop

\debdem
%$\hR = S^{-1}\circ R$ gives
\be
\eqalign{
(\hR^T.\vW_2,\vW_1)_G
= & (\hR.\vW_1,\vW_2)_G \quad\hbox{(definition of the transposed)}\cr
= &(S^{-1}.R.\vW_1,\vW_2)_G \quad\hbox{(definition of $\hR$)} \cr
%= &(R.\vW_1,\vW_2)_G
= & (R.\vW_1,\vW_2)_G \quad\hbox{($S$ is the algebraic identity)} \cr
= &(R^T.\vW_2,\vW_1)_g \quad\hbox{(definition of $R^T$)} \cr
= & (R^{-1}.\vW_2,\vW_1)_g \quad\hbox{(\cf~\eref{eqFeRUu0} )} \cr
= & (\hR^{-1}.\vW_2,\vW_1)_G\quad\hbox{($S$ is the algebraic identity)},
}
\ee
true for all $\vW_1,\vW_2\in\RRntz$, thus $\hR^T=\hR^{-1}$ in $(\RRntz,\dd_G)$.
And~\eref{eqhR} and~\eref{eqFeRU2} give~\eref{eqFeRUS}. 
\findem

\debexe
Let $D=\diag(\alpha_i)$, let $(\va_i)$ be a Euclidean basis in~$\RRntz$,
let $P$ be the transition matrix from~$(\va_i)$ to~$(\vc_i)$, so $[C]_{|\va}=P.D.P^{-1}$;
Prove $[U]_{|\va} = P.\sqrt D.P^{-1}$. Case $(\va_i)=(\vE_i)$ is a $\dd_g$-orthonormal basis?

\debrep
The $n$ equations \eref{eqCeigval} (for $j=1,...,n$), read as the matrix equation $[C]_{|\va}.P = P.D$ since $[\vc_j]_{\va}$ is the $j$-th column of~$P$.
And he $n$ equations \eref{eqUeigval} (for $j=1,...,n$), read as the matrix equation $[U]_{|\va}.P = P.\sqrt D$ since $[\vc_j]_{\va}$ is the $j$-th column of~$P$.
\finrep
\finexe

\debrem
\def\tR{\tilde R}
Instead of $\hR\in\calL(\RRntz;\RRntz)$, \cf~\eref{eqhR}, you may prefer to consider $\tilde R_0\in\calL(\RRnt;\RRnt)$ defined by
$R=\tilde R_0 \circ S$, \ie, $\tilde R_0=R\circ S^{-1}$.
\finrem
% debrem See remark~\ref{rem2secCG1},4-: $F$ is the main ingredient, and $U$ is the main ingredient, not $U^2=C$. \finrem

\comment{
\debexe
\eref{eqCeigval} shows that the $\alpha_j$ and $\vu_j$ depend on~$\dd_G$ and~$\dd_g$.
Suppose: 
$\dd_G$ and~$\dd_g$ are Euclidean dot products, \eg,
$\dd_G$ defined by an English observer and his foot, and 
$\dd_g$ defined by a French observer and his metre. Do the $\alpha_j$ and $\vu_j$ still depend on~$\dd_G$ and~$\dd_g$?

\debrep
$C.\vu_j = \alpha_j\vu_j$ gives 
$([G]_{\vE})^{-1}.([F]_{|\vE,\ve})^T.[g]_{|\ve}.[F]_{|\vE,\ve}.[\vu_j]_{|\vE}
= \alpha_j[\vu_j]_{|\vE}$, so the $\alpha_j$ and $[\vu_j]_{|\vE}$ are eigenvalues and eigenvectors
of the matrix $([G]_{\vE})^{-1}.([F]_{|\vE,\ve})^T.[g]_{|\ve}.[F]_{|\vE,\ve}$ (eigenvalue problem in~$\vRRn$
with its canonical basis).

Choose $(\vE_i)$ a $\dd_G$-orthonormal basis in~$\RRntz$ and $(\ve_i)$ a $\dd_g$-orthonormal basis in~$\RRnt$. 
Then the $\alpha_j$ and $[\vu_j]$ are eigenvalues and ``eigenvectors''
of the matrix $([F]_{|\vE,\ve})^T.[F]_{|\vE,\ve}$.
They are independent of~$\dd_G$ and~$\dd_g$, but are a priori dependent on the choice of the basis;
Closer look:
\eref{eqFtztnew}, that is
$[F]_{|\vE,\vbd} = P^{-1}.[F]_{|\vE,\vbu}$,
gives
$([F]_{|\vE,\vbu})^T.[F]_{|\vE,\vbu}
= ([F]_{|\vE,\vbd})^T.P^T.P.[F]_{|\vE,\vbd}
$ where $P$ is the transition matrix from $(\vbu)$ to~$(\vbd)$.
Thus with $(\vbu)$ and~$(\vbd)$ $\dd_g$-Euclidean matrices,
we have $P^T=P^{-1}$, thus $([F]_{|\vE,\vbu})^T.[F]_{|\vE,\vbu}
= ([F]_{|\vE,\vbd})^T.[F]_{|\vE,\vbd}$ ($ \eqnote [C]_{|\vE}$ in this Euclidean framework) is independent
of any choice of Euclidean basis in~$\RRnt$.
\finrep
\finexe
}

%%%%%%%%%%%%%%%%%%%%%%%%%%%%%%%%%%%%%%%%%%%%%%%%%%%%%%%%%%%%%%%%%%%%%%%%%%%%%%%%%%%

\subsubsection{$F=V.R$ (left polar decomposition)}
\label{secrpd2}

Same steps than for the right polar decomposition, but with pull-backs (with $F^{-1}$ instead of~$F$).

Let $\pt=\Phitzt(\ptz)\in\Omegat$,
let $\uubtzt(\pt) := \Ftzt(\ptz) \circ (\Ftzt)^T(\pt) \in\calL(\RRnt;\RRnt)$, written
$\uub=F \circ F^T$ (the left Cauchy--Green deformation tensor also called the Finger tensor). The endomorphism $b$ being symmetric definite positive: 
%Relative to a , 
$\exists \beta_1,...,\beta_n\in\RR_+^*$ (the eigenvalues), $\exists \vd_1,...,\vd_n\in\RRnt$ (associated eigenvectors), such that, for all $i=1,...,n$,
\be
\label{eqCeigval3}
\uub.\vd_i = \beta_i \vd_i, \qand \hbox{$(\vd_i)$ is a $\dd_g$-orthonormal basis in~$\RRnt$}.
\ee
Then, define the unique endomorphism $V \in\calL(\RRnt;\RRnt)$, called the left stretch tensor, by, for all $i=1,...,n$,
\be
\label{eqUeigval4}
V.\vd_i = \sqrt{\beta_i} \vd_i, \qand V \eqnote \sqrt \uub.
\ee
(Full notation: $V^\tz_{t,Gg}(\pt) = \sqrt{\uubtzt(\pt)_{Gg}}$.)
% the left stretch tensor at~$\pt$ between $\tz$ and~$t$ relative to~$\dd_G$ and~$\dd_g$.)
Then define the linear map $R_\ell\in\calL(\RRntz;\RRnt)$ by 
\be
\label{eqFeVR}
R_\ell := V^{-1} \circ F \eqnote V^{-1}.F ,
\ee
so that
\be
\label{eqFeVR2}
\boxed{F=V\circ R_\ell} \eqnote V.R_\ell, \quad\hbox{called the left polar decomposition of~$F$}.
\ee

\debprop
We have: 1-
\be
\label{eqFeVR3}
\uub=V\circ V \eqnote V^2,\quad \hbox{$V$ is symmetric definite positive}, \quad R_\ell^{-1}=R_\ell^T.
\ee
And the left polar decomposition $F=V\circ R$ is unique.

2- $R_\ell=R$ and $V = R.U.R^{-1}$ (so $U$ and $V$ are similar), thus $U$ and $V$ have the same eigenvalues, \ie, $\alpha_i=\beta_i$ for all~$i$, and %you can choose the eigenvectors such that 
$\vd_i = R.\vc_i$ for all~$i$ gives a relation between eigenvectors.
\finprop

\debdem
1- Use $F^{-1}$ and $\uub^{-1}=(F^{-1})^T.(F^{-1})$, instead of~$F$ and $C=F^T.F$, to get $F^{-1} = R_\ell^{-1}.U_{\ell}^{-1}$, \cf~\eref{eqFeRU}; Thus $F=U_\ell.R_\ell$; Then name $U_\ell=V$ to get~\eref{eqFeVR2} and~\eref{eqFeVR3}.

2- $V.R_\ell = F = R.U = (R.U.R^{-1}).R$,
thus, by uniqueness of the right polar decomposition, $V = R.U.R^{-1}$ (so $U$ and $V$ are similar) and $R_\ell=R$. Thus,
with~\eref{eqCeigval3},
$\beta_i \vd_i = V.\vd_i = R.U.(R^{-1}.\vd_i)$, thus with $\vc_i=R^{-1}.\vd_i$, then $(\vc_i)$ is an orthonormal basis in~$\RRntz$
and
$\beta_i R.\vc_i = R.U.\vc_i = \alpha_i R.\vc_i$ gives $\beta_i = \alpha_i$ and the $\vc_i$ are eigenvectors of~$U$, for all~$i$.
\findem

%For covariant objectivity you can write the left polar shifted decomposition.

% debrem See remark~\ref{rem2secCG1},4-: $F$ is the main ingredient, and $U$ is the main ingredient, not $U^2=C$. \finrem

%%%%%%%%%%%%%%%%%%%%%%%%%%%%%%%%%%%%%%%%%%%%%%%%%%%%%%%%%%%%%%%%%%%%%%%%%%%%%%%%%%%

\subsection{Linear elasticity: A Classical ``tensorial'' approach}
\label{secrem2secCG1}

%%%%%%%%%%%%%%%%%%%%%%%%%%%%%%%%%%%%%%%%%%%%%%%%%%%%%%%%%%%%%%%%%%%%%%%%%%%%%%%%%%%
\comment{
\subsubsection{Introduction: Remarks}
%\label{secCG1}

The introduction of the Cauchy strain tensor $C=F^T \circ F$ raises questions:

\debrem
\label{rem2secCG1}
The linearization of $E={C-I \over 2} = {F^T.F-I \over 2}$ (the Green--Lagrange tensor) gives $\uueps={F+F^T\over 2}-I$ (the infinitesimal strain tensor). Steps:

$\bullet$  Start with the linear map $F=d\Phitzt(P)$, %see~\S~\ref{secFT},

$\bullet$  Introduce Euclidean dot products to define $F^T$, %see~\S~\ref{secFT},

$\bullet$  Create the quadratic type map $C=F^T.F$, %see~\S~\ref{secC}, 
then the quadratic type map $E={C-I \over 2}$, %see \S~\ref{secGL},

$\bullet$  Linearize $E$ to get back to a linear result: the~$\uueps$ expression. %, see \S~\ref{seceps}.
\\
So: From the linear map $F=d\Phi$ you get back to a ``linear'' $\uueps$... after having ``squared''~$F$ , to get~$C$ and $E=\demi(C-I)$ which you linearize... % (the $\demi$ introduced since your ``squared'' to begin with)... 
And you end with a kind of first order Taylor expansion of the motion~$\Phi$ (the $F$ term is expected), but with an unwanted~$F^T$ (in~$\uueps$)... and, moreover, $\uueps$ is not a function (is not a tensor), \cf~remark~\ref{remuuepsnaf}.
%We mainly get back to the linear $F=d\Phi$ (your starting point) and to  %(The Lie derivative approach is exempt of this drawback).
%In 1-D, or for an elongation motion where $[F]$ is diagonal, we have $F^T=F$, thus $C=F^2$ and $E={F^2-I\over 2}$ (quadratic) and $\uueps = F-I$ (linear). So, to get $\uueps$ (the linear part of~$(F{-}I)^2$), you first square the linear operator $F$ (to get~$C$) and then linearize to get back to~$F$... %See~\S~\ref{secrem2secCG1} for example.
\finrem

\debrem
It is simple to compose and use the differentials along a trajectory (linear map are composed):
$F^{t_1}_{t_2}\circ F^{\tz}_{t_1} = F^{t_0}_{t_2}$, \cf~\eref{eqcompf02}.
It is more difficult to use the Cauchy--Green deformation tensors~$C$ (quadratic type map):
$C^{t_1}_{t_2}\circ C^{\tz}_{t_1} = (F^{t_1}_{t_2})^T\circ F^{t_1}_{t_2}\circ (F^{t_0}_{t_1})^T\circ F^{t_0}_{t_1}$ is not of the type $C^{t_0}_{t_2}$ (the composition $C^{t_1}_{t_2}\circ C^{\tz}_{t_1}$ does not seem to be very used, many descriptions with~$C$ starting from a ``rest state'').
%(With the dot notation due to linearity, $F^{t_1}_{t_2}.F^{\tz}_{t_1} = F^{t_0}_{t_2}$, and $C^{t_1}_{t_2}.C^{\tz}_{t_1} = (F^{t_1}_{t_2})^T.F^{t_1}_{t_2}.(F^{t_0}_{t_1})^T.F^{t_0}_{t_1}$.)
\finrem

The above remarks end with the question: Can we start a linear classical theory without~$\uueps$? Proposal for a~yes:

%$\bullet$ It induces the notation $\sigma_{ij}$ for the stress tensor, which type is undecided: ${1\choose 1}$, ${0\choose 2}$, or ${2\choose 0}$, because of the hidden use of the Riesz representation theorem; Or it can induce the misuse of the Einstein convention if duality notations seem to be used.
%And therefore the impossible visual distinction between covariance and contravariance when components are used.

}

%%%%%%%%%%%%%%%%%%%%%%%%%%%%%%%%%%%%%%%%%%%%%%%%%%%%%%%%%%%%%%%%%%%%%%%%%%%%%%%%%%%

\subsubsection{Classical approach (``isometric objectivity''), and an issue}
\label{secCF}

%Cauchy's approach is extremely useful: It has produced most of the computations for elastic materials over the past two hundred years;
%(It however raises questions, see remark~\ref{rem2secCG1} and next paragraphs.) 
With the infinitesimal strain ``tensor'' (which is not a tensor but a matrix)
\be
\label{eqe01}
\uueps = {F+F^T\over 2} -I,
\ee
the homogeneous isotropic elasticity constitutive law reads (matrix equation for the stress)
\be
\label{eqe1}
(\uusigma(\Phi) = ) \quad \uusigma = \lambda \Tr( \uueps)I + 2\mu \uueps,
\ee
where $\lambda,\mu$ are the Lamé coefficients and $\uusigma$ is the Cauchy stress ``tensor''.

{\bf Issue:} Recall: Adding $F$ and $F^T$ to make $\uueps$ is functionally a mathematical nonsense since
$F:\RRntz\rar\RRnt$ and $F^T:\RRnt\rar\RRntz$
and $I$ is some identity operator: $\uusigma$ is not a tensor.
%(more precisely$F=\Ftzt(\ptz):\TptzOmegatz \rar \TptOmegat$ and $F^T = F^T(\pt) :\TptOmegat\rar\TptzOmegatz$).
%So what could be the set of definition for $\uueps$? 
In particular the meaning of $\Tr(\uueps)$ is questionable (since $\uueps$ is not an endomorphism
and $\Tr(\uueps)$ means $\Tr([\uueps])={\Tr([F])+\Tr([F^T])\over 2} - n$), 
as well as the meaning of $\uueps.\vn = \demi(F.\vn + F^T.\vn)$, or the meaning of
\be
\label{eqe12}
\uusigma.\vn = \lambda \Tr(\uueps)\vn + 2\mu \uueps.\vn
\ee
%for some vector $\vn$,
since $\vn$ has to be defined at $(\tz,\ptz)$ for $F$ and at $(t,\pt)$ for $F^T$.
(Cauchy's approach: $\vn$ is defined at~$(t,\pt)$.)

So, despite the eventual claims, neither $\uueps$ nor $\uusigma$ are tensors (they don't have a functional meaning): They only have a questionable matrix meaning (observer dependent) $[\uueps]:={[F]{+}[F]^T\over 2} -[I]$ and
\be
[\uusigma] = \lambda \Tr([\uueps])[I] + 2\mu [\uueps],
\qand [\uusigma].[\vn] = \lambda \Tr( [\uueps])[\vn] + 2\mu [\uueps].[\vn].
\ee
%with problematic interpretations.

\debrem
\label{rempitfall}
%Another pitfall:
To justify the name ``tensor'' applied to~$\uueps$, you may read: ``For small displacements the Eulerian variable~$\pt$ and the Lagrangian variable~$\ptz$ can be confused'': $\pt \simeq \ptz$ (so $\Omegatz$ and~$\Omegat$ are ``almost equal'',
so $F(\ptz)+F^T(\pt)$ can be considered). Which means that you use the zero-th order Taylor expansions $\pt=\Phitzptz(t) = \ptz+o(1)$. But then, you can\textslbf{not} also use the first (or higher) order Taylor expansion in following calculations, \eg\ you cannot use velocities...
\finrem

%%%%%%%%%%%%%%%%%%%%%%%%%%%%%%%%%%%%%%%%%%%%%%%%%%%%%%%%%%%%%%%%%%%%%%%%%%%%%%%%%%%

\subsubsection{A functional (tensorial) formulation (``isometric objectivity'')}
\label{secFfeps}

Question: Can the constitutive law \eref{eqe1} be modified into a tensorial expression (a functional expression)?
Proposal for a~yes:

\vspace{-1ex}
\begin{enumerate}[itemsep=1pt,wide] %, 
%\begin{enumerate}[noitemsep,wide] %itemsep=1ex, 
%\item Start with .

\item Consider the ``right polar decomposition'' $F=R. U$ where $U\in\calL(\RRntz;\RRntz)$, \cf~\eref{eqFeRU}.
%Thus $\Phitzt(\ptz{+}h\vW)=\Phitzt(\ptz) +h\,R.U(\ptz).\vW+o(h)$, 
The Green Lagrange tensor $E={C{-}I\over2}$ (endomorphism in~$\RRntz$) then reads, with~\eref{eqFeRUu0},
\be
\label{eqtuuepstzt0}
E= {U^2 {-} I_\tz \over 2} ={(U{-}I_\tz)^2 +2(U - I_\tz) \over 2}
\ee
(the Green--Lagrange tensor is independent of the rotation~$R$),
thus, with $U-I_\tz=O(h)$ (small deformation approximation), we get the modified infinitesimal strain tensor at $\ptz\in\Omegatz$
\be
\label{eqtuuepstzt}
\boxed{\tuueps = U{-}I_\tz}\in\calL(\RRntz;\RRntz),
\ee
endomorphism in~$\RRntz$ (to compare with~$\uueps$ which is not a function, \cf\ the previous~\S).
(Full notation $\tuueps^\tz_{t,Gg}(\ptz) = \UtztGg(\ptz){-}I_\tz(\ptz)$ in $\calL(\RRntz;\RRntz)$.)
%And an endomorphism is naturally canonically identified with a tensor: \eref{eqtuuepstzt} is a functional and tensorial expression.
%To compare with~\eref{eqe01} (matrix meaning).
And, for all $\vW\in\RRntz$ we get
\be
\label{eqtuuepstzt2}
\boxed{\tuueps.\vW = U.\vW - \vW = R^{-1}.\vw-\vW}\; \in\RRntz, \qwhen \vw=F.\vW %=R.U.\vW
\hbox{ (push-forward)}.
\ee
{\bf Interpretation}: % of~\eref{eqtuuepstzt2}:
From $\vw=F.\vW=R.U.\vW \in \RRnt$ (the deformed by the motion), remove the ``rigid body rotation'' to get $R^{-1}.\vw = U.\vW\in\RRntz$, to which you remove the initial~$\vW$ to obtain $\tuueps.\vW\in\RRntz$.
In~particular $||\tuueps.\vW||_G = ||(U{-}I_\tz).\vW||_G$ measures the relative elongation undergone by~$\vW$.
%(to compare with $\uueps.\vW$ which is not functionally defined).
%\Eg, with~\eref{eqUeigval}, $\tuueps.\vc_j = (\sqrt{\alpha_j}{-}1) \vc_j\in \RRntz$.
And you can then apply $R$ to get back into~$\RRnt$ at~$\pt$:
\be
\label{eqtuuepstzt3}
R.(\tuueps.\vW) = F.\vW - R.\vW = \vw-R.\vW\; \in\RRnt, \qwhen \vw=F.\vW=\hbox{ (push-forward)}.
\ee

\def\sph{{\rm sph}}
\def\dev{{\rm dev}}
\def\deform{{\rm def}}

\item Then, %instead of the Piola--Kirchhoff tensor,
at $\ptz\in\Omegatz$, consider the stress tensor $\tSigma(\Phi) \eqnote \tSigma\in \calL(\RRntz;\RRntz)$ (functionally well) defined by
\be
\label{eqSigmaobj}
\boxed{\tSigma = \lambda \Tr(\tuueps)I_\tz + 2\mu \tuueps}
= \lambda \Tr(U{-}I_\tz)I_\tz + 2\mu (U{-}I_\tz).
\ee
(The trace $\Tr(\tuueps)$ is well defined since $\tuueps$ is an endomorphism.)
Then for any $\vW\in\RRntz$ you get in $\RRntz$, at $\ptz\in\Omegatz$,
\be
\tSigma.\vW
= \lambda  \Tr(\tuueps)\vW +  2\mu \tuueps.\vW
= \lambda \Tr(U{-}I_\tz)\vW + 2\mu (U.\vW{-}\vW)
%= 3\kappa \Tr(\tuueps)\vW + 2\mu \tuueps_{\dev}.\vW
\ee
(functionally well defined in~$\RRntz$). %, vectorial stress which is independent of any rotation~$R$.
%In particular, $\tSigma.\vc_j = ((\sumin\alpha_i {-}n) + 2\mu (\alpha_j{-}1)).\vc_j$.
%\item
Then rotate and shift with $R$ to get into~$\RRnt$ at~$\pt$,
\be
\eqalign{
R.\tSigma.\vW
= &\lambda  \Tr(\tuueps)R.\vW +  2\mu R.\tuueps.\vW
= \lambda  \Tr(U{-}I_\tz)R.\vW +  2\mu R.(U{-}I_\tz).\vW \cr
= & \lambda  \Tr(U{-}I_\tz)R.\vW +  2\mu (F - R).\vW, \cr
= & \lambda  \Tr(U{-}I_\tz)R.\vW +  2\mu (\vw - R.\vW), \qwhere \vw= F.\vW=R.U.\vW.
}
%= 3\kappa \Tr(\tuueps)\vW + 2\mu \tuueps_{\dev}.\vW
\ee
You have defined the two point tensor (functionally well defined)
\be
%(R\circ \tSigma \eqnote)\quad  
R.\tSigma
 = \lambda  \Tr(\tuueps)R +  2\mu R.\tuueps\; \in \calL(\RRntz;\RRnt).
\ee

\item
Then you can propose the constitutive law with the stress tensor (the symmetric endomorphism) in~$\RRnt$ given by
\be
\label{eqe2}
(\tuusigma(\Phi)=) \quad\boxed{\tuusigma = R \circ \tSigma \circ R^{-1}} \eqnote R.\tSigma.R^{-1} \;\in \calL(\RRnt;\RRnt).
\ee
(Functionally well defined.) % (while $\uusigma$ is not, \cf~\S~\ref{secCF}).
And, for all vector fields $\vw$ defined in~$\Omegat$, you get the (functionally well defined) vector field
\be
\label{eqe3}
\tuusigma.\vw 
=  R.\tSigma.R^{-1}.\vw \;\;\in\RRnt.
\ee
{\bf Interpretation} of~\eref{eqe2}-\eref{eqe3}: Shift and rigid rotate backward by applying~$R^{-1}$, apply the elastic stress law with~$\Sigma$ which corresponds to a rotation free motion (Noll's frame indifference principle), then shift and rigid rotate forward by applying~$R$.

Detailed expression for~\eref{eqe2}-\eref{eqe3}:
With $\Tr(R.\tuueps.R^{-1}) = \Tr(\tuueps)$ (see exercise~\ref{exetrrer}), we get, at $(t,\pt)$,
\be
\label{eqe2ts}
\eqalign{
\tuusigma
=  \lambda \Tr(\tuueps)\,I_t + 2\mu R.\tuueps.R^{-1} %= n\kappa \Tr(\tuueps) I_t + 2\mu R.\tuueps_{dev}.R^{-1} 
= & \lambda \Tr(U{-}I_\tz)\,I_t + 2\mu R.(U{-}I_\tz).R^{-1} \cr
= & \lambda \Tr(U{-}I_\tz)\,I_t + 2\mu (F.R^{-1}{-}I_t). \cr
}
\ee
%where $\tuueps = U-I_\tz = R^{-1}.F-I_\tz$. 
And for any $\vw\in\RRnt$, and with $\vw=R.\vW$, you get %the tensorial expression
\be
\eqalign{
\tuusigma.\vw
=  \lambda  \Tr(\tuueps)\,\vw + 2\mu R.\tuueps.\vW 
%=  n\kappa\Tr(\tuueps)\vw + 2\mu R.\tuueps_{dev}.\vW, \qwhen \vw=R.\vW, 
= & \lambda  \Tr(U{-}I_\tz)\,\vw + 2\mu R.(U{-}I_\tz).\vW 
%=  n\kappa\Tr(U{-}I_\tz)\vu + 2\mu F.\vW, 
\cr
= & \lambda  \Tr(U{-}I_\tz)\,\vw + 2\mu (R.U.R^{-1}.\vw{-}\vw)
%=  n\kappa\Tr(U{-}I_\tz)\vw + 2\mu F.R^{-1}.\vw
. \cr
}
\ee
To compare with the classical functionally meaningless~\eref{eqe12}.
\end{enumerate}

\debrem
Doing so, you avoid the use of the Piola--Kirchhoff tensors.
\finrem

\debexe
\label{exetrrer}
Prove: $\Tr(R.\tuueps.R^{-1}) = \Tr(\tuueps)= \sum_i(\alpha_i{-}1)$.
(NB: $\tuueps$ is an endomorphism in~$\RRntz$ while $R.\tuueps.R^{-1}$ is an endomorphism in~$\RRnt$.)

\debrep
$\det_{|\ve}(R.\tuueps.R^{-1} - \lambda I_t)
=\det_{|\ve}(R.(\tuueps{-}\lambda I_\tz).R^{-1}) = \det_{|\vE,\ve}(R).\det_{|\vE}(\tuueps{-}\lambda I).\det_{|\ve,\vE}(R^{-1}) = \det_{|\vE}(\tuueps{-}\lambda I)$ for all Euclidean bases $(\vE_i)$ and~$(\ve_i)$ in~$\RRntz$ and~$\RRnt$.
(With $L=\tuueps$ and components, $\Tr(R.L.R^{-1})
= \sum_i (R.L.R^{-1})^i_i
= \sum_{ijk} R^i_j L^j_k (R^{-1})^k_i
%= \sum_{ijk} (R^{-1})^k_i R^i_j L^j_k(R^{-1})k_i
= \sum_{jk} (R^{-1}.R)^k_j L^j_k
= \sum_{jk} \delta^k_j L^j_k
= \sum_{j}  L^j_j
= \Tr(L)$.)
\finrep
\finexe

\debexe
Elongation in $\RR^2$ along the first axis : origin~$O$, same Euclidean basis~$(\vE_1,\vE_2)$ and Euclidean dot product at all time,
$\xi>0$, $t\ge\tz$, $L,H>0$, $P\in[0,L]\times[0,H]$, $[\ora{OP}]_{|\vE} = \pmatrix{X_0 \cr Y_0}$, and
$[\ora{O\Phitzt(P)}]_{|\vE} = \pmatrix{X_0+\xi(t{-}\tz)X_0 \cr Y_0} = \pmatrix{X_0(\kappa{+}1) \cr Y_0} = \pmatrix{x \cr y}=[\ora{Op}]_{|\vE}$,
where $\kappa=\xi(t{-}\tz)>0$ for $t>\tz$.

1- Give $F$, $C$, $U=\sqrt{C}$ and $R=F.U^{-1}$. Relation with the classical expression ?

2- Spring $\ora{OP}=\ora{Oc_\tz} (s) = X_0\vE_1 + Y_0\vE_2 + s\vW$,
\ie\ $[\ora{OP}]_{|\vE} = [\ora{Oc_\tz}]_{|\vE} = \pmatrix{X_0 {+} sW_1 \cr Y_0 {+} sW_2}_{|\vE}$ with $s\in[0,L]$ and $\vW =W_1\vE_1 + W_2\vE_2$.
Give the deformed spring, \ie\ give $p=c_t(s)=\Phitzt(c_\tz(s))$, and $\vc_t{}'$, and the stretch ratio.

\debrep
1- $[F]=[d\Phi]=\pmatrix{\kappa{+}1 & 0 \cr 0 & 1}$, same Euclidean dot product and basis at all time,
thus $[F^T]=[F]^T=[F]$, then
$[C]=[F^T].[F]=[F]^2 = \pmatrix{(\kappa{+}1)^2 & 0 \cr 0 & 1}$,
thus $[U]=[F]=\pmatrix{\kappa{+}1 & 0 \cr 0 & 1}$, thus $[R]=[I]$.
All the matrices are given relative to the basis~$(\vE_i)$, thus $F,C,U,R$ (\eg, $C.\vE_1=(\kappa{+}1)^2\vE_1$ and $C.\vE_2 = \vE_2$).

Since $R=I$ and $[\uueps]=[\tuueps]$, \eref{eqe2ts} gives the usual result
$[\uusigma] = \lambda \Tr( [\uueps])I + 2\mu [\uueps]$, cf~\eref{eqe1} (matrix meaning).

2- $\ora{Oc_t(s)}=\ora{O\Phitzt(c_\tz(s))}=\pmatrix{(X_0 {+} sW_1)(\kappa{+}1) \cr Y_0 {+} sW_2}_{|\vE}$, thus $\vc_t{}' (s)=\pmatrix{W_1(\kappa{+}1) \cr W_2}_{|\vE}$,
stretch ration ${W_1^2(\kappa{+}1)^2 + W_2^2 \over W_1^2 + W_2^2}$ at $(t,\pt)$.
\finrep
\finexe

\debexe
\label{exess1}
Simple shear in $\RR^2$ :
$[\ora{O\Phitzt(P)}]_{|\vE} = \pmatrix{X+\xi(t{-}\tz)Y \cr Y} \eqnote \pmatrix{X+\kappa Y \cr Y} =\pmatrix{x \cr y}=[\ora{Op}]_{|\vE}$. Same questions, and moreover give the eigenvalues of~$C$.

\debrep
1- $[F]=\pmatrix{1 & \kappa \cr 0 & 1}$, 
$[C] = \pmatrix{1 & 0 \cr \kappa & 1}.\pmatrix{1 & \kappa \cr 0 & 1}=\pmatrix{1 & \kappa \cr \kappa & \kappa^2{+}1}$. Eigenvalues: 
$\det(C-\lambda I)=\lambda^2-(2{+}\kappa^2)\lambda+1$.
Discriminant $\Delta = (2{+}\kappa^2)^2-4=\kappa^2(\kappa^2{+}4)$.
Eigenvalues $\alpha_\pm = \demi(2{+}\kappa^2 \pm \kappa\sqrt{\kappa^2{+}4})
%= \demi(2 + \kappa^2 \pm \kappa^2\sqrt{1{+}{4\over \kappa^2}})
$.
(We check that $\alpha_\pm>0$.)
Eigenvectors $\vv_\pm$(main directions of deformations) given by $(1{-}\alpha_\pm)x+\kappa y=0$,
\ie, $ y=\demi(\kappa\pm\sqrt{\kappa^2{+}4})x$,  
thus, \eg,
$\vv_\pm 
=\pmatrix{
2 \cr 
\kappa \pm \sqrt{\kappa^2{+}4} \cr
}
$.
(We check that $\vv_+ \perp \vv_-$.)
With $P$ the transition matrix from $(\vE_1,\vE_2)$ to $({\vv_+\over ||\vv_+||},{\vv_-\over ||\vv_-||})$ and $D= {\rm diag}(\alpha_+,\alpha_-)$
we get $C = P.D.P^{-1}$ (with $P^{-1}=P^T$ since here $({\vv_+\over ||\vv_+||},{\vv_-\over ||\vv_-||})$
is an orthonormal basis), thus $U=P.\sqrt D.P^{-1}$ (we check that $U^T=U$ and $U^2=C$).
And $R = F.U^{-1}$.

2- $\ora{Oc_t(s)}=\ora{O\Phitzt(c_\tz(s))}=\pmatrix{(X_0 {+} sW_1)+\kappa(Y_0 {+} sW_2) \cr Y_0 {+} sW_2}$, thus $[\vc_t{}' (s)]=\pmatrix{W_1+\kappa W_2 \cr W_2}$.
Stretch ratio ${(W_1+\kappa W_2)^2 + W_2^2 \over W_1^2 + W_2^2}$ at $(t,\pt)$.
\finrep
\finexe

%%%%%%%%%%%%%%%%%%%%%%%%%%%%%%%%%%%%%%%%%%%%%%%%%%%%%%%%%%%%%%%%%%%%%%%%%%%%%%%%%%%

\comment{

\def\tf{\tilde f}
\subsubsection{Example of a 1-D spring in a 3-D problem}

Consider a spring $\Obj$ which particles $\Pobj$ are located at~$t$ at $\tPhi(t,\Pobj)$.
Let $\vutz\in\vRRt$.
Suppose that at $\tz$ the particles of the spring lie along the straight segment $c_\tz:[s_1,s_2] \rar \RRt$ given by
\be
\label{eqctz}
c_\tz (s) = c_\tz (s_1) + s\vutz\eqnote \qtz(s), \qand \vc_\tz{}' (s)=\vutz.
\ee
(\Ie, if $\Pobj$ is a particle of the spring then $\exists s\in[s_1,s_2]$ \st\ $c_\tz (s)=\tPhi(\tz,\Pobj)$, \cf\ figure~\ref{figpf}.)

Then consider an affine motion $\Phitzt$, so $\Phitzt(Q_2)=\Phitzt(Q_1) + F.\ora{Q_1Q_2}$ for all $Q_1,Q_2$.
%(here $d\Phitzt(Q)\eqnote F$ is independent of~$Q$). 
Then $c_\tz$ is transformed by the motion into the curve $c_t=\Phitzt\circ c_\tz$:
The particles $\Pobj$ are located at~$t$ along the straight segment $c_t:[s_1,s_2] \rar \RRt$ given by
\be
\label{eqctz2}
c_t (s) %= \tPhi(t,\Pobj) 
= \Phitzt(c_\tz(s))
= \Phitzt(c_\tz(s_1)) + F.\ora{c_\tz(s_1)c_\tz(s)}
\eqnote \qt(s), \qand
\vc_t{}' (s) = F.\vutz = \vutzs.
\ee
And the stretch ratio at~$t$ at $\qt(s)$ (in the direction~$\vutz$) is 
\be
\lambdasr(\qt(s)) = {||F.\vutz||_g \over ||\vutz||_g}=\lambdasr
\ee
(independent of~$s$),
an implicit Euclidean dot product $\dd_g$ being understood:
Dilation if $\lambdasr \ge1$ and contraction if $\lambdasr \le 1$ (old length
$L_\tz= \int_{s_1}^{s_2} ||\vc_\tz{}' (s)||_g \,ds = (s_2{-}s_1)||\vutz||_g$,
new length $L_t= \int_{s_1}^{s_2} ||\vc_t{}' (s)||_g \,ds = \lambdasr (s_2{-}s_1)||\vutz||_g$).

\def\vdeltaftzt{\vec{\delta f^\tz_t}}
\def\Wtzt{\delta f^\tz_t}

\debrem
1-D spring \eref{eqctz}-\eref{eqctz2}. Suppose that $||\vutz||=1$ and, at $\qt(s)=c_t(s)$,
define the density of force to be
$\vdeltaftzt(\qt(s),\vutz)
=-k(\Utzt-I_\tz)(\qt(s)).\vutz $.
Thus $\vdeltaftzt$ derives from the potential $\Wtzt=-k\Phitzt$, that is,
$\vdeltaftzt(\qt(s),\vutz) = d\Wtzt(\qt(s)).\vutz = -k\, d\Phitzt(\qt(s)).\vutz$.

And $\int_{s=s_1}^{s_2} -k ||d\Phitzt(\qt(s)).\vutz||\,ds
= \int_{s=s_1}^{s_2} -k ||{\pa \Phitzt\circ c_\tz \over ds}(s)||\,ds
= \int_{s=s_1}^{s_2} -k ||{\pa c_t \over ds}(s)||\,ds
= -k L_t
$, to be compared with~$-kL_\tz$ to get $-k (L_t{-}L_\tz)=$
the force applied by of the spring on its end: Usual scalar result.
\finrem

}

%%%%%%%%%%%%%%%%%%%%%%%%%%%%%%%%%%%%%%%%%%%%%%%%%%%%%%%%%%%%%%%%%%%%%%%%%%%%%%%%%%%

\subsubsection{Second functional formulation: With the Finger tensor}

\def\ttuueps{{\tilde{\tilde\uueps}}}

The above approach uses the push-forward: It uses~$F$, \ie\ you arrive with your memory. You may prefer to use the pull-back, \ie\ use $F^{-1}$ (you remember the past which is Cauchy's point of view): Then you use $F^{-1} = R^{-1}.V^{-1}$ the right polar decomposition of $F^{-1}$, %, see the Euler--Almansi approach~\S~\ref{seceulal}.
and you consider the tensor
\be
\label{eqVmu}
\ttuueps_t = V^{-1}{-}I_t \; \in\calL(\RRnt;\RRnt),
\ee
and 
\be
\label{eqVmu2}
(\uusigma_t(\Phi)=)\quad \uusigma_t = \lambda \Tr(\ttuueps_t)I_t + 2\mu\ttuueps_t,
 \qand
\uusigma_t.\vn_t = \lambda \Tr(\ttuueps_t)\,\vn_t + 2\mu\ttuueps_t.\vn_t.
\ee
(Quantities functionally well defined: Give a tensorial approach). % for any $\vn_t\in\RRnt$ since $\ttuueps_t : \RRnt\rar\RRnt$.

%%%%%%%%%%%%%%%%%%%%%%%%%%%%%%%%%%%%%%%%%%%%%%%%%%%%%%%%%%%%%%%%%%%%%%%%%%%%%%%%%%%

\subsection{Elasticity with a covariant objective approach?}
\label{secelaoa}

\def\vT{{\vec T}}
\def\vwts{\vec w_{t*}}

In~\S~\ref{secrem2secCG1} you need to start with Euclidean dot products, so from the start the result can't be covariant objective. Can you start without Euclidean dot products to set up general laws?
Proposal: 

\medskip
Hypothesis: The Cauchy stress $\vw$ is a Eulerian vector field.
\comment{
, % (defined $\bigcup_{t\in[t_1,t_2]} (\{t\}\times\Omegat)$,
and, at~$\tau$ at~$\ptau$, the stress $\vw(\tau,\ptau)$ is a function of the 
$\Fttau(\tau,\pt).\vw(t,\pt) \eqnote \vwts(\tau,\ptau)$ (the push-forward).
}

Then we could use the (covariant objective) Lie derivative which characterizes the rate of stress,
\cf~\S~\ref{secdlcv} and~\ref{secdlcdf}: With a particle $\Pobj\in\Obj$, with $\vv(\tau,\ptau) = {\pa \tPhi \over \pa \tau}(\tau,\Pobj)$ its Eulerian velocity at $\tau$ at $\ptau = \tPhi(\tau,\Pobj)$,
the Lie derivative of a Eulerian vector field~$\vw$ along~$\vv$ is, at $(t,\pt)$,
\be
\label{eqLieelas}
\eqalign{
\calL_\vv \vw(t,\pt)
= & \lim_{\tau\rar t} { \vw(\tau,\ptau)-\vwts(\tau,\ptau) \over \tau-t}
%= \lim_{\tau\rar t} { \vw(\tau,\ptau)-\vw_{\tau,t*}(\tau,\ptau) \over \tau-t} 
%\cr
=  ({\pa \vw \over \pa t} + d\vw.\vv - d\vv.\vw)(t,\pt). \cr
}
\ee
Hence the proposal, with the virtual power principle to measure the rate of stress (see https://arxiv.org/abs/2208.10780v1 for a full description).
%(following Germain: ``To know the weight of a suitcase you have to move it'').

\def\pint{p_{\rm int}}

\medskip
1- Hypotheses:
1.1- Suppose that $n$ Eulerian vector fields~$\vw_j$ (``force fields''), $j=1,...,n$, enable to characterize a material.
(In fact, for elasticity problems it could be better %to follow the usual characterization of the materials
to replace vector fields $\vw_j$ with  1-forms~$\alpha_i$ to characterize the work.)
% (At $(t,\pt)$, $n$ forces $\vw_j(t,\pt)$ are enough to describe the ``whole state of stress''.)

1.2- 
With a basis $(\ve_i)$ chosen in~$\RRnt$, with $(e^i)$ its (covariant) dual basis in~$\RRnts$, assume that the internal power density at $(t,\pt)$ is given by (at first order):
\be
\label{eqpintlie}
\pint(\vv) = \sumjn e^j.\calL_\vv \vw_j = \sumjn e^j.({\pa \vw_j\over \pa t}+d\vw_j.\vv - d\vv.\vw_j).
\ee
(At second order you can add second order Lie derivatives as $\calL_\vv(\calL_\vv \vw_j)$,
similarly for higher orders.)

2- %Up to now the approach is qualitative (objective, observer independent).
Then, so that this $\pint$ satisfies the frame invariance hypothesis, %in a Galilean referential, 
choose a Euclidean dot product~$\dd_g$ in~$\RRnt$; Then, first, the internal power has to vanish if $d\vv=0$, thus we are left with
\be
\label{eqpintlie2}
\pint(\vv) = -\sumjn  e^j.d\vv.\vw_j = -\uutau \odd d\vv,
\qwhere \uutau = \sumjn \vw_j \otimes e^j,
\ee
defined at~$t$.
(The $j$-th column of~$[\uutau]_{|\ve}$ is~$[\vw_j]_{|\ve}$.)
And, second, the internal power vanishes if $d\vv + d\vv^T=0$ (rotation), thus we are left with
\be
\label{eqpintlie3}
\pint(\vv) = -\uutau \odd {d\vv + d\vv^T \over 2} = -\uusigma \odd d\vv 
\qwhere \uusigma = {\uutau + \uutau^T \over 2},
\ee
this $\pint(\vv) = -\uusigma \odd d\vv$ being the usual expression of the internal power at first order.
%Then, \eg\ for homogeneous isotropic elasticity, you can use~\eref{eqVmu2} (a functionally well defined tensor), or~\eref{eqe12} (matrix approach).

%For details, see \eg\ \verb+https://www.isima.fr/leborgne/IsimathMeca/PpvObj.pdf+.

\debexa
\eref{eqpintlie} may be applied to orthotropic elasticity, \eg\ for a material which fibers at some time~$\tz$ are along~$\ve_1$, in a 2-D case for simplicity: 1- With an elongation type motion $(\Phi_e)$ given by
$[(F_e)(\ptz)]=[d(\Phi_e)(\ptz)]=\pmatrix{1{+}\alpha_{11}(\ptz) & 0 \cr 0 & 1{-}\alpha_{22}(\ptz)}$
you measure the Young moduli in the directions~$\ve_1$ and~$\ve_2$;
2- And with a shear type motion given by
$[(F_s)(\ptz)]=[d(\Phi_s)(\ptz)]=\pmatrix{1 & \gamma_{12} \cr 0 & 1}$
you measure the shear modulus. %=[R_s].[U_s]$.

For more complex material, you may need more vectors~$\vw_j$ to describe the constitutive law, that is, \eref{eqpintlie} may be considered with $\sumim e^j.\calL_\vv \vw_j$ with $m>n$. %(And toward crystallography?)
\finexa

\debrem
The Lie approach is different from the usual classic approach:

1- The classic approach looks for an order two stress tensor $[\uusigma]$ as a function of the deformation gradient~$[F]$, \cf~\eref{eqe1}.

2- The Lie approach begins with the internal power (which measures forces), \cf~\eref{eqpintlie}, which then enable to build $\uutau$ and the~$\uusigma$ (the stress tensor), \cf~\eref{eqpintlie2}-\eref{eqpintlie3}.
%In a first order approximation (``small deformations''), there seems to be little interest.

\Eg, application to visco-elasticity: With the Lie approach, you automatically use Lie derivative of vector fields (and/or of differential forms), instead of Lie derivative of order 2 tensor fields (which does not seem to give good result, see \eg\ the Maxwell visco-elastic type laws, as well as footnote\footref{footrem} page~\pageref{footrem}).
\finrem

\comment{
%Here $F=R.U$ with $[R]=[I]$ and $[F]=[U]$, 
$[\tuusigma] = \lambda \Tr([F_e]{-}[I]) I+ 2\mu ([F_e]{-}[I])
=\pmatrix{\lambda(\alpha_{11}{-}\alpha_{22})+2\mu\alpha_{11} & 0 \cr
 0 & \lambda(\alpha_{11}{-}\alpha_{22})-2\mu\alpha_{22}}$.
Then choose $[\vw_1] = [\tuusigma].[\ve_1]$ (first column) and $[\vw_2] = [\tuusigma].[\ve_2]$ (second column).

Consider a shear type motion given by
$[(F_s)(\ptz)]=[d(\Phi_s)(\ptz)]=\pmatrix{1 & \gamma_{12} \cr 0 & 1}=[R_s].[U_s]$.
\\
$[\uutau] = \lambda \Tr([U_s]{-}[I]) I+ 2\mu R_s([U_s]{-}[I]).R_s]^{-1}
%=\pmatrix{0 & 2\mu\gamma_{12} \cr  0 & 0}
$.
Then choose $[\vw_1] = [\tuusigma].[\ve_1]$ (first column) and $[\vw_2] = [\tuusigma].[\ve_2]$ (second column).
}

\comment{
\begin{figure}[!h]
\qquad\qquad\qquad\includegraphics[width=0.5\textwidth]{knitElastic.png}
\caption{Elongation, from 
Geometry and Elasticity of a Knitted Fabric,
Samuel Poincloux 1 Mokhtar Adda-Bedia Frédéric Lechenault, LPS - Laboratoire de Physique Statistique de l'ENS,
https://hal.sorbonne-universite.fr/hal-01833356/document.
}\label{figelongation}
\end{figure}
}

\comment{
\begin{figure}[!h]
\qquad\qquad\qquad\includegraphics[width=0.7\textwidth]{elongation.png}
\caption{Elongation.}
%From https://eduscol.education.fr/sti/sites/eduscol.education.fr.sti/files/ressources/pedagogiques/6689/ 6689-modelisation-du-comportement-des-composites1-3-lelasticite-anisotrope-ensps.pdf
\end{figure}
}

%%%%%%%%%%%%%%%%%%%%%%%%%%%%%%%%%%%%%%%%%%%%%%%%%%%%%%%%%%%%%%%%%%%%%%%%%%%%%%%%%%%

\section{Displacement}

%%%%%%%%%%%%%%%%%%%%%%%%%%%%%%%%%%%%%%%%%%%%%%%%%%%%%%%%%%%%%%%%%%%%%%%%%%%%%%%%%%

\subsection{The displacement vector $\vcalU$}
\label{secvcalU}

In~$\RRn$, let $\pt = \Phitzt(\ptz)$. Then the bi-point vector
\be
\label{eqdvut0}
\vcalUtzt(\ptz) = \Phitzt(\ptz)-I_\tz(\ptz) = \pt-\ptz = \ora{\ptz \pt}
\ee
is called the displacement vector at~$\ptz$ relative to $t_0$ and~$t$.
This defines the map
\be
\label{eqdvut1}
\vcalUtzt :
\left\{\eqalign{
\Omegatz & \rar \vRRn \cr
\ptz & \rar \vcalUtzt(\ptz) \eqdef \pt-\ptz = \ora{\ptz \pt} \qwhen \pt = \Phitzt(\ptz).
}\right.
\ee

\debrem
\label{remU}
$\vcalUtzt(\ptz)$ doesn't define a vector field (it is not tensorial),
because $\vcalUtzt(\ptz) = \pt-\ptz = \ora{\ptz\pt}$ is a bi-point vector which is neither in~$\RRntz$ or in~$\RRnt$
since $\ptz\in \Omegatz$ and $\pt\in\Omegat$ (it requires time and space ubiquity gift). In particular, it makes no sense on a non-plane surface (manifold).
%If you prefer, in~\eref{eqdvut1}, you need the image space to be~$\vRRn$ to use time ubiquity (the image cannot be just~$\RRnt$ or~$\RRntz$). 
More at \S~\ref{secappgd}.
\finrem

\debrem
For elastic solids in~$\RRn$,
the function $\vcalUtz$ is often considered to be the unknown (to be computed);
But the ``real'' unknown is the motion~$\Phitz$.
And it is sometimes confused with the extension of a spring $1$-D case; But see figure~\ref{figpf} where
$||\vw_\tz(\ptz)||$ represents the initial length and  $||\vw_{\tz*}(t,\pt)||$ represents the current length of the spring, while the length of the displacement vector $\vcalUtzt=\pt-\ptz$ can be very long for a very small elongation $||\vw_{\tz*}(t,\pt)||-||\vw_\tz(\ptz)||$ of the spring.
\finrem

%%%%%%%%%%%%%%%%%%%%%%%%%%%%%%%%%%%%%%%%%%%%%%%%%%%%%%%%%%%%%%%%%%%%%%%%%%%%%%%%%%

\subsection{The differential of the displacement vector}

%What follows is mathematically meaningless, but it is used in the literature.

The differential of~$\vcalUtzt$ at~$\ptz$ is
\be
\label{eqdvut0d}
d\vcalUtzt(\ptz) = d\Phitzt(\ptz) - I_\tz = \Ftzt(\ptz) - I_\tz, \qwritten d\vcalU = F-I,
\ee
thus isn't defined as a function,
because $\Ftzt(\ptz):\RRntz\rar\RR$ while $I_\tz:\RRntz\rar\RRntz$. 

So $d\vcalUtzt(\ptz)$ as to be understood as a matrix: With
\be
[\vcalUtzt(\ptz)] = [\ora{\ptz\pt}] = [\ora{O\Phitzt(\ptz)}] - [\ora{O\ptz}],
\ee
relative to an origin~$O$ and a unique basis at all time,
compute $[d\vcalUtzt(\ptz)] = [d\Phitzt(\ptz)]-I$, abusively written $d\vcalU = d\Phi - I$.
Then, with $\vW\in\RRntz$ ,
\be
\label{eqdlU}
d\vcalU.\vW = F.\vW - \vW ,
\quad \hbox{which means}\quad  = [\Ftzt(\ptz)].[\vW] - [\vW].
\ee
%which compares the transported vector $\vwtzs(\pt)$ (the push-forward by the flow)  with the original vector~$\vw_\tz(\ptz)$ at~$\tz$.
Thus we have defined (matrix meaning)
\be
\vcalUtz :
\left\{\eqalign{
[\tz,T] \times \Omegatz & \rar \vRRn \cr
(t,\ptz) & \rar \vcalUtz(t,\ptz) := \vcalUtzt(\ptz),
}\right.
\qand
\vcalUtzptz :
\left\{\eqalign{
[\tz,T] & \rar \vRRn \cr
t & \rar \vcalUtzptz(t) := \vcalUtzt(\ptz).
}\right.
\ee

%%%%%%%%%%%%%%%%%%%%%%%%%%%%%%%%%%%%%%%%%%%%%%%%%%%%%%%%%%%%%%%%%%%%%%%%%%%%%%%%%%%

\subsection{Deformation ``tensor'' $\protect\uueps$ (matrix),  bis}

\eref{eqdvut0d} gives (matrix meaning)
\be
\Ftzt(\ptz) = I_\tz + d\vcalUtzt(\ptz)  , \qwritten F = I + d\vcalU.
\ee
Therefore, Cauchy--Green deformation tensor $C=F^T.F$ reads, in short, (matrix meaning)
\be
C = I + d\vcalU + d\vcalU^T + d\vcalU^T.d\vcalU  \quad \hbox{(matrix meaning)},
\ee
\ie\ $[\Ctzt(\ptz)] = [I_\tz] + [d\vcalUtzt(\ptz)] + [d\vcalUtzt(\ptz)]^T + [d\vcalUtzt(\ptz)]^T.[d\vcalUtzt(\ptz)]$.

Thus the Green--Lagrange deformation tensor $E = {C-I \over 2}$, \cf~\eref{eqdefGSV}, reads, in short, (matrix meaning)
\be
\label{eqdvut9f}
%\Etzt(\ptz) = {d\vcalUtzt(\ptz) + d\vcalUtzt(\ptz)^T + d\vcalUtzt(\ptz)^T.d\vcalUtzt(\ptz)\over 2}, \qwritten
E = {d\vcalU + d\vcalU^T \over 2} + \demi d\vcalU^T.d\vcalU  \quad \hbox{(matrix meaning)}.
\ee
Thus the deformation tensor~$\uueps$, \cf~\eref{equuepsm}, reads (matrix meaning)
\be
\label{eqdvut9f2}
\uueps
%= {d\Ftz + d\Ftz^T \over 2} - I  + \demi d\vcalU^T.d\vcalU
= E - \demi (d\vcalU)^T.d\vcalU ,
\ee
with $\uueps$ the ``linear part'' of~$E$ (small displacements: we only used the first order derivative $d\Phitzt$).

%%%%%%%%%%%%%%%%%%%%%%%%%%%%%%%%%%%%%%%%%%%%%%%%%%%%%%%%%%%%%%%%%%%%%%%%%%%%%%%%%%%

\subsection{Small displacement hypothesis, bis}
\label{secsmh}

(Usual introduction.)
Let $\pt=\Phitzt(\ptz)$, $\vW_i\in\vRRntz$, $\vw_i(\pt) = \Ftzt(\ptz).\vW_i(\ptz) \in \RRnt$ (the push-forwards), written $\vw_i = F.\vW_i$.
Then define (matrix meaning)
\be
\vDelta_i \eqdef \vw_i - \vW_i = d\calU.\vW_i, \qand
||\vDelta||_\infty = \max(||\vDelta_1||_\RRn,||\vDelta_2||_\RRn).
\ee
Then the small displacement hypothesis reads (matrix meaning):
\be
\label{eqvD}
||\vDelta||_\infty= o(||\vW||_\infty).
\ee
Thus
$\vw_i = \vW_i + \vDelta_i$ (with $\vDelta_i$ ``small'') and the hypothesis $\dd_g= \dd_G$ (same inner dot product at~$\tz$ and~$t$) give
$$
(\vw_1,\vw_2)_G - (\vW_1,\vW_2)_G
= (\vDelta_1,\vW_2)_G + (\vDelta_2,\vW_1)_G + (\vDelta_1,\vDelta_2)_G.
$$
So~\eref{eqdvut9f2} gives
$
2(E.\vW_1,\vW_2)_G = 2(\uueps.\vW_1,\vW_2)_G + (d\vcalU^T.d\vcalU.\vW_1,\vW_2)_G,
$
And~\eref{eqvD} gives
\be
\label{eqdvut91}
(E.\vW_1,\vW_2)_G = (\uueps.\vW_1,\vW_2)_G + O(||\vDelta||_\infty^2),
\ee
so $\Etzt$ is approximated by~$\uuepstzt$, that is, $\Etzt \simeq \uuepstzt$ (matrix meaning).

\comment{
%%%%%%%%%%%%%%%%%%%%%%%%%%%%%%%%%%%%%%%%%%%%%%%%%%%%%%%%%%%%%%%%%%%%%%%%%%%%%%%%%%

\subsection{Hypothèse des Petites Pertubations (petites déformations)}

\noindent
{\bf Approche élémentaire ``bipoint''.}
Soit $Q\in\Omegatz$ dans un voisinage de~$P$ et soit
$\qt = \Phitzt(Q) \in \Omegat$.

Comme $\Phitzt(Q) = \Phitzt(P) + d\Phitzt(P).(Q-P) + o(Q-P)$,
i.e. $\ora{\pt\qt} = \Ftzt(P).\ora{PQ} + o(\ora{PQ})$, on~a :
\be
\label{eqai}
\ora{\pt\qt} - \ora{PQ} = (\Ftzt(P) - I).\ora{PQ} + o(\ora{PQ})
=  d\vcalUtzt(P).\ora{PQ} + o(\ora{PQ}).
\ee
Négligeant le petit~$o$, on note :
\be
\label{eqvdd}
\eqalign{
\vDeltatzt (P,\ora{PQ})
= &\ora{\pt\qt} - \ora{PQ}
=  (\Ftzt(P) - I).\ora{PQ} \cr
= & d\vcalUtzt(P).\ora{PQ}.
}
\ee

\noindent
{\bf Approche vectorielle, \cf~\S~\ref{secsa}.}
Soit $\alpha : s\in[-\eps,\eps] \rar \Omegatz$ une courbe dans~$\Omegatz$ avec $P=\alpha(0)$,
et soit $\beta = \Phitzt\circ\alpha  : s\in[-\eps,\eps] \rar \Omegat$ sa transformée par le mouvement,
donc avec $\pt=\beta(0)$.

Soit $\vW_P = \alpha'(0) \in\RRntz$ le vecteur tangent en~$P$ à $\Im\alpha \subset\Omegatz$,
et soit $\vw_\pt = \beta'(0) \in\RRnt$ le vecteur tangent en~$\pt$ à $\Im\beta \subset\Omegat$.
Donc, \cf~\eref{defrapvEivei2z}, $\vw_\pt$ est le push-forward de~$\vW_P$ par $\Phitzt$ en~$P$ :
\be
\label{eqai1}
(\vw^\tz_t(\pt) = ) \quad \vw_\pt = d\Phitzt(P).\vW_P = \Ftzt(P).\vW_P, \qqn \vw = F.\vW.
\ee
On note :
\be
\label{eqvdd2}
\eqalign{
\vDeltatzt(P,\vW_P)
=  &\vw_\pt - \vW_P
=  (\Ftzt(P) - I).\vW_P \cr
= & d\vcalUtzt(P).\vW_P \qquad (=\vw(\pt) - \vW_P),
}
\ee
qu'on préfèrera à~\eref{eqvdd} (on utilise ici de ``vrais vecteurs'' et on ne néglige pas un petit~o).

\debdef
On dit qu'on travaille entre $\Omegatz$ et~$\Omegat$ dans le cadre de l'hypothèse des petites perturbations (HPP),
ou hypothèse des petites déformations, 
lorsque, pour $t_1>t$, pour tout $t\in[\tz,t_1]$, pour tout $P\in\Omegatz$ et tout $\vW_P \in\RRntz$ on~a :
\be
||\vw^\tz_t(\pt)||^2 = ||\Ftzt(P).\vW_P||^2 = ||\vW_P||^2 + o(||\vW_P||^2),
\ee
i.e. le tenseur de Cauchy
$\Ctzt = \Ftzt^T.\Ftzt$ (endomorphisme dans~$\RRntz$) à toutes ses valeurs propres
proches de~$1$.
C'est exprimé par les mécaniciens sous la forme :
\be
\label{eqvdd2p}
\vDeltatzt(P,\vW_P) = o(\vW_P), \qie \vw(\pt) = \vW(P) + o(\vW_P).
\ee
(Les mécaniciens disent que les notation eulériennes et lagrangiennes peuvent être confondues dans le cadre HPP.)
\findef

\debrem
\label{remhpp}
Dans ce cadre HPP, \eref{eqvdd2} et~\eref{eqdPhi2} indiquent que
$\vDeltatzt(P,\vW_P) = O(t-\tz)$ :
c'est une autre manière de définir l'hypothèse des petites déformations entre les instants $\tz$ et~$t$ :
quand $t$ est proche de~$\tz$.
\finrem
}

%%%%%%%%%%%%%%%%%%%%%%%%%%%%%%%%%%%%%%%%%%%%%%%%%%%%%%%%%%%%%%%%%%%%%%%%%%%%%%%%%%%

\subsection{Displacement vector with differential geometry}
\label{secappgd}

\def\Stzt{{S^\tz_t}}
\def\tS{{\tilde S}}
\def\tStzt{{\tilde{S^\tz_t}}}
\def\RRntzP{{\vec\RR^n_{\tz,P}}}
\def\RRntpt{{\vec\RR^n_{t,\pt}}}
\def\vwpt{{\vw_\pt}}

%%%%%%%%%%%%%%%%%%%%%%%%%%%%%%%%%%%%%%%%%%%%%%%%%%%%%%%%%%%%%%%%%%%%%%%%%%%%%%%%%%%

\subsubsection{The shifter}

We give the steps, see Marsden--Hughes~\cite{marsden-hughes}.
The complexity introduced is due to the small displacement hypothesis applied to the Green--Lagrange tensor $E={F^T.F - I \over 2}$ which linearization gives $\uueps={F+F^T\over 2}-I$
(the classical approach ``squares the motion'' to get~$E$, then ``linearizes''~$E$ ... to get back to~$F$... with a spurious~$F^T$).

%Voir poly ``Tenseurs...'' pour les définitions précises.

Let $P\in\Omegatz$, $\vW_P \in\RRntz$, $\pt = \Phitzt(P)\in\Omegat$, and $\vw_\pt = \Ftzt(P).\vW_P \in \RRnt$ (push-forward).

\comment{
To compare $\vwpt$ with $\vWP$,
an observer starts from $P$ at~$\tz$, and moves along the trajectory $\PhitzP$ with~$\vWP$
to be at $\pt$ at~$t$ with the vector $\vwpt$ (the push-forward).
}

\medskip
\noindent
{\bf $\bullet$ Affine case $\RRn$ (continuum mechanics):}
With $p_t=\Phitzt(P)$, the shifter is:
\be
\label{eqdefshift}
\tStzt : 
\left\{\eqalign{
\Omegatz \times \RRntz & \rar  \Omegat \times \RRnt \cr
(P,\vZ_P) & \rar \tStzt(P,\vZ_P) = (\pt,\Stzt(\vZ_P))\qwith  \Stzt(\vZ_P) = \vZ_P.
%\quad\hbox{quand}\quad p_t=\Phitzt(P),
}\right.
\ee
(The vector is unchanged but the time and the application point have changed: A real observer has no ubiquity gift).
So:
\be
\Stzt \in \calL(\RRntz ; \RRnt)\qand
[\Stzt]_{|\ve} = I \hbox{ identity matrix},
\ee
the matrix equality being possible after the choice of a unique basis at $\tz$ and~at~$t$.
And (simplified notation)
$
%\label{eqdefshiftn}
\tStzt(P,\vZ_P) \eqnote \Stzt(\vZ_P)
$.
Then the deformation tensor $\uueps$ at~$P$ can be defined by
\be
\uuepstzt(P).\vZ(P) = {(\Stzt)^{-1}(\Ftzt(P).\vZ(P)) + \Ftzt(P)^T.(\Stzt(P).\vZ(P)) \over 2} - \vZ(P),
\ee
in short: $\uueps.\vZ = {(\Stzt)^{-1}(F.\vZ) + F^T.(\Stzt.\vZ) \over 2} - \vZ)$.

%\medskip
\noindent
{\bf $\bullet$ In a manifold:} $\Omega$ is a manifold
(like a surface in~$\RR^3$ from which we cannot take off).
%\ Misner, Thorne et Wheeler \cite{misner-thorne-wheeler} pour l'illustration à l'aide d'une fourmi se déplaçant sur la surface d'une pomme.
Let $T_P\Omegatz$ be the tangent space à~$P$ (the fiber at~$P$), and
$T_\pt\Omegat$ be the tangent space à~$\pt$ (the fiber at~$\pt$).
In general $T_P\Omegatz \ne T_\pt\Omegat$ (\eg\ on a sphere ``the  Earth'').
The bundle (the union of fibers) at~$\tz$ is $T\Omegatz = \bigcup_{P \in\Omegatz}(\{P\} \times T_P\Omegatz)$,
and the bundle at~$t$ is $T\Omegat = \bigcup_{\pt \in\Omegat}(\{\pt\} \times T_\pt\Omegat)$.
Then the shifter
\be
\label{eqdefshift2}
\tStzt : 
\left\{\eqalign{
T\Omegatz & \rar  T\Omegat \cr
(P,\vZ_P) & \rar \tStzt(P,\vZ_P) = (\pt,\Stzt(\vZ_P)), 
%\quad\hbox{quand}\quad p_t=\Phitzt(P),
}\right.
\ee
is defined such that $\vZ_P\in T_P\Omegatz$ ``as little distorted as possible'' along a path.
\Eg, on a sphere, if the path is a geodesic,
if $\theta_\tz$ is the angle between $\vZ_P$ and the tangent vector to the geodesic at~$P$,
then $\theta_\tz$ is also the angle between $\Stzt(\vZ_P)$ and the tangent vector to the geodesic at~$\pt$,
and $\Stzt(\vZ_P)$ has the same length than~$\vZ_P$
(at constant speed in a car you think the geodesic is a straight line, although $\Stzt(\vZ_P) \ne \vZ_P$: the Earth is not flat).

%On note $\RRntz=T_P$ l'espace tangent en~$P$ à~$\tz$, et $\RRnt = T_\pt$ l'espace tangent en~$\pt$ à~$t$, et on note $\calL(T_P,T_\pt)$ l'ensemble des applications linéaires de $T_P$ dans~$T_\pt$.

%%%%%%%%%%%%%%%%%%%%%%%%%%%%%%%%%%%%%%%%%%%%%%%%%%%%%%%%%%%%%%%%%%%%%%%%%%%%%%%%%%%

\subsubsection{The displacement vector}

(Affine space framework, $\Omegatz$ open set in~$\RRn$.)
Let $P\in\Omegatz$, $\vW_P \in\RRntz$, $\pt = \Phitzt(P)\in\Omegat$,
%and $\vw_\pt = \Ftzt(P).\vW_P \in \RRnt$ (push-forward).
and $d\Phitzt = \Ftzt \in \calL(\RRntz ; \RRnt)$. Define
\be
\label{eqdvut2}
\delta\tilde\vcalUtzt :
\left\{\eqalign{
\Omegatz \times \RRntz & \rar \Omegat \times \calL(\RRntz;\RRnt) \cr
(P,\vZ_P) & \rar \delta\tilde\vcalUtzt(P,\vZ_P) = (\pt , \delta\vcalUtzt(\vZ_P)) 
\qwith \delta\vcalUtzt(\vZ_P) = (\Ftzt - \Stzt).\vZ_P. \cr
}\right.
\ee
Then $\delta\tilde\vcalUtzt = \Ftzt - \Stzt$ is a two-point tensor. And
\be
\label{eqdvut3}
\eqalign{
\Ctzt
= & (\Ftzt)^T.\Ftzt = (\delta\calU^\tz_t+ S^\tz_t)^T.(\delta\calU^\tz_t +S^\tz_t) \cr
= & I + (S^\tz_t)^T.\delta\calU^\tz_t + (\delta\calU^\tz_t)^T.S^\tz_t + (\delta\calU^\tz_t)^T.\delta\calU^\tz_t,
}
\ee
since $(S^\tz_t)^T. S^\tz_t=I$ identity in~$T\Omegatz$: Indeed,
$((S^\tz_t)^T.S^\tz_t.\vA, \vB)_\RRn = (S^\tz_t.\vA, S^\tz_t.\vB)_\RRn = (\vA,\vB)_\RRn$,
\cf~\eref{eqdefshift}, for all $\vA,\vB$.
Then the Green--Lagrange tensor is defined on~$\Omegatz$ by
\be
\label{eqdefGSV2}
E^\tz_t=\demi(\Ctzt-I_\tz)= {(S^\tz_t)^T.\delta\calU^\tz_t + S^\tz_t.(\delta\calU^\tz_t)^T\over 2} +\demi (\delta\calU^\tz_t)^T.\delta\calU^\tz_t,
\ee
to compare with~\eref{eqdefGSV}.

%Ainsi $\delta\vcalUtzt(P)$ est une application linéaire de $T^\tz_P \rar T^t_\pt$.

%%%%%%%%%%%%%%%%%%%%%%%%%%%%%%%%%%%%%%%%%%%%%%%%%%%%%%%%%%%%%%%%%%%%%%%%%%%%%%%%%%%
%%%%%%%%%%%%%%%%%%%%%%%%%%%%%%%%%%%%%%%%%%%%%%%%%%%%%%%%%%%%%%%%%%%%%%%%%%%%%%%%%%%

%\section{Determinants and endomorphisms in $\RRn$}
\section{Determinants}
\label{secdetend}

\def\Al{{A\!\ell}}
\def\Bl{{B\!\ell}}
\def\Jtz{{J^\tz}}
\def\Jtzt{{J^\tz_t}}

%%%%%%%%%%%%%%%%%%%%%%%%%%%%%%%%%%%%%%%%%%%%%%%%%%%%%%%%%%%%%%%%%%%%%%%%%%%%%%%%%%%

\subsection{Alternating multilinear form}
\label{remvolume}

Let $E$ be a vector space, and let
$\calL(E,...,E;\RR) \eqnote\calL(E^n;\RR)$  be the set of multilinear forms,
%(the set $\calL^0_n(E)$ of uniform ${0\choose n}$ tensors), 
\ie\ $m\in \calL(E^n;\RR)$ iff
\be
m(...,\vx+\lambda\vy, ...) = m(...,\vx, ...) + \lambda m(...,\vy, ...)
\ee
for all $\vx,\vy\in E$ and all $\lambda\in \RR$ and for all ``slot''.
In particular, $m(\lambda_1 \vx_1,..., \lambda_n \vx_n)
= (\prod_{i=1,...,n} \lambda_i)\; m( \vx_1,..., \vx_n)$,
for all $\lambda_1,...,\lambda_n\in \RR$ and all $\vx_1,...,\vx_n \in E$.

\debdef
\label{defnlina}
If $n=1$ then a $1$-alternating multilinear function is a linear form, also called a $1$-form.
If $n\ge2$ then $\Al : 
\left\{\eqalign{
E^n & \rar \RR \cr
(\vv_1,...,\vv_n) & \rar \Al(\vv_1,...,\vv_n) \cr
}\right\} \in \calL(E^n;\RR)
$
is a $n$-alternating multilinear form iff, for all $\vu,\vv\in E$,
\be
\label{eqdefnlina}
\Al(...,\vu,...,\vv,...) = - \Al(...,\vv,...,\vu,...),
\ee
the other elements being unchanged.
If $n=1$, the set of $1$-forms is $\Omega^1(E) = E^*$. If $n\ge 2$,
the set of $n$-alternating multilinear forms is
\be
\Omega^n(E) = \{m\in\calL(E^n;\RR) : m=\Al \hbox{ is alternating}\}.
\ee
\findef

If $\Al,\Bl \in \Omega^n(E)$ and $\lambda\in\RR$
then $\Al+\lambda \Bl \in \Omega^n(E)$ thanks to the linearity for each variable.
Thus $\Omega^n(E)$ is a vector space, sub-space in $(\calF(E^n;\RR),+,.)$.

%%%%%%%%%%%%%%%%%%%%%%%%%%%%%%%%%%%%%%%%%%%%%%%%%%%%%%%%%%%%%%%%%%%%%%%%%%%%%%%%%%%

\subsection{Leibniz formula}

Particular case $\dim E{=}n$. % and $\Omega^n(E)$.
Let $\Al \in \Omega^n(E)$ (a $n$-alternating multilinear form).
Recall (see \eg\ Cartan~\cite{cartanh}):

1- A permutation $\sigma:[1,n]_\NN \rar[1,n]_\NN$ is a bijective map (\ie\ one-to-one and onto); Let $S_n$ be the set of permutations of $[1,n]_\NN$.

2- A transposition $\tau:[1,n]_\NN \rar[1,n]_\NN$ is a permutation that exchanges two elements, that is,
$\exists i,j$ \st\ $\tau(...,i,...,j,...) = (...,j,...,i,...)$,
the other elements being unchanged.

3- A permutation is a composition of transpositions (theorem left as an exercise, of see Cartan).
And a permutation is even iff the number of transpositions is even,
and a permutation is odd iff the number of transpositions is odd. Based on:
The parity (even or odd) of a permutation is an invariant.

4- The signature $\eps(\sigma)=\pm1$ of a permutation $\sigma$ is $+1$ if $\sigma$ is even,
and is $-1$ if $\sigma$ is odd.

%(These definitions are based on the fact that the parity (even or odd) of a permutation is an invariant.)

\begin{prop}[Leibniz formula]
\label{propleibniz}
Let $\Al  \in \Omega^n(E)$.
Let $(\ve_i)_{i=1,...,n}\eqnote(\ve_i)$ be a basis in~$E$.
%$(\ve_i)$ being a basis in~$E$, %and let $c := \Al(\ve_1,...,\ve_n)$ ($\in \RR$).
For all vectors $\vv_1,...,\vv_n \in E$, %and let the $v^i_j$ be the components of~$\vv_j$,
with $\vv_j = \sumin v^i_j \ve_i$ for all~$j$,
\be
\label{eqfleibniz0}
\Al(\vv_1,...,\vv_n)
= c \sum_{\sigma\in S_n} \eps(\sigma) \prod_{i=1}^n v^{\sigma(i)}_i
= c \sum_{\tau\in S_n} \eps(\tau) \prod_{i=1}^n v^i_{\tau(i)}
\quad\hbox{(with $c := \Al(\ve_1,...,\ve_n)$)}
.
\ee
Thus if $c=\Al(\ve_1,...,\ve_n)$ is known, then $\Al$ is known. Thus $\dim(\Omega^n(E))=1$.
\\
(Classic not.: $\vv_j = \sumin v_{ij} \ve_i$,
$\Al(\vv_1,...,\vv_n)
= c \sum_{\sigma\in S_n} \eps(\sigma) \prod_{i=1}^n v_{\sigma(i),i}
= c \sum_{\tau\in S_n} \eps(\tau) \prod_{i=1}^n v_{i,\tau(i)}$.)
\finprop

\debdem
Let $F := \calF([1,n]_\NN;[1,n]_\NN) \eqnote [1,n]_\NN^{[1,n]_\NN}$ 
be the set of functions $i : 
\left\{\eqalign{
[1,n]_\NN & \rar [1,n]_\NN \cr
k  & \rar i_k=i(k) \cr
}\right\}
$.
$\Al$~being multilinear,
$\Al(\vv_1,...,\vv_n)
= \sum_{j_1=1}^n v^{j_1}_1\Al( \ve_{j_1},\vv_2,...,\vv_n)
$ (``the first column'' development).
By recurrence we get
$\Al(\vv_1,...,\vv_n)
= \sum_{j_1,...,j_n=1}^n v^{j_1}_1...  v^{j_n}_n \Al( \ve_{j_1},...,\ve_{j_n})
= \sum_{j\in F} \prod_{k=1}^n v^{j(k)}_k \Al( \ve_{j(1)},...,\ve_{j(n)})$.

And $\Al( \ve_{i_1},...,\ve_{i_n})\ne0$
iff  $i : k\in \{1,...,n\} \rar  i(k) = i_k \in \{1,...,n\}$ is one-to-one (thus bijective).
Thus
$\Al(\vv_1,...,\vv_n)
= \sum_{\sigma\in S_n} \prod_{i=1}^n v^{\sigma(i)}_i \Al(\ve_{\sigma(1)},...,\ve_{\sigma(n)})
= \sum_{\sigma\in S_n} \eps(\sigma) \prod_{i=1}^n v^{\sigma(i)}_i \Al(\ve_1,...,\ve_n)$,
which is the first equality in~\eref{eqfleibniz0}.
Then
$\sum_{\sigma\in S_n} \eps(\sigma) \prod_{i=1}^n v^{\sigma(i)}_i
=\sum_{\sigma\in S_n} \eps(\sigma) \prod_{i=1}^n v^{\sigma(\sigma^{-1}(i))}_{\sigma^{-1}(i)}$
since $\sigma$ is bijectif, 
thus
$\sum_{\sigma\in S_n} \eps(\sigma) \prod_{i=1}^n v^{\sigma(i)}_i
=\sum_{\tau\in S_n} \eps(\tau^{-1}) \prod_{i=1}^n v^i_{\tau(i)}
$, thus the second equality in~\eref{eqfleibniz0} since $\eps(\tau)^{-1} = \eps(\tau)$.
(See Cartan~\cite{cartanh}.)
\findem

%%%%%%%%%%%%%%%%%%%%%%%%%%%%%%%%%%%%%%%%%%%%%%%%%%%%%%%%%%%%%%%%%%%%%%%%%%%%%%%%%%%

\subsection{Determinant of vectors}

\debdef
$(\ve_i)_{i=1,...,n}$ being a basis in~$E$, the alternating multilinear form $\det_{|\ve} \in \Omega^n(E)$ defined by
\be
\label{eqdettrans}
\det_{|\ve}(\ve_1,...,\ve_n)=1
\ee
is called the determinant relative to~$(\ve_i)$.
And, with prop.~\ref{propleibniz} (here $c=1$), %$\det_{|\ve}$ is given by
\be
\label{eqdettrans2}
\det_{|\ve}(\vv_1,...,\vv_n)
= \sum_{\sigma\in S_n} \eps(\sigma) \prod_{i=1}^n v^{\sigma(i)}_i
= \sum_{\tau\in S_n} \eps(\tau) \prod_{i=1}^n v^i_{\tau(i)}
\ee
is called the determinant of the vectors $\vv_i$ relative to~$(\ve_i)$. And we write
\be
%\dim(\Omega^n(E)) = 1, \qwith 
\Omega^n(E) = \Vect\{\det_{|\ve}\}\quad\hbox{(the 1-D vector space spanned by $\det_{|\ve}$)}.
\ee

\findef

Thus, if $\Al\in \Omega^n(E)$ then
\be
\label{eqald}
\Al = \Al(\ve_1,...,\ve_n)\,\det_{|\ve},
\ee
thus if $(\vb_i)$ is another basis then
\be
\label{eqpropdet}
\exists c\in\RR,\quad \det_{|\vb} = c\det_{|\ve}, \qwith  c=\det_{|\vb}(\ve_1,...,\ve_n).
\ee

\debexe
%\label{exechvol}
Change of measuring unit: If $(\va_i)$ is a basis and $\vb_j = \lambda \va_j$ for all~$j$, prove 
\be
\label{eqexechvol}
\forall j=1,...,n, \quad\vb_j = \lambda \va_j \quad  \Longrightarrow \quad \det_{|\va} =\lambda^n \det_{|\vb}
\ee
 (relation between volumes relative to a change of measuring unit in the Euclidean case).

\debrep
$\ds \det_{|\va}(\vb_1,...,\vb_n)
= \det_{|\va}(\lambda\va_1,...,\lambda\va_n)
\mathop{=}^{multi}_{linear} \lambda^n \det_{|\va}(\va_1,...,\va_n)
\mathop{=}^{\eref{eqdettrans}} \lambda^n
\mathop{=}^{\eref{eqdettrans}} \lambda^n \det_{|\vb}(\vb_1,...,\vb_n)
$. %gives $\ds \det_{|\va} =\lambda^n \det_{|\vb}$.
\finrep
\finexe

\debprop
$\det_{|\ve}(\vv_1,...,\vv_n) \ne 0$ iff $(\vv_1,...,\vv_n)$ is a basis,
or equivalently, $\det_{|\ve}(\vv_1,...,\vv_n) = 0$ iff $\vv_1,...,\vv_n$ are linearly dependent.
\finprop

\debdem
If one of the $\vv_i$ is $=\vec0$ then $\det_{|\ve}(\vv_1,...,\vv_n) = 0$ (multilinearity),
and if $\sumin c_i\vv_i=0$ and one of the $c_i\ne0$ and then a $\vv_i$ is a linear combination of the others
thus $\det_{|\ve}(\vv_1,...,\vv_n)=0$ (since $\det_{|\ve}$ is alternate).
Thus $\det_{|\ve}(\vv_1,...,\vv_n) \ne 0$ $\Rightarrow$ the $\vv_i$ are independent.
And if the $\vv_i$ are independent then
$(\vv_1,...,\vv_n)$ is a basis, thus $\det_{|\vv}(\vv_1,...,\vv_n)=1 \ne 0$,
with $\det_{|\vv}=c\det_{|\ve}$, thus $\det_{|\ve}(\vv_1,...,\vv_n) \ne 0$.
\findem

\debexe
\label{exedet2}
In~$\RR^2$. Let $\vv_1=\sum_{i=1}^2 v_1^i\ve_i$ and $\vv_2 = \sum_{j=1}^2 v_2^j\ve_j$ (duality notations). Prove:
\be
\det_{|\ve}(\vv_1,\vv_2)=v_1^1 v_2^2 - v_1^2 v_2^1.
\ee

\debrep
Development relative to the first column (linearity used for the first vector $\vv_1= v_1^1\ve_1 + v_1^2\ve_2$):
$\det_{|\ve}(\vv_1,\vv_2)
=\det_{|\ve}(v_1^1\ve_1+v_1^2\ve_2,\vv_2)
= v_1^1 \det_{|\ve}(\ve_1,\vv_2) + v_1^2\det_{|\ve}(\ve_2,\vv_2)$.
Thus (linearity used for the second vector  $\vv_2= v_2^1\ve_1 + v_2^2\ve_2$):
$\det_{|\ve}(\vv_1,\vv_2)
%= v_1^1 \det(\ve_1,v_2^1\ve_1+v_2^2\ve_2) + v_1^2 \det(\ve_2,v_2^1\ve_1+v_2^2\ve_2)
=0+v_1^1v_2^2 \det(\ve_1,\ve_2) + v_1^2v_2^1 \det(\ve_2,\ve_1)+0
=v_1^1 v_2^2 - v_1^2 v_2^1$.
\finrep
\finexe

\debexe
\label{exedet3}
In~$\RR^3$, with $\vv_j=\sum_{i=1}^3 v_j^i \ve_i$, prove:
\be
\det(\vv_1,\vv_2,\vv_3)=\sum_{i,j,k=1}^3\eps_{ijk}v_1^i v_2^j v_3^k,
\ee
where $\eps_{ijk} = \demi(j{-}i)(k{-}j)(k{-}i)$, \ie\
$\eps_{ijk}=1$ if $(i,j,k)=(1,2,3)$, $(3,1,2)$ or $(2,3,1)$ (even signature),
$\eps_{ijk}=-1$ if $(i,j,k)=(3,2,1)$, $(1,3,2)$ and $(2,1,3)$ (odd signature),
and $\eps_{ijk}=0$ otherwise.

\debrep
Development relative to the first column (as in exercise~\ref{exedet2}). 

Result
$=v^1_1v^2_2v^3_3 + v^1_2v^2_3v^3_1 + v^1_3v^2_1v^3_2 - v^3_1v^2_2v^1_3 - v^3_2v^2_3 v^1_1 - v^3_3v^2_1v^1_2$.
\finrep
\finexe

%%%%%%%%%%%%%%%%%%%%%%%%%%%%%%%%%%%%%%%%%%%%%%%%%%%%%%%%%%%%%%%%%%%%%%%%%%%%%%%%%%%

\subsection{Determinant of a matrix}
\label{secdetmat}

\def\Mij{{M^i{}_j}}
\def\Nij{{N^i{}_j}}

%%%%%%%%%%%%%%%%%%%%%%%%%%%%%%%%%%%%%%%%%%%%%%%%%%%%%%%%%%%%%%%%%%%%%%%%%%%%%%%%%%%

%\subsubsection{Definition}

Let $M = [M_{ij}]_{i=1,...,n \atop j=1,...,n}$ be a $n^2$ real matrix.
Let $\vRRn = \RR\times...\times \RR$ (Cartesian product $n$-times) with its canonical basis $(\vE_i)$. 
Let $\vv_j\in\vRRn$, $\vv_j = \sumin M_{ij}\vE_i$; So
$M=\pmatrix{[\vv_1]_{|\vE},...,[\vv_n]_{|\vE}}$.

\debdef
\label{defdetM}
\def\vm{{\vec m}}
The determinant of the matrix $M=\pmatrix{[\vv_1]_{|\vE},...,[\vv_n]_{|\vE}}$ is
\be
\label{eqdefdetM}
\det(M) :=\det_{|\vE}(\vv_1,...,\vv_n)  .
\ee
\findef

%%%%%%%%%%%%%%%%%%%%%%%%%%%%%%%%%%%%%%%%%%%%%%%%%%%%%%%%%%%%%%%%%%%%%%%%%%%%%%%%%%%

%\subsubsection{Determinant of a transposed matrix}

\debprop
Let $M^T$ be the transposed matrix, \ie, $(M^T)_{ij} = M_{ji}$ for all $i,j$. Then
\be
\label{eqMT}
\det (M^T) = \det (M).
\ee
\finprop

\debdem
$\ds \det [M_{ij}]
=  \det_\vE(\vv_1,...,\vv_n)
\mope^{\eref{eqdettrans2}}  \sum_{\sigma\in S_n} \eps(\sigma) \prod_{i=1}^n v^{\sigma(i)}_i
=  \sum_{\tau\in S_n} \eps(\tau) \prod_{i=1}^n v^i_{\tau(i)}
= \det [M_{ji}]$.
\findem

\comment{
\debdem
$\ds \det [M_{ij}]
=  \det_\vE(\vv_1,...,\vv_n)
\mope^{\eref{eqdettrans2}}  \sum_{\sigma\in S_n} \eps(\sigma) \prod_{i=1}^n v_{\sigma(i),i}
=  \sum_{\tau\in S_n} \eps(\tau) \prod_{i=1}^n v_{i,\tau(i)}
= \det [M_{ji}]$.
\findem
}

%%%%%%%%%%%%%%%%%%%%%%%%%%%%%%%%%%%%%%%%%%%%%%%%%%%%%%%%%%%%%%%%%%%%%%%%%%%%%%%%%%%

\subsection{Volume}

\begin{definition} %[Algebraic volume]
Let $(\ve_i)$ be a Euclidean basis. % (\ie, \st\ $(\ve_i,\ve_j)_g = \delta_{ij}$ for all~$i,j$).
Consider a parallelepiped in~$\RRn$ which sides are vectors $\vv_1,...,\vv_n$; Its algebraic volume relative to~$(\ve_i)$ is
\be
\label{eqvolalg}
\hbox{algebraic volume} = \det_{|\ve}(\vv_1,...,\vv_n).
\ee
And its volume relative to~$(\ve_i)$ is (non negative)
\be
\label{eqvolalg2}
\hbox{volume} = \bigl|\det_{|\ve}(\vv_1,...,\vv_n)\bigr|.
\ee
\findef

\Eg, if $n=1$ and $\vv = v^1\ve_1$, then $\det_{|\ve}(\vv) = v^1$ is the algebraic length of~$\vv$ (relative to the unit of measurement given by~$\ve_1$).
And $|\det_{|\ve}(\vv)| = |v^1|$ is the length of~$\vv$ (the norm of~$\vv$).
(The volume function $(\vv_1,...,\vv_n) \rar \bigl|\det_{|\ve}(\vv_1,...,\vv_n)\bigr|$ is \textbf{\textsl{not}} a multilinear form, because the absolute value function is not linear.)
\Eg, if $n=2$ or~$3$, see exercises~\ref{exedet2}-\ref{exedet3}.

\mn
{\bf Notation.}
Let $(\ve_i)$ be a Cartesian basis and $(e^i)=(dx^i)$ be the dual basis. Then, \cf~Cartan~\cite{cartan},
\be
\det_{|\ve} \eqnote e^1 \wedge ... \wedge e^n = dx^1 \wedge ... \wedge dx^n.
\ee
And, for integration, the volume element (non negative) uses a Euclidean basis $(\ve_i)$ and is
\be
\label{eqvolalgdo}
d\Omega(\vx) = |\det_{|\ve}| = |dx^1\wedge...\wedge dx^n| \eqnote dx^1... dx^n.
\ee
Thus the volume of a parallelepiped at~$\vx$ which sides are given by $\delta x_1\vu_1,...,\delta x_n\vu_n$ is
$d\Omega(\vx)(\delta x_1\vu_1,...,\delta x_n\vu_n)=|\delta x_1...\delta x_n|\,|\det_{|\ve}(\vu_1,...,\vu_n)|$; 
Thus the volume of a polygonal domain $\Omega=\bigcup_{i=1}^N P_i$ where $P_i$ is a parallelepiped which sides are given by $\delta x_{i,1}\vu_{i,1},...,\delta x_{i,n}\vu_{i,n}$ is
\be
|\Omega|=\sum_{i=1}^N |\det_{|\ve}(\vu_{i,1},...,\vu_{i,n})|\delta x_{i,1}...\delta x_{i,n}.
\ee
And thus (Riemann approach), the volume of a regular domain~$\Omega$ is written
\be
|\Omega|=\int_\Omega d\Omega %=\int_{\vx\in\Omega} d\Omega(\vx)
=\int_{\vx\in\Omega} |\det_{|\ve}(\vu_{i,1},...,\vu_{i,n})|\;dx^1... dx^n.
\ee
In particular, since any regular volume~$\Omega$ can be approximated with cubes as small as wished,
$|\Omega|
=\sum_{i=1}^N |\delta x_{i,1}...\delta x_{i,n}\det_{|\ve}(\ve_1,...,\ve_n)|
=\sum_{i=1}^N |\delta x_{i,1}...\delta x_{i,n}|
$ gives
\be
|\Omega|=\int_\Omega d\Omega %=\int_{\vx\in\Omega} d\Omega(\vx)
=\int_{\vx\in\Omega}\;dx^1... dx^n.
\ee

\debexe
Let $\Psi:\vq=(q_1,...,q_n)\in[a_1,b_1] \times ...\times [a_n,b_n] \rar \vx=(x_1=\Psi_1(\vq),...,x_n=\Psi_n(\vq))\in\Omega$ be a parametric description of a domain~$\Omega$;
Prove 
\be
d\Omega(\vx) = |J_\Psi(\vq)| \, dq^1...dq^n \quad (= |\det_{|\ve}(\vp_1(\vx),...,\vp_n(\vx))|\, dq^1...dq^n),
\ee
where
$(\vp_i(x))=({\pa \Psi \over \pa q_i}(\vq))$ is the parametric basis at $\vx=\Psi(\vq)$ and 
$J_\Psi(\vq) = \det_{|\ve}[d\Psi(\vq)]_{|\ve}$
is the Jacobian matrix of~$\Psi$ at~$\vq$. And thus $|\Omega|=\int_{\vq} |J_\Psi(\vq)| \, dq^1...dq^n$.

\debrep
Polar coordinates for illustration purpose (immediate generalization): Consider the disk~$\Omega$ parametrized with the polar coordinate system $\Psi:\vq=(\rho,\theta) \in[0,R]\times [0,2\pi] \rar \vx=(x=\rho\cos\theta, y=\rho\sin\theta)\in\RR^2$
where a Euclidean basis $(\ve_1,\ve_2)$ has been used in~$\RR^2$ (so $\vx = \rho\cos\theta \ve_1 + \rho\sin\theta \ve_2$).
The associated polar basis at $\vx=\Psi(\vq)$ is
$(\vp_1(\vx)={\pa\Psi\over \pa \rho}(\rho,\theta), \vp_2(\vx)={\pa\Psi\over \pa \theta}(\rho,\theta))$,
so $[\vp_1(\vx)]_{|\ve} = \pmatrix{\cos\theta \cr \sin\theta}$ and 
$[\vp_2(\vx)]_{|\ve} = \pmatrix{-\rho\sin\theta \cr \rho \cos\theta}$.
Thus $\det_{|\ve}(\vp_1(\vx),\vp_2(\vx))=\rho$ ($>0$ here), thus
$d\Omega = |\rho|\, d\rho d\theta = \rho\, d\rho d\theta$.
% with the usual notation $d\rho:=p^1$ and $d\theta:=p^2$ where $(p^1(\vx),p^2(\vx))$ is the dual basis of~$(\vp_1(\vx),\vp_2(\vx))$.
Thus the volume is $|\Omega|= \int_{\vx\in\Omega} d\Omega= \int_{\rho=0}^R \int_{\theta=0}^{2\pi} \rho\,d\rho d\theta$ ($=\pi R^2$).
\finrep
\finexe

\debexe
What is the ``volume element'' on a regular surface~$\Sigma$ in~$\RR^3$, called the ``surface element''?
\def\vt{\vec t}

\debrep
Let $(\ve_1,\ve_2,\ve_3)$ be a Euclidean basis in~$\RR^3$.
We need a regular parametric description
$\Psi:(u,v)\in [a_1,b_2]\times [a_2,b_2] \rar \vx=\Psi(u,v)=x_1(u,v)\ve_1+...+x_3(u,v)\ve_3$
of the geometric surface $\Sigma=\Im(\Psi)$.
Thus $\vt_1(\vx)={\pa \Psi\over \pa u}(u,v)$ and $\vt_2(\vx)={\pa \Psi\over \pa v}(u,v)$ are tangent vectors
%which define a basis in the tangent vector plane 
at~$\Sigma$ at $\vx=\Psi(u,v)$. % (since $\Psi$ is supposed to be regular).
Hence a normal unit vector is $\vn(\vx)={\vt_1(\vx)\wedge \vt_2(\vx) \over ||\vt_1(\vx)\wedge \vt_2(\vx)||}$, and thus
$\det_{|\ve}(\vt_1,\vt_2,\vn)=||\vt_1(\vx)\wedge \vt_2(\vx)||$ is the area of the parallelogram which sides are given by $\vt_1$ and~$\vt_2$ (volume with height~$1$). Thus the surface element at $\vx=\Psi(u,v)$ is
$d\Sigma(\vx) = ||{\pa \Psi\over \pa u}(u,v)\wedge {\pa \Psi\over \pa v}(u,v)||\,dudv$. Thus
$|\Sigma|=\int_{\vx\in\Sigma} d\Sigma(\vx)=\int_{u=a_1}^{b_1}\int_{v=a_2}^{b_2}||{\pa \Psi\over \pa u}(u,v)\wedge {\pa \Psi\over \pa v}(u,v)||\,dudv$.
\finrep
\finexe

%%%%%%%%%%%%%%%%%%%%%%%%%%%%%%%%%%%%%%%%%%%%%%%%%%%%%%%%%%%%%%%%%%%%%%%%%%%%%%%%%%%

\subsection{Determinant of an endomorphism}

%%%%%%%%%%%%%%%%%%%%%%%%%%%%%%%%%%%%%%%%%%%%%%%%%%%%%%%%%%%%%%%%%%%%%%%%%%%%%%%%%%%

\subsubsection{Definition and basic properties}

\debdef
The determinant of an endomorphism $L\in \calL(E;E)$ relative to a basis~$(\ve_i)$ is
\be
\label{eqdefdetL}
\tdet_{|\ve}(L)
\eqdef \det_{|\ve}(L.\ve_1,...,L.\ve_n).
\ee
This define $\tdet_{|\ve}:\calL(E;E) \rar \RR$.
(If the context is not ambiguous, then $\tdet_{|\ve}\eqnote \det_{|\ve}$.)
\findef

\debprop
\label{propdetLv}
Let $L\in \calL(E;E)$.

1- If $L=I$ the identity, then $\tdet_{|\ve}(I) = 1$ for all basis $(\ve_i)$.

2- For all $\vv_1,...,\vv_n\in E$,
\be
\label{eqdetLv}
\det_{|\ve}(L.\vv_1,...,L.\vv_n) = \tdet_{|\ve}(L)\det_{|\ve}(\vv_1,...,\vv_n).
\ee

3- If $L.\ve_j = \sumin L_{ij}\ve_i$, \ie\ $[L]_{|\ve} = [L_{ij}]$, then
\be
\label{eqdetLv3}
\tdet_{|\ve}(L) = \det([L]_{|\ve}) = \det([L_{ij}]).
\ee

4- For all $M\in \calL(E;E)$, and with $M \circ  L \eqnote  M . L$ (thanks to linearity),
\be
\label{eqddetLM}
\tdet_{|\ve}( M . L) = \tdet_{|\ve}( M ) \tdet_{|\ve}( L)
= \tdet_{|\ve}( L. M ) . %\qand \tdet_{|\ve}( M \circ  L) \eqnamed \tdet_{|\ve}( M .  L).
\ee

5- $L$ is invertible iff $\tdet_{|\ve}( L)\ne 0$.

6- If $ L$ is invertible then
\be
\label{eqdetinvL}
\tdet_{|\ve}( L^{-1}) = {1\over \tdet_{|\ve}( L)}.
\ee

7- If $\dd_g$ is an inner dot product in~$E$ and $L^T_g$ is the $\dd_g$ transposed of~$L$
(\ie, $(L^T_g\vw,\vu)_g = (\vw,L.\vu)_g$ for all $\vu,\vw\in E$) then %(result independent of~$\dd_g$)
\be
\label{eqdetLv4}
\tdet_{|\ve}(L^T_g) = \tdet_{|\ve}(L).
\ee

8- If $(\ve_i)$ and $(\vb_i)$ are two $\dd_g$-orthonormal bases in~$\RRnt$ (\eg\ two Euclidean basis for the same measuring unit), then $\det_{|\vb} = \pm\det_{|\ve}$.
\finprop

\debdem
1- $\tdet_{|\ve}(I)
\mathop{=}^{\eref{eqdefdetL}} \det_{|\ve}(I.\ve_1,...,I.\ve_n)
\mathop{=}^{\eref{eqdettrans}} \det_{|\ve}(\ve_1,...,\ve_n)=1$, true for all basis.

2- Let $m:(\vv_1,...,\vv_n) \rar m(\vv_1,...,\vv_n) := \det_{|\ve}(L.\vv_1,...,L.\vv_n)$:
It is a multilinear alternated form, since $L$ is linear;
Thus $m\mope^{\eref{eqald}} m(\ve_1,...,\ve_n)\det_{|\ve}$;
With $m(\ve_1,...,\ve_n) \mathop{=}^{\eref{eqdefdetL}} \tdet_{|\ve}(L)$,
% with $m \mathop{=}^{\eref{eqdefdetL}} m(\ve_1,...,\ve_n)\det_{|\ve}$,
thus~\eref{eqdetLv}.

3- Apply~\eref{eqdefdetM} with $M=[L]_{|\ve}$ to get~\eref{eqdetLv3}.

4-
$
%\tdet_{|\ve}( M . L) =
\det_{|\ve} (( M . L).\ve_1,...,( M . L).\ve_n)
=\det_{|\ve} ( M .( L.\ve_1),..., M .( L.\ve_n))
\mope^{\eref{eqdetLv}}\tdet_{|\ve} ( M )\det_{|\ve}( L.\ve_1,..., L.\ve_n)
%=\tdet_{|\ve} ( M )\tdet_{|\ve}( L)
%=\det_{|\ve}( L)\det_{|\ve} ( M )
$.

5- If $ L$ is invertible, then
$1=\tdet_{|\ve}(I)
=\tdet_{|\ve}( L. L^{-1})
=\tdet_{|\ve}( L)\tdet_{|\ve}( L^{-1})$,
thus $\tdet_{|\ve}( L) \ne 0$.

If $\tdet_{|\ve}( L) \ne 0$ then 
$\det_{|\ve}( L.\ve_1,..., L.\ve_n)\ne 0$,
thus $( L.\ve_1,..., L.\ve_n)$ is a basis,
thus $ L$ is invertible.

6- \eref{eqddetLM} gives 
$1= \tdet_{|\ve}(I) = \tdet_{|\ve}( L^{-1}. L) = \tdet_{|\ve}( L).\tdet_{|\ve}( L^{-1})$,
thus~\eref{eqdetinvL}.

7- %\eref{eqaltltbb00} 
$(L^T_g\vw,\vu)_g = (\vw,L.\vu)_g$ gives $[g]_{|\ve}.[L^T_g]_{|\ve} = ([L]_{|\ve})^T.[g]_{|\ve}$,
thus $\det([g]_{|\ve})\det([L^T_g]_{|\ve})=\det(([L]_{|\ve})^T)\det([g]_{|\ve})$, and $\det([g]_{|\ve})\ne0$ (exercise), thus~\eref{eqdetLv4}.

8- Let $\calP$ be the change of basis endomorphism from $(\ve_i)$ to~$(\vb_i)$,
and $P$ be the transition matrix from~$(\ve_i)$ to~$(\vb_i)$.
Both basis being $\dd_g$-orthonormal, $P^T.P=I$, thus $\det(P) = \pm1=\tdet_{|\ve}(\calP)$.
And $\det_{|\ve}(\vb_1,...,\vb_n) = \det_{|\ve}(\calP.\ve_1,...,\calP.\ve_n)
=\tdet_{|\ve}(\calP)\det_{|\ve}(\ve_1,...,\ve_n)
%=\tdet_{|\ve}(\calP)
=\tdet_{|\ve}(\calP)\det_{|\vb}(\vb_1,...,\vb_n)
$, thus $\det_{|\ve} =\tdet_{|\ve}(\calP)\det_{|\vb}=\pm\det_{|\vb}$.
\findem

\debdef
Two $\dd_g$-orthonormal bases $(\ve_i)$ and $(\vb_i)$ have the same orientation iff $\det_{|\vb} = +\det_{|\ve}$.
\findef

\debexe
Prove $\tdet_{|\ve}(\lambda L) = \lambda^n\tdet_{|\ve}(L)$.

\debrep
$\ds \tdet_{|\ve}(\lambda L)
= \det_{|\ve}(\lambda L.\ve_1,...,\lambda L.\ve_n)
= \lambda^n \det_{|\ve}( L.\ve_1,..., L.\ve_n)
= \lambda^n \tdet_{|\ve}( L) 
$.
\finrep
\finexe

\comment{
\debexe
%\label{exedev1erc}
In $\RR^2$, and the component expressions in the basis~$(\ve_i)$,
%Develop relative to the first column to 
check that
$\det_{|\ve}( L.\vv_1, L.\vv_2) = \det_{|\ve}( L)\det_{|\ve}(\vv_1,\vv_2)$.

\debrep
Let $\det=\det_{|\ve}$.
Let $\vc_j = L^1_j\ve_1+L^2_j\ve_2$ and $\vv_j = v_j^1\ve_1 + v_j^2\ve_2$.
Thus $ L.\vv_j = v_j^1\vc_1 + v_j^2\vc_2$.
Then (development relative to the first column)
$\det_{|\ve}( L.\vv_1, L.\vv_2)
= \det_{|\ve} (v_1^1\vc_1 + v_1^2 \vv_2, L.\vv_2)
= v_1^1 \det_{|\ve}(\vc_1, L.\vv_2) + v_1^2 \det_{|\ve}(\vc_2, L.\vv_2)$.
And $\det_{|\ve}(\vc_1, L.\vv_2) = v_2^1\det_{|\ve}(\vc_1,\vc_1) + v_2^2\det_{|\ve}(\vc_1,\vc_2)
= 0+ v_2^2\det_{|\ve}(\vc_1,\vc_2) = v_2^2\det_{|\ve}( L)$,
and $\det_{|\ve}(\vc_2, L.\vv_2) = v_2^1\det_{|\ve}(\vc_2,\vc_1) + v_2^2\det_{|\ve}(\vc_2,\vc_2)
= v_2^1\det_{|\ve}(\vc_2,\vc_1)+0 = -v_2^1\det_{|\ve}( L)$.
Thus $\det_{|\ve}( L.\vv_1, L.\vv_2)=v_1^1v_2^2\det_{|\ve}( L) - v_1^2v_2^1\det_{|\ve}( L)= \det_{|\ve}(\vv_1,\vv_2)\det( L)$.
\finrep
\finexe
}

%%%%%%%%%%%%%%%%%%%%%%%%%%%%%%%%%%%%%%%%%%%%%%%%%%%%%%%%%%%%%%%%%%%%%%%%%%%%%%%%%%%

\subsubsection{The determinant of an endomorphism is objective}

\debprop
Let $(\va_i)$ and $(\vb_i)$ be bases in~$E$. The determinant of an endomorphism $L  \in \calL(E;E)$ is objective (observer independent, here basis independent):
%that is, if $(\va_i)$ and $(\vb_i)$ are bases in~$E$ then
\be
\label{eqobjdet}
(\det([L]_{|\va})=)\quad
\tdet_{|\va}( L ) = \tdet_{|\vb}( L )
\quad (=\det([L]_{|\vb})).
\ee
{\bf NB:} But the determinant of $n$ vectors is {\bf not} objective, \cf~\eref{eqpropdet}
(compare the change of basis formula for vectors $[\vw]_{|\vb}=P^{-1}.[\vw]_{|\va}$ with the change of basis formula for endomorphisms $[ L ]_{|\vb}=P^{-1}.[ L ]_{|\va}.P$).
\finprop

\debdem
Let $(\va_i)$ and $(\vb_i)$ be bases in~$E$, and $P$ be the transition matrix from~$(\va_i)$ to~$(\vb_i)$.
The change of basis formula $[L]_{|\vb} = P^{-1}.[L]_{|\va}.P$ and~\eref{eqddetLM} give
$\det([L]_{|\vb}) = \det(P^{-1})\det([L]_{|\va})\det(P) = \det([L]_{|\va})$, thus \eref{eqdetLv3} gives~\eref{eqobjdet}.
\findem

\debexe
%\label{exedetab}
Let $(\va_i)$ and $(\vb_i)$ be bases in~$E$,
and $\calP \in \calL(E;E)$ be the change of basis endomorphism from $(\va_i)$ to~$(\vb_i)$
(\ie, $\calP.\va_j = \vb_j$ for all~$j$). Prove
\be
\label{eqdefvaln0}
\det_{|\va}(\vb_1,...,\vb_n) = \tdet_{|\va}(\calP), 
\qthus \det_{|\va}= \tdet_{|\va}(\calP)\det_{|\vb},
\qie \det_{|\vb}=  { \det_{|\va} \over\tdet_{|\va}(\calP)} ,
\ee

\debrep
$\ds \det_{|\va}(\vb_1,...,\vb_n)
= \det_{|\va}(\calP.\va_1,...,\calP.\va_n)
\mathop{=}^{\eref{eqdetLv}}  \tdet_{|\va}(\calP)\det_{|\va}(\va_1,...,\va_n)
= \tdet_{|\va}(\calP) \,1
= \tdet_{|\va}(\calP)\det_{|\vb}(\vb_1,...,\vb_n)
$, thus~\eref{eqdefvaln0} and $\det_{|\va}= \tdet_{|\va}(\calP)\det_{|\vb}$ and $\det_{|\va}(\vv_1,...,\vv_n) = \tdet_{|\va}(\calP)\det_{|\vb}(\vv_1,...,\vv_n)$.
\comment{
%Thus $\det_{|\va}(\vv_1,...,\vv_n) \mathop{=}^{\eref{eqdefvaln0}_2} \tdet_{|\va}(\calP)\det_{|\vb}(\vv_1,...,\vv_n)$.

Then prop.~\ref{propdefP0i0} gives $[\calP]_{|\va} = [\calP]_{|\vb}$, thus $\det([\calP]_{|\va})=\det([\calP]_{|\vb})$,
thus~\eref{eqdetLv3} gives~\eref{eqobjdet}; Or, use~\eref{eqdefvaln0} to get
$\ds \tdet_{|\vb}(\calP)
\mathop{=}^{\eref{eqdefdetL}}\det_{|\vb}(\calP.\vb_1,...,\calP.\vb_n)
\mathop{=}^{\eref{eqdefvaln0}}{ \det_{|\va}(\calP.\vb_1,...,\calP.\vb_n) \over \tdet_{|\va}(\calP) }
\mathop{=}^{\eref{eqdetLv}}{ \tdet_{|\va}(\calP) \det_{|\va}(\vb_1,...,\vb_n) \over \tdet_{|\va}(\calP) }
=\det_{|\va}(\vb_1,...,\vb_n)$
$\mathop{=}^{\eref{eqdefvaln0}} \tdet_{|\va}(\calP)
$, thus~\eref{eqdefvaln3}.
Thus 
$$
\eqalign{
\tdet_{|\va}( L )
\mathop{=}^{\eref{eqdefdetL}} & \det_{|\va}( L .\va_1,..., L .\va_n)
\mathop{=}^{\eref{eqdefvaln0}} \tdet_{|\va}(\calP)\det_{|\vb}( L .\va_1,...,L .\va_n)
\mathop{=}^{\eref{eqdefvaln3}} \tdet_{|\vb}(\calP)\tdet_{|\vb}( L )\det_{|\vb}(\va_1,...,\va_n) \cr
= & \tdet_{|\vb}(\calP)\tdet_{|\vb}( L )\det_{|\vb}(\calP^{-1}.\vb_1,...,\calP^{-1}.\vb_n))
=\tdet_{|\vb}(\calP)\tdet_{|\vb}( L )\tdet_{|\vb}(\calP^{-1}) %\det_{|\vb}(\vb_1,...,\vb_n))
=\tdet_{|\vb}( L ),
}
$$
\ie~\eref{eqobjdet}.
}
\finrep
\finexe

%%%%%%%%%%%%%%%%%%%%%%%%%%%%%%%%%%%%%%%%%%%%%%%%%%%%%%%%%%%%%%%%%%%%%%%%%%%%%%%%%%%

\subsection{Determinant of a linear map}

(Needed for the deformation gradient $\Ftzt(P) = d\Phitzt(P) : \RRntz \rar \RRnt$.) % which is not an endomorphism.)

Let $A$ and $B$  be vector spaces, $\dim A = \dim B = n$,
and $(\va_i)$ and $(\vb_i)$ be bases in~$A$ and~$B$.

%%%%%%%%%%%%%%%%%%%%%%%%%%%%%%%%%%%%%%%%%%%%%%%%%%%%%%%%%%%%%%%%%%%%%%%%%%%%%%%%%%%

\subsubsection{Definition and first properties}

\debdef
The determinant of a linear map $L \in \calL(A;B)$ relative to the bases $(\va_i)$ and $(\vb_i)$ is
\be
\label{eqdetal}
\tdet_{|\va,\vb}(L) \eqdef \det_{|\vb}(L.\va_1,...,L.\va_n). % \eqnote J_{|\va,\vb}(L),
\ee
(And $\tdet_{|\va,\vb}(L) \eqnote \det(L)$ if the bases are implicit.)
\findef

Thus, \eref{eqdefdetM} gives, with $L.\va_j = \sumin L_{ij}\vb_i$, \ie\ $[L]_{|\va,\vb}=[L_{ij}]$:
\be
\tdet_{|\va,\vb}(L) = \det([L]_{|\va,\vb}) = \det([L_{ij}]).
\ee

\debprop
Let $\vu_1,...,\vu_n \in A$.
Then
\be
\label{eqddetLw}
\det_{|\vb}(L.\vu_1,...,L.\vu_n)
= \tdet_{|\va,\vb}(L)\,\det_{|\va}(\vu_1,...,\vu_n)
.
\ee
\finprop

\debdem
$m:(\vu_1,...,\vu_n)\in A^n \rar m(\vu_1,...,\vu_n) := \det_{|\vb}(L.\vu_1,...,L.\vu_n) \in \RR$
is a multilinear alternated form, since $L$ is linear;
And $m(\va_1,...,\va_n)
=\det_{|\vb}(L.\va_1,...,L.\va_n)
\mathop{=}^{\eref{eqdetal}} \tdet_{|\va,\vb}(L)
= \tdet_{|\va,\vb}(L)\det_{|\va}(\va_1,...,\va_n)$.
Thus $m = \tdet_{|\va,\vb}(L)\det_{|\va}$, \cf~\eref{eqpropdet}, thus~\eref{eqddetLw}.
\findem

\debcor
Let $A,B,C$ be vector spaces such that $\dim A = \dim B = \dim C = n$. Let $(\va_i)$, $(\vb_i)$, $(\vc_i)$ be bases in~$A,B,C$.
Let $L : A \rar B$ and $M : B \rar C$ be linear. Then, with $M \circ  L \eqnote  M . L$ (thanks to linearity),
\be
\label{eqdML0}
\tdet_{|\va,\vc}(M. L) = \tdet_{|\va,\vb}(L) \tdet_{|\vb,\vc}(M) .
\ee
\fincor

\debdem
$\ds\tdet_{|\va,\vc}(M. L)
= \det_{|\vc}(M. L.\va_1),...,M. L.\va_n))
%= \det_{|\vc}((M. L).\va_1),...,(M. L).\va_n))
%= \det_{|\vc}(M.(L.\va_1),...,M.(L.\va_n))
= \tdet_{|\vb,\vc}(M) \det_{|\vb}(L.\va_1,...,L.\va_n)
= \tdet_{|\vb,\vc}(M)\tdet_{|\va,\vb}(L) $.
\findem

%%%%%%%%%%%%%%%%%%%%%%%%%%%%%%%%%%%%%%%%%%%%%%%%%%%%%%%%%%%%%%%%%%%%%%%%%%%%%%%%%%%

\subsubsection{Jacobian of a motion, and dilatation}
\label{secdilv}

Let $\tPhi$ be a motion, let $\tz,t\in\RR$, let $\Phitzt$ be the associated motion, let $\Ftzt(\ptz) := d\Phitzt(\ptz) : \RRntz \rar \RRnt$
the deformation gradient at $\ptz\in\Omegatz$ relative to~$\tz$ and~$t$, \cf~\eref{eqdefFtf}.
Let $(\vE_i)$ be a Euclidean basis in~$\vRRntz$ and $(\ve_i)$ be a Euclidean basis in~$\vRRnt$ for all $t\ge\tz$,
and $[\Ftzt(\ptz)]_{|\vE,\ve}=[F_{ij}(\ptz)]$, \ie, $\Ftzt(\ptz).\vE_j=\sumijn F_{ij}(\ptz) \ve_i$ for all~$j$.
%and $[\Ftzt(\ptz)]_{|\vE,\ve}=[\FiJ(\ptz)]$, \ie, $\Ftzt(\ptz).\vE_J=\sumijn \FiJ(\ptz) \ve_i$ for all~$J$ (we use Marsden--Hughes duality notations).

\debdef
The ``volume dilatation'' at~$\ptz$, relative to the Euclidean bases 
$(\vE_i)$ in~$\vRRntz$ and $(\ve_i)$ in~$\vRRnt$, is
\be
\label{eqdilv}
J_{|\vE,\ve}(\Phitzt)(\ptz) := \tdet_{|\vE,\ve}(\Ftzt(\ptz)) 
\quad (=\det_{|\ve}(\Ftzt(\ptz).\vE_1,...,\Ftzt(\ptz).\vE_n) = \det([F_{ij}(\ptz)])),
\ee
usually written $J_{|\vE,\ve}:=\det([F]_{|\vE,\ve})$ (or simply $J=\det(F)$ when everything is implicit).
\findef

So, at~$\tz$ at~$\ptz$,
$(\ptz,\vE_1,...,\vE_n)$ is a unit parallelepiped which volume is~$1$ relative to the unit of measurement chosen in~$\RRntz$,
and, at~$t$ at $\pt=\Phitzt(\ptz)$, $J_{|\vE,\ve}(\Phitzt)(\ptz) =\det_{|\ve}(\Ftzt(\ptz).\vE_1,...,\Ftzt(\ptz).\vE_n)$ is the volume of the parallelepiped $(\pt,\Ftzt(\ptz).\vE_1,...,\Ftzt(\ptz).\vE_n)$ relative to the unit of measurement chosen in~$\RRnt$.

\mn
\def\td{{t_2}}%
Interpretation: With $\td>\tu\ge\tz$, and $[\ve_i)$ is the basis at~$\tu$ and~$\td$:

$\bullet$ Dilatation if  $J_{|\vE,\ve}(\Phi^\tz_\td)(\ptz)>J_{|\vE,\ve}(\Phi^\tz_\tu)(\ptz)$ (volume increase),

$\bullet$ contraction if $J_{|\vE,\ve}(\Phi^\tz_\td)(\ptz)<J_{|\vE,\ve}(\Phi^\tz_\tu)(\ptz)$ (volume decrease), and

$\bullet$ incompressibility if $J_{|\vE,\ve}(\Phi^\tz_\td)(\ptz)=J_{|\vE,\ve}(\Phi^\tz_\tu)(\ptz)$ for all~$t$ (volume conservation).
\\
In particular, if $(\ve_i)=(\vE_i)$ then $J_{|\ve,\ve}(\Phitztz)(\ptz) =1$, and if $t>\tz$, then

$\bullet$ Dilatation if  $J_{|\ve,\ve}(\Phitzt)(\ptz)>1$ (volume increase),

$\bullet$ contraction if $J_{|\ve,\ve}(\Phitzt)(\ptz)<1$ (volume decrease), and

$\bullet$ incompressibility if $J_{|\ve,\ve}(\Phitzt)(\ptz)=1$ for all~$t$ (volume conservation).

\debexe
Let $(\vE_i)$ be a Euclidean basis in~$\RRntz$, and let $(\va_i)$ and $(\vb_i)$ be two Euclidean bases in~$\RRnt$ for the same Euclidean dot product~$\dd_g$.
Prove: %$\det_{|\ve}(F.\vE_1,...,F.\vE_n) = \pm\det_{|\vb}(F.\vE_1,...,F.\vE_n)$, \ie, 
\be
J_{|\vE,\va}(\Phitzt(P))=\pm J_{|\vE,\vb}(\Phitzt(P)).
\ee

\debrep
$P$ being the transition matrix from $(\va_i)$ to~$(\vb_i)$, $\det(P) = \pm1$ here.
And~\eref{eqFtztnew} gives $[F]_{|\vE,\va} = P.[F]_{|\vE,\vb}$,
thus $\det([F]_{|\vE,\va}) = \pm \det([F]_{|\vE,\vb})
$, thus $\det_{|\va}(F.\vE_1,...,F.\vE_n) = \pm \det_{|\vb}(F.\vE_1,...,F.\vE_n)$.
\finrep
\finexe

%%%%%%%%%%%%%%%%%%%%%%%%%%%%%%%%%%%%%%%%%%%%%%%%%%%%%%%%%%%%%%%%%%%%%%%%%%%%%%%%%%%

\subsubsection{Determinant of the transposed}

Let $(A,\dd_g)$ and $(B,\dd_h)$ be finite dimensional Hilbert spaces. %, $\dim A = \dim B = n$.
Let $L \in \calL(A;B)$ (a linear map).
Recall: The transposed $L_{gh}^T\in\calL(B;A)$ is defined by, for all $\vu\in A$ and all $\vw\in B$, \cf~\eref{eqseccpgdd0}
\be
\label{eqdetMT}
(L_{gh}^T.\vw,\vu)_g \eqdef (\vw,L.\vu)_h.
\ee
Let $(\va_i)$ be a basis in~$A$ and $(\vb_i)$ be a basis in~$B$.
Then
\be
\label{eqdetMT2}
\tdet([L^T_{gh}]_{|\vb,\va}) = \det([L]_{|\va,\vb}){\det([\dd_g]_{|\va}) \over \det([\dd_h]_{|\vb})}.
\ee
Indeed, \eref{eqdetMT} gives
$[\dd_g]_{|\va}.[L_{gh}^T]_{|\vb,\va} = ([L]_{|\va,\vb})^T.[\dd_h]_{|\vb}$.

%In particular, if $L$ is an endomorphism and $(\va_i)=(\vb_i)$ and $\dd_g=\dd_h$, then we recover~\eref{eqdetLv4}.

%%%%%%%%%%%%%%%%%%%%%%%%%%%%%%%%%%%%%%%%%%%%%%%%%%%%%%%%%%%%%%%%%%%%%%%%%%%%%%%%%%%
%%%%%%%%%%%%%%%%%%%%%%%%%%%%%%%%%%%%%%%%%%%%%%%%%%%%%%%%%%%%%%%%%%%%%%%%%%%%%%%%%%%

\subsection{Dilatation rate}
\label{secdilrate}

A unique Euclidean basis~$(\ve_i)$ at all time is chosen, and $\dd_g$ is the associated inner dot product.
 
%%%%%%%%%%%%%%%%%%%%%%%%%%%%%%%%%%%%%%%%%%%%%%%%%%%%%%%%%%%%%%%%%%%%%%%%%%%%%%%%%%%

\subsubsection{${\pa \Jtz\over\pa t}(t,\ptz)=\Jtz(t,\ptz)\,\dvg\vv(t,\pt)$}

A regular motion $\tPhi$ is considered, \cf~\eref{eqdeftPhi0}, and
the Eulerian velocity is $\vv(t,\pt) = {\pa \tPhi \over \pa t}(t,\Pobj)$ at $\pt= \tPhi(t,\Pobj)$.
Let $\tz$ be given; The associated motion $\Phitz$ is given by
$\Phitz(t,\ptz)=\tPhi(t,\Pobj)\eqnote \pt$ when $\ptz=\tPhi(\tz,\Pobj)$,
~\eref{eqdefPhi}, and is supposed to be at least~$C^2$;
The Lagrangian velocity is $\vV(t,\ptz) = {\pa \Phitz \over \pa t}(t,\ptz)$,
and the Eulerian velocity satisfies $\vv(t,\pt) = {\pa \Phitz \over \pa t}(t,\ptz)$ when $\pt = \Phitz(t,\ptz)$, \cf~\eref{eqremvVnoncv}. Let $\Ftz(t,\ptz) = d\Phitz(t,\ptz) = \Ftzt(\ptz)=d\Phitzt(\ptz)$,
and consider the Jacobian
\be
\Jtzt(\ptz)=\det_{|\ve} (\Ftzt(\ptz))=\Jtz(t,\ptz),
\ee

\deblem
${\pa \Jtz\over\pa t}(t,\ptz)$ satisfies, with $\pt=\Phitzt(\ptz)$,
\be
\label{eqdjodt}
{\pa \Jtz\over\pa t}(t,\ptz) = \Jtz(t,\ptz)\,\dvg\vv(t,\pt)
\ee
(value to be considered at $t$ at~$\pt$).
In particular, $\tPhi$ is incompressible iff $\dvg\vv(t,\pt)=0$.
\finlem

\debdem
Let $\calO$ be a origin in~$\RRn$.
Let $\ora{\calO \Phitz} = \sumin \Phi^i\ve_i$, $\vVtz = \sumin V^i \ve_i$, $\vv = \sumin v^i \ve_i$,
$\Ftz.\vE_j = d\Phitz.\vE_j = \sumin {\pa\Phi^i\over\pa X^j}\ve_i$.
Let $[\Ftz]_{|\vE,\ve} \eqnote F$, $\Jtz\eqnote J$ and
$[d\Phi^i]_{|\vE} = \pmatrix{{\pa\Phi^i \over \pa X^1} &...& {\pa\Phi^i \over \pa X^n}} \eqnote d\Phi^i$  (row matrix).
Thus $J=\det F=\det\pmatrix{d\Phi^1 \cr \vdots \cr d\Phi^n}$, thus (a determinant is multilinear)
$$
{\pa J\over\pa t}
= \det\pmatrix{\ds{\pa(d\Phi^1)\over\pa t} \cr d\Phi^2 \cr \vdots \cr d\Phi^n}
+\ldots 
+ \det\pmatrix{d\Phi^1 \cr \vdots \cr d\Phi^{n-1} \cr \ds{\pa(d\Phi^n)\over\pa t})}.
%= \det\pmatrix{\ds dV^1 \cr d\Phi^2 \cr \vdots \cr d\Phi^n} +\ldots  + \det\pmatrix{d\Phi^1 \cr \vdots \cr d\Phi^{n-1} \cr \ds dV^n}.
$$
With $\Phitz$ $C^2$, thus
$\ds {\pa(d\Phi^i)\over\pa t}(t,\ptz) 
\mope^{\hbox{\footnotesize Swhartz}} d({\pa\Phi^i\over\pa t})(t,\ptz) 
= dV^i(t,\ptz) = dv^i(t,\pt).F(t,\ptz)$% since $V^i(t,\ptz) = v^i(t,\pt)= v^i(t,\Phitz(t,\ptz))$
, \cf~\eref{eqvVvv}.
\comment{
And with the dual basis $(\pi_{ej})$ and $(\pi_{Ek})$ of the basis~$(\ve_j)$ and $(\vE_k)$ we have
$v^i=\sumjn {\pa v^i \over \pa x^j}\pi_{ej}$ and 
$F = \sumjkn {\pa \Phi^j\over \pa X^k}\ve_j\otimes \pi_{Ek}$ (with tensorial notations for simple calculations)
thus
$\ds {\pa(d\Phi^i)\over\pa t}(t,\ptz)
=\sumjkn {\pa v^i \over \pa x^j}(t,\pt){\pa \Phi^j\over \pa X^k} \pi_{Ek}
=\sumjn {\pa v^i \over \pa x^j}(t,\pt) d\Phi^j(t,\ptz)$
}
Thus
$$
\det\pmatrix{\ds{\pa(d\Phi^1)\over\pa t} \cr d\Phi^2 \cr \vdots \cr d\Phi^n}
= \det\pmatrix{\ds\sumin {\pa v^1 \over \pa x^i} d\Phi^i \cr d\Phi^2 \cr \vdots \cr d\Phi^n}
\mope^{\hbox{\footnotesize $\det$ is}}_{\hbox{\footnotesize alternating}} \det\pmatrix{\ds{\pa v^1 \over \pa x^1} d\Phi^1 \cr d\Phi^2 \cr \vdots \cr d\Phi^n}
= {\pa v^1 \over \pa x^1}\det\pmatrix{ d\Phi^1 \cr d\Phi^2 \cr \vdots \cr d\Phi^n}
= {\pa v^1 \over \pa x^1}J
$$
Idem for the other terms, thus
$$
{\pa J\over\pa t}(t,\ptz)
={\pa v^1\over\pa x^1}(t,\pt)\;J(t,\ptz)+\ldots+{\pa v^n\over\pa x^n}(t,\pt)\;J(t,\ptz)
 = \dvg\vv(t,\pt)\;J(t,\ptz),
$$
i.e.~\eref{eqdjodt}.
\findem

\debdef
$\dvg\vv(t,\pt)$ is the dilatation rate.
\findef

\comment{
\debexe
Dans~$\RR^3$, démontrer~\eref{eqdjodt} par un calcul direct.

\debrep
On a (déterminant d'une matrice 3*3), notant $X^4{=}X^1$ et $X^5{=}X^2$ :
$$
   J(t,P)
=\sum_{j=1}^3{\pa\phi^1\over\pa X^j}{\pa\phi^2\over\pa X^{j+1}}{\pa\phi^3\over\pa X^{j+2}}
-\sum_{j=1}^3{\pa\phi^1\over\pa X^{j+2}}{\pa\phi^2\over\pa X^{j+1}}{\pa\phi^3\over\pa X^j},
$$
d'où : %(on utilise la multilinéarité sans s'en rendre compte) :
$$
\eqalign{
   {\pa J\over\pa t}(t,P)
=&\sum_{j=1}^3
  ({\pa V^1\over\pa X^j}{\pa\phi^2\over\pa X^{j+1}}{\pa\phi^3\over\pa X^{j+2}}
 +{\pa\phi^1\over\pa X^j}{\pa V^2\over\pa X^{j+1}}{\pa\phi^3\over\pa X^{j+2}}
 +{\pa\phi^1\over\pa X^j}{\pa\phi^2\over\pa X^{j+1}}{\pa V^3\over\pa X^{j+2}})\cr
-&\sum_{j=1}^3
  ({\pa V^1\over\pa X^{j+2}}{\pa\phi^2\over\pa X^{j+1}}{\pa\phi^3\over\pa X^j}
 +{\pa\phi^1\over\pa X^{j+2}}{\pa V^2\over\pa X^{j+1}}{\pa\phi^3\over\pa X^j}
 +{\pa\phi^1\over\pa X^{j+2}}{\pa\phi^2\over\pa X^{j+1}}{\pa V^3\over\pa X^j}).
\cr}
$$
D'autre part, ayant $V^i(t,P)=v^i(t,p)=v^i(t,\Phi(P))$, on~a :
$$
   {\pa V^i\over\pa X^j}=\sum_{k=1}^3{\pa v^i\over\pa x^k}{\pa \phi^k\over\pa X^j}.
$$
Sachant que
$\dvg\vv(t,p)=({\pa v^1\over\pa x}+{\pa v^2\over\pa y}+{\pa v^3\over\pa z})(t,p)$,
on obtient~\eref{eqdjodt}.
\finrep
\finexe
}

%%%%%%%%%%%%%%%%%%%%%%%%%%%%%%%%%%%%%%%%%%%%%%%%%%%%%%%%%%%%%%%%%%%%%%%%%%%%%%%%%%%

\subsubsection{Leibniz formula}

\begin{prop}[Leibniz formula]
Under regularity assumptions (\eg\ hypotheses of the Lebesgue theorem to be able to derive under~$\int$) we have
\be
\label{eqdjodt2}
\eqalign{
{d\over dt}\Bigl(\int_{\pt\in\Omegat} f(t,\pt)\,d\Omega_t \Bigr)
 = & \int_{\pt\in\Omegat} \bigl({Df\over Dt}+f\,\dvg\vv\bigr)(t,\pt)\,d\Omega_t \cr
= & \int_{\pt\in\Omegat} \bigl({\pa f\over \pa t}+ d f.\vv+f\,\dvg(\vv)  \bigr)(t,\pt)\,d\Omega_t\cr
 = & \int_{\pt\in\Omegat} \bigl({\pa f\over \pa t}+\dvg(f\vv)\bigr)(t,\pt)\,d\Omega_t.\cr
}
\ee
\finprop

\debdem
Let 
$$
\eqalign{
Z(t) \eqdef \int_{p\in\Omegat} f(t,p)\,d\Omegat
%= & \int_{P\in\Omegatz} f(t,\Phitzt(P))\,|\Jtzt(P)|\,d\Omegatz \cr
= & \int_{P\in\Omegatz} f(t,\Phitz(t,P))\,\Jtz(t,P)\,d\Omegatz.
}
$$
(The Jacobian is positive for a regular motion.) Then (derivation under~$\int$)
$$
\eqalign{
Z'(t) %\eqnote {d\over dt}\Bigl(\int_{p\in\Omega_t} f(t,p)\,d\Omega_t\Bigr)
%= & {d\over dt}\Bigl(\int_{P\in\Omegatz} f(t,\Phitz(t,P))\,\Jtz(t,P)\,d\Omegatz\Bigr) \cr
= & \int_{P\in\Omega_{t_0}} {Df\over D t}(t,\pt)\,\Jtz(t,P)    + f(t,\pt){\pa \Jtz\over\pa t}(t,P)\,d\Omega_{t_0} \cr
= & \int_{P\in\Omega_{t_0}} ({Df\over D t}(t,\pt)\
   + f(t,\pt)\, \dvg\vv(t,\pt))\Jtz(t,P)\,d\Omega_{t_0},
}
$$
thanks to~\eref{eqdjodt}.
And $\dvg(f\vv)= d f.\vv+f\,\dvg\vv$ gives~\eref{eqdjodt2}.
\findem

\debcor
With $(\vu,\vw)_g\eqnote \vu\bcdot \vw$ (in the given Euclidean framework),
\be
\label{eqdjodt3}
{d\over dt}\int_{\Omega_t} f(t,\pt)\,d\Omega_t 
 = \int_{\Omega_t} {\pa f\over \pa t}(t,\pt)\,d\Omega_t
   + \int_{\pa\Omega_t}(f\vv \bcdot \vn)(t,\pt)\,d\Gamma_t,
\ee
sum of the temporal variation within~$\Omega_t$ and the flux through the surface~$\pa\Omega_t$.
\fincor

\debdem
Apply% the Gauss formula to
~\eref{eqdjodt2}$_3$.
\findem

%%%%%%%%%%%%%%%%%%%%%%%%%%%%%%%%%%%%%%%%%%%%%%%%%%%%%%%%%%%%%%%%%%%%%%%%%%%%%%%%%%%

\subsection{$\pa J / \pa F = J\,F^{-T}$}

%%%%%%%%%%%%%%%%%%%%%%%%%%%%%%%%%%%%%%%%%%%%%%%%%%%%%%%%%%%%%%%%%%%%%%%%%%%%%%%%%%%

\subsubsection{Meaning of ${\pa \det \over \pa M_{ij}}$?}
\label{secpadet}

\def\tZ{\tilde Z}

Let $\calM_{nn}=\{M= [M_{ij}]\in \RR^{n^2}\}$ be the set of $n*n$ matrices, and consider the function
\be
Z:=\det:\left\{\eqalign{
\calM_{nn} &\rar \RR \cr
M=[M_{ij}] & \rar Z(M) := \det(M)=\det([M_{ij}]). \cr
}\right.
\ee

Question: What does ${\pa Z \over \pa M_{ij}}(M)$ mean?

Answer: It is the ``standard meaning'' of a directional derivative ${\pa f \over \pa x_i}(\vx)=df(\vx).\ve_i$% of a function $f:\vec{\RR^k} \rar\RR$
... where here %$k=mn$, 
$f=Z$, thus $\vx \eqnote M$ is a matrix (a vector in~$\calM_{nn}$), and $(\ve_i)$ is the canonical basis $(m_{ij})$ in~$\calM_{nn}$ (all the elements of the matrix $m_{ij}$ vanish but the element at intersection of line~$i$ and column~$j$ which equals~$1$). So:
\be
\label{eqpaZpaMij}
{\pa Z \over \pa M_{ij}}(M) := dZ(M).m_{ij} = \lim_{h\rar0} {Z(M+hm_{ij}) - Z(M) \over h} \;\;(\in\RR).
\ee

%%%%%%%%%%%%%%%%%%%%%%%%%%%%%%%%%%%%%%%%%%%%%%%%%%%%%%%%%%%%%%%%%%%%%%%%%%%%%%%%%%%

\subsubsection{Calculation of ${\pa \det \over \pa M_{ij}}$}

\debprop
\be
\label{eqpaJpaF0}
\forall i,j,\;\;{\pa Z \over \pa M_{ij}}(M) = Z(M)\,(M^{-T})_{ij},
\qwritten 
{\pa Z \over \pa M} = Z\,M^{-T}.
\ee
%(Warning: $\tJ$ is a function of~$F$, while $J$ is a function of~$\ptz$, both depending on~$(\vE_i)$ and~$(\ve_i)$.)
\finprop

\debdem
${\pa Z \over \pa M_{ij}}(M) 
:= \lim_{h\rar0} {\det(M+hm_{ij}) - \det(M) \over h}
$;
The development of the determinant $\det(M+hm_{ij})$ relative to the column~$j$ gives
\be
\det(M+h[m_{ij}]) = \det(M) + h \, c_{ij}
\ee
where $c_{ij}$ is the $(i,j)$-th cofactor of~$M$;
Thus ${\pa Z \over \pa M_{ij}}(M) = \lim_{h\rar0} {Z(M+hm_{ij}) - Z(M) \over h} = c_{ij}$;
And since $M^{-1} = {1\over \det(M)}[c_{ij}]^T$, \ie\ $[c_{ij}] = \det(M)M^{-T}$, we get
${\pa Z \over \pa M_{ij}}(M) = \det(M)(M^{-T})_{ij}$, \ie~\eref{eqpaJpaF0}.
\findem

%%%%%%%%%%%%%%%%%%%%%%%%%%%%%%%%%%%%%%%%%%%%%%%%%%%%%%%%%%%%%%%%%%%%%%%%%%%%%%%%%%%

\subsubsection{$\partial J / \partial F = J\,F^{-T}$ usually written $[{\pa J \over \pa F_{ij}}] = J\,F^{-T}$}

Setting of~\S~\ref{secdilrate}: With $F:=d\Phi(\ptz)$ we have
$F.\vE_j = \sumin F_{ij} \ve_i$ where $F_{ij}={\pa\Phi_i \over \pa X_j}(\ptz)$, % for all~$i,j$. 
and
\be
J_{\Phi,\ptz,\vE,\ve} \eqnote 
J:
\left\{\eqalign{
\calL(\RRntz;\RRnt) & \rar \RR \cr
F & \rar J(F):= %\tdet_{|\vE,\ve}(F)
\det([F_{ij}]) \quad (=\det([{\pa\Phi_i \over \pa X_j}(\ptz)]),
}\right. %\qso J(F(\ptz)) :=J(\ptz)
\ee
%for all $\ptz\in\Omegatz$.
so, $J(F)$ is the Jacobian $\tdet_{|\vE,\ve}(d\Phi(\ptz))$ of~$\Phi$ at~$\ptz$ relative to~$(\vE_i)$ and~$(\ve_i)$.
Thus~\eref{eqpaJpaF0} gives:

\debcor
\be
\label{eqpaJpaF}
\forall i,j,\;\;{\pa J \over \pa F_{ij}}(F) = J(F)\,([F]^{-T})_{ij},
\qwritten 
{\pa J \over \pa F} = J\,F^{-T}.
\ee
%(Warning: $\tJ$ is a function of~$F$, while $J$ is a function of~$\ptz$, both depending on~$(\vE_i)$ and~$(\ve_i)$.)
\fincor

%%%%%%%%%%%%%%%%%%%%%%%%%%%%%%%%%%%%%%%%%%%%%%%%%%%%%%%%%%%%%%%%%%%%%%%%%%%%%%%%%%%

\subsubsection{Interpretation of ${\pa J \over \pa F_{ij}}$?}
\label{secmean} 
%\label{remmean} 

The first derivations into play are along the directions~$\vE_j$ at~$\tz$: The $F_{ij}={\pa\Phi_i \over \pa X_j}(\ptz) := d\Phi_i(\ptz).\vE_j$.

\medskip
Question: ${\pa J \over \pa F_{ij}}$ is the usual notation for a directional derivative, \cf~\S~\ref{secpadet}.
So ${\pa J \over \pa F_{ij}}$ is the derivative in which direction?

\medskip
Answer:
1- ``Identify'' $F\in\calL(\RRntz;\RRnt)$ with the tensor $\tF \in \calL(\RRnts,\RRntz;\RR)$
given by $\tF(\ell,\vU)=\ell.(F.\vU)$;
So, if $F.\vE_j = \sumin F_{ij} \ve_i$ then $\tF = \sumijn F_{ij}\ve_i \otimes \pi_{Ej}$,
relative to a basis $(\vE_i)$ and its covariant dual basis $(\pi_{Ei})$ in~$\RRntz$ and a basis $(\ve_i)$ and~$\RRnt$.
%, \ie\ the $F_{ij}$ are the components of~$\tF$ relative to the basis $(\ve_i \otimes \pi_{Ej})$.

2- Define the function
$\tdet_{\vE,\ve}=\tJ:
\left\{\eqalign{
\calL(\RRnts,\RRntz;\RR) &\rar \RR \cr
\tF &\rar \tJ(\tF): = J(F) = \det_{\vE,\ve}(F) = \det([F_{ij}]) \cr
}\right\}$;

3- Then it is meaningful to differentiate $\tJ$ along the direction $\ve_i \otimes \pi_{Ej}\in\calL(\RRnts,\RRntz;\RR)$ to get
\be
{\pa \tJ \over \pa F_{ij}}(\tF)
:= \lim_{h\rar0} {\tJ_{|\ve,\vE}(\tF+h\ve_i \otimes \pi_{Ej}) - \tJ_{|\ve,\vE}(\tF) \over h}
\quad (\eqnote {\pa J \over \pa F_{ij}}(F));
\ee
This is a derivation in both directions $\pi_{Ej}$ in~$\RRntz$ (past at~$\ptz$) \textsl{\textbf{and}} $\ve_i$ in~$\RRnt$ (present at~$\pt$).
What does this derivative %along $\ve_i \otimes \pi_{Ej}$ 
mean?
(The answer is unknown to the author.)

\comment{
And then there is a derivation in both directions $\ve_i$ and $\pi_{E_j}$ where $(\pi_{E_j})$ is the (covariant) dual basis of~$(\vE_i)$:
\be
{\pa J \over \pa F_{ij}}(F)
= \lim_{h\rar0} {\det([{\pa\Phi_i \over \pa X_j}(\ptz)]_{|\vE,\ve} + h [\ve_i\otimes \pi_{E_j}]_{|\vE,\ve} - \det([{\pa\Phi_i \over \pa X_j}(\ptz)]_{|\vE,\ve}) \over h},
\ee 
which makes the interpretation of ${\pa J \over \pa F_{ij}}$ not obvious.
}

\comment{
%%%%%%%%%%%%%%%%%%%%%%%%%%%%%%%%%%%%%%%%%%%%%%%%%%%%%%%%%%%%%%%%%%%%%%%%%%%%%%%%%%%

\subsubsection{Problem ${\pa  \over \pa F^i_J}$ of with tensorial notations}

Let $(\vE_i)$ and $(\ve_i)$ be bases in~$\RRntz$ and~$\RRnt$, let $P\in\Omegatz$, let $o\in\RRn$ be an origin,
let $\Phi = \Phitzt$, let $\ora{o\Phi(P)} = \sumin \Phi^i(P)\ve_i$,
let $F_P.\vE_j = \sumin F^i_j(P) \ve_i = \sumin {\pa\Phi^i \over \pa X^j}(P) \ve_i$ for all~$j$. Consider a function, \cf~\eref{eqdetal},
\be
\tJ_P:=
\left\{\eqalign{
\calL(\RRntz;\RRnt) & \rar \RR \cr
F_P & \rar \tJ_P(F_P) = \tdet_{|\vE,\ve}(F_P) = \det_{|\ve} (F_P.\vE_1,...,F_P.\vE_n) = \det([F^i_j(P)]). % = \det([F^i_J(P)]).
}\right.
\ee

Let $\tF_P
\in \calL(\RRnts,\RRntz;\RR)$ be the bilinear map that naturally represents the linear map $F_P\in\calL(\RRntz;\RRnt)$, that is, $\tF_p(\ell,\vu)=\ell.F.\vu$ for all $(\ell,\vu)\in\RRnts\times \RRntz$, \cf~\eref{eqdefcalJ}. Thus
$\tF_P= \sumijn F_{ij}(P) \ve_i\otimes \pi_{Ej}
$, where $(\pi_{Ei})$ is the dual basis of~$(\ve_i)$, and $[\tF]_{|\vE,\ve} = [F]_{|\vE,\ve}$. Then let
\be
\tJ_P: 
\left\{\eqalign{
\calL(\RRntz,\RRntz;\RR) & \rar \RR \cr
\tF_P & \rar \tJ_P(\tF_P):= J_P(F_P). % = \tdet_{|\vE,\ve}(\tF) = \det_{|\ve} (F.\vE_1,...,F.\vE_n) = \det([F^i_J]),
}\right.
\ee
Thus we have
\be
\label{eqdJF}
{\pa \tJ_P \over \pa F_{ij}}(\tF_P)
:= d\tJ_P(\tF_P).(\ve_i \otimes \pi_{Ej})
=\lim_{h\rar0} {\tdet_{|\vE,\ve}(\tF_P+h\ve_i \otimes \pi_{Ej}) - \tdet_{|\vE,\ve}(\tF_P) \over h}.
%\eqnote {\pa J \over \pa F_{ij}}(F_P).
\ee
(Duality notations: ${\pa \tJ_P \over \pa F^i_J}$.) And let
\be
[{\pa \tJ \over \pa \tF}(\tF)]_{|\vE,\ve} := [{\pa \tJ_P \over \pa F_{ij}}(\tF_P)]
\ee

\debprop
\be
\label{eqpaJpaF}
[{\pa \tJ \over \pa \tF}]_{|\vE,\ve} = \tJ\,[\tF]_{|\vE,\ve}^{-T}, % \eqnamed {\pa J \over \pa F},
\qwritten {\pa J \over \pa F} = J\,F^{-T}.
\ee
\finprop

\debdem
${\pa \tJ \over \pa F_{ij}}(\tF) %= d\tJ(\tF)(\ve_i \otimes \pi_{Ej})
= \lim_{h\rar0} {\tdet(\tF+h\ve_i \otimes \pi_{Ej}) - \tdet(\tF) \over h}
$, and the developments of the determinants relative to the column~$j$ give
$\tdet(\tF+h\ve_i \otimes \pi_{Ej}) - \tdet(\tF)
= h \, c_{ij}$ with $c_{ij}$ the cofactor of $[F]:=[\tF]_{|\vE,\ve}=[F]_{|\vE,\ve}$;
%Since $[c] = \tdet(F)[\tF]^{-T}$, we get
And $[\tF]^{-1} = {1\over \tdet(\tF)}[c]^T$, thus
${\pa \tJ \over \pa F_{ij}}(\tF) = c_{ij} = \tdet(\tF)([\tF]^{-T})_{ij}$, thus \eref{eqpaJpaF}.
\findem

}

\comment{
\debrem
\label{rempaFij}
The derivation ${\pa J \over \pa F^i_J}(F)$, \cf~\eref{eqdJF}, is a derivation where you derive at the same time in a direction at~$\tz$ (in the past) and in a direction at~$t$... 
However, a direct application of the definition of the differential~$dJ$ indicates
that $dJ(L).M 
=\lim_{h\rar0} {\tdet_{|\vE,\ve}(L+hM) - \tdet_{|\vE,\ve}(F) \over h}
=\lim_{h\rar0} {\det_{\ve}((L+hM).\vE_1,...,(L+hM).\vE_n) - \det_{|\ve} (L.\vE_1,...,L.\vE_n) \over h}
$

\finrem
}

\comment{

Then, given a motion $\Phitzt\eqnamed \Phi$, its Jacobian $\Jtzt \eqnamed J$ is, with~\eref{eqhWP1},
\be
\label{eqhWP1}
J(\Phi) = \hW(d\Phi) :
\left\{\eqalign{
\Omegatz & \rar \RR \cr
P & \rar J(\Phi)(P) = \hW(d\Phi)(P) = \tdet(d\Phi(P)) = \tdet(F(P)) . %\qwhen F:=\Ftzt(P)=d\Phitzt(P).
}\right.
\ee

And the notation ${\pa J \over \pa F^i_J}(P)$ means the partial derivative of~$\tJ$ in the direction $\ve_i \otimes E^J$:
\be
{\pa \tJ \over \pa F^i_J}(F)  := d\tJ(F).(\ve_i\otimes E^J)
=\lim_{h\rar0} {\tJ(F+h\ve_i\otimes E^J) - \tJ(F) \over h}.
\ee
}

\comment{
%%%%%%%%%%%%%%%%%%%%%%%%%%%%%%%%%%%%%%%%%%%%%%%%%%%%%%%%%%%%%%%%%%%%%%%%%%%%%%%%%%%

\subsection{Deformation tensor $C$ and dilatation}

By definition the deformation tensor~$C$ is the endomorphism in~$\vRRntz$ defined by,
for all $\vU,\vW\in\vRRntz$,
\be
(C.\vU,\vW)_\gtz \eqdef (F.\vU,F.\vW)_\gt, \qie C=F^T_{\gtz\gt}.F.
\ee

\debcor
Hypotheses: 
$\gt$ is a Euclidean dot product in~$\RRnt$.
And $\gtz$ is the pull-back of~$\gt$ by~$\Phitzt$, that is, for all $\vWu,\vWd$,
\be
\gtz(\vWu,\vWd) = \gt(F.\vWu,F.\vWd).
\ee

On suppose que $\det F>0$ (conserve l'orientation), \cf~\eref{eqdilv}.
Soit $\dd_\gt$ un produit scalaire dans~$\RRnt$.
Soit $\dd_\gtz$ le produit scalaire dans~$\RRntz$ défini par (pull-back), pour tout $i,j$ :
\be
(\vE_i,\vE_j)_\gtz \eqdef (F.\vE_i,F.\vE_j)_\gt.
\ee
avec $C=F_\gt^T.T$ le tenseur des déformations.)
Alors pour les matrices relativement à la base~$(\vE_i)$ on~a
$[h]=[g].[F^T_g].[F]$
où $[h] = [\dd_h]$ est la matrice du produit scalaire~$\dd_h$ dans la base~$(\vE_i)$,
et $[F^T_g] = [g]^{-1}.[F]^T.[g]$. Donc :
\be
\label{eqhgT}
[h] = [g].[F]^T.[F],
\ee
et :
\be
\label{eqhgT2}
\det[h] = (\det [g])(\det F)^2.
\ee
En particulier si $\dd_g$ est le produit scalaire euclidien associé à la base euclidienne~$(\vE_i)$
alors $\det[h] = (\det F)^2 =$ le coefficient de dilatation volumique au carré, \cf~\eref{secdilv}.
\fincor
%Dans~$\vRRn$, soit une base euclidienne~$(\vE_i)$ and son produit scalaire euclidien~$\dd_g$ associé.

C'est la démarche adoptée pour le tenseur des déformations de Cauchy $C=F^T.F$ \cf~\eref{eqdefC}-\eref{eqdefCt2} :
si $\dd_\gt$ a également un sens dans~$\RRntz$, alors, avec $F^T=F_\gt^T$ défini par
$(F^T_\gt.\vy,\vx)_\gt = (\vy,F.\vx)_\gt$ pour tout $\vx,\vy$,
on~a $(\vE_i,\vE_j)_\gtz = (C.\vE_i,\vE_j)_\gt$.
Donc ici $\dd_\gtz$ est la métrique ``pull-back'' de la métrique euclidienne~$\dd_\gt$.

}

%%%%%%%%%%%%%%%%%%%%%%%%%%%%%%%%%%%%%%%%%%%%%%%%%%%%%%%%%%%%%%%%%%%%%%%%%%%%%%%%%%%
%%%%%%%%%%%%%%%%%%%%%%%%%%%%%%%%%%%%%%%%%%%%%%%%%%%%%%%%%%%%%%%%%%%%%%%%%%%%%%%%%%%

\section{Transport of volumes and areas}

%%%%%%%%%%%%%%%%%%%%%%%%%%%%%%%%%%%%%%%%%%%%%%%%%%%%%%%%%%%%%%%%%%%%%%%%%%%%%%%%%%%

\def\vUP{{\vU_P}}
\def\vWP{{\vW_P}}
\def\vUuP{{\vU_{1P}}}
\def\vUdP{{\vU_{2P}}}
\def\vUnP{{\vU_{nP}}}
\def\vNP{{\vN_P}}
\def\vup{{\vu_p}}
\def\vwp{{\vw_p}}
\def\vuup{{\vu_{1p}}}
\def\vudp{{\vu_{2p}}}
\def\vunp{{\vu_{np}}}
\def\vnp{{\vn_p}}

Here $\RRn=\RRt$ the usual affine space.
Let $\tz,t\in\RR$, and $\Phitzt : \RR\times \Omegatz \rar \Omegat$, see~\eref{eqdefPhi}.
Let $F_P = d\Phitzt(P)$.
Let $\dd_g$ be a Euclidean dot product in~$\vRRn$ (English, French...),
with $||.||_g$ the associated norm.

%Let $(\ve_i)$ be a unique Euclidean basis at all time (so the same in~$\RRntz$ and~$\RRnt$), \eg\ made with a foot of with a metre,
% (which means $\ve_i = \Stzt.\vE_i$, \cf\ the shifter~\eref{eqdefshift}),

%%%%%%%%%%%%%%%%%%%%%%%%%%%%%%%%%%%%%%%%%%%%%%%%%%%%%%%%%%%%%%%%%%%%%%%%%%%%%%%%%%%

\subsection{Transformed parallelepiped}

The Jacobian of~$\Phitzt$ at~$P$ relative to a $\dd_g$-Euclidean bases is defined in~\eref{eqdilv}: With $F_P = \Ftzt(P)$, % and $\det_{|\ve} \eqnote \det$,
\be
\label{eqJ0}
J_P = J(P) := \det_{|\ve}(\Ftzt(P).\vE_1,...,\Ftzt(P).\vE_n)), \qand J_P>0
\ee
the motion being supposed regular.
%So, $J_P$ is the volume of the parallelepiped $(F_P.\ve_1,....,F_P.\ve_n)$.
Thus, if $(\vUuP,...,\vUnP)$ is a parallelepiped at $P$ at~$\tz$, if $\vu_{ip} = F_P.\vU_{iP}$,
then $(\vuup,...,\vunp)$ is a parallelepiped at $p$ at~$t$ which volume is
\be
\label{eqfcv0}
\det_{|\ve}(\vuup,...,\vunp) = J_P\;\det_{|\ve}(\vUuP,...,\vUnP).
\ee

%%%%%%%%%%%%%%%%%%%%%%%%%%%%%%%%%%%%%%%%%%%%%%%%%%%%%%%%%%%%%%%%%%%%%%%%%%%%%%%%%%%

\subsection{Transformed volumes}

Riemann integrals and~\eref{eqfcv0} give the change of variable formula: For any regular function $f:\Omegat\rar\RR$,
\be
\label{eqcvdi}
%((d\Omegat(f)=)\quad 
\int_{\pt\in\Omegat} f(\pt)\, d\Omega_t
= \int_{P\in\Omegatz} f(\Phitzt(P))\, |J(P)|\,d\Omega_{t_0}.
\ee
(See~\eref{eqvolalgdo}: $d\Omegat$ is a positive measure: It is not a multilinear form.)
In particular,
\be
|\Omegat| = \int_{\pt\in\Omegat}  d\Omegat(\pt) = \int_{P\in\Omegatz}  |J(P)|\,d\Omegatz(P).
%\qwritten \int_\Omegat d\Omegat = \int_\Omegatz |J|\,d\Omegatz.
\ee
(With $J(P)>0$ for regular motions.)

\comment{
Then the volume element is defined to be the determinant, that is,
with $(e^i)=(dx^i)$ the dual basis of the basis~$(\ve_i)$,
define, at all time~$t$,
\be
d\Omegat = \det = dx^1\wedge...\wedge dx^n.
\ee
}

%%%%%%%%%%%%%%%%%%%%%%%%%%%%%%%%%%%%%%%%%%%%%%%%%%%%%%%%%%%%%%%%%%%%%%%%%%%%%%%%%%%

\subsection{Transformed parallelogram}

Consider two independent vectors $\vUuP,\vUdP\in\RRntz$ at~$\tz$ at $P$,
and the vectors $\vuup = F_P.\vUuP$ and $\vudp = F_P.\vUdP$ at~$t$ at $p=\Phitzt(P)$. Since $\Phitzt$ is a diffeomorphism, $\vuup$ and $\vudp$ are independent.

Then choose a Euclidean dot product~$\dd_g$ (English, French...) to be able to use the vectorial product, \cf~\eref{eqw1}, the same at all time~$t$. %, and a unit normal vector (relative to the chosen unit of measurement).
Then the areas of the parallelograms are
\be
||\vUuP\wedge \vUdP||_g \qand ||\vuup\wedge \vudp)||_g,
\ee
and unit normal vectors to the quadrilaterals are
\be
\vNP={\vUuP\wedge\vUdP\over||\vUuP\wedge\vUdP||_g}\;\in\RRntz,\qand
\vnp={\vuup\wedge\vudp\over||\vuup\wedge\vudp||_g}\;\in\RRnt.
\ee

\debprop
\label{proptransaire00}
If $\vuup =  F_P.\vUuP$ and $\vudp =  F_P.\vUdP$, then
\be
\label{eqtransaire00}
\vuup\wedge \vudp = J_P\,F_P^{-T}.\bigl(\vUuP\wedge \vUdP\bigr),
\qand ||\vuup\wedge \vudp||_g = J_P\,||F_P^{-T}.(\vUuP\wedge \vUdP)||_g,
\ee
since $J_P>0$ (for regular motions), and
\be
\label{eqrelvnvN}
\vnp= {F_P^{-T}.\vNP\over ||F_P^{-T}.\vNP||_g} \quad ( \ne F_P.\vNP \hbox{ in general}).
\ee
%And $\vnp \ne F_P.\vNP$ in general.
\finprop

\debdem
Let $\vWP\in\vRRntz$ and $\vwp=F_P.\vWP$.
Then the volume of the parallelepiped $(\vuup,\vudp,\vwp)$ is
\be
\label{eqtransaire001}
\eqalign{
(\vuup\wedge\vudp,\vwp)_g
= & \det(\vuup,\vudp,\vwp)
=  \det(F_P.\vUuP,F_P.\vUdP,F_P.\vWP)
=  \det(F_P)\det(\vUuP,\vUdP,\vWP) \cr
= & J_P\,(\vUuP\wedge \vUdP,\vWP)_g 
=  J_P\,(\vUuP\wedge \vUdP,F_P^{-1}.\vwp)_g
=  J_P\,(F_P^{-T}.(\vUuP\wedge \vUdP),\vwp)_g,
}
\ee
for all~$\vwp$, thus~\eref{eqtransaire00},
thus ${\vuup\wedge \vudp\over ||\vuup\wedge \vudp||_g}
 = {J_P\,F_P^{-T}.(\vUuP\wedge \vUdP) \over J_P ||F_P^{-T}.(\vUuP\wedge \vUdP)||_g},
$
 thus~\eref{eqrelvnvN}.
\findem

\def\vrtz{{\vr_\tz}}
\def\vs{{\vec s}}
\def\vsigma{{\vec\sigma}}
\def\vSigma{{\vec\Sigma}}
\def\vSigmaP{{\vec\Sigma_P}}
\def\SigmaP{{\Sigma_P}}
\def\sigmap{{\sigma_p}}
\def\vsigmap{{\vsigma_p}}
\def\Stz{{S_\tz}}
\def\vT{{\vec T}}
\def\vTuP{{\vec T_{1P}}}
\def\vTdP{{\vec T_{2P}}}
\def\vt{{\vec t}}
\def\vtup{{\vec t_{1p}}}
\def\vtdp{{\vec t_{2p}}}

%%%%%%%%%%%%%%%%%%%%%%%%%%%%%%%%%%%%%%%%%%%%%%%%%%%%%%%%%%%%%%%%%%%%%%%%%%%%%%%%%%%

\subsection{Transformed surface}
\label{secSurfs}

%%%%%%%%%%%%%%%%%%%%%%%%%%%%%%%%%%%%%%%%%%%%%%%%%%%%%%%%%%%%%%%%%%%%%%%%%%%%%%%%%%%
\comment{
\subsubsection{Parametrized surface in~$\RRn$}

A parametric surface in~$\RRt$ and the associated geometric surface are
\be
\label{eqdefr00}
\Psi:
\left\{\eqalign{
[a,b]\times[c,d] & \rar\RRt \cr
(u,v) & \rar P =\Psi(u,v), %= \pmatrix{x_{t_0}(u,v) \cr y_{t_0}(u,v) \cr z_{t_0}(u,v)}.
}\right\} \qand S = \Im(\Psi)\subset\RRt.
\ee
$u,v$ are the parameters, $P$ is the associated position in our affine space $\RRt$.
\Eg, the sphere in~$\RRt$ with $u=\theta$ and $v=\phi$.

So a parametrized surface is generated by the curves
$\Psi_v : u\in [a,b] \rar \Psi_v(u) = \Psi(u,v) \in \vRRt$ (for each~$v$),
and by the curves 
$\Psi_u : v\in [c,d] \rar \Psi_u(v) = \Psi(u,v) \in \vRRt$ (for each~$v$).
\Eg, on the sphere, $\Psi_v$ is a parallel and $\Psi_u$ is a meridian.
With an origin $O$ chosen in~$\RRt$:
The associated vector parametrized surface in~$\vRRt$ is the function $\vr = \ora{O\Psi}$, that is,
\be
\label{eqdefr0}
\vr:
\left\{\eqalign{
[a,b]\times[c,d] & \rar\vRRt \cr
(u,v) & \rar \ora{OP} = \ora{O\Psi(u,v)} = \vr(u,v). % = \pmatrix{x(u,v) \cr y(u,v) \cr z(u,v)}
}\right.
\ee
(With a basis $(\ve_1,\ve_2,\ve_3)$ in~$\vRRt$, $[\vr(u,v)]_{|\ve} = \pmatrix{x(u,v) \cr y(u,v) \cr z(u,v)}$.)

Let $(\vE_1,\vE_2)$ be the canonical basis in the space $\RR\times \RR \supset[a,b]\times[c,d] = \{(u,v)\}$ of parameters.

The surface $\Psi$ is supposed to be regular  (and so $\vr$ is regular), that is, $\psi$ is $C^1$ and
the tangents vectors
$\vTuP$ and $\vTdP$ at~$P$ are independent at all $P = O+\vr(u,v) \in S$ , that is,
\be
\left.\eqalign{
& \vTuP:= d\Psi(u,v).\vE_1 = d\vr(u,v).\vE_1 \eqnamed {\pa \vr\over\pa u}(u,v) , \cr
& \vTdP := d\Psi(u,v).\vE_2 = d\vr(u,v).\vE_2 \eqnamed {\pa \vr\over\pa v}(u,v),
}\right\}  \qand \vTuP \wedge \vTdP \ne \vec0.
\ee

Then choose a Euclidean dot product~$\dd_g$ (English, French...) in~$\vRRt$.
Then the vectorial area element $d\vSigmaP$ and the scalar area element at~$P$ at $S=\Im(\vr)$ relative to~$\vr$ are, by definition,
\be
\eqalign{
&d\vSigmaP := {\pa \vr\over\pa u}(u,v) \wedge {\pa \vr\over\pa v}(u,v)\,du\,dv
\qquad (= \vTuP \wedge \vTdP\,du\,dv), \cr
& d\SigmaP := ||{\pa \vr\over\pa u}(u,v) \wedge {\pa \vr\over\pa v}(u,v)||\,du\,dv
\qquad (= ||\vTuP \wedge \vTdP||\,du\,dv)
.
}
\ee
And a unit normal vector $\vNP$ at $S$ at~$P$ is
\be
\vNP 
:= {\vTuP\wedge \vTdP \over ||\vTuP\wedge \vTdP||_g}
\quad (= {{\pa \vr\over\pa u}(u,v) \wedge  {\pa \vr\over\pa v}(u,v)
\over  ||{\pa \vr\over\pa u}(u,v) \wedge {\pa \vr\over\pa v}(u,v)||_g}).
\ee
(The other one is $-\vNP$.)

So the area of the parallelogram $(\vTuP , \vTdP)$ is
\be
%\hbox{area of }(\vTuP , \vTdP) =
||\vTuP\wedge \vTdP||_g
= |(\vTuP \wedge \vTdP,\vNP)_g|
=|\det(\vTuP , \vTdP,\vNP)|
,
\ee
and we have
\be
d\SigmaP = d\vSigmaP \bcdot \vNP := (d\vSigmaP,\vNP)_g := (\vTuP \wedge \vTdP,\vNP)_\RRt\;du\,dv.
\ee
And the area of $\Stz$ is by definition
\be
|S| = \int_{P\in \Stz} d\SigmaP
:=\int_{u=a}^b\int_{v=c}^d ||{\pa \vr\over\pa u}(u,v)\wedge{\pa \vr\over\pa v}(u,v)||_g\,du\,dv
.
\ee
}

%%%%%%%%%%%%%%%%%%%%%%%%%%%%%%%%%%%%%%%%%%%%%%%%%%%%%%%%%%%%%%%%%%%%%%%%%%%%%%%%%%%

\subsubsection{Deformation of a surface}

A parametrized surface $\Psitz$ in~$\Omegatz$ and the associated geometric surface $S_\tz$ are defined by
\be
\Psitz:
\left\{\eqalign{
[a,b]\times[c,d] & \rar\Omegatz \cr
(u,v) & \rar P =\Psitz(u,v) %= \pmatrix{x_{t_0}(u,v) \cr y_{t_0}(u,v) \cr z_{t_0}(u,v)}.
}\right\} \qand S_\tz = \Im(\Psitz)\subset\Omegatz.
\ee
(It is also represented after a choice of an origin $O$ by the vector valued parametrized surface
$\vrtz = \ora{O\Psi_\tz}$.)
\comment{, that is,
\cf~\eref{eqdefr0},
\be
\vrtz:
\left\{\eqalign{
[a,b]\times[c,d] & \rar\vRRt \cr
(u,v) & \rar \ora{OP} = \ora{O\Psi_\tz(u,v)} = \vrtz(u,v) = \pmatrix{x_{t_0}(u,v) \cr y_{t_0}(u,v) \cr z_{t_0}(u,v)}.
}\right.
\ee
}

The transformed parametric surface is $\Psit := \Phitzt\circ\Psitz$ and the associated geometric surface is~$S_t$:
\be
\label{eqdefr10}
\Psit:=\Phitzt\circ\Psitz:
\left\{\eqalign{
[a,b]\times[c,d] & \rar\Omegatz \cr
(u,v) & \rar p =\Psit(u,v) = \Phitzt(\Psitz(u,v)) = \Phitzt(P) %= \pmatrix{x_{t_0}(u,v) \cr y_{t_0}(u,v) \cr z_{t_0}(u,v)}.
}\right\} \qand S_t = \Phitzt(S_\tz). % = \Im(\Psit)
\ee
(After a choice of an origin $O$, the associated vector valued parametrized surface is $\vr_t = \ora{O\Psi_t}$.)
\comment{, that is, with~\eref{eqdefr0},
\be
\label{eqdefr}
\vr_t:=\ora{O\Psi_t} :
\left\{\eqalign{
[a,b]\times[c,d] & \rar\vRRt \cr
(u,v) & \rar \ora{Op} = \ora{O\Psi_t(u,v)} = \vr_t(u,v) = \pmatrix{x_t(u,v) \cr y_t(u,v) \cr z_t(u,v)}.
}\right.
\ee
}

Let $(\vE_1,\vE_2)$ be the canonical basis in the space $\RR\times \RR \supset[a,b]\times[c,d] = \{(u,v)\}$ of parameters.

The surface $\Psitz$ is supposed to be regular, that is, $\Psitz$ is $C^1$ and,
for all $P = \Psitz(u,v) \in \Stz$, the tangents vectors
$\vTuP$ and $\vTdP$ at~$P$ are independent, that is,
\be
\left.\eqalign{
& \vTuP:= d\Psitz(u,v).\vE_1 %= d\vrtz(u,v).\vE_1 
%=\pa_1\Psitz(u,v) 
\eqnote {\pa \Psitz\over\pa u}(u,v) , \cr
& \vTdP := d\Psitz(u,v).\vE_2 %= d\vrtz(u,v).\vE_2
%=\pa_2\Psitz(u,v) 
\eqnote {\pa \Psitz\over\pa v}(u,v),
}\right\}  \qand \vTuP \wedge \vTdP \ne \vec0.
\ee
And the tangent vectors at~$S_t$ at $p= \Phitzt(P)$ at~$t$ are
\be
\left\{\eqalign{
& \vtup := d\Psi_t(u,v).\vE_1 %= d\vr_t(u,v).\vE_1 
%=\pa_1\Psi_t(u,v) 
= {\pa \Psi_t\over\pa u}(u,v), \qso \vtup = F_P.\vTuP \quad(= d\Phitzt(P).{\pa \Psitz\over\pa u}(u,v)), \cr
& \vtdp := d\Psi_t(u,v).\vE_2 %= d\vr_t(u,v).\vE_2 
%=\pa_2\Psi_t(u,v) 
= {\pa \Psi_t\over\pa v}(u,v), \qso \vtdp = F_P.\vTdP \quad(= d\Phitzt(P).{\pa \Psitz\over\pa v}(u,v)).
}\right.
\ee
These vectors are independent since $\Phitzt$ is a diffeomorphism and~$\Psitz$ is regular.
In facts, we used tangent vectors to curves and their push-forwards, \cf\ figure~\ref{figpf} and~\S~\ref{secsa}.

%%%%%%%%%%%%%%%%%%%%%%%%%%%%%%%%%%%%%%%%%%%%%%%%%%%%%%%%%%%%%%%%%%%%%%%%%%%%%%%%%%%

\subsubsection{Euclidean dot product and unit normal vectors}

Then choose a Euclidean dot product~$\dd_g$ (English, French...), to be able to use the vectorial product, \cf~\eref{eqw1}, the same at all time~$t$.
Then the scalar area elements $d\SigmaP$ at~$P$ at $S_\tz$ relative to~$\Psitz$, and
$d\sigmap$ at~$p$ at $S_t$ relative to~$\Psi_t$, are
\be
\left\{\eqalign{
&d\SigmaP := ||{\pa \Psitz\over\pa u}(u,v) \wedge {\pa \Psitz\over\pa v}(u,v)||_g\,du\,dv
\quad (= ||\vTuP\wedge \vTdP||_g\,du\,dv), \cr
& d\sigmap := ||{\pa \Psi_t\over\pa u}(u,v) \wedge {\pa \Psi_t\over\pa v}(u,v)||_g\,du\,dv
\quad (= ||\vtup\wedge \vtdp||_g\,du\,dv).
%\quad (=||\vTuP\wedge \vTdP||_g\,du\,dv).
%\eqnote  d\vSigmaP \bcdot \vNP,
}\right.
\ee
And the areas of $\Stz$ and $S_t$ are
\be
\left\{\eqalign{
&|S_{t_0}| = \int_{P\in \Stz} d\SigmaP
:=\int_{u=a}^b\int_{v=c}^d ||{\pa \Psitz\over\pa u}(u,v)\wedge{\pa \Psitz\over\pa v}(u,v)||_g\,du\,dv, \cr
&|S_t| = \int_{p\in S_t} d\sigmap
:=\int_{u=a}^b\int_{v=c}^d ||{\pa \Psi_t\over\pa u}(u,v)\wedge{\pa \Psi_t\over\pa v}(u,v)||_g\,du\,dv.
}\right.
\ee
(See~\eref{eqvolalgdo}: $d\SigmaP$ and $d\sigmap$ are positive measures: They are not multilinear forms.)

And the unit normal vectors $\vNP$ at $S_\tz$ at~$P$ at~$\tz$ and $\vnp$ at $S_t$ at~$p$ at~$t$ are
\be
\left\{\eqalign{
&\vNP  = {{\pa \Psitz\over\pa u}(u,v)\wedge{\pa \Psitz\over\pa v}(u,v) \over ||{\pa \Psitz\over\pa u}(u,v)\wedge{\pa \Psitz\over\pa v}(u,v)||_g}
\quad (= {\vTuP\wedge \vTdP \over ||\vTuP\wedge \vTdP||_g}) \cr
&\vnp = {{\pa \Psi_t\over\pa u}(u,v)\wedge{\pa \Psi_t\over\pa v}(u,v) \over ||{\pa \Psi_t\over\pa u}(u,v)\wedge{\pa \Psi_t\over\pa v}(u,v)||_g}
\quad (= {\vtup\wedge \vtdp \over ||\vtup\wedge \vtdp||_g}=.
%\quad (= {{\pa \Psitz\over\pa u}(u,v) \wedge  {\pa \Psitz\over\pa v}(u,v)
%\over  ||{\pa \Psitz\over\pa u}(u,v) \wedge {\pa \Psitz\over\pa v}(u,v)||_g}).
}\right.
\ee
Then the vectorial area elements $d\vSigmaP$ at~$P$ at $S_\tz=\Im(\Psitz)$ relative to~$\vr_\tz$
and $d\vsigmap$ at $p$ at $S_t=\Im(\Psi_t)$ relative to~$\Psi_t$ are
\be
\left\{\eqalign{
&d\vSigmaP :=\vNP\, d\SigmaP = {\pa \Psitz\over\pa u}(u,v) \wedge {\pa \Psitz\over\pa v}(u,v)\,du\,dv
\quad (= \vTuP \wedge \vTdP\,du\,dv) \cr
&d\vsigmap := \vnp\, d\sigmap = {\pa \Psi_t\over\pa u}(u,v) \wedge {\pa \Psi_t\over\pa v}(u,v)\,du\,dv
\quad (= \vtup \wedge \vtdp\,du\,dv). \cr
}\right.
\ee
(Useful to get the flux through a surface: $\int_\Gamma \vf \bcdot \vn\,d\sigma = \int_\Gamma \vf \bcdot d\vsigma$.)

(NB: $d\vSigmaP$ and $d\vsigmap$ are  not multilinear since $d\SigmaP$ and $d\sigmap$ are not.)

%%%%%%%%%%%%%%%%%%%%%%%%%%%%%%%%%%%%%%%%%%%%%%%%%%%%%%%%%%%%%%%%%%%%%%%%%%%%%%%%%%%

\subsubsection{Relations between surfaces}

$\vtup\wedge \vtdp = J_P\,F_P^{-T}.(\vTuP\wedge \vTdP)$, \cf~\eref{eqtransaire00}, gives
\be
\label{eqtransaire}
{\pa \Psi_t\over\pa u}(u,v) \wedge {\pa \Psi_t\over\pa v}(u,v)
= J_P\,F_P^{-T}.({\pa \Psitz\over\pa u}(u,v) \wedge {\pa \Psitz\over\pa v}(u,v)).
\ee
This gives the relation between vectorial and scalar area elements,
\be
\label{eqtransaire1}
\vn\,d\sigma_p = d\vsigma_p 
%= J_P\,||F_P^{-T}.\vNP||_g\vn_p\,d\SigmaP
%= J_P\,F_P^{-T}.\vNP\,d\SigmaP
= J_P\,F_P^{-T}.d\vSigmaP
= J_P\,F_P^{-T}.\vNP\,d\SigmaP
,\qand
d\sigma_p = J_P\,||F_P^{-T}.\vNP||_g\,d\SigmaP 
.
\ee
(Check with~\eref{eqrelvnvN}.)

%%%%%%%%%%%%%%%%%%%%%%%%%%%%%%%%%%%%%%%%%%%%%%%%%%%%%%%%%%%%%%%%%%%%%%%%%%%%%%%%%%%

\subsection{Piola identity}

Reminder:
Let $M = [M^i_j]$ be a $3*3$ matrix function.
%$M= \sumijn M^i_j\ve_i\otimes e^j$, 
We use the usual divergence in continuum mechanics (non objective) given by
$\dvg M := \pmatrix{
{\pa M^1_1\over\pa X^1}+{\pa M^1_2\over\pa X^2}+{\pa M^1_3\over\pa X^3}\cr
{\pa M^2_1\over\pa X^1}+{\pa M^2_2\over\pa X^2}+{\pa M^2_3\over\pa X^3}\cr
{\pa M^3_1\over\pa X^1}+{\pa M^3_2\over\pa X^2}+{\pa M^3_3\over\pa X^3}\cr
}
= \pmatrix{
\sumjn {\pa M^1_j\over\pa X^j}\cr
\sumjn {\pa M^2_j\over\pa X^j}\cr
\sumjn {\pa M^3_j\over\pa X^j}\cr
}$, \cf~\eref{eqdivmec}.
And if $\Cof(M)$ is the matrix of cofactors
(in~$\RRt$: $\Cof(M)^i_j = M^{i+1}_{j+1} M^{i+2}_{j+2} - M^{i+1}_{j+2} M^{i+2}_{j+1}$),
then
$ M^{-1}={1\over \det M} \Cof(M)^T$, \ie,
\be
\label{eqcof}
(\det M)M^{-1} = \Cof(M)^T.
\ee

The framework being Euclidean, we use a Euclidean basis and the associated matrix, and thus (matrix meaning)
\be
\label{eqcof2}
%(JF^{-T})(P) = 
J(P)F(P)^{-T} =  \Cof(F(P)) \eqnote \Cof(F)(P) ,\qwritten JF^{-T} = \Cof(F).
\ee

\begin{prop}[Piola identity]
In~$\RRt$, we have
\be
\label{eqidpiola}
\dvg(JF^{-T})(P) = 0 ,
\qie \forall i,\; \sumjn {\pa \Cof(F)^i_j\over\pa X^j}(P) = 0.
\ee
Also written $\sumjn {\pa \over \pa X^j} (J {\pa X^i \over \pa x^j}) = 0$...
NB: \eref{eqidpiola} is a just a matrix computation since we used the divergence of a matrix (we used components relative to a given basis).
\finprop

\debdem
We are in~$\RRt$, thus
$\Cof(F)^i_j =F^{i+1}_{j+1} F^{i+2}_{j+2} - F^{i+1}_{j+2} F^{i+2}_{j+1}$,
and $F=[d\Phi_t] = [{\pa\phi^i\over\pa X^j}]$, that is, $F^i_j={\pa\phi^i\over\pa X^j}$. Thus
$$
  {\pa \Cof(F)^i_j\over\pa X^j}
 ={\pa^2\phi^{i+1}\over\pa X^j\pa X^{j+1}}{\pa\phi^{i+2}\over\pa X^{j+2}}
 + {\pa\phi^{i+1}\over\pa X^{j+1}}{\pa^2\phi^{i+2}\over\pa X^j\pa X^{j+2}}
 -{\pa^2\phi^{i+1}\over\pa X^j\pa X^{j+2}}{\pa\phi^{i+2}\over\pa X^{j+1}}
 - {\pa\phi^{i+1}\over\pa X^{j+2}}{\pa^2\phi^{i+2}\over\pa X^j\pa X^{j+1}}.
$$
Thus, for all $i=1,2,3$, we get $\sumjn {{\pa \Cof(F)^i_j\over\pa X^j}} = 0$
(the terms cancel each other out two by two).
\findem

%%%%%%%%%%%%%%%%%%%%%%%%%%%%%%%%%%%%%%%%%%%%%%%%%%%%%%%%%%%%%%%%%%%%%%%%%%%%%%%%%%%

\subsection{Piola transformation}

\def\Piola{{\rm Piola}}
\def\omegatz{{\omega_\tz}}
\def\omegat{{\omega_t}}

Let $\vu$ be a vector field in~$\Omegat$.
The goal is to find a vector field $\vU_\Piola$ in~$\Omegatz$ \st\
for all open subset $\omegat\subset\Omegat$ with $\omegatz=\Phitzt^{-1}(\omegat) \subset\Omegatz$,
\be
\label{eqdefpiola4}
\int_{\pa\omegatz} \vU_\Piola \bcdot \vN\,d\Sigma
=\int_{\pa\omegat} \vu \bcdot \vn \,d\sigma,
\ee
%(which means $\int_{P\in\pa\omegatz} (\vU_\Piola(P) , \vN(P))_g\,d\Sigma(P) =\int_{p\in\pa\omegat} (\vu(p) , \vn(p))_g \,d\sigma(p)$).
or
\be
\int_\omegatz \dvg(\vU_\Piola) \, d\Omegatz
= \int_\omegat \dvg(\vu)\, d\Omegat,
\ee
%(which means $\int_{P\in \omegatz} \dvg(\vU_\Piola)(P) \, d\Omegatz = \int_{p\in\omegat} \dvg(\vu)(p)\, d\Omegat$), and
\ie
\be
\int_{P\in \omegatz} \dvg(\vU_\Piola)(P) \, d\Omegatz
= \int_{P\in \omegatz} \dvg(\vu)(\Phitzt(P))\,J(P)\, d\Omegatz.
\ee
(The motion is supposed to be regular, so $J(P)>0$). Thus we want
\be
\dvg\vU_\Piola(P) = J(P)\,\dvg\vu(p) \qwhen p=\Phitzt(P).
\ee

\debdef
The Piola transform is the map
\be
\label{eqdefpiola}
\left\{\eqalign{
C^\infty(\Omega_t;\RRn) & \rar C^\infty(\Omega_{t_0};\RRn) \cr
\vu & \rar \vU_\Piola , \quad\ \boxed{\vU_\Piola(P)\eqdef  J(P) F(P)^{-1}.\vu(p) } 
  \qwhen    p=\Phitzt(P).
}\right.
\ee
(So $\vU_\Piola(P) =  J(P) \Phi^*(\vu)(P)$ where $\Phi^*(\vu)(P) = F(P)^{-1}.\vu(p) =$ the pull-back with~$\Phi=\Phitzt$.)
\findef

\debprop
With $p=\Phitzt(P)$, $\vU_\Piola = \sumin U_\Piola^i \ve_i$ and $\vu = \sumin u^i \ve_i$ we get
\be
\label{eqdefpiola2}
\dvg\vU_\Piola(P) = J(P)\,\dvg\vu(p),\qie
\sumin {\pa U_\Piola^i \over \pa X^i}(P) = J(P) \sumin {\pa u^i \over \pa x^i}(p).
\ee
\finprop

\debdem
$\dvg(\uutau.\vw) \equalref{eqdtw0} \tdvg(\uutau).\vw + \uutau \odd d\vw$ gives
$\dvg((JF^{-1}).(\vu\circ\Phitzt))(P) = (\dvg(JF^{-T})(P),\vu(p))_g + (J(P)F(P)^{-1}) \odd (d\vu(p).F(P))
\equalref{eqidpiola} 0 + J(P) (F(P).F(P)^{-1})\odd d\vu(p) = J(P) \,I\odd d\vu(p) = J(P)\dvg\vu(p)$,
which gives~\eref{eqdefpiola2}.
\findem

%%%%%%%%%%%%%%%%%%%%%%%%%%%%%%%%%%%%%%%%%%%%%%%%%%%%%%%%%%%%%%%%%%%%%%%%%%%%%%%%%%%
%%%%%%%%%%%%%%%%%%%%%%%%%%%%%%%%%%%%%%%%%%%%%%%%%%%%%%%%%%%%%%%%%%%%%%%%%%%%%%%%%%%

\section{Work and power}

\def\courbe{{c}}
\def\tPhiP{{\tilde\Phi_\Pobj}}
\def\Jtzt{{J^\tz_t}}
\def\Jtut{{J^\tu_t}}
\def\Jtz{{J^\tz}}
\def\WtzT{{W^\tz_T}}
\def\calPa{{{\cal P}_a}}
\def\calPe{{{\cal P}_e}}
\def\calPi{{{\cal P}_i}}
\def\vUtzt{{\vU^\tz_t}}
\def\vUtz{{\vU^\tz}}

%%%%%%%%%%%%%%%%%%%%%%%%%%%%%%%%%%%%%%%%%%%%%%%%%%%%%%%%%%%%%%%%%%%%%%%%%%%%%%%%%%%

\subsection{Definitions}

%%%%%%%%%%%%%%%%%%%%%%%%.%%%%%%%%%%%%%%%%%%%%%%%%%%%%%%%%%%%%%%%%%%%%%%%%%%%%%%%%%%%

\subsubsection{Work}

(Thermodynamic like approach.)
The elementary work is a differential form $\alpha$, \eg\ $ \alpha = dU$ (internal energy density), $\alpha= \delta W=$ (elementary work).
Consider a regular curve $\courbe : t \in [\tz,T] \rar \courbe(t) \in \RRn$.
And let $\vv(t,\courbe(t)) := %{d\courbe\over dt}(t) 
\vcp(t)$.
The work of~$\alpha$ along the curve~is
\be
\label{eqfdt}
\eqalign{
\int_\courbe \alpha
:= &\int_{t=\tz}^T \alpha(t,\courbe(t)).\vcp(t)\,dt
\eqnote \int_{t=\tz}^T \alpha.d\vc
\cr
= & \int_{t=\tz}^T \alpha(t,\courbe(t)).\vv(t,\courbe(t))\,dt
\eqnote \int_{t=\tz}^T \alpha.\vv\,dt.
%\quad \mathop{=}^{\eg} \int_\courbe \delta W =  \WtzT(\alpha,\courbe).
}
\ee
\Eg, $\WtzT(\alpha,\courbe) = \int_\courbe \delta W=$ work along~$c$ of the differential form $\alpha=\delta W$.

Then consider an object~$\Obj$ and its motion $\tPhi:(t,\Pobj)\rar p(t)=\tPhi(t,\Pobj)=\tPhi_\Pobj(t) \in \RRn$, %\cf~\eref{eqdeftPhi0},
the curves $c_\Pobj=\tPhi_\Pobj : t \in [\tz,T] \rar p(t)=\tPhi_\Pobj(t) \in \RRn$,
and the Eulerian velocities $\vv(t,p(t)) = \tPhi_\Pobj{}'(t)$.
The work for $\Obj$ and a Eulerian differential form~$\alpha$ along~$\tPhi$ is the sum of work of $\alpha$ of all particles, formally $\sum_{\Pobj\in\Obj} (\int_{c_\Pobj} \alpha_\Pobj)$. So with the associated motion
$\Phitz(t,\ptz) = \tPhi(t,\Pobj)=p(t) = \Phitzptz(t)$ when $\ptz = \tPhi(\tz,\Pobj)$, and with $\Omega_t = \tPhi(t,\Obj)$,
\be
\label{eqfdt23}
\WtzT(\tPhi)
:= \int_{\ptz\in\Omegatz}\int_{t=\tz}^T \alpha(t,\Phitzptz(t)).\vv(t,\Phitzptz(t)) \,dt\,d\Omegatz
= \int_{t=\tz}^T \int_{\pt\in\Omegat}  \alpha(t,\pt)).\vv(t,\pt)\,d\Omegat \,dt.
\ee
(The last equality if Fubini theorem can be applied, \eg\ if $\alpha$ is $C^0$ and $\Phitz$ is~$C^1$, $\Obj$ being bounded.)

\debexe
If $\alpha$ is a stationary and exact differential form, $\alpha = dU$, then prove that
\be
\int_\courbe dU
= U(\courbe(T)) - U(\courbe(\tz)) \eqnote \Delta U
\ee
only depends on the extremities $\courbe(\tz)$ and~$\courbe(T)$ of the curve~$c$.

\debrep
$
\int_\courbe dU
= \int_{t=\tz}^T dU(\courbe(t)).\vcp(t)\,dt
= \int_{t=\tz}^T {d(U \circ \courbe)\over dt}(t)\,dt
= [U \circ \courbe]_\tz^T
= U(\courbe(T)) - U(\courbe(\tz))$.
%Ici $\WtzT(\courbe)$ est indépendant d'un changement de base spatial, \cf~\eref{eqdefP1b}.
\finrep
\finexe

Remark (continuum mechanics): An observer chooses a Euclidean dot product $\dd_g = .\bcdotg. = .\bcdot.$
(if $\dd_g$ is imposed and implicit).
And if he chooses to represent a linear form $\alpha_t(p_t)$ with its $\dd_g$-Riesz representation vector $\vf_t(p_t)$ (observer dependent),
\cf~\eref{eqsecemp}, then
\be
\int_\courbe \alpha
= \int_{t=\tz}^T \alpha(t,\courbe(t)).\vcp(t)\,dt
= \int_{t=\tz}^T \vf\bcdot d\vc = \int_{t=\tz}^T \vf\bcdot \vv \,dt.
\ee

%%%%%%%%%%%%%%%%%%%%%%%%%%%%%%%%%%%%%%%%%%%%%%%%%%%%%%%%%%%%%%%%%%%%%%%%%%%%%%%%%%%

\subsubsection{And its associated power density}

Definition: %for an object~$\Obj$ and its motion~$\tPhi$,
The power density of a differential form~$\alpha$
along~$\tPhi$ is the Eulerian function
\be
\label{eqpcla0}
\psi := \alpha.\vv : 
\left\{\eqalign{
\bigC=\bigcup_{t\in[\tz,T]}(\{t\} \times \Omegat) & \rar \RR \cr
(t,p) & \rar \psi(t,p)=\alpha(t,p).\vv(t,p).
}\right.
\ee
And the power at~$t$ is
\be
\label{eqpow}
\calP_t(\tPhi)=\calP(t,\tPhi)
:= \int_{p\in\Omegat} \psi(t,p)\,d\Omegat
= \int_{p\in\Omegat} \alpha_t(p).\vv_t(p)\,d\Omegat
\eqnote \calP(t,\vv_t) = \calP_t(\vv_t).
\ee
%Here $\sum$ is meant for a finite number of points, while $\sum\eqnote \int$ (Riemann notation) is meant for a continuous material.

Remark: With a Euclidean dot product~$\dd_g$, then with the $\dd_g$-Riesz representation vector $\vf$ of~$\alpha$
(observer dependent) we get
\be
\label{eqpcla0b}
\psi=\vf\bcdot\vv, \qie 
\psi(t,p) = \vf(t,p) \bcdotg \vv(t,p) \quad (= (\vf(t,p),\vv(t,p))_g),
\ee
which gives
$
\calP(t,\tPhi)
%:=  %\sum_{\pt\in\Omegat} \psi(t,\pt)
%= \sum_{\pt\in\Omegat} \alpha(t,\pt).\vv(t,\pt)
:= \int_{p\in\Omegat} \vf(t,p) \bcdot \vv(t,p)\,d\Omegat.
$

%%%%%%%%%%%%%%%%%%%%%%%%%%%%%%%%%%%%%%%%%%%%%%%%%%%%%%%%%%%%%%%%%%%%%%%%%%%%%%%%%%%

\subsection{Piola--Kirchhoff tensors}
\label{secWPi}

%%%%%%%%%%%%%%%%%%%%%%%%%%%%%%%%%%%%%%%%%%%%%%%%%%%%%%%%%%%%%%%%%%%%%%%%%%%%%%%%%%%

%\subsubsection{Double objective contraction}

Consider a regular Eulerian velocity field $\vv$, so $d\vv$ is an endomorphism (identified with a ${1\choose1}$ tensor).
Then we need another endomorphism~$\uutau$ (identified with a ${1\choose1}$ to get the objective double contraction
\be
\uutau \odd d\vv := \Tr(\uutau.d\vv),
\ee
which means $\uutau(t,p) \odd d\vv(t,p) := \Tr(\uutau(t,p).d\vv(t,p))$.

Quantification: With a basis $(\ve_i)$ at~$t$ and 
$\uutau = \sum_{ij} \tau^i_j \ve_i\otimes e^j$ and $d\vv = \sum_{jk} v^j_{|k} \ve_j\otimes e^k$ (the endomorphisms have been written like ${1\choose1}$ tensors for calculation purpose), $\uutau.d\vv = \sum_{ijk} \tau^i_j v^j_{|k}\ve_i\otimes e^k$ and
\be
\label{eqPKdc}
\uutau\odd d\vv = \sumijn \tau^i_j v^j_{|i}  \quad \hbox{(objective value)},
\ee
see~\eref{eqdefSoddT}. Then choose a Euclidean dot product $\dd_g$, to be able to use the double matrix product
\be
\uutau : d\vv := \sumijn \tau^i_j v^i_{|j} = [\uutau]_{|\ve}{}^T : [d\vv]_{|\ve} \quad \hbox{(subjective value)}.
\ee

%%%%%%%%%%%%%%%%%%%%%%%%%%%%%%%%%%%%%%%%%%%%%%%%%%%%%%%%%%%%%%%%%%%%%%%%%%%%%%%%%%%

\subsubsection{Objective internal power for the stress: function of~$d\vv$}

{\bf Usual hypothesis for the internal stress in a material:} At first order, 
%at~$t$ at $p\in\Omegat$, 
the power density is of the type
\be
\label{eqttPK}
\psi = \uutau \odd d\vv, \qie \psi(t,p) =  \uutau(t,p) \odd d\vv(t,p),\;\;\forall (t,p)\in\bigC,
\ee
thus the power at~$t$ is
\be
\label{eqttPK00}
\calP_t(\vv_t)
= \int_{p\in\Omegat} \psi(t,p)\,d\Omegat
= \int_{p\in\Omegat} \uutau_t(p) \odd d\vv_t(p)\,d\Omegat.
\ee

\comment{
\debexe
$\RRn$ Euclidean. If $S,T \in \Tuuo$, prove:
\be
%\label{eqoddsy}
{S+S^T \over 2} \odd T = {S+S^T \over 2} \odd {T+T^T \over 2} .
\ee
(If $S=S^T$ then ${S+S^T \over 2} \odd T = S\odd T$.)

\debrep
%With bases,  $S\odd T^T = \sum_{ik} S^i_k T^i_k = \sum_{ik} S^k_i T^i_k = \sum_{ik} S^i_k T^k_i = S\odd T$.
Since ${S+S^T \over 2}$ symmetric, we get ${S+S^T \over 2} \odd T = {S+S^T \over 2} \odd T^T$,
hence~\eref{eqoddsy}.
\finrep
\finexe
}

%%%%%%%%%%%%%%%%%%%%%%%%%%%%%%%%%%%%%%%%%%%%%%%%%%%%%%%%%%%%%%%%%%%%%%%%%%%%%%%%%%%

\subsubsection{The first Piola--Kirchhoff tensor}
\label{secPK}

The Piola--Kirchhoff approach consists in transforming Eulerian quantities into Lagrangian quantities
to refer to the initial configuration.
\eref{eqttPK00} gives
\be
\label{eqttPK2}
\calP_t(\vv_t)
= \int_{P\in\Omegatz} \psi_t(\Phitzt(P))\, |\Jtzt(P)|\,d\Omegatz
= \int_{P\in\Omegatz} \uutau_t(\Phitzt(P)) \odd d\vv_t(\Phitzt(P))\;\Jtzt(P)\,d\Omegatz
\ee
(the Jacobian $\Jtzt(P) = \det(\Ftzt(P))$ of~$\Phitzt$ at~$P$ is positive for a regular motion).
The Lagrangian velocity $\vVtz(t,P) = \vv_t(\Phitzt(P))$ satisfies
$d\vVtzt(P) = d\vv_t(p_t).\Ftzt(P)$ where $p_t = \Phitz(t,P)$.
Thus
\be
\uutau_t(\pt) \odd d\vv_t(\pt)
= \uutau(\pt) \odd (d\vVtzt(P).\Ftzt(P)^{-1}) 
= (\Ftzt(P)^{-1}.\uutau(\pt))\odd d\vVtzt(P).
\ee
Quantification: Choose a basis and a Euclidean dot product $\dd_g$,
% Thus \eref{eqPKdc} gives $\uutau_t(\pt) \odd d\vv_t(\pt) = (\uutau(\pt)^T.\Ftzt(P)^{-T}): d\vVtzt(P)$,
thus
\be
\calP_t(\vv_t)  = \int_{P\in\Omegatz} (
\underbrace{\Jtzt(P)\uutau(\pt)^T.\Ftzt(P)^{-T}}_{\PKtzt(P)}) : d\vVtzt(P)\,d\Omegatz.
\ee

\debdef
The first Piola--Kirchhoff (two point) tensor at $P\in\Omegatz$, relative to $\tz$, $t$ and a basis~$(\ve_i)$,
is the linear map $\PKtzt(P) \in \calL(\RRntz;\RRnt)$ defined by %with $\pt = \Phitzt(P)$ by
\be
\label{eqtPKsig}
\PKtzt(P) = \Jtzt(P)\, \uusigma_t(\Phitzt(P)).\Ftzt(P)^{-T} , \qwhere \uusigma = \uutau^T,
\ee
abusively written
\be
\label{eqtPKsig2}
\boxed{\PK = J\,\uusigma.F^{-T}}.
\ee
\findef

Hence
\be
\label{eqtPKpuis}
\calP_t(\vv_t)  = \int_{\Omegatz} \PKtzt(P) : d\vVtzt(P)\,d\Omegatz.
\ee

\comment{
\debrem
For any vector $\vN \in \RRntz$, \eref{eqtPKsig2} gives
\be
\PK.\vN = J\,\uusigma.F^{-T}.\vN = J\,\uusigma.\vn \qwhen \vn=F^{-T}.\vN \quad(\hbox{\ie\ } \vN=F^T.\vn).
\ee
And the fundamental law $\dvg\uusigma(t,p) + \vf(t,p) = \rho(t,p) \vgamma(t,p)$ postulated in~$\Omegat$ will be transformed into, in~$\Omegatz$, %, with $p_t=\Phitzt(P)$,
\be
\label{eqtPKpuis2}
\dvg\PKtzt(P) + \Jtzt(P) \vf^\tz_t(P) = \rho_J(t,P) {\pa^2\Phitz \over \pa t^2}(t,P).
\ee
Indeed, $d\vV = \sumiJn {\pa V^i \over \pa X^J}\ve_i \otimes E^J$
and $\PK = \sumiJn \PK^i_J\ve_i \otimes E^J$ give
$\calP_t(\vv_t)  = \sumiJn\int_\Omegatz \PK^i_J{\pa V^i \over \pa X^J}\,d\Omegatz
= - \sumiJn\int_\Omegatz {\pa \PK^i_J \over \pa X^J}\, V^i\,d\Omegatz
+ \sumiJn \int_{\Gamma_\tz} \PK^i_J \, V^i\, N_J\,d\Gamma_\tz
$;
Then with
$\dvg\PK := \sumin(\sumJn {\pa \PK^i_J \over \pa X^J})\ve_i$
and $\rho_J(t,P) := \Jtzt(P)\,\rho(t,p_t)$,
and $\vf^\tz(t,P) = \vf(t,p_t)$, we get~\eref{eqtPKpuis2}.
\finrem
}

\debrem
The Piola--Kirchhoff tensor is not that easy to master:
Everything is quite simple in a Eulerian framework (the configuration at~$t$ where the laws are expressed to begin with), but then everything is made more complicated when expressed in an initial configuration (at~$\tz$)...
%for in the end (one more complication) to get back to the current configuration (at~$t$).
So, when the Piola--Kirchhoff tensor is used to introduce the Lie derivatives (Eulerian type),
it makes the Lie derivative quite a mysterious mathematical object, see footnote page~\pageref{footrem}.
%Since the Piola--Kirchhoff is classic, let us introduce it, that is, let us make simple results become difficult.
%(Gabrio Piola died in 1850, before the Lie derivative exists).
\finrem

\debrem
\label{rempbve}
Continuation of the remark:
With the pull-backs, \eref{eqttPK2} reads ($\Jtzt(P)$ being positive)
\be
\calP(t,\tPhi)
=\int_{\pt \in\Omegat}\psi_t(\pt)\,d\Omegat
= \int_{P\in\Omegatz} \bigl((\Phitzt)^*\psi_t\bigr)(P)\, \bigl((\Phitzt)^* d\Omegat\bigr),
\ee
since $((\Phitzt)^* d\Omegat) = \Jtzt(P)\,d\Omegatz$ and $((\Phitzt)^*\psi_t)(P) = \psi_t(\pt)$ (scalar valued functions).
It gives the Piola--Kirchhoff tensor (pull-back to the initial configuration)
since
%\Eg, with $\psi = \alpha.\vv$ we have $ ((\Phitzt)^*\alpha_t)(P) = \alpha_t(\Phitzt(P)).\Ftzt(P)$,
%and $((\Phitzt)^*\vv_t)(P) = \Ftzt(P)^{-1}.\vv_t(\Phitzt(P))$,
%and $(\Phitzt)^*(\alpha_t.\vv_t) = (\Phitzt)^*\alpha_t.(\Phitzt)^*(\vv_t)$, \cf~\eref{eqdefpfalpha0},
%thus we get
$(\Phitzt)^*(\alpha_t.\vv_t)(\pt) = (\alpha_t.\vv_t)(\Phitzt(P)) = \alpha_t(\Phitzt(P)).\vv_t(\Phitzt(P))$.
\comment{
Reminder: The pull-back $((\Phitzt)^*\vv_t)(P) = \Ftzt(P)^{-1}.\vv_t(\Phitzt(P))$
of the Eulerian velocity is \textslbf{not} the Lagrangian velocity $\vVtzt(P)=\vv_t(\Phitzt(P))$
(unless $\Ftzt(P)=I$).
Besides, a Lagrangian velocity doesn't define a vector field, \cf~\eref{eqdefV99} and \S~\ref{secbrtpt},
contrary to the pull-back of a Eulerian velocity.
}
\finrem

%%%%%%%%%%%%%%%%%%%%%%%%%%%%%%%%%%%%%%%%%%%%%%%%%%%%%%%%%%%%%%%%%%%%%%%%%%%%%%%%%%%

\subsubsection{The second Piola--Kirchhoff tensor}
\label{secPK2}

The first Piola--Kirchhoff tensor~$\PK$ may confuse Eulerian and Lagrangian variables, linear maps and
endomorphisms... %, type ${1\choose1}$ tensors and any kind of second order tensors...
And~$\PK(\ptz)$ is not symmetric:
It can't be since $\PK(\ptz)\in \calL(\RRntz;\RRnt)$ is not an endomorphism.
% (even if $\uutau$ is symmetric, see \eg\ Marsden--Hughes~\cite{marsden-hughes} if not convinced).
%(its matrix can be).
To get a symmetric tensor, the second Piola--Kirchhoff tensor is defined: % (endomorphism in $\calL(\RRntz;\RRntz)$).

\debdef
The second Piola--Kirchhoff tensor is the endomorphism $\SKtzt(P) \in \calL(\RRntz;\RRntz)$ defined by, in short,
\be
\label{eqSK}
\SK = F^{-1}.\PK = JF^{-1}.\uusigma.F^{-T}.
\ee
Full notation: $\SKtzt(P) = (\Ftzt(P))^{-1}.\PKtzt(P) = \Jtzt(P) (\Ftzt(P))^{-1}.\uusigma_t(p).(\Ftzt(P))^{-T}$.
\findef

So if $\uusigma_t(p) \in \calL(\RRnt;\RRnt)$ is symmetric then $\SKtzt(P) \in \calL(\RRntz;\RRntz)$ is symmetric.

\medskip
Thus, with the pull-back of the endomorphism $d\vv_t\in\calL(\RRnt,\RRnt)$:
\be
((\Phitzt)^* d\vv_t)(P) = \Ftzt(P)^{-1}.d\vv_t(\pt).\Ftzt(P) , %= \Ftzt(P)^{-1}.d\vVtzt(P),
\ee
and with $d\vv_t(\pt) = d\vVtzt(P).\Ftzt(P)^{-1}$ and $\uusigma_t(p)$ symmetric (so $\SKtzt$ is symmetric),
\be
\eqalign{
\calP_t(\vv_t)
= & \int_{\Omegatz} \PKtzt : d\vVtzt\,d\Omegatz
= \int_{\Omegatz} (\Ftzt.\SKtzt) : d\vVtzt\,d\Omegatz
= \int_{\Omegatz} ([\Ftzt].[\SKtzt]) : [d\vVtzt]^T\,d\Omegatz
 \cr
%= & \int_{\Omegatz} \Jtzt\,(\Ftzt^{-1}.\uusigma_t.\Ftzt^{-T}) \odd (\Ftzt^T.d\vVtzt)\;d\Omegatz, \cr
%= & \int_{P\in\Omegatz} \Jtzt(P)\,(\Ftzt(P)^{-1}.\uutau_t(\pt)) \odd (d\vVtzt(P))\;d\Omegatz, \cr
= & %\int_{\Omegatz} \SKtzt \odd (d\vVtzt^T.\Ftzt)\,d\Omegatz
\int_{\Omegatz} \SKtzt : ((d\vVtzt)^T.\Ftzt)\,d\Omegatz
= \int_{\Omegatz} 
\SKtzt : ({\Ftzt^T.d\vVtzt + d(\vVtzt)^T.\Ftzt\over 2})\;d\Omegatz. \cr
}
\ee
%(And we have ${\Ftzt(P)^T.d\vVtzt(P) + d\vVtzt(P)^T.\Ftzt(P)\over 2} = {\Ftzt(P)^T.(d\vv(t,p(t) + d\vv(t,p(t)^T).\Ftzt(P)\over 2} $.)
%où $\Jtzt(P)\,(\Ftzt(P)^{-1}.\uutau_t(\pt).\Ftzt(P))$ n'est pas le second tenseur de Piola--Kirchhoff.

\debrem
It is a ``chosen time derivative'' of $\SK(t) = J(t)F(t)^{-1}.\uusigma(t).F(t)^{-T}$
that leads to some kind of Lie derivative as explain in books in continuum mechanics, as in footnote page~\pageref{footrem}.
\finrem

\comment{
Ou si on préfère utiliser $\uutau = \uusigma^T$,
on pose $\tSK=\SK^T = \tPK.F^{-T} = JF^{-1}.\uutau.F^{-T}$, et donc $\tPK = \tSK.F^T$.

Si on considère le pull-back $\Phi^*(d\vv) = F^{-1}.d\vv.F$,
on~a $d\vv = F.\Phi^*(d\vv).F^{-1}$,
donc $d\vV = F.\Phi^*(d\vv)$,
donc $\tPK \odd d\vV
= \tPK \odd (F.\Phi^*(d\vv))
= (\tPK.F)\odd\Phi^*(d\vv)
= (\tSK.F^T.F)\odd\Phi^*(d\vv)
= (\tSK.C):\Phi^*(d\vv)
$.
Mais on rappelle que le produit de deux matrices symétriques n'est pas une matrice symétrique
en général, donc $\tSK.C$ n'est pas symétrique en général.
}

%%%%%%%%%%%%%%%%%%%%%%%%%%%%%%%%%%%%%%%%%%%%%%%%%%%%%%%%%%%%%%%%%%%%%%%%%%%%%%%%%%%

\subsection{Classical hyper-elasticity: $\partial W / \partial F$}

\def\hW{{\widehat W}}
$E$ and $F$ are finite dimensional spaces, $\dim E=n$, $\dim F=m$.

% Marsden et Hughes $\omega$-lemma p 187.)
%We use Marsden and Hughes notations. (One of the difficulties is in the notations.)

%%%%%%%%%%%%%%%%%%%%%%%%%%%%%%%%%%%%%%%%%%%%%%%%%%%%%%%%%%%%%%%%%%%%%%%%%%%%%%%%%%%

\subsubsection{Definition}

Reminder: Consider a function
\be
\label{eqhWP0}
\hW : 
\left\{\eqalign{
\calL(E;F) & \rar \RR \cr
L & \rar \hW(L)
}\right.
\ee
Its differential
$d\hW : 
\left\{\eqalign{
\calL(E;F) & \rar \calL(\calL(E;F);\RR) \cr
L & \rar d\hW(L)
}\right\}
$
is defined at~$L$ by, in a direction $M$, \cf~\eref{eqdifff9},
\be
d\hW(L)(M) = \lim_{h\rar0} {\hW(L+hM) - \hW(L) \over h}
\eqnote {\pa \hW \over \pa L}(L)(M).
\ee
Also written $d\hW(L)(M) = d\hW(L).M$ since $d\hW(L)$ is linear.
%, but warning: This dot is not the dot of the matrix product (no bases have been introduced), so, if in doubt, stick to the notation $d\hW(L)(M)$.

\debexa
\label{exapaW2}
$\hW:F\in\calL(\RRntz;\RRnt) \rar \hW(F)\in\RR$ (real valued function),
with $F:=\Ftzt(\ptz) %=d\Phitzt(P)\in\calL(\RRntz;\RRnt)
$ 
the deformation gradient at~$\in\Omegatz$ at~$t$ at~$\ptz$. Thus % in the following;
$d\hW(F).M = \lim_{h\rar0} {\hW(F+hM) - \hW(F) \over h}
\eqnote {\pa \hW \over \pa F}(F).M \in \RR$ is the derivative of~$\hW$ at~$F$ in a direction $M\in\calL(\RRntz;\RRnt)$.
\finexa

\debexa
\label{exapaW1}
$m=n$, endomorphisms $L\in\calL(\vRRn;\vRRn)$, and $\hW(L):=\Tr(L)$ (the trace).
%For the classical hyper-elasticity, $L\in\calL(\RRntz;\RRnt)$.
Here %: $\hW(L)=\Tr(L)$ give
$d\Tr(L)(M) %= d\hW(L)(M) 
= \lim_{h\rar0} {\Tr(L+hM) - \Tr(L) \over h}
= \Tr(M)$ (the trace is linear), thus
$d\Tr(L) =\Tr$ for all~$L$.
\finexa

%%%%%%%%%%%%%%%%%%%%%%%%%%%%%%%%%%%%%%%%%%%%%%%%%%%%%%%%%%%%%%%%%%%%%%%%%%%%%%%%%%%

\subsubsection{Expression with bases (quantification): The $\pa W / \pa L_{ij}$}

Let $(\va_i)$ and $(\vb_i)$ be bases in~$E$ and~$F$, with $(\piai)$ the (covariant) dual basis of~$(\va_i)$.
Let $(\calL_{ij})_{i=1,...,m \atop j=1,...,n} = (\vb_i\otimes \piaj)$ be the associated basis in $\calL(E;F)$, \ie\ the $\calL_{ij}$ are defined by $\calL_{ij}.\va_\ell = \delta_{j\ell}\vb_i$ for all $i=1,...,m$ and $j,\ell=1,...,n$, \cf~prop.~\ref{propLab} (all the elements of the matrix $[\calL_{ij}]_{|\va,\vb}$ vanish except the element at the intersection of row~$i$ and column~$j$ which equals~$1$). %, and $\calL_{ij}\eqnote \vb_i\otimes a^j$).
Let $L \in \calL(E;F)$. The derivation of~$\hW$ at~$L$ in a direction~$\calL_{ij}$ is
\be
\label{eqdhWij}
d\hW(L).\calL_{ij}
= \lim_{h\rar0} {\hW(L+h \calL_{ij}) - \hW(L) \over h}
\eqnote {\pa \hW \over \pa L_{ij}}(L)
\ee
(usual notation).
\comment{
 and $L=\sumijn L_{ij}\calL_{ij}$ (\ie\ the $L_{ij}$ are the components of~$L$ in the basis~$(\calL_{ij})$, so $L.\va_j=\sumin L_{ij}\vb_i$ for all~$j=1,...,n$ and $[L]_{|\va,\vb}=[L_{ij}])$. Then,
derivation in a direction~$\calL_{ij}$,
\be
\label{eqdhWij}
d\hW(L).\calL_{ij}
= \lim_{h\rar0} {\hW(L+h \calL_{ij}) - \hW(L) \over h}
\eqnote {\pa \hW \over \pa L_{ij}}(L)
\ee
(usual notation giving the derivation of~$\hW$ at~$L$ in the direction $\calL_{ij}$).
}
Associated matrix relative to the chosen bases:
\be
\label{eqdhWij2}
[d\hW(L)]_{|\calL_{ij}}
:= [{\pa \hW \over \pa L_{ij}}]_{i=1,...,m \atop j=1,...,n}
 \eqnote [d\hW(L)]_{|\va,\vb}
\quad (\eqnote [d\hW(L)_{ij}] )
%\quad (\eqnote [d\hW(L)_{ij}]_{i=1,...,m \atop j=1,...,n} )
.
\ee
So if $M=\sumim \sumjn M_{ij}\calL_{ij}$ %\in \calL(\vRRn;\vRRm)$ 
then
$d\hW(L)(M) 
= \sum_{ij} M_{ij}\, d\hW(L)(\calL_{ij})$ since $d\hW(L)$ is linear, so
\be
\label{eqdhWij3}
d\hW(L)(M) 
= \sumijn {\pa \hW \over \pa L_{ij}}(L) M_{ij} =  [d\hW(L)]_{|\va,\vb} : [M]_{|\va,\vb},
\ee
double matrix contraction. Duality notations: $d\hW(L)(M) = \sum_{ij} {\pa \hW \over \pa L^i_j}(L) M^i_j$. 
%(it is in fact $[d\hW(L)]_{|\vc} . [M]_{|\vc}$ where $(\vc_k)_{k=1,...,mn}$ is the basis $(\vb_i\otimes \piaj)$ in~$\calL(E;F)$ written as one set of the $nm$ elements $c_k$ where $k=k(i,j)=m(i-1)+j$.) 

\debrem
The notation $[M]_{|\vE,\ve} : [d\hW(L)]_{|\vE,\ve}$ is just a matrix product, since $M=\calL(\vRRn;\vRRm)$ and $d\hW(L) \in \calL(\calL(\vRRn;\vRRm);\RR)$ are different kinds of mathematical objects.
\finrem

\debexa
Continuing example~\ref{exapaW1} with $(\ve_i)=(\vE_i)$: Then
$\hW(L)=\Tr(L)$ gives
$
d\hW(L).M =\Tr(M) = \sum_{i} M_{ii}
$, thus ${\pa \hW \over \pa L_{ij}}(L)= \delta_{ij}$ for all $i,j$, thus
$[d\hW(L)]_{|\ve} =[I] = [{\pa \Tr \over \pa L_{ij}}(L)]$ (identity matrix), and we recover
$d\Tr(L)(M)=[{\pa \Tr \over \pa L_{ij}}(L)]:[M]=[I]:[M] = \sumin M_{ii} = \Tr(M)$.
\finexa

\debexa
Continuing example~\ref{exapaW2}:
The meaning of the derivation ${\pa \hW \over \pa F_{ij}}$ is intriguing:
It is a derivation in the direction $\calL_{ij} \eqnote \ve_i \otimes \pi_{Ej}$, where $(\ve_i)$ is a basis at $p=\Phitzt(P)$ in~$\RRnt$ and $(\pi_{Ej})$ is the dual basis of a basis~$(\vE_j)$ at $P$ in~$\RRntz$, %\cf~\eref{eqdhWij},
\ie\ ${\pa \hW \over \pa F_{ij}}(F)=d\hW(F).\calL_{ij} \eqnote d\hW(F).(\ve_i \otimes \pi_{Ej})$ is a derivation
% in~$\RRntz$ and in~$\RRnt$ 
``at the same time'' in the directions $\ve_i$ (at $(t,p)$) and~$\pi_{Ej}$ (at $(\tz,P)$),
where $F$ stands for $\Ftzt(P)$.
\finexa

\comment{
%%%%%%%%%%%%%%%%%%%%%%%%%%%%%%%%%%%%%%%%%%%%%%%%%%%%%%%%%%%%%%%%%%%%%%%%%%%%%%%%%%%

\subsubsection{Notation $D_2W = \partial W_P / \partial F$}

Let $\Omega$ be an open set in~$\RRn$. Let $P \in \Omega$ and let
\be
\label{eqhWP}
\hW_P : 
\left\{\eqalign{
\calL(\vRRm;\vRRn) & \rar \RR \cr
L & \rar \hW_P(L)
}\right\}
\ee
be regular (see example~\ref{exapaW0}).
Its differential
$d\hW_P : 
\left\{\eqalign{
\calL(\vRRm;\vRRn) & \rar \calL(\calL(\vRRm;\vRRn);\RR) \cr
L & \rar d\hW_P(L)
}\right\}
$
is defined by, at~$L$ in a direction $M$,
\be
d\hW_P(L)(M) = \lim_{h\rar0} {\hW_P(L+hM) - \hW_P(L) \over h}
\eqnote {\pa \hW_P \over \pa L}(L)(M).
\ee

Full functional notation: Let
$\hW : 
\left\{\eqalign{
\Omega \times \calL(\vRRm;\vRRn) & \rar \RR \cr
(P, L) & \rar \hW(P,L) := \hW_P(L)
}\right\}
$, supposed to be $C^1$. Then its differential at~$(P,L)$ relative to the second variable (here $L$) is named
\be
D_2\hW(P,L) := d\hW_P(L) = {\pa \hW_P \over \pa L}(L) \eqnote {\pa \hW \over \pa L}(P,L) ,
\ee
so, for all $M\in\calL(\vRRn;\vRRm)$,
\be
D_2\hW(L)(M) = {\pa \hW \over \pa L}(P,L)(M) = \lim_{h\rar0} {\hW(P,L+hM) - \hW(P,L) \over h}.
\ee

}

%%%%%%%%%%%%%%%%%%%%%%%%%%%%%%%%%%%%%%%%%%%%%%%%%%%%%%%%%%%%%%%%%%%%%%%%%%%%%%%%%%%

\subsubsection{Motions and $\omega$-lemma}

\def\oOmegatz{\overline{\Omegatz}}
\def\oOmegat{\overline{\Omegat}}

Generalization of~\eref{eqhWP0} to $C^1$ functions, with $\UE$ open subset in a affine space which associated vector space is~$E$, % (non homogenous):
\be
\hW:\left\{\eqalign{
\UE \times \calL(E;F) &\rar \RR \cr
(P,L) & \rar \hW(P,L).\cr
}\right.
\ee
At $P$, let $\hW_P(L):=\hW(P,L)$.
The differential $d\hW_P(L) \eqnote \pa_2\hW(P,L)$
%(relative to the second variable of~$\hW$) 
in a direction $M \in \calL(E;F)$ is
\be
\pa_2\hW(P,L).M:= d(\hW_P)(L).M
= \lim_{h\rar 0} {\hW(P,L+hM) - \hW(P,L) \over h}
\eqnote {\pa\hW \over \pa L}(P,L).M.
\ee
With a motion $\Phi:=\Phitzt:\Omegatz\rar\Omegat$
%and with $F(P):=\Ftzt(P)= d\Phitzt(P)\in\calL(\RRntz;\RRnt)$ the deformation gradient between $\tz$ and~$t$ at $P\in\Omegatz$, 
define
\be
f:\left\{\eqalign{
C^1(\oOmegatz;\oOmegat) & \rar C^0(\oOmegatz;\RR) \cr
\Phi & \rar f(\Phi) := \hW(.,d\Phi(.)),
}\right.
\ee
a function of~$\Phi$ which only depends on its first (covariant) gradient; So, for all $P\in\Omegatz$,
\be
f(\Phi)(P) = \hW(P,d\Phi(P))\in \RR.
\ee
(This kind of relation is generally deduced after application of the frame invariance principle, and the hypothesis of dependence on only the first order derivative $d\Phi=F$.)

\begin{lemme}[$\omega$-lemma] For all $\Phi,\Psi \in C^1(\oOmegatz;\oOmegat)$,
\be
\label{eqolemma}
\boxed{df(\Phi)(\Psi) = \pa_2\hW(.,d\Phi)(d\Psi) \eqnote {\pa \hW \over \pa F}(.,d\Phi)(d\Psi)},
\ee
\ie,  for all $P\in\Omegatz$,
$
df(\Phi)(\Psi)(P) = {\pa \hW \over \pa F}(P,d\Phi(P))(d\Psi(P)).
$

With bases $(\vE_i)$ and $(\ve_i)$ in~$\RRntz$ and~$\RRnt$ %an origin $o_t\in\RRn$ at~$t$,
and $d\Psi.\vE_j = \sumin {\pa \Psi_i \over \pa X_j} \ve_i$, we get
\be
df(\Phi)(\Psi)
= \sumijn {\pa \hW \over \pa F_{ij}}(.,d\Phi){\pa \Psi_i \over \pa X_j}(.)
\eqnote [{\pa \hW \over \pa F_{ij}}]:[{\pa \Psi_i \over \pa X_j}].
%= \sumiJn {\pa \hW \over \pa F^i_J}(F_\Psi)^i_J
%\quad ( = \sumiJn{\pa \hW \over \pa ({\pa \Psi^i \over \pa X^J})}{\pa \Psi^i \over \pa X^J}).
\ee
Marsden notations: 
$df(\Phi)(\Psi)=\sumiJn {\pa \hW \over \pa F^i_J}{\pa \Psi^i \over \pa X^J} = [{\pa \hW \over \pa F^i_J}]:[{\pa \Psi^i \over \pa X^J}]$.
\finlem

\debdem
$C^1(\oOmegatz;\oOmegat)$ is a vector space, so %$df(\Phi)\in \calL(C^1(\oOmegatz;\oOmegat);C^0(\oOmegatz;\oOmegat))$ for any $\Phi\in C^1(\oOmegatz;\oOmegat)$ is given by, \cf~\eref{defdifff9},
$
df(\Phi)(\Psi)
= \lim_{h\rar 0} {f(\Phi {+} h\Psi) - f(\Phi) \over h} \in C^0(\oOmegatz;\oOmegat)
$, \ie, for any $P\in\oOmegatz$ we have
$
df(\Phi)(\Psi)(P)
= \lim_{h\rar 0} {f(\Phi {+} h\Psi)(P) - f(\Phi)(P) \over h} 
= \lim_{h\rar 0} {\hW(P,d\Phi(P){+}h \,d\Psi(P)) - \hW(P,d\Phi(P)) \over h}
= d\vW_P(d\Psi(P)
$, \ie~\eref{eqolemma}
\findem

%%%%%%%%%%%%%%%%%%%%%%%%%%%%%%%%%%%%%%%%%%%%%%%%%%%%%%%%%%%%%%%%%%%%%%%%%%%%%%%%%%%

\subsubsection{Application to classical hyper-elasticity: $\PK = \pa W / \pa F$}

\def\hPK{{\widehat{\PK}}}
\def\hPKtzt{{\widehat{\PKtzt}}}
\def\hSK{{\widehat{\SK}}}
\def\hSKtzt{{\widehat{\SKtzt}}}

Let $\dd_g$ be a unique Euclidean dot product in~$\RRnt$ at all times~$t$,
and let $(\vE_i)$ and $(\ve_i)$ be Euclidean bases at $\tz$ and~at~$t$.
%Thus we can consider the transposed $F^T$ and the Jacobian $J(P)=\det_{|\vE,\ve}(F(P))$.
Let $\uusigma_t(p)=$ the Cauchy stress tensor at~$t$ at $p=\Phitzt(P)$.

Let $\PK(P) = J(P)\,\uusigma_t(p).F^{-T}(P)=$
the first Piola--Kirchhoff (two point) tensor at~$P$, \cf~\eref{eqtPKsig}.
Since $\PK$ depends on~$\Phi$, the full notation is
$\PK = \PK(\Phi)$ given by
\be
\PK(\Phi)(P) 
=J(P)\,\uusigma_t(\Phi(P)).d\Phi(P)^{-T}.
\ee

\debdef
If there exists a function $\hPK$ such that $\PK$ reads 
\be
\PK(\Phi)(P) = \hPK(P,d\Phi(P))
\ee
then $\hPK$ is called a constitutive function.
(First order hypothesis: $\hPK$ only depends on $d\Phi=F$ the first order derivative of~$\Phi$.)
\findef

\def\hWtzt{\hW^\tz_t}

\debdef
The material is hyper-elastic iff there exists a function 
$\hW : 
\left\{\eqalign{
\Omegatz \times \calL(\RRntz;\RRnt) & \rar \RR \cr
(P, L) & \rar \hW(P,L)
}\right\}
$
such that
\be
(\PK(\Phi)=) \qquad \hPK(.,d\Phi) = {\pa \hW \over \pa F}(.,d\Phi), \qwritten \hPK = {\pa \hW \over \pa F},
\ee
that is, $\hPK(P,F(P)) = {\pa \hW \over \pa F}(P,F(P))$ for all $P\in\Omegatz$, where $F=d\Phi$.
%(A full notation is $\hW^\tz_t$ with $\hPKtzt(P,d\Phi(P)) = {\pa \hWtzt \over \pa F}(P,d\Ftzt(P))$.)
\findef

With bases $(\vE_I)$ and $(\ve_i)$ in~$\RRntz$ and~$\RRnt$, and $(E^I)$ the dual basis of~$(\vE_I)$,
and $\PK = \sumiJn \PK^i_J \ve_i \otimes E^J$,
\be
[\PK(\Phi)]_{|\vE,\ve} = [\hPK(.,F)]_{|\vE,\ve} = [{\pa \hW \over \pa F}(.,F)]_{|\vE,\ve},
\qie [\PK^i_J] = [{\pa \hW \over \pa F^i_J}(.,F)].
\ee
Thus, for any (virtual) motion $\Psi : \Omegatz \rar \Omegat$, with~\eref{eqolemma} and~\eref{eqdhWij3},
\be
\hPK(d\Phi)(d\Psi)
= {\pa \hW \over \pa F}(d\Phi)(d\Psi)
\qquad
= \sum_{iJ} {\pa \hW \over \pa F^i_J}(F){\pa \Psi^i \over \pa X^J}
\eqnote [\hPK]: [d\Psi],
\ee
that is,
$\hPK(d\Phi)(d\Psi)(P)=\sum_{iJ} {\pa \hW \over \pa F^i_J}(P,\Ftzt(P)){\pa \Psi^i \over \pa X^J}(P)$ for all $P\in\Omegatz$.

\def\tW{{\tilde W}}

\debexe
With a unique Euclidean dot product $\dd_g$ both in~$\RRntz$ and~$\RRnt$,
with Euclidean bases $(\vE_i) \in \RRntz$ and~$(\ve_i) \in \RRnt$, and with $(E^i)$ the dual basis of~$(\vE_i)$,
with $C=F^T.F$, prove  (derivation in the direction $\ve_i \otimes E^J$):
\be
\label{eqdCF}
\eqalign{
{\pa C \over \pa F^i_J}(F)
=  \sum_K F^i_K \vE_J\otimes E^K + \sum_K F^i_K \vE_K \otimes E^J
\quad (=dC(F).(\ve_i \otimes E^J))
%\quad \scriptstyle  (:= dC(F)(\ve_i \otimes E^J) = \lim_{h\rar0} {C(F + h\ve_i \otimes E^J) - C(F) \over h})
%\quad (\hbox{\footnotesize  $\ds := dC(F)(\ve_i \otimes E^J) = \lim_{h\rar0} {C(F + h\ve_i \otimes E^J) - C(F) \over h}$} )
.
}
\ee
\be
{\pa \sqrt C \over \pa F} (F)
= \demi \bigl(\sqrt{C(F)}\,\bigr)^{-1}.{\pa C \over \pa F}(F).
\ee
\be
{\pa \sqrt C \over \pa C} = \demi (\sqrt C)^{-1}.
\ee

\debrep
Let $F=\sum_{iJ} F^i_J \ve_i \otimes E^J$,
so $F^T = \sum_{Ij} (F^T)^I_j \vE_I \otimes e^j = \sum_{Ij} F^j_I \vE_I \otimes e^j$,
%that is $[F]_{|\vE,\ve} = [F^i_J]$,
and
$C
= \sum_{IJ} C^I_J \vE_i \otimes E^J = F^T.F
= \sum_{IJ} \sum_k (F^T)^I_k F^k_J \vE_I \otimes E^J
= \sum_{IJ} \sum_k F^k_I F^k_J \vE_I \otimes E^J = C(F)$,
so $C^I_J = \sum_k F^k_I F^k_J = C^I_J(F)$.
And
\be
\eqalign{
C(F + h\ve_i \otimes E^J)
= & (F + h\ve_i \otimes E^J)^T.(F + h\ve_i \otimes E^J)
=  (F^T + h\vE_J \otimes e^i).(F + h\ve_i \otimes E^J) \cr
= & C(F) + h\, (\vE_J \otimes e^i).F + h\,F^T.(\ve_i \otimes E^J) + h^2\,\vE_J \otimes E^i \cr
= & C(F) + h\,(\sum_K F^i_K \vE_J\otimes E^K + \sum_K (F^T)^K_i \vE_K \otimes E^J) + h^2\,\vE_J \otimes E^J \cr
}
\ee
Thus~\eref{eqdCF}. And
$C(F + h\ve_i \otimes E^J) - C(F)
=(\sqrt C(F + h\ve_i \otimes E^J) + \sqrt C(F)).(\sqrt C(F + h\ve_i \otimes E^J) - \sqrt C(F))
$ gives
$
dC(F)(\ve_i \otimes E^J) = 2\sqrt C(F).d\sqrt C(F)(\ve_i \otimes E^J)
$.
Thus ${\pa \sqrt C \over \pa F^i_J} (F)
= \demi (\sqrt{C(F)})^{-1}.{\pa C \over \pa F^i_J}(F)$.

$(C+h\ve_i\otimes e^j) - C
= (\sqrt{C+h\ve_i\otimes e^j} + \sqrt C).(\sqrt{C+h\ve_i\otimes e^j} - \sqrt C)
$, divided by~$h$,
gives
$\ve_i\otimes e^j
= 2\sqrt C.\lim_{h\rar0}{\sqrt{C+h\ve_i\otimes e^j} - \sqrt C \over h}
= 2\sqrt C.d\sqrt C.(\ve_i\otimes e^j)
$,
thus $L = 2\sqrt C.(d\sqrt C.L)$ for all $L$ (linearity of $d\sqrt C$),
thus $d\sqrt C.L = \demi (\sqrt C)^{-1}.L$.
\finrep
\finexe

%%%%%%%%%%%%%%%%%%%%%%%%%%%%%%%%%%%%%%%%%%%%%%%%%%%%%%%%%%%%%%%%%%%%%%%%%%%%%%%%%%%

\subsubsection{Corollary (hyper-elasticity): $\SK = \pa W / \pa C$}

With the symmetry of the second Piola--Kirchhoff tensor $\SK = F^{-1}.\PK$,
we deduce $\SKtzt(\Phitzt)(P) = \hSKtzt(P,\Ftzt(P))$ (constitutive function).
And we deduce the existence of a function 
$\tW : 
\left\{\eqalign{
\Omegatz \times \calL(\RRntz;\RRntz) & \rar \RR \cr
(P, L) & \rar \tW(P,L)
}\right\}
$
such that, 
\be
\hSKtzt(.,C) = %\pa_2\tW(.,C) \eqnote 
{\pa \tW \over \pa C}(.,C).
\ee
(See Marsden and Hughes~\cite{marsden-hughes} for details and the thermodynamical hypotheses required.)

%\debrem
%With $C=F^T.F$, the ${\pa \tW \over \pa C}$ give results relative to derivation in~$\RRntz$. However derivations relative to derivation in~$\RRnt$ would be closer to Cauchy's approach, \ie\ derivations $ {\pa \over \pa b_{ij}}$ where $\uub$ is the Finger tensor.
\comment{
The previous issues remain: Derivation at~$\tz$ (the tensor $C$ is defined in~$\RRntz$
and the derivations ${\pa \tW \over \pa C^I_J}(.,C) := d\vW.(\vE_I \otimes E^J)$ are considered)
while a derivation at~$t$ could be more connected to Cauchy's approach which starts with considerations at~$t$, see theorem~\ref{sectcC}.
} 
%\finrem

\comment{
Indeed,
1- the frame invariance principle gives that $\hW_P(Q.F) = \hW_P(F)$ for all rotation $Q\in\calL(\RRnt;\RRnt)$ (change of orthonormal basis at~$t$),
2- with $F=R_F.U_F$, matrix product unique with $U_F = \sqrt C \in \calL(\RRntz;\RRntz)$ symmetric positive-definite (square root of~$C$) and $R_F \in \calL(\RRntz;\RRnt)$ satisfying $R_F^T.R_F=I_\tz$,
we get $\hW_P(Q.R_F.U_F) = \hW_P(R_F.U_F) = \hW_P(R_F.\sqrt{C(F)})$,
3- the frame invariance principle also gives the symmetry of the stress, which yields the existence
of $\tW$ such that $\hW_P(R_F.\sqrt{C(F)}) = \hW_P(\sqrt{C(F)})\eqnamed\tW_P(C(F))$.
(See Marsden and Hughes for details.)
}

%%%%%%%%%%%%%%%%%%%%%%%%%%%%%%%%%%%%%%%%%%%%%%%%%%%%%%%%%%%%%%%%%%%%%%%%%%%%%%%%%%%
\comment{
\subsection{Toward an objective formulation: Lie approach}

We cannot know $\vec T(t,\pt)$ by simply looking at the material:
``to know the weight of a suitcase you have to lift it'', says Germain~\cite{germain},
and it is the virtual power principle that is used to deduce the forces acting on a body.
The classical approach (first order) states the existence of a tensor~$\uusigma$ such that the virtual internal power reads,
for all (virtual mathematically admissible) $\vv$,
\be
\label{eqpv0}
\calP_i(\vv) = -\int_{\Omegat} \uusigma : d\vv \,d\Omega.
%=\int_{\Omegat} \dvg\uusigma_t.\vw_t \,d\Omegat - \int_{\pa\Omegat} (\uusigma_t.\vn_t).\vw_t \,d\Gammat.
\ee
This approach is stated in a Galilean Euclidean framework with a Euclidean basis~$(\ve_i)$.
%Unfortunately, with the non-objective double dot used in $\uusigma:d\vw$, the value $\calP_i(\vw)$ depends on the observer (English? French?), see~example~\ref{exaissuedd}: $\calP_i$ is quantitative not qualitative. Indeed the computation $\uusigma:d\vv$ requires a Euclidean dot product (in foot? in metre?).

The Lie approach  proposes %, \eg\ for Newtonian fluids, 
to state the existence of $n$ vectors fields $\va_i$ (like the vectors $\vt$ used by Cauchy to build~$\uusigma$) to obtain, in a Galilean Euclidean framework with a Euclidean basis~$(\ve_i)$ and its dual basis~$(e^i)$,
\be
\label{eqpv0lie}
\calP_i(\vv) = \int_{\Omegat} e^i.\calL_\vv \va_i \,d\Omegat
\ee
(see https://arxiv.org/abs/2208.10780v1).

For these first order approximations, in a Galilean Euclidean framework, both formulas~\eref{eqpv0} and~\eref{eqpv0lie} give the same results. %, cf~\S~\ref{secelaoa} 
%(we get $\uusigma = -\sumjn \va_j\otimes e^j = -\sumijn \sigma^i_j \ve_j\otimes e^j$ in a Euclidean Galilean framework).
% (or of $n$ differential forms

At second order, the classic approach proposes to add Lie derivatives $\calL_\vv(\uusigma_2)$ of second order tensors~$\uusigma_2$ to~\eref{eqpv0}, \eg\ the Jaumann derivative $\calL_\vv \uusigma_2 = {D \uusigma_2 \over Dt} - d\vv.\uusigma_2 + \uusigma_2.d\vv$, \cf~\eref{eqJaumann};

While the Lie approach proposes to add \eg\ second order Lie derivatives of vector fields
(not derivative of second order tensors).
}

\comment{
%%%%%%%%%%%%%%%%%%%%%%%%%%%%%%%%%%%%%%%%%%%%%%%%%%%%%%%%%%%%%%%%%%%%%%%%%%%%%%%%%%%

\subsection{Hyper-elasticity and Lie derivative}
\label{secheLd}

\def\pow{{\rm pow}}
\def\powt{{\rm pow}_t}
\def\alphait{{\alpha^i_t}}
\def\zetait{{\zeta^i_t}}
\def\alphaitz{{\alpha^i_\tz}}
\def\betaitz{{\beta^i_\tz}}
\def\betajtz{{\beta^j_\tz}}
\def\vatj{{\va_{t,j}}}
\def\vauj{{\va_{u,j}}}
\def\vatzj{{\va_{\tz,j}}}
\def\vWtzptz{\vW^\tz_\ptz}

We look for a ``stored energy function'' in the actual state.
Thus we apply a motion to measure the work or the power (to be able to quantified forces).
Let $\vw$ be a (virtual) Eulerian velocity; Then the Lie Derivative $\calL_\vw$ seems to be an adequate tool
to measure the rate of stress along~$\vw$,
approach proposed in www.isima.fr/leborgne/IsimathMeca/PpvObj.pdf;  %\verb+https://www.isima.fr/leborgne/IsimathMeca/PpvObj.pdf+
Instead of a Cauchy stress tensor as a starting point,

\medskip
{\bf Hypothesis: First order  approximation}: $n$~Eulerian differential forms $\alpha_i$, $i=1,...,n$, characterize the elastic material and at~$t$ in a referential $(O,(\ve_i))$ in~$\RRn$, along a virtual motion which Eulerian velocity field is~$\vw$, the density of virtual power due to~$\vw$ is
$\pow(\vw)(t,\pt) = \sumjn \calL_\vw\alpha_i(t,\pt). \ve_i$, where
$\calL_\vw\alpha={\pa \alpha\over \pa t} + d\alpha.\vw+\alpha.d\vw$ is the Lie derivative of a differential form~$\alpha$.
(So it is the Lie derivative of differential forms that is considered, not the Lie derivative of second order tensors.)
%(full notation: $\calL_\vw\alpha(t,\pt)={\pa \alpha\over \pa t}(t,\pt) + d\alpha(t,\pt).\vw(t,\pt)+\alpha(t,\pt).d\vw(t,\pt)$).

\medskip
Corollary: Then in a Galilean framework, with a Cartesian basis~$(\ve_i)$, its dual basis~$(e^i)$,
and the usual hypothesis (isometric objectivity = independence of any rigid body motion), the power reduces to
\be
\pow(\vw)= \sumin \alpha_i.d\vw.\ve_i = -\uutau\odd d\vw, \qwhere \uutau = -\sumin \ve_i \otimes \alpha_i,
\ee
\ie, $e^i.\uutau = -\alpha_i$.
That is, for all $\pt\in\Omegat$,
$\pow(\vw)(t,\pt)= \sumin \alpha_i(t,\pt).d\vw(t,\pt).\ve_i = -\uutau(t,\pt)\odd d\vw(t,\pt)$
where $\uutau(t,\pt) = -\sumin \ve_i \otimes \alpha_i(t,\pt)$.
With components, $\vw= \sumkn w^k\ve_k$ gives
$d\vw = \sumikn {\pa w^k \over \pa x^i}\ve_k \otimes e^i$ and
$d\vw.\ve_i = \sumkn {\pa w^k \over \pa x^i}\ve_k$, and
$\alpha^i = \sumjn \alpha^i_j e^j$ gives
$\pow(\vw)=\alpha^i.d\vw.\ve_i = \sumjn \alpha^i_j {\pa w^j \over \pa x^i}$, thus
$\uutau = - \sumijn \alpha^i_j\ve_i\otimes e^j$
(the $i$-th row of~$[\uutau]_{|\ve} =[\alpha^i_j]$ is~$[\alpha^i]_{|\ve}$).
%so $\powt(\vw_t) = \sumijn \alpha^i_j {\pa w^j \over \pa x^i}$.

\debexa
Isotropic homogeneous elastic material hypothesis: With $(\vE_i)$ a Cartesian basis in~$\RRntz$,
\be
\alpha_i(t,\pt)=\alpha_i(\tz,\ptz).(\Ftzt)^{-1}(\pt),\qwhere \alpha_i(\tz,\ptz)=2\mu E^i.
\ee
And, with $(\ve_i)$ a Cartesian basis in~$\RRnt$, $(\Ftzt)^{-1}(\pt).\ve_j = \sumin (F^{-1})^i_j \vE_i$  gives
$[\uutau(t,\pt)]_{|\ve}=- 2\mu[(\Ftzt)^{-1}(\pt)]_{|\ve,\vE}$.

Non isotropic, \eg, $\alpha_1(\tz,\ptz)=10\mu E^i$ and $\alpha_2(\tz,\ptz)=20\mu E^i$.
\finexa

\comment{
\medskip
Hypothesis:
Suppose that the material is such that, 
$\exists \beta^1_\tz,..., \beta^n_\tz$ differential forms defined in~$\RRntzs$ and a function $\zeta_t$ \st, for all $t$ and all $i=1,...,n$, %and all motions $\Phitzt : \Omegatz\rar\Omegat$
\be
\alpha^i_t(\pt) = \zetait(\pt,(\betajtz(\ptz))_{j=1,...,n},\Ftzt(\ptz)) \qwhen \pt = \Phitzt(\ptz).
\ee
when $\pt=\Phitzt(\ptz)$. 
(The $\beta^i$ are related to strain while the $\alpha^i$ are related to stress.)

\Eg, if $\alphait$ is the result of deformation (push-forward), 
\be
\alphait(\pt) = \alphaitz(\ptz).d\Phitzt(\ptz)^{-1} = \alphaitz(\ptz).\Htzt(\pt),
\ee
where $\Htzt(\pt)=(\Ftzt(\ptz))^{-1}$, \cf~\eref{eqHtz}, then $\uutau_t=-\sumin \ve_i\otimes ( \alphaitz.\Htzt)$, \ie,
\be
\uutau_t(\pt) = \uutau_\tz(\ptz) . \Htzt(\pt) 
\qwhere \uutau_\tz(\ptz)  = -\sumin \ve_i \otimes \alphaitz(\ptz).
\ee
%(two different motions can start from~$\Omegatz$ and end at~$\Omegat$ with the same~$\Ftzt$).
With components: 
$\alphaitz = \sumkn (\alphaitz)_k E^k$ and 
$H = \sumjkn H^k_j \vE_k \otimes e^j$ give
$\alphait = \sumjkn (\alphaitz)_k H^k_j e^j$
and
$\uutau_\tz
= -\sumikn (\alphaitz)_k \ve_i \otimes E^k
$ and
$\uutau_t = -\sumijkn (\alphaitz)_k H^k_j \ve_i\otimes e^j$.
And
\be
\eqalign{
\powt(\vw_t)(\pt)
= & \sumin \alphaitz(\ptz).\Htzt(\pt).d\vw_t.\ve_i
= \sumin (\ve_i \otimes \alphaitz(\ptz)) \odd (\Htzt(\pt).d\vw_t) \cr
= & -\uutau_\tz(\ptz) \odd (\Htzt(\pt).d\vw_t).
}
\ee

And the Cauchy stress vector $\uutau_t(\pt).\vn_t(\pt)$ in a direction $\vn_t(\pt)$ is 
\be
\uutau_t.\vn_t
= \sumin (\alphait.\vn_t)\ve_i
= \sumin (\alphaitz.\Htzt.\vn_t)\ve_i
= \sumin (\alphaitz.\vN_{H\tz})\ve_i, 
\qwhere \vN_{H\tz}:=\Htzt.\vn_t.
\ee
Consider a trajectory $c=\Phitzptz:u\in[\tz,t] \rar c(u) = \Phitzptz(u) \in\RRt$, and its tangent vector ${d\vc \over du}(u)\eqnamed \vn(u,\pu)$ at $\pu=c(u)$.
The work of the Cauchy stress vector along this trajectory is
\be
W(\Phitzptz)
=\int_{u=\tz}^t \vT(u,\Phitzptz(u)) \bcdotg {d\Phitzptz \over du}(u)\,du
=\int_{u=\tz}^t \vT(u,\Phitzptz(u)) \bcdotg\vWtzptz(u)\,du.
\ee

The work along a trajectory $\Phitzptz : u \rar \pu=\Phitzptz(u)$ is, here with
$\vn(u,\pu)={d\Phitzptz \over d u}(u)=\vWtzptz(u)$ for all~$u$,
\be
\eqalign{
W(\Phitzt)
= &- \sumjn \int_{u=\tz}^t  \sumjn  n^j(u,\pu) \va_j(u,\pu) \bcdotg\vWtzptz(u)\,du \cr
= &-\sumjn \int_{u=\tz}^t  (\vWtz)^j(u,\ptz)(\Ftz(u,\ptz).\vajtz(\ptz))\bcdotg \vWtz(u,\ptz)\,du \cr
= &-\sumjn\vajtz(\ptz)\bcdotg  (\int_{u=\tz}^t (\vWtzptz)^j(u) \Ftzptz(u)^T.\vWtzptz(u)\,du) \cr
}
\ee

\debrem
Special case of a isotropic material: With $\alphaitz=\mu E^i$ for all~$i$
we get $\alphait=\mu E^i.\Htzt = \mu \sumjn H^i_j e^j$, and $\uutau_t =-\mu \sumijn H^i_j \ve_i \otimes e^j = -\mu \Htzt$,
and $\uutau_t.\vn_t=-\mu \Htzt.\vn_t$.
\finrem
}
}

%%%%%%%%%%%%%%%%%%%%%%%%%%%%%%%%%%%%%%%%%%%%%%%%%%%%%%%%%%%%%%%%%%%%%%%%%%%%%%%%%%%
%%%%%%%%%%%%%%%%%%%%%%%%%%%%%%%%%%%%%%%%%%%%%%%%%%%%%%%%%%%%%%%%%%%%%%%%%%%%%%%%%%%

\section{Conservation of mass}

Let $\rho(t,\pp)=\rho_t(p)$ be the (Eulerian) mass density at~$t$ at~$p\in\Omegat$, supposed to be $>0$;
The mass $m(\omegat)$ of a subset $\omegat \subset\Omegat$ %=\tPhi(t,\Obj)$ 
is
\be
%\label{eqrhot}
m(\omegat)=\int_{p\in\omegat}\rho_t(p)\,d\omegat.
\ee
%Let $\omegat = \Phitzt(\omegatz)$.

\medskip
\noindent{\bf Conservation of mass principle} (no loss nor production of particles): For all $\omega_{t_0}\subset\Omega_{t_0}$ and all~$t$,
% transformé en le sous-domaine $\omega_t=\Phi_t(\omega_{t_0})\subset \Omega_t$, on~a :
\be
\label{eqpcm}
m(\omega_t)  = m(\omega_{t_0}),\qie
  \int_{p\in\omega_t}\rho_t(p)\,d\omega_t=\int_{P\in\omegatz}\rho_\tz(P)\,d\omega_{t_0}.
\ee

\debprop
If~\eref{eqpcm} then, with $\Jtzt(P)= \det(d\Phitzt(P))$ (positive Jacobian the motion being supposed regular) and $p=\Phitzt(P)$,
\be
\label{eqconvmasse}
\rho_t(p)={\rho_\tz(P)\over \Jtzt(P)}.
\ee
%En particulier, si $\Jtzt(P)\le1$ (contraction), alors $\rho_t(p)\ge\rho_\tz(P)$ (la densité augmente), et si $\Jtzt(P)\ge1$ (dilatation), alors $\rho_t(p)\le\rho_\tz(P)$ (la densité diminue).
\finprop

\debdem
The change of variable formula gives
$$
  \int_{p\in\omegat}\rho_t(p)\,d\omega_t
=\int_{P\in\omega_\tz} \rho_t(\Phitzt(P))\,\Jtzt(P)\,d\omega_\tz,
$$
thus~\eref{eqpcm} gives $\rho_t(\Phitzt(P))\Jtzt(P)=\rho_\tz(P)$.
\findem

\debprop
$\vv=\vv(t,\pt)$ being the Eulerian velocity at $(t,\pt)\in\RR\times \Omegat$, \eref{eqpcm} gives
\be
\label{eqdrodt}
{D\rho\over Dt}+\rho\,\dvg\vv=0, \qie
{\pa\rho\over\pa t}+\dvg(\rho\,\vv)=0.
\ee
%And let $\vec J=\rho\vv$, called the (Eulerian) mass flow (beware of notation: This is not the Jacobian).
Thus, for all $\omegat\subset\Omegat$,
\be
\label{eqdrodt2}
  \int_{\omega_t} {\pa\rho\over\pa t} \,d\omega_t = -\int_{\pa\omega_t} \rho\vv.\vn\,d\sigma_t.
\ee
\finprop

\debdem
\eref{eqpcm} gives
$
  {d\over dt}(\int_{p(t)\in\omega_t}\rho(t,p(t))\,d\omega_t)=0,
$
and Leibniz formula~\eref{eqdjodt2} applied for all~$\omegat$ gives~\eref{eqdrodt}.
Then the Green formula
$\int_{\Omega_t}\dvg(\rho\,\vv)\,d\Omega_t = \int_{\pa\Omega_t} \rho\vv.\vn\,d\sigma_t$
gives~\eref{eqdrodt2}.
\findem

\debexe
Use~\eref{eqconvmasse} to prove~\eref{eqdrodt}.

\debrep
$J(t,P)\rho(t,\Phi(t,P))=\rho_\tz(P)$ give, with $\pt=\Phi(t,P)$,
$$
   {\pa J\over\pa t}(t,P)\,\rho(t,\pt)
 + J(t,P)\,\Bigl({\pa\rho\over\pa t}(t,\pt)
          +d\rho(t,\pt).d\Phi(t,P)\Bigr)=0.
$$
Thus 
${\pa J\over\pa t}(t,P)=J(t,P)\,\dvg\vv(t,p)$, \cf~\eref{eqdjodt},
gives~\eref{eqdrodt}.
\finrep
\finexe

%%%%%%%%%%%%%%%%%%%%%%%%%%%%%%%%%%%%%%%%%%%%%%%%%%%%%%%%%%%%%%%%%%%%%%%%%%%%%%%%%%%
%%%%%%%%%%%%%%%%%%%%%%%%%%%%%%%%%%%%%%%%%%%%%%%%%%%%%%%%%%%%%%%%%%%%%%%%%%%%%%%%%%%

\section{Balance of momentum}

\def\paomegat{{\pa\omega_t}}
\def\omegatz{{\omega_\tz}}
\def\omegat{{\omega_t}}

%%%%%%%%%%%%%%%%%%%%%%%%%%%%%%%%%%%%%%%%%%%%%%%%%%%%%%%%%%%%%%%%%%%%%%%%%%%%%%%%%%%

\subsection{Framework}

$\tPhi:[\tz,T]\times\Obj\rar\RRn$ is a regular motion,
$\Omegat = \tPhi(t,\Obj)$, $\Gammat = \pa\Omegat$ (the boundary),
$\vv$ is the Eulerian velocity field,
$\omegat$ is a regular sub domain in~$\Omegat$ and $\paomegat$ is its boundary.

An observer chooses a Euclidean basis~$(\ve_i)$ (\eg\ made with the foot or the metre)
and call $\dd_g$ the associated Euclidean dot product.
And $\vn(t,p)=\vn_t(p)$ is the outer unit normal at~$t$ at $p\in\paomegat$.
%, and $\vn_t(p) = \sumin n^i_t(p)\ve_i$.

\comment{
We will use the ``double matrix contraction'', that is, \cf~\eref{eqttg},
\be
\label{eqpcqm0}
[\sigma_{ij}] : [\tau_{ij}] = \sumijn \sigma_{ij}\tau_{ij}.
\ee
(Non objective: see~\eref{eqS2pT2}-\eref{eqS2pT3}.)
And $\uusigma=[\sigma_{ij}]$ gives the column matrix
$\dvg\uusigma = \pmatrix{
\sum_j{\pa\sigma_{1j}\over \pa x^j} \cr \vdots \cr \sum_j{\pa\sigma_{nj}\over \pa x^j}}$,
written $\dvg\uusigma = \sumijn {\pa\sigma_{ij}\over \pa x^j}\ve_i$,
\cf~\eref{eqdivmec}.
}

All the functions are assumed to be regular enough to validate the following calculations.%
\def\paomega{{\pa\omega}}
\def\dext{{d_{ext}}}
\def\vk{{\vec k}}

\comment{
We will use the  Gauss--Green--Ostrogradsky theorem: 
\be
\label{eqGGO}
\int_{p\in\omega} \dvg \vk(p)\,d\Omega  = \int_{p\in\paomega} \vk(p)\bcdot\vn(p)\,d\Gamma.
\ee
}

%On dispose la vitesse eulérienne donnée par $\vv(t,\pt)= {\pa \tPhi \over \pa t}(t,\Pobj)$ quand $\pt=\tPhi(t,\Pobj)$.
Let $\rho :
\left\{\eqalign{
\bigcup_{t\in[\tz,T]}(\{t\} \times \Omegat) &\rar  \RR \cr
(t,\pt) &\rar \rho(t,\pt) \cr
}\right\}
$
(a mass density),
let $\vf:
\left\{\eqalign{
\bigcup_{t\in[\tz,T]}(\{t\} \times \Omegat) &\rar  \vRRn \cr
(t,\pt) &\rar \vf(t,\pt) \cr
}\right\}
$
(a body force density),
and
let $\vec T:
\left\{\eqalign{
\bigcup_{t\in[\tz,T]}(\{t\} \times \paomegat \times \vRRnt) &\rar  \vRRn \cr
(t,\pt,\vn(\pt)) &\rar \vec T(t,\pt,\vn(\pt)) \cr
}\right\}
$
(a surface force density) defined for any regular subset $\omegat\subset\Omegat$.

%%%%%%%%%%%%%%%%%%%%%%%%%%%%%%%%%%%%%%%%%%%%%%%%%%%%%%%%%%%%%%%%%%%%%%%%%%%%%%%%%%%

\subsection{Master balance law}

\debdef
The balance of momentum is satisfied by $\rho$, $\vf$ and~$\vec T$ iff, for all regular open subset $\omegat$ in~$\Omegat$,
\be
\label{eqp1}
{d\over dt}(\int_\omegat \rho \vv\,d\Omegat)
 = \int_\omegat \vf \,d\Omegat + \int_{\pa\omega_t} \vec T_\paomegat\,d\Gammat
\quad\hbox{(master balance law)}.
\ee
\findef
(It is in fact a linearity hypothesis, see theorem~\ref{thm1emeloiC}.)

%(Full notation: $\ds {d\over dt}(\int_{\pt\in\omegat} \rho(t,\pt) \vv(t,\pt)\,d\Omegat)  = \int_{\pt\in\omegat} \vf(t,\pt) \,d\Omegat + \int_{\pt\in\paomegat} \vec T_\paomegat(t,\pt)\,d\Gammat$.)
Thus, with~\eref{eqdjodt2},
\be
\label{eqp12}
  \int_\omegat {D(\rho \vv)\over Dt}+\rho \vv\,\dvg\vv\;d\Omegat
 = \int_\omegat \vf \,d\Omegat + \int_{\pa\omega_t} \vec T_\paomegat\,d\Gammat.
\ee
And with the conservation of mass hypothesis, \cf~\eref{eqdrodt}, we get
\be
\label{eqp13}
  \int_\omegat \rho {D\vv\over Dt}\;d\Omegat
 = \int_\omegat \vf \,d\Omegat + \int_{\pa\omega_t} \vec T_\paomegat\,d\Gammat,
\ee
with  ${D\vv\over Dt}=\vgamma = $ the Eulerian acceleration.

%%%%%%%%%%%%%%%%%%%%%%%%%%%%%%%%%%%%%%%%%%%%%%%%%%%%%%%%%%%%%%%%%%%%%%%%%%%%%%%%%%%

\subsection{Cauchy theorem $\vec T = \protect\uusigma.\vn$ (stress tensor $\protect\uusigma$)}
\label{sectcC}

\begin{theorem} [Cauchy first law: Cauchy stress tensor]
\label{thm1emeloiC}
If the master balance law~\eref{eqp1} is satisfied, then
$\vec T$ is linear in~$\vn$, that is, there exists a Eulerian endomorphism
$\uusigma$, identified to a Eulerian tensor $\uusigma\in \Tuuot$,
called the Cauchy stress tensor, \st\ on all~$\paomegat$, in short
\be
\label{eq1lc}
\vec T = \uusigma.\vn, %\quad \hbox{au sens matriciel}\quad T_i = \sumijn \sigma_{ij}n_j,
\ee
where $\vn$ is the unit outward normal to~$\paomegat$ (\ie,
$\vec T(t,\pt) = \uusigma(t,\pt).\vn(t,\pt)$ for all $t$ and $\pt\in\paomegat$).
\finthm

The proof is based on:

\deblem
\label{lemC}
Let
$\phi :
\left\{\eqalign{
\bOmega & \rar \RR \cr
p & \rar \phi(p)
}\right\}
 \in C^1(\bOmega;\RR)$
and
$\psi :
\left\{\eqalign{
\bOmega \times \vRRt & \rar \RR \cr
(p,\vw) & \rar \psi(p,\vw)
}\right\}
\in C^1(\bOmega,\vRRt;\RR)$. If % (more precisely, $\psi : \overline{T\Omega}\rar\RR$).
\be
%\label{eqlem1lcd}
\forall \omega \hbox{ open in }\Omega,\;
\int_{p\in\omega} \phi(p)\,d\Omega  = \int_{p\in\paomega} \psi(p,\vn(p))\,d\Gamma
\ee
(no dependence on the curvature or on higher derivatives since at any $p\in\paomega$, $\psi$ only depends on $\vn(p)$),
then
\be
\label{eqlem1lcd}
\exists \vk \in C^1(\bOmega;\vRRt)\hbox{ \st\ } \psi = (\vk,\vn)_g,
\hbox{ and } \phi= \dvg\vk,
\ee
\ie\ $\psi$ depends linearly on~$\vn$, and $\phi$ is a divergence.
%(The last equality with the divergence formula~\eref{eqipp}).
% that is, at $p\in\paomega$, if $\psi$ only depends on~$\vn$ (no dependence on the curvature or on higher derivatives) then \eref{eqlem1lcd}$_1$ implies $\exists \vk \in C^1(\bOmega;\vRRt)$ \st\ $\phi(p)= \dvg\vk(p)$. % and $\psi(p,\vn(p)) = (\vk(p),\vn(p))_g \eqnote \vk_p\bcdot\vn_p$.
\finlem

\def\paomega{{\pa\omega}}
\debdem (Lemma~\ref{lemC}.)
(This proof is standard: We recall it.)
Let $p\in\Omega \subset\RRt$.
Consider the tetrahedral defined by its vertices
$p$, $p+(h_1,0,0)$, $p+(0,h_2,0)$ and $p+(0,0,h_3)$, with $h_i>0$ for all~$i$.
(On each face of a tetrahedron, the unit normal vector is uniform.)
Let $\Sigma_1$ the side which outer unit normal is $-\vE_1$:
It area is $\sigma_1=\demi h_2h_3$ (square triangle).
Idem for $\Sigma_2$ and $\Sigma_3$.
Let $\Sigma$ be the fourth side: its area is
$\sigma= \demi\sqrt{h_2^2h_3^2+h_3^2h_1^2+h_1^2h_2^2}$ and
its outer unit normal is $\vn={1\over 2\sigma}(h_2h_3,h_3h_1,h_1h_2)$ (see exercise~\ref{exoaire4}),
that is $\vn=(n_1,n_2,n_3)$ with $n_i = {\sigma_i\over \sigma}$ pour $i=1,2,3$.
The volume of the tetrahedral is ${1\over6}h_1h_2h_3 \eqnote \ell^3$. %, où $\ell$ à la dimension d'une longueur.
Let $M %=||\phi||_\infty 
:= \sup_{p\in\bOmega} |\phi(p)|$; We have $M<\infty$, since $\phi$ is continuous in~$\bOmega$. Then \eref{eqlem1lcd} give
\be
\label{eq1lcd2}
M\ell^3 \ge  |\int_{\pa\omega_t} \psi(p,\vn(p))\,d\Gamma|, \qso
\int_{\pa\omega_t} \psi(p,\vn(p))\,d\Gamma  = O(\ell^3).
\ee
And $\psi$ being continuous, the mean value theorem applied on~$\Sigma_i$ gives: %, pour $i=1,...,4$,
There exists $p_i\in\Sigma_i$ \st
$$
\int_{\Sigma_i}\psi(p,\vn(p))\,d\Gamma= \sigma_i \psi(p_i,\vn_i).
$$
Thus
$$
\int_{\pa\omega_t}\psi(p,\vn(p))\,d\Gamma=\Bigl(\sigma_1 \psi(p_1,-\vE_1)+\sigma_2 \psi (t,p_2,-\vE_2)
  +\sigma_3 \psi(p_3,-\vE_3)+\sigma \psi(p_4,\vn)\Bigr).
$$
Then, $\Psi$ being continous, \eref{eq1lcd2} gives
\be
\label{eqaux}
 \sigma_1 \psi(p_1,-\vE_1)
+\sigma_2 \psi(p_2,-\vE_2)
+\sigma_3 \psi(p_3,-\vE_3)
+\sigma \psi(p_4,\vn)   = O(\ell^3).
\ee
We flatten the tetrahedron on the $yz$ face by taking $h_2=h_3 \eqnote h$ and $h_1=h^2$;
Thus $\sigma_1=\demi h^2$, $\sigma_2=o(h^2)$, $\sigma_3=o(h^2)$,
$\sigma \sim \sigma_1$, $\ell^3 = {1\over 6} h^4$, with $\vn \sim -\vn_1 = \vE_1$
and $p_i\sim p$; Then
\be
\label{eqaux2}
   \psi(p,-\vE_1)+ \psi(p,+\vE_1)=0.
\ee
Idem with $xz$ and $xy$.
And for a fixed tetrahedron with $h_1,h_2,h_3$ given,
consider the smaller tetrahedron with $\eps h_1,\eps h_2, \eps h_3$.
Then as $\eps\rar0$ \eref{eqaux} with~\eref{eqaux2} give
$$
\eqalign{
   \psi(p,\vn) = & -{\sigma_i \over \sigma} \psi(p,-\vE_1)
	-{\sigma_2 \over \sigma} \psi(p,-\vE_2)
	-{\sigma_3 \over \sigma} \psi(p,-\vE_3)
 = \sum_{i=1}^3 n_i \psi(p,\vE_i),\cr
% & = & n_1 \psi(p,\vE_1)+n_2 \psi(p,\vE_2)+n_3 \psi(p,\vE_3),\cr
}
$$
since $n_i= {\sigma_i \over \sigma}$ pour $i=1,2,3$.
The same steps can be done for any (inclined) tetrahedron (or apply a change of variable to get back to the above tetrahedron).
Thus $\psi_p$ is a linear map in~$\vn_p$,
that is, there exists a linear form $\alpha_p$ \st\ $\psi_p(\vn_p) = \alpha_p.\vn_p$ for any $p\in\paomega$.
And the Riesz representation theorem gives:
$\exists \vk_p$ \st\ $\alpha_p.\vn_p = (\vk_p,\vn_p)_g \eqnote \vk_p\bcdot\vn_p$.
\findem

\debdem (Theorem.)
Apply Lemma~\ref{lemC} component by component with
$\vec\phi = \rho {D\vv\over Dt} - \vf = \sumin \phi^i\ve_i$,
\cf~\eref{eqp13}.
\findem

\comment{
Se plaçant dans une base euclidienne $(\ve_i)$ de~$\RRn$,
pour une fonction différentiable
$\uusigma:p\in\RRn\rar  = \sumijn \sigma^i_j \ve_i\otimes e^j$
(considérée comme un tenseur~${1\choose1}$)
de matrice $[\sigma^i_j(p)]$,
on appelle divergence la fonction vectorielle
$\dvg\uusigma : p\in\RR\rar \dvg\uusigma(p)\in\RRn$ donnée par, \cf~\eref{eqdivmec} :
\be
\label{eqdivmec2}
\dvg\uusigma=\pmatrix{
{\pa\sigma^1_1\over\pa x^1}+\ldots+{\pa\sigma^1_n\over\pa x^n}
= \sumjn {\pa\sigma^1_j\over\pa x^j}\cr
\vdots\cr
{\pa\sigma^n_1\over\pa x^1}+\ldots+{\pa\sigma^n_n\over\pa x^n}
= \sumjn {\pa\sigma^n_j\over\pa x^j}\cr
}
= \sumijn {\pa\sigma^i_j\over\pa x^j}\ve_i,
\ee
divergence des <<vecteurs lignes>> de~$\uusigma$.
}

\debcor
With $\dvg\uusigma \eqdef \sumin (\sumjn {\pa\sigma_{ij}\over \pa x_j})\ve_i$
(definition of ``the matrix divergence'' see~\eref{eqdivmec}),
\be
\label{eqcorp120}
\left\{\eqalign{
& \vf + \dvg\uusigma = \rho{D \vv\over Dt} \quad \hbox{in }\Omegat, \cr
& \uusigma.\vn = \vec T \quad\hbox{on }\Gammat
}\right.
\ee
(matrix meaning).
(With duality notations, 
$\dvg\uusigma \eqdef \sumin (\sumjn {\pa\sigma^i_j\over \pa x^j})\ve_i$.)
\fincor

\debdem
Apply the divergence Formula to~\eref{eqp13}.
\findem

\debexe
\label{exoaire4}
Consider a triangle $T$ in~$\RRt$ which vertices are $A=(h_1,0,0)$, $B=(0,h_2,0)$, $C=(0,0,h_3)$.
Prove that $\vn = (h_2h_3,h_3h_1,h_1h_2)$ is orthogonal to~$T$
and that $\sigma=\demi\sqrt{h_2^2h_3^2+h_3^2h_1^2+h_1^2h_2^2}$ is its area.

\debrep
Consider the parametric surface $\vr(t,u)=A + t \vec{AB} + u \vec{AC}$ for $t,u\in[0,1]$
describing the triangle. Thus
$\vn = {\pa \vr\over \pa t} \wedge {\pa \vr\over \pa u} =  \vec{AB} \wedge \vec{AC}
=\pmatrix{-h_1\cr h_2 \cr 0} \wedge \pmatrix{-h_1\cr 0 \cr h_3}
=\pmatrix{h_2h_3 \cr h_3h_1 \cr h_1h_2}$ is orthonormal.
And $d\sigma = || {\pa \vr\over \pa t} \wedge {\pa \vr\over \pa u}||dudt
=\sqrt{h_2^2h_3^2+h_3^2h_1^2+h_1^2h_2^2}dudt$.
Thus $\sigma = \int_{t=0}^1\int_{u=0}^1 d\sigma=\sqrt{h_2^2h_3^2+h_3^2h_1^2+h_1^2h_2^2}$
is twice the aera of the triangle.
\finrep
\finexe

% \\ \verb+http://www.isima.fr/leborgne/IsimathMeca/PpvObj.pdf+.

\comment{

On ne peut pas connaître $\vec T(t,\pt)$ en se contentant de regarder le matériau :
``pour connaître le poids d'une valise il faut la soulever'', dixit Germain.
Exemple pour le travail des forces extérieures : on mesure un travail dû à un déplacement
le long d'une trajectoire (une courbe paramétrée)
$\vc_\pt : \tau\in[t,t{+}\eps] \rar q_\tau=\vc_\pt(\tau)\in\RRn$ avec $\vc_\pt(t)=\pt$.
Et de manière classique ce travail est quantifié (mesuré) à l'aide d'un produit scalaire euclidien~$\dd_g$ :
\be
\label{eqtvx2}
\calT(\vc)
= \int_{\Im\vc} (\vec T , d\vell)_g
%= \int_{\qt\in \Im\vc} (\vec T(q_t) , d\vell(q_t))_g
\eqdef \int_{\tau=t}^{t{+}\eps} (\vec T(\tau,\vc_\pt(\tau)) , \vc_\pt{}'(\tau))_g\,d\tau
\simeq \eps (\vec T(t,\pt), \vcp(t))_g,
\ee
avec l'approximation de Riemann à gauche.
%Et $(\vec T(\vc_x(t)) , \vcxp(t))_g \eqnote \vec T(\vc_x(t)) \cdot \vcxp(t)$.
On déduit ainsi~$\vec T(t,\pt)$, et on applique le \S~précédent pour en déduire le tenseur de Cauchy~$\uusigma$.

%\medskip
Sans produit scalaire donné a priori,
on suit la démarche de la thermodynamique et le travail est décrit
à l'aide d'une forme différentielle $T^\flat$, où donc $T^\flat(t,\vc_\pt(t))\in \RRns$, par :
\be
\label{eqtvx3}
\calT(\vc) = \int_{\Im\vc} T^\flat.d\vell
\eqdef \int_{\tau=t}^{t{+}\eps} T^\flat(\tau,\vc_\pt(\tau)).\vc_\pt{}'(\tau)\,d\tau.
%\simeq \eps T^\flat(t,\pt). \vcp(t),
\ee
%où on a utilisé d'approximation de Riemann à gauche.

\comment{
Une fois~\eref{eqtvx3} obtenu (valide pour un anglais ou un français),
on décide d'introduire une unité de mesure et un produit scalaire euclidien
(impossible en thermodynamique, possible ici)
et on dispose alors d'un vecteur normal unitaire à une surface
(dépend du choix de l'unité de mesure).
}

Si on suit cette démarche pour le travail des forces intérieurs
alors l'approche de Cauchy consiste à considérer que les efforts sont caractérisés par
un tenseur $\uutau \in \Tuuu$, et la densité de puissance caractérisée par $\uutau \odd d\vv$.
On peut alors introduire une unité euclidienne, une base euclidienne $(\ve_i)$
et un produit scalaire euclidien~$\dd_g$ associés.
On pose $\uusigma = \uutau^T$, et $\uutau \odd d\vv = \uusigma : d\vv$
redonne l'approche classique.

}

%%%%%%%%%%%%%%%%%%%%%%%%%%%%%%%%%%%%%%%%%%%%%%%%%%%%%%%%%%%%%%%%%%%%%%%%%%%%%%%%%%%
%%%%%%%%%%%%%%%%%%%%%%%%%%%%%%%%%%%%%%%%%%%%%%%%%%%%%%%%%%%%%%%%%%%%%%%%%%%%%%%%%%%

\section{Balance of moment of momentum}

\debdef
The balance of moment of momentum is satisfied by $\rho$, $\vf$ and~$\vec T$ iff for all regular sub-open set $\omegat\subset \Omegat$
\be
\label{eqp2}
  {d\over dt}\int_\omegat \rho\, \ora{\calO M}\wedge\vv\,d\Omegat
 = \int_\omegat \rho\,\ora{\calO M}\wedge \vf \,d\Omegat
    + \int_{\pa\omega_t} \ora{\calO M}\wedge\vec T\,d\Gammat,
\ee
equality called the master balance of moment of momentum law.
(This excludes \eg\ Cosserat continua materials.)
\findef

\debthm (Cauchy second law.)
\label{thm2emeloiC}
If the master balance law  (so $\vec T=\uusigma.\vn$)
and the master balance of moment of momentum law are satisfied then $\uusigma$ is symmetric.
\finthm

\debdem (Standard proof.)
Let $\vx= \ora{\calO M} = \sum_i x_i\vE_i$,
and $\vec T = \sum_i T_i\vE_i = \uusigma.\vn = \sum_{ij} \sigma_{ij} n_j \vE_i$.
Then (first component)
$(\vx\wedge\vec T)_1
= x_2T_3 - x_3T_2
= x_2(\sigma_{31} n_1 + \sigma_{32} n_2 + \sigma_{33} n_3)
- x_3(\sigma_{21} n_1 + \sigma_{22} n_2 + \sigma_{23} n_3)
= (x_2\sigma_{31}-x_3\sigma_{21})n_1
+ (x_2\sigma_{32}-x_3\sigma_{22})n_2
+ (x_2\sigma_{33}-x_3\sigma_{23})n_3
$.
Thus
$\int_{\pa\omega_t} (\vx\wedge\vec T)_1\,d\Gammat
= \int_{\omega_t}
  {\pa (x_2\sigma_{31}-x_3\sigma_{21}) \over \pa x_1}
+ {\pa (x_2\sigma_{32}-x_3\sigma_{22}) \over \pa x_2}
+ {\pa (x_2\sigma_{33}-x_3\sigma_{23}) \over \pa x_3}
\,d\Omegat
= \int_{\omega_t}
  x_2 (\dvg\uusigma)_3
+ x_3 (\dvg\uusigma)_2 + \sigma_{32} - \sigma_{23}
\, d\omegat
$.
%Les autres composantes par permutation circulaire.

\eref{eqcorp120} gives $\rho {D\vv\over Dt} - \vf = \dvg\uusigma$,
thus $\vx\wedge(\rho\vgamma - \vf) = \vx\wedge\dvg\uusigma$,
so the first component of $\vx\wedge(\rho\vgamma - \vf)$ is
$ x_2 (\dvg\uusigma)_3 - x_3(\dvg\uusigma)_2
%= x_2 ({\pa \sigma_{31} \over \pa x_1} + {\pa \sigma_{32} \over \pa x_2} + {\pa \sigma_{33} \over \pa x_3})
%- x_3 ({\pa \sigma_{21} \over \pa x_1} + {\pa \sigma_{22} \over \pa x_2} + {\pa \sigma_{23} \over \pa x_3})
$, \cf~\eref{eqcorp120}.
Thus~\eref{eqp2} gives $\int_{\omega_t} \sigma_{32} - \sigma_{23}\, d\omegat = 0$.
True for all $\omegat$, thus $\sigma_{32} - \sigma_{23} = 0$.
Idem for the other components: $\uusigma$ is symmetric.
\findem

%%%%%%%%%%%%%%%%%%%%%%%%%%%%%%%%%%%%%%%%%%%%%%%%%%%%%%%%%%%%%%%%%%%%%%%%%%%%%%%%%%%
%%%%%%%%%%%%%%%%%%%%%%%%%%%%%%%%%%%%%%%%%%%%%%%%%%%%%%%%%%%%%%%%%%%%%%%%%%%%%%%%%%%

\section{Uniform tensors in $\calL^r_s(E)$}
\label{secunift}

Uniform tensors enable to define without ambiguity the ``objective contraction rules''.
Uniform tensors are scalar valued multilinear functions acting on both vectors and linear forms.

NB: In classical mechanics courses, what is called a ``tensor'' generally not a tensor but a matrix.
\Eg\ you may encounter the expression ``Euclidean tensor'' which means:
% which is self-contradictory, since a tensor exists independently of any Euclidean dot product; \Eg\ a wooden stick %, which exists independently of any observer, is represented by a vector~$\vec{AB}$ by all observers, \cf~\S~\ref{secRfc},1-, independent of any unit of measurement such as the foot or metre, therefore independent of any inner dot product. So, if you come accross the expression ``Euclidean tensor'' It means: 
The matrix representation of ``something'' with respect to a Euclidean basis (based on the foot, metre,...) chosen by some observer. (An ``Euclidean tensor'' is a non-sense, \eg\ can you define a ``Euclidean vector''?)

%%%%%%%%%%%%%%%%%%%%%%%%%%%%%%%%%%%%%%%%%%%%%%%%%%%%%%%%%%%%%%%%%%%%%%%%%%%%%%%%%%%

\subsection{Tensorial product and multilinear forms}

Let $A_1,...,A_n$ be $n$ finite dimension vector spaces. %, $\dim(A_i)=d_i\in\NNs$. 
And $A_i^*=\calL(A_i;\RR)$ the set of linear forms.

%%%%%%%%%%%%%%%%%%%%%%%%%%%%%%%%%%%%%%%%%%%%%%%%%%%%%%%%%%%%%%%%%%%%%%%%%%%%%%%%%%%

\subsubsection{Tensorial product of functions}

Let $f_1:A_1\rar \RR$, ..., $f_n:A_n\rar \RR$ be $n$ functions.
Their tensorial product is the function
$f_1\otimes...\otimes f_n : A_1 \times...\times A_n \rar \RR$ defined by (separate variable function)
\be
\label{eqdefpt}
%\forall (\vx_1,...,\vx_n) \in A_1\times...\times A_n,\quad
(f_1\otimes... \otimes f_n)(\vx_1,...,\vx_n) = f_1(\vx_1)...f_n(\vx_n).
\ee
(\Eg, $n=2$ and $A_1=A_2=\RR$ and $(\cos\otimes \sin)(x,y)=\cos(x) \sin(y)$.)

%%%%%%%%%%%%%%%%%%%%%%%%%%%%%%%%%%%%%%%%%%%%%%%%%%%%%%%%%%%%%%%%%%%%%%%%%%%%%%%%%%%

\subsubsection{Tensorial product of linear forms: multilinear forms}

Let $\calL(A_1,...,A_n ; \RR)$ be the set of $\RR$-multilinear forms on the Cartesian product
$A_1\times ...\times A_n$, that is, the set of the functions
$M : A_1 \times ... \times A_n \rar \RR$ \st,
for all $i=1,...,n$, all $\vx_i,\vy_i\in A_i$ and all $\lambda\in\RR$,
\be
M(...,\vx_i+\lambda \vy_i,...) = M(...,\vx_i,...) + \lambda\; M(...,\vy_i,...),
\ee
the other variables being unchanged. 

Definition: An elementary tensor is multilinear form $M=\ell_1\otimes .... \otimes \ell_n$, with $\ell_i\in A_i^*$ for all~$i$; So
\be
\forall (\vx_i)_{i\in\NNs}\in \prod_{i=1}^n A_i,\quad
(\ell_1\otimes... \otimes \ell_n)(\vx_1,...,\vx_n) 
%= \ell_1(\vx_1)...\ell_n(\vx_n)
= (\ell_1.\vx_1)...(\ell_n.\vx_n)\;\in\RR.
\ee
(The dot in $\ell_i.\vx_i$ is {\bf not} an inner dot product: It is the duality ``outer product'' $\ell_i.\vx_i:=\ell_i(\vx_i)$, \cf~\eref{eqellpv}.)

%%%%%%%%%%%%%%%%%%%%%%%%%%%%%%%%%%%%%%%%%%%%%%%%%%%%%%%%%%%%%%%%%%%%%%%%%%%%%%%%%%%

\comment{
\subsubsection{With bases}

\def\vaij{{\va_{(i)j}}}
\def\vauj{{\va_{(1)j}}}
\def\vanj{{\va_{(n)j}}}
\def\vauji{{\va_{(1)j_i}}}
\def\vauju{{\va_{(1)j_1}}}
\def\vanji{{\va_{(n)j_i}}}
\def\vanjn{{\va_{(n)j_n}}}
\def\aij{{a_{(i)}^j}}
\def\auj{{a_{(1)}^j}}
\def\anj{{a_{(n)}^j}}

Let $(\vaij)_{j=1,...,d_i}$ be a basis in~$A_i$, $i=1,...,n$,
and $(\aij)_{j=1,...,d_i}$ be the dual bases;
Then a multilinear form~$M$ is known iff the values $M(\vauju,...,\vanjn)$
are known for all $j_1,...,j_n$.
Indeed,  if $\vv_1=\sum_j v_1^j \vauj$, ..., $\vv_n=\sum_j v_n^j \vanj$,
then (multilinearity)
$M(\vv_1,...,\vv_n)
= \sum_{j_1,...,j_n} v_1^{j_1}... v_n^{j_n} M(\vauju,...,\vanjn)$.
Therefore $(a_{(1)}^{j_1}\otimes...\otimes a_{(n)}^{j_n})_{j_1 = 1,...,d_1,... \atop j_n = 1,...,d_n}$
is a basis in $\calL(A_1,...,A_n ; \RR)$, and $M\in \calL(A_1,...,A_n ; \RR)$ reads
\be
M = \sum_{j_1=1}^{d_1}...\sum_{j_n=1}^{d_n} M_{j_1,...,j_n} a_{(1)}^{j_1}\otimes...\otimes a_{(n)}^{j_n}
\qwhere  M_{j_1,...,j_n} = M(\va_{(1)j_1},...,\va_{(n)j_n}),
\ee
since 
$\sum\sum(M_{j_1,...,j_n}a_{(1)}^{j_1}\otimes...\otimes a_{(n)}^{j_n})(\vauju,...,\vanjn)
=\sum\sum M_{j_1,...,j_n}a_{(1)}^{j_1}(\vauju)...a_{(n)}^{j_n}(\vanjn)
=\sum\sum M_{j_1,...,j_n}\delta^{j_1}_{j_1}...\delta^{j_n}_{j_n}
= M_{j_1,...,j_n}
$.
}

%%%%%%%%%%%%%%%%%%%%%%%%%%%%%%%%%%%%%%%%%%%%%%%%%%%%%%%%%%%%%%%%%%%%%%%%%%%%%%%%%%%

\subsection{Uniform tensors in $\calL^0_s(E)$}
\label{secut0s}

Let $E$ be a real vector space, with $\dim(E)=n\in\NNs$.
In this section we consider the first overlay on~$E$ made of
multilinear forms $M$ on~$E$, called the uniform tensors of type $0$~$s$ or of type~${0 \choose s}$.

\Eg,  $M\in\calL^0_1(E)$ a linear form, $M\in\calL^0_2(E)$ an inner dot product,  $M\in\calL^0_n(E)$ a determinant...

Notations for quantification purposes:  $(\ve_i)$ is a basis in~$E$,
$(\pi_{ei})$ is its (covariant) dual basis (basis in $\Es=\calL(E;\RR)$),
$(\pa_i)$ is its bidual basis (basis in $\Ess=\calL(E^*;\RR)$).

%%%%%%%%%%%%%%%%%%%%%%%%%%%%%%%%%%%%%%%%%%%%%%%%%%%%%%%%%%%%%%%%%%%%%%%%%%%%%%%%%%%

\subsubsection{Definition of type ${0 \choose s}$ uniform tensors}
\label{seczs}

$\calL^0_0(E) := \RR$, and if $s\in\NNs$ then
\be
\calL^0_s(E)
:= \calL(\underbrace{E \times ... \times E}_{s\;\rm times} ; \RR)
\ee
is called the set of uniform tensors of type~${0 \choose s}$ on~$E$.

%%%%%%%%%%%%%%%%%%%%%%%%%%%%%%%%%%%%%%%%%%%%%%%%%%%%%%%%%%%%%%%%%%%%%%%%%%%%%%%%%%%

\subsubsection{Example: Type ${0\choose1}$ uniform tensor = linear forms}

A type~${0\choose1}$ uniform tensor is an element of $\calL^0_1(E) = \calL(E;\RR) = \Es$:
It is a linear form $\ell\in \calL^0_1(E) = E^*$.

\mn
{\bf Quantification:}
With $\ell_i := \ell(\ve_i)$ we have, \cf~\eref{eqxi2},
\be
\ell = \sumin \ell_i \pi_{ei} ,\qand
[\ell]_{|\pi_e} = \pmatrix{\ell_1 & ... & \ell_n} \eqnote [\ell]_{|\ve}
\ee
(row matrix for a linear form).
Duality notations: $(e^i)$ is the covariant dual basis and $\ell = \sumin \ell_i e^i$.

Thus, if $\vv\in E$, $\vv = \sumin v_i\ve_i$,
then $\vv$ is represented by
$[\vv]_{|\ve} = \pmatrix{v_1 \cr \vdots \cr v_n}$ (column matrix for a vector), and
the matrix calculation rules give
\be
\ell(\vv) = [\ell]_{|\ve}.[\vv]_{|\ve}
= \pmatrix{\ell_1 & ... & \ell_n}. \pmatrix{v_1 \cr \vdots \cr v_n} = \sumin \ell_i v_i \eqnote \ell.\vv.
\ee
%the result being objective.
Duality notations: $\vv = \sumin v^i\ve_i$ and $\ell(\vv)=\sumin \ell_iv^i$, and Einstein's convention is satisfied.

%%%%%%%%%%%%%%%%%%%%%%%%%%%%%%%%%%%%%%%%%%%%%%%%%%%%%%%%%%%%%%%%%%%%%%%%%%%%%%%%%%%

\subsubsection{Example: Type ${0\choose2}$ uniform tensor}

A type~${0\choose2}$ uniform tensor is an element of $\calL^0_2(E) = \calL(E,E;\RR)$:
It is a bilinear form $T\in \calL(E,E;\RR)$. %\Eg, an inner dot product is a ${0\choose2}$ uniform tensor.

\mn
{\bf Quantification:}
Let $T_{ij} := T(\ve_i,\ve_j)$. Then, with $\vv=\sumin v_i\ve_i$ and $\vw=\sumin w_i\ve_i$,
\be
T(\vv,\vw)= \sumijn T_{ij} v_i w_j = [\vv]_{|\ve}^T.[T]_{|\ve}.[\vw]_{|\ve}, 
\qie T = \sumijn T_{ij} \pi_{ei} \otimes \pi_{ej}.
\ee
Duality notations: $T(\vv,\vw)= \sumijn T_{ij} v^i w^j$, 
 and Einstein's convention is satisfied.
%The $T_{ij}$ are the components of~$T$, and $[T]_{|\ve} = [T_{ij}]$ is the matrix of~$T$ relative to the basis $(\ve_i)$.

An elementary uniform tensor in $\calL^0_2(E)$ is a tensor $T=\ell \otimes m$, where $\ell,m\in\Es$.
And so, for all $\vv,\vw\in E$,
\be
\label{eqexat02}
(\ell\otimes m)(\vv,\vw) =  (\ell.\vv)(m.\vw).
\ee
%Thus, with a basis $(e_i)$, $(\ell\otimes m)(\vv,\vw) = (\sumin \ell_i v^i)(\sumin m_i w^i)$ when $\vv=\sumin v^i\ve_i$, $\vw=\sumin w^i\ve_i$, $\ell = \sumin \ell_i e^i$ and $m = \sumin m_i e^i$.

%%%%%%%%%%%%%%%%%%%%%%%%%%%%%%%%%%%%%%%%%%%%%%%%%%%%%%%%%%%%%%%%%%%%%%%%%%%%%%%%%%%

\subsubsection{Example: Determinant}

The determinant is a alternating ${0\choose n}$ uniform tensor, \cf~\eref{eqdefnlina}.

%%%%%%%%%%%%%%%%%%%%%%%%%%%%%%%%%%%%%%%%%%%%%%%%%%%%%%%%%%%%%%%%%%%%%%%%%%%%%%%%%%%

\subsection{Uniform tensors in $\calL^r_s(E)$}
\label{secut}

In this section we consider an over-overlay on~$E$:
The multilinear forms acting on both vectors ($\in E$) and functions $\in \Es$ (linear forms).

%%%%%%%%%%%%%%%%%%%%%%%%%%%%%%%%%%%%%%%%%%%%%%%%%%%%%%%%%%%%%%%%%%%%%%%%%%%%%%%%%%%

\subsubsection{Definition of type ${r\choose s}$ uniform tensors}
\label{secrsu}

Let $r,s\in\NN$ \st\ $r+s\ge 1$. The set of multilinear forms
\be
\label{eqdefLrsE}
\calL^r_s(E)
:= \calL(\underbrace{E^* \times ... \times E^*}_{r\;\rm times}, \underbrace{E \times ... \times E}_{s\;\rm times} ; \RR)
\ee
is called the set of uniform tensors of type~${r \choose s}$ on~$E$.
%And $\calL^0_0(E) := E$.

The case $r=0$ has been considered at~\S~\ref{secut0s}.

When $r\ge1$, a tensor $T\in \calL^r_s(E)$ is a functional:
Its domain of definition contains a set of functions (the set $\Es = \calL(E;\RR)$).

%%%%%%%%%%%%%%%%%%%%%%%%%%%%%%%%%%%%%%%%%%%%%%%%%%%%%%%%%%%%%%%%%%%%%%%%%%%%%%%%%%%

\subsubsection{Example: Type ${1\choose0}$ uniform tensor: Identified with a vector}

A uniform ${1\choose0}$ tensor is a element $T\in \calL^1_0(E) = \calL(\Es;\RR) = \calL(\calL(E;\RR);\RR) = \Ess$.
With the natural canonical isomorphism
\be
\label{eqisomcan00}
\calJ : 
\left\{\eqalign{
E & \rar \Ess = \calL^1_0(E) \cr
\vw & \rar \calJ(\vw) = w,\quad\hbox{defined by}\quad w(\ell) \eqdef \ell(\vw), \quad \forall \ell\in\Es,
}\right.
\ee
\cf~\eref{eqisomcan0} and prop.~\ref{propJnat},
\be
\label{eqwevw0}
w \eqnote \vw ,\qso w.\ell \eqnote \vw.\ell \quad(= \ell.\vw).
\ee
So a ${1\choose0}$ type uniform tensor $w$ is identified (natural canonical) to the vector $\vw=\calJ^{-1}(w)$.

\mn
{\bf Interpretation: } $\Ess$ is the set of directional derivatives.
Indeed, if $\calE$ is an affine space, if $E$ is the associated vector space,
if $p\in\calE$, and if $f$ is a differentiable function at~$p$,
then $w.df(p) \mope^{\eref{eqisomcan00}} df(p).\vw$ is the directional derivative along~$\vw$.

Remark: In differential geometry, $w.df$ is written $\vw(f)$, so $\vw(f)(p) := df(p).\vw$, the definition of a vector being a directional derivative.
%(As a matter of fact, in differential geometry a vector field can be defined as being an element of~$\Ess$, that is a field of directional derivative.)

\mn
{\bf Quantification:} For all~$i,j$,
\be
\pa_i.\pi_{ej} = \delta_{ij} = \pi_{ej}.\ve_i,\qthus \pa_i=\calJ(\ve_i) \eqnote \ve_i.
\ee
Duality notations: $\pa_i.e^j = \delta^j_i = e^j.\ve_i$.
\Eg, if $f$ is a $C^1$ function then $df(p) = \sumin f_{|i}(p)\,\pi_{ei}$ ($= \sumin f_{|i}(p)\,e^i$) and
\be
\pa_i(df(p)) = df(p).\ve_i  = f_{|i}(p) \eqnote \pa_i(f)(p) \eqnote \ve_i(f)(p). % = {\pa f \over \pa x^i}(p)
\ee
%(In a Cartesian basis in~$\RRn$, then $(e^i) \eqnote (dx^i)$, and  the usual notation is $f_{|i} = {\pa f \over \pa x^i}$ and $df(p)=\sumin {\pa f \over \pa x^i}(p)dx^i$.)
% Un tenseur uniforme de type~${1\choose0}$ est un élément de $\calL^1_0(E) = \calL(\Es;\RR) = \Ess$, donc $\calL^1_0(E)\simeq E$ grâce à~$\calJ$, \cf~\eref{eqisomcan00}.
%(donc identification d'un vecteur $\vw$ et de l'opérateur de dérivation dans la direction~$\vw$).

%%%%%%%%%%%%%%%%%%%%%%%%%%%%%%%%%%%%%%%%%%%%%%%%%%%%%%%%%%%%%%%%%%%%%%%%%%%%%%%%%%%

\subsubsection{Example: Type ${1\choose1}$ uniform tensor}
\label{eqicetu}

An elementary uniform tensor in $\calL^1_1(E)$ is a tensor $T=  u \otimes \beta$,
where $u\in \Ess$ and $\beta\in \Es$.
And, with $\vu = J^{-1}(u)\in E$, \cf~\eref{eqisomcan00}, 
we also write $T= \vu \otimes \beta$. Thus, for all $\ell\in\Es$ and $\vw\in E$
\be
(u\otimes \beta)(\ell,\vw) = u(\ell)\beta(\vw) = \ell(\vu)\beta(\vw)
\eqnote \vu(\ell)\beta(\vw) \eqnote (\vu\otimes \beta)(\ell,\vw).
\ee

%And any tensor $T\in \calL^1_1(E)$ is a sum of elementary tensors.
%(The identification with an endomorphism is done in~\S~\ref{secicetu2}.)

\noindent
{\bf Quantification:}
Let $T(\pi_{ei},\ve_j)$. So 
\be
\label{eqTuuu}
T = \sumijn T_{ij}\; \ve_i \otimes \pi_{ej} , \qand [T]_{|\ve} = [T_{ij}],
%T = \sumijn T^i_j\; \pa_i \otimes e^j \eqnote \sumijn T^i_j \ve_i \otimes e^j, \qwith [T]_{|\ve} = [T^i_j],
\ee
$[T]_{|\ve} = [T_{ij}]$ being the matrix of~$T$ relative to the basis~$(\ve_i)$.
Duality notations: $T(e^i,\ve_j) = T^i{}_j$, $[T]_{|\ve} = [T^i{}_j]$, $T= \sumijn T^i{}_j \ve_i\otimes e^j$, and
Einstein's convention is satisfied.

Thus with $\ell\in\Es$, $\ell = \sumin \ell_i e^i \in \Es$, and~$\vw\in E$, $\vw = \sumin w^i\ve_i \in E$, \eref{eqTuuu} gives
\be
\label{eqTuuub}
T(\ell,\vw)
= \sumijn T_{ij} \ve_i(\ell) \pi_{ej}(\vw)
= \sumijn T_{ij} \ell_i w_j
=[\ell]_{|\ve}.[T]_{|\ve}.[\vw]_{|\ve}
\ee
($[\ell]_{|\ve}$ is a row matrix). Duality notations:
$T(\ell,\vw) = \sumijn T^i{}_j \ell_i w^j$ and Einstein convention is satisfied.

%%%%%%%%%%%%%%%%%%%%%%%%%%%%%%%%%%%%%%%%%%%%%%%%%%%%%%%%%%%%%%%%%%%%%%%%%%%%%%%%%%%

\subsubsection{Example: Type ${1\choose2}$ uniform tensor}

The same steps are applied to any tensor.
\Eg, if $T\in \calL^1_2(E)$, then with duality notations, $T^i{}_{\!jk} = T(e^i,\ve_j,\ve_k)$ and
\be
T= \sumijkn T^i{}_{\!jk} \ve_i \otimes e^j \otimes e^k,\qand T(\ell,\vu,\vw) = \sumijkn T^i{}_{\!jk} \ell_i u^j w^k.
\ee
%\Eg, $d^2\vv = \sumijkn v^i_{|jk}\ve_i \otimes e^j \otimes e^k$, and $d^2\vv(\vu,\vw) = \sumijkn v^i_{|jk}u^jw^k\ve_i $.

%%%%%%%%%%%%%%%%%%%%%%%%%%%%%%%%%%%%%%%%%%%%%%%%%%%%%%%%%%%%%%%%%%%%%%%%%%%%%%%%%%%

\subsection{Exterior tensorial products}

Let $T_1 \in \calL^{r_1}_{s_1}(E)$ and $T_2 \in \calL^{r_2}_{s_2}(E)$.
Their tensorial product is the tensor $T_1 \otimes T_2 \in \calL^{r_1+r_2}_{s_1+s_2}(E)$
defined by
\be
\label{eqdefpt0}
(T_1 \otimes T_2)(\ell_{1,1},...,\ell_{2,1},...,\vu_{1,1},...,\vu_{2,1},...)
\eqdef T_1(\ell_{1,1},...,\vu_{1,1},...) T_2(\ell_{2,1},...,\vu_{2,1},...).
\ee
Particular case: with $\lambda \in \calL^0_0(E)=\RR$ and $T \in \calL^r_s(E)$,
\be
\label{eqdefpt1}
\lambda \otimes T = T\otimes \lambda \eqdef \lambda T \in \calL^r_s(E).
\ee

\debexa
let $T_1,T_2 \in \calL^1_1(E)$. Quantification:
Let $T_1 = \sumijn (T_1)^i_j \ve_i\otimes e^j$ and
let $T_2 = \sumkmn (T_2)^k_m \ve_k\otimes e^m$; Then
$T_1 \otimes T_2 = \sumijkmn (T_1)^i_k (T_2)^j_m \ve_i\otimes \ve_j \otimes e^k\otimes e^m \in \calL^2_2(E)$.
\finexa

\debrem
\label{remad}
Alternative definition:
$T_1 \tilde\otimes T_2 := \sumijkmn (T_1)^i_j(T_2)^k_m \ve_i\otimes e^j \otimes \ve_k\otimes e^m
\in \calL(\Es,E,\Es,E;\RR)$.
And we get back to the previous definition thanks to the natural canonical isomorphism
$\tilde J : \calL(\Es,E,\Es,E;\RR) \rar \calL(\Es,\Es,E,E;\RR) = \calL^2_2(E)$
defined by
$ \tilde J(\tilde T) = T$ where $T(\ell,m,\vv,\vw) = \tilde T(\ell,\vv,m,\vw)$.
\finrem

%%%%%%%%%%%%%%%%%%%%%%%%%%%%%%%%%%%%%%%%%%%%%%%%%%%%%%%%%%%%%%%%%%%%%%%%%%%%%%%%%%%

\subsection{Contractions}
\label{secAtco}

%%%%%%%%%%%%%%%%%%%%%%%%%%%%%%%%%%%%%%%%%%%%%%%%%%%%%%%%%%%%%%%%%%%%%%%%%%%%%%%%%%%

\subsubsection{Contraction of a linear form with a vector}

Let $\ell \in \calL^0_1(E) = \Es$ and $\vw\in E$.
Their contraction is the value
\be
\label{eqclv}
\ell(\vw)  \mope^{\hbox{\footnotesize linearity}} \ell.\vw  \eqnote \vw.\ell . %= \vw(\ell)
\ee
And with a basis $(\ve_i)$ and its dual basis $(\pi_{ei})$, $\ell=\sumin \ell_i \pi_{ei}$ and $\vw=\sumin w_i\ve_i$ give
\be
\label{eqclv0}
\ell.\vw = \sumin \ell_i w_i = [\ell]_{|\ve}.[\vw]_{|\ve} = \sumin w_i\ell_i  = \vw.\ell = \Tr(\vw\otimes \ell),
\ee
where $\Tr$ is the objective trace operator $\Tr : \calL(E;E) \simeq \calL^1_1(E) \rar\RR$
(defined by $\Tr(\ve_i\otimes \pi_{ej}) = \delta^i_j$).
% (after using the natural canonical isomorphism $\calL^1_1(E)\simeq \calL(E;E)$).
%This result~\eref{eqclv0} is objective.
Duality notations: $\ell.\vw = \sumin \ell_i w^i$, and Einstein convention is satisfied.

\debexe
Use the change of coordinate formulas to prove that the computation $\ell.\vw$ in~\eref{eqclv0}
gives a result independent of the basis.

\debrep
Let $P$ be the change of basis matrix.
So $[\vw]_\new = P^{-1}.[\vw]_\old$ and $[\ell]_\new = [\ell]_\old.P$, \cf~\eref{eqdefP1},
thus $[\ell]_\new.[\vw]_\new
= ([\ell]_\old.P).(P^{-1}.[\vw]_\old)
= [\ell]_\old.(P.P^{-1}).[\vw]_\old
= [\ell]_\old[\vw]_\old$ ($=\ell.\vw$).
\finrep
\finexe

\comment{
En particulier le réel $\ell.\vw = \vw.\ell$ ne dépend
pas du choix de la base (avec les formules de changement de base
on~a $[\ell]_\new = [\ell]_\old.P$ et $[\vw]_\new = P^{-1}.[\vw]_\old$, \cf~\eref{eqdefP1},
donc $[\ell]_\new.[\vw]_\new = [\ell]_\old.P.P^{-1}.[\vw]_\old = [\ell]_\old[\vw]_\old$).
%Et $\Tr(\vw\otimes \ell) = \Tr(\sumijn w^i\ell_j \ve_i\otimes e^j = \sumin w^i\ell_i$ (trace).
}

%%%%%%%%%%%%%%%%%%%%%%%%%%%%%%%%%%%%%%%%%%%%%%%%%%%%%%%%%%%%%%%%%%%%%%%%%%%%%%%%%%%

\subsubsection{Contraction of a ${1\choose 1}$ tensor and a vector}

Let $\ell\in E^*$ and $\vu\in E$.
The contraction of the elementary tensor $\vw\otimes \ell\in \calL^1_1(E)$ with~$\vu$ is defined by:
\be
\label{eqobjcL}
%(\vw\otimes \ell).\vu = 
(\vw\otimes \underbrace{\ell).\vu}_{\makebox[.1cm]{\footnotesize\rm contraction}} = (\ell.\vu)\vw.
\ee
Thus, if $(\ve_i)$ is a basis in~$E$ and $(\pi_{ei})$ is the dual basis, 
and $T = \sumijn T_{ij} \ve_i\otimes \pi_{ej} \in \calL^1_1(E)$ and $\vu = \sumjn u_j\ve_j\in E$, then
\be
\label{eqobjcL3}
T= \sumijn T_{ij} \ve_i\otimes e^j \quad \Longrightarrow \quad 
T.\vu = \sumijn T_{ij} u_j^j\ve_i
\ee
because $\pi_{ej}(\vu)=u_j$. Duality notations: $T.\vu = \sumijn T^i_j u^j\ve_i$.

Then, with the natural canonical isomorphism ($ \calL^1_1(E) =$) $\calL(E,E^*;\RR) \simeq \calL(E;E)$, see~\eref{propeqtJ2f},
any endomorphism $L\in \calL(E;E)$ defined by $L.\ve_j = \sumin L_{ij}\ve_i$ can be written, for calculation purpose,
\be
\tL = \sumijn L_{ij}\ve_i\otimes \pi_{ej} \eqnote L, \quad\hbox{which means}\quad  
L.\vu \mope^{\eref{eqobjcL}} \sumin L_{ij} u_j\ve_i
\ee
when $\vu=\sum_i u_j\ve_j$,
since $\pi_{ej}(\vu)=u_j$. Duality notations: $L = \sumijn \Lij\ve_i\otimes e^j$.

%%%%%%%%%%%%%%%%%%%%%%%%%%%%%%%%%%%%%%%%%%%%%%%%%%%%%%%%%%%%%%%%%%%%%%%%%%%%%%%%%%%

\subsubsection{Contractions of uniform tensors}

More generally, the contraction of two tensors, if meaningful, is defined thanks to~\eref{eqclv}:
Let $T_1 \in \calL^{r_1}_{s_1}(E)$, $T_2 \in \calL^{r_2}_{s_2}(E)$,
$\ell \in \Es$ and $\vu \in E$.
%, and consider $T_1 \otimes \ell \in \calL^{r_1}_{s_1+1}(E)$ and $\vu \otimes T_2 \in \calL^{r_2+1}_{s_2}(E)$.
%together with $\calJ$,  \cf~\eref{eqisomcan00}.

\debdef
The objective contraction of $T_1 \otimes \ell \in \calL^{r_2}_{s_2+1}(E)$
and $\vu \otimes T_2 \in \calL^{r_2+1}_{s_2}(E)$ is the tensor
$(T_1\otimes \ell) . (\vu\otimes T_2) \in\calL^{r_1+r_2}_{s_1+s_2}$ given by
\be
\label{eqdefptc1}
(T_1\otimes \underbrace{\ell) . (\vu}_{\makebox[.1cm]{\footnotesize\rm contraction}}\otimes T_2) \eqdef (\ell . \vu)\,T_1 \otimes T_2.
\ee
In particular $(T_1\otimes \ell) . \vu = (\ell . \vu)\,T_1$ (as in~\eref{eqobjcL}),
and $\ell . (\vu\otimes T_2) = (\ell.\vu)\,T_2$.

And the objective contraction of $T_1 \otimes \vu \in \calL^{r_2+1}_{s_2}(E)$
and $\ell \otimes T_2 \in \calL^{r_2}_{s_2+1}(E)$ is the tensor
$(T_1\otimes \vu) . (\ell\otimes T_2) \in\calL^{r_1+r_2}_{s_1+s_2}$ given by
\be
(T_1\otimes \vu) . (\ell\otimes T_2) = (\vu.\ell)\,T_1 \otimes T_2
\quad (=(\ell.\vu)\,T_1 \otimes T_2).
\ee
\findef

Quantification with a basis~$(\ve_i)$, examples to avoid cumbersome notations:

\debexa
Let $T\in \calL^1_1(E) = \calL^1_{0+1}(E)$, $T = \sumijn T^i_j \ve_i \otimes e^j$.
With $\vw \in E\sim \Ess = \calL^1_0(E)$, $\vw = \sumjn w^j\ve_j$,
\eref{eqdefptc1} gives $T.\vw \in \calL^1_0(E)\sim E$ and
\be
\label{eqTvw}
T.\vw = \sumijn T^i_j w^j \ve_i, \qie [T.\vw]_{|\ve} = [T]_{|\ve}.[\vw]_{|\ve} \quad\hbox{(column matrix)}.
\ee
(Einstein's convention is satisfied.)
Indeed,	
$T.\vw
= \sumijkn T^i_j w^k(\ve_i \otimes e^j).\ve_k
= \sumijkn T^i_j w^k\ve_i (e^j.\ve_k)
= \sumijkn T^i_j w^k\ve_i (\delta^j_k)
= \sumijn T^i_j w^j\ve_i$.
With $\ell\in\Es=\calL^0_1(E)$, $\ell = \sumin \ell_i e^i$,
\eref{eqdefptc1} gives $\ell.T\in \calL^0_1(E) = \Es$ and
\be
\ell.T = \sumijn \ell_i T^i_j e^j, \qie [\ell.T]_{|\ve} = [\ell]_{|\ve}.[T]_{|\ve} \quad \hbox{(row matrix)}.
\ee
(Einstein's convention is satisfied.)
Indeed $\ell.T = (\sumin \ell_i e^i).(\sumjkn T^k_j \ve_k \otimes e^j)
= \sumijkn \ell_i T^k_j (e^i.\ve_k) e^j %= \sumijkn  \ell_k T^i_j \delta^i_k e^j 
= \sumijn \ell_i T^i_j e^j$.
\finexa

\debexa
Let $S,T \in \calL^1_1(E)$, $S = \sumikn S^i_k \ve_i \otimes e^k$
and $T = \sumjkn T^k_j \ve_k \otimes e^j$. Then
\be
\label{eqST}
S.T = \sumijkn S^i_k T^k_j \ve_i \otimes e^j , \qie
[S.T]_{|\ve} %= [\sumkn S^i_k T^k_j]_{i=1,...,n \atop j=1,...,n}
= [S]_{|\ve}.[T]_{|\ve}
\ee
(Einstein's convention is satisfied.)
Indeed 
$S.T = (\sumikn S^i_k \ve_i \otimes e^k).(\sum_{j,m=1}^n T^m_j \ve_m \otimes e^j)
= \sumijkmn S^i_k T^m_j \ve_i (e^k.\ve_m) \otimes e^j
%= \sumijkmn S^i_k T^m_j \delta^k_m \ve_i \otimes e^j
= \sumijkn S^i_k T^k_j \ve_i \otimes e^j$.
%$[S.T]_{|\ve} = [\sumkn S^i_k T^k_j]_{i=1,...,n \atop j=1,...,n} = [S]_{|\ve}.[T]_{|\ve}$.
\finexa

\debexa
Let $T\in \calL^1_2(E)$, $T = \sumijkn T^i_{jk} \ve_i \otimes e^j \otimes e^k$,
and $\vu,\vw\in E\sim \calL^1_0(E)$, $\vw = \sumin w^i \ve_i$ and $\vu = \sumin u^i \ve_i$. Then
\be
T.\vw = \sumijkn T^i_{jk}w^k \ve_i \otimes e^j  \in \calL^1_1(E),\qand
(T.\vw).\vu =  \sumijkn T^i_{jk}w^k u^j \ve_i \eqnote T(\vu,\vw).
\ee
(Einstein's convention is satisfied.)
So $[T.\vw]_{|\ve} = [\sumkn T^i_{jk}w^k]_{i=1,...,n \atop j=1,...,n}$.
And with $\ell\in \Es$, $\ell = \sumin \ell_i e^i$,
\be
((T.\vw).\vu).\ell = \sumijkn T^i_{jk}w^k u^j \ell_i = T(\ell,\vu,\vw) = \ell.T(\vu,\vw) = \ell.(T.\vw).\vu.
\ee
\finexa

%%%%%%%%%%%%%%%%%%%%%%%%%%%%%%%%%%%%%%%%%%%%%%%%%%%%%%%%%%%%%%%%%%%%%%%%%%%%%%%%%%%

\subsubsection{Objective double contractions of uniform tensors}

\debdef
Let $S,T\in\calL^1_1(E)$. %, identified with the endomorphisms still denoted $S,T\in\calL(E;E)$.
And let $(\ve_i)$ be a basis in~$E$, $(e^i)$ its dual basis, $S=\sumijn S^i_j \ve_i\otimes e^j$ and $T=\sumijn T^i_j \ve_i\otimes e^j$.
The double objective contraction $S\odd T$ of $S$ and~$T$ is defined by
\be
\label{eqdefSoddT}
S \odd T = \sumijn S^i_j T^j_i.
\ee
(Einstein convention is satisfied.)
\findef

\debprop
\label{propSoddT}
$S\odd T$ defined in \eref{eqdefSoddT} is an invariant:
It is the trace $\Tr(L_S \circ L_T)$ of the endomorphisms $L_S,L_T\in\calL(E;E)$ naturally canonically associated
to~$S$ and~$T$ (given by $\ell.L_S.\vu := S(\ell,\vu)$ and $\ell.L_T.\vu := T(\ell,\vu)$ for all $(\vu,\ell)\in E\times E^*$). So the real value $\sumijn S^i_j T^j_i$ has the same real value regardless of the chosen basis~$(\ve_i)$.
(Which is not the case of the term to term matrix multiplication $S:T=\sumijn S^i_j T^i_j$, see next~\S~\ref{secnoct} and example~\ref{exaissuedd}.)
\finprop

\debdem
Let $(\va_i)$ and $(\vb_i)$ be two bases
and $P= [P^i_j]$ be the transition matrix from $(\va_i)$ to $(\vb_i)$, \ie, $\vb_j = \sumin P^i_j \va_i$ for all~$j$.
Let $Q= [Q^i_j] := P^{-1}$.
Then $b^i = \sumin Q^i_j a^i$.
Let $S= \sum_{ij} (S_a)^i_j\va_i\otimes a^j= \sum_{ij} (S_b)^i_j\vb_i\otimes b^j$.
So $[(S_b)^i_j] = P^{-1}.[(S_a)^i_j].P$ (change of basis formula for ${1\choose1}$ tensors identified with endomorphisms), \ie\
$(S_b)^i_j = \sum_{k m} Q^i_k  (S_a)^k_m P^m_j$ for all $i,j$.
Idem with~$T$.
Thus
$\sum_{i,j}(S_b)^i_j (T_b)^j_i
= \sum_{i,j,k ,m,\alpha,\beta}
Q^i_k  (S_a)^k_m P^m_j Q^j_\alpha (T_a)^\alpha_\beta P^\beta_i
= \sum_{i,j,k ,m,\alpha,\beta}
(S_a)^k_m (T_a)^\alpha_\beta P^\beta_i Q^i_k   P^m_j Q^j_\alpha 
= \sum_{k ,m,\alpha,\beta}
(S_a)^k_m (T_a)^\alpha_\beta \delta^\beta_k   \delta^m_\alpha 
= \sum_{k ,m}
(S_a)^k_m (T_a)^m_k
$.
\findem

\debdef
More generally, the objective double contractions $S\odd T$ of uniform tensors,
is obtained by applying the objective simple contraction twice consecutively, when applicable.
\findef

\Eg, $ T_1 \otimes \ell_ {1,1} \otimes \ell_ {1,2} $ and
$ \vu_ {2,1} \otimes \vu_ {2,2} \otimes T_2 $ give
\be
\label{eqdefptc2}
\eqalign{
(T_1\otimes \ell_{1,1}\otimes \underbrace{\ell_{1,2}) .(\vu_{2,1}}_{\hbox{\footnotesize first}}\otimes \vu_{2,2}\otimes T_2)
= & (\ell_{1,2} .\vu_{2,1})(T_1\otimes \underbrace{\ell_{1,1})\otimes (\vu_{2,2}}_{\hbox{\footnotesize second}}\otimes T_2) \cr
= & (\ell_{1,2} . \vu_{2,1})(\ell_{1,1} . \vu_{2,2})\, T_1 \otimes T_2. \cr
}
\ee

\debexa
\label{exaverstriple}
Let $S \in \calL^1_2(E)$, $T \in \calL^2_1(E)$,
$S = \sumijkn S^i_{jk} \ve_i \otimes e^j \otimes e^j$,
$T = \sum_{\alpha,\beta,\gamma=1}^n T^{\alpha\beta}_\gamma \ve_\alpha \otimes \ve_\beta \otimes e^\gamma$.
Then
\be
\label{eqToddS3}
S . T = \sum_{i,j,k,\beta,\gamma=1}^n S^i_{jk} T^{k\beta}_\gamma \ve_i \otimes e^j\otimes \ve_\beta \otimes e^\gamma, \qand
S \odd T = \sum_{i,j,k,\gamma=1}^n S^i_{jk} T^{kj}_\gamma \ve_i \otimes e^\gamma.
\ee
(Einstein's convention is satisfied.)
\finexa

\debexe
If $S\in\calL(E,F;\RR)$, $T\in \calL(F,G;\RR)$ and $U\in\calL(G,E;\RR)$ then prove
\be
\label{eqalg}
S \odd (T.U)
= (S.T) \odd U
= (U.S) \odd T \quad \hbox{(circular permutation)}.
\ee

\debrep
If $S = \sum S^i_j \va_i \otimes b^j$,
$T = \sum T^i_j \vb_i \otimes c^j$
and $U = \sum U^i_j \vc_i \otimes a^j$,
then $T . U = \sum T^i_k U^k_j \vb_i \otimes a^j$,
thus %$S . (T.U) = \sum S^i_m T^m_k U^k_j \va_i \otimes  a^j$ et
$S \odd (T.U) = \sum S^i_m T^m_k U^k_i$,
and
$S . T = \sum S^i_k T^k_j \va_i \otimes c^j$,
so $(S.T) \odd U = \sum S^i_k T^k_mU^m_i$.
And the second equality thanks to the symmetry of~$\odd$, \ie\
$(S.T) \odd U = U \odd (S.T) = (U.S) \odd T$
with the previous calculation.
\finrep
\finexe

We define in the same way the triple objective contraction (apply the simple contraction three times consecutively).
\Eg, with~\eref{eqToddS3} we get
\be
S \otd T = \sum_{i,j,k=1}^n S^i{}_{jk} T^{kj}{}_i.
\ee
(Einstein's convention is satisfied.)

%%%%%%%%%%%%%%%%%%%%%%%%%%%%%%%%%%%%%%%%%%%%%%%%%%%%%%%%%%%%%%%%%%%%%%%%%%%%%%%%%%%

\subsubsection{Non objective double contraction: Double matrix contraction}
\label{secnoct}

The engineers often use the double matrix contraction of second order tensors defined by (term to term multiplication):
If $S=[S_{ij}]=[S^i_j]$ and $T=[T_{ij}]=[T^i_j]$ then
\be
\label{eqS2pT}
S:T := \sumijn S_{ij}T_{ij} = \sumijn S^i_jT^i_j \eqnote \Tr(S.T^T).
\ee
Einstein's convention is \textslbf{not} satisfied, and the result is observer dependent for associated endomorphism:
%(in particular the transpose $T^T$ depends on some inner dot product to be defined, and the writing $\Tr(S.T^T)$ corresponds to a matrix multiplication). 

\debexa
\label{exaissuedd}
%{\bf Issue:}
Let $(\ve_i)$ be a basis, let $S\in\calL(E;E)$ given by $[S]_{\ve} = \pmatrix{ 0 & 4 \cr 2 & 0}$ 
(so $S.\ve_1=2 \ve_2$ and $S.\ve_2=4\ve_1$). Then the double matrix contraction~\eref{eqS2pT} gives
\be
\label{eqS2pT2}
S:S =[S]_{\ve}:[S]_{\ve} = 4*4+2*2=20.
\ee
Change of basis: let $\vb_1=\ve_1$ and $\vb_2=2\ve_2$.
The transition matrix from~$(\ve_i)$ to~$(\vb_i)$ is $P=\pmatrix{ 1 & 0 \cr 0 & 2}$.
Thus $[S]_{\vb}
= P^{-1}.[S]_{\ve} . P
= \pmatrix{ 1 & 0 \cr 0 & {1\over 2}}.\pmatrix{0 & 8 \cr 2 & 0}
=\pmatrix{0 & 8 \cr 1 & 0}$. Thus
\be
\label{eqS2pT3}
S:S =[S]_{\vb}:[S]_{\vb}= 8*8 + 1*1 = 65 \ne 20.
\ee
To be compared with the double objective contraction:
$[S]_{\ve}\odd [S]_{\ve}=4*2+2*4= 16 = [S]_{\vb}\odd [S]_{\vb} = S\odd S $ (observer independent result = objective result).

So it is absurd to use $S:S$ (double matrix contraction) if you need objectivity:
Recall that the foot is the international vertical unit in aviation, and thus the use of the double \textbf{objective} contraction is vital, while the use of the double matrix contraction can be fatal (really).
Also see the Mars climate orbiter probe crash.
\finexa

\comment{
\debcor
Let $S,T\in\calL^1_1(E)$.
The double objective contraction $S\odd T$ is invariant, \cf~prop.~\ref{propSoddT}, while
the double matrix contraction~\eref{eqS2pT} is \underline{\bf not} invariant:
It depends on the choice of the basis (quite annoying), see \eref{eqS2pT2}-\eref{eqS2pT3}.
(Hence the double objective contraction $S\odd T$ is suitable to get an objective virtual power principle.)
\fincor
}

\debexe
Let $S \in \calL^0_2(E)$ (\eg\ a metric), let $(\va_i)$ be a Euclidean basis in \foot,
and let $(\vb_i) = (\lambda\va_i)$ be the related euclidean basis in \metre\ (change of unit).
Give $[S]_{|\va}:[S]_{|\va}$ and $[S]_{|\vb}:[S]_{|\vb}$ and compare.
(The simple and double objective contractions are impossible here since $S$ and $T$ are not compatible.)

\debrep
Let $S = \sumijn S_{a,ij} a^i \otimes a^j = \sumijn S_{b,ij} b^i \otimes b^j$.
Since $(\vb_i) = (\lambda\va_i)$ we have $b^i = {1\over \lambda} a^i$.
Thus $\sumijn S_{a,ij} a^i \otimes a^j =  \sumijn S_{a,ij}\lambda^2 b^i \otimes b^j$,
thus $ \lambda^2 S_{a,ij} = S_{b,ij}$.
Thus
\be
[S]_{|\vb}:[S]_{|\vb} = \sumijn (S_{b,ij})^2
= \lambda^4 \sumijn (S_{a,ij})^2 = \lambda^4 [S]_{|\va}:[S]_{|\va},
\ee
with $\lambda^4 \ge 100$: Quite a difference isn't it?
\finrep
\finexe

\comment{
\debexe
Prove: $S:T$ is left unchanged for a change of orthonormal basis from $(\ve_i)$ to $(\vb_i)$.

\debrep
Here the transition matrix $P$ from $(\ve_i)$ to~$(\vb_i)$ is orthonormal, \ie\ $P^T=P^{-1}$, thus
 $P.P^T=I$
and $Q^T.Q=I$ where $Q=P^{-1}$. %, \ie\ $\sumkn P^i_kP^j_k=\delta^{ij}$ for all $i,j$.

And $[S]_{|\vb} = Q.[S]_{|\ve}.P
= [\sum_{\alpha,\beta=1}^n Q^i_\alpha S^\alpha_{e,\beta} P^\beta_j]_{i=1,...,n \atop j=1,...,n}$,
and $[T]_{|\vb} = Q.[T]_{|\ve}.P 
= [\sum_{\gamma,\delta=1}^n Q^i_\gamma T^\gamma{e,\delta} P^\delta_j]_{i=1,...,n \atop j=1,...,n}$.
Thus 
$\ds [S]_{|\vb}:[T]_{|\vb}
= \sum_{ij}
(\sum_{\alpha\beta}Q^i_\alpha S^\alpha_{e,\beta} P^\beta_j)
(\sum_{\gamma\delta}Q^i_\gamma T^\gamma_{e,\delta} P^\delta_j)
= \sum_{ij\alpha\beta\gamma\delta}
Q^i_\alpha Q^i_\gamma P^\beta_j P^\delta_j S^\alpha_{e,\beta} T^\gamma_{e,\delta}
= \sum_{ij\alpha\beta\gamma\delta}
(Q^T)^\alpha_i Q^i_\gamma P^\delta_j (P^T)^j_\beta S^\alpha_{e,\beta} T^\gamma_{e,\delta}
= \sum_{\alpha\beta\gamma\delta}
\delta^\alpha_\gamma \delta^\delta_\beta S^\alpha_{e,\beta} T^\gamma_{e,\delta}
= \sum_{\alpha\beta}S^\alpha_{e,\beta} T^\alpha_{e,\beta}
=[S]_{|\ve}:[T]_{|\ve}
$.
\finrep
\finexe
}

\comment{
The following definition may also be found:
\be
S:T = \sumijn S_{ij}T^{ij}.
\ee
This one can be justified if another double contraction is defined by applying the
simple contraction twice with jumps.
However, their usual definition explains that $S:T$ is defined by $S:T := S\odd T^T$,
\ie\ they need the transpose. But if they consider $T$ as linear map, then
the transposed needs an inner dot product to be defined, and the use of an inner dot product leads to the loss of objectivity.
(See~\S~\ref{secsdce} and~\eref{eqJdL4} for endomorphisms.)
(See~\S~\ref{secdmcno} for double matrix contraction.)
}

%\medskip

\comment{
%%%%%%%%%%%%%%%%%%%%%%%%%%%%%%%%%%%%%%%%%%%%%%%%%%%%%%%%%%%%%%%%%%%%%%%%%%%%%%%%%%%

\subsection{Endomorphism and tensorial notation}

\subsubsection{Endomorphism identified to a 1 1 uniform tensor}
\label{secicetu2}

We have the natural canonical isomorphism, \cf~\eref{propeqtJ2f},
\be
\label{eqtJ1_e}
\calJ_2 :
\left\{\eqalign{
\calL(E;E) & \rar \calL(\Es,E;\RR) \cr
L & \rar  T_L = \calJ_2(L)
}\right\}, \qwith \forall (\vv,m)\in E\times E^*,\quad T_L(m,\vv) \eqdef m.L.\vv.
\ee
Thus, for direct calculations (and for nothing else, in particular not with transposed!), we can write
\be
L  \eqnote T_L \quad (=\calJ_2(L)).
\ee
Details: Let $(\ve_i)$ be a basis and $(e^i)$ be its dual basis.
Let $L \in \calL(E;E)$ (an endomorphism), $[L^i_j]=[L]_{|\ve}$, i.e.
\be
\label{eqdefptc12}
L .\ve_j = \sumijn L^i_j \ve_i, \qie
\ve_i.L.\ve_j = L^i_j \mope^{\eref{eqtJ1_e}} T_L(e^i,\ve_j).
\ee
Thus
\be
\label{eqTuuuL}
T_L = \sumijn L^i_j \ve_i \otimes e^j \eqnote L  \quad (=\calJ_2(L)).
\ee
Thus, if $\vw\in E$ and $\vw=\sumjn w^j\ve_j$, then (computation)
\be
\label{eqTuuuL2}
L.\vw = T_L.\vw \mope^{\hbox{\footnotesize contraction}} \sumijn L^i_j w^j \ve_i,\qand [L.\vw]_{\ve} = [L]_{\ve}.[\vw]_{\ve} .
\ee
Indeed, 
$(\sumijn L^i_j \ve_i \otimes e^j).(\sumkn w^k\ve_k)
= \sumijn L^i_jw^k \ve_i (e^j.\ve_k)
= \sumijn L^i_jw^k \ve_i \delta^j_k = \sumijn L^i_j w^j \ve_i$.

\comment{
\debexe
Soit $L$ un endomorphisme sur~$E$. Soit $(\va_i)$ une base de~$E$, soit $\lambda>0$,
et soit $(\vb_i) = (\lambda \va_i)$ (base~$E$).
Soit $M=[L]_{|\va}$ et $N=[L]_{|\vb}$. Montrer $M=N$.

\debrep
$N=P^{-1}.M.P$ avec $P=\lambda I$. Cqfd.
\finrep
\finexe
}

%%%%%%%%%%%%%%%%%%%%%%%%%%%%%%%%%%%%%%%%%%%%%%%%%%%%%%%%%%%%%%%%%%%%%%%%%%%%%%%%%%%

\subsubsection{Simple and double objective contractions of endomorphisms}
\label{secsdce}

The simple contraction of $L\in\calL(E;E)$ and $M\in\calL(E;E)$
is the composed endomorphism 
\be
L{\circ} M \eqnote L.M \quad(\hbox{called a simple contraction}).
\ee
So, if $L.\ve_j = \sumin L^i_j \ve_i$ and $M.\ve_j = \sumin M^i_j \ve_i$ then
$(L.M).\ve_j := (L\circ M).\ve_j = L(M.\ve_j)
=L.\sumkn M^k_j\ve_k
= \sumikn L^i_k M^k_j \ve_i$, \ie\ $[L.M]_{|\ve} = [L]_{|\ve}.[M]_{|\ve} = [L \circ M]_{|\ve}$.

And the double objective contraction of endomorphisms is % with~\eref{eqdefSoddT}:
\be
\label{eqJdL4}
L\odd M := \Tr(L.M) \quad (=\Tr(L \circ M)).
\ee

Remark: For computations purposes with a chosen basis $(\ve_i)$ in~$E$,
with $T_L=\calJ_2(L) = \sumikn L^i_k \ve_i \otimes e^k$ and $T_M=\calJ_2(M)= \sumjkn M^k_j \ve_k \otimes e^j$, \cf~\eref{eqtJ1_e}, then $L.M$ is also the notation of the contraction $T_L.T_M$:
\be
T_L.T_M \mope^{\eref{eqST}} \sumijkn L^i_kM^k_j \ve_i \otimes e^j = T_{L\circ M},\qand
[T_L.T_M]_{|\ve} = [L]_{|\ve}.[M]_{|\ve} = [L.M]_{|\ve},
\ee
and
\be
T_L \odd T_M = \sumijn L^i_jM^j_i = \Tr(L.M) = L\odd M .
\ee
(Objective, and the Einstein's convention is satisfied.)

}

%%%%%%%%%%%%%%%%%%%%%%%%%%%%%%%%%%%%%%%%%%%%%%%%%%%%%%%%%%%%%%%%%%%%%%%%%%%%%%%%%%%

\subsection{Kronecker (contraction) tensor, trace}
\label{seckroten}
%%%%%%%%%%%%%%%%%%%%%%%%%%%%%%%%%%%%%%%%%%%%%%%%%%%%%%%%%%%%%%%%%%%%%%%%%%%%%%%%%%%

%\subsubsection{Kronecker contraction tensor}

\debdef
\label{defkro}
The Kronecker tensor is the ${1\choose1}$ uniform tensor $\uu\delta \in \calL^1_1(E)$ defined by
\be
\forall (\ell,\vu) \in \Es \times E,\quad \uu\delta(\ell,\vu) \eqdef \ell.\vu.
\ee
And the Kronecker symbols relative to a basis $(\ve_i)$ are the reals defined by, calling $(\pi_{ei})$ the dual basis,
\be
\label{eqexpdb0}
\delta_{ij} := \delta(\pi_{ei},\ve_j)
= \left\{\eqalign{
& 1 \;\hbox{if } i=j, \cr
& 0 \;\hbox{if } i \ne j, \cr
}\right\}
\qie \uu\delta := \sum_{i=1}^n \pi_{ei}\otimes e^i,\quad [\uu\delta]=[\delta_{j}]=[I]
\ee
(identity matrix whatever the basis).
Duality notations: $\delta^i_j := \delta(e^i,\ve_j)$, $\uu\delta := \sum_{i=1}^n \ve_i\otimes e^i$ and $[\uu\delta]=[\delta^i_j]$.
\findef

\debdef
The trace of a ${1\choose1}$ uniform tensor $T \in \calL^1_1(E)$ is
\be
\tilde\Tr(T) = \uudelta \odd T \quad (=\Tr(L_T))
\ee
(with the natural canonical isomorphism $T\in\calL^1_1(E) \simeq L_T\in\calL(E;E)$ given by $T(\ell,\vv) := \ell.L_T.\vv$).
\findef

Thus $\tilde\Tr(T) %= (\sumijn \delta^i_j\ve_i \otimes e^j)\odd (\sumjkn T^k{}_\ell \ve_k \otimes e^\ell)
=\sumin T^i{}_i$. 

In particular $\tilde\Tr(\uudelta)=n$,
and $\tilde\Tr(\vv\otimes \ell) = \sum_i v^i\ell_i = \ell.\vv$ when
$\vv=\sum_i v^i\ve_i$ and $\ell= \sum_j \ell_j e^j$.

%%%%%%%%%%%%%%%%%%%%%%%%%%%%%%%%%%%%%%%%%%%%%%%%%%%%%%%%%%%%%%%%%%%%%%%%%%%%%%%%%%%

\section{Tensors in $\Trsu$}
\label{secrtbase}

%%%%%%%%%%%%%%%%%%%%%%%%%%%%%%%%%%%%%%%%%%%%%%%%%%%%%%%%%%%%%%%%%%%%%%%%%%%%%%%%%%%
%\setcounter{subsubsection}{-1}

\subsection{Introduction, module, derivation}
\label{secmodule}

\def\CIU{C^\infty(U;\RR)}
\def\CuU{C^1(U;\RR)}

Let $A$ and $B$ be any sets, and let $\calF(A;B)$ be the set of functions $A\rar B$.
The ``plus'' inner operation and the ``dot'' outer operation are defined by,
for all $f,g\in \calF(A;B)$, all $\lambda\in\RR$ and all $p\in A$,
\be
\label{eqfpg}
\left\{\eqalign{
&(f+g)(p) \eqdef f(p) + g(p),\qand\cr
&(\lambda .f)(p) \eqdef \lambda \,f(p), \quad
\lambda .f \eqnote \lambda f.
}\right.
\ee
$( \calF(A;B),+,.,\RR)$ is thus a vector space on the field~$\RR$ (see any elementary course) called $ \calF(A;B)$.

But the field~$\RR$ is ``too small'' to define a tensor which can be seen as ``a linear tool that satisfies the change of coordinate system rules'': %\Eg, the derivation $d$ is linear, but does not satisfies the rules:

\debexa
\label{remunpeuplus}
{\bf Fundamental counter-example: Derivation.} %(the $\RR$-linearity is not sufficient). 
Let $U$ be an open set in~$\RRn$.
The derivation $d:\vw \in C^1(U;\vRRn) \rar d\vw \in C^0(U;\calL(\vRRn;\vRRn))$ is $\RR$-linear: In particular $d(\lambda \vw) = \lambda (d\vw)$ for all $\lambda\in\RR$... 

...but $d$ doesn't satisfy the change of coordinate system rules, %(because of the term $\sum_{\mu\nu} Q^i_\lambda P^\mu_j (dP^\lambda_k.\va_\mu)$ i
see~\eref{exochrist2}.
%(the ``change of coordinate system formulas'' result from an algebraic calculation, while a derivation calculation is part of functional analysis).

So a derivation it \textslbf{not} a tensor (it is a ``spray'', see Abraham--Marsden~\cite{abraham-marsden}).

In fact, one requirement for $T$ to be a tensor is, \eg\ with $T=\vw$ a vector field: %:\Gamma(U)\rar \RR$:
For all $\phi\in \CIU$, and all $\vw\in\Gamma(U)$ ($C^\infty$-vector field),
\be
\label{eqflinearity}
T(\phi\vw) = \phi\,T(\vw).
\ee
While
\be
d(\phi\vw) \ne \phi\,d(\vw), \quad \hbox{because} \quad d(\phi\vw) = \phi\,d\vw + d\phi.\vw.
\ee
Thus the elementary $\RR$-linearity requirement ``$T.(\lambda \vw) = \lambda (T.\vw)$ for all $\lambda\in\RR$
is not sufficient to characterize a tensor: The $\RR$-linearity has to be replaced by the $\CIU$-linearity, \cf~\eref{eqflinearity}.

Thus we will have to replace a real vector space $(V,+,.,\RR)$ over the field~$\RR$ with the ``module'' $(V,+,.,\CIU)$ over the ring~$\CIU$, which mainly amounts to consider~\eref{eqfpg} for all $\lambda=\phi\in\CIU$.
Remark: The use of a module is very similar to the use of a vector space, but for the use of the inverse:
all real $\lambda\ne0$ has a multiplicative inverse in~$\RR$ (namely ${1\over \lambda}$), but a function $f\in \CIU$
\st\ ``$f\ne0$ and $f$ vanishes at one point'' doesn't have a multiplicative inverse in~$\CIU$.
\finexa

\comment{
{\bf So:} we will replace the vector space $(C^1(A;B),+,.,\RR)$ built with the field~$\RR$
(where $(\lambda .T)(p):=\lambda T(p)$ for all $\lambda\in\RR$) 
by the ``module'' $(C^1(A;B),+,.,\CIU)$ build with the ring $\CIU$
(where $(f.T)(p):=f(p)T(p)$ for all $f\in \CIU$. 

}

%%%%%%%%%%%%%%%%%%%%%%%%%%%%%%%%%%%%%%%%%%%%%%%%%%%%%%%%%%%%%%%%%%%%%%%%%%%%%%%%%%%

\subsection{Field of functions and vector fields}
\label{seccfcv}

{\bf Framework of classical mechanics:}
$U$ is an open set in an affine space~$\calE$ which  associated vector is~$E$.
And the definition of tensors is done at a fixed time~$t$ (concerns the space variables).
As before, the approach is first qualitative, then quantitative with a basis $(\ve_i(p))$
and its dual basis $(\piei(p))=(e^i(p))$,
at any $p\in\calE$.
%(More generally $U$ is an open set in a differentiable manifold).

%%%%%%%%%%%%%%%%%%%%%%%%%%%%%%%%%%%%%%%%%%%%%%%%%%%%%%%%%%%%%%%%%%%%%%%%%%%%%%%%%%%

\subsubsection{Field of functions}

\def\tf{{\tilde f}}
\def\tg{{\tilde g}}

Let $f\in \CIU$ be a  function. %(sufficiently) regular function (for mechanical calculations).
The associated function field is
\be
\label{eqcf0}
\tf :
\left\{\eqalign{
U &\rar U\times \RR \cr
p & \rar \tf(p) := (p;f(p)),
}\right.
\ee
and $p$ is called the base point.
So $\Im \tf = \{(p;f(p)) : p\in U\}$ is the graph of~$f$.
%(A field of functions is of Eulerian type, not Lagrangian.)
Definition:
\be
\Tzzu := \{\tf : f\in\CIU \} = \{\hbox{field of functions}\}=  \hbox{the set of ${0\choose 0}$ type tensor on~$U$},
\ee
or the set of tensors of order~$0$ on~$U$.
Abusive short notations (to lighten the writings): 
\be
%\left\{\eqalign{
 \tf(p) \eqnote f(p) , \qand 
 T^0_0(U) \eqnote  \CIU,
%}\right.
\ee
but keep the base point in mind (no ubiquity gift).

In $\Tzzu$,
the internal sum is defined by, for all $\tf,\tg\in \Tzzu$ with $\tf(p) = (p;f(p))$ and $\tg(p) = (p;g(p))$,
\be
%\label{eqram}
(\tf+\tg)(p) := (p;(f+g)(p)) \quad (= (p;f(p)+g(p))),
\ee
and the external multiplication on the ring~$ \CIU$ is defined by, for all $\phi\in \CIU$,
\be
\label{eqram}
(\phi\tf)(p) := (p; (\phi f)(p)) \quad ( = (p; \phi(p) f(p)))
\ee
(the base point~$p$ remains unchanged).
%\eref{eqram}~models the actual computation made by an observer located at~$p$ (no gift of ubiquity).
Thus $(\Tzzu,+,.)$ is a module over the ring~$\CIU$.

%%%%%%%%%%%%%%%%%%%%%%%%%%%%%%%%%%%%%%%%%%%%%%%%%%%%%%%%%%%%%%%%%%%%%%%%%%%%%%%%%%%

\subsubsection{Vector fields}

\def\tvw{{\tilde\vw}}
Let $\vw\in C^\infty(U,E)$ be a vector valued function
(at least Lipschitzian, to get integral curves, \cf~Cauchy--Lipschitz theorem).
The associated vector field is
\be
\label{eqdefcvg}
\tvw :
\left\{\eqalign{
U &\rar U\times E \cr
p & \rar \tvw(p) = (p;\vw(p)).
}\right.
\ee
So $\Im \tvw = \{(p;\vw(p)) : p\in U\}$ is the graph of~$\vw$,
and the definition of~$\tvw$ tells that the vector $\vw(p)$ has to be drawn at~$p$ (the base point).
%(A vector field is of Eulerian type, not Lagrangian.)
Abusive short notation:
\be
\tvw(p) \eqnote \vw(p)\quad\hbox{instead of }\tvw(p) = (p;\vw(p)).
\ee
It lightens the notations, but keep the base point in mind.
Let
\be
\label{eqdefGU}
\Gamma(U) := \hbox{ the set of vector fields on~$U$}.
\ee

\medskip
\noindent
{\bf More precisely,} we will use the following full definition of vector fields
(see \eg\ Abraham--Marsden~\cite{abraham-marsden}):
A~vector field is built from tangent vectors to curves.
It makes sense on non planar surfaces, and more generally on differential manifolds.

%\Eg, see~\S~\ref{secsa} for a fundamental example in mechanics.

%Avec une base $(\ve_i(p))$ en~$p$ on notera $\vw(p) = \sumin w^i(p)\ve_i(p)$ et $\vw = \sumin w^i\ve_i$.

%%%%%%%%%%%%%%%%%%%%%%%%%%%%%%%%%%%%%%%%%%%%%%%%%%%%%%%%%%%%%%%%%%%%%%%%%%%%%%%%%%%

\subsection{Differential forms}

%%%%%%%%%%%%%%%%%%%%%%%%%%%%%%%%%%%%%%%%%%%%%%%%%%%%%%%%%%%%%%%%%%%%%%%%%%%%%%%%%%%

\comment{
\subsubsection{Exact differential forms}

If $f:U\subset E\rar\RR$ is~$C^1$, then its differential $df : U \rar \Es$ is called an ``exact differential form''
(example of the density of the internal energy in thermodynamics).
Recall: If $p\in U$ then $df(p)\in \Es = \calL(E;\RR)$ is the linear form defined on~$E$ by,
for all $\vu\in E$, $df(p).\vu := \lim_{h\rar0} {f(p+h\vu) - f(p) \over h}$.
%(And if $U$ is a non planar surface, then  $df(p).\vu := \lim_{h\rar0} {f(c_p(h)) - f(c_p(0)) \over h}$ where $c_p: h\rar c_p(h)\in U$ is a regular curve \st\ $c_p(0) = p$ and $c_p{}'(h)=\vu$; If $U$ were planar: $c_p(h)=p+h\vu + o(h)$ and $c_p(h)=p+h\vu$ can be chosen.)

An ``exact differential form'' is a particular case of a ``differential form'' (example of the elementary heat or work in thermodynamics):
}

%%%%%%%%%%%%%%%%%%%%%%%%%%%%%%%%%%%%%%%%%%%%%%%%%%%%%%%%%%%%%%%%%%%%%%%%%%%%%%%%%%%

%\subsubsection{Differential forms}

\def\talpha{{\tilde\alpha}}

The basic concept is that of vector fields.
A first over-layer is made of differential forms (which ``measure vector fields''): % which are ``functions defined on vector fields \st'':

\debdef
Let $\alpha
\left\{\eqalign{
U & \rar E^* \cr
p & \rar \alpha(p)\cr
}\right\}
$ 
(so $\alpha(p)$ is a linear form at~$p$).
The associated differential form (also called a $1$-form) is ``the field of linear forms'' defined by
\be
\label{eqchampfd}
\talpha :
\left\{\eqalign{
U &\rar U\times \Es \cr
p & \rar \talpha(p) = (p;\alpha(p)) \quad\hbox{( = ``a pointed linear form at $p$'')}.
}\right.
\ee
And $p$ is called the base point, and $\Im \talpha = \{(p;\alpha(p)) : p\in U\}$ is the graph of~$\alpha$.
\findef

Thus, if $\talpha \in \Omega^1(U)$ (differential form) and $\tvw\in\Gamma(U)$ (vector field), then $\talpha.\tvw  \in \Tzzu$ (field of scalar valued functions) satisfies
\be
\talpha.\tvw:
\left\{\eqalign{
U & \rar U\times \RR \cr
p& \rar (\talpha.\tvw)(p) = (p;(\alpha.\vw)(p)) = (p;\alpha(p).\vw(p)) \;\in U \times \RR.
}\right.
\ee
Short notation:
\be
\talpha(p) \eqnote \alpha(p),\quad\hbox{instead of } \talpha(p) = (p;\alpha(p)),
\ee
but keep the base point in mind. And
\be
\Omega^1(U) := \hbox{ the set of differential forms~$U$}.
\ee

\comment{
\debrem
Thermodynamics: Let $U$ be the internal energy. Then its differential $dU$ is a (exact) differential form (first principle of thermodynamic).
The elementary work $w= \delta W$ is a differential form which is not exact in general,
\ie\ it doesn't derive from a potential in general (\eg\ because of friction losses),
and the loss $q:= dU - \delta W \eqnote \delta Q $ is called the elementary heat, thus $q= \delta Q$ is a not an exact differential form either.
%However their sum is exact:  $ \delta W+\delta Q = dU$ = first principle of thermodynamic.
\finrem
}

\comment{
%%%%%%%%%%%%%%%%%%%%%%%%%%%%%%%%%%%%%%%%%%%%%%%%%%%%%%%%%%%%%%%%%%%%%%%%%%%%%%%%%%%

\subsubsection{Covariance and contravariance}

\debdef
A vector field $\vw$ is said to be contravariant.

A differential form $\alpha$ (a $1$-form),
which is a function acting on the vector fields~$\vw$, is said to be covariant.
\findef

(See Misner, Thorne, Wheeler~\cite{misner-thorne-wheeler} box~2.1:
``Without it [the distinction between covariance and contravariance],
one cannot know whether a vector is meant or the very different () object that is a 1-form.'')
}

%%%%%%%%%%%%%%%%%%%%%%%%%%%%%%%%%%%%%%%%%%%%%%%%%%%%%%%%%%%%%%%%%%%%%%%%%%%%%%%%%%%

\subsection{Tensors}
\label{sectrs}

\def\tT{{\tilde T}}

A second over-layer is introduced with the tensors with are ``functions defined on vector fields and on differential forms'' (which ``measure vector fields and differential forms'').

Let $r,s\in\NN$, $r{+}s\ge 1$, and let $T :
\left\{\eqalign{
U & \rar \calL^r_s(E) \cr
p & \rar T(p)
}\right\}
$
(so~$T(p)$ is a~uniform ${r\choose s}$ tensor for each~$p$, \cf~\eref{secrsu}).
And consider the associated function
\be
\label{eqdeftT}
\tT :
\left\{\eqalign{
U &\rar U \times \calL^r_s(E) \cr
p & \rar \tT(p)=(p;T(p))
}\right.
\ee
Abusive short notation:
\be
\tT(p) \eqnote T(p)\quad\hbox{instead of } \tT(p) = (p;T(p)),
\ee
but keep the base point in mind.

\debdef
\label{deftT} 
(Abraham--Marsden~\cite{abraham-marsden}.)
$\tT$ is a tensor of type~${r\choose s}$ iff $T$ is $\CIU$-multilinear (not only $\RR$-multilinear), \ie,
for all $f \in \CIU$, all $z_1,z_2$ vector field or differentiable form where applicable, and all $p\in U$,
\be
\label{eqdeftT1}
\left\{\eqalign{
& T(p)(...,z_1(p)+z_2(p),...) = T(p)(...,z_1(p),...) + T(p)(...,z_2(p),...), \qand \cr
& T(p)(...,f(p)z_1(p),...) = f(p)\,T(p)(...,z_1(p),...), \cr
}\right.
% \hbox{ et}\cr & T(p)(...,f(p)z(p),...) =  f(p)T(p)(...,z(p),...),
\ee
written in short
\be
\label{eqdeftT2}
\left\{\eqalign{
& T(...,z_1+z_2,...) = T(...,z_1,...) + T(...,z_2,...), \qand \cr
& T(...,fz_1,...) = f\,T(...,z_1,...). \cr
}\right.
\ee
\findef

And
\be
\Trsu := \hbox{ the set of ${r\choose s}$ type tensors on~$U$}.
\ee
(Recall: $\ds \Tzzu := \CIU$ the set of function fields, \cf~\eref{eqcf0}.)

%(In manifolds  we have $T(p)\in \calL^r_s(T_pU)$ and $\tT : U \rar \bigcup_{p\in\Omega} (\{p\} \times \calL^r_s(T_pU))$.)

\debrem
Definition in differential geometry lessons: A tensor is a section of a certain bundle over a manifold.
For classical mechanics, definition~\ref{deftT} gives an equivalent definition.
\finrem

\comment{
\debrem
This definition of tensors (or tensor fields) enables to exclude the derivation operators which are
not tensors, \cf~example~\ref{remunpeuplus}.
\finrem
}

%%%%%%%%%%%%%%%%%%%%%%%%%%%%%%%%%%%%%%%%%%%%%%%%%%%%%%%%%%%%%%%%%%%%%%%%%%%%%%%%%%%

\subsection{First Examples}
%%%%%%%%%%%%%%%%%%%%%%%%%%%%%%%%%%%%%%%%%%%%%%%%%%%%%%%%%%%%%%%%%%%%%%%%%%%%%%%%%%%

\subsubsection{Type ${0\choose1}$ tensor = differential forms}
\label{sectzu}

If $T\in\Tzuu$ then $T(p)\in \Es$, so $T=\alpha\in\Omega^1(U)$ is a differential form: $\Tzuu \subset\Omega^1(U)$.

Converse: Does a differential form $\alpha\in\Omega^1(U)$ defines a ${0\choose1}$ type tensor on~$U$?
Yes: We have to check~\eref{eqdeftT1}, which is trivial.
So $\alpha\in\Tzuu$, so $\Omega^1(U) \subset \Tzuu$. 

Thus
\be
\label{eqchampfd2}
\Tzuu = \Omega^1(U).
\ee
%Et dans une base on notera $T = \sumin T_i e^i$.

%%%%%%%%%%%%%%%%%%%%%%%%%%%%%%%%%%%%%%%%%%%%%%%%%%%%%%%%%%%%%%%%%%%%%%%%%%%%%%%%%%%

\subsubsection{Type ${1\choose0}$ tensor (identified to a vector field)}
\label{sectuz}

Let $T\in\Tzuu$,
so $T(p)\in \calL^1_0(E) = \calL(E^*;\RR)=\Ess$ for all $p\in U$.
Thus, thanks to the natural canonical isomorphism $\Ess\simeq E$, $T(p)$ can be identified to a vector,
thus $\Tzuu \subset \Gamma(U)$.

Converse: Does a vector field $\vw\in\Gamma(U)$ defines a ${1\choose0}$ type tensor on~$U$?
Yes: We have to check~\eref{eqdeftT1}, which is trivial.
So $\Gamma(U) \subset \Tuzu$. 

Thus
\be
\label{eqwevw}
\Tuzu \simeq \Gamma(U).
\ee
%Et dans une base on notera $T = \sumin T^i \ve_i$.

%%%%%%%%%%%%%%%%%%%%%%%%%%%%%%%%%%%%%%%%%%%%%%%%%%%%%%%%%%%%%%%%%%%%%%%%%%%%%%%%%%%

\subsubsection{A metric is a ${0\choose2}$ tensor}

Let $T \in \Tzdu$,
so $T(p)\in \calL^0_2(E)$ for all $p\in U$,
and $T(\vu,\vw)\in\Tzzu$ for all $\vu,\vw\in\Gamma(U)$.

\comment{
Et, avec une base $(\ve_i(p))$ en~$p$, on notera $T(p) = \sumijn T_{ij}(p) e^i(p) \otimes e^j(p)$
et $T = \sumijn T_{ij} e^i \otimes e^j$.
Ainsi $T(\vv,\vw) = \sumijn T_{ij}v^iw^i$ quand $\vv= \sumin v^i\ve_i$ et $\vw= \sumin w^i\ve_i$, et la convention d'Einstein est respectée.
}

\debdef
A metric $g$ on~$U$ is a ${0\choose2}$ type tensor on~$U$ such that, for all $p \in E$,
$g(p) \eqnote g_p$ is an inner dot product on~$E$.
\findef

%\Eg, $C^\flat$ is a metric, which is \st\ $C^\flat(p)$ is not a Euclidean dot product in general. \Eg, a Riemannian metric~$g$ is \st\ $g(p)$ is a Euclidean dot product at each~$p$.

%%%%%%%%%%%%%%%%%%%%%%%%%%%%%%%%%%%%%%%%%%%%%%%%%%%%%%%%%%%%%%%%%%%%%%%%%%%%%%%%%%%

\subsection{${1\choose1}$ tensor, identification with fields of endomorphisms}
\label{eqicet}

\def\tL{{\tilde L}}

Let $T \in \Tuuu$,
so $T(p)\in \calL^1_1(E)$ for all $p\in U$,
and $T(\alpha,\vw) \in \Tzzu$ for all $\alpha\in\Omega^1(U)$ and $\vw\in\Gamma(U)$
(so $T(p)(\alpha(p),\vw(p)) \in \RR$ for all~$p$).

The associated field of endomorphisms on~$U$ is
$
\tL_T :
\left\{\eqalign{
U &\rar U \times \calL(E;E) \cr
p & \rar \tL_T(p) = (p,L_T(p))
}\right\}
$
where $L_T(p)$ is identified with $T(p)$ thanks to the natural canonical isomorphism
$\calL(E;E) \simeq \calL(E^*,E;\RR)=\calL^1_1(E)$
given by
\be
\label{eqLidT}
\forall \ell\in \Es,\; \forall \vw\in E,\; \quad
\ell.(L_T(p).\vw) = T(p)(\ell,\vw).
\ee

\comment{
Conversely, with the natural canonical isomorphism (in fact with its inverse),
a field of endomorphisms can be written as a ${1\choose1}$ tensor.

\debexa
%\label{exadvwce}
If $\vw$ be a $C^1$ vector field on~$U$, then $d\vw$ is a field of endomorphisms on~$U$
(we have  $d\vw(p)\in\calL(E;E)$ for all $p\in U$). And $d\vw$ can be written as a ${1\choose1}$ tensor.
So, with a basis $(\ve_i)$ and with $w^i_{|j} := e^i.d\vw.\ve_j$, we have $d\vw.\ve_j= \sumin w^i_{|j}\ve_i$,
and we can can write
$d\vw \simeq T_{d\vw} = \sumijn w^i_{|j} \ve_i \otimes e^j$, for calculations purpose with the contraction rule~\eref{eqobjcL}.
\finexa
}

\comment{
%%%%%%%%%%%%%%%%%%%%%%%%%%%%%%%%%%%%%%%%%%%%%%%%%%%%%%%%%%%%%%%%%%%%%%%%%%%%%%%%%%%

\subsection{Example: Type ${2\choose0}$ tensor...}

Same steps to define ${2\choose0}$ tensors, or more generally ${r\choose s}$ tensors.
}

%%%%%%%%%%%%%%%%%%%%%%%%%%%%%%%%%%%%%%%%%%%%%%%%%%%%%%%%%%%%%%%%%%%%%%%%%%%%%%%%%%%

\subsection{Unstationary tensor}

Let $t\in[t_1,t_2] \subset \RR$.
Let $(T_t)_{t\in[t_1,t_2]}$ be a family of ${r \choose s}$ tensors, \cf~\eref{eqdeftT}.
Then $T : t \rar T(t):=T_t$ is~called an unstationary tensor.
And the set of unstationary tensors is also noted $\Trsu$.
\Eg, a Eulerian velocity field is a ${1\choose 0}$ unstationary vector field.

%%%%%%%%%%%%%%%%%%%%%%%%%%%%%%%%%%%%%%%%%%%%%%%%%%%%%%%%%%%%%%%%%%%%%%%%%%%%%%%%%%%
%%%%%%%%%%%%%%%%%%%%%%%%%%%%%%%%%%%%%%%%%%%%%%%%%%%%%%%%%%%%%%%%%%%%%%%%%%%%%%%%%%%

\section{Differential, its eventual gradients, divergences}

%%%%%%%%%%%%%%%%%%%%%%%%%%%%%%%%%%%%%%%%%%%%%%%%%%%%%%%%%%%%%%%%%%%%%%%%%%%%%%%%%%%

\subsection{Differential}
\label{seccontree1}

\def\aff{{a\!f\!\!f}}
\def\UE{U}
\def\pE{{p}}
\def\qE{{q}}

The definition of the differential of a function is observer independent: All observers have the same definition
(qualitative: no man made tool required, like a basis or an inner dot product).

%%%%%%%%%%%%%%%%%%%%%%%%%%%%%%%%%%%%%%%%%%%%%%%%%%%%%%%%%%%%%%%%%%%%%%%%%%%%%%%%%%%

\subsubsection{Framework}

Classical Framework: $\calE$ are $\calF$ affine spaces associated with vector spaces $E$ and~$F$, and $||.||_E$ and $||.||_F$ are norms in~$E$ and~$F$ such that $(E,||.||_E)$ and~$(F,||.||_F)$ are complete (we need ``limit that stay in the space as $h\rar0$'', ). 
%(If $E$ and $F$ are infinite dimensional, we suppose that , \ie\ complete normed vector spaces.)
$\UE$ is an open set in $\calE$, % and~$\calF$,
and $\Phi :
\left\{\eqalign{
\UE & \rar \calF \cr
\pE & \rar \pF=\Phi(\pE)
}\right\}
$ is a function. % (\ie, $\Phi\in \calF(E;F)$).
If applicable, $\calE$ and/or $\calF$ can be replaced by $E$ and/or~$F$. 
(The definitions can be generalized to manifolds.)
Reminder:

\debdef
Let $\pE\in\UE$. The function $\Phi$ is said to be continuous at~$p$ iff $\ds \Phi(q) \mrar_{\qE\rar p} \Phi(\pE)$
relative to the considered norms, \ie, $||\Phi(q) - \Phi(p)||_F \mrar_{||q-p||_E\rar0}0$, also written (Landau notation): Near~$p$,
\be
\Phi(\qE) = \Phi(\pE) + o(1),
\ee
called ``the zero-th order Taylor expansion of~$\Phi$ near~$p$''.
In other words:

 $\forall \eps>0$, $\exists \eta>0$ \st\
$\forall \qE\in \calE$ \st\ $||\qE-\pE||_E<\eta$ we have $||\Phi(\qE)-\Phi(\pE)||_F<\eps$.
\\
And $C^0(\UE;\calF)$ is the set of functions that are continuous at all $\pE\in\UE$.
%And $\Phi\in C^0(E;F)$ iff $\Phi$ is continuous at all $\pE\in\calE$.
\findef

%%%%%%%%%%%%%%%%%%%%%%%%%%%%%%%%%%%%%%%%%%%%%%%%%%%%%%%%%%%%%%%%%%%%%%%%%%%%%%%%%%%

\subsubsection{Directional derivative and differential (observer independent)}
\label{secddad}

Let $\pE\in\UE$, $\vu\in E$, and let $f:\RR \rar \calF$ defined by
\be
\label{eqdifff90}
f(h) :=\Phi(\pE + h\vu)
\ee

\debdef
The function $\Phi$ is differentiable at $\pE$ in the direction~$\vu$ iff $f$ is  derivable at~$0$,
\ie\ iff the limit
$f'(0)=\lim_{h\rar0}{\Phi(\pE + h\vu) - \Phi(\pE) \over h} \eqnote d\Phi(\pE)(\vu)$ exists in~$F$, \ie\ iff, near~$p$,
\be
\label{eqdifff9}
\Phi(\pE + h\vu) = \Phi(\pE) + h\,d\Phi(\pE)(\vu) + o(h),
\ee
equation called the first order Taylor expansion of~$\Phi$ at~$\pE$ in the direction~$\vu$
(it is the first order Taylor expansion of~$f$ near~$p$).

Then $d\Phi(\pE)(\vu)$ is called the directional derivative of $\Phi$ at~$\pE$ in the direction~$\vu$.

And if, for all $\vu\in E$, $d\Phi(\pE)(\vu)$ exists (in $F$) then $\Phi$ is called Gâteaux differentiable at~$\pE$. 
\findef

\debexe
Prove: If $\Phi$ is Gâteaux differentiable at~$\pE$ then $d\Phi(\pE)$ is homogeneous, \ie, $d\Phi(\pE)(\lambda\vu) = \lambda\,d\Phi(\pE)(\vu)$ for all $\vu\in E$ and all $\lambda\in\RR$.

\debrep
$ %d\Phi(\pE)(\lambda\vu) = 
\lim_{h\rar0}{\Phi(\pE + h(\lambda\vu)) - \Phi(\pE) \over h}
= \lambda \lim_{h\rar0}{\Phi(\pE + \lambda h\vu) - \Phi(\pE) \over \lambda h}
= \lambda \lim_{k\rar0}{\Phi(\pE + k\vu) - \Phi(\pE) \over k}
%= \lambda\,d\Phi(\pE)(\vu)
$.
\finrep
\finexe

\debdef
\label{defdifff9}
If $\Phi$ is Gateaux differentiable and if moreover $d\Phi(p)$ is linear and continuous at~$p$,
then $\Phi$ is said to be differentiable at~$p$ (or Fréchet differentiable at~$p$). So
\be
\Phi(\qE ) = \Phi(\pE) + h\,d\Phi(\pE).\ora{\pE\qE} + o(||\ora{\pE\qE}||_E),
\ee
since then $d\Phi(p)(\vu) \eqnote d\Phi(p). \vu$ for all $\vu\in E$ (linearity).
%; And $\sup_{||\vu||_E=1}||d\Phi(p). \vu||_F <\infty$ (continuity = bounded for a linear map).

And the affine function $\aff_{\!\!\pE}: \qE \rar  \aff_{\!\!\pE}(\qE) :=\Phi(\pE) + d\Phi(\pE).\ora{\pE\qE}$ is the affine approximation of~$\Phi$ at~$\pE$.
(So, the graph of~$\aff_{\!\!\pE}$ is the tangent plane of $\Phi$ at~$\pE$.)
\findef

\debdef
$\Phi:\UE\rar \calF$ is said to be differentiable in~$\UE$ iff $\Phi$ is differentiable at all $\pE\in\UE$.
Then its differential is the map
\be
\label{eqdefdPsi}
d\Phi :
\left\{\eqalign{
\UE & \rar \calL(E;F) \cr
\pE & \rar d\Phi(\pE).
}\right.
\ee
And $C^1(\UE;\calF)$ is the set of differentiable functions $\psi$ such that $d\Phi \in C^0(\UE;\calL(E;F))$.

And $C^2(\UE;\calF)$ is the set of differentiable functions $\psi$ such that $d\Phi \in C^1(\UE;\calL(E;F))$.

...
And $C^k(\UE;\calF)$ is the set of differentiable functions $\psi$ such that $d\Phi \in C^{k-1}(\UE;\calL(E;F))$....
\findef

\debprop
The differentiation (or derivation) operator $d: 
\left\{\eqalign{
C^1(\UE;\calF) & \rar C^0(\UE;\calL(E;F)) \cr
\Phi & \rar d\Phi \cr
}\right\}
$
is $\RR$-linear (``a derivation is linear'').
\finprop

\debdem
$d(\Phi + \lambda\Psi)(\pE).\vu
= \lim_{h\rar0}{(\Phi + \lambda\Psi)(\pE{+}h\vu) - (\Phi + \lambda\Psi)(\pE) \over h}
= \lim_{h\rar0}{\Phi(\pE{+}h\vu)-\Phi(\pE) + \lambda\Psi(\pE{+}h\vu) - \lambda\Psi(\pE) \over h}
= \lim_{h\rar0}{\Phi(\pE{+}h\vu)-\Phi(\pE) \over h} + \lambda\lim_{h\rar0}{\Psi(\pE{+}h\vu) - \Psi(\pE) \over h}
= d\Phi(\pE).\vu + \lambda d\Psi(\pE).\vu
= (d\Phi(\pE) + \lambda d\Psi(\pE)).\vu
$ for all $p$ and~$\vu$,
thus  $d(\Phi + \lambda\Psi) = d\Phi + \lambda d\Psi$ for all $\lambda\in\RR$ and $\Phi,\Psi\in C^1(\UE;\calF)$.
\findem

\debexe
Prove: if $f\in C^1(\UE;\RR)$ (scalar values) and $\Phi \in C^1(\UE;\calF)$ then, for all $\vu\in E$,
\be
d(f\Phi).\vu = (df.\vu)\Phi + f(d\Phi.\vu)
\ee
(and we also write $d(f\Phi) = \Phi \otimes df + f\, d\Phi$ for a use with contraction rules).

\debrep
%Contraction rule: $(\Phi \otimes df).\vu := \Phi(df.\vu) =(df.\vu)\Phi$. And
\be
\label{eqderprod}
\eqalign{
d(f\Phi)(\pE).\vu
= &\lim_{h\rar0} {f(\pE{+}h\vu)\Phi(\pE{+}h\vu) - f(\pE)\Phi(\pE) \over h} \cr
= &\lim_{h\rar0} {f(\pE{+}h\vu)\Phi(\pE{+}h\vu) - f(\pE)\Phi(\pE{+}h\vu) \over h}
+  {f(\pE)\Phi(\pE{+}h\vu) - f(\pE)\Phi(\pE) \over h} \cr
= &\lim_{h\rar0} {f(\pE{+}h\vu) - f(\pE) \over h}(\Phi(\pE)+o(1))
+ \lim_{h\rar0} f(\pE){\Phi(\pE{+}h\vu) - \Phi(\pE) \over h} \cr
= &(df(\pE).\vu)\Phi(\pE) + f(\pE)(d\Phi(\pE).\vu).
}
\ee
Tensorial writing: $d(f\Phi).\vu = (\Phi \otimes df).\vu + (f\, d\Phi).\vu$,
thanks to the contraction rule which gives
$(\Phi \otimes df).\vu + (f\, d\Phi).\vu = \Phi(df.\vu)+ f (d\Phi.\vu)$.
\finrep
\finexe

\debrem
In differential geometry, the definition of a tangent map is defined by, with definition~\ref{defdifff9}:
\be
T\Phi :
\left\{\eqalign{
\UE \times E & \rar \calF \times F \cr
(\pE,\vu) & \rar T\Phi(\pE,\vu) = (\Phi(\pE),d\Phi(\pE).\vu). \cr
}\right.
\ee
The two points $\pE$ (input) and $\Phi(\pE)$ (output) are the base points, and
the two vectors $\vu$ (input) and $d\Phi(\pE).\vu$ (output) are the initial vector and its push-forward by~$\Phi$. % (considered at~$\Phi(p)$).
\finrem

%%%%%%%%%%%%%%%%%%%%%%%%%%%%%%%%%%%%%%%%%%%%%%%%%%%%%%%%%%%%%%%%%%%%%%%%%%%%%%%%%%%

\subsubsection{Notation for the second order Differential}
\label{secasod}

Let $\Phi\in C^2(\UE;\calF)$; 
Thus $d\Phi \in C^1(\UE;\calL(E;F))$, thus $d(d\Phi) \in C^0(\UE;\calL(E;\calL(E;F)))$;
So, for $p\in U$ and $\vu\in E$, we have
$
d(d\Phi)(\pE).\vu = \lim_{h\rar0}{d\Phi(\pE + h\vu) - d\Phi(\pE) \over h}\;\in \calL(E;F)
$, and, with $\vv \in E$ we have $(d(d\Phi)(\pE).\vu).\vv \in F$.

\debdef
The bilinear map $d^2\Phi(\pE) \in \calL(E,E;F)$ is defined by % (bilinear map $E\times E \rar F$) by 
\be
d^2\Phi(\pE)(\vu,\vv) = (d(d\Phi)(\pE).\vu).\vv,
\ee
thanks to the natural canonical isomorphism $L\in \calL(E;\calL(E;F)) \leftrightarrow T_L\in \calL(E,E;F)$ given by
$T_L(\vu_1,\vu_2) := (L.\vu_1).\vu_2$ for all $\vu_1,\vu_2\in E$; Thus $L\eqnote T_L$, thus $d(d\Phi) \eqnote d^2\Phi(\pE)\in \calL(E,E;F)$.
\findef

This gives the usual second order Taylor expansion
of~$\Phi$ (supposed~$C^2$) near~$p$ in the direction~$\vu$:
\be
\Phi(\pE + h\vu) = \Phi(\pE) + h\,d\Phi(\pE).\vu + {h^2\over 2}\,d^2\Phi(\pE)(\vu,\vu) + o(h^2)
\ee
(=the second order Taylor expansion of $f:h\rar f(h) = \Phi(\pE + h\vu)$ near $h=0$, \cf~\eref{eqdifff90}).

And Schwarz's theorem tells that $d^2\Phi(p)$ is symmetric when $\Phi$ is~$C^2$, \ie\ $d^2\Phi(\pE)(\vu,\vv)=d^2\Phi(\pE)(\vv,\vu)$.

\comment{
Moreover $\calL(E,E;F) \simeq \calL(F^*,E,E;\RR)$, thus $d^2\Phi(\pE)\simeq T(\pE) \in \calL(F^*,E,E;\RR)$ where $T(\pE)(\ell,\vu,\vv)=\ell.d^2\Phi(\pE)(\vu,\vv)$. 
%Thus, when $F=E$, $d^2\Phi$ can be considered to be a ${1\choose2}$ tensor.

\debexa
If $E=F=\RRn$ and $\vu\in \Gamma(U)$ is a vector field in $U$ an open set in~$\RRn$, then, for $p\in U$, %with natural canonical isomorphisms,

$\vu(p)\in\vRRn \simeq \vec\RRn^{**} = \calL^1_0(\vRRn)$, so $\vu \in \Tuzu$,

$d\vu(p)\in \calL(\vRRn;\vRRn) \simeq \calL(\RRns,\RRn;\RR) = \calL^1_1(\vRRn)$, so $\vu \in \Tuuu$,

$d^2\vu(p) \in \calL(\vRRn,\vRRn;\vRRn)\simeq \calL(\RRns,\vRRn,\vRRn;\RR) = \calL^1_2(\vRRn)$, so $d^2\vu \in \Tudu$, ...,

and $d^k\vu \in T^1_k(U)$ for any $k\in \NN$.
\finexa
}

%%%%%%%%%%%%%%%%%%%%%%%%%%%%%%%%%%%%%%%%%%%%%%%%%%%%%%%%%%%%%%%%%%%%%%%%%%%%%%%%%%%

\subsection{A basis and the $j$-th partial derivative}

\debdef
Let $\Phi \in C^1(\UE;\calF)$, $\vu\in \Gamma(\UE)$ (a vector field), $p\in U$.
The derivative of $\Phi$ at~$p$ along $\vu$ is defined by
\be
\pa_\vu \Phi(\pE) := d\Phi(\pE).\vu(\pE)  %\eqnote {\pa \Phi \over \pa \ve_j}(\pE)
\quad (= \lim_{h\rar0}{\Phi(\pE + h\vu(\pE)) - \Phi(\pE) \over h}\;\in F).
\ee
This defines the directional derivative operator along~$\vu$:
\be
\pa_\vu:
\left\{\eqalign{
C^1(\UE;\calF) &\rar C^0(\UE;F) \cr
\Phi & \rar \pa_\vu (\Phi):= d\Phi.\vu, \qie \pa_\vu (\Phi)(\pE):= d\Phi(\pE).\vu(\pE).
}\right.
\ee
(And $\pa_\vu (\Phi)(\pE) \eqnote \vu(\Phi)(\pE)$ in differential geometry thanks to $E\simeq \Ess$ which gives $\pa_\vu\simeq \vu$.)
\findef

In particular, if $(\ve_i(\pE))$ is a basis at $\pE$, then the $j$-th partial derivative of~$\Phi$ at~$\pE$
is $\pa_{\ve_j} \Phi(\pE) \eqnote \pa_j \Phi(\pE)$ (the derivative along~$\ve_j$), and the $j$-th directional derivative operator is
\be
\label{eqpajpsi}
\pa_j:
\left\{\eqalign{
C^1(\UE;\calF) &\rar C^0(\UE;F) \cr
\Phi & \rar \boxed{\pa_j \Phi := d\Phi.\ve_j}, \qie \pa_j (\Phi)(\pE):= d\Phi(\pE).\ve_j(\pE) .
}\right.
\ee
(In differential geometry $\pa_j \Phi \eqnote \ve_j(\Phi)$, so $\ve_j (\Phi)(\pE):= d\Phi(\pE).\ve_j(\pE)$.)

%Particular case of a flat surface: There exists a Cartesian basis $(\ve_i)$ (the same basis at all points), and the $i$-th partial derivative at a point~$\pE$ is often written $\pa_i f(\pE) = {\pa f \over \pa x_i}(\pE)$.

\comment{
Particular case of Cartesian coordinates, \ie\ $(\ve_i)$ is a Cartesian basis in~$E$, $(\vb_i)$ is a Cartesian basis in~$F$
(both bases are uniform), and $O_\calE$ is an origin in~$\calE$: 
If $x_i$ is the usual name of the $i$-th coordinate, \ie\ $\ora{O\pE}= \sum_i x_i \ve_i$,
and if $\ora{O\Phi(\pE)}= \sum_i \Phi_i(\pE) \vb_i$, 
then $d\Phi(\pE).\vu(\pE) = \sum_i (d\Phi_i(\pE).\vu(\pE)) \vb_i$ (since the $\vb_i$ are uniform);
And if $\vu(\pE) = \sum_j u_j(\pE) \ve_i$, 
then $d\Phi(\pE).\vu = \sum_{ij} u_j(\pE) (d\Phi_i(\pE).\ve_j) \vb_i$, and
\be
\pa_j \Phi_i(\pE) = d\Phi_i(\pE).\ve_j \eqnote {\pa\Phi_i\over \pa x_j}(\pE).
\ee
(With duality notations, $\pa_j \Phi_j(\pE) \eqnote {\pa\Phi^i\over \pa x^j}(\pE)$.)
%(This notation ${\pa\Phi\over \pa x_j}(\pE)$ can be ambiguous, see \eg\ \S~\ref{secrnna}.)
}

%%%%%%%%%%%%%%%%%%%%%%%%%%%%%%%%%%%%%%%%%%%%%%%%%%%%%%%%%%%%%%%%%%%%%%%%%%%%%%%%%%%

\subsection{Application 1: Scalar valued functions}
\label{secdifffavs}

\def\pE{p}

%%%%%%%%%%%%%%%%%%%%%%%%%%%%%%%%%%%%%%%%%%%%%%%%%%%%%%%%%%%%%%%%%%%%%%%%%%%%%%%%%%%

\subsubsection{Differential of a scalar valued function (objective)}

Here $\ds \Phi\eqnote f :
\left\{\eqalign{
\UE & \rar \RR \cr
\pE & \rar f(\pE)
}\right\}
$
is a~$C^1$ scalar valued function,
so $df\in \Omega^1(U) \cap C^0(U;E^*)$ (a $C^0$ differential form).
So $df(p)\in E^*$ for all $p\in U$, and $df(p).\vu = \lim_{h\rar0} {f(\pE+h\vu) - f(\pE) \over h}\in \RR$ for all $\vu\in E$.

\debexe
Prove: If $f,g\in C^1(\UE;\RR)$ then (derivative of a product)
\be
\label{eqdfg}
d(fg) = (df)g + f(dg), %\qthus (fg)_{|i} = f_{|i} g + fg_{|i}, \; \forall i=1,...,n,
\ee
i.e., $d(fg).\vw= (df.\vw)g + f(dg.\vw)$ for all $\vw\in\Gamma(\UE)$.

\debrep
$\lim_{h\rar0} {f(\pE{+}h\vw)g(\pE{+}h\vw) - f(\pE)g(\pE) \over h}
= \lim_{h\rar0} {f(\pE{+}h\vw)g(\pE{+}h\vw) - f(\pE)g(\pE{+}h\vw) \over h}
+ \lim_{h\rar0} {f(\pE)g(\pE{+}h\vw) - f(\pE)g(\pE) \over h}
= \lim_{h\rar0} {f(\pE{+}h\vw) - f(\pE) \over h}(g(\pE)+o(1))
+ \lim_{h\rar0} f(\pE){g(\pE{+}h\vw) - g(\pE) \over h}
$, calculation that only requires the first order (affine) approximation of $f$ and $g$:
We get the same result as with the affine functions
$f(x)=a_0 + a_1 x$ and $g(x)=b_0+b_1x$, which give $(fg)(x)= a_0b_0 + (a_0b_1{+} a_1b_0) x + a_1b_1 x^2$, and then
$(fg)'(x) = a_0b_1{+} a_1b_0 + 2a_1b_1 x$, which is indeed equal to $(f'g+fg')(x)=a_1(b_0{+} b_1 x) + (a_0{+} a_1 x)b_1$.
\finrep
\finexe

%%%%%%%%%%%%%%%%%%%%%%%%%%%%%%%%%%%%%%%%%%%%%%%%%%%%%%%%%%%%%%%%%%%%%%%%%%%%%%%%%%%

\subsubsection{Quantification}

Let $(\ve_i(p))$ be a basis at~$p$. So
$\pa_j f(\pE) \equalref{eqpajpsi} df(\pE).\ve_j(\pE)$ ($= \lim_{h\rar0} {f(\pE+h\ve_j(\pE)) - f(\pE) \over h}$),
and we write
\be
\pa_j f(\pE) \eqnote f_{|j}(\pE).  %= {\pa f \over \pa \ve_j}(\pE) 
\ee
So, with $(\pi_{ei}(\pE))$ the dual basis  of the basis~$(\ve_i(\pE))$,
and with $f_{|j}(\pE):= \pi_{ei}(\pE).df(\pE)$ ($j$-th component of $df(\pE)$ in the basis~$(\pi_{ei}(\pE))$),
we have %$df(\pE) = \sumjn f_{|j}(\pE)\,\pi_{ei}(\pE)$, so 
\be
\label{eqpaif0}
df = \sumjn f_{|j}\pi_{ej}, \qand [df(\pE)]_{|\ve} = \pmatrix{f_{|1}(\pE)&...&f_{|n}(\pE)} \quad\hbox{(row matrix)}.
\ee
So $df.\vu = \sumjn f_{|j} u_j = [df]_{|\ve}.[\vu]_{|\ve}$ when $\vu(p) = \sum_i u_i(p)\ve_i(p)$.
In particular with a Cartesian basis, $(\pi_{ei}(\pE))\eqnote (dx_j)$, and $df = \sumjn  {\pa f\over \pa x_j}dx_j$.
%(unmissable in thermodynamics).

Duality notations: $\piei = e^i$, $\vu = \sumjn u^j \ve_j$, $df = \sumjn f_{|j}\,e^j$, $df.\vu = \sumjn f_{|j}u^j$, and with a Cartesian basis, $\piei=dx^i$ and $df = \sumjn {\pa f\over \pa x^j}dx^j$.

%And Cartesian basis: $e^i\eqnote dx^i$, and $df = \sumjn {\pa f \over \pa x^j}\,dx^j$, and $df.\vu = \sumjn {\pa f \over \pa x^j} u^j$.

\debexe
Prove: $(fg)_{|j} = f_{|j}\,g + f\,g_{|j}$ when $f,g:U\rar\RR$ are $C^1$ scalar valued functions.

\debrep
Apply~\eref{eqderprod}: here $d(fg)=g\, df + f \, dg$, \ie\
 $d(fg).\ve_j = (df.\ve_j)\,g + f\,(dg.\ve_j)$ for all~$j$.
\finrep
\finexe

And $df(p)\in E^*$ satisfies the covariant change of basis formula for linear forms, \ie, if $(\va_i(p))$ and $(\vb_i(p))$ are two bases at~$p$ and $P(p)$ is the transition matrix from $(\va_i(p))$ to~$(\vb_i(p))$, then
$[df(p)]_{|\vb} \equalref{eqdefP1} [df(p)]_{|\va}.P(p)$, or in short:
\be
\label{eqdfvavbP}
[df]_{|\vb} = [df]_{|\va}.P  \quad(\hbox{covariance formula}).
\ee

%%%%%%%%%%%%%%%%%%%%%%%%%%%%%%%%%%%%%%%%%%%%%%%%%%%%%%%%%%%%%%%%%%%%%%%%%%%%%%%%%%%

\subsubsection{Gradients (subjective) associated with a differential through inner dot products}
\label{seccontree2}

Let $f\in C^1(\UE;\RR)$ (a $C^1$ scalar valued function). Choose (subjective) an inner dot product $\dd_g$ in~$E$.

\debdef
The conjugate gradient $\vgrad_g f(\pE)$ of $f$ at $\pE\in\UE$ relative to~$\dd_g$, also called the $\dd_g$-conjugate gradient of $f$ at $p$,
is the $\dd_g$-Riesz representation vector of the linear form $df(\pE)\in E^*$:
\be
\vgrad_g f(\pE) := \vR_g(df(\pE)).
\ee
\Ie, the vector $\vgrad_g f(\pE)\in E$ is characterized by, \cf~\eref{eqrtr},
\be
\label{eqdifff2}
\forall\vu\in E,\quad 
\boxed{df(\pE).\vu = (\vgrad_g f(\pE),\vu)_g} =\vgrad_g f(\pE) \bcdotg\;\vu.
\ee
\findef

\noindent
{\bf Fundamental:} An English observer with his Euclidean dot product $\dd_a$ in foot and a French observer with his Euclidean dot product $\dd_b$ in metre 
have the same differential $df$ (defined independently of any unit of measurement); But do \textslbf{not} have the same gradient:
\be
\label{eqrtr20abg}
\vgrad_b f \equalref{eqrtr20} \lambda^2 \vgrad_a f \qwith \lambda^2>10.
\ee
Quite different vectors isn't it?
The ``gradient vector'' strongly depends on the chosen inner dot product. And to forget this fact leads to accidents like the crash of the Mars Climate Orbiter probe, \cf~remark~\ref{remMCOC}.

\mn
{\bf Subjective first order Taylor expansion: }
If an inner dot product~$\dd_g$ exists and is used, then the first order Taylor expansion~\eref{eqdifff9} gives
\be
f(\pE + h\vu)
= f(\pE) + h\,(\vgrad_g f(\pE),\vu)_g + o(h)
\qquad (= f(\pE) + h\,\vgrad_g f(\pE) \bcdotg \vu + o(h))
.
\ee

\mn
{\bf Fundamental once again} (we insist):

$\bullet$ An inner dot product does not always exist (as a meaningful tool), see~\S~\ref{seccontree} (thermodynamics),
thus, for a $C^1$ function, a gradient does not always exists (contrary to a differential).

$\bullet$ $df(\pE)$ is a linear form (covariant) while $\vgrad_g f(\pE)$ is a vector (contravariant).
In particular the change of basis formulas differ, \cf~\eref{eqdefP1}:
\be
[df]_{|\new} = [df]_{|\old}.P, \quad \hbox{while}\quad
[\vgrad_g]_{|\new} = P^{-1}.[\vgrad_g]_{|\old}.
\ee

$\bullet$  $df$ cannot be identified $\vgrad f$ (with one?) (Recall; there is no natural canonical isomorphims between $E$ and~$E^*$.)
The differential $df$ is also called the ``covariant gradient'',
and any of its associated gradient vectors is also called the ``contravariant gradient relative to an inner dot product''.

\mn
{\bf Isometric Euclidean framework:}
%If $\dd_g$ is a Euclidean dot product (\eg\ chosen by an English-foot or French-metre observer), then $\vgrad_g f(\pE)$ is called the gradient of $f$ at $\pE\in\UE$ relative to~$\dd_g$ (or the $\dd_g$- gradient). 
If one Euclidean dot product can be imposed to all observers (foot? metre?) then $\vgrad_g f \eqnote \vgrad f = \vec\nabla f$ and~\eref{eqdifff2} is written $df.\vu=\vgrad f \bcdot\vu = \vec\nabla f \bcdot \vu$ (isometric framework).

\debexe
Cartesian basis $(\ve_i)$ and $\dd_g$ given by $[g]_{[\ve} = \pmatrix{1&0\cr0&2}$. Give
$[df]_{|\ve}$ and $[\vgradg f]_{|\ve}$.

\debrep
$[df]_{|\ve}= \pmatrix{{\pa f \over \pa x_1} & {\pa f \over \pa x_2}}$ (row matrix)
and \eref{eqdifff2} gives $[\vgradg f]_{|\ve} = \pmatrix{{\pa f \over \pa x_1} \cr \demi{\pa f \over \pa x_2}}$ (column matrix $\ne [df]^T$).
\finrep
\finexe

%%%%%%%%%%%%%%%%%%%%%%%%%%%%%%%%%%%%%%%%%%%%%%%%%%%%%%%%%%%%%%%%%%%%%%%%%%%%%%%%%%%

\subsection{Application 2: Coordinate system basis and Christoffel symbols}
\label{secCsb}

\def\Upar{U_{\!\rm par}}
\def\piAi{\pi_{\!Ai}}
\def\piAj{\pi_{\!Aj}}

(Necessary when dealing with covariance.)

%%%%%%%%%%%%%%%%%%%%%%%%%%%%%%%%%%%%%%%%%%%%%%%%%%%%%%%%%%%%%%%%%%%%%%%%%%%%%%%%%%%

\subsubsection{Coordinate system, and coordinate system basis}

Consider a (open) set $\Upar=\{\vq\in ]a_1,b_1[\times ... \times ]a_n,b_n[\}$, called the set of parameters,
in the Cartesian space~$\RRn$,
consider an open set $U\subset \RRn$, called the set of geometric positions,
and consider a $C^2$-diffeomorphism $\Psi: \vq\in \Upar \rar p\in U$, called a coordinate system.

Let $(\va_i)$ the canonical basis of the parameter space, let $\vq=\sum_i q_i\va_i\in \Upar$ (the $q_i$ are called the parameters). 
\Eg, see the polar coordinate system at~\S~\ref{secrempfcb} where $\vq=(q_1,q_2)=(r,\theta)$.

$\Psi$ being a diffeomorphism, at any $p=\Psi(\vq) \in U$ the vectors
\be
\label{eqveiPsi}
\va_{i*}(p) :=d\Psi(\vq).\va_i
\ee
make a basis in~$E$ at $p$, and $(\va_{i*}(p))$ is called the coordinate system basis at~$p$.
Its dual basis at~$p$ is made of the linear forms $dq_i(p)$, so where, for all~$i,j$,
\be
dq_i(p).\va_{j*}(p) =\delta^j_i.
\ee
Duality notations: $dq^i(p).\va_{j*}(p)=\delta^i_j$ for all $i,j$.

%%%%%%%%%%%%%%%%%%%%%%%%%%%%%%%%%%%%%%%%%%%%%%%%%%%%%%%%%%%%%%%%%%%%%%%%%%%%%%%%%%%

\subsubsection{Parametric expression of the differential of a scalar valued function}

With a coordinate system~$\Psi$, a scalar valued function 
$f:
\left\{\eqalign{
U & \rar\RR \cr
p & \rar f(p)
}\right\}
$
defined in~$U$ can be described with the function
$g=f\circ\Psi:
\left\{\eqalign{
\Upar & \rar\RR \cr
\vq & \rar g(\vq):=f(p) \;\hbox{when}\; p=\Psi(\vq)
}\right\}
$
defined in~$\Upar$, and $g$ is called the parametric expression of~$f$. Thus
\be
dg(\vq) = df(p).d\Psi(\vq) \qwhen p=\Psi(\vq),
\ee
in particular,
\be
\label{eqpafcoor0}
{\pa g \over \pa q_j}(\vq) := dg(\vq).\va_j = df(p).d\Psi(\vq).\va_j = df(p).\va_{j*}(p)
\eqnote {\pa f \over \pa q_j}(p).
\ee
{\bf Warning, pay attention:} $f$ is a function of~$p$, not a function of~$\vq$, and the notations
${\pa f \over \pa q_j}(p)$ means $:={\pa (f\circ\Psi) \over \pa q_j}(\vq)$ when $p=\Psi(\vq)$, and nothing else.

Thus with $(dq_j(p))$ the dual basis of the coordinate basis~$(\va_{i*}(p))$ at~$p$,
\be
\label{eqpafcoor}
df(p) \equalref{eqpafcoor0} \sumjn {\pa f \over \pa q_j}(p)\, dq_j(p).
\ee
Duality notations: $df(p) = \sum_j {\pa f \over \pa q^j}(p)\, dq^j(p)$.

\debrem
Pay attention to the notations that could contradict themselves: % if you don't pay attention to the context:

1- In~$\Upar$ the dual basis $(\piai)$ of the Cartesian basis~$(\va_i)$ is a uniform basis (independent of~$\vq$)...
and is (almost) never written $(dq_i)$...

2- Indeed, $(dq_i(p))$ is the name reserved for the dual basis of~$(\va_{i*}(p))$ in the geometric space... Mind the notations!
\Eg\ for polar coordinates $(dq_1(p),dq_2(p))=(dr(p),d\theta(p))$ is the dual basis of the polar coordinate system basis $(\va_{1*}(p),\va_{2*}(p))$ at~$p$, \cf~\eref{secrempfcb}.
\finrem

\debexe
Bases $(\va_i)$ and $(\vb_i)$ at~$p$.
A vector $\vx$ is expressed as $\vx=\sum_i x_{a,i}\va_i = \sum_i x_{b,i}\vb_i$.
%Suppose (change of unit \cf~\eref{eqexa0a}). 
Prove:
\be
\label{eqddfvbva}
\vb_i=\lambda\va_i, \;\; \forall i\;\; \Longrightarrow\;\;
\boxed{{\pa f \over \pa x_{b,i}} = \lambda {\pa f \over \pa x_{a,i}}}
\qor {\pa f \over \pa x_{a,i}}  = {\pa f \over \pa (\lambda  x_{a,i})}.
\ee
(Change of unit formula.)
Duality notations: ${\pa f \over \pa x_b^j} = \lambda {\pa f \over \pa x_a^j}$.

\debrep
$df(p).\vb_j(p) = \lambda df(p).\va_j(p)$ (linearity of $df(p)$) reads~\eref{eqddfvbva}.
(Or $[df]_{|\vb} = [df]_{|\va}.P$ with $P=\lambda I$ here.) 
%thus, with $df(\pE) = \sumjn{\pa f \over \pa x_{a,j}}\, dx_j = \sumjn{\pa f \over \pa x_{b,j}}\, dx_j$ as written by $A$ and~$B$,
%the last notation being valid  $[\vx]_{|\vb} = P^{-1}.[\vx]_{|\va} = {1\over \lambda} [\vx]_{|\va}$ gives $x_a^j = \lambda  x_b^j$.
\finrep
\finexe

\debexe
\label{exepxpx}
$[df]_{|\vb} = [df]_{|\va}.P$, \cf~\eref{eqdfvavbP}, \ie\ ${\pa f \over \pa x_b^j} = \sumin  {\pa f \over \pa x_a^i} P^i_j$ is also noted
\be
\label{eqdfvavbP3}
{\pa f \over \pa x_b^j} = \sumin  {\pa f \over \pa x_a^i}{\pa x_a^i \over \pa x_b^j}.
\ee
Why?

\debrep
{\bf Quick answer.}
$[\vx]_{|\vb} = P^{-1}.[\vx]_{|\va}$,
\ie\ $[\vx]_{|\va} = P.[\vx]_{|\vb}$,
which means $[\vx]_{|\va}([\vx]_{|\vb}) = P.[\vx]_{|\vb}$,
\ie
\be
\label{exepxpx1}
\pmatrix{ x_a^1(x_b^1,...,x_b^n) \cr \vdots \cr x_a^1(x_b^1,...,x_b^n) }
= \pmatrix{ \sumjn P^1_j x_b^j \cr \vdots \cr \sumjn P^n_j x_b^j},
\qthus
{\pa x_a^i\over \pa x_b^j}(x_b^1,...,x_b^n) = P^i_j,\;\; \forall i,j.
\ee
Thus~\eref{eqdfvavbP3} means
\be
\label{eqdfvavbP4}
{\pa f \over \pa x_b^j}(\pE) = \sumin  {\pa f \over \pa x_a^i}(\pE){\pa x_a^i \over \pa x_b^j}(x_b^1,...,x_b^n)
\qthus =\sumin  {\pa f \over \pa x_a^i}(\pE)\,P^i_j,
\ee
as given in~\eref{eqdfvavbP}.

\noindent
{\bf Detailed answer.}
%\eref{eqdfvavbP4} is not satisfactory since $\pE$ and the $x_a^i$ and $x_b^i$ should be related: What are the relations?
Let $O$ be a point (origin) in~$\UE$. If $\pE\in \UE$,
let $\vx=\ora{O\pE} = \sumin x_a^i\va_i = \sumin x_b^i\vb_i$.

This define the function $[\vx]_{|\va} : [\vx]_{|\vb} \rar [\vx]_{|\va}([\vx]_{|\vb})$,
and we have $[\vx]_{|\va}([\vx]_{|\vb}) = P.[\vx]_{|\vb}$ (change of basis formula). %$x_a^i = x_a^i(x_b^1,...,x_b^n)$

Then let $f_a,f_b: \RR^n \rar \RR$ be defined by
$
f_a([\vx]_{|\va}) :=  f(\pE)$ and $%f_b(x_b^1,...,x_b^1) = 
f_b([\vx]_{|\vb}) := f(\pE)
$.
(NB: $f_a$ and $f_b$ don't have the same definition domain: They are different).

Thus $f_a([\vx]_{|\va}) = f_b([\vx]_{|\vb})$ ($=f(p)$) when $[\vx([\vx]_{|\vb})]_{|\va} = P.[\vx]_{|\vb}$, so
$(f_a\circ[\vx]_{|\va})([\vx]_{|\vb}) = f_b([\vx]_{|\vb})$.
Thus ${\pa (f_a\circ[\vx]_{|\va}) \over \pa x_b^i}([\vx]_{|\vb}) 
= {\pa f_b \over \pa x_b^i}([\vx]_{|\vb})$,
thus the meaning of~\eref{eqdfvavbP3} is
\be
\label{eqdfvavbP5}
\sumjn {\pa f_a \over \pa x_a^j}([\vx]_{|\va}){\pa x_a^j \over \pa x_b^i}([\vx]_{|\vb})
= {\pa f_b \over \pa x_b^i}([\vx]_{|\vb}).
\ee
%This is the meaning of~\eref{eqdfvavbP3}.

{\bf Question:} Why did we introduce $f_a$ and $f_b$ (and not just keep~$f$)?

{\bf Answer:} Because a vector is not just a collection of components (is not just a matrix),
and $\ora{O\pE}$ cannot be reduced to a matrix of components (which one: $[\vx]_{|\va}$? $[\vx]_{|\vb}$?).
Here $f$ is a function acting on a point~$\pE$ (independent of a referential), while $f_a$ and $f_b$ are functions acting on matrices (dependent on the choice of a referential):
The domain of definitions are different, so the functions $f$, $f_a$ and $f_b$ are different.
\finrep
\finexe

%%%%%%%%%%%%%%%%%%%%%%%%%%%%%%%%%%%%%%%%%%%%%%%%%%%%%%%%%%%%%%%%%%%%%%%%%%%%%%%%%%%

\subsubsection{Christoffel symbols}

We use duality notations for readability and usage.

\debdef
In a coordinate system basis $(\ve_i(p))$ in~$E$ (previously called $(\va_{i*}(p)$),
the Christoffel symbol $\gamma^i_{jk}(p)\in \RR$ are the components of the vector $d\ve_k(p).\ve_j(p)$,
\ie\ $d\ve_k(p).\ve_j(p) = \sumkn \gamma_{jk}^i(p) \ve_i(p)$, so
\be
\label{eqdefCs}
\boxed{d\ve_k.\ve_j = \sumin \gamma^i_{jk} \ve_i} , \qor
d\ve_j.\ve_i = \sumkn \gamma^k_{ij} \ve_k.
\ee
(So, with $(e^i(p))$ the dual basis of~$(\ve_i(p))$, $\gamma^i_{jk} := e^i.d\ve_k.\ve_j$, and, for calculations with contractions, $d\ve_k = \sum_{ij} \gamma^i_{jk} \ve_i \otimes e^j$.)

(The Christoffel symbols vanish in a Cartesian framework.)

(Differential geometry in manifolds: $\nabla_{\ve_j}\ve_k = \sumin \gamma^i_{jk} \ve_i$, \ie\ the $\gamma^i_{jk} = e^i.\nabla_{\ve_j}\ve_k$ are the component of the connection~$\nabla$, the usual connection in a surface in~$\RRn$ being the Riemannian connection, in which case $\nabla_{\ve_j}\ve_k$ is the orthogonal projection of $d\ve_k.\ve_j$ on the surface relative to a Euclidean dot product.)
\findef

\Eg\ for the polar coordinate system, see remark~\ref{rempfcb}, $d\ve_2.\ve_2=-r\ve_1$, thus $\gamma^1_{22}=-r$
and $\gamma^2_{22}=0$.

%\Eg\ for a Cartesian coordinate system, $\gamma^i_{jk}=0$ for all $i,j,k$ since the $\ve_i$ are independent of~$p$.

\debexe
Prove: If $(\ve_i(p))$ is the coordinate system basis of a $C^2$ coordinate system, then:
\be
\label{eqChrsym}
\forall i,j,\; d\ve_i.\ve_j = d\ve_j.\ve_i\quad (={\pa^2\Psi \over \pa q^i\pa q^j}), 
\qand \forall i,j,k,\;  \gamma^k_{ji}=\gamma^k_{ij}
\quad\hbox{(symmetry for lower indices)}.
\ee

\debrep
$\ve_i(p) = (\ve_i\circ \Psi)(\vq) \equalref{eqveiPsi} d\Psi(\vq).\va_i$ gives
$d(\ve_i\circ \Psi)(\vq).\va_j = d(d\Psi(\vq).\va_i).\va_j$,
thus
$d\ve_i(\Psi(\vq)).d\Psi(\vq).\va_j %= d^2\Psi(\vq)(\va_i,\va_j) 
= {\pa{\pa\Psi \over \pa q^i}\over \pa q^j}
= {\pa{\pa\Psi \over \pa q^j}\over \pa q^i}$ (Schwarz theorem since $\Psi$ is~$C^2$)
$= d\ve_j.\ve_i \eqnote {\pa^2\Psi \over \pa q^j\pa q^i}$,
thus $\sumkn \gamma^k_{ij} \ve_k=\sumkn \gamma^k_{ji} \ve_k$.
\finrep
\finexe

\debexe
\label{execgammaijk0}
Consider two coordinate system bases $(\va_i(p))$ and~$(\vb_i(p))$ at~$p$,
$P(p)=[P^i_j(p)]$ the transition matrix from $(\va_i(p))$ to~$(\vb_i(p))$, and $Q=P^{-1}$.
Using the generic notation $d\ve_k.\ve_j = \sumin \gamma^i_{jk,e} \ve_i$, prove the change of basis formula for the Christoffel symbols:
\be
\label{exochrist2}
\gamma_{jk,b}^i
=  \sum_{\lambda,\mu,\nu=1}^n  Q^i_\lambda P^\mu_j P^\nu_k \gamma_{\mu\nu,a}^\lambda
+ \sum_{\lambda,\mu=1}^n Q^i_\lambda P^\mu_j (dP^\lambda_k.\va_\mu) \quad (
= \sum_{\lambda,\mu,\nu=1}^n Q^i_\lambda P^\mu_j P^\nu_k \gamma_{\mu\nu,a}^\lambda
+ \sum_{\lambda=1}^n Q^i_\lambda (dP^\lambda_k.\vb_j)).
\ee
(Because of the term $\sum_{\mu\nu} Q^i_\lambda P^\mu_j (dP^\lambda_k.\va_\mu)$, a derivation is not a tensor.)
%(The term $\sum_{\mu\nu} Q^i_\lambda P^\mu_j (dP^\lambda_k.\va_\mu)$ is not algebraic, so a derivation is not a tensor.)
%(A connection is not a tensor: The last term with~$dP$ doesn't exist in the change of basis formula for $1\choose2$ tensors.)
\comment{
\be
\label{exochrist2}
\gamma_{ij,b}^k
=  \sum_{\lambda,\mu,\nu=1}^n P^\lambda_i P^\mu_j Q^k_\nu \gamma_{\lambda\mu,a}^\nu
+ \sum_{\lambda,\nu=1}^n P^\lambda_i Q^k_\nu (dP^\nu_j.\va_\lambda) \quad (
= \sum_{\lambda,\mu,\nu=1}^n P^\lambda_i P^\mu_j Q^k_\nu \gamma_{\lambda\mu,a}^\nu
+ \sum_{\nu=1}^n Q^k_\nu (dP^\nu_j.\vb_i))
\ee
}

\debrep
$\vb_k(p) = \sum_\nu P^\nu_k(p)\va_\nu(p)$ gives
$d\vb_k.\vb_j = \sum_{\nu} (dP^\nu_k.\vb_j)\va_\nu + \sum_{\nu} P^\nu_k (d\va_\nu.\vb_j)
= \sum_{\mu\nu} P^\mu_j (dP^\nu_k.\va_\mu)\va_\nu + \sum_{\mu\nu} P^\nu_k P^\mu_j (d\va_\nu.\va_\mu)
$;
And $b^i=\sum_\lambda Q^i_\lambda a^\lambda$, thus
$$
\gamma^i_{jk,b} =b^i.d\vb_k.\vb_j
= \sum_{\lambda\mu\nu} Q^i_\lambda P^\mu_j (dP^\nu_k.\va_\mu) a^\lambda.\va_\nu
+ \sum_{\lambda\mu\nu} Q^i_\lambda P^\mu_j P^\nu_k a^\lambda.(d\va_\nu.\va_\mu)
=  \sum_{\lambda\mu} Q^i_\lambda P^\mu_j (dP^\lambda_k.\va_\mu)
+ \sum_{\lambda\mu\nu} Q^i_\lambda P^\mu_j P^\nu_k \gamma^\lambda_{\mu\nu,a},
$$
thus~\eref{exochrist2}.
\finrep
\finexe

%%%%%%%%%%%%%%%%%%%%%%%%%%%%%%%%%%%%%%%%%%%%%%%%%%%%%%%%%%%%%%%%%%%%%%%%%%%%%%%%%%%

\subsection{Application 3: Differential of a vector field}

Here $F=E=\vRRn$,
$\Phi \eqnote \vw \in \Gamma(U)$ is a vector field.
Thus $d\vw(p) \in \calL(E;E)$ and $d\vw.\vu$ is a vector field in~$E$ for all $\vu\in\Gamma(U)$, given by
$(d\vw.\vu)(p) = d\vw(p).\vu(p) = \lim_{h\rar0}{\vw(p + h\vu(p)) - \vw(p) \over h}\in E$.

\mn
{\bf Quantification:}
$(\ve_i(p))$ is a basis at~$p$ in~$E$. Call $w_i(p)\in\RR$ the components of $\vw(p)$, \ie\ $\vw(p) = \sumin w_i(p) \ve_i(p)$.
And call $w_{i|j}(p)$ the components of $d\vw(p)$ (endomorphism in~$E$):
\be
\label{eqOFPsid}
\vw = \sumin w_i \ve_i, \quad
d\vw.\ve_j = \sumin w_{i|j} \ve_i,
\quad [d\vw]_{|\ve} = [w_{i|j}] \quad\hbox{(Jacobian matrix)}.
\ee
And tensorial notations for calculations with contractions: $(\pi_{ei}(p))$ being the dual basis,
\be
d\vw =  \sumijn w_{i|j} \ve_i\otimes \pi_{ej}.
\ee
Duality notations:  $\vw = \sumin w^i \ve_i$, 
$d\vw.\ve_j =  \sumijn w^i_{|j} \ve_i$, $[d\vw]_{|\ve} = [w^i_{|j}]$,
and $d\vw =  \sumijn w^i_{|j} \ve_i\otimes e^j$.

\mn
{\bf In a Cartesian basis:}
Here $(\ve_i)$ is uniform,
so $\vw(p) = \sumin w_i(p) \ve_i$ gives $d\vw(p).\ve_j = \sumin (dw_i(p).\ve_j) \ve_i$,
thus \eref{eqOFPsid} gives
\be
\label{eqdPsi}
w_{i|j} = {\pa w_i \over \pa x_j}(p) \eqnote w_{i,j},
\qso [d\vw]_{|\ve} = [ {\pa w_i \over \pa x_j}].
\ee
Duality notations: $w^i_{|j} = {\pa w^i \over \pa x^j}$
and $[d\vw]_{|\ve} = [ {\pa w^i \over \pa x^j}]$.

\mn
{\bf In a coordinate system basis:}
With the coordinate system described in \S~\ref{secCsb} and the duality notations for readability (and usage).
$\vw(p) = \sumin w^i(p) \ve_i(p)$ gives, for all~$j$,
\be
\label{eqdPsi2}
d\vw.\ve_j = \sumin (dw^i.\ve_j)\ve_i + \sumin w^i (d\ve_i.\ve_j)
\quad ( = \sumin w^i_{|j} \ve_i).
\ee
(Tensorial notations to be used with contractions:
$d\vw = \sum_i \ve_i \otimes dw^i + \sum_i w^i\,d\ve_i = \sum_{ij} w^i_{|j} \ve_i\otimes e^j$.)

And $\sum_i w^i (d\ve_i.\ve_j) 
\equalref{eqdefCs} \sum_{ik} w^i \gamma^k_{ji} \ve_k = \sum_{ik} w^k \gamma^i_{jk} \ve_i$,
thus, for all~$i,j$,
%$d\vw.\ve_j = \sum_i (dw^i.\ve_j)\ve_i + \sum_{ik} w^i \gamma^k_{ji} \ve_k = \sum_i (dw^i.\ve_j)\ve_i + \sum_{ik} w^k \gamma^i_{jk} \ve_i$, thus
\be
\label{eqdPsi2b}
%d\vw.\ve_j %= \sumin (dw^i.\ve_j + \sumkn w^k \gamma^i_{jk})\ve_i
%= \sumin w^i_{|j}\ve_i, \qwhere 
\boxed{w^i_{|j} = {\pa w^i \over \pa q^j} + \sumkn w^k \gamma^i_{jk} }
\qwhere {\pa w^i \over \pa q^j} := dw^i.\ve_j.
\ee
(${\pa w^i \over \pa q^j} := dw^i.\ve_j$ is the derivation along the $j$-th coordinate line of the scalar valued function~$w^i$).
% Thus the Jacobian matrix of~$\vw$ relative to $(\ve_i)$ is $[d\vw]_{|\ve} = [ w^i_{|j}] = [{\pa w^i \over \pa q^j} + \sumkn w^k \gamma^i_{jk}]. %\quad\hbox{(Jacobian matrix)}.$

(In particular, if $\vw=\ve_\ell = \sum_i \delta^i_\ell \ve_i$, 
we recover $d\ve_\ell.\ve_j = \sum_i 0 \ve_i + \sum_{ik}\delta^k_\ell \gamma^i_{jk} \ve_i= \sum_i\gamma^i_{j\ell} \ve_i$, 
\cf~\eref{eqdefCs}.)

\debexe
\label{execgammaijk}
With exercise~\ref{execgammaijk0}, and $\vw=\sumin u^i\va_i = \sum_n v^i\vb_i$,
check with calculations ($d\vw$ is an endomorphism defined independently of any basis): %, when , by computation with components:
\be
\label{eqgamdw}
[d\vw]_{|\vb} = P^{-1}.[d\vw]_{|\va}.P, \qie v^i_{|j} = \sumkln Q^i_k u^k_{|\ell} P^\ell_j.
\ee
%\goodbreak\penalty-500

\debrep
$\vb_j = \sum_\ell P^\ell_j \va_\ell$ for all~$i$, $Q=P^{|1}$, and $[\vw]_{|\vb} = Q.[\vw]_{|\va}$ reads 
$v^i = \sum_k Q^i_k u^k$ for all~$i$.
%, and $d\vw.\va_j=\sum_i (du^i.\va_j)\va_i$ and $d\vw.\vb_j=\sum_i (dv^i.\vb_j)\vb_i$; 

Cartesian basis:
$dv^i.\vb_j
= d(\sum_k Q^i_k u^k).(\sum_\ell P^\ell_j \va_\ell)
= \sum_{k\ell} Q^i_k (du^k.\va_\ell) P^\ell_j$, qed (here the $Q^i_k$ are uniform \ie\ independent of~$p$).

Coordinate system basis:
$v^i = \sum_\lambda Q^i_\lambda u^\lambda$
gives $dv^i.\vb_j = \sum_\lambda (dQ^i_\lambda.\vb_j) u^\lambda + \sum_\lambda Q^i_\lambda (du^\lambda.\vb_j)$; Thus
$$
\eqalign{
v^i_{|j}
\equalref{eqdefCs} & dv^i.\vb_j + \sum_k v^k \gamma^i_{jk,b} %+ \sum_k v^k b^i.(d\vb_k.\vb_j)
\cr
= & \sum_{\lambda\mu} u^\lambda P^\mu_j(dQ^i_\lambda.\va_\mu) + \sum_{\lambda\mu} Q^i_\lambda P^\mu_j (du^\lambda.\va_\mu)
 + \sum_{k\lambda\mu\nu} (Q^k_\lambda u^\lambda) Q^i_\nu P^\mu_j (dP^\nu_k.\va_\mu)
+ \sum_{k\omega\lambda\mu\nu} (Q^k_\omega u^\omega) Q^i_\lambda P^\mu_j P^\nu_k \gamma^\lambda_{\mu\nu,a}
\cr
}
$$
And $Q^k_\omega P^\lambda_k = \delta^\lambda_\omega$ gives $(dQ^k_\omega.\va_\mu) P^\lambda_k + Q^k_\omega(dP^\lambda_k.\va_\mu)=0$,
thus the third term reads
$$
\sum_{k\lambda\mu\nu} u^\lambda Q^i_\nu P^\mu_j Q^k_\lambda (dP^\nu_k.\va_\mu)
= - \sum_{k\lambda\mu\nu} u^\lambda Q^i_\nu P^\mu_j P^\nu_k (dQ^k_\lambda.\va_\mu)
= - \sum_{\lambda\mu} u^\lambda P^\mu_j (dQ^i_\lambda.\va_\mu),
$$
which cancels the first term:
Thus
$v^i_{|j} = \sum_{\lambda\mu} Q^i_\lambda P^\mu_j (du^\lambda.\va_\mu) 
+ \sum_{\lambda\mu\nu}  u^\nu Q^i_\lambda  P^\mu_j \gamma^\lambda_{\mu\nu} = \sum_{\lambda\mu} Q^i_\lambda u^i_{|j} P^\mu_j$, \ie~\eref{eqgamdw}.
\finrep
\finexe

%%%%%%%%%%%%%%%%%%%%%%%%%%%%%%%%%%%%%%%%%%%%%%%%%%%%%%%%%%%%%%%%%%%%%%%%%%%%%%%%%%%

\subsection{Application 4: Differential of a differential form}

Here $F=\RR$, $\Phi \eqnote \ell \in \Omega^1(U)$ (differential form) supposed~$C^1$, $p\in U$, so $\ell(p)\in E^*$.
Its differential at~$p$ in a direction $\vu$ is
$d\ell(p).\vu
= \lim_{h\rar0} {\ell(p+h\vu) - \ell(p)\over h} \;\in\Es.
$
And $(d\ell(p).\vu).\vv = \lim_{h\rar0} {\ell(p+h\vu).\vv - \ell(p).\vv\over h} \;\in\RR$
for all $\vu,\vv\in E$.
 
\mn
{\bf Quantification:}
$(\pi_{ei}(p)$ its the dual basis. 

Call $\ell_i(p)\in\RR$ the components of $\ell(p)$, \ie\ $\ell(p) = \sumin \ell_i(p) \pi_{ei}(p)$.
And call $\ell_{i|j}(p)$ the components of $d\ell(p) \in \calL(E;E^*)$:
\be
\ell = \sumin \ell_i \pi_{ei}, \quad
d\ell.\ve_j = \sumin \ell_{i|j} \pi_{ei},
\quad [d\ell]_{|\ve} = [\ell_{i|j}].
\ee
Tensorial notations, to be used with contractions:
$
d\ell =  \sumijn \ell_{i|j} \pi_{ei}\otimes \pi_{ej}.
$

Duality notations: %$\pi_{ei}(p)=e^i(p)$, 
$\ell = \sum_i \ell_i e^i$, $d\ell.\ve_j = \sumin \ell_{i|j} e^i$, $[d\ell]_{|\ve} = [\ell_{i|j}]$, and
$d\ell =  \sumijn \ell_{i|j} e^i\otimes e^j$.

\mn
{\bf In a Cartesian basis:}
Here $(\ve_i)$ is uniform, so 
\be
\ell_{i|j} %= d\ell_i(p).\ve_j \eqnote  
= {\pa \ell_i \over \pa x_j}(p) \eqnote \ell_{i,j},
\qso [d\ell]_{|\ve} = [ {\pa \ell_i \over \pa x_j}].
\ee
Duality notations: $\ell_{i|j} = d\ell_i.\ve_j={\pa \ell_i \over \pa x^j}$ and $[d\ell]_{|\ve} = [ {\pa \ell_i \over \pa x^j}]$.

\mn
{\bf In a coordinate system basis:} With duality notations and Christoffel symbols:
\be
\label{eqdeijdual}
\boxed{de^i.\ve_j = -\sumkn \gamma^i_{jk} e^k}.
%d\ell.\ve_j =  \sumin \ell_{i|j}\,e^i \qwhere \ell_{i|j} = \sumjn d\ell_i.\ve_j + \sumkn \ell_k \gamma^k_{ji},
\ee
Indeed,
$e^i.\ve_k = \delta^i_k$ gives $(de^i.\ve_j).\ve_k + e^i.(d\ve_k.\ve_j)=0$,
thus $(de^i.\ve_j).\ve_k = -e^i.\sum_\ell \gamma^\ell_{jk}\ve_\ell = -\gamma^i_{jk}$. Thus
\be
\label{eqdeijdual2}
\boxed{\ell_{i|j} = {\pa \ell_i \over \pa q^j} - \sumkn \ell_k \gamma^k_{ji} }
\qwhere {\pa \ell_i \over \pa q^j}(p) := d\ell_i(p).\ve_i(p).
\ee
Indeed,
$\ell = \sum_i \ell_i e^i$ gives
$d\ell.\ve_j
= \sum_i (d\ell_i.\ve_j) e^i + \sum_i \ell_i (de^i.\ve_j)
= \sum_i (d\ell_i.\ve_j) e^i - \sum_{ik} \ell_i \gamma^i_{jk} e^k
$.

%%%%%%%%%%%%%%%%%%%%%%%%%%%%%%%%%%%%%%%%%%%%%%%%%%%%%%%%%%%%%%%%%%%%%%%%%%%%%%%%%%%

\subsection{Application 5: Differential of a 1 1 tensor}

Consider a $C^1$ ${1\choose1}$ tensor $\uutau:
\left\{\eqalign{
 \UE &\rar \calL(\Es,E;\RR) \cr
p &\rar \uutau(p)  \cr
}\right\}
$. %So $\uutau(p)\in\calL(\Es,E;\RR)$ for all $p \in \UE$.
Its differential
$d\uutau:
\left\{\eqalign{
 \UE &\rar \calL(E; \calL(\Es,E;\RR)) \cr
p &\rar d\uutau(p)  \cr
}\right\}
$
is defined by $d\uutau(p).\vu = \lim_{h\rar0} {\uutau (p+h\vu) - \uutau(p) \over h} \in \calL(\Es,E;\RR)$,
so $(d\uutau(p).\vu)(\ell,\vv) = \lim_{h\rar0} {\uutau (p+h\vu)(\ell,\vv) - \uutau(p)(\ell,\vv) \over h}$
($\in\RR$), for all $\vu,\vv\in E$ and $\ell\in E^*$.

\mn
{\bf Quantification} (duality notations):
Basis $(\ve_i(p))$ in~$E$ at~$p$, dual basis $(e^i(p))$,
call $\tau^i_j(p)$ the components of $\uutau(p)$,
call $\tau^i_{j|k}(p)$ the components of $d\uutau(p)$:
\be
\label{eqdifftau}
\uutau = \sum_{ij} \tau_{ij} \ve_i \otimes e^j,\quad
\boxed{d\uutau.\ve_k = \sumijn \tau^i_{j|k} \ve_i\otimes e^j}.
%, \qand d\uutau \eqnote \sumijkn (\tau_e)^i_{j|k} \ve_i\otimes e^j \otimes e^k \in \Tudu.
\ee
Tensorial notations, to be used with contractions:
%With natural canonical isomorphisms, we can write (for computation purposes with contractions rules)
$
d\uutau = \sumijkn \tau^i_{j|k} \ve_i\otimes e^j \otimes e^k.
$
\\
(Classical notations: $\uutau = \sum_{ij} \tau_{ij} \ve_i \otimes \pi_{ej}$,
$d\uutau.\ve_k = \sum_{ij} \tau_{ij|k} \ve_i\otimes \pi_{ej}$, and
$d\uutau = \sum_{ijk} \tau_{ij|k} \ve_i \otimes \pi_{ej} \otimes \pi_{ek}$.)

\mn
{\bf Cartesian basis:} $d\uutau(p).\ve_k = \sum_{ij} (d\tau^i_j(p).\ve_k) \ve_i \otimes e^j$,
so
\be
\tau^i_{j|k} %=  d\tau^i_j.\ve_k 
= {\pa \tau^i_j\over \pa x^k} \eqnote \tau^i_{j,k} \quad (:= d\tau^i_j.\ve_k).
\ee
%$ %\label{eqtauijkd1} \tau^i_{j|k} = {\pa \tau^i_j\over \pa x^k} \quad(\hbox{also written} = \tau^i_{j,k}).$

\noindent
{\bf Coordinate system basis:} $\uutau(p)  = \sumijn \tau^i_j(p) \ve_i(p) \otimes e^j(p)$ gives, for all~$k$,
\be
\eqalign{
d\uutau.\ve_k 
= & \textstyle \sum_{ij} (d\tau^i_j.\ve_k) \ve_i \otimes e^j
+ \sum_{ij} \tau^i_j (d\ve_i.\ve_k) \otimes e^j 
+ \sum_{ij} \tau^i_j \ve_i \otimes (de^j.\ve_k) \cr
= & \textstyle \sum_{ij} (d\tau^i_j.\ve_k) \ve_i \otimes e^j
+ \sum_{ij\ell} \tau^i_j \gamma^\ell_{ki} \ve_\ell \otimes e^j
- \sum_{ij\ell} \tau^i_j \gamma^j_{k\ell} \ve_i \otimes e^\ell \cr
= & \textstyle \sum_{ij} (d\tau^i_j.\ve_k) \ve_i \otimes e^j
+ \sum_{ij\ell} \tau^\ell_j \gamma^i_{k\ell} \ve_i \otimes e^j
- \sum_{ij\ell} \tau^i_\ell \gamma^\ell_{kj} \ve_i \otimes e^j \cr
}
\ee
thus %, with ${\pa \tau^i_j \over \pa q^k}:=d\tau^i_j.\ve_k$,
\be
\label{eqtauijkd2}
\boxed{\tau^i_{j|k} = {\pa \tau^i_j \over \pa q^k}
+ \sumln \tau^\ell_j \gamma^i_{k\ell}
- \sumln \tau^i_\ell \gamma^\ell_{kj}} \qwhere {\pa \tau^i_j \over \pa q^k}:=d\tau^i_j.\ve_k.
\ee
(We have the $+$ sign from vector fields, \cf~\eref{eqdPsi2b}, and the $-$ sign from differential forms, \cf~\eref{eqdeijdual2}.)
%(The $\gamma$s vanish in a Cartesian basis.)

\debexe
If $\vu\in E$, $\ell \in E^*$ then for the elementary ${1\choose1}$ tensor $\uutau = \vu\otimes \ell$ prove:
\be
\label{eqcaldutvek}
d(\vu\otimes \ell).\ve_k = (d\vu.\ve_k) \otimes \ell + \vu \otimes (d\ell.\ve_k), \qand
(\vu\otimes \ell)^i_{j|k} = u^i_{|k}\ell_j + u^i \ell_{j|k},
\ee
when $\vu=\sum_i u^i\ve_i$,
$\ell = \sum_j \ell_j e^j$,
$d\vu.\ve_k = \sum_{i} u^i_{|k} \ve_i$,
$d\ell.\ve_k = \sum_j \ell_{j|k} e^j$.

\debrep
$\uutau= \vu\otimes \ell = \sum_{ij} \tau^i_j \ve_i \otimes e^j$.
 where $\tau^i_j=u^i\ell_j$, and 
$d\uutau.\ve_k = \sumijn \tau^i_{j|k} \ve_i\otimes e^j$ where
$\tau^i_{j|k} = (u^i\ell_j)_{|k} = u^i_{|k}\ell_j + u^i \ell_{j|k}=(\vu\otimes \ell)^i_{j|k}$. Thus
(similar to the derivation of a product):
$$
\eqalign{
d(\vu\otimes \ell)(p).\ve_k(p)
%= \lim_{h\rar0} {\uutau(p{+}h\ve_k(p)) - \uutau(p) \over h}
= & \lim_{h\rar0} {(\vu\otimes \ell)(p{+}h\ve_k(p)) - (\vu\otimes \ell)(p) \over h}
= \lim_{h\rar0} {\vu(p{+}h\ve_k(p))\otimes \ell(p{+}h\ve_k(p)) - \vu(p) \otimes \ell(p) \over h} \cr
= & \lim_{h\rar0} {\vu(p{+}h\ve_k(p))\otimes \ell(p{+}h\ve_k(p)) - \vu(p{+}h\ve_k(p))\otimes \ell(p) \over h}
+ \lim_{h\rar0}{ \vu(p{+}h\ve_k(p))\otimes \ell(p) - \vu(p) \otimes \ell(p) \over h} \cr
= & \lim_{h\rar0} (\vu(p{+}h\ve_k(p))\otimes ({ \ell(p{+}h\ve_k(p)) -  \ell(p) \over h})
+ \lim_{h\rar0} ({\vu(p{+}h\ve_k(p)) - \vu(p) \over h})\otimes \ell(p) \cr
= & \vu(p)\otimes (d\ell(p).\ve_k(p)) + (d\vu(p).\ve_k(p)) \otimes \ell(p),
}
$$
thus~\eref{eqcaldutvek}$_1$.
Which gives $d(\vu\otimes \ell).\ve_k 
= (\sum_{i} u^i \ve_i) \otimes (\sum_j \ell_{j|k} e^j)
+ (\sum_{i} u^i_{|k} \ve_i)\otimes (\sum_j\ell_j  e^j)
$, thus~\eref{eqcaldutvek}$_2$.
\finrep
\finexe

%%%%%%%%%%%%%%%%%%%%%%%%%%%%%%%%%%%%%%%%%%%%%%%%%%%%%%%%%%%%%%%%%%%%%%%%%%%%%%%%%%%

\subsection{Divergence of a vector field: Invariant}

%%%%%%%%%%%%%%%%%%%%%%%%%%%%%%%%%%%%%%%%%%%%%%%%%%%%%%%%%%%%%%%%%%%%%%%%%%%%%%%%%%%

%\subsubsection{Definition}

$\Gamma(\UE)$ is the set of $C^1$ vector fields in~$\UE$, and $\Tr:\calL(E;E)\rar\RR$ is the trace operator.

\debdef
The divergence operator is
\be
\dvg := \Tr \circ d : 
\left\{\eqalign{
\Gamma(\UE) & \rar C^0(\UE;\RR) \cr
\vw & \rar \dvg\vw := \Tr(d\vw), \qie \dvg\vw(p):=\Tr(d\vw(p)).
}\right. %\quad\hbox{donc pour } p\in \Omega,\quad \dvg\vw(p)\eqdef \Tr(d\vw(p)),
\ee
(So $\dvg\vw(p) =$ trace of the endomorphism~$d\vw(p)$).
%i.e. pour $p\in \Omega$ le réel $\dvg\vw(p)$ est la trace de la différentielle de~$\vw$ en~$p$.
%If $\vw$ is unstationary, $\vw:(t,p)\rar \vw(t,p)$, then, for $t$ fixed, $\vw_t(p):=\vw(t,p)$, and $\dvg\vw(t,p) \eqdef \dvg\vw_t(p)$.
\findef

$\Tr$ and~$d$ are linear, hence $\dvg = \Tr \circ d$ is $\RR$-linear (composed of two $\RR$-linear maps).

\debprop
The divergence of a vector field is objective (is an invariant): Same value for all observers (objective quantity) intrinsic to~$\vw$.
\finprop

\debdem
The differential and the trace are objective.
(Or computation: $\vw = \sum_i u^i\va_i = \sum_i v^i \vb_i$ gives $v^i_{|j} = \sum_{k\ell} Q^i_k u^k_{|\ell} P^\ell_j$, see~\eref{eqgamdw}, thus
$\sum_i v^i_{|i} = \sum_{ik\ell} P^\ell_i Q^i_k u^k_{|\ell} 
= \sum_{k\ell} \delta^\ell_k u^k_{|\ell} = \sum_k u^k_{|k}$.)
%(Or direct calculation, see next exercises~\ref{exediverg1}-\ref{exediverg}.)
\findem

\noindent
{\bf Quantification:} $\vw\in\Gamma(\UE)$, $(\ve_i)$ is a basis, $\vw = \sumin w_i\ve_i$ with classical notations, and $w_{i|j}(p)$ are the components of the vector $d\vw(p).\ve_j(p)$ in the basis~$(\ve_i(p))$. Thus
\be
\label{eqdvgw}
%d\vw.\ve_j = \sumin w_{i|j}\ve_i,  \quad [d\vw]_{|\ve} = [w_{i|j}], \quad 
\boxed{\dvg\vw = \sumin w_{i|i}}.
\ee
Duality notations: 
$\vw =\sumin w^i\ve_i$, $d\vw.\ve_j = \sumin w^i_{|j}\ve_i$, $[d\vw]_{|\ve} = [w^i_{|j}]$, $\dvg\vw = \sumin w^i_{|i}$.

\mn
{\bf Cartesian basis} $(\ve_i)$ (classical notations): $dw_i.\ve_j \eqnote {\pa w_i\over \pa x_j}$ and
%Explicit expressions for the $w_{i|j} = w^i_{|j}$:
\be
\label{eqCh1c}
w_{i|j} = {\pa w_i\over \pa x_i}, \qthus \dvg\vw = \sumin {\pa w_i\over \pa x_i}. %, \qso \dvg\ve_i=0.
\ee

\noindent
{\bf Coordinate system basis} $(\ve_i)$ (duality notations): With the Christoffel symbols, \cf~\eref{eqdefCs}, \eref{eqdPsi2b} gives
\be
\label{eqCh1d}
w^i_{|i} = {\pa w^i\over \pa q^i} + \sumin w^k \gamma^i_{ik}, \qthus
\dvg\vw = \sumin {\pa w^i\over \pa q^i} + \sumikn w^k \gamma^i_{ik}.
%\qso \dvg\ve_i=0.
\ee

\debexe
Prove:
\be
\label{eqdwbsc3}
\dvg(f\vw) = df.\vw + f\,\dvg\vw.
\ee

\debrep
$d(f\vw) = \vw \otimes df + f\,d\vw$, % (with the contraction rule and $d(f\vw).\vu = (df.\vu)\vw + f(d\vw.\vu)$), 
thus $\Tr(d(f\vw))=\Tr(\vw \otimes df) + \Tr(f\,d\vw)
= df.\vw + f\,\Tr(d\vw)$. Use a coordinate system if you prefer.
\finrep
\finexe

\comment{
\debexe
\label{exediverg1}
With components, prove that $\dvg\vw$ is invariant  by a change of Cartesian bases,

\debrep
Cartesian bases $(\va_i)$ and~$(\vb_i)$, $\vb_j = \sum_i P_{ij}\va_i$, 
generic notations $df.\va_i = {\pa f \over \pa x_i}$ and $df.\vb_i = {\pa f \over \pa y_i}$,
and $\vw(p) = \sum_i u_i(p)\va_i = \sum_i v_i(p)\vb_i$.
Thus
%$d\vw.\va_k = \sum_\ell (du_\ell.\va_k) \va_\ell = \sum_\ell {\pa u_\ell\over \pa x_k}\va_\ell$ and 
%$d\vw.\vb_j = \sum_i (dv_i.\vb_j) \vb_i = \sum_i {\pa v_i\over \pa y_j}\vb_i$; And
%$\sum_i {\pa v_i\over \pa y_j}\vb_i$
${\pa v_i\over \pa y_j} = d\vw.\vb_j
= \sum_k P_{kj}\,d\vw.\va_k
= \sum_{k\ell} P_{kj} {\pa u_\ell\over \pa x_k}\va_\ell
= \sum_{k\ell i} P_{kj} {\pa u_\ell\over \pa x_k} Q_{i\ell}\vb_i
$
where $Q=P^{-1}$, thus
${\pa v_i\over \pa y_j} = \sum_{k\ell} P_{kj} Q_{i\ell} {\pa u_\ell\over \pa x_k}$;
Hence 
$\dvg_b\vw = \sum_i {\pa v_i\over \pa y_i} 
= \sum_{ik\ell} P_{ki}Q_{i\ell} {\pa u_\ell\over \pa x_k} 
= \sum_{k\ell} \delta_{k\ell} {\pa u_\ell\over \pa x_k} 
= \sum_{k} {\pa u_k\over \pa x_k} = \dvg_a\vw
$, true for all~$\vw$, thus $\dvg_b=\dvg_a$, thus $\dvg_b=\dvg_a\eqnote\dvg: \Tuuu \rar \RR$ is objective.
\finrep
\finexe
\debexe
\label{exediverg}
With components, prove that $\dvg\vw$ is invariant by a change of coordinate system bases.

\debrep
Coordinate system bases $(\va_i(p))$ and~$(\vb_i(p))$ at~$p$, $\vb_j(p)=\sum_i P^i_j(p)\va_i(p)$, $P=[P^i_j]$, $Q=P^{-1} = [Q^i_j]$, and $\vw(p) = \sum_i u^i(p)\va_i = \sum_i v^i(p)\vb_i$,
thus $[\vw]_{|\vb} = Q.[\vw]_{|\va}$, which reads $v^i = \sum_j Q^i_j u^j$. We get
\be
\eqalign{
\sum_i dv^i.\vb_i
= &\sum_{ij} (dQ^i_j.\vb_i) u^j + \sum_{ij} Q^i_j (du^j.\vb_i)
= \sum_{ijk} P^k_i (dQ^i_j.\va_k) u^j + \sum_{ijk} P^k_i Q^i_j(du^j.\va_k)  \cr
= & %\sum_{ijk} P^k_i (dQ^i_j.\va_k) u^j + \sum_{kj} \delta^k_j (du^j.\va_j) 
\sum_{ijk} P^k_i (dQ^i_j.\va_k) u^j + \sum_{j} (du^j.\va_j) \cr
}
\ee
And
$\gamma_{ik,b}^i
\equalref{exochrist2} \sum_{\lambda\mu\nu} Q^i_\nu P^\mu_k P^\lambda_i \gamma_{\lambda\mu,a}^\nu
+ \sum_{\lambda\mu} Q^i_\mu P^\lambda_i (dP^\mu_k.\va_\lambda)
$ gives
\be
\eqalign{
\sum_{ik} v^k \gamma_{ik,b}^i
= &\sum_{ikj\lambda\mu\nu} Q^k_j u^j Q^i_\nu P^\mu_k P^\lambda_i \gamma_{\lambda\mu,a}^\nu
+ \sum_{ikj\lambda\mu} Q^k_j u^j Q^i_\mu P^\lambda_i (dP^\mu_k.\va_\lambda) \cr
= &\sum_{j\lambda\mu\nu} \delta^\mu_j \delta^\lambda_\nu \gamma_{\lambda\mu,a}^\nu u^j
+ \sum_{kj\lambda\mu} Q^k_j \delta^\lambda_\mu (dP^\mu_k.\va_\lambda) u^j 
= \sum_{j\lambda} \gamma_{\lambda j,a}^\lambda u^j
+ \sum_{kj\lambda} Q^k_j (dP^\lambda_k.\va_\lambda) u^j \cr
}
\ee
Thus %, with $v^i = \sum_j Q^i_j u^j$
\be
\eqalign{
\dvg_b\vw 
= &  \sumin dv^i.\vb_i + \sumikn v^k \gamma^i_{ik,b} 
=  \sum_{ijk} P^k_i (dQ^i_j.\va_k) u^j + \sum_{j} (du^j.\va_j)
+ \sum_{j\lambda} \gamma_{\lambda j,a}^\lambda u^j
+ \sum_{kj\lambda} Q^k_j (dP^\lambda_k.\va_\lambda) u^j \cr
= & \dvg_a\vw  + \sum_{ijk} P^k_i (dQ^i_j.\va_k) u^j + \sum_{ijk} Q^i_j (dP^k_i.\va_k) u^j. \cr
}
\ee
And $\sum_{i}P^k_i Q^i_j  = (P.Q)^k_j = \delta^k_j$ gives
$\sum_{i}d(P^k_i Q^i_j).\va_k = 0 = \sum_{i}(dP^k_i.\va_k) Q^i_j + \sum_i P^k_i(dQ^i_j.\va_k)$, thus
$\sum_{i}(dP^k_i.\va_k) Q^i_j u^j + \sum_i P^k_i(dQ^i_j.\va_k) u^j = 0$,
thus
$\dvg_b\vw =\dvg_a \vw$.
True for all~$\vw$, thus $\dvg_b=\dvg_a$, thus $\dvg_b=\dvg_a\eqnote\dvg: \Tuuu \rar \RR$ is objective.
\comment{
And \eref{eqdwbsc3} gives
$\dvg_b\vw
= \dvg_b(\sum_j v^j\vb_j)
= \sum_j dv^j.\vb_j + \sum_j v^j \dvg_b\vb_j
= \sum_i d(\sum_i Q^j_i u^i).\vb_j + \sum_j (\sum_i Q^j_i u^i) \dvg_b\vb_j
$,
thus
\be
\eqalign{
\dvg_b\vw 
= & \sum_{ij} (dQ^j_i.\vb_j) u^i
+ \sum_{ij} Q^j_i(du^i.\vb_j)
+ \sum_{ij} Q^j_i u^i \dvg_b\vb_j \cr
= & \sum_{ijk} P^k_j (dQ^j_i.\va_k) u^i 
+ \sum_{ijk} P^k_j Q^j_i(du^i.\va_k)
+ \sum_{ij\alpha\beta} Q^j_i u^i P^\alpha_j \gamma_{\beta\alpha,a}^\beta
+ \sum_{ijk} Q^j_i u^i (dP^k_j.\va_k) \cr
= & \sum_{i} du^i.\va_i
+ \sum_{i\beta}   u^i \gamma_{\beta i,a}^\beta
+ \sum_{ijk} P^k_j (dQ^j_i.\va_k) u^i 
+ \sum_{ijk} Q^j_i u^i (dP^k_j.\va_k) \cr
= & \dvg_a\vw
+ \sum_{ijk} u^i (P^k_j (dQ^j_i.\va_k) +  Q^j_i (dP^k_j.\va_k))
= \dvg_a\vw + \sum_{ijk} u^i d(P^k_j Q^j_i).\va_k, \cr
}
\ee
and $\sum_{jk}P^k_j Q^j_i  = \sum_k (P.Q)^k_i = (PQ)^i_i=1$, thus $\sum_{ijk} u^i d(P^k_j Q^j_i).\va_k=0$,
 true for all~$\vw$, thus $\dvg_b=\dvg_a$, and $\dvg_b=\dvg_a\eqnote\dvg$ is objective.
}
\finrep
\finexe
}

\debrem
\label{remda}
If $\alpha$ is a differential form, if $(\ve_i)$ is a basis and $(e^i)$ its dual basis,
and if $\alpha = \sumin \alpha_i e^i$, then $d\alpha = \sumin \alpha_{i|j} e^i \otimes e^j$,
with $\alpha_{i|j}:= \ve_i.d\alpha.\ve_j$.
Here it is impossible to define an objective trace of $d\alpha$ like $\sumin \alpha_{i|i}$: The result depends on the choice of the basis (the Einstein convention is not satisfied, and \eg\ with a Euclidean basis the result depends on the choice of unit of length: Foot? Meter?). Thus the objective (or intrinsic) divergence of a differential form is a nonsense.
\finrem

%%%%%%%%%%%%%%%%%%%%%%%%%%%%%%%%%%%%%%%%%%%%%%%%%%%%%%%%%%%%%%%%%%%%%%%%%%%%%%%%%%%

\subsection{Objective divergence for 1 1 tensors}
\label{secdiv}

To create an objective divergence for a second order ${1\choose1}$ tensor $\uutau \in \Tuuu$, in~\eref{eqdifftau} we have to contract an admissible index with the ``differential index~$k$'', So, no choice: Contract $i$ and~$k$ to get $\tdvg\uutau  := \sumijn \tau^i_{j|i} e^j$. 
Let us start with:
%(so that English and French observers can work together with their foot and metre)

\debdef
\label{defdvgul}
Let $\vu\in\Gamma(\UE)$ and $\ell\in\Omega^1(\UE)$ be $C^1$.
The objective divergence of the elementary ${1\choose1}$ tensor $\vu\otimes\ell \in \Tuuu$ is the differential form
$\tdvg(\vu\otimes \ell)\in \Omega^1(\UE)$ defined by
\be
\label{eqdvgul}
\tdvg(\vu\otimes \ell) = (\dvg\vu)\ell +  d\ell.\vu,
\ee
\ie\ defined by $\tdvg(\vu\otimes \ell).\vw = (\dvg\vu)(\ell.\vw) +  (d\ell.\vw)\vu$ for all $\vw\in E$.
And the objective divergence operator 
$\tdvg:
\left\{\eqalign{
\Tuuu &\rar \Omega^1(\UE) \cr
\uutau & \rar \tdvg\uutau \cr
}\right\}
$ is the linear map defined on elementary tensors with~\eref{eqdvgul}.
%any tensor being a sum of elementary tensors.
\findef

\noindent
{\bf Quantification:} 
$(\ve_i)$ is a basis, $(e^i)$ its dual basis,
$\vu= \sum_i u^i \ve_i$,
$\ell = \sum_j \ell_j e^j$.
Thus $\vu\otimes \ell
%= \sum_{ij} (\vu\otimes \ell)^i_j \ve_i\otimes e^j
= \sum_{ij} u^i\ell_j \ve_i\otimes e^j
$,
and \eref{eqcaldutvek}-\eref{eqdvgul} give
\be
\label{eqdeftdvgtau0}
\tdvg(\vu\otimes \ell)
= \sumijn (u^i_{|i}\ell_j + \ell_{j|i} u^i) e^j.
%\quad (= \sumijn (\vu\otimes \ell)^i_{j|i} e^j).
\ee
So for an elementary tensor $\uutau = \vu\otimes \ell$, $\uutau = \sum_{ij}\tau^i_j \ve_i\otimes e^j$,
and $d\uutau.\ve_k = \sum_{ijk} \tau^i_{j|k} \ve_i\otimes e^j$
and $\tdvg(\uutau)=\sum_{ij}\tau^i_{j|i} e^i$, with
$\tau^i_{j|k} = u^i_{|k}\ell_j + u^i \ell_{j|k}$, here with $\tau^i_j = u^i\ell_j$,
and $\tau^i_{j|k} = u^i_{|k}\ell_j + u^i \ell_{j|k}$, so
$\tau^i_{j|i} = u^i_{|i}\ell_j + u^i \ell_{j|i}$. 

Thus, by linearity of~$\tdvg$, for all tensors $\uutau\in\Tuuu$, we have with~\eref{eqdifftau}:
\be
\label{eqdeftdvgtau}
\boxed{\tdvg\uutau = \sumijn \tau^i_{j|i} e^j}, \qie
[\tdvg\uutau]_{|\ve} = \pmatrix{\sum_i\tau^i_{1|i} & ... \sum_i\tau^i_{n|i}}
\ee
(row matrix since $\tdvg\uutau$ is a differential form). %(See exercise~\ref{exetdvgtau}.)
\Ie, we have contracted $i$~and~$k$ in~\eref{eqdifftau}.

(Classical notations: $\tdvg\uutau  := \sumijn \tau_{ij|i} \pi_{ej}$, \ie\ $[\tdvg\uutau]_{|\ve} = \pmatrix{\sum_i\tau_{i1|i} & ... \sum_i\tau_{in|i}}$.)

\comment{
\debrem
\eref{eqdeftdvgtau} can be taken to be the definition of the objective divergence of a ${1\choose1}$ tensor $\uutau\in\Tuuu$
given in a basis $(\ve_i)$ by $\uutau= \sum_{ij}\tau^i_j \ve_i\otimes e^j$.
\finrem
}

So %, with Cartesian basis and coordinate system basis, \eref{eqtauijkd1}-\eref{eqtauijkd2} give
\be
\label{eqdvguutaucoor}
\eqalign{
&\hbox{Cartesian bases: } 
\tdvg\uutau = \sumijn {\pa \tau^i_j\over \pa x^i} e^j , \cr 
&\hbox{Coord. sys. bases: } 
\tdvg\uutau = \sumijn \bigl({\pa \tau^i_j\over \pa q^i}
+ \sumkn \tau^k_j \gamma^i_{ik}
- \sumkn \tau^k_i \gamma^i_{kj} \bigr) e^j.
}
\ee
Indeed: With $\uutau = \sum_j(\sum_i \tau^i_j\ve_i)\otimes e^j
= \sum_j \vw_j \otimes e^j$ 
where $\vw_j=\sum_i \tau^i_j\ve_i$, the linearity of~$\tdvg$ gives
$\tdvg\uutau = \sum_j \tdvg(\vw_j \otimes e^j)$; Thus, with~\eref{eqdvgul}:
1- 
Cartesian basis:
$\dvg\vw_j = \sum_i {\pa \tau^i_j \over \pa x^i} $
and $de^j=0$ give
$\tdvg\uutau = \sum_j\sum_i {\pa \tau^i_j \over \pa x^i}e^j = \sum_{ij} \tau^i_{j,i} e^j$, thus~\eref{eqdvguutaucoor}$_1$; And
2- 
Coordinate system basis:
$\dvg\vw_j 
\equalref{eqCh1d} \sum_i {\pa \tau^i_j \over \pa q^i} + \sum_{ik} \tau^k_j \gamma^i_{ik}$
and $de^j.\vw_j
= \sum_k \tau^k_j de^j.\ve_k
= \sum_k \tau^k_j (-\sum_i \gamma^j_{ki} e^i)$,
thus $\sum_j de^j.\vw_j
= -\sum_{ijk} \tau^k_j \gamma^j_{ki} e^i
= -\sum_{ijk} \tau^k_i \gamma^i_{kj} e^j
$, thus~\eref{eqdvguutaucoor}$_2$.

\debexe
Prove: If $f\in C^1(\UE;\RR)$ and $\uutau = \sumijn \tau^i_j \ve_i\otimes e^j\in \Tuuu \cap C^1$ then
\be
\label{eqdefdvi3c}
\tdvg(f\uutau) = df.\uutau + f\,\tdvg \uutau.
\ee

\debrep
%$df = \sumin f_{|i} e^i$, $\uutau = \sumijn \tau^i_j \ve_i\otimes e^j$,
%$d\uutau = \sumijkn \tau^i_{j|k} \ve_i\otimes e^j \otimes e^k$, so on the one hand 
$f\uutau = \sum_{ij} f\tau^i_j \ve_i\otimes e^j$ gives
$d(f\uutau)
= \sum_{ijk} (f\tau^i_{j})_{|k} \ve_i\otimes e^j \otimes e^k
=\sum_{ijk} (f_{|k} \tau^i_{j} + f\tau^i_{j|k}) \ve_i\otimes e^j \otimes e^k$, thus
$\tdvg(f\uutau) = \sum_{ij} (f_{|i} \tau^i_j + f\tau^i_{j|i}) e^j$;
And
$df.\uutau + f\,\tdvg \uutau
= \sum_{ij} f_{|i}\tau^i_{j}  e^j + f\, \sum_{ij} \tau^i_{j|i} e^j$.
\finrep
\finexe

\debexe
Prove: If $\uutau\in \Tuuu$ and $\vw\in\Gamma(\UE)$ then
\be
\label{eqdtw0}
\boxed{\dvg(\uutau.\vw) = \tdvg(\uutau).\vw + \uutau \odd d\vw}.
\ee

\debrep
$\uutau = \sum_{ij} \tau^i_j \ve_i\otimes e^j$ and $\vw=\sum_i w^i\ve_i$ give
$\uutau.\vw = \sum_{ij} \tau^i_j w^j \ve_i$, thus
$\dvg(\uutau.\vw) = \sum_{ij} \tau^i_{j|i} w^j + \tau^i_j w^j_{|i}$.
\finrep
\finexe

\debexe
If $\uutau \in \Tuuu$ check with component calculations (since $\tdvg(\uutau)$ is objective):
\be
\label{eqptdvg2}
[\tdvg(\uutau)]_{|b} = [\tdvg(\uutau)]_{|a}.P \quad (\hbox{covariance formula}),
\ee
where $P$ is the transition matrix from a basis~$(\va_i)$ to a basis~$(\vb_i)$.

\debrep
Let $\uutau = \sum_{ij} \sigma^i_j \va_i \otimes a^j = \sum_{ij} \tau^i_j \vb_i \otimes b^j$,
so $\tau^i_j = \sum_{\lambda\mu} Q^i_\lambda \sigma^\lambda_\mu P^\mu_j$.

1- Cartesian bases: $\sum_i \tau^i_{j|i} = \sum_i d\tau^i_j.\vb_i
= \sum_i d(\sum_{\lambda\mu} Q^i_\lambda \sigma^\lambda_\mu P^\mu_j).(\sum_\nu P^\nu_i.\va_\nu)
= \sum_{i\lambda\mu \nu} Q^i_\lambda P^\mu_j P^\nu_i (d \sigma^\lambda_\mu.\va_\nu)
= \sum_{\lambda\mu \nu} \delta^\nu_\lambda P^\mu_j (d \sigma^\lambda_\mu.\va_\nu)
= \sum_{\lambda\mu} P^\mu_j (d \sigma^\lambda_\mu.\va_\lambda)
= \sum_{\mu}(\sum_{\lambda} \sigma^\lambda_{\mu|\lambda}) P^\mu_j
$ as desired.

2- Coordinate system bases:
$\sum_i \tau^i_{j|i} \equalref{eqdvguutaucoor}
\sum_i d\tau^i_j.\ve_i + \sum_{i\ell} \tau^\ell_j \gamma^i_{i\ell,b} - \sum_{i\ell} \tau^i_\ell \gamma^\ell_{ij,b}$
(with $j$ fixed); With
$$
\eqalign{
\sum_{i} (d\tau^i_j.\vb_i)
= &
\sum_{i\lambda\mu} Q^i_\lambda\, (d\sigma^\lambda_\mu.\vb_i)\, P^\mu_j
+ \sum_{i\lambda\mu} (dQ^i_\lambda.\vb_i)\, \sigma^\lambda_\mu\, P^\mu_j
+ \sum_{i\lambda\mu} Q^i_\lambda\, \sigma^\lambda_\mu\, (dP^\mu_j.\vb_i)
\cr
= &
\sum_{i\lambda\mu\nu} Q^i_\lambda P^\mu_j P^\nu_i (d\sigma^\lambda_\mu.\va_\nu)
+ \sum_{i\lambda\mu\nu} \sigma^\lambda_\mu\, P^\mu_j P^\nu_i (dQ^i_\lambda.\va_\nu)
+ \sum_{i\lambda\mu\nu} \sigma^\lambda_\mu\, Q^i_\lambda P^\nu_i (dP^\mu_j.\va_\nu)
\cr
= &
\sum_{\lambda\mu} P^\mu_j (d\sigma^\lambda_\mu.\va_\lambda)
- \sum_{i\lambda\mu\nu} \sigma^\lambda_\mu\, P^\mu_j Q^i_\lambda (dP^\nu_i.\va_\nu)
+ \sum_{\lambda\mu} \sigma^\lambda_\mu\, (dP^\mu_j.\va_\lambda)
\cr
}
$$
since $P^\nu_i Q^i_\lambda = \delta^\nu_\lambda$ gives
$P^\nu_i (dQ^i_\lambda.\va_\nu) - Q^i_\lambda (dP^\nu_i.\va_\nu)$.
And, with \eref{exochrist2},
\be
\eqalign{
\sum_{i\ell} \tau^\ell_j \gamma^i_{i\ell,b}
= &
\sum_{i\ell}(\sum_{\lambda\mu} Q^\ell_\lambda \sigma^\lambda_\mu P^\mu_j)
(\sum_{\alpha\beta\omega}
Q^i_\alpha P^\beta_i P^\omega_\ell \gamma_{\beta\omega,a}^\alpha
+ \sum_{\alpha\beta} Q^i_\alpha P^\beta_i (dP^\alpha_\ell.\va_\beta))
\cr
= & 
\sum_{\lambda\mu\alpha} \sigma^\lambda_\mu  P^\mu_j  \gamma_{\alpha\lambda,a}^\alpha
+ \sum_{\ell\lambda\mu\alpha}  \sigma^\lambda_\mu Q^\ell_\lambda P^\mu_j (dP^\alpha_\ell.\va_\alpha),
}
\ee
and
\be
\eqalign{
- \sum_{i\ell} \tau^i_\ell \gamma^\ell_{ij,b}
= & -
\sum_{i\ell}  (\sum_{\lambda\mu} Q^i_\lambda \sigma^\lambda_\mu P^\mu_\ell)
(\sum_{\alpha\beta\omega} P^\alpha_i P^\beta_j Q^\ell_\omega \gamma_{\alpha\beta,a}^\omega
+ \sum_{\alpha\omega} P^\alpha_i Q^\ell_\omega (dP^\omega_j.\va_\alpha))
\cr
= & -
\sum_{\lambda\mu\beta} \sigma^\lambda_\mu P^\beta_j  \gamma_{\lambda\beta,a}^\mu
- \sum_{\lambda\mu} \sigma^\lambda_\mu   (dP^\mu_j.\va_\lambda).
}
\ee
Thus
$
\sum_i \tau^i_{j|i} = 
\sum_{\lambda\mu} P^\mu_j (d\sigma^\lambda_\mu.\va_\lambda)
+ \sum_{\lambda\mu\alpha} \sigma^\lambda_\mu  P^\mu_j  \gamma_{\alpha\lambda,a}^\alpha
- \sum_{\lambda\mu\beta} \sigma^\lambda_\mu P^\beta_j  \gamma_{\lambda\beta,a}^\mu
= \sum_{\lambda\mu} P^\mu_j \sigma^\lambda_{\mu|\lambda}
$ as desired.\finrep
\finexe

\comment{
The differential form $\tdvg_e(\uutau)$ relative to~$(\ve_i)$, \cf~\eref{eqdefdvi3},
does not depends on the chosen basis~$(\ve_i)$, that is, if $(\va_i)$ and $(\vb_i)$ are bases then
\be
[\tdvg_e(\uutau)]_{|\va} = [\tdvg_e(\uutau)]_{|\vb}.P, \qand \tdvg_e(\uutau) \eqnote \tdvg\uutau (\in \Tzuu).
\ee
\be
\label{eqptdvg}
\tdvg_a(\uutau) = \tdvg_b(\uutau) \eqnote \tdvg\uutau \in \Tzuu,
\ee
that is, for all $\vw\in\Gamma(\UE)$, and $\vw=\sumjn (w_e)^j\ve_j$,
\be
\tdvg_a(\uutau).\vw = \tdvg_b(\uutau).\vw \eqnote \tdvg(\uutau).\vw =  \sumijn (\tau_e)^i_{j|i}(w_e)^j.
\ee
That is, if $P=[P^i_j]$ is the transition matrix from $(\va_i)$ to~$(\vb_i)$, then
\be
\label{eqptdvg2}
%[\tdvg_b(\uutau)]_{|b} = [\tdvg_a(\uutau)]_{|a}.P,\qs 
[\tdvg(\uutau)]_{|b} = [\tdvg(\uutau)]_{|a}.P.
\ee
Thus,  the ``objective divergence'' $\tdvg\uutau$, of a ${1\choose1}$ tensor~$\uutau$, is objective.
(Einstein convention is satisfied.)
}

\comment{
\debexa Suite.
Le laplacien d'une fonction $f$ $C^2$ est défini dans un cadre euclidien
avec une base euclidienne par (\cf~\S~suivant) :
\be
\label{eqdefDelta}
\Delta f = \sumjn f_{|jj} = \sumjn {\pa ^2 f \over \pa (x^j)^2},
\ee
valeur qui dépend de l'unité de mesure choisie (on divise par une longueur au carré : dans quelle unité ?).
Donc $\tdvg(d\vw)$ (divergence objective) n'est pas le laplacien
$\Delta\vw = \sumijn w^i_{|jj}\ve_i = \sumijn \Delta w^i\ve_i$, \cf~\eref{eqtdvgdvq}
(on~a $\Delta\vw \eqnote \dvg_e(\vgrad\vw)$ dans le cadre cartésien euclidien \cf~\S~suivant).
\finexa
}

%%%%%%%%%%%%%%%%%%%%%%%%%%%%%%%%%%%%%%%%%%%%%%%%%%%%%%%%%%%%%%%%%%%%%%%%%%%%%%%%%%%

\subsubsection{Divergence of a 2 0 tensor}

Let $\uutau \in \Tdzu$ and $\uutau = \sumijn \tau^{ij}\ve_i \otimes \ve_j$,
thus $d\uutau = \sumijkn \tau^{ij}_{\;\;|k} \ve_i \otimes \ve_j \otimes e^k$;
Then two objective divergences may be defined: by contracting $k$ with~$i$,
or $k$ with~$j$. (The Einstein convention is then satisfied.)

%%%%%%%%%%%%%%%%%%%%%%%%%%%%%%%%%%%%%%%%%%%%%%%%%%%%%%%%%%%%%%%%%%%%%%%%%%%%%%%%%%%

\subsubsection{Divergence of a 0 2 tensor}

Let $\uutau = \sumijn \tau_{ij} e^i \otimes e^j \in \Tzdu$.
Thus $d\uutau = \sumijkn \tau_{ij|k} e^i \otimes e^j \otimes e^k$,
and there are no indices to contract to satisfy Einstein convention:
There is no objective divergence of 0 2 tensors.

%%%%%%%%%%%%%%%%%%%%%%%%%%%%%%%%%%%%%%%%%%%%%%%%%%%%%%%%%%%%%%%%%%%%%%%%%%%%%%%%%%%

\subsection{Euclidean framework and ``classic divergence'' of a tensor (subjective)}
\label{secdivcla}

\def\dvga{\dvg_{\!a}}
\def\dvgb{\dvg_{\!b}}
\def\dvge{\dvg_{\!e}}

Let $\uusigma$ be a $C^1$ tensor of order 2 of any kind.
An observer chooses a (Cartesian) Euclidean basis~$(\ve_i)$ and call $\dd_g$ the associated Euclidean dot product.
And he calls $\sigma_{ij}$ the components of~$\uusigma$,
\eg\ writes $\uusigma.\ve_j = \sum_i \sigma_{ij}\ve_i$.

\debdef (Usual divergence in classical mechanics.)
The divergence $\dvg \uusigma$ of~$\uusigma$ relative to the basis~$(\ve_i)$,
is the column matrix (it is not a vector)
\be
\label{eqdivmec}
%\hbox{si}\quad [\uusigma]_{|\ve} = [\sigma^i_j] \qalors
\dvge\uusigma
%= \pmatrix{\sumjn \sigma^1_{j|j} \cr \vdots \cr \sumjn \sigma^n_{j|j} }
= \pmatrix{\sumjn {\pa \sigma_{1j} \over \pa x^j} \cr \vdots \cr 
\sumjn {\pa \sigma_{nj} \over \pa x^j}} \eqnote \dvg\uusigma \quad\hbox{(a matrix)}.
%\eqnote [\dvg_e\uusigma].
\ee
(Take the divergences of the ``row vectors''  of $[\uusigma]_{|e}= [\sigma_{ij}]$
to make the ``column vector'' $[\dvg_e \uusigma]$.)
\findef

\debprop
\label{propdiv}
The ``so called vector'' $\dvg\uusigma$, in~\eref{eqladvgm}, is not a vector: It
does not satisfy the change of basis formula:
If $(\va_i)$ and $(\vb_i)$ are bases, if $P$ is the transition matrix from~$(\va_i)$ to~$(\vb_i)$, if $[\uusigma]_{|\va} = [A_{ij}]$ and $[\uusigma]_{|\vb} = [B_{ij}]$,
with the divergence of $\uusigma$ relative to~$(\va_i)$ and $(\vb_i)$ called $\dvga\uusigma$ and $\dvgb\uusigma$,
then neither a contravariant nor a covariant change of basis formula applies in general:
\be
\label{eqdivmec2}
\hbox{neither}\quad[\dvgb \uusigma]_{|\vb} \ne P^{-1}.[\dvga \uusigma]_{|\va}\quad \hbox{nor}\quad
[\dvgb\uusigma]_{|\vb}^T = [\dvga\uusigma]_{|\va}^T.P
\ee
(compare with~\eref{eqptdvg2}). So $\dvg \uusigma$ as given in~\eref{eqladvgm} is neither a contravariant vector nor a covariant vector (it is just a matrix which depends on an observer).
% Recall: The introduction of a Euclidean basis (it is the case here) and the implicit use of the associated Euclidean dot product produce ``subjective'' results, \ie\ results which depend on an observer.
\finprop

\debdem
Consider the simple case $\vb_i=\lambda \va_i$, for all $i$, $\lambda >1$: %, \cf~\eref{eqexa0a}.
Transition matrix $P=\lambda I$, and $P^{-1} = {1\over \lambda}I$.

For a ${1\choose1}$ tensor: 
$\uusigma = \sum_{ij} (\sigma_b)^i_j \vb_i \otimes b^j = \sum_{ij} (\sigma_a)^i_j \va_i \otimes a^j$, $[\uusigma]_{|\vb} = P^{-1}.[\uusigma]_{|\va}.P = {1\over\lambda}.[\uusigma]_{|\va}.\lambda = [\uusigma]_{|\va}$,
\ie\ $(\sigma_a)^i_j = (\sigma_b)^i_j$ for all $i,j$.
Thus \eref{eqladvgm} gives
$\dvgb\uusigma = \sum_{ij} (d(\sigma_b)^i_j.\vb_j) \vb_i
%= \sum_{ij} (d(\sigma_a)^i_j.\vb_j) \vb_i
= \sum_{ij} (d(\sigma_a)^i_j.(\lambda\va_j)) (\lambda\va_i)
%= \lambda^2 \sum_{ij} (d(\sigma_a)^i_j.\va_j) \va_i
= \lambda^2 \dvga\uusigma.
$
Thus $[\dvgb\uusigma]_{|\vb} \ne P^{-1}.[\dvgb\uusigma]_{|\va}$ and $[\dvgb\uusigma]_{|\vb}^T \ne [\dvga\uusigma]_{|\va}^T.P$.

For a ${0\choose2}$ tensor: 
$\uusigma = \sum_{ij} \sigma_{b,ij} b^i \otimes b^j = \sum_{ij} \sigma_{a,ij} a^i \otimes a^j$, and
$[\uusigma]_{|\vb} = P^T.[\uusigma]_{|\va}.P = \lambda^2[\uusigma]_{|\va}$,
\ie\ $\sigma_{b,ij} = \lambda^2\sigma_{a,ij}$ for all $i,j$.
Thus \eref{eqladvgm} gives
$\dvgb\uusigma = \sum_{ij} (d\sigma_{b,ij}.\vb_j) \vb_i
= \lambda^2 \sum_{ij} (d\sigma_{a,ij}.(\lambda\va_j)) (\lambda\va_i)
%= \lambda^4 \sum_{ij} (d\sigma_{a,ij}.\va_j) \va_i
= \lambda^4 \dvga\uusigma.
$
Thus $[\dvgb\uusigma]_{|\vb} \ne P^{-1}.[\dvgb\uusigma]_{|\va}$ and $[\dvgb\uusigma]_{|\vb}^T \ne [\dvga\uusigma]_{|\va}^T.P$.

For a ${2\choose0}$ tensor: 
$\uusigma = \sum_{ij} \sigma_b^{ij} \vb_i \otimes \vb_j = \sum_{ij} \sigma_a^{ij} \va_i \otimes \va_j$, and
$[\uusigma]_{|\vb} = P^{-T}.[\uusigma]_{|\va}.P^{-1} = {1\over \lambda^2}[\uusigma]_{|\va}$,
\ie\ $\sigma_b^{ij} = {1\over \lambda^2}\sigma_a^{ij}$ for all $i,j$.
Thus \eref{eqladvgm} gives
$\dvgb\uusigma = \sum_{ij} (d\sigma_b^{ij}.\vb_j) \vb_i
= {1\over \lambda^2} \sum_{ij} (d\sigma_a^{ij}.(\lambda\va_j)) (\lambda\va_i)
%=  \sum_{ij} (d\sigma_a^{ij}.\va_j) \va_i
=  \dvga\uusigma.
$
Thus $[\dvgb\uusigma]_{|\vb} \ne P^{-1}.[\dvgb\uusigma]_{|\va}$ and $[\dvgb\uusigma]_{|\vb}^T \ne [\dvga\uusigma]_{|\va}^T.P$.
\findem

Remark: \eref{eqdivmec} can be written
\be
\label{eqladvgm}
\dvg\uusigma = \sum_{ij} {\pa \sigma_{ij} \over \pa x_j}\vE_i
\ee
where $(\vE_i)$ is the canonical basis in $\Mnu$ the space of $n*1$ column vectors.

%%%%%%%%%%%%%%%%%%%%%%%%%%%%%%%%%%%%%%%%%%%%%%%%%%%%%%%%%%%%%%%%%%%%%%%%%%%%%%%%%%%
%%%%%%%%%%%%%%%%%%%%%%%%%%%%%%%%%%%%%%%%%%%%%%%%%%%%%%%%%%%%%%%%%%%%%%%%%%%%%%%%%%%

\section{Natural canonical isomorphisms}
\label{secisnn}

\def\calZ{{\cal Z}}
\def\calLic{{\calL_i}}

%%%%%%%%%%%%%%%%%%%%%%%%%%%%%%%%%%%%%%%%%%%%%%%%%%%%%%%%%%%%%%%%%%%%%%%%%%%%%%%%%%%

\subsection{The adjoint of a linear map}
%\label{secinterMs}

Setting of~\S~\ref{secadjlm}: $E$ and $F$ are vector spaces, $\Es=\calL(E;\RR)$ and $\Fs=\calL(F;\RR)$ are their dual spaces,
and the adjoint of a linear map $\calP\in\calL(E;F)$ is the linear map $\calP^* \in \calL(F^*;E^*)$ canonically defined by
\be
\label{eqdefcalIs}
\forall \ell \in \Fs,\quad
\calP^*(\ell) := \ell \circ \calP, \qwritten \calP^*.\ell = \ell.\calP
\ee
(dot notations $\calP^*(\ell) \eqnote \calP^*.\ell$ and $\ell \circ \calP \eqnote \ell.\calP$ since $\ell$ and $\calP^*$ are linear),
\ie, for all $(\ell,\vu)\in \Fs \times E$,
\be
\calP^*(\ell)(\vu) = \ell(\calP(\vu)), \qwritten (\calP^*.\ell).\vu = \ell.\calP.\vu.
\ee
Interpretation: If $\calP$ is the push-forward of vector fields, then $\calP^*$ is the pull-back of differential forms, see~remark~\ref{rempb}.
In particular, it will be interpreted with $\calP\in\calLic(E;F)$ (linear and invertible = a change of observer).

\comment{
And if $(\va_i)$ is a basis in~$E$ and $[\calP]_{|\va} = [P_{ij}]$,
that is, $\calP.\va_j = \sumin P_{ij}\va_i$ for all~$j$, then with $(\pi_{ai})$ the dual basis of~$(\va_i)$
(that is $\pi_{ai}\in E^*=\calL(E;\RR)$ and $\pi_{ai}(\va_j)=\delta_{ij}$ for all~$i,j$), we have, \cf~\eref{defadjLq},
\be
[\calP^*]_{|\pi_a} = ([\calP]_{|\va})^T,
\ee
that is, if $\calP^*.\pi_{aj} = \sumin (\calP^*)_{ij}\pi_{ai}$ for all~$j$, then $(\calP^*)_{ij} = P_{ji}$ for all $i,j$.
Moreover, if $\vb_j=\calP.\va_j$ (and $(\vb_j)$ is a new basis), then we have $[\calP]_{|\vb} =[\calP]_{|\va}$ ($= [P_{ij}]$), see~\eref{eqdefP0i}, thus $[\calP^*]_{|\pi_a}=[\calP^*]_{|\pi_b}$.
}

%%%%%%%%%%%%%%%%%%%%%%%%%%%%%%%%%%%%%%%%%%%%%%%%%%%%%%%%%%%%%%%%%%%%%%%%%%%%%%%%%%%

\subsection{An isomorphism $E\simeq E^*$ is never natural (never objective)}
\label{secEEsnonnat}

%%%%%%%%%%%%%%%%%%%%%%%%%%%%%%%%%%%%%%%%%%%%%%%%%%%%%%%%%%%%%%%%%%%%%%%%%%%%%%%%%%%

%\subsubsection{Definition}

Two observers $A$ and~$B$ consider a linear map $L \in \calL(E;E^*)$;
Let $\calP \in \calL(E;E)$ be the change of observer endomorphism.
Willing to work together, $A$ and $B$ (``naturally'') consider the diagram
\be
\label{eqdiagnc}
\eqalign{
E\;\mrar^{\ds\; L } \;& E^* \qquad\qquad\leftarrow\hbox{considered by observer $A$}
\cr
\noalign{\vskip-6pt}
\calP \downarrow \qquad\quad &\;\;\uparrow \calP^* \cr
\noalign{\vskip-4pt}
E\;\mrar_{\ds\; L} \;& E^* \qquad\qquad\leftarrow\hbox{considered by observer $B$}
\cr
}
\ee

\debdef
%\label{defdiagnats}
(Spivak~\cite{spivak}.) A linear map $L\in\calL(E;E^*)$ is natural iff the diagram~\eref{eqdiagnc} commutes for all~$\calP \in \calL(E;E)$:
\be
\hbox{$L \in \calL(E;E^*)$ is natural} \;\; \Longleftrightarrow \;\; \forall \calP\in\calL(E;E),\;\; \calP^*\circ L \circ \calP = L.
\ee
(In that case, if $A$ computes $L.\vu$ with the top line of the diagram, if $B$ computes with the bottom line of the diagram, then they can easily check their results since here $L.\vu=(\calP^*\circ L \circ \calP).\vu$.)
\findef

\noindent
{\bf Question:} Does there exist an endomorphism~$L$ such that the diagram~\eref{eqdiagnc} commutes for all change of observers? That is, do we have
\be
\label{eqquestionnat}
\exists? L\in\calL(E;E),\;\forall \calP\in\calLic(E;E),\quad \calP^*\circ L \circ \calP = L\; ?
\ee
{\bf Answer:} Always {\large \textslbf{no}} (if $L\ne 0$):

\debthm
\label{propspivak}
A (non-zero) linear map $L\in\calL(E;E^*)$ is not natural: If $L\in\calL(E;E^*)-\{0\}$, then
\be
\label{eqpropspivak}
\exists\calP\in \calLic(E;E) \qst    L \ne \calP^*\circ L \circ\calP.
\ee
\finthm

\debdem (Spivak~\cite{spivak}.) %p.148
\def\vRR{\vec\RR}%
It suffices to prove this proposition for $E = \vRR$. Let $L\in \calL(\vRR;(\vRR)^*)$, $L\ne0$.

Let $(\va_1)$ be a basis in~$\vRR$ (chosen by~$A$). %, and let $(\pi_{a1})$ be the dual basis (basis in~$(\vRR)^*$).
Let $(\vb_1)$ be a basis in~$\vRR$ (chosen by~$B$). %, and let $(\pi_{b1})$ be the dual basis (basis in~$(\vRR)^*$).

Consider $\calP\in \calLic(\vRR;\vRR)$ defined by $\calP(\va_1)=\vb_1$ (change of observer),
and let $\lambda\in\RR$ \st\ $\vb_1= \lambda\va_1$. %, with $\lambda\ne1$ (so $\calP\ne I$).
Then 
\eref{eqdefcalIs} gives
$\calP^*(\ell)(\va_1) := \ell(\calP(\va_1)) = \ell(\vb_1) = \ell(\lambda\va_1) =\lambda  \ell(\va_1)$,
thus $\calP^*(\ell)= \lambda  \ell$ for all $\ell\in (\vRR)^*$.

Thus $\calP^*(L (\calP(\va_1)))
= \calP^*(L(\lambda\va_1))
= \lambda \calP^*(L(\va_1))
=\lambda^2 L(\va_1)
\ne L(\va_1)
$ when $\lambda^2\ne 1$. %Thus~\eref{eqpropspivak}: 
\Eg, $\calP=2 I$ gives $L \ne \calP^*\circ L \circ\calP$ ($=4L$), thus~\eref{eqpropspivak}:
A (non-zero) linear map $E\rar E^*$ cannot be natural. 
\comment{
Thus $(\calP^*\circ L \circ\calP)(\va_1)
= (\calP^*\circ L)(\lambda\va_1)
= \lambda \calP^*(L.\va_1))
=\lambda^2 L.\va_1
\ne L.\va_1
$ when $\lambda\ne 1$. %Thus~\eref{eqpropspivak}: 
\Eg, $\calP=2 I$ gives $L \ne \calP^*\circ L \circ\calP$ ($=4L$), thus~\eref{eqpropspivak}:
A (non-zero) linear map $E\rar E^*$ cannot be natural. 
}
\findem

%%%%%%%%%%%%%%%%%%%%%%%%%%%%%%%%%%%%%%%%%%%%%%%%%%%%%%%%%%%%%%%%%%%%%%%%%%%%%%%%%%%

%\subsubsection{Illustrations (two fundamental examples)}

\debexa
\label{exaisnonn0}
Consider $E$ \st\ $\dim E=1$, and consider the linear map $L\in\calL(E;E^*)$ which sends a basis $(\va_1)$ onto its dual basis $(\pi_{a1})$, so $L$ is defined by $L.\va_1:=\pi_{a1}$.

Question: If $(\vb_1)$ is another basis, $\lambda\ne\pm 1$ and $\vb_1 = \lambda \va_1$ (change of unit of measurement), does $L.\vb_1=\pi_{b1}$, \ie\ does $L$ also sends~$(\vb_1)$ onto its dual basis?

Answer: No. Indeed,
$\vb_1= \lambda\va_1$ gives $\pi_{b1}={1\over \lambda} \pi_{a1}$, thus
$L.\vb_1 = \lambda L.\va_1 = \lambda \pi_{a1} = \lambda^2 \pi_{b1} \ne \pi_{b1}$ since $\lambda^2\ne 1$.
In words: $L$ is not natural, \cf~\eref{eqpropspivak}.

A different presentation: 
Let $L_A$ and $L_B$ be defined by $L_A.\va_j=\pi_{aj}$ and $L_B.\vb_j=\pi_{bj}$ for all~$j$.
And suppose that $\vb_j=\lambda \va_j$ for all~$j$.
Then, $L_A.\vb_j = \lambda L_A.\va_j = \lambda \pi_{aj} = \lambda^2\pi_{bj} = \lambda^2 L_B.\vb_j \ne L_B.\vb_j$ when $\lambda^2\ne 1$, that is, $L_A\ne L_B$ when $\lambda^2\ne 1$: An operator that sends a basis onto its dual basis is not natural.
\finexa

\debexa
\label{exaisnonn2}
Let $\dd_g$ be an inner dot product in $E=\vRRn$. Let $\vR_g\in\calL(E^*;E)$ be the Riesz representation map, that is, defined by $\vR_g(\ell) = \vell_g$ where $\vell_g$ is defined by
$(\vell_g,\vv)_g=\ell.\vv$ for all $\vv\in\vRRn$, cf~\eref{eqJg}. 

Question: Is $\vR_g$~natural?

Answer: No: Consider the diagram
$\Bigl(
\eqalign{
E^*\;\mrar^{\ds \vR_g } \;& E\cr
\noalign{\vskip-6pt}
\calP^* \downarrow \qquad\quad  &\;\uparrow \calP \cr
\noalign{\vskip-4pt}
E^*\;\mrar_{\ds \vR_g } \;& E\cr
}\Bigr)
$
with $\calP=\lambda I$, $\lambda\ne\pm1$. % (change of unit of measurement).
Then $\calP^* = \lambda I$, and $\calP.\vR_g.\calP^*.\ell = \lambda^2 \vR_g.\ell \ne \vR_g.\ell$ gives $\calP.\vR_g.\calP^*\ne \vR_g$:
So $\vR_g$ is not natural, \cf~\eref{eqpropspivak}.
(You may prefer to consider the diagram~\eref{eqdiagnc} with $L=\vR_g^{-1}$.)

A different presentation: 
Consider two distinct Euclidean dot products $\dd_g$ and $\dd_h$ (\eg, built with a foot and built with a metre).
So $\dd_h = \lambda^2 \dd_g$ with $\lambda^2\ne1$.
Let $\vR_g,\vR_h\in\calL(\RRns;\RRn)$ be the Riesz operators relative to~$\dd_g$ and~$\dd_h$, that is $\vR_g.\ell=\vell_g$ and $\vR_h.\ell=\vell_h$ are given by $\ell.\vv = (\vell_g,\vv)_g = (\vell_h,\vv)_h$ for all $\vv\in\vRRn$.
We have $\vell_h = \lambda^2\vell_g$, \cf~\eref{eqfcvvr3}, thus $\vR_h = \lambda^2 \vR_g\ne \vR_g$ since $\lambda^2\ne  1$:
A Riesz representation operator is not natural (it is observer dependent).
\finexa

%%%%%%%%%%%%%%%%%%%%%%%%%%%%%%%%%%%%%%%%%%%%%%%%%%%%%%%%%%%%%%%%%%%%%%%%%%%%%%%%%%%

\subsection{Natural canonical isomorphism $E \simeq E^{**}$}
\label{secEEssnat}

\def\calPss{{\calP^{**}}}

%%%%%%%%%%%%%%%%%%%%%%%%%%%%%%%%%%%%%%%%%%%%%%%%%%%%%%%%%%%%%%%%%%%%%%%%%%%%%%%%%%%

%\subsubsection{Framework and definition}

Two observers $A$ and $B$ consider the same linear map $L\in\calL(E;\Ess)$ (where $\Ess=(E^*)^* = \calL(\Es;\RR)$).
%And more generally, $A$ considers a linear map 
Willing to work together, they (``naturally'') consider the diagram
\be
\label{eqdiagncss}
\eqalign{
E\;\mrar^{\ds L } \;& \Ess \qquad\qquad\leftarrow\hbox{considered by observer $A$}
\cr
\noalign{\vskip-6pt}
\calP \downarrow \qquad\quad &\;\;\downarrow \calPss \cr
\noalign{\vskip-4pt}
E\;\mrar_{\ds L } \;& \Ess \qquad\qquad\leftarrow\hbox{considered by observer $B$}
\cr
}
\ee
where $\calP \in \calL(E;E)$ is a linear diffeomorphism, $\calP^*\in \calL(E^*;E^*)$ its adjoint,
given by $\calP^*(\ell) = \ell\circ \calP$ \cf~\eref{eqdefcalIs}, 
and $\calPss \in\calLic (\Ess;\Ess)$ the adjoint of~$\calP^*$, thus
given by $\calPss(u) = u \circ\calP^*$ for all $u\in \Ess$ \cf~\eref{eqdefcalIs},
\ie\ $\calPss$ is given by, for all $(\ell,u)\in E^*\times \Ess$,
\be
\label{defadjLP2}
(\calPss(u))(\ell) = u(\ell\circ \calP), \qie (\calPss.u).\ell = u.(\ell.\calP).
\ee

\comment{
\debdef
(Spivak~\cite{spivak}.)
Let $L\in\calL(E;\Ess)$.
If the diagram~\eref{eqdiagncss} is commutative, then $L$ is said to be natural:
\be
\hbox{$L \in \calL(E;E^*)$ is natural} \;\; \Longleftrightarrow \;\; 
\forall \calP\in\calL(E;E),\;\; \calPss\circ L = L \circ \calP.
\ee

%: In that case, for all~$\calP$, both path $(\calPss)^{-1}\circ L\circ \calP$ and $L$ give the same result.
\findef
}

%%%%%%%%%%%%%%%%%%%%%%%%%%%%%%%%%%%%%%%%%%%%%%%%%%%%%%%%%%%%%%%%%%%%%%%%%%%%%%%%%%%

%\subsubsection{The Theorem}

\noindent
{\bf Question:} Does there exist a linear map $L\in \calL(E;\Ess)$ that is  natural? 
%(The applications being linear, do we have $\calP^*. L . \calP = L$ for all~$\calP$?)

\noindent
{\bf Answer:} Yes (particular case of the next proposition):

\debprop
\label{propJnat}
The canonical isomorphism
\be
\label{eqisomcan0}
\calJ_E: 
\left\{\eqalign{
E & \rar \Ess \cr
\vu & \rar u=\calJ_E(\vu) \quad\hbox{defined by}\quad
\calJ_E(\vu)(\ell) := \ell.\vu,\quad \forall \ell\in \Es,
}\right.
\ee
is natural, that is, $F$ being another finite dimensional vector space, the diagram
\be
\eqalign{
E\;\mrar^{\ds\calJ_E } \;& \Ess %\qquad\qquad\leftarrow\hbox{considered by observer $A$}
\cr
\noalign{\vskip-6pt}
\calP \downarrow \qquad\quad &\;\;\downarrow \calPss \cr
\noalign{\vskip-4pt}
F\;\mrar_{\ds\calJ_F } \;& \Fss %\qquad\qquad\leftarrow\hbox{considered by observer $B$}
\cr
}
\qwritten
\eqalign{
E\;\mrar^{\ds\calJ } \;& \Ess %\qquad\qquad\leftarrow\hbox{considered by observer $A$}
\cr
\noalign{\vskip-6pt}
\calP \downarrow \qquad\quad &\;\;\downarrow \calPss \cr
\noalign{\vskip-4pt}
F\;\mrar_{\ds\calJ } \;& \Fss %\qquad\qquad\leftarrow\hbox{considered by observer $B$}
\cr
}
\ee
commutes for all $\calP\in\calL(E;F)$, \ie
\be
\label{eqpropJnat0}
\forall \calP \in \calL(E;F),\quad \calPss \circ \calJ_E = \calJ_F \circ \calP,
\quad\hbox{and we write}\quad E \simeq \Ess.
\ee
Thus we can use the unambiguous notation (observer independent)
\be
\label{eqpropJnat}
\calJ(\vu) \eqnote \vu, \qand
\calJ(\vu).\ell \eqnote \vu.\ell \quad (=\ell.\vu).
\ee
(And $u=\calJ(\vu)$ is the derivation operator in the direction~$\vu$.)
\finprop

\debdem
(Spivak~\cite{spivak}.)
It is trivial that $\calJ_E$ is linear and bijective ($E$ is finite dimensional): It is an isomorphism.
Then
$\ds (\calP^{**} \circ \calJ_E(\vu))(\ell)
\mathop{=}^{\eref{defadjLP2}} \calJ_E(\vu)(\ell.\calP)
\mathop{=}^{\eref{eqisomcan0}} (\ell\circ \calP)(\vu)
= \ell(\calP(\vu))
\mathop{=}^{\eref{eqisomcan0}}\calJ_F(\calP(\vu))(\ell)
$, for all $\ell\in F^*$ and all $\vu\in E$,
thus $\calP^{**}\circ \calJ_E(\vu) = \calJ_F(\calP(\vu))$, for all $\vu\in E$, thus $\calP^{**} \circ \calJ_E = \calJ_F\circ \calP$.
\findem

\debprop
\label{propJnat2}
(Characterization of~$\calJ_E$.) $\calJ_E$ sends any basis $(\va_i)$ onto its bidual basis.
(Expected, since $\calJ_E(\vu)$ is the directional derivative in the direction~$\vu$, whatever~$\vu$.)
\finprop

\debdem
Let $(\va_i)$ be a basis and $(\pi_{ai})$ be its dual basis (defined by $\pi_{ai}.\va_j = \delta_{ij}$ for all $i,j$).
Then~\eref{eqisomcan0} gives
$\calJ_E(\va_j).\pi_{ai} = \pi_{ai}.\va_j = \delta_{ij}$ for all $i,j$,
thus $(\calJ_E(\va_j))$ is the dual basis of~$(\pi_{ai})$, \ie, is the bidual basis of~$(\va_i)$;
True for all basis: $\calJ_E(\vb_j).\pi_{bi} = \pi_{bi}.\vb_j = \delta_{ij}$ for all $i,j$.
\findem

\comment{
%%%%%%%%%%%%%%%%%%%%%%%%%%%%%%%%%%%%%%%%%%%%%%%%%%%%%%%%%%%%%%%%%%%%%%%%%%%%%%%%%%%

\subsubsection{Illustration}

Change of observer for propositions~\ref{propJnat}-\ref{propJnat2}:
Let $L_A,L_B\in\calL(E;\Ess)$ be defined by $L_A.\va_i = \pa_{ia}$ and $L_B.\vb_i = \pa_{ib}$,
where $(\pa_{ia})$ and $(\pa_{ib})$ are the bidual bases of the bases $(\va_i)$ and~$(\vb_i)$.
%Let $(\va_i)$ be a basis chosen by~$A$, let $(\pa_{ia})$ be the bidual basis;
%Let $(\vb_i)$ be a basis chosen by~$B$, let $(\pa_{ib})$ be the bidual basis;
%Let $L_A,L_B\in\calL(E;\Ess)$ be defined by $L_A.\va_i = \pa_{ia}$ and $L_B.\vb_i = \pa_{ib}$;
And let $\calP$ given by $\calP.\va_j=\vb_j=\sumin P_{ij}\va_i$ for all~$j$.

1- Do we have $L_B=L_A$ for all~$\calP$? Yes: On the one hand
$(L_B.\vb_j).\ell = \pa_{bj}.\ell = \ell.\vb_j = \ell.\calP.\va_j$,
and on the other hand
$L_A.\vb_j = L_A.(\calP.\va_j) = (\sumin P_{ij}L_A.\va_i) = \sumin P_{ij} \pa_{ai}$
gives
$(L_A.\vb_j).\ell = \sumin  P_{ij} \pa_{ai}.\ell = \sumin  P_{ij} \ell.\va_i = \ell.\calP.\va_j$, for all~$\ell$ and all~$j$, thus $L_B.\vb_j = L_A.\vb_j$ for all~$i$.
Thus $L_A=L_B\eqnamed L$.

2- Do we have $\calPss\circ L = L \circ \calP$? Yes, since here we have $L=\calJ$.
}

\comment{
%%%%%%%%%%%%%%%%%%%%%%%%%%%%%%%%%%%%%%%%%%%%%%%%%%%%%%%%%%%%%%%%%%%%%%%%%%%%%%%%%%%

\subsubsection{Generalization}

Name $\calJ_E$ the canonical isomorphism defined in~\eref{eqisomcan0}.
Let $F$ be another finite dimensional vector space. Then the diagram
\be
\eqalign{
E\;\mrar^{\ds\calJ_E } \;& \Ess %\qquad\qquad\leftarrow\hbox{considered by observer $A$}
\cr
\noalign{\vskip-6pt}
\calP \downarrow \qquad\quad &\;\;\downarrow \calPss \cr
\noalign{\vskip-4pt}
F\;\mrar_{\ds\calJ_F } \;& \Fss %\qquad\qquad\leftarrow\hbox{considered by observer $B$}
\cr
}
\ee
commutes for all $\calP\in\calL(E;F)$: % (application \eg\ with $E=\RRntz$, $F=\RRnt$, and $\calP=\Ftzt(P)$).
%Similar proof:
$\ds (\calP^{**}. \calJ_E(\vu)).\ell
\mathop{=}^{\eref{defadjLP2}} \calJ_E(\vu).(\ell.\calP)
\mathop{=}^{\eref{eqisomcan0}} (\ell.\calP).\vu
= \ell.(\calP.\vu)
\mathop{=}^{\eref{eqisomcan0}}\calJ_F(\calP.\vu). \ell
$, for all $\ell\in F^*$ and all $\vu\in E$,
thus $\calP^{**}. \calJ_E(\vu) = \calJ_F(\calP.\vu)$, for all $\vu\in E$, thus $\calP^{**} \circ \calJ_E = \calJ_F\circ \calP$. %, for all~$\calP$.
}

%%%%%%%%%%%%%%%%%%%%%%%%%%%%%%%%%%%%%%%%%%%%%%%%%%%%%%%%%%%%%%%%%%%%%%%%%%%%%%%%%%%

\subsection{Natural canonical isomorphisms $\calL(E;F) \simeq \calL(F^*,E;\RR) \simeq \calL(E^*;F^*)$}
\label{secal}

\def\calIP{{\calI_\calP}}
\def\calIPEF{{\calI_{\calP_{E\!F}}}}
\def\calIPFE{{\calI_{\calP_{F\!E}}}}
\def\tcalP{\tilde\calP}
\def\tcalIP{\tilde\calIP}
\def\calJEF{{\calJ_{\!E\!F}}}
\def\calJAB{{\calJ_{\!A\!B}}}
\def\ZAB{Z_{\!A\!B}}
\def\ZBA{Z_{\!B\!A}}
\def\calKAB{{\cal K}_{\!\!A\!B}}
\def\calKEF{{\cal K}_{\!\!E\!F}}

$E,F,A,B$ are finite dimensional vector spaces.
Consider the canonical isomorphism
\be
\label{eqdefJ2g}
\calJEF :
\left\{\eqalign{
\calL(E;F) & \rar \calL(\Fs,E;\RR)  \cr
L & \rar \tL = \calJEF(L)\quad\hbox{where}
 \quad \tL(\ell,\vu) := \ell.L.\vu, \quad \forall (\ell,\vu)\in F^* \times E .
}\right.
\ee
Let $\calP_1 \in\calLic(E;A)$ and $\calP_2 \in\calL(F;B)$, and consider the diagram
\be
\label{eqdiagcalj2g}
\eqalign{
\calL(E;F)\;\mrar^{\ds \calJEF } \;& \calL(F^*,E;\RR) \cr
\noalign{\vskip-4pt}
\calIP \downarrow \qquad\quad  \hphantom{\;\mrar_{\ds L } \;} & \qquad\qquad \downarrow \tcalIP\cr
\noalign{\vskip-2pt}
\calL(A;B)\;\mrar_{\ds \calJAB } \;&  \calL(B^*,A;\RR)\cr
}
\ee
where
\be
\label{eqdiagcalj2g2}
\calIP(L) = \calP_2.L.\calP_1^{-1} \qand 
\tcalIP(\tL)(b,\va) =  \tL(b.\calP_2,\calP_1^{-1}.\va)\quad \forall(b,\va)\in B^*\times A.
\ee
($\calIP$ and $\tcalIP$ are the push-forwards for linear maps $L\in\calL(E;F)$ and for bilinear forms $\tL\in\calL(F^*,E;\RR)$.)

\debprop
\label{propeqtJ2f}
The canonical isomorphism $\calJEF$ is natural, that is, the diagram~\eref{eqdiagcalj2g} commutes for all
$\calP_1\in\calLic(E,A)$ and all $\calP_2\in\calL(F,B)$:
\be
\label{eqtJ2f}
%\forall (\calP_1,\calP_2)\in \calLic(E;\tE)\times\calL(F;\tF), \quad
\tcalIP \circ\calJEF = \calJAB \circ \calIP, \qand \calL(E;F) \mathop{\simeq}^{natural}  \calL(F^*,E;\RR).
\ee
Thus $\ds \calL(E^*;F^*) \mathop{\simeq}^{natural} \calL(E;F)$.
\finprop

\debdem
$\ds \calJAB(\calIP(L))(b,\va) $
$\ds \mathop{=}^{\eref{eqdefJ2g}} b.\calIP(L).\va$
$\ds \mathop{=}^{\eref{eqdiagcalj2g2}} b.(\calP_2.L.\calP_1^{-1}).\va$
$= (b.\calP_2).L.(\calP_1^{-1}.\va)$
$\ds \mathop{=}^{\eref{eqdefJ2g}} \calJEF(L)(b.\calP_2,\calP_1^{-1}.\va)$
$\ds \mathop{=}^{\eref{eqdiagcalj2g2}}\tilde\calIP(\calJEF(L))(b,\va)
$, true for all $L\in\calL(E;F)$, $b\in B^*$, $\va\in A$, thus~\eref{eqtJ2f}.

Thus $\ds \calL(\Es;\Fs) 
{\mathop{\simeq}^{\eref{eqtJ2f}}} \calL((\Fs)^*,\Es;\RR) 
{\mathop{\simeq}^{\eref{eqpropJnat0}}} \calL(F,\Es;\RR) 
{\mathop{\simeq}^{\eref{eqtJ2f}}} \calL(\Ess;F)
{\mathop{\simeq}^{\eref{eqpropJnat0}}} \calL(E;F)
$.
\findem

Consider the canonical isomorphism (defines the transposed of a bilinear map)
\be
\calKEF: 
\left\{\eqalign{
\calL(E,F;\RR) & \rar \calL(F,E;\RR) \cr
T & \rar \calKEF(T)
}\right\} , \quad \calKEF(T)(\vu,\vv) := T(\vv,\vu),\quad \forall (\vu,\vv)\in E \times F,
\ee
and $\ZAB\in \calL(E,F;\RR) \rar \calL(A,B;\RR)$ defined by 
$\ZAB(T)(\va,\vb) :=  T(\calP_1^{-1}.\va,\calP_2^{-1}.\vb)$ for all $(\va,\vb)\in A\times B$.

\debprop
The canonical isomorphism $\calKEF$ is natural: 
For all $(\calP_1,\calP_2)\in \calLic(E;A)\times\calL(F;B)$,
the diagram
$
\eqalign{
\calL(E,F;\RR)\;\mrar^{\ds \calKEF } \;& \calL(F,E;\RR) \cr
\noalign{\vskip-4pt}
\ZAB\; \downarrow \qquad\quad  \hphantom{\;\mrar_{\ds L } \;} & \qquad\quad \downarrow \ZBA\cr
\noalign{\vskip-2pt}
\calL(A,B;\RR)\;\mrar_{\ds \calKAB } \;&  \calL(B,A;\RR)\cr
}
$
commutes: $\ds \calL(E,F;\RR) \mathop{\simeq}^{natural} \calL(F,E;\RR)$.
\finprop

\debdem
$\calKEF(\ZAB(T))(\vb,\va)
=\ZAB(T)(\va,\vb)
=T(\calP_2^{-1}.\vb,\calP_1^{-1}.\va)
$
and 
$\ZBA(\calKEF(T))(\va,\vb)=\calKEF(T)(\calP_1^{-1}.\va,\calP_2^{-1}.\vb)
=T(\calP_2^{-1}.\vb,\calP_1^{-1}.\va)
$,
thus $\calKAB\circ \ZAB = \ZBA\circ \calKEF$.
\findem

%%%%%%%%%%%%%%%%%%%%%%%%%%%%%%%%%%%%%%%%%%%%%%%%%%%%%%%%%%%%%%%%%%%%%%%%%%%%%%%%%%%

\subsection{Natural canonical isomorphisms $\calL(E;\calL(E;F)) \simeq \calL(E,E;F) \simeq \calL(F^*,E,E;\RR)$}

For application to the second order derivative $d(d\vu)\simeq d^2\vu$ and, with $\vu\in\Tuzu$,  the notation $d\vu\in \Tuuu$, then $d^2\vu\in \Tudu$, ..., $d^k\vu\in T^1_k(U)$, ...

Consider the canonical isomorphism
\be
\calJ_{12E}: 
\left\{\eqalign{
\calL(E;\calL(E;F)) & \rar \calL(E,E;F) \cr
T_1 & \rar T_2 = \calJ_{12E}(T_1)
}\right\} , \quad 
\calJ_{12E}(T_1)(\vu_1,\vu_2) %=T_2(\vu_1,\vu_2) 
:= T_1(\vu_1).\vu_2\in F,\quad \forall \vu_1,\vu_2\in E,
\ee
and the canonical isomorphism
\be
\calJ_{23E}: 
\left\{\eqalign{
\calL(E,E;F) & \rar \calL(F^*,E,E;\RR) \cr
T_2 & \rar \calJ_{23E}(T_2)=T_3
}\right\} , \quad T_3(\ell,\vu,\vv) := \ell.T_2(\vu_1,\vu_2),\quad \forall \vu_1,\vu_2\in E,\;\forall \ell\in F^*.
\ee

\debprop
$\calJ_{12}$ and $\calJ_{23}$ are natural. Thus $\calJ_{23} \circ \calJ_{12}$ is natural.
\finprop

\def\YAB{Y_{\!A\!B}}

\debdem
1- We have to prove that the following diagram commutes: 
\be
\eqalign{
\calL(E;\calL(E;F))\;\mrar^{\ds \calJ_{12E} } \;& \calL(E,E;F) \cr
\noalign{\vskip-4pt}
\ZAB \downarrow \qquad\qquad\qquad  & \qquad \downarrow \YAB\cr   % \hphantom{\;\mrar_{\ds L } \;}
\noalign{\vskip-2pt}
\calL(A;\calL(A;B))\;\mrar^{\ds  \calJ_{12A}} \;& \calL(A,A;B) \cr
}
\qwhere
\eqalign{
& \ZAB(T_1)(\va_1).\va_2 :=  T_1(\calP_1^{-1}.\va_1).(\calP_1^{-1}.\va_2), \cr
& \YAB(T_2)(\va_1,\va_2) = T_2(\calP_1^{-1}.\va_1,\calP_1^{-1}.\va_2),\cr
}
\ee
(the ``push-forwards)
for all $\va_1,\va_2\in A$ and $L_{AB}\in \calL(A;B)$.

Let $T_1\in \calL(E;\calL(E;F))$. We have

$
%(\calJ_{12A} \circ\ZAB(T_1))(\va_1,\va_2) = \ZAB(T_1)(\va_1).\va_2 =T_1(\calP_1^{-1}.\va,\calP_1.L_{AB}.\calP_2^{-1})
\calJ_{12A} (\ZAB(T_1))(\va_1).\va_2
= \ZAB(T_1)(\va_1).\va_2
= T_1(\calP_1^{-1}.\va_1).(\calP_1^{-1}.\va_2)
$, and

$
\YAB(\calJ_{12E}(T_1))(\va_1,\va_2)
= \calJ_{12E}(T_1)(\calP_1^{-1}.\va_1,\calP_1^{-1}.\va_2)
= T_1(\calP_1^{-1}.\va_1).(\calP_1^{-1}.\va_2)
$, 

thus $\calJ_{12A}\circ \ZAB  = \YAB \circ \calJ_{12E}$, thus $\calJ_{12}$ is natural.

2- We have to prove that the following diagram commutes: 
\be
\eqalign{
\calL(E,E;F)\;\mrar^{\ds \calJ_{23E} } \;& \calL(F^*,E,E;\RR) \cr
\noalign{\vskip-4pt}
\ZAB \downarrow \qquad\qquad\qquad  & \qquad \downarrow \YAB\cr   % \hphantom{\;\mrar_{\ds L } \;}
\noalign{\vskip-2pt}
\calL(A,A;B)\;\mrar^{\ds \calJ_{23A} } \;& \calL(B^*,A,A;\RR) \cr
}
\qwhere
\eqalign{
& \ell_B.\ZAB(T_2)(\va_1,\va_2) := (\ell_B.\calP_2). T_2(\calP_1^{-1}.\va_1,\calP_1^{-1}.\va_2), \cr
& \YAB(T_3)(\ell_B,\va_1,\va_2) = T_3(\ell_B.\calP_2,\calP_1^{-1}.\va_1,\calP_1^{-1}.\va_2),\cr
}
\ee
(the ``push-forwards)
for all $\va_1,\va_2\in A$ and $\ell_B\in B^*$.

Let $T_2\in \calL(E,E;F)$. We have 

$\calJ_{23A}(\ell_B,\ZAB(T_2)(\va_1,\va_2))
= \ell_B.\ZAB(T_2)(\va_1,\va_2)
= (\ell_B.\calP_2).T_2(\calP_1^{-1}.\va_1,\calP_1^{-1}.\va_2)
$, and

$\YAB(\calJ_{23A}(T_2))(\ell_B,\va_1,\va_2)
=\calJ_{23A}(T_2)(\ell_B.\calP_2,\calP_1^{-1}.\va_1,\calP_1^{-1}.\va_2)
= \ell_B.\calP_2.T_2(\calP_1^{-1}.\va_1,\calP_1^{-1}.\va_2)
$

thus $\calJ_{23A}\circ \ZAB  = \YAB \circ \calJ_{23E}$, thus $\calJ_{23}$ is natural.
\findem

%%%%%%%%%%%%%%%%%%%%%%%%%%%%%%%%%%%%%%%%%%%%%%%%%%%%%%%%%%%%%%%%%%%%%%%%%%%%%%%%%%%

\section{Distribution in brief: A covariant concept}

We refer to the books of Laurent Schwartz for a full description.
In continuum mechanics, with $\Omega$ an open set in~$\RRn$ and
for the %infinite dimensional 
space of the finite energy functions~$\Ldo$
and its sub-spaces, a distribution gives a covariant formulation for the virtual power, as used by Germain.

%%%%%%%%%%%%%%%%%%%%%%%%%%%%%%%%%%%%%%%%%%%%%%%%%%%%%%%%%%%%%%%%%%%%%%%%%%%%%%%%%%%

\subsection{Definitions}

\def\DO{{\calD(\Omega)}}
\def\DR{{\calD(\RR)}}
\def\DRn{{\calD(\RRn)}}
\def\DPO{{\calD'(\Omega)}}
\def\DPR{{\calD'(\RR)}}
\def\DPRn{{\calD'(\RRn)}}
\def\Huo{{H^1(\Omega)}}
\def\Hu{{H^1}}
\def\Hmu{{H^{-1}}}
\def\Ld{{L^2}}
\def\Lp{{L^p}}
\def\Li{{L^\infty}}
\def\LpO{{L^p(\Omega)}}
\def\LiO{{L^\infty(\Omega)}}
\def\supp{{\rm supp}}
\def\xz{{x_0}}
\def\Ldon{{\Ldo^n}}
\def\Huz{{H^1_0}}
\def\Huzo{{H^1_0(\Omega)}}
\def\Hmuo{{H^{-1}(\Omega)}}

Usual notations:
Let $p\in[1,\infty[$ (\eg\ $p=2$ for finite energy functions), and let
\be
\LpO := \{f:\Omega\rar\RR : \int_\Omega |f(x)|^p\,d\Omega < \infty \} \qand ||f||_p= (\int_\Omega |f(x)|^p\,d\Omega)^{1\over p},
\ee
the space of functions such that $|f|^p$ is Lebesgue integrable with $||.||_p$ its usual norm. Then $(\LpO,||.||_\Lp)$ is a Banach space (a complete normed space). And let
\be
\LiO:=\{f:\Omega\rar\RR : \sup_{x\in \Omega}( |f(x)|) < \infty \}, \qand ||f||_\infty= \sup_{x\in \Omega}( |f(x)|),
\ee
the space of Lebesgue measurable bounded functions with $||.||_\infty$ its usual norm. Then $(\LiO,||.||_\Li)$ is a Banach space (a complete normed space).

\debdef
If $f\in\calF(\Omega;\RR)$, then its support is the set
\be
\supp(f) := \overline{\{x\in \Omega : f(x)\ne 0\}} = \hbox{the closure of } \{x\in \Omega : f(x)\ne 0\}
\ee
= the set where it is interesting to study~$f$.
\findef

The closure is required:
\Eg, if $\Omega=]0,2\pi[$ and $f(x)= \sin x$ for all $x\in\omega$, then $%\{f=0\}=
\{x\in \Omega : f(x)\ne 0\}=]0,\pi[\cup]\pi,2\pi[$;
And the point~$\pi$ is a point of interest since $\sin$ varies in its vicinity: $f'(\pi)=\cos(\pi)=-1\ne0$;
So it is the closure $\supp(f) := \overline{]0,\pi[\cup]\pi,2\pi[}=[0,2\pi]$ that is considered.
%: In particular $\sin $ vanishes at~$\pi$ but the point~$\pi$ is a point of interest since $\sin$ varies in its vicinity: $\sin'(\pi)=\cos(\pi)=-1\ne0$.

\debdef (Schwartz notation, $D$ being the letter after~$C$:)
Let
\be
\DO := C^\infty_c(\Omega;\RR) = \{\phi \in C^\infty(\Omega;\RR) \hbox{ \st\ $\supp(\phi)$ is compact in $\Omega$}\}.
\ee
\findef

\Eg, $\Omega=\RR$, $\phi(x) := e^{-{1\over 1-x^2}}$ if $x\in]-1,1[$ and $\phi(x):=0$ elsewhere:
$\phi\in\DR$ with $\supp(\phi) = [-1,1]$.

\medskip
And $\DO$ is a vector space %, subspace of~$C^\infty(\Omega;\RR)$,
which is dense in $(\LpO,||.||_\Lp)$ for any $p\in[1,\infty[$.

\debdef
\label{defT}
A distribution in~$\Omega$ is a linear $\DO$-continuous\footnote{
The $\DO$-continuity of~$T$ is defined by: 1- A sequence $(\phi_n)_\NNs$ in~$\DO$ converges in~$\DO$ towards a function $\phi\in\DO$
iff there exists a compact $K\subset\Omega$ \st\ $\supp(\phi_n)\subset K$ for all~$n$, and
%$||\phi-\phi_n||_\infty\matrarrow_{n\rar\infty}0$, and 
$||{\pa^k\phi\over \pa x_{i_1}...\pa x_{i_k}}-{\pa^k\phi_n\over \pa x_{i_1}...\pa x_{i_k}}||_\infty\matrarrow_{n\rar\infty}0$ for all $k\in\NN$ and all $i_j$; %, and for all derivations $\pa^k$, $k\in\NN$.
2- $T$ is continuous at $\phi\in\DO$ iff $\ds T(\phi_n) \mrar_{n\rar\infty} T(\phi)$ for any sequence $\ds (\phi_n)_\NN \in \DO^\NN \matrarrow_{n\rar\infty}\phi$ in~$\DO$.
} function
\be
\label{eqdefT}
T:\left\{\eqalign{ 
\DO & \rar \RR \cr 
\phi & \rar T(\phi) \eqnote \la T,\phi\ra
} \right.
\ee
%(see remark~\ref{remdocont}). 
The space of distribution in~$\Omega$ is named $\DPO$ (the dual of~$\DO$).

The notation $\la T,\phi \ra_{\DPO,\DO} = \la T,\phi \ra$ is the ``duality bracket'' = the ``covariance--contravariance bracket'' between a linear function $T \in \DPO$ and a vector $\phi\in\DO$.
\findef

\debdef
Let $f\in \LpO$. The regular distribution $T_f\in\DPO$ associated to~$f$ is defined by
\be
\label{eqTf}
T_f(\phi) := \int_\Omega f(x)\phi(x)\,d\Omega,\quad \forall \phi\in\DO.
\ee
So $T_f$ is a measuring instrument with density~$dm_f(x)=f(x)\,d\Omega$, \ie\ $T_f(\phi) := \int_\Omega \phi(x)\,dm_f(x)$.
\findef

\debdef
Let $x_0\in\RRn$. The Dirac measure at~$\xz$ is the distribution $\ds T \eqnote \delta_\xz \in\DPR$ defined by,
for all $\phi\in\DR$,
\be
\delta_\xz(\phi)=\phi(\xz), \qie   \la \delta_\xz,\phi\ra=\phi(\xz).
\ee
\findef

And $\delta_\xz$ is not a regular distribution ($\delta_\xz$ is not a density measure): There is no integrable function~$f$ such that $T_f=\delta_\xz$.
Interpretation: $\delta_\xz$ corresponds to an ideal measuring device: The precision is perfect at~$\xz$ (gives the exact value $\phi(\xz)$ at~$\xz$). In real life $\delta_\xz$ is the ideal approximation of $T_{f_n}$ where $f_n$ is \eg\ given by $f_n(x)= n 1_{[\xz,\xz+{1\over n}]}$ (drawing):
For all $\phi\in\DO$, $T_{f_n}(\phi) \mrar_{n\rar\infty} \delta_\xz(\phi) = \phi(\xz)$.

\medskip
\noindent
{\bf Generalization of the definition:}  In~\eref{eqdefT} $\DO = C^\infty_c(\Omega;\RR)$
is replaced by $C^\infty_c(\Omega;\vRRn)$. So if you consider a basis~$(\ve_i)$ then $\vphi \in C^\infty_c(\Omega;\vRRn)$ reads $\vphi = \sumin \phi^i\ve_i$ with $\phi^i\in \DO$ for all~$i$.

\debexa
Power: Let $\alpha:\Omega\rar \Tzuo$ be a differential form. Then
$P=T_\alpha$ defined by $P(\vv) = \int_\Omega \alpha.\vv\,d\Omega$ gives the virtual power associated to~$\alpha$ relative to the vector field~$\vv$ (mechanics and thermodynamics).
%\Eg, if $\alpha$ is exact, $\alpha = df$, then $P(\vv) = \int_\Omega df.\vv\,d\Omega$.
\finexa

%\debrem \label{remdocont}  \finrem

%%%%%%%%%%%%%%%%%%%%%%%%%%%%%%%%%%%%%%%%%%%%%%%%%%%%%%%%%%%%%%%%%%%%%%%%%%%%%%%%%%%

\subsection{Derivation of a distribution}

Let $O$ be a point in $\RRn$ (an origin). If $p\in\RRn$ and if $(\ve_i)$ is a basis in~$\vRRn$, let $\vx=\ora{Op}= \sumin x_i\ve_i$.

\debdef
The derivative ${\pa T\over \pa x_i}$ of a distribution $T\in \DPO$ is the distribution $\in\DPO$ defined by, for all $\phi\in\DO$,
\be
\label{eqTfp}
{\pa T \over \pa x_i}(\phi) := - T({\pa\phi\over \pa x_i}), \qie %\quad\hbox{\ie, with the duality notation,}\quad
\la{\pa T \over \pa x_i},\phi\ra := - \la T,{\pa\phi\over \pa x_i}\ra.
\ee
(${\pa T \over \pa x_i}$ is indeed a distribution: Easy check.)
\findef

\debexa
If $T=T_f$ is a regular distribution with $f\in C^1(\Omega)$, then ${\pa (T_f) \over \pa x_i} = T_{({\pa f \over \pa x_i})}$.
Indeed, for all $\phi\in\DO$,
${\pa (T_f) \over \pa x_i}(\phi) = - T_f({\pa\phi\over \pa x_i})
= - \int_\Omega f(x) {\pa\phi\over \pa x_i}\,d\Omega
= + \int_\Omega {\pa f\over \pa x_i}\phi(x) \,d\Omega + \int_\Gamma 0\,d\Gamma
$, since $\phi$ vanishes on $\Gamma=\pa \Omega$ (the support of~$\phi$ is compact in~$\Omega$), thus
${\pa (T_f) \over \pa x_i}(\phi)
= T_{({\pa f \over \pa x_i})}(\phi)
$ for all $\phi\in\DO$.
\finexa

\debexa
\def\urp{{1_{\RR_+}}}
\def\Hz{{H_0}}
Consider the Heaviside function (the unit step function) $\Hz:=\urp$ and the associated distribution
$T=T_\Hz$. Then $\la (T_\Hz)',\phi\ra := -\la T_\Hz , \phi'\ra
= -\int_\Omega \Hz(x)\phi'(x)\,dx = -\int_0^\infty \phi'(x)\,dx = \phi(0) = \la\delta_0,\phi\ra$
for any $\phi\in\DR$, thus $(T_\Hz)' = \delta_0$. Written $\Hz'=\delta_0$ in~$\DPO$,
which is not in a equality between functions, because $\Hz$ is not derivable at~$0$ as a function, and $\delta_0$ is not a function; It is equality between distributions: The notation $\Hz'$ can only be used
to compute $\Hz'(\phi)$ ($=\la \Hz',\phi\ra:= -\la \Hz,\phi'\ra$).
% And $\delta_0'$ satisfies $\delta_0'(\phi) = -\phi'(0)$ (easy check).
\finexa

%%%%%%%%%%%%%%%%%%%%%%%%%%%%%%%%%%%%%%%%%%%%%%%%%%%%%%%%%%%%%%%%%%%%%%%%%%%%%%%%%%%

\subsection{Hilbert space $\Huo$}

%(See \eg\ Brezis~\cite{brezis} for proof.) 

%%%%%%%%%%%%%%%%%%%%%%%%%%%%%%%%%%%%%%%%%%%%%%%%%%%%%%%%%%%%%%%%%%%%%%%%%%%%%%%%%%%

\subsubsection{Motivation}

\def\hh{{\Lambda}}

%Issue with $\RRn=\RR$: 
Consider the hat function
$\hh(x)
\left\{\eqalign{
=&x+1 \hbox{ if } x\in[-1,0], \cr
=&1-x \hbox{ if } x\in[0,1], \cr
=& 0  \hbox{ otherwise }
}\right\}
$ (drawing). 
When applying the finite element method, it is well-known that, if you use integrals (if you use the virtual power principle which makes you compute average values), then you can consider the derivative of the hat function $\hh$ as if it was the usual derivative, \ie\ at the points where the usual computation of $\Lambda'$ is meaningful, that is,
\be
\label{eqhhp}
\hh'(x)
\left\{\eqalign{
=&1 \hbox{ if } x\in]-1,0[, \cr
=&{-}1 \hbox{ if } x\in]0,1[, \cr
=& 0  \hbox{ if } x\in \RR-\{-1,0,1\}
}\right.
\ee (drawing). %Remember: $\Lambda'$ is undefined at $-1,0,1$. 

\medskip
\noindent
{\bf Problem:} $\Lambda'$ is not defined at $-1,0,1$ (the function $\hh$ is not derivable at $-1,0,1$);

\noindent
{\bf Question:}
So does the ``usual'' computation $I=\int_\RR \Lambda'(x)\phi(x)\,dx$ with~\eref{eqhhp} gives the good result? (This is not a trivial question: \Eg, with $H_0=1_{\RR_+}$ instead of~$\Lambda$, we would get the absurd result $H_0'=0$,
absurd since $H_0'=\delta_0$.)
%In particular, $\hh'$ is no the classic derivative of~$\hh$ since $\hh$ is not derivable at all points.

\medskip
\noindent
{\bf Answer:} Yes:

1- Consider $T_\hh$ the regular distribution associated to~$\Lambda$, \cf~\eref{eqTf};

2- Then consider $(T_\hh)'$, \cf~\eref{eqTfp}: We get 
$\ds \la (T_\hh)',\phi\ra
\mathop{=}^{\eref{eqTfp}} -\la T_\hh, \phi'\ra
= -\int_\RR \hh(x)\phi'(x)\,dx 
= -\int_{-1}^0 \hh(x)\phi'(x)\,dx - \int_0^1 \hh(x)\phi'(x)\,dx
= +\int_{-1}^0 1_{]-1,0[}(x)\phi(x)\,dx + \int_0^1 1_{]0,1[}\phi(x)\,dx$,
for any $\phi\in\calD(\RR)$; 

3- Thus $(T_\hh)' = T_f$ where $f=1_{]-1,0[} + 1_{]0,1[}$, that is $(T_\hh)'$ is a regular distribution, And its is named $f=\hh'$ within the distribution framework, \ie, for computations $\la \hh',\phi\ra:=\la (T_\hh)',\phi\ra$ with $\phi\in\DR$ (value $= \int_\RR f(x)\phi(x)\,dx$).

%%%%%%%%%%%%%%%%%%%%%%%%%%%%%%%%%%%%%%%%%%%%%%%%%%%%%%%%%%%%%%%%%%%%%%%%%%%%%%%%%%%

\subsubsection{Definition of $\Huo$}

The space $C^1(\Omega;\RR)$ is too small in many applications (\eg, for the $\Lambda$ function above); We need a larger space where the functions are ``derivable is a weaker sense'' which is the distribution sense.
Consider a basis in~$\RRn$:
%And $\Huo$ will be that space (of once derivable functions in the Lebesgue integral framework):

\debdef
The Sobolev space $\Huo$ is the subspace of $\Ldo$ restricted to functions whose generalized derivatives are in~$\Ldo$:
\be
\label{eqdefHuo}
\Huo = \{v\in\Ldo : {\pa v \over \pa x_i}\in\Ldo,\; \forall i=1,...,n \}.
\ee
Usual shortened notation: $\Huo = \{v\in\Ldo : \vgrad v \in\Ldon \}$.
\findef

So to check that $v\in \Huo$, even if ${\pa v \over \pa x_i}$ does not exists in the classic way
(see the above hat function~$\Lambda$), you have to:

1- Consider its associated regular distribution $T_v$,

2- Compute ${\pa T_v \over \pa x_i}$ in~$\DPO$, 

3- And if, for all~$i$, there exists $f_i\in\Ldo$ \st\ ${\pa T_v \over \pa x_i}= T_{f_i}$,
then $v\in\Huo$.

4- And then $f_i$ is noted ${\pa v \over \pa x_i}$ when used within the Lebesgue integrals
$%\int_\Omega f_i(x)\phi(x)\,dx \eqnote 
\int_\Omega {\pa v \over \pa x_i}(x)\phi(x)\,dx$ with $\phi\in\DO$.
\\
\Eg, $\hh \in H^1(\RR)$ since $(T_\Lambda)'=T_f$ with $f=1_{]-1,0[} + 1_{]0,1[} \in L^2(\RR)$; And $(T_\Lambda)' \eqnote \Lambda'$ ($=f$) in the distribution context (integral computations).

%%%%%%%%%%%%%%%%%%%%%%%%%%%%%%%%%%%%%%%%%%%%%%%%%%%%%%%%%%%%%%%%%%%%%%%%%%%%%%%%%%%

%\subsubsection{The Hilbert space $\Huo$}
\medskip
Let $\dd_\Ld$ and~$||.||_\Ld$ be the usual inner dot product and norm in~$\Ldo$, \ie
\be
(u,v)_\Ld = \int_\Omega u(x)v(x)\,d\Omega, \qand ||v||_\Ld = \sqrt{(v,v)_\Ld} = ( \int_\Omega v(x)^2\,d\Omega)^\demi.
\ee
$(\Ldo,\dd_\Ld)$ is a Hilbert space. Then define, for all $u,v \in\Huo$,
\be
(u,v)_\Hu 
=(u,v)_\Ld + \sumin ({\pa u \over \pa x_i},{\pa v \over \pa x_i})_\Ld,
 \qand ||v||_\Hu = (v,v)_\Hu^\demi.
\ee
Then $(\Huo,\dd_\Hu)$ is a Hilbert space (Riesz--Fisher theorem).

With a Euclidean dot product $\dd_g$ in~$\vRRn$ and a $\dd_g$-orthonormal basis, 
\be
(u,v)_\Hu = (u,v)_\Ld + (\vgrad u, \vgrad v)_\Ld.
\ee

%%%%%%%%%%%%%%%%%%%%%%%%%%%%%%%%%%%%%%%%%%%%%%%%%%%%%%%%%%%%%%%%%%%%%%%%%%%%%%%%%%%

\subsubsection{Subspace $\Huzo$ and its dual space $\Hmuo$}

%$\Omega$ is bounded and t
The boundary $\Gamma = \pa\Omega$ of~$\Omega$ is supposed to be regular. Let
\be
\Huzo := \{v\in\Huo : v_{|\Gamma} = 0\}.
\ee
Then $(\Huzo,\dd_\Hu)$ is a Hilbert space. 

More generally (without any regularity assumption on~$\Gamma$),  $\Huzo:=\overline{\DO}^\Hu=$ the closure of~$\DO$ in $(\Huo,||.||_\Hu)$: This closure of~$\DO$ in~$\Huo$ enables the use of the distribution framework.

Notation : The dual space of~$\Huzo$ is the space
\be
\Hmuo = (\Huzo)' =\calL(\Huzo;\RR)
\ee
equipped with the (usual) norm
$\ds ||T||_\Hmu
:= \sup_{||v||_\Huz=1} |T(v)|$.
%= \sup_{||v||_\Huz=1} |\la T, v\ra|$. 
And (duality bracket), if $v\in\Huzo$ and $T\in\Hmuo$ then
\be
T(v) \eqnote \la T,v \ra_{\Hmu,\Huz} \eqnote \la T,v \ra.
\ee

\debthm
({\bf Characterization of $\Hmuo = (\Huzo)'$}.) A distribution $T$ is in~$\Hmuo$ iff
\be
\label{eqdhuzo}
\exists (f,\vg)\in\Ldo \times \Ldon \qst T = f - \dvg\vg \;\;(\in\DPO),
\ee
that is, for all $v\in\Huzo$, %
\be
\label{eqdhuzo2}
\la T, v \ra_{\Hmu,\Huz}
= \int_\Omega fv\,d\Omega + \int_\Omega dv.\vg\,d\Omega.
%= \int_\Omega fv\,d\Omega + \int_\Omega \vgrad v \bcdot \vg\,d\Omega,
\ee
And if $\Omega$ is bounded then we can choose $f=0$.
If moreover $\vg \in \Huo^n$ then % (we need the unit exterior normal~$\vn(x)$),
\be
\label{eqdhuzo3}
\la T,v\ra_{\Hmu,\Huz}
= \int_\Omega f(x)v(x)\,dx - \int_\Omega \dvg\vg(x)v(x)\,dx.
\ee
(In fact we only need $\vg \in H_\dvg(\Omega) = \{\vg\in\Ldon : \dvg\vg\in\Ldo\}$.)
\finthm

\debproof
\Eg, see Brezis~\cite{brezis}.
\finproof

For boundary value problems with Neumann boundary conditions, 
we then need $(\Huo)'$ the dual space of~$\Huo$.
% , that is the space of linear functions $T:\Huo \rar \RR$ such that $\ds \sup_{||v||_\Huo=1} |T(v)|<\infty$. And then $T(v)\eqnamed\la T,v\ra_{(\Hu)',\Hu}$.
% where $T \in (\Huo)'$ and $v\in\Huo$. 
Characterization of $(\Huo)'$: We still have~\eref{eqdhuzo2}, but we have to replace~\eref{eqdhuzo} or~\eref{eqdhuzo3} by, with a Euclidean dot product in~$\vRRn$, see Brezis~\cite{brezis},
\be
\la T,v\ra_{(\Hu)',\Hu}
= \int_\Omega f(x)v(x)\,dx - \int_\Omega \dvg\vg(x)v(x)\,dx + \int_\Gamma \vg(x)\bcdot\vn(x) \,v(x)\,dx.
\ee

%\section{Connections, Riemannian curvature tensor} https://perso.isima.fr/~gileborg/IsimathMeca/Rcurv.pdf

%\section{Thermodynamics} https://perso.isima.fr/~gileborg/IsimathMeca/Thermo.pdf

%%%%%%%%%%%%%%%%%%%%%%%%%%%%%%%%%%%%%%%%%%%%%%%%%%%%%%%%%%%%%%%%

\end{document}